%% file: main-new-layout-2.tex
\documentclass{scrbook}

\usepackage{multicol}
\usepackage{epsfig}
\usepackage{makeidx}
\usepackage{latexsym}

\usepackage{tikz}
\usepackage{latexsym}
\usepackage{amssymb}
\usepackage{amsfonts}
\usepackage{amsmath}
\usepackage{stmaryrd}
\usepackage{multirow}
\usepackage{hyperref}
\usepackage{dsfont,paralist,graphicx}

\usepackage[numbers]{natbib}

\usepackage{chngcntr}
\counterwithout{equation}{chapter}
\counterwithin*{equation}{section}

\usepackage[OT2,OT1]{fontenc}
\newcommand\cyr
{
\renewcommand\rmdefault{wncyr}
\renewcommand\sfdefault{wncyss}
\renewcommand\encodingdefault{OT2}
\normalfont
\selectfont
}
\DeclareTextFontCommand{\textcyr}{\cyr}

\usetikzlibrary{shapes.geometric,arrows}
\tikzset{grow=up,node/.style={draw,circle,scale=0.3},every label/.style={scale=0.8}}

\usepackage{caption}
\Roman{chapter}

\newtheorem{df}{Definition}[section]
\newtheorem{lm}[df]{Lemma}
\newtheorem{ex}[df]{Example}
\newtheorem{thm}[df]{Theorem}
\newtheorem{cor}[df]{Corollary}

\def\pr{\noindent{\bf Proof.\ }}
\def\epr{\hfill$\Box$}

\newcommand{\bA}{\mathbf A}
\newcommand{\bB}{\mathbf B}
\newcommand{\bC}{\mathbf C}
\newcommand{\bF}{\mathbf F}
\newcommand{\bN}{\mathbf N}
\newcommand{\bT}{\mathbf T}
\newcommand{\ba}{\mathbf a}
\newcommand{\bb}{\mathbf b}
\newcommand{\bc}{\mathbf c}
\newcommand{\bd}{\mathbf d}
\newcommand{\bp}{\mathbf p}
\newcommand{\bq}{\mathbf q}
\newcommand{\br}{\mathbf r}
\newcommand{\bs}{\mathbf s}
\newcommand{\bt}{\mathbf t}
\newcommand{\bu}{\mathbf u}

\newcommand{\cA}{\mathcal A}
\newcommand{\cB}{\mathcal B}
\newcommand{\cC}{\mathcal C}

\newcommand{\cDF}{\mathcal{DF}}
\newcommand{\cDG}{\mathcal{DG}}
\newcommand{\cDR}{\mathcal{DR}}
\newcommand{\cF}{\mathcal F}
\newcommand{\cG}{\mathcal G}
\newcommand{\cH}{\mathcal H}
\newcommand{\cK}{\mathcal K}
\newcommand{\cLDF}{\mathcal{LDF}}
\newcommand{\cLDR}{\mathcal{LDR}}
\newcommand{\cLF}{\mathcal{LF}}
\newcommand{\cLNDF}{\mathcal{LNDF}}
\newcommand{\cLNDR}{\mathcal{LNDR}}
\newcommand{\cLNF}{\mathcal{LNF}}
\newcommand{\cLNR}{\mathcal{LNR}}
\newcommand{\cLR}{\mathcal{LR}}
\newcommand{\cR}{\mathcal R}
\newcommand{\cS}{\mathcal S}
\newcommand{\cL}{\mathcal L}

\newcommand{\gA}{\mathfrak{A}}
\newcommand{\gB}{\mathfrak{B}}
\newcommand{\gC}{\mathfrak{C}}
\newcommand{\gD}{\mathfrak{D}}
\newcommand{\gp}{\mathfrak{p}}

\newcommand{\abs}{\mathrm{abs}}
\newcommand{\Alg}{\mathrm{Alg}}
\newcommand{\CF}{\mathrm{CF}}
\newcommand{\Comb}{\mathrm{Comb}}
\newcommand{\DGSDT}{\mathrm{DGSDT}}
\newcommand{\dom}{\mathrm{dom}}
\newcommand{\ET}{\mathrm{ET}}
\newcommand{\fork}{\mathrm{fork}}
\newcommand{\GSDH}{\mathrm{GSDH}}
\newcommand{\GSDT}{\mathrm{GSDT}}
\newcommand{\gsm}{\mathrm{gsm}}
\newcommand{\hg}{\mathrm{hg}}
\newcommand{\Loc}{\mathrm{Loc}}

\newcommand{\nroot}{\mathrm{nroot}}
\newcommand{\Par}{\mathrm{Par}}
\newcommand{\ph}{\mathrm{ph}}
\newcommand{\ppath}{\mathrm{path}}
\newcommand{\range}{\mathrm{range}}
\newcommand{\RE}{\mathrm{RE}}
\newcommand{\rel}{\mathrm{rel}}
\newcommand{\Rec}{\mathrm{Rec}}
\newcommand{\Reg}{\mathrm{Reg}}
\newcommand{\rest}{\mathrm{rest}}
\newcommand{\rres}{\mathrm{res}}
\newcommand{\rroot}{\mathrm{root}}
\newcommand{\sub}{\mathrm{sub}}
\newcommand{\Surf}{\mathrm{Surf}}
\newcommand{\yd}{\mathrm{yd}}

\makeindex

\begin{document}

\pagestyle{empty}

\begin{center} \huge{\bf TREE AUTOMATA}
\end{center}

\vspace{10mm}

\begin{center}
\begin{tabular}{cc}
FERENC G\'ECSEG & MAGNUS STEINBY \\[2mm]
Bolyai Institute & Department of Mathematics\\
J\'ozsef Attila University  &       University of Turku\\
Szeged, Hungary & Turku, Finland
\end{tabular}
\end{center}

\newpage

$\ $
\newpage


\cleardoublepage
\phantomsection
\addcontentsline{toc}{chapter}{Prefaces}
 
\pagestyle{headings}
\pagenumbering{roman}

\cleardoublepage
\chapter*{PREFACES}
\label{Prefaces}
\input{Preface_Magnus}

\input{Preface}

\cleardoublepage

\pagestyle{headings}
\pagenumbering{arabic}
\setcounter{page}{1} 


\tableofcontents


\cleardoublepage
\phantomsection
\addcontentsline{toc}{chapter}{Notes to the reader}
\chapter*{NOTES TO THE READER}
\label{Notes-to-the-reader}
\input{Notes-to-the-reader}


\chapter{PRELIMINARIES}
\label{Chapter.1}
\input{Prologue.1}

\section[Sets, relations and mappings]{SETS, RELATIONS AND MAPPINGS}
\label{Section.1.1}
\input{Section.1.1}

\section[Universal algebras]{UNIVERSAL ALGEBRAS}
\label{Section.1.2}
\input{Section.1.2}

\section[Terms, polynomial functions and free algebras]{TERMS, POLYNOMIAL FUNCTIONS AND FREE ALGEBRAS}
\label{Section.1.3}
\input{Section.1.3}

\section[Lattices]{LATTICES}
\label{Section.1.4}
\input{Section.1.4}

\section[Finite recognizers and regular languages]{FINITE RECOGNIZERS AND REGULAR LANGUAGES}
\label{Section.1.5}
\input{Section.1.5}

\section[Grammars and context-free languages]{GRAMMARS AND CONTEXT-FREE LANGUAGES}
\label{Section.1.6}
\input{Section.1.6}

\section[Sequential machines]{SEQUENTIAL MACHINES}
\label{Section.1.7}
\input{Section.1.7}

\section[References]{REFERENCES}
\label{Section.1.8}
\input{Section.1.8}


\chapter{TREE RECOGNIZERS AND RECOGNIZABLE FORESTS}
\label{Chapter.2}
\input{Prologue.2}
\section[Trees and forests]{TREES AND FORESTS}
\label{Section.2.1}
\input{Section.2.1}

\section[Tree recognizers]{TREE RECOGNIZERS}
\label{Section.2.2}
\input{Section.2.2}

\section[Regular tree grammars]{REGULAR TREE GRAMMARS}
\label{Section.2.3}
\input{Section.2.3}

\section[Operations on forests]{OPERATIONS ON FORESTS}
\label{Section.2.4}
\input{Section.2.4}

\section[Regular expressions. Kleene's theorem]{REGULAR EXPRESSIONS. KLEENE'S THEOREM}
\label{Section.2.5}
\input{Section.2.5}

\section[Minimal tree recognizers]{MINIMAL TREE RECOGNIZERS}
\label{Section.2.6}
\input{Section.2.6}

\section[Algebraic characterizations of recognizability]{ALGEBRAIC CHARACTERIZATIONS OF RECOGNIZABILITY}
\label{Section.2.7}
\input{Section.2.7}

\section[A Medvedev-type characterization]{A MEDVEDEV-TYPE CHARACTERIZATION}
\label{Section.2.8}
\input{Section.2.8}

\section[Local forests]{LOCAL FORESTS}
\label{Section.2.9}
\input{Section.2.9}

\section[Some basic decision problems]{SOME BASIC DECISION PROBLEMS}
\label{Section.2.10}
\input{Section.2.10}

\section[Deterministic R-recognizers]{DETERMINISTIC R-RECOGNIZERS}
\label{Section.2.11}
\input{Section.2.11}

\section[Exercises]{EXERCISES}
\label{Section.2.12}
\input{Section.2.12}

\section[Notes and references]{NOTES AND REFERENCES}
\label{Section.2.13}
\input{Section.2.13}


\chapter{CONTEXT-FREE LANGUAGES AND TREE  RECOGNIZERS}
\label{Chapter.3}
\input{Prologue.3}

\section[The yield function]{THE YIELD FUNCTION}
\label{Section.3.1}
\input{Section.3.1}

\section[Context-free languages and recognizable forests]{CONTEXT-FREE LANGUAGES AND RECOGNIZABLE FORESTS}
\label{Section.3.2}
\input{Section.3.2}

\section[Further results and applications]{FURTHER RESULTS AND APPLICATIONS}
\label{Section.3.3}
\input{Section.3.3}

\section[Another way to recognize CF languages]{ANOTHER WAY TO RECOGNIZE CF LANGUAGES}
\label{Section.3.4}
\input{Section.3.4}

\section[Exercises]{EXERCISES}
\label{Section.3.5}
\input{Section.3.5}

\section[Notes and references]{NOTES AND REFERENCES}
\label{Section.3.6}
\input{Section.3.6}


\chapter{TREE TRANSDUCERS AND TREE TRANSFORMATIONS}
\label{Chapter.4}
\input{Prologue.4}

\section[Basic concepts]{BASIC CONCEPTS}
\label{Section.4.1}
\input{Section.4.1}

\section[Some classes of tree transformations]{SOME CLASSES OF TREE TRANSFORMATIONS}
\label{Section.4.2}
\input{Section.4.2}

\section[Compositions and decompositions of tree transformations]{COMPOSITIONS AND DECOMPOSITIONS OF TREE TRANSFORMATIONS}
\label{Section.4.3}
\input{Section.4.3}

\section[Tree transducers with regular look-ahead]{TREE TRANSDUCERS WITH REGULAR LOOK-AHEAD}
\label{Section.4.4}
\input{Section.4.4}

\section[Generalized syntax directed translators]{GENERALIZED SYNTAX DIRECTED TRANSLATORS}
\label{Section.4.5}
\input{Section.4.5}

\section[Surface forests]{SURFACE FORESTS}
\label{Section.4.6}
\input{Section.4.6}

\section[Auxiliary concepts and results]{AUXILIARY CONCEPTS AND RESULTS}
\label{Section.4.7}
\input{Section.4.7}

\section[The hierarchies of tree transformations, surface forests and\\ transformational languages]{THE HIERARCHIES OF TREE TRANSFORMATIONS, SURFACE FORESTS AND TRANSFORMATIONAL LANGUAGES}
\label{Section.4.8}
\input{Section.4.8}

\section[The equivalence of tree transducers]{THE EQUIVALENCE OF TREE TRANSDUCERS}
\label{Section.4.9}
\input{Section.4.9}

\section[Exercises]{EXERCISES}
\label{Section.4.10}
\input{Section.4.10}

\section[Notes and references]{NOTES AND REFERENCES}
\label{Section.4.11}
\input{Section.4.11}


\cleardoublepage
\phantomsection
\addcontentsline{toc}{chapter}{Bibliography}
\bibliographystyle{plain}
\input{bib}

\cleardoublepage
\phantomsection
\addcontentsline{toc}{chapter}{Index}
\input{main-new-layout-2.ind}


\chapter{APPENDIX}
\label{Chapter.5}
\input{Appendix}

\end{document}

%% file: Preface_Magnus.tex
\section*{Preface to the Second Edition}

When the present book was written in the early 1980s, the theory of tree automata, tree languages and tree transformations was young but already  quite extensive. Our aim was to give a systematic and mathematically sound exposition of some central parts of this subject. The presentation uses universal algebra in the spirit of J.~R.~B\"uchi and J.~B.~Wright from whose ideas of automata as algebras tree automata once emerged. That the algebraic formalism encourages and supports precise definitions and rigorous proofs may explain why the book has remained a general reference for many mathematically minded workers in the field ever since its publication in 1984. Unfortunately, it has long been out of print and hard to obtain.

Soon after the regrettable death of Ferenc G\'ecseg in October 2014, Zolt\'an F\"ul\"op (Szeged) and Heiko Vogler (Dresden) proposed a reissue of this book. Akad\'emiai Kiad\'o, the original publisher, did not find the project feasible but gave us free hands to proceed on our own. Professor G\'ecseg's family also willingly endorsed the idea. Since the book did not exist in any electronic form, the whole text had to be retyped in Latex. For this exacting task F\"ul\"op and Vogler quickly assembled a highly qualified team that, besides themselves, included
\mbox{Johanna} \mbox{Bj\"orklund} (Ume{\aa}),
Frank \mbox{Drewes} (Ume{\aa}),
Zsolt \mbox{Gazdag} (Budapest),
\mbox{Eija} \mbox{Jurvanen} (Turku),
\mbox{Andreas} \mbox{Maletti} (Stuttgart),
\mbox{George} \mbox{Rahonis} (Thessaloniki),
Kai \mbox{Salomaa} (Kingston, Ontario), and
\mbox{S\'andor} \mbox{V\'agv\"olgyi} (Szeged). Professors F\"ul\"op and Vogler also undertook the overall management of the work. The generous contributions of all these individuals are acknowledged with many thanks.

From the very beginning it was decided that this new edition should be true to the original one. In particular, the terminology was preserved even in cases in which some alternative terms have become prevailing. However, a few mistakes were corrected and a couple of obscure passages were clarified.

Of course, the book was never claimed to offer a complete presentation of its subject matter. In fact, some important topics were totally left out. It was hoped that the extensive bibliography, fairly complete up to around 1982, and the notes and references at the end of each chapter would, at least partly, make up for the shortcomings. Now, over thirty years later, the incompleteness is naturally even more obvious. Much progress has been made in already established areas and many new topics have emerged. Some of the new work is strongly motivated by applications, old or new. No book of this size could do justice to all these developments. Instead, we have to trust that the matters presented here still belong to the core of the theory and are worth studying by anyone who wants to work in this field. Moreover, to account for more recent contributions and lines of research, an appendix has been added to the book. In it several topics are briefly surveyed and some relevant  references are given to help an interested reader get started on them. I thank Heiko Vogler and Zolt\'an F\"ul\"op for some important additions to the bibliography.\\

\noindent Turku   \hfill Magnus Steinby\\
August 2015

%% file: Preface.tex

\section*{Preface to the Original Edition}
The purpose of this book is to give a mathematically rigorous presentation of the theory of tree automata, recognizable forests, and tree transformations. Apart from its intrinsic interest this theory offers some new perspectives to various parts of mathematical linguistics. It has also been applied to some decision problems of logic, and it provides tools for syntactic pattern recognition. We have not even tried to discuss all aspects of the subject or any of the applications, but enough central material has been included to give the reader a firm basis for further studies. Being relatively new and very manyfaceted, the field still lacks a uniform widely accepted formalism. We have chosen the language of universal algebra as our vehicle of presentation. However, we have not assumed that the reader is familiar with universal algebra; the preparatory sections in Chapter~\ref{Chapter.1} should make the book self-contained in this respect. On the other hand, it is natural to assume that anyone interested in such a book has some general mathematical training and some knowledge of finite automata and formal languages.

The book consists of four chapters, a bibliography and an index. The first chapter contains an exposition of the necessary universal algebra and lattice theory, as well as a quick review of finite automata and formal languages. We also recommend some books on these subjects. In Chapter~\ref{Chapter.2} trees, forests, tree recognizers, tree grammars, and some operations on forests are introduced. Several characterizations and closure properties of recognizable forests are presented. Chapter~\ref{Chapter.3} is devoted to the connections between recognizable forests and context-free languages. Chapter~\ref{Chapter.4} deals with tree transducers and tree transformations. Chapters~\ref{Chapter.2}--\ref{Chapter.4} contain some exercises. Each of these chapters is concluded with some historical and bibliographical comments. We also point out some topics not discussed in the book. We have tried to make the Bibliography as complete as possible. Of course, it has not always been easy to decide whether a given item should be included or not.

We want to thank our colleagues and the staffs at our institutions for the good working atmosphere in which this book was written. Dr.~Andr\'{a}s \'{A}d\'{a}m and Professor Istv\'{a}n Pe\'{a}k gave the text a careful scrutiny. We gratefully acknowledge their many remarks. We are also indebted to Dr.~Zolt\'{a}n \'{E}sik for his very helpful comments on Chapter~\ref{Chapter.4}. We wish to express our warmest thanks to Mrs.~Piroska Folberth for performing very competently the difficult task of typing the manuscript. Also, we want to thank our wives and daughters for their support and for putting so gracefully up with the inconveniences inevitably caused by our undertaking.

The writing of the book has involved several trips between Turku and Szeged. We gratefully acknowledge the financial support provided by the Academy of Finland, the Hungarian Academy of Sciences, the J\'{a}nos Bolyai Mathematical Society, the University of Szeged, and the University of Turku. Our work was also furthered by a possibility for the first-named author to spend a term at the Tampere University of Technology. For this thanks are due Professor Timo Lepist\"{o}.

%% file: Notes-to-the-reader.tex

Within each section, there is one counter which is incremented by each of the environments definition, lemma, theorem, corollary, and example.
The end of a proof or an example is indicated by the mark $\Box$. It appears immediately after a theorem, lemma or corollary if this is not followed by a proof. The references to the literature are by the author(s)
and the number with which the publication occurs in the Bibliography. In a few cases we refer to a book mentioned at the end of Chapter~\ref{Chapter.1}.

%% file: Prologue.1.tex

In this chapter we shall review some basic concepts and results from the theories of automata, formal languages, and universal algebras. It is reasonable to assume that a potential reader of this book already knows something about automata and formal languages. On the other hand, we do not presuppose any knowledge of universal algebra. These two assumptions suggested the styles and extents of the following seven sections.

Section~\ref{Section.1.1} (Sets, relations and mappings) may be skimmed through for terminology and notation.

Sections~\ref{Section.1.2} and~\ref{Section.1.3} present the required universal algebraic concepts and results. These are not many, but they should be mastered well as the very basic concepts of the theory of tree automata are defined in terms of universal algebra. We have tried to make the book self-contained in this respect, but a reader who wants to pursue further the algebraic aspects of the theory should certainly consult one of the references on universal algebra.

The lattice theory presented in Section~\ref{Section.1.4} is less important here, and the reading of this section may be postponed until needed.

Sections~\ref{Section.1.5}, \ref{Section.1.6} and~\ref{Section.1.7} survey some of the most essential facts about finite recognizers, regular languages, context-free grammars, and (generalized) sequential machines. A reader less familiar with these matters would do wisely to look up these subjects in some of the references given at the end of the chapter.

%% file: Section.1.1.tex
The set theory needed here is very elementary and most of our set theoretic notation is well-known. However, a few conventions should be pointed out:
\begin{itemize}
  \item[(i)] $A \subseteq B$ means that the set $A$ is a subset of the set $B$. Proper inclusion is denoted by $A \subset B$.
  \item[(ii)] $\emptyset$ denotes the empty set.
  \item[(iii)] $|A|$ denotes the cardinality of the set $A$.
  \item[(iv)] The {\em power set}\index{set!power} of a set $A$, i.e., the set of all subsets of $A$, is denoted by $\gp A$.
  \item[(v)] The union of a family $(A_i \mid i \in I)$ of subsets (indexed by $I$) of some set is written as $\bigcup(A_i \mid i \in I)$. Similarly, $\bigcap(A_i \mid i \in I)$ is the intersection.
  \item[(vi)] The set $\{ x \in A \mid P_1(x), \dots, P_k(x) \}$ of all elements $x$ in $A$ with the properties $P_1$, \dots, $P_k$ may also be written as $\{x\mid P_1(x), \dots, P_k(x) \}$ when $A$ is understood from the context. We shall use this notation in the following more general form, too. Suppose $f(x_1, \dots, x_m)$ is an object defined in some way in terms of the objects $x_1$, \dots, $x_m$. Then
      $$
          \{ f(x_1, \dots, x_m) \mid P(x_1, \dots, x_m) \}
      $$
      is the set of all such objects constructed from objects $x_1$, \dots, $x_m$ satisfying the condition $P(x_1, \dots, x_m)$. Furthermore, we use
      $$
          \{ f_1(x_1, \dots, x_m), \dots, f_k(x_1, \dots, x_m) \mid P(x_1, \dots, x_m) \}
      $$
      as a short form for the union
      $$
          \{ f_1(x_1, \dots, x_m) \mid P(x_1, \dots, x_m) \} \cup \dots \cup \{ f_k(x_1, \dots, x_m) \mid P(x_1, \dots, x_m) \}.
      $$
  \item[(vii)] If there is no danger of confusion, we may write simply $a$ for the one-element set $\{a\}$. Of course, we should not write $\emptyset$ for $\{\emptyset\}$.
\end{itemize}

Sometimes we employ some notation from logic as abbreviations:
\begin{itemize}
  \item[(i)] ``$(\forall x \in A) \, P(x)$'' states that $P(x)$ holds for all $x \in A$.
  \item[(ii)] ``$(\exists x \in A) \, P(x)$'' states that there exists an $x$ in $A$ such that $P(x)$ holds.
  \item[(iii)] ``$P \Longrightarrow Q$'' means that $Q$ holds if $P$ holds.
  \item[(iv)] ``$P \Longleftrightarrow Q$'' states that the conditions $P$ and $Q$ are equivalent, i.e., both of them hold or then neither one holds.
  \item[(v)] ``$P \wedge Q$'' is the statement that both $P$ and $Q$ hold. Similarly, ``$P \vee Q$'' states that at least one of $P$ and $Q$ holds.
\end{itemize}

The numbers dealt with here are always integers and mostly even non-negative integers. When we write ``\dots~for all $n\ge 1$'' we mean, in fact, ``\dots~for all integers $n\ge 1$''. The set of all integers is denoted by ${\mathbf Z}$, the set of the natural numbers 1, 2, \dots\ by ${\mathbf N}$, and the set of all non-negative integers by ${\mathbf N_{\mathbf 0}}$.

Let $A$ and $B$ be sets and $\varrho \subseteq A \times B$ a (binary) {\em relation}\index{relation} from $A$ to $B$. The fact that $(a,b) \in \varrho$ ($a \in A$, $b \in B$) is also expressed by writing $a \varrho b$ or $a \equiv b \,(\varrho)$. The opposite case may be expressed by $a\!\! \not\! \varrho \, b$ or by $a \not\equiv b \,(\varrho)$. For any $a \in A$, we put
$$
    a\varrho = \{ b \in B \mid a \varrho b \}.
$$
This notation is extended to subsets of $A$:
$$
    A_1\varrho = \bigcup ( a\varrho \mid a \in A_1 )
    \quad\text{for $A_1 \subseteq A$}.
$$
The {\em converse}\index{converse of relation} of $\varrho$ is the relation
$$
   \varrho^{-1} = \{ (b,a) \mid (a,b) \in \varrho \} \subseteq B \times A.
$$
Obviously,
$$
    b\varrho^{-1} = \{ a \in A \mid a \varrho b \}
$$
and
$$
    B_1\varrho^{-1} = \{ a \in A \mid (\exists b \in B_1) a \varrho b \}
$$
for all $b \in B$ and $B_1 \subseteq B$. The {\em domain}\index{domain!of relation} of $\varrho$ is the subset $\dom(\varrho) = B\varrho^{-1}$ of $A$, and its {\em range}\index{range of!relation} is the subset $\range(\varrho) = A\varrho$ of $B$.

The {\em product}\index{product!of relations} or {\em composition}\index{composition of!relations} of two relations $\varrho \subseteq A \times B$ and $\tau \subseteq B \times C$ is the relation
$$
    \varrho \circ \tau = \{ (a,c) \mid (\exists b \in B) a\varrho b \tau c \}
    \subseteq A \times C.
$$
In this definition we used the short form $a \varrho b \tau c$ to express the fact that $a \varrho b$ and $b \tau c$. Often we write $\varrho \tau$ for $\varrho \circ \tau$. The product of relations is associative. We note also the equality $(\varrho \circ \tau)^{-1} = \tau^{-1} \circ \varrho^{-1}$.

Consider now (binary) relations on a set $A$, i.e.\ subsets of $A \times A$. These include the {\em diagonal relation}\index{relation!diagonal} $\delta_A = \{ (a,a) \mid a \in A \}$ and the {\em total relation}\index{relation!total} $\iota_A = A \times A$. For any relation $\varrho$ on $A$ we define the {\em powers}\index{power of!relation} $\varrho^n$ ($n \ge 0$) with respect to the product of relations:
\begin{align*}
  &1^{\circ}\quad \varrho^0 = \delta_A \quad\text{and} \\
  &2^{\circ}\quad \varrho^{n+1} = \varrho^n \circ \varrho \quad\text{for $n\ge 0$}.
\end{align*}

The relation $\varrho \subseteq A \times A$ is called
\begin{itemize}
  \item[(a)] {\em reflexive}\index{relation!reflexive} if $\delta_A \subseteq \varrho$,
  \item[(b)] {\em symmetric}\index{relation!symmetric} if $\varrho^{-1} \subseteq \varrho$,
  \item[(c)] {\em antisymmetric}\index{relation!antisymmetric} if $\varrho \cap \varrho^{-1} \subseteq \delta_A$ and
  \item[(d)] {\em transitive}\index{relation!transitive} if $\varrho^2 \subseteq \varrho$.
\end{itemize}

The intersection of any reflexive relations (on a given $A$) is reflexive, and the intersection of transitive relations is transitive. Thus there exists for every $\varrho \subseteq A \times A$  a unique minimal reflexive, transitive relation $\varrho^*$ containing $\varrho$. It is called the {\em reflexive, transitive closure}\index{reflexive transitive closure} of $\varrho$. One verifies easily that
$$
    \varrho^* = \delta_A \cup \varrho \cup \varrho^2 \cup \varrho^3 \cup \dots,
$$
i.e., for any $a$, $b \in A$ we have $a \varrho^* b$ iff
$$
    a = a_1 \, \varrho \, a_2 \, \varrho \, a_3 \, \dots \, a_{n-1} \, \varrho \, a_n = b
$$
for some $n \ge 1$ and $a_1$, \dots, $a_n \in A$.

A relation on $A$ is called an {\em equivalence relation}\index{relation!equivalence} on $A$, if it is reflexive, symmetric and transitive. The set of all equivalence relations on $A$ is denoted by $E(A)$. Clearly, $\delta_A \in E(A)$  and $\iota_A \in E(A)$. Let $\varrho$ be an equivalence relation on $A$. The {\em $\varrho$-class}\index{rho@$\rho$-class} (or the {\em equivalence class}\index{class!equivalence} modulo $\varrho$) of an element $a \in A$ is the set $a\varrho$. Obviously, $a \varrho b$ iff $a\varrho = b\varrho$. We shall also write $a/\varrho$ for $a\varrho$ and extend this notation to subsets $A_1 \subseteq A$ and $n$-tuples $\ba = (a_1, \dots, a_n)$  of elements of $A$ ($n \ge 1$): $A_1/\varrho = \{ a/\varrho \mid a \in A_1 \}$ and $\ba/\varrho = (a_1/\varrho, \dots, a_n/\varrho)$. The {\em quotient set}\index{set!quotient} of $A$ modulo $\varrho$ is $A/\varrho$. Obviously, $A/\varrho$ is a partition on $A$, that is, every element of $A$ belongs to exactly one $\varrho$-class. On the other hand, every partition on $A$ can be obtained this way as the quotient set from a unique equivalence relation and there is a natural one-to-one correspondence between the partitions on $A$ and $E(A)$. The cardinality of $A/\varrho$ is called the {\em index}\index{index of equivalence relation} of $\varrho \in E(A)$. If $|A/\varrho|$ is finite, we say that $\varrho$ is of {\em finite index}. We say that $\varrho \in E(A)$ {\em saturates}\index{relation!saturating a subset} the subset $H \subseteq A$ if $H\varrho = H$, i.e., if $H$ is the union of some $\varrho$-classes.

A {\em mapping}\index{mapping} or a {\em function}\index{function} from a set $A$ to a set $B$ is a triple $(A,B,\varphi)$, where $\varphi \subseteq A \times B$ is a relation such that for every $a \in A$ there exists exactly one $b \in B$ satisfying $a \varphi b$. As usual we write $\varphi\colon A \to B$ and say that $\varphi$ is a mapping from $A$ to $B$. If $a \varphi b$ ($a \in A$, $b \in B$), $b$ is called the {\em image}\index{image} of $a$ and $a$ an {\em inverse image}\index{image!inverse} of $b$. This is expressed by writing $b = a\varphi$, $b= \varphi(a)$ or $\varphi\colon a \mapsto b$. For a subset $A_1$ of $A$ we also use the two notations $A_1\varphi$ and $\varphi(A_1)$ for the set $\{ a\varphi \mid a \in A_1 \}$. The converse $\varphi^{-1}$ of $\varphi$ is always defined as a relation ($\subseteq B \times A$), but it is usually not a mapping from $B$ to $A$. Again, $\varphi^{-1}(B_1)$  will sometimes be used instead of $B_1\varphi^{-1}$ when $B_1 \subseteq B$. Note that $\dom(\varphi) = A$ and $\range(\varphi) \subseteq B$. The set of all mappings from $A$ to $B$ is denoted by $B^A$.

The {\em composition}\index{composition of!mappings} or {\em product}\index{product!of mappings} of two mappings $\varphi\colon A \to B$ and $\psi\colon B \to C$ is the mapping
$$
    \varphi\psi\colon A \to C
$$
where $\varphi\psi$ is the product of $\varphi$ and $\psi$ as relations. Clearly, $a\varphi\psi = (a\varphi)\psi$ for all $a \in A$.

The {\em restriction}\index{restriction of!mapping} of a mapping $\varphi\colon  A \to B$ to a subset $C$ of $A$ is the mapping
$$
    \varphi|C\colon C \to B
$$
where $\varphi|C = \varphi \cap (C \times B)$. If $\psi\colon C \to B$ is obtained from $\varphi\colon A \to B$ as the restriction of $\varphi$ to $C$, i.e., $C \subseteq A$ and $\psi = \varphi|C$, then we say also that $\varphi$ is an {\em extension}\index{extension of mapping} of $\psi$ to $A$.

The {\em kernel}\index{kernel of mapping} $\varphi\varphi^{-1}$ of a mapping $\varphi\colon A \to B$ is an equivalence relation on $A$ and \linebreak $a_1 \equiv a_2  \,(\varphi\varphi^{-1})$ iff $a_1\varphi = a_2\varphi$ ($a_1$, $a_2 \in A$). On the other hand, one can associate with every $\theta \in E(A)$ a mapping
$$
    \theta^{\natural}\colon A \to A/\theta, \quad a \mapsto a\theta, \quad (a \in A)
$$
such that the kernel of $\theta^{\natural}$ is $\theta$. This $\theta^{\natural}$ is called the {\em natural mapping}\index{mapping!natural} associated with $\theta$.

A mapping $\varphi\colon A \to B$ is called
\begin{itemize}
  \item[(i)] {\em injective}\index{mapping!injective} (or an {\em injection}\index{injection}), if $\varphi\varphi^{-1} = \delta_A$,
  \item[(ii)] {\em surjective}\index{mapping!surjective} (or a {\em surjection}\index{surjection}), if $\range(\varphi) = B$, and
  \item[(iii)] {\em bijective}\index{mapping!bijective} (or an {\em bijection}\index{bijection}), if it is injective and surjective.
\end{itemize}

If $\varphi\colon A \to B$ is surjective, one says also that $\varphi$ is a mapping of $A$ {\em onto}\index{mapping!onto} $B$. It is obvious that the natural mapping $\theta^{\natural}$ is always surjective ($\theta \in E(A)$). The diagonal relation of a set $A$ defines the {\em identity mapping}\index{mapping!identity} $A \to A$, $a \mapsto a$ ($a \in A$). It is denoted by $1_A$.

We shall also meet partial mappings, that is, mappings for which the image of some elements may be undefined. A {\em partial mapping}\index{mapping!partial} from $A$ to $B$ is defined by a relation $\varphi \subseteq A \times B$ such that $|a\varphi| \le 1$ for all $a \in A$. Again, we write $\varphi\colon A \to B$. If $a\varphi = \emptyset$, then we say that $\varphi$ is {\em undefined\/}\index{mapping!undefined for an element} for $a$ ($a \in A$). The notations and terminology introduced above for mappings apply to partial mappings, too, although $\dom(\varphi)$ may be a proper subset of $A$ when $\varphi\colon A \to B$ is a partial mapping.

It is convenient to think of the elements of a cartesian product $A_1 \times \dots \times A_n$ as $n$-tuples $(a_1, \dots, a_n)$ with $a_1 \in A_1$, \dots, $a_n \in A_n$. We adopt the definition of an ordinal number $n$ as the set of all ordinals smaller that $n$: $0 = \emptyset$, $1 = \{0\}$, $2 = \{0,1\}$ etc.\ and, in general, $n = \{0, 1, \dots, n-1\}$. Then $A_1 \times \dots \times A_n$ can also be defined as the set of all mappings
$$
    \varphi\colon n \to A_1 \cup \dots \cup A_n
$$
such that $i\varphi \in A_{i+1}$ for $i=0$, 1, \dots, $n-1$. Of course, we may identify such a $\varphi$ with the $n$-tuple $(0\varphi, 1\varphi, \dots, (n-1)\varphi)$. Now the cartesian power $A^n = A \times \dots \times A$ ($n$ times) is the set of all mappings $\varphi\colon n \to A$. In particular, $A^0 = \{\emptyset\}$ since $\emptyset$ is the only mapping from $\emptyset$ to $A$. Note that the notation $A^n$ is consistent with our earlier notation $B^A$ for the set of all mappings from $A$ to $B$.

We shall also need countably infinite sequences of elements. Let $\omega = \{0, 1, 2, \dots \}$ be the smallest infinite ordinal and $A$ any set. The elements of $A^{\omega}$ are called {\em $\omega$-sequences}\index{omega@$\omega$-sequence}. Thus an $\omega$-sequence of elements of $A$ is a mapping
$$
    \varphi\colon \omega \to A
$$
which we may also write as
$$
    (0\varphi, 1\varphi, \dots, n\varphi, \dots)_{n < \omega}.
$$

We conclude the section by considering operations. These are special mappings and are among the most fundamental concepts of algebra. Let $m \ge 0$. An {\em $m$-ary operation}\index{operation!m-ary@$m$-ary} on a set $A$ is a mapping from $A^m$ to $A$. If $\varphi\colon A^m \to A$ is an $m$-ary operation on $A$, then $\varphi$ assigns to every $m$-tuple $(a_1, \dots, a_m)$ of elements of $A$ a unique element of $A$ which we write as $\varphi(a_1, \dots, a_m)$. The number $m$ is called the {\em arity}\index{arity of!operation} or the {\em rank}\index{rank of!operation} of $\varphi$. Most operations encountered in the usual algebraic systems (groups, rings, lattices etc.) have rank 0, 1 or 2. A few comments on these special cases:
\begin{itemize}
  \item[(i)] A 0-ary operation $\varphi\colon \{\emptyset\} \to A$ is completely determined by its only image $\varphi(\emptyset)$, and often $\varphi$ is given simply by naming this element. Note that here $\emptyset$ may also be seen as the empty sequence of elements, and often one writes $\varphi(\;)$, or just $\varphi$, for $\varphi(\emptyset)$.
  \item[(ii)] When $m=1$, we have a mapping from $A$ to itself. Such operations are called {\em unary}\index{operation!unary}.
  \item[(iii)] An operation of rank 2 is called a {\em binary operation}\index{operation!binary}. For example, the addition and the multiplication in a ring are binary operations. In most such concrete examples one uses the {\em infix notation}\index{infix notation} for binary operations. Thus it is customary to write the ring operations in the form $a+b$ and $a \cdot b$ instead of $+(a,b)$ and $\cdot(a,b)$, respectively.
\end{itemize}

A {\em partial $m$-ary operation}\index{operation!partial m-ary@partial $m$-ary} on a set $A$ is a partial mapping from $A^m$ to $A$. For any partial $m$-ary operation $\varphi\colon A^m \to A$ and subset $B$ of $A$ we have a partial mapping
$$
    \varphi|B\colon B^m \to B,
$$
where $\varphi|B = \varphi \cap (B^m \times B)$. If $\varphi$ is an operation and $B$ is {\em closed}\index{subset!closed with respect to operation} with respect to $\varphi$, i.e., $\varphi(a_1, \dots, a_m) \in B$ whenever $a_1$, \dots, $a_m \in B$, then $\varphi|B$ is an $m$-ary operation on $B$ called the {\em restriction}\index{restriction of!operation} of $\varphi$ to $B$. Often the same symbol is used to denote an operation and its restrictions.

Suppose we are given a set $A$, $k$ $m$-ary operations $\varphi_1$, \dots ,$\varphi_k$ on $A$ and a $k$-ary operation $\psi$ on $A$ ($m$, $k \ge 0$). The {\em composition}\index{composition of!operations} of $\varphi_1$, \dots, $\varphi_k$ with $\psi$ is the $m$-ary operation $\psi(\varphi_1, \dots, \varphi_k)$ defined so that
$$
   \psi(\varphi_1, \dots, \varphi_k)(a_1, \dots, a_m) =
   \psi(\varphi_1(a_1, \dots, a_m), \dots, \varphi_k(a_1, \dots, a_m))
$$
for all $a_1$, \dots, $a_m \in A$. Note that the possibilities $k=0$ or $m=0$ are included. If $k=0$, then the composition is an $m$-ary operation with the constant image $\psi(\emptyset)$. If $m=0$, then the composition is a 0-ary operation with the single value $\psi(\varphi_1(\emptyset), \dots, \varphi_k(\emptyset))$.

Let $\varphi$ be an $m$-ary operation on a set $A$ and $A_1$, \dots, $A_m$ any subsets of $A$. Then we write
$$
    \varphi(A_1, \dots, A_m) = \{ \varphi(a_1, \dots, a_m) \mid a_1 \in A_1, \dots, a_m \in A_m \}.
$$
Thus $\varphi$ is extended to an $m$-ary operation on the power set $\gp A$. In general, there is no need to introduce a new notation for this extension.

%% file: Section.1.2.tex
In this and the next section some concepts and results from universal algebra are surveyed. Universal algebra is an extensive field of mathematics, but we need really just certain basic parts of it. On the other hand, a good grasp of the material of these sections is essential to an understanding of the rest of the book.

Generally speaking, an {\em algebra} (or a {\em universal algebra}\index{Algebra!universal}) is a set together with a set of operations on this set. There may be a finite or an infinite number of operations, but we insist that they all are {\em finitary}\index{operation!finitary}, i.e., the ranks are finite as in the definition of operations given in the previous section. As a first example we consider the algebra of subsets of a given set $U$. In the power set $\gp U$ we have several naturally defined operations. For example, there is a binary operation $\cup$ that forms the union $A \cup B$ of any two $A$, $B \in \gp U$. Similarly, we have the binary operation $\cap$ that forms the intersection of two subsets of $U$. A unary operation is obtained if we map every $A \in \gp U$ to its complement $A^c = U - A$. Furthermore, we introduce two 0-ary operations, one that has $\emptyset$ and one that has $U$ as its image. Of course, an infinite number of operations could be defined on $\gp U$, but if we restrict ourselves to those defined above, we get the algebra
$$
    (\gp U,\, \cup,\, \cap,\, {}^c,\, \emptyset,\, U)
$$
with two binary, one unary and two 0-ary operations. Note that we get such an algebra for each set $U$. In fact, all of these algebras can be viewed as special instances of a general class of algebras known as {\em Boolean algebras}\index{Algebra!Boolean}.

The example brings forth an important point. In algebra, and this will be the case here, too, one is generally not interested just in individual algebras, but rather in whole classes of algebras. Algebras in such a class are all ``similar'' in the sense that there is a natural correspondence between the operations of any two algebras of the class. Such a correspondence of operations is needed when one defines any concept, such as homomorphisms or direct products, involving more than one algebra. For example, the multiplications of any two groups correspond to each other, and a homomorphism of groups should preserve the multiplication. We shall now introduce a convenient vehicle to define such a class of similar algebras.

\begin{df}\label{Definition.1.2.1}\rm An {\em operator domain}\index{domain!operator} is a set $\Sigma$ together with a mapping
$$
    r\colon\Sigma \to {\mathbf N}_{\mathbf 0}
$$
that assigns to every $\sigma \in \Sigma$ an {\em arity}\index{arity of!operator}, or {\em rank}\index{rank of!operator}, $r(\sigma)$. For any $m \ge 0$,
$$
    \Sigma_m = \{ \sigma \in \Sigma \mid r(\sigma) = m \}
$$
is the set of the {\em $m$-ary operators}\index{operator} (or {\em operational symbols}\index{operational symbol}).
\end{df}

From now on $\Sigma$ is an operator domain. The mapping $r$ is usually not mentioned, but we denote by $r(\Sigma)$ the set of all $m \ge 0$ such that $\Sigma_m \ne \emptyset$. One can write $\Sigma$ as the disjoint union $\Sigma_0 \cup \Sigma_1 \cup \Sigma_2 \cup \dots$ from which the empty sets will be omitted.

\begin{df}\label{Definition.1.2.2}\rm A {\em $\Sigma$-algebra}\index{Sigma-algebra@$\Sigma$-algebra, see algebra} $\cA$ is a pair consisting of a nonempty set $A$ (of elements of $\cA$) and a mapping that assigns to every operator $\sigma \in \Sigma$ an $m$-ary operation
$$
    \sigma^{\cA}\colon A^m \to A,
$$
where $m$ is the arity of $\sigma$. The operation $\sigma^{\cA}$ is called the {\em realization}\index{realization of!operator} of $\sigma$ in $\cA$. The mapping $\sigma \mapsto \sigma^{\cA}$ will not be mentioned explicitly, but we write $\cA = (A,\Sigma)$. The \text{$\Sigma$-algebra} $\cA$ is {\em finite}\index{Algebra!finite} if $A$ is finite, and it is of {\em finite type}\index{Algebra!of finite type} if $\Sigma$ is finite. When $\Sigma$ is not specified, or not emphasized, we speak simply about {\em ``algebras''}\index{Algebra}. An algebra with just one element is called {\em trivial}\index{Algebra!trivial}.

In general,  $\cA = (A,\Sigma)$, $\cB = (B,\Sigma)$ and ${\cC} = (C,\Sigma)$, possibly equipped with subscripts, will be $\Sigma$-algebras. The realizations of an operator $\sigma \in \Sigma$ in these algebras are denoted by $\sigma^{\cA}$, $\sigma^{\cB}$ and $\sigma^{\cC}$, respectively.
\end{df}

In the previous example of subset algebras we would have $\Sigma = \Sigma_0 \cup \Sigma_1 \cup \Sigma_2$ with (for example) $\Sigma_0 = \{0,1\}$, $\Sigma_1 = \{\neg\}$ and $\Sigma_2 = \{\wedge,\vee\}$. The algebra of the subsets of a set $U$ is then the $\Sigma$-algebra $\cA$, where $A = \gp U$ and the operators are realized as follows: $0^{\cA} = \emptyset$, $1^{\cA} = U$, $\neg^{\cA} = {}^c$ (complement in $U$), $\wedge^{\cA} = \cap$ (intersection) and $\vee^{\cA} = \cup$ (union).

Note that the possibility $m = 0$ is not excluded when we consider generally an $m$-ary operation. For $\sigma \in \Sigma_0$ one often writes $\sigma^{\cA}$ instead of $\sigma^{\cA}(\phantom{x})$ or $\sigma^{\cA}(\emptyset)$ (this involves the harmless confusion of a 0-ary operation and its value). When $\Sigma = \{\sigma_1, \dots, \sigma_k\}$ is finite, one usually writes $\cA = (A, \sigma_1, \dots, \sigma_k)$ instead of $\cA = (A,\Sigma)$.

We introduce now several concepts related to algebras.

\begin{df}\label{Definition.1.2.3}\rm The  $\Sigma$-algebra $\cB$ is a {\em subalgebra}\index{subalgebra} of the $\Sigma$-algebra $\cA$ if $B \subseteq A$ and $\sigma^{\cB} = \sigma^{\cA}|B$ for all $\sigma \in \Sigma$.
\end{df}

If $\cB$ is a subalgebra of $\cA$, then $B$ is a {\em closed subset}\index{subset!closed} of $\cA$, i.e., $\sigma^{\cA}(b_1, \dots, b_m) \in B$ for all $\sigma \in \Sigma_m$ ($m \ge 0$) and $b_1$, \dots, $b_m \in B$. For every nonempty closed subset $B$ of $\cA$, there is exactly one way to realize the operators on $B$ in such a way that we get a subalgebra $\cB$ of $\cA$: obviously every $\sigma^{\cB}$ should be the restriction $\sigma^{\cA}|B$ of the corresponding operation of $\cA$ to $B$. Hence, a subalgebra is completely determined by its set of elements and one may call this subset a subalgebra.  If $\sigma$ is a 0-ary operator, then every subalgebra of $\cA$ contains the element $\sigma^{\cA}$. If $\Sigma_0$ is empty, then $\emptyset$ is a closed subset, but we do not count it among the subalgebras.

It is easy to see that the intersection of any family of closed subsets of a given algebra $\cA$ is again closed. Thus we have for any $H \subseteq A$ a unique minimal closed subset containing $H$:
$$
    [H] = \bigcap ( B \mid H \subseteq B \subseteq A, B \text{ closed} ).
$$
If $H \ne \emptyset$ or $\Sigma_0 \ne \emptyset$, then $[H]$ is also nonempty and thus a subalgebra. It is called the {\em subalgebra generated\/}\index{subalgebra!generated by a set} by $H$. If $\Sigma_0 = \emptyset$, then $[\emptyset] = \emptyset$. A {\em generating set\/}\index{generating set}\index{set!generating} of $\cA$ is a subset $H \subseteq A$ such that $[H] = A$ and $\cA$ is said to be {\em finitely generated\/}\index{Algebra!finitely generated} if it has a finite generating set. It is clear that every finite algebra is finitely generated.

\begin{df}\label{Definition.1.2.4}\rm A {\em homomorphism}\index{homomorphism!of algebra} from a $\Sigma$-algebra $\cA$ to a $\Sigma$-algebra $\cB$ is a mapping $\varphi\colon A \to B$ such that for all $m \ge 0$, $\sigma \in \Sigma_m$ and $a_1$, \dots, $a_m \in A$,
$$
    \sigma^{\cA}(a_1, \dots, a_m)\varphi =
    \sigma^{\cB}(a_1\varphi, \dots, a_m\varphi).
$$
We write then $\varphi\colon \cA \to \cB$. This homomorphism is called
\begin{itemize}
\item[(a)] an {\em epimorphism}\index{epimorphism!of algebra}, if $\varphi$ is surjective,
\item[(b)] a {\em monomorphism}\index{monomorphism of!algebra}, if $\varphi$ is injective, and
\item[(c)] an {\em isomorphism}\index{isomorphism of!algebras}, if $\varphi$ is bijective.
\end{itemize}
\end{df}
If there exists an epimorphism from $\cA$ to $\cB$, then $\cB$ is said to be an {\em epimorphic image}\index{image!epimorphic} of $\cA$. A monomorphism is also called an {\em embedding}\index{embedding of!algebra}. If there is an isomorphism from $\cA$ to $\cB$, then $\cA$ and $\cB$ are {\em isomorphic} and we write $\cA \cong \cB$. Homomorphisms are often also called {\em morphisms}\index{morphism}.

If $\cA \cong \cB$, then $\cA$ and $\cB$ are the same algebra from the abstract point of view. An easy computation shows that the composition $\varphi\psi$ of two homomorphisms $\varphi\colon \cA \to \cB$ and $\psi\colon \cB \to {\cC}$ is a homomorphism from $\cA$ to ${\cC}$.

A homomorphism is a mapping that is compatible with the operations of the algebras. For example, let $\cA = ({\mathbf Z},+)$ be the algebra of the integers with the usual addition as the only operation, $n \ge 1$ and $\cB = ({\mathbf Z}_n,+)$ the algebra where ${\mathbf Z}_n = \{0,1, \dots, n-1\}$ and the sum is formed modulo $n$. Then the mapping $\varphi\colon {\mathbf Z} \to {\mathbf Z}_n$ that maps every $a \in {\mathbf Z}$ to its remainder $r_n(a)$ modulo $n$ ($0 \le r_n(a) < n$) is an epimorphism from $\cA$ to $\cB$. Of course, the homomorphisms defined in group theory, lattice theory etc.\ provide further general examples.

The proof of the following lemma is straightforward and thus it is omitted.

\begin{lm}\label{Lemma.1.2.5} Let $\varphi\colon \cA \to \cB$ be a homomorphism. If ${\cC}$ is a subalgebra of $\cA$, then $C\varphi$ is a subalgebra of $\cB$. If ${\mathcal D}$ is a subalgebra of $\cB$ and $D\varphi^{-1}$ is nonempty, then $D\varphi^{-1}$ is a subalgebra of $\cA$. \epr
\end{lm}

The following lemma contains an important observation.

\begin{lm}\label{Lemma.1.2.6} Let $\varphi\colon \cA \to \cB$ and $\psi\colon \cA \to \cB$ be two homomorphisms and $H$ a generating set of $\cA$. If $\varphi|H = \psi|H$, then $\varphi = \psi$. In other words, a homomorphism is completely determined by its restriction to a generating set.
\end{lm}

\pr Let $C = \{a\in A\mid a\varphi = a\psi\}$. Then $H \subseteq C$ by the assumption. If $m\ge 0$, $\sigma \in \Sigma_m$ and $a_1$, \dots, $a_m \in C$, then $\sigma^{\cA}(a_1, \dots, a_m) \in C$:
$$
    \sigma^{\cA}(a_1, \dots, a_m)\varphi =
    \sigma^{\cB}(a_1\varphi, \dots, a_m\varphi) =
    \sigma^{\cB}(a_1\psi, \dots, a_m\psi) = \sigma^{\cA}(a_1, \dots, a_m)\psi.
$$
Hence $C$ is closed and we get $C=A$. This implies $\varphi = \psi$.
\epr

\vspace{\baselineskip}

We define now two concepts closely related to homomorphisms, namely congruences and quotient algebras.

\begin{df}\label{Definition.1.2.7}\rm A {\em congruence}\index{congruence!ofSigma-algebra@of $\Sigma$-algebra} (relation\index{relation!congruence}) of $\cA$ is an equivalence relation on $A$ which is invariant with respect to all operations $\sigma^{\cA}$ ($\sigma \in \Sigma$). A relation $\varrho \subseteq A \times A$ is said to be {\em invariant}\index{relation!invariant with respect to operation} with respect to an $m$-ary operation $f\colon A^m \to A$ if
$$
    f(a_1, \dots, a_m) \equiv f(b_1, \dots, b_m)~(\varrho)
$$
for all elements $a_1$, \dots, $a_m$, $b_1$, \dots, $b_m \in A$ such that
$$
    a_1 \equiv b_1,\; \dots,\; a_m \equiv b_m~(\varrho).
$$
The set of all congruences of an algebra $\cA$ is denoted by $C(\cA)$.
\end{df}

Every algebra $\cA$ has at least the trivial congruences $\delta_A$ and $\iota_A$. For $\varrho \in C(\cA)$, the $\varrho$-class $a\varrho$ of an element $a \in A$ is also called a {\em congruence class}\index{class!congruence} (modulo $\varrho$). The partition $A/\varrho$ of $A$ defined by the congruence classes is {\em compatible}\index{compatible partition} in the sense that for all $m \ge 0$, $\sigma \in \Sigma_m$ and $a_1\varrho$, \dots, $a_m\varrho \in A/\varrho$ there is a class $a\varrho$ such that
$$
    \sigma^{\cA}(a_1\varrho, \dots, a_m\varrho) \subseteq a\varrho.
$$
Obviously, we can choose $a = \sigma^{\cA}(a_1, \dots, a_m)$. It is also easy to see that an equivalence relation $\varrho \in E(A)$ is a congruence of $\cA$ only in case $A/\varrho$ is a compatible partition. In fact, in automata theory it is usual to deal with compatible partitions (also called SP partitions) rather than with congruences, but both concepts convey the same idea.

The fact that $A/\varrho$ is a compatible partition for any $\varrho \in C(\cA)$ also justifies the following definition; the operations are well-defined.

\begin{df}\label{Definition.1.2.8}\rm The {\em quotient algebra}\index{Algebra!quotient} $\cA/\varrho = (A/\varrho,\Sigma)$ of a $\Sigma$-algebra $\cA$ by a congruence $\varrho \in C(\cA)$ is defined as follows. For any $m\ge 0$, $\sigma \in \Sigma_m$ and $a_1$, \dots, $a_m \in A$ we put
$$
    \sigma^{\cA/\varrho}(a_1\varrho, \dots, a_m\varrho) = \sigma^{\cA}(a_1, \dots, a_m)\varrho.
$$
\end{df}

The definition of $\sigma^{\cA/\varrho}$ may be explained as follows. To compute $\sigma^{\cA/\varrho}(a_1\varrho, \dots, a_m\varrho)$ one takes a representative from each of the $\varrho$-classes, say $a_1$, \dots, $a_m$, computes $\sigma^{\cA}$ for the representatives and forms then the $\varrho$-class of the resulting element.

Homomorphisms, congruences and quotient algebras are closely related to each other as the following three theorems show.

\begin{thm}\label{Theorem.1.2.9} For any $\varrho \in C(\cA)$, the natural mapping $\varrho^{\natural}\colon a \mapsto a\varrho$ is an epimorphism $\cA \to \cA/\varrho$ (the {\em natural homomorphism}\index{homomorphism!natural}).
\end{thm}

\pr We know that $\varrho^{\natural}$ is a surjection from $A$ to $A/\varrho$ so it suffices to verify that it is a homomorphism: for all $m\ge 0$, $\sigma \in \Sigma_m$ and $a_1$, \dots, $a_m \in A$,
\begin{align*}
    \sigma^{\cA}(a_1, \dots, a_m)\varrho^{\natural}
    = \text{}&\sigma^{\cA}(a_1, \dots, a_m)\varrho
    = \sigma^{\cA/\varrho}(a_1\varrho, \dots, a_m\varrho) = \\
    &  \sigma^{\cA/\varrho}(a_1\varrho^{\natural}, \dots, a_m\varrho^{\natural}). \tag*{\epr}
\end{align*}

\begin{thm}\label{Theorem.1.2.10} The kernel $\varphi\varphi^{-1}$ of any homomorphism $\varphi\colon\cA \to \cB$ is a congruence of $\cA$.
\end{thm}

\pr Consider any $m\ge 0$, $\sigma \in \Sigma_m$ and elements $a_1$, \dots, $a_m$,  $a'_1$, \dots, $a'_m \in A$ such that
$$
    a_1 \equiv a'_1,\; \dots,\; a_m \equiv a'_m~(\varphi\varphi^{-1}).
$$
Then $a_1\varphi = a'_1\varphi$, \dots, $a_m\varphi = a'_m\varphi$, which implies $\sigma^{\cA}(a_1, \dots, a_m)\varphi \!=\! \sigma^{\cB}(a_1\varphi, \dots, a_m\varphi) \!=\!$
$\sigma^{\cB}(a'_1\varphi, \dots, a'_m\varphi) \!=\! \sigma^{\cA}(a'_1, \dots, a'_m)\varphi$.
This means that $\sigma^{\cA}(a_1, \dots, a_m) \equiv \sigma^{\cA}(a'_1, \dots, a'_m)$ $(\varphi\varphi^{-1})$ as required.
\epr

\begin{thm}\label{Theorem.1.2.11} Every epimorphic image of an algebra $\cA$ is isomorphic to some quotient algebra of $\cA$.
\end{thm}

\pr Let $\varphi\colon\cA \to \cB$ be an epimorphism and $\theta = \varphi\varphi^{-1}$ its kernel. We claim that $\cB \cong \cA/\theta$. The required isomorphism $\cA/\theta \to \cB$ is shown to be given by
$$
    \psi\colon a\theta \mapsto a\varphi \qquad(a \in A).
$$
For any $a_1$, $a_2 \in A$,
\begin{align*}
    a_1\theta\psi = a_2\theta\psi \quad &\text{iff} \quad a_1\varphi = a_2\varphi \\
                                        &\text{iff} \quad a_1 \equiv  a_2~(\theta).
\end{align*}
This shows that $\psi$ is well-defined (i.e., $a\theta\psi$ is independent of the choice of the representative $a \in A$ of the $\theta$-class $a\theta$) and injective. Since $\varphi$ is surjective, it is clear that $\psi$ is surjective, too. It remains to be shown that $\psi$ is a homomorphism. Let $m \ge 0$, $\sigma \in \Sigma_m$ and $a_1$, \dots, $a_m \in A$. Then
\begin{align*}
    \sigma^{\cA/\theta}(a_1\theta, \dots, a_m\theta)\psi
    &= \sigma^{\cA}(a_1, \dots, a_m)\theta\psi \\
    &= \sigma^{\cA}(a_1, \dots, a_m) \varphi \\
    &= \sigma^{\cB}(a_1\varphi, \dots, a_m\varphi) \\
    &= \sigma^{\cB}(a_1\theta\psi, \dots, a_m\theta\psi). \tag*{\epr}
\end{align*}

\vspace{\baselineskip}

Taken together, Theorems \ref{Theorem.1.2.9} and \ref{Theorem.1.2.11} say that the epimorphic images of an algebra are exactly its quotient algebras (when one does not distinguish between isomorphic algebras).

Next, direct products of algebras are introduced. We may restrict ourselves to the case of a finite number of factors.

\begin{df}\label{Definition.1.2.12}\rm The {\em direct product}\index{direct product of!algebras} of two $\Sigma$-algebras $\cA$ and $\cB$ is the $\Sigma$-algebra
$$
    \cA \times \cB = (A \times B, \Sigma),
$$
where the operations are defined so that
$$
    \sigma^{\cA \times \cB}((a_1,b_1), \dots, (a_m,b_m)) = (\sigma^{\cA}(a_1, \dots, a_m), \sigma^{\cB}(b_1, \dots, b_m))
$$
for all $m\ge 0$, $\sigma \in \Sigma_m$ and $(a_1,b_1)$, \dots, $(a_m,b_m) \in A \times B$. The $k^{\rm th}$ ($k \ge 0$) {\em direct power}\index{direct power of algebra} $\cA^k$ of the $\Sigma$-algebra $\cA$ is defined inductively:
\begin{itemize}
  \item[(i)] $\cA^0 = (\{\emptyset\},\Sigma)$ is the trivial $\Sigma$-algebra.
  \item[(ii)] $\cA^{k+1} = \cA^k \times \cA$ for all $k\ge 0$.
\end{itemize}
\end{df}

It is easy to see that direct products are associative in the sense that $(\cA \times \cB) \times \cC \cong \cA \times (\cB \times \cC)$ for all $\cA$, $\cB$ and $\cC$. Both of these products can be written simply as $\cA \times \cB \times \cC$  and their elements may be identified with the triples $(a,b,c)$  with $a \in A$, $b \in B$ and $c \in C$. More generally, one can define the direct product $\cA_1 \times \dots \times \cA_k$ of $k$ ($k \ge 0$) \text{$\Sigma$-algebras} as an algebra with $A_1 \times \dots \times A_k$ as its set of elements and operations performed componentwise. It is easy to see that the projections
$$
    \pi_i\colon A_1 \times \dots \times A_k \to A_i, \quad
    (a_1, \dots, a_k) \mapsto a_i
$$
($i = 1, \dots, k$) are epimorphisms from $\cA_1 \times \dots \times \cA_k$ to the respective factor algebras $\cA_i$. Hence, every factor in a direct product is an epimorphic image of the direct product.

We shall also need the following, perhaps, less usual, way to construct a new algebra from a given one.

\begin{df}\label{Definition.1.2.13}\rm The {\em subset algebra\/}\index{Algebra!subset} (or {\em power algebra}\index{Algebra!power}) $\gp\cA = (\gp A, \Sigma)$ of a $\Sigma$-algebra $\cA$ is defined as follows. If $m\ge 0$, $\sigma \in \Sigma_m$ and $H_1$, \dots, $H_m \in \gp A$, then put
$$
    \sigma^{\gp\cA}(H_1, \dots, H_m) = \sigma^{\cA}(H_1, \dots, H_m).
$$
\end{df}

Note that the singleton sets $\{a\}$ ($a \in A$) form in $\gp\cA$ a subalgebra isomorphic to $\cA$. If $\Sigma_0 = \emptyset$, $\gp\cA$ has the trivial subalgebra $\{\emptyset\}$.

We conclude this section with a simple example illustrating these constructions.

\begin{ex}\label{Example.1.2.15}\rm Suppose $\Sigma$ consists of one binary operator $\sigma$ and a nullary operator $\gamma$. Let $\cA = (\{a,b\},\Sigma)$ be a $\Sigma$-algebra such that $\gamma^{\cA} = a$ and $\sigma^{\cA}(a,a) = \sigma^{\cA}(a,b) = \sigma^{\cA}(b,a) = a$,  $\sigma^{\cA}(b,b) = b$. Consider first the direct power $\cA^2 = \cA \times \cA$. If we
write $aa$ for $(a,a)$ etc., then $\gamma^{\cA \times \cA} = aa$ and $\sigma^{\cA \times \cA}$ is given by the following multiplication table:
$$
    \begin{array}[h]{c|cccc}
      \sigma^{\cA \times \cA} & aa & ab & ba & bb \\ \hline
      aa & aa & aa & aa & aa \\
      ab & aa & ab & aa & ab \\
      ba & aa & aa & ba & ba \\
      bb & aa & ab & ba & bb
    \end{array}
$$
Let us now construct the subset algebra. The value of the 0-ary operation is $\gamma^{\gp\cA} = \{a\}$ and the operation $\sigma^{\gp\cA}$ is given by the table below.
$$
    \begin{array}[h]{c|cccc}
      \sigma^{\gp\cA} & \emptyset & \{a\} & \{b\} & \{a,b\} \\ \hline
      \emptyset & \emptyset & \emptyset & \emptyset & \emptyset \\
      \{a\} & \emptyset & \{a\} & \{a\} & \{a\} \\
      \{b\} & \emptyset & \{a\} & \{b\} & \{a,b\} \\
      \{a,b\} & \emptyset & \{a\} & \{a,b\} & \{a,b\}
    \end{array}
$$\epr
\end{ex}

%% file: Section.1.3.tex
The concepts ``term'' and ``polynomial function'' are all-important in our modelling of the
theory
of tree automata. Let us consider an introductory example. An expression like $(x+y)(y+z)$, such
expressions are called terms, represents in a natural manner a function of the three variables
$x$, $y$, and $z$. Two things should be pointed out here. First of all, the term defines such a
function in any algebra with operations denoted by the operators appearing in the term. In our
case it could define, for example, a mapping ${\mathbf Z^{\mathbf 3}} \to {\mathbf Z}$ or a
mapping ${\mathbf R^{\mathbf 3}} \to {\mathbf R}$ depending on whether the addition and
multiplication are interpreted as those of integers or those of real numbers. Generally speaking,
the terms are determined by the operator domain, but they define operations in all algebras with
that operator domain. Secondly, we note that the term not only defines a function, but it also
describes a way to compute its values from the values of the variables once the operations of
the
algebra in question are known. In fact, algebras can be viewed as devices that evaluate terms.
When we interpret (in Chapter~\ref{Chapter.2}) terms as trees, the step from algebras to tree automata is not
long.

From now on, $X$ will be a set disjoint from the operator domain $\Sigma$. The elements of $X$ are
called {\em variables}\index{variable}. Other symbols used for sets of variables are $Y$ and $Z$.

\begin{df}\label{Definition.1.3.1}\rm The set $F_{\Sigma}(X)$ of {\em $\Sigma$-terms in $X$}\index{Sigma-term in X@$\Sigma$-term in $X$}, or
{\em $\Sigma X$-terms}\index{SigmaX-term@$\Sigma X$-term} for short, is defined as follows:


\begin{itemize}
  \item[(i)] $X \subseteq F_{\Sigma}(X)$,
  \item[(ii)] $\sigma(t_1, \dots, t_m) \in F_{\Sigma}(X)$ whenever $m\ge 0$, $\sigma \in \Sigma_m$
    and $t_1$, \dots, $t_m \in F_{\Sigma}(X)$, and
  \item[(iii)] every $\Sigma X$-term can be obtained by applying the rules (i) and (ii) a finite
    number of times.
\end{itemize}
\end{df}

If $\sigma$ is a 0-ary operator, then we get by rule (ii) the $\Sigma X$-term $\sigma(\phantom{a})$.
It is convenient to write just $\sigma$ for such a term. Then the definition of $F_{\Sigma}(X)$ may
be reformulated as follows.

\bgroup
\let\oldthedf\thedf
\edef\thedf{\oldthedf'}%
\begin{df}\label{Definition.1.3.1'}\rm The set $F_{\Sigma}(X)$ of $\Sigma X$-terms is defined
as follows:
\begin{itemize}
  \item[(i)] $X \cup \Sigma_0 \subseteq F_{\Sigma}(X)$,
  \item[(ii)] $\sigma(t_1, \dots, t_m) \in F_{\Sigma}(X)$ whenever $m > 0$, $\sigma \in \Sigma_m$
    and $t_1$, \dots, $t_m \in F_{\Sigma}(X)$, and
  \item[(iii)] every $\Sigma X$-term can be obtained by applying the rules (i) and (ii) a finite
    number of times.
\end{itemize}
\end{df}
\let\thedf\oldthedf
\egroup
\addtocounter{df}{-1}%

When $\Sigma$ and $X$ are unspecified or unemphasized, we shall speak simply about {\em terms}\index{term}.
The inductive definition of $F_{\Sigma}(X)$ suggests a useful method to deal with terms. It
could be called {\em term induction}\index{induction!term}. If we want to define a property or quantity $c(t)$ for every
$\Sigma X$-term $t$, it suffices
\begin{itemize}
  \item[(i)] to define $c(t)$ for all $t \in X$, and then
  \item[(ii)] to give a rule how to determine $c(\sigma(t_1, \dots, t_m))$ in terms of $\sigma$
  ($\in \Sigma_m$) and $c(t_1)$, \dots, $c(t_m)$ $(m \ge 0)$.
\end{itemize}

Sometimes the variation suggested by Definition \ref{Definition.1.3.1'} is more convenient: in (i)
one defines $c(t)$ for $t \in \Sigma_0$, too, but in (ii) one can then restrict oneself to values
$m > 0$. {\em Proofs by term induction} can be modelled according to the same pattern.

Note that $F_{\Sigma}(X)$ is empty iff $\Sigma_0 = X = \emptyset$. Since we do not want to consider
this uninteresting case separately every time, we shall tacitly assume that always $\Sigma_0 \cup X
\ne \emptyset$.

\begin{ex}\label{Example.1.3.2}\rm Let $\Sigma = \Sigma_0 \cup \Sigma_1 \cup \Sigma_2$, where
$\Sigma_0 = \{\mu\}$, $\Sigma_1 = \{\tau\}$ and $\Sigma_2 = \{\sigma\}$. If $X = \{x,y,z\}$, then
$x$, $z$, $\mu$, $\tau(z)$, $\tau(\mu)$, $\sigma(z,\tau(\mu))$ and $t =
\sigma(x,\sigma(z,\tau(\mu)))$ are some examples of $\Sigma X$-terms.\epr
\end{ex}

A $\Sigma X$-term $t$ is evaluated in a given $\Sigma$-algebra as follows. First we assign a value
$x\alpha \in A$ to every variable $x \in X$. Then the operations of $\cA$ are applied to these
elements as indicated by the form of $t$. For example, given a mapping $\alpha\colon X \to A$, the $t$ of
the previous example would yield the element
$$
    \sigma^{\cA}(x\alpha,\sigma^{\cA}(z\alpha,\tau^{\cA}(\mu^{\cA}))).
$$
Of course, the result depends on the choice of $\alpha$, too. This evaluation process can be
formalized as follows.

\begin{df}\label{Definition.1.3.3}\rm With every $\Sigma$-algebra $\cA$ and $\Sigma X$-term $t$
we associate a mapping
$$
    t^{\cA}\colon A^X \to A
$$
as follows: for any $\alpha\colon X \to A$
\begin{itemize}
  \item[(i)] $x^{\cA}(\alpha) = x\alpha$ ($x \in X$) and
  \item[(ii)] $t^{\cA}(\alpha) = \sigma^{\cA}(t_1^{\cA}(\alpha), \dots, t_m^{\cA}(\alpha))$ when
    $t = \sigma(t_1, \dots, t_m)$ ($m \ge 0$, $\sigma \in \Sigma_m$, $t_1$, \dots, $t_m \in
    F_{\Sigma}(X)$). The mappings $t^{\cA}$ are called the {\em polynomial functions}\index{function!polynomial} of $\cA$
    in variables $X$ and their set is denoted by $P_X(\cA)$.
\end{itemize}
\end{df}

It may seem strange that the polynomial functions $t^{\cA} \in P_X(\cA)$ are evaluated on
mappings from $X$ to $A$, but this is, in fact, just a modification of the usual way to express
polynomial functions. When one writes the value of a polynomial function in the form
$p(a_1, \dots, a_n)$, a given order of the variables is assumed, say $X = \{x_1, \dots,x_n\}$,
and the $n$-tuple $(a_1, \dots, a_n)$ is just a convenient way to give the mapping
$\alpha\colon X \to A$ such that $x_i\alpha = a_i$ ($i=1,$ \dots, $n$).

In a sense, the polynomial functions of an algebra $\cA$ are the operations one can derive by
composition from the basic operations $\sigma^{\cA}$ ($\sigma \in \Sigma$) of $\cA$, and they
share many properties with these. This is exemplified by the following four lemmas.

\begin{lm}\label{Lemma.1.3.4} If $\cB$ is a subalgebra of the $\Sigma$-algebra $\cA$ and
$\alpha\colon X \to A$ a mapping such that $X\alpha \subseteq B$, then $t^{\cA}(\alpha) \in B$ for all
$t \in F_{\Sigma}(X)$. \epr
\end{lm}

The lemma states, in other words, that subalgebras are closed with respect to polynomial functions.
The proof is a simple exercise in term induction quite similar to that of the next lemma which
expresses formally the fact that congruences are invariant with respect to polynomial functions.

\begin{lm}\label{Lemma.1.3.5} Let $\theta$ be a congruence of the $\Sigma$-algebra $\cA$ and
$\alpha\colon X \to A$, $\beta\colon X \to A$ two mappings such that
$$
    x\alpha \equiv x\beta~(\theta) \qquad\text{for all $x \in X$}.
$$
Then $t^{\cA}(\alpha) \equiv t^{\cA}(\beta)\,(\theta)$ for all $t \in F_{\Sigma}(X)$.
\end{lm}

\pr We proceed by term induction on $t$. If $t = x \in X$, then
$$
    t^{\cA}(\alpha) = x\alpha \equiv x\beta = t^{\cA}(\beta)~(\theta).
$$
Let $t = \sigma(t_1, \dots, t_m)$ and suppose
$$
    t_i^{\cA}(\alpha) \equiv t_i^{\cA}(\beta)~(\theta)
    \qquad\text{for all $i=1$, \dots, $m$}.
$$
Then also
$$
    t^{\cA}(\alpha) = \sigma^{\cA}(t_1^{\cA}(\alpha), \dots, t_m^{\cA}(\alpha)) \equiv
    \sigma^{\cA}(t_1^{\cA}(\beta), \dots, t_m^{\cA}(\beta)) = t^{\cA}(\beta)~(\theta)
$$
as $\theta$ is a congruence. Here the possibility $m=0$ can be allowed as a trivial special case.
\mbox{\ }\epr

\begin{lm}\label{Lemma.1.3.6} Let $\varphi\colon \cA \to \cB$ be a homomorphism of $\Sigma$-algebras.
Then
$$
    t^{\cA}(\alpha)\varphi = t^{\cB}(\alpha\varphi)
$$
for each mapping $\alpha\colon X \to A$ and each $\Sigma X$-term $t$. \epr
\end{lm}

\begin{lm}\label{Lemma.1.3.7} Let $\cA$ and $\cB$ be $\Sigma$-algebras, and
$\alpha\colon X \to A$ and $\beta\colon X \to B$ any mappings. If we define a mapping
$\gamma\colon X \to A \times B$ by putting
$$
    x\gamma = (x\alpha,x\beta) \qquad\text{for all $x \in X$},
$$
then
\begin{equation*}
    t^{\cA \times \cB}(\gamma) = (t^{\cA}(\alpha), t^{\cB}(\beta))
    \qquad\text{for all $t \in F_{\Sigma}(X)$}. \tag*{\epr}
\end{equation*}
\end{lm}

Lemmas \ref{Lemma.1.3.6} and \ref{Lemma.1.3.7} can easily be verified by term induction.

The subalgebra generated by a subset can also be described in terms of polynomial functions.

\begin{lm}\label{Lemma.1.3.8} For any subset $X$ of a  $\Sigma$-algebra $\cA$ we have $[X] =
\{ t^{\cA}(\alpha_X) \mid t \in F_{\Sigma}(X) \}$, where $\alpha_X = 1_A|X$, i.e., $\alpha_X$
is the mapping from $X$ to $A$ such that $x\alpha_X = x$ for all $x \in X$.
\end{lm}

\pr Denote $\{ t^{\cA}(\alpha_X) \mid t \in F_{\Sigma}(X) \}$ by $C$. For every $x \in X$, $x =
x\alpha_X = x^{\cA}(\alpha_X) \in C$. Hence $X \subseteq C$. Also, $C$ is closed under the
operations of $\cA$:
$$
    \sigma^{\cA}(t_1^{\cA}(\alpha_X), \dots, t_m^{\cA}(\alpha_X))
    = \sigma(t_1, \dots, t_m)^{\cA}(\alpha_X) \in C
$$
for all $m \ge 0$, $\sigma \in \Sigma_m$ and $t_1$, \dots, $t_m \in F_{\Sigma}(X)$. Lemma
\ref{Lemma.1.3.4} implies that $C \subseteq B$ for every subalgebra $\cB$ which contains $X$.
Hence $C = [X]$. Note that the result is true even if $\Sigma_0 = X = \emptyset$. In this case
$[X] = \emptyset$.
\epr

\vspace{\baselineskip}

We shall now turn to the $\Sigma$-algebra formed by the $\Sigma X$-terms.

\begin{df}\label{Definition.1.3.9}\rm The $\Sigma$-algebra $\cF_{\Sigma}(X) =
(F_{\Sigma}(X),\Sigma)$ defined so that
$$
    \sigma^{\cF_{\Sigma}(X)}(t_1, \dots, t_m) = \sigma(t_1, \dots, t_m)
$$
for all $m \ge 0$, $\sigma \in \Sigma_m$ and $t_1$, \dots, $t_m \in F_{\Sigma}(X)$, is called the
{\em $\Sigma X$-term algebra}\index{Algebra!SigmaX-term@$\Sigma X$-term} or the {\em free $\Sigma$-algebra generated by $X$}.
\end{df}

We shall first account for the name ``free algebra''.

\begin{df}\label{Definition.1.3.10}\rm Let $K$ be a class of $\Sigma$-algebras. A $\Sigma$-algebra
$\cF = (F, \Sigma)$ is said to be {\em freely generated\/}\index{Algebra!freely generated over a class} over $K$ by a subset $X \subseteq F$, if
the following conditions are satisfied:
\begin{itemize}
  \item[(i)] $\cF \in K$.
  \item[(ii)] $X$ generates $\cF$.
  \item[(iii)] Every mapping $\alpha\colon X \to A$ of $X$ into any algebra $\cA$ in $K$ has an
    extension to a homomorphism $\widehat{\alpha}\colon \cF \to \cA$.
\end{itemize}

If these conditions are satisfied for some subset $X$ of $F$, then $\cF$ is called a {\em free
algebra}\index{Algebra!free} over $K$ (with $|X|$ generators), and $X$ is called a {\em free generating set}\index{set!free generating}\index{generating set!free}.
\end{df}

A well-known example is provided by the free semigroup $X^+$ generated by a set (alphabet) $X$.
The elements of $X^+$ are all the finite nonempty strings of elements of $X$. The product of two
such strings $u$ and $v$ is simply their concatenation $uv$. The associativity of this product is
obvious and thus $X^+$ is a semigroup. As every string $u \in X^+$ is obtained by concatenating
individual elements of $X$, it is clear that $X$ generates $X^+$. To prove that $X^+$ is freely
generated by $X$ over the class of all semigroups we consider any semigroup $S$ and mapping
$\alpha\colon X \to S$. The required (unique) homomorphism
$$
    \widehat{\alpha}\colon X^+ \to S
$$
is obtained by putting
$$
    (x_1x_2 \dots x_k)\widehat{\alpha} =
    (x_1\alpha) \cdot (x_2\alpha) \cdot \ldots \cdot (x_k\alpha)
$$
for all $x_1x_2\dots x_k \in X^+$ (products to the right are formed in $\cS$).

Free semigroups are considered later again, but we return now to our term algebras.

\begin{thm}\label{Theorem.1.3.11} The $\Sigma X$-term algebra $\cF_{\Sigma}(X)$ is freely
generated by $X$ over the class of all $\Sigma$-algebras.
\end{thm}

\pr That $X$ generates $\cF_{\Sigma}(X)$ is quite obvious when we compare the definitions of
$F_{\Sigma}(X)$ and $\cF_{\Sigma}(X)$, but it follows also from the useful observation that
\begin{equation*}\tag{*}
    t^{\cF_{\Sigma}(X)}(\alpha_X) = t
    \qquad\text{for all $t \in F_{\Sigma}(X)$}
\end{equation*}
(where $\alpha_X = 1_{F_{\Sigma}(X)}|X$). The proof of (*) goes again by term induction. Let $\cA$
be any $\Sigma$-algebra and $\alpha\colon X \to A$ any mapping. We claim that the mapping
$$
    \widehat{\alpha}\colon F_{\Sigma}(X) \to A, \quad
    t \mapsto t^{\cA}(\alpha) \quad (t \in F_{\Sigma}(X))
$$
is the required homomorphism. For every $x \in X$, $x\widehat{\alpha} = x^{\cA}(\alpha) = x\alpha$.
Hence, $\widehat{\alpha}|X = \alpha$. It remains to be verified that $\widehat{\alpha}$ is a
homomorphism. Indeed,
\begin{align*}
    \sigma^{\cF_{\Sigma}(X)}(t_1, \dots, t_m)\widehat{\alpha}
        &= \sigma(t_1, \dots, t_m)^{\cA}(\alpha) \\
        &= \sigma^{\cA}(t_1^{\cA}(\alpha), \dots, t_m^{\cA}(\alpha)) \\
        &= \sigma^{\cA}(t_1\widehat{\alpha}, \dots, t_m\widehat{\alpha})
\end{align*}
for all $m \ge 0$, $\sigma \in \Sigma_m$ and $t_1$, \dots, $t_m \in F_{\Sigma}(X)$.
\epr

\vspace{\baselineskip}

We add a few general comments on free algebras. First of all, one should note that the homomorphic
extension $\widehat{\alpha}\colon \cF \to \cA$ of a mapping $\alpha\colon X \to A$ ($\cA \in K$) is unique.
This follows from Lemma \ref{Lemma.1.2.6}. Free algebras over a given class do not always exist,
but when they do, they are determined up to isomorphism by the cardinality of the free generating
set. This is stated formally in the following lemma.

\begin{lm}\label{Lemma.1.3.12} Any two algebras freely generated over the same class of algebras
by sets of the same cardinality are isomorphic.
\end{lm}

\pr Suppose $\cA$ and $\cB$ both are free over the same class $K$ and that they have free
generating sets $X$ and $Y$, respectively, such that $|X| = |Y|$. Then there is a bijection
$\alpha\colon X \to Y$. The converse of it, $\beta = \alpha^{-1}$, defines a bijection from $Y$ to $X$.
Now there exist morphisms
$$
    \widehat{\alpha}\colon \cA \to \cB
    \quad\text{and}\quad
    \widehat{\beta}\colon \cB \to \cA
$$
such that $\widehat{\alpha}|X = \alpha$ and $\widehat{\beta}|Y = \beta$. But then
$$
    \widehat{\alpha}\widehat{\beta}\colon \cA \to \cA
    \quad\text{and}\quad
    \widehat{\beta}\widehat{\alpha}\colon \cB \to \cB
$$
are homomorphisms such that $\widehat{\alpha}\widehat{\beta}|X = 1_X$ and
$\widehat{\beta}\widehat{\alpha}|Y = 1_Y$. This means by Lemma \ref{Lemma.1.2.6} that
$\widehat{\alpha}\widehat{\beta} = 1_A$ and $\widehat{\beta}\widehat{\alpha} = 1_B$. Hence,
$\widehat{\alpha}$ and $\widehat{\beta}$ are isomorphisms inverse to each other. This implies
$\cA \cong \cB$.
\epr

\vspace{\baselineskip}

Lemma \ref{Lemma.1.3.12} allows us to speak about {\em the} algebra freely generated over a class $K$
by a set $X$.

We shall fix the notation $\widehat{\alpha}$ used above for the rest of the book: for any $\!\cA$ and
$\alpha\colon \!X \!\to\! A$, $\widehat{\alpha}\colon \cF_{\Sigma}(X) \to \cA$ is the homomorphism such that
$\widehat{\alpha}|X = \alpha$. To evaluate a $\Sigma X$-term $t$ in a $\Sigma$-algebra $\cA$ for a
given assignment $\alpha\colon X \to A$ of values to the variables amounts to the computation of
$t\widehat{\alpha}$. Indeed, we showed in the proof of Theorem \ref{Theorem.1.3.11} that
$t^{\cA}(\alpha) = t\widehat{\alpha}$ for all $\cA$, $\alpha$ and $t$.

The polynomial functions in variables $X$ of an algebra $\cA$ are the mappings one can get from the ``projections'' $x^{\cA}$ ($x \in X$) by iterated compositions with the basic operations $\sigma^{\cA}$ ($\sigma \in \Sigma$). If the generating set of functions is enlarged by the set of all constant mappings ($c \in A$)
$$
    \gamma_c\colon A^X \to A, \quad \alpha \mapsto c \quad (\alpha \in A^X),
$$
then we get, in general, a larger class of functions. These are called algebraic functions. We shall need just the unary (i.e., 1-place) algebraic functions and these only are defined below. In this special case $X$ is a singleton $\{x\}$  and we may identify any mapping $\alpha\colon X \to A$ with the element $x\alpha \in A$. Then the unary algebraic functions can be defined simply as certain mappings from $A$ to $A$.

\begin{df}\label{Definition.1.3.13}\rm The set of {\em unary algebraic functions}\index{function!unary algebraic} $\Alg_1(\cA)$ of a $\Sigma$-algebra $\cA$ is defined as follows:
\begin{itemize}
  \item[(i)] $1_A \in \Alg_1(\cA)$.
  \item[(ii)] For every $c \in A$, $\Alg_1(\cA)$ contains the {\em constant mapping}\index{mapping!constant} $\gamma_c\colon A \to A$, $a \mapsto c$ ($a \in A$).
  \item[(iii)] The composition $\sigma^{\cA}(f_1, \dots, f_m)$ is in $\Alg_1(\cA)$ whenever $m \ge 0$, $\sigma \in \Sigma_m$ and $f_1$, \dots, $f_m \in \Alg_1(\cA)$.
  \item[(iv)] All members of $\Alg_1(\cA)$ are obtained by the rules (i)--(iii).
\end{itemize}
\end{df}

The constant mapping $\gamma_c$ ($c \in A$) is usually denoted simply by $c$. It is intuitively clear from Definition \ref{Definition.1.3.13} that every $f \in \Alg_1(\cA)$ can be represented by an expression similar to the terms that gave the polynomial functions. Let $X = A \cup \{x\}$ ($x \notin A$). Following the inductive form of Definition \ref{Definition.1.3.13} we associate with every $f \in \Alg_1(\cA)$ a $\Sigma X$-term $t_f$ as follows:
\begin{itemize}
  \item[(i)] $t_{1_A} = x$.
  \item[(ii)] $t_c = c$ for all $c$ ($ = \gamma_c$) ($c \in A$).
  \item[(iii)] If $f = \sigma^{\cA}(f_1, \dots, f_m)$, then $t_f = \sigma(t_{f_1}, \dots, t_{f_m})$.
\end{itemize}

It is now an easy task to verify that the following lemma holds.

\begin{lm}\label{Lemma.1.3.14} For every $f \in \Alg_1(\cA)$ there exists a term $t_f \in F_{\Sigma}(A \cup x)$ such that, for all $a \in A$,
$$
    f(a) = t_f^{\cA}(\alpha_a)
$$
when $\alpha_a$ is the mapping such that $\alpha_a|A = 1_A$ and $x\alpha_a = a$. \epr
\end{lm}

The assignment $\alpha_a$ depends on $a \in A$ only. We may think of $t_f$ as a $\Sigma X$-term for a suitable $X$, in which all variables, save $x$, have been assigned constant values from $A$. In other words, the unary algebraic functions are obtained from polynomial functions by fixing the values of some variables. It is now obvious, in view of Lemma \ref{Lemma.1.3.5}, that congruences of $\cA$  are invariant with respect to unary algebraic functions. The converse of this observation holds also. In fact, it can be stated in a stronger form in terms of the special unary algebraic functions introduced in the following definition.

\begin{df}\label{Definition.1.3.15}\rm A mapping $f\colon A \to A$ is called an {\em elementary translation}\index{translation!elementary} of the $\Sigma$-algebra $\cA$, if there exist an $m > 0$, a $\sigma \in \Sigma_m$, a $j$ ($1 \le j \le m$) and elements $c_1$, \dots, $c_{j-1}$, $c_{j+1}$, \dots, $c_m \in A$ such that
$$
    f(a) = \sigma^{\cA}(c_1, \dots, c_{j-1},a,c_{j+1}, \dots, c_m)
    \quad\text{for all $a \in A$}.
$$
The set of all elementary translations of $\cA$ is denoted by $\ET(\cA)$.
\end{df}

It is obvious that $\ET(\cA) \subseteq \Alg_1(\cA)$.

\begin{lm}\label{Lemma.1.3.16} An equivalence relation $\theta \in E(A)$ is a congruence of $\cA$ iff $\theta$ is invariant with respect to all elementary translations of $\cA$.
\end{lm}

\pr Suppose $a \equiv b \,(\theta)$ implies $f(a) \equiv f(b)\,(\theta)$ for all $a$, $b \in A$ and $f \in \ET(\cA)$. Consider any $m > 0$, $\sigma \in \Sigma_m$ and elements $a_1$, \dots, $a_m$, $b_1$, \dots, $b_m \in A$ such that $a_1 \equiv b_1$, \dots, $a_m \equiv b_m \,(\theta)$. Define the following $m$ elementary translations:
$$
    f_j(\xi) = \sigma^{\cA}(b_1, \dots, b_{j-1},\xi,a_{j+1}, \dots, a_m)
    \quad(j = 1, \dots, m).
$$
Then
\begin{align*}
    \sigma^{\cA}(a_1, a_2, \dots, a_m)
                 &= f_1(a_1) \equiv f_1(b_1) ~(\theta) \\
                 &= f_2(a_2) \equiv f_2(b_2) ~(\theta) \\
                 &\vdots \\
                 &= f_m(a_m) \equiv f_m(b_m) ~(\theta) \\
                 &= \sigma^{\cA}(b_1, b_2, \dots, b_m).
\end{align*}
Hence $\sigma^{\cA}(a_1, \dots, a_m) \theta \sigma^{\cA}(b_1, \dots, b_m)$ and we have verified that $\theta \in C(\cA)$. The converse is obvious.
\epr

%% file: Section.1.4.tex
We shall need a few facts from lattice theory, and these are quickly surveyed here.

\begin{df}\label{Definition.1.4.1}\rm Let $A$ be a set. A relation $\varrho \subseteq A \times A$
is called a {\em partial ordering}\index{ordering!partial} of $A$, if
\begin{itemize}
  \item[(1)] $\delta_A \subseteq \varrho$ ($\varrho$ is reflexive),
  \item[(2)] $\varrho \cap \varrho^{-1} \subseteq \delta_A$ ($\varrho$ is antisymmetric), and
  \item[(3)] $\varrho\varrho \subseteq \varrho$ ($\varrho$ is transitive).
\end{itemize}
If $\varrho$ is a partial ordering of $A$, then $(A,\varrho)$ is called a {\em poset}\index{poset}.
\end{df}

The usual symbol for a partial ordering is $\le$. Often a set $A$ is called a poset when a certain partial ordering of $A$ is understood.

An example of a poset is $(\gp S, \subseteq)$, where $S$ is a set and $\subseteq$ the usual subset relation in the power set $\gp S$. Another simple example is $({\mathbf N}, \le)$ where $\le$ is the ``less than or equal'' -relation of natural numbers.  This $\le$ is a {\em total ordering}\index{ordering!total}, which means that any two elements of the poset are {\em  comparable}\index{comparable elements}, i.e., either $a \le b$ or $b \le a$ holds for any two elements $a$ and $b$. A poset $(A, \le)$ in which $\le$ is a total ordering is called a {\em chain}\index{chain}.

Let $(A, \le)$  be a poset and $a$, $b \in A$. We may write $a \ge b$ when $b \le a$, $a < b$ when $a \le b$ and $a \ne b$, and $a > b$ when $a \ge b$ and $a \ne b$. Clearly $\ge$ is a partial ordering and the poset $(A, \ge)$ is said to be {\em dual\/}\index{poset!dual} to $(A, \le)$. Each one of the relations $\ge$, $<$ and $>$ determines $\le$ completely.

An element $a \in A$ is an {\em upper bound}\index{bound!upper} of a subset $H \subseteq A$ if $b \le a$ for all $b \in H$. An upper bound $a$ of $H \subseteq A$ is the {\em least upper bound}\index{bound!least upper}, or the {\em supremum}\index{supremum}, of $H$, if $a \le c$ for all upper bounds $c$ of $H$. {\em Lower bounds}\index{bound!lower} and {\em greatest lower bounds\index{bound!greatest lower} (infimums\index{infimum})\/} are defined similarly. The least upper bound and the greatest lower bound of a subset $H$ are denoted, respectively, by $\bigvee H$  and $\bigwedge H$. In case of an indexed family $(a_i \mid i \in I)$ of elements the notations $\bigvee(a_i \mid i \in I)$ and $\bigwedge(a_i \mid i \in I)$ may be used.

An element $c \in A$ is a {\em zero element}\index{element!zero} of the poset $A$ if $c \le a$ for every $a \in A$. If a poset has a zero element, it is unique and usually it is denoted by 0. Similarly, the {\em unit element\/}\index{element!unit} 1, is defined by the condition that $a \le 1$ for all $a \in A$. Clearly, $\bigwedge A$ exists iff the poset has a zero element 0, and then $\bigwedge A = 0$. Similarly, $\bigvee A$ exists, and then equals 1, iff $A$ has a unit element 1.

\begin{df}\label{Definition.1.4.2}\rm A poset $(A, \le)$ is a {\em lattice}\index{lattice}, if $\bigvee \{a,b\}$ and $\bigwedge\{a,b\}$ exist for all $a$, $b \in A$. It is a {\em complete lattice}\index{lattice!complete}, if
$\bigvee H$ and $\bigwedge H$ exist for all subsets $H$ of $A$.
\end{df}

In a lattice one usually writes $a \vee b$ and $a \wedge b$ for $\bigvee \{a,b\}$ and $\bigwedge\{a,b\}$, respectively. The element $a \vee b$ is also called the {\em join}\index{join} of $a$ and $b$, and $a \wedge b$ is the {\em meet}\index{meet} of $a$ and $b$. It is easy to see that $\bigvee H$ and $\bigwedge H$ exist for every finite, nonempty subset $H$ of a lattice. However, $\bigvee \emptyset$ exists only in case the lattice has a zero element 0. Then $\bigvee \emptyset = 0$. Similarly, $\bigwedge \emptyset$ exists iff the lattice has a unit element 1; then $\bigwedge \emptyset = 1$.

The following lemma follows directly from the definitions of the join and the meet.

\begin{lm}\label{Lemma.1.4.3} If $(A,\le)$ is a lattice then $\wedge$ and $\vee$ satisfy the following identities:
\begin{itemize}
  \item[\rm (L1)] $x \wedge x =x$, $x \vee x =x$ (idempotence).
  \item[\rm (L2)] $x \wedge y = y \wedge x$, $x \vee y = y \vee x$ (commutativity).
  \item[\rm (L3)] $x \wedge (y \wedge z) = ( x \wedge y) \wedge z$, $x \vee (y \vee z) = ( x \vee y) \vee z$ (associativity).
  \item[\rm (L4)] $x \wedge (x \vee y) = x$, $x \vee (x \wedge y) = x$ (absorption). \epr
\end{itemize}
\end{lm}

The identities (L1)--(L4) are characteristic of lattices in the following sense. If $(A,\wedge,\vee)$ is an algebra with two binary operations that satisfy these identities, then $(A,\le)$ is a lattice when $\le$ is defined so that
$$
    a \le b \quad\text{iff}\quad a \wedge b = a \quad(a, b \in A).
$$
In this lattice $\bigvee\{a,b\} = a \vee b$ and $\bigwedge\{a,b\} = a \wedge b$ for all $a$, $b \in A$. In lattice theory lattices are usually defined and considered in parallel both as posets and as algebras. The two aspects of the theory complement each other.

The following lemma is often useful when one wants to show that a certain poset is a complete lattice.

\begin{lm}\label{Lemma.1.4.4} A poset $(A,\le)$ is a complete lattice, if $\bigwedge H$ exists for each subset $H \subseteq A$. \\ \mbox{\ } \epr
\end{lm}

Note that the existence of $\bigwedge \emptyset = 1$ should also be ascertained when Lemma \ref{Lemma.1.4.4} is used. We shall now apply the lemma to an important example. Let $A$ be a set. It is easy to see that the intersection $\bigcap (\varepsilon_i \mid i \in I)$ of any equivalence relations $\varepsilon_i$ ($i \in I$) of $A$ is again in $E(A)$. This means that
$$
    \bigwedge (\varepsilon_i \mid i \in I) = \bigcap (\varepsilon_i \mid i \in I)
$$
always exists in the poset $(E(A), \subseteq)$. (In particular, $\bigwedge \emptyset = \iota_A$.) Hence, we get

\begin{lm}\label{Lemma.1.4.5} For each set $A$, $(E(A), \subseteq)$ is a complete lattice. \epr
\end{lm}

In general, the union of equivalence relations is not an equivalence relation. For any $H \subseteq E(A)$, $\bigvee H$ is the intersection of all equivalence relations which contain the union $\bigcup H$. A more useful description of the supremum is given in the following lemma.

\begin{lm}\label{Lemma.1.4.6} Let $H \subseteq E(A)$ and $a$, $b \in A$. Then $a \equiv b \,(\bigvee H)$ iff there exist an $n \ge 0$, $\varepsilon_1$, \dots, $\varepsilon_n \in H$ and $a_1$, \dots, $a_{n-1} \in A$ such that
\begin{equation*}
    a \, \varepsilon_1 \, a_1 \, \varepsilon_2 \, a_2 \, \dots \, a_{n-1} \, \varepsilon_n \, b.   \tag*{\epr}
\end{equation*}
\end{lm}

The lemma may be used to prove the following important fact.

\begin{thm}\label{Theorem.1.4.7} For any algebra $\cA = (A,\Sigma)$, $C(\cA)$ forms a {\em complete sublattice}\index{complete sublattice} of $(E(A),\subseteq)$, that is to say, $\bigvee H \in C(\cA)$ and $\bigwedge H \in C(\cA)$ whenever $H \subseteq C(\cA)$. \epr
\end{thm}

The {\em direct product}\index{direct product of!posets} $(L_1 \times \dots \times L_n, \le)$ of posets $(L_1,\le)$, \dots, $(L_n,\le)$ is a poset when we define $\le$ in $L_1 \times \dots \times L_n$ so that
$$
    (a_1, \dots, a_n) \le (b_1, \dots, b_n)
    \quad\text{iff}\quad
    a_i \le b_i \quad\text{for all $i=1$, \dots, $n$.}
$$
If the $(L_i,\le)$'s are lattices, then the direct product is also a lattice in which
\begin{align*}
    &(a_1, \dots, a_n) \vee (b_1, \dots, b_n) = (a_1 \vee b_1, \dots, a_n \vee b_n)
\intertext{and}
    &(a_1, \dots, a_n) \wedge (b_1, \dots, b_n) = (a_1 \wedge b_1, \dots, a_n \wedge b_n).
\end{align*}

An {\em ideal}\index{ideal} of a lattice $(A, \le)$ is a nonempty subset $I$ of $A$ such that, for all $a$, $b \in A$,
\begin{itemize}
    \item[(1)] $a$, $b \in I$ implies $a \vee b \in I$, and
    \item[(2)] $a \le b \in I$ implies $a \in I$.
\end{itemize}

A {\em dual ideal}\index{ideal!dual} of a lattice $(A, \le)$ is a nonempty subset $D$ of $A$ such that, for all $a$, $b \in A$,
\begin{itemize}
    \item[(1')] $a$, $b \in D$ implies $a \wedge b \in D$, and
    \item[(2')] $a \ge b \in D$ implies $a \in D$.
\end{itemize}

General examples are provided by the
\begin{itemize}
    \item[(i)] {\em principal ideal\/}\index{ideal!principal} $(a] = \{ x \in A \mid x \le a \}$ generated by an element $a \in A$, and by the
    \item[(ii)] {\em principal dual ideal\/}\index{ideal!principal dual} $[a) = \{ x \in A \mid x \ge a \}$ generated by an element $a \in A$.
\end{itemize}

Let $A$ and $B$ be posets. A mapping $\varphi\colon A \to B$ is said to be {\em isotone}\index{mapping!isotone}, if
$$
    (\forall a_1, a_2 \in A) \, a_1 \le a_2 \Longrightarrow a_1\varphi \le a_2 \varphi.
$$
Suppose now that $A$ and $B$ are complete lattices. The mapping $\varphi$ is {\em $\omega$-continuous}\index{mapping!omega-continuous@$\omega$-continuous}, if
$$
    \bigvee (a_i \mid i\ge 0) \varphi = \bigvee (a_i \varphi \mid i \ge 0)
$$
for every ascending $\omega$-sequence
$$
    a_0 \le a_1 \le a_2 \le \dots
$$
of elements $a_i \in A$ ($0 \le i < \omega$). An $\omega$-continuous mapping is always isotone, but the converse is false.

Let $A$ be a poset and $\varphi\colon A \to A$ a mapping. An element $a \in A$ is a {\em fixed-point}\index{fixed-point} of $\varphi$, if $a\varphi = a$. It is the {\em least fixed-point}\index{fixed-point!least} of $\varphi$, if all other fixed-points of $\varphi$ are above it. Of course, there can be at most one least fixed-point. A well-known theorem by A.~Tarski states that every isotone mapping in a complete lattice has a fixed-point. For $\omega$-continuous mappings the following stronger result holds.

\begin{thm}\label{Theorem.1.4.8} Let $(A,\le)$ be a complete lattice and $\varphi\colon A \to A$ an $\omega$-continuous mapping. Then
$$
    [\varphi] = \bigvee ( 0\varphi^i \mid i \ge 0)
$$
is the least fixed-point of $\varphi$.
\end{thm}

\pr Since $\varphi$ is isotone, $0 \le 0\varphi$ implies
$$
    0 \le 0\varphi \le 0\varphi^2 \le 0\varphi^3 \le \dots.
$$
By $\omega$-continuity, we get now
$$
    [\varphi]\varphi = \bigvee(0\varphi^{i+1} \mid i \ge 0)
     = \bigvee(0\varphi^i \mid i \ge 0) = [\varphi].
$$
For any fixed-point $a$ of $\varphi$, $0 \le a$ implies
$$
    0\varphi \le a\varphi = a,
$$
and in general by induction on $i \ge 0$, $0\varphi^i \le a$. Hence $[\varphi] \le a$, and $[\varphi]$ is the least fixed-point of $\varphi$. \epr

%% file: Section.1.5.tex

In this section several basic concepts and facts from the theory of finite automata are reviewed. For many readers there is probably nothing really new. The presentation is quite telegraphic and proofs are sketched at most. Much of the material will be generalized to tree automata in Chapter~\ref{Chapter.2}, and the present section is intended mainly as an outline of the proper background scenery.

An \emph{alphabet}\index{alphabet} is a finite nonempty set of symbols which are called \emph{letters}\index{letter}. We shall usually use the letters $X$, $Y$ and $Z$ to indicate alphabets. A finite string of letters from an alphabet $X$ is called an \emph{$X$-word} \index{xword@$X$-word, see word} or a \emph{word}\index{word} over $X$. Consider an arbitrary $X$-word
\begin{equation*}
w = x_1 x_2 \ldots x_n \quad (n\geq0,~x_1, \ldots, x_n \in X).
\end{equation*}
Here $x_i = x_j$ is possible even for $i \neq j$. If $n=0$, then $w$ is the \emph{empty word}\index{word!empty} which is denoted by $e$. The \emph{length}\index{length of!word} of $w$ is $n$ and we write it $|w|$. Obviously, $|w| = 0$ iff $w=e$. The set of all $X$-words is denoted by $X^*$, and the set of all nonempty $X$-words is denoted by $X^+$. The letters of an alphabet are viewed as indivisible symbols. This means, in particular, that for any $m \geq 0$, $n \geq 0$ and $x_1, \ldots, x_m, \; y_1, \ldots, y_n \in X$,
\begin{equation*}
x_1 x_2 \ldots x_m = y_1 y_2 \ldots y_n
\end{equation*}
holds just in case $m=n$ and $x_i= y_i$ for all $i= 1, ..., m$. Letters are considered words of length $1$. Hence, we may write $X \subset X^+ \subset X^*$ and $X^* = X^+ \cup e$.

In Section 3 we noted that $X^+$ is the free semigroup generated by $X$, when the product of two words is defined to be their catenation. Similarly, $X^*$ is the \emph{free monoid}\index{monoid!free} generated by $X$. The identity element is the empty word: $ew = we = w$ for each $w \in X^*$.

A \emph{language}\index{language} over $X$, or an \emph{$X$-language}, \index{xlang@$X$-language, see language} is simply a subset of $X^*$. An $X$-language is \emph{$e$-free}\index{language!e-free@$e$-free} if it does not include the empty word. Of course, formal language theory concerns itself with such languages only that can be specified in some effective manner.

A \emph{family of languages}\index{family of languages} $\cL$ is defined by indicating for each alphabet the set $\cL(X)$ of $X$-languages belonging to the family. For example, $\cL(X)$ could consist of all languages recognized by automata of a given type with input alphabet $X$. If $L \in \cL(X)$, one may write just $L \in \cL$. Two families of languages $\cK$ and $\cL$ are equal, which we write $\cK = \cL$, if $\cK(X)= \cL(X)$ for every alphabet $X$. Similarly, the inclusion $\cK \subseteq \cL$ means that $\cK(X) \subseteq \cL(X)$ for every $X$.

One way to specify a language $L \subseteq X^*$ is to give an automaton that can examine any given $X$-word and then tell whether the word is in $L$ or not. Such automata are called \emph{recognizers}\index{recognizer}. The most basic type of recognizers is the following:

\begin{df}
\label{Definition.1.5.1}
\rm
An \emph{$X$-recognizer} \index{xrecog@$X$-recognizer, see recognizer} (also called a \emph{Rabin-Scott recognizer}\index{recognizer!Rabin-Scott}) $\bA$ consists of
\begin{enumerate}[(1)]
\item a finite (nonvoid) set $A$ of \emph{states}\index{state!of recognizer},
\item the \emph{input alphabet}\index{input alphabet of!recognizer} $X$,
\item a \emph{next-state function}\index{next-state function of!recognizer} $\delta \colon A \times X \rightarrow A$,
\item an \emph{initial state}\index{initial state of!recognizer} $a_0 \in A$, and
\item a set $A' \subseteq A$ of \emph{final states}\index{final state of!recognizer}.
\end{enumerate}
 We write $\bA = (A, X, \delta, a_0, A')$.
\end{df}

If the $X$-recognizer $\bA$ of Definition~\ref{Definition.1.5.1} is in state $a~(\in\hspace{-.5ex}A)$ and receives the input $x~(\in\hspace{-.5ex}X)$, it enters state $\delta(a, x)$ and remains in this state until it reads the next input letter. The next-state function is extended to a function
\begin{equation*}
\hat{\delta} \colon A \times X^* \rightarrow A
\end{equation*}
as follows:
\begin{itemize}
\item[$1^\circ\,$] $\hat{\delta}(a, e) = a$ for each $a \in A$, and
\item[$2^\circ\,$] $\hat{\delta}(a, wx) = \delta(\hat{\delta}(a, w), x)$ for all $a \in A,$ $w \in X^*$ and $x \in X$.
\end{itemize}

We will omit the cap from $\hat{\delta}$. For any $a \in A$ and $w \in X^*$, $\delta(a, w)$ is the state of $\bA$ when it has read the whole input word $w$, from left to right, and the state in the beginning was $a$. As a language recognizer $\bA$ operates as follows. The word $w$ to be tested for membership is entered to $\bA$ so that the state of $\bA$ initially is $a_0$. Now $w$ is \emph{accepted}\index{word!accepted by recognizer} by $\bA$ if $\delta(a_0, w)$ is a final state. Otherwise $w$ is said to be \emph{rejected} by $\bA$. The \emph{language recognized}\index{language!recognized by recognizer} by $\bA$ consists of all $X$-words accepted by $\bA$, i.e., it is the $X$-language
\begin{equation*}
 L(\bA) = \{w \in X^* \mid \delta(a_0,w) \in A'\}.
\end{equation*}
An $X$-language $L$ is called \emph{recognizable}\index{language!recognizable}, if there exists an $X$-recognizer $\bA$ such that $L=L(\bA)$. The family of recognizable languages is denoted by $\Rec$, and $\Rec~X$ denotes the set of all recognizable $X$-languages.

In the definition of $X$-recognizers the finiteness of the state set is essential. Otherwise, every $X$-language would be recognizable.

We shall now prepare for the first of the many characterizations of recognizable languages.

The \emph{product}\index{product!of languages} of two $X$-languages $U$ and $V$ is the $X$-language
\begin{equation*}
 UV = \{uv \mid u \in U, v \in V\}.
\end{equation*}
The product is associative:
\begin{equation*}
 U(VW) = (UV)W \quad \textrm{ for all}~U, V, W \subseteq X^*.
\end{equation*}
%
Furthermore,
\begin{equation*}
U\emptyset = \emptyset U = \emptyset \quad \textrm{and} \quad U\{e\} = \{e\}U = U
\end{equation*}
for every $X$-language $U$.

The \emph{powers}\index{power of!language} $U^n~(n \geq 0)$ of an $X$-language $U$ are defined inductively:
\begin{equation*}
\begin{array}{ll}
1^\circ & U^0 = \{e\} \textrm{ and }\\
2^\circ & U^n = U^{n-1}U \textrm{ for } n>0.
\end{array}
\end{equation*}
By means of the powers we may define the \emph{iteration}\index{iteration} of $U$
\begin{equation*}
U^* = \bigcup (U^n \mid n \geq 0).
\end{equation*}
Excluding $U^0$, we get the language
\begin{equation*}
U^+ = \bigcup (U^n \mid n \geq 1).
\end{equation*}
Clearly, $U^* = U^+ \cup \{e\}$, and $U^+ = U^*$ iff $e \in U$. A word $w \in X^*$ belongs to $U^*$ iff it can be expressed in the form $w=u_1 u_2 \ldots u_n$, where $n \geq 0$ and $u_1,\ldots u_n \in U$.

Note that $X^n$ is the set of all $X$-words of length $n~(n \geq 0)$ and the set $X^*$ of all $X$-words really is the iteration of $X$ (when $X$ is viewed as the set of $X$-words of length 1).

Union, product and iteration are called the \emph{regular language operations}.

\begin{df}
\label{Definition.1.5.2}
\rm
The set $\Reg~X$ of \emph{regular}\index{language!regular} $X$-languages is the smallest set $R$ such that
\begin{equation*}
\begin{array}{ll}
1^\circ & \emptyset \in R  \textrm{ and } \{x\} \in R \textrm{ for each } x \in X, \textrm{ and }\\
2^\circ & U,\,V \in R \textrm{ implies } U \cup V,\, UV,\, U^* \in R.
\end{array}
\end{equation*}
\end{df}

Regular languages are also called rational languages. All finite languages are regular. Hence $\Reg~X$ is the smallest set of $X$-languages containing the finite $X$-languages which is closed under the three regular operations.

The form of Definition~\ref{Definition.1.5.2} implies that every regular $X$-language can be represented by a \emph{regular expression}\index{regular expression} which shows how the language is obtained from $\emptyset$ and the languages $\{x\}$ by forming unions, products and iterations.

\begin{ex}
\label{Example.1.5.3}
\rm
Let $X = \{x, y\}$. Some members of $\Reg~X$ are $\emptyset$, $\{x\}$, $\{y\}$, $\{xy\}= \{x\}\{y\}$, $\{xy, yy\}= \{x\}\{y\} \cup \{y\}\{y\} = (\{x\} \cup \{y\})\{y\}$ and
\begin{equation*}
U = \{x^i y^j \mid i \geq 1, j \geq 0\} \cup \{yx^{2k} \mid k \geq 0 \}.
\end{equation*}
A possible regular expression for the language $U$ would be $\eta = (x(x)^* (y)^*)+(y(xx)^*)$ (usually `+' is used for union). If we agree on the usual hierarchy of regular operations (first iterations, then products, and unions last), then some parentheses can be omitted and $\eta$ becomes $xx^*y^*+y(xx)^*$. The language $U$ is
recognized by the $X$-recognizer defined by the state graph of Fig~\ref{Figure.1.1} (the initial state is $a_0$ and the final states are $a$, $b$ and $c$). 
\end{ex} \epr

The following theorem is one of the cornerstones of finite automaton theory.

\begin{thm}
\label{Theorem.1.5.4}
(S. C. Kleene 1956)
$\Rec = \Reg$. \epr
\end{thm}

The theorem is effective in the following sense. There are algorithms to construct a recognizer for any regular language given by a regular expression. Conversely, a regular expression representing $L(\bA)$ can be found for any given recognizer $\bA$.

\begin{figure}[h]
\input{Figure.1.1}
\caption{\label{Figure.1.1}}
\end{figure}

Kleene's theorem implies also that the family $\Rec$ is closed under the regular operations. We shall present some more closure properties of the family $\Rec$.

\begin{thm}
\label{Theorem.1.5.5}
Let $X$ and $Y$ be arbitrary alphabets.
\begin{itemize}
\item[\emph{(a)}] If $U,\,V \in \Rec~X$, then $U \cap V,~U-V \in \Rec~X$.
\item[\emph{(b)}] If $U$ is a recognizable $X$-language, then so is its \emph{mirror image}\index{mirror image} (or \emph{reversal})
\begin{equation*}
\textrm{mi}(U) = \{x_n \ldots x_2 x_1 \mid n \geq 0,\; x_1 x_2 \ldots x_n \in U\; (x_i \in X)\}.
\end{equation*}
\item[\emph{(c)}] If $U$ and $V$ are recognizable $X$-languages, then so are the \emph{quotient languages}\index{language!quotient}
\begin{equation*}
U^{-1}V = \{w \in X^* \mid uw = v \textrm{ for some } u \in U, v \in V\}
\end{equation*}
and
\begin{equation*}
UV^{-1} = \{w \in X^* \mid wv = u \textrm{ for some } u \in U, v \in V\}.
\end{equation*}
\item[\emph{(d)}] Let $\varphi \colon X^* \rightarrow Y^*$ be a homomorphism (of monoids). If $U \in \Rec~X$, then $U\varphi \in \Rec~Y$. If $V \in \Rec~Y$, then $V\varphi^{-1}\in \Rec~X$.
\item[\emph{(e)}] If $U \in \Rec~X$ and $\varphi \colon \gp X^* \rightarrow \gp Y^*$ is such a \emph{substitution mapping}\index{mapping!substitution} that $x\varphi \in \Rec~Y$ for all $x \in X$, then $U\varphi \in \Rec~Y$. \epr
\end{itemize}
\end{thm}

Recall that a mapping $\varphi \colon \gp X^* \rightarrow \gp Y^*$ is a substitution, if
\begin{equation*}
\begin{array}{ll}
1^\circ & \{e\}\varphi = \{e\},\\
2^\circ & \{wx\}\varphi = (w\varphi)(x\varphi) \textrm{ for all } w \in X^*,~x \in X, \textrm{ and }\\
3^\circ & U\varphi = \bigcup (u\varphi \mid u \in U) \textrm{ for all } U \subseteq X^*.
\end{array}
\end{equation*}
%
Obviously, the substitution is completely defined when the languages $x \varphi$ $(x \in X)$ are given. Extended to mappings of languages, homomorphisms $\varphi \colon X^* \rightarrow Y^*$ are special substitutions for which every $x\varphi$ $(x \in X)$ consists of exactly one word.

Often it is convenient to allow a recognizer to be nondeterministic. In a \emph{\text{nondeterministic} $X$-recognizer}\index{recognizer!nondeterministic} $\bA = (A, X, \delta, A_0, A')$ the next-state function is a mapping
\begin{equation*}
\delta \colon A \times X \rightarrow \gp A.
\end{equation*}
Also, the recognizer has a set $A_0 \subseteq A$ of initial states. If $\bA$ receives in state $a$ the input letter $x$, then it may enter any one of the states in $\delta(a , x)$. The operation of $\bA$ may be started in any initial state $a_0 \in A_0$. A word $w = x_1 x_2 \ldots x_n~(n \geq 0, x_1, \ldots, x_n \in X)$ is accepted by $\bA$ if there is such a choice of states $a_0, a_1, \ldots, a_n$ that
\begin{enumerate}[(i)]
\item $a_0 \in A_0$,
\item $a_i \in \delta(a_{i-1},x_i)$ for all $i = 1, \ldots, n$, and
\item $a_n \in A'$.
\end{enumerate}

The mapping $\delta$ extends to a mapping
\begin{equation*}
\hat{\delta} \colon\gp A \times X^* \rightarrow \gp A
\end{equation*}
as follows:
\begin{itemize}
\item[$1^\circ$] $\hat{\delta}(H,e)=H$ for all $H \subseteq A$, and
\item[$2^\circ$] $\hat{\delta}(H,wx)=\bigcup(\delta(a,x)\mid a \in \hat{\delta}(H,w))$ for all $H\subseteq A, w \in X^*$ and $ x \in X$.
\end{itemize}

Obviously, $\hat{\delta}(H, w)$ is the set of states $\bA$ may reach under the input word $w$ from at least one state in $H$. The \emph{language recognized} by $A$ can now be defined formally as
\begin{equation*}
L(\bA) = \{w \in X^* \mid \hat{\delta} (A_0, w) \cap A' \neq \emptyset\}.
\end{equation*}

Every $X$-recognizer may be interpreted as a nondeterministic $X$-recognizer $\bA$, where $A_0$ and the sets $\delta(a, x)$ all are singletons. On the other hand, every nondeterministic $X$-recognizer $\bA$ may be turned into the equivalent $X$-recognizer
\begin{equation*}
\bB = (\gp A, X, \hat{\delta}, A_0, A''),
\end{equation*}
where $A'' = \{H \in \gp A \mid H \cap A' \neq \emptyset\}$; this is the well-known ``subset construction''\index{subset construction}. Hence, a language can be recognized by a nondeterministic recognizer iff it is recognizable in our original sense of the word.

Now we recall some algebraic characterizations of $\Rec$.

An equivalence relation $\varrho$ on a semigroup $\cS$ is a \emph{right congruence}\index{congruence!right}, if $a \varrho b$ implies $ac \varrho bc$ for all $a, b, c \in S$. Every $X$-recognizer $\bA = (A, X, \delta, a_0, A')$ defines a right congruence $\varrho_{\bA}$ of the free monoid $X^*$ as follows:
\begin{equation*}
u \equiv v~(\varrho_{\bA}) \quad\textrm{iff}\quad \delta (a_0, u) = \delta(a_0, v)\quad(u, v \in X^*).
\end{equation*}
The index of $\varrho_{\bA}$ is at most $|A|$ and
\begin{equation*}
L(\bA) = \bigcup (u \varrho_{\bA} \mid u \in X^*, \delta(a_0, u) \in A').
\end{equation*}
This shows that every recognizable $X$-language is saturated by a right congruence of $X^*$ of finite index.

Suppose now that the $X$-language $L$ is saturated by a right congruence $\varrho$ of $X^*$ of finite index. The $X$-recognizer
\begin{equation*}
\bA = (X^*/\varrho, X, \delta, e \varrho, L/\varrho),
\end{equation*}
where $\delta$ is defined by the condition
\begin{equation*}
\delta(u\varrho, x) = (ux)\varrho \quad (u \in X^*, x \in X),
\end{equation*}
is then well-defined and
\begin{equation*}
\delta(e\varrho, u) = u\varrho
\end{equation*}
for each $u \in X^*$. This implies $L(\bA) = L \in \Rec~X$. Among all right congruences of $X^*$ saturating a given $X$-language there is a greatest one which is called the \emph{Nerode congruence}\index{Nerode congruence of!language} of $L$. We denote it by $\varrho_L$ and it can be defined by the condition that
\begin{equation*}
u \equiv v~(\varrho_L) \quad\textrm{iff}\quad (\forall w \in X^*)~(uw \in L \Leftrightarrow vw \in L)
\end{equation*}
for all $u, v \in X^*$. From these observations it is easy to construct a proof for the following theorem.

\begin{thm}
\label{Theorem.1.5.6}
(A. Nerode 1957). For any $X$-language $L$ the following three conditions are equivalent:
\begin{itemize}
\item[\emph{(1)}] $L \in  \Rec~X$.
\item[\emph{(2)}] $L$ is saturated by a right congruence of $X^*$ of finite index.
\item[\emph{(3)}] The Nerode congruence $\varrho_L$ is of finite index. \epr
\end{itemize}
\end{thm}

There is a similar characterization which uses congruences of $X^*$. Every $X$-recognizer $\bA$ defines a congruence $\theta_{\bA}$ of $X^*$ of finite index which saturates $L(\bA)$:
\begin{equation*}
u \equiv v~(\theta_{\bA}) \quad\textrm{iff}\quad (\forall a \in A)~\delta(a, u) = \delta(a, v).
\end{equation*}

If $L \subseteq X^*$ is saturated by a congruence, then a recognizer for $L$ can be constructed as above in the case of right congruences. The greatest congruence $\theta_L$ saturating $L$ is called the \emph{syntactic congruence}\index{congruence!syntactic} of $L$. It may be defined by the condition that
\begin{equation*}
u \equiv v~(\theta_L) \quad\textrm{iff}\quad (\forall w, w' \in X^*)~(wuw' \in L \Leftrightarrow wvw' \in L)
\end{equation*}
for all $u, v \in  X^*$.

\begin{thm}
\label{Theorem.1.5.7}
(J. R. Myhill 1957). For every $X$-language $L$ the following three conditions are equivalent:
\begin{itemize}
\item[\emph{(1)}] $L \in \Rec~X$.
\item[\emph{(2)}] $L$ is saturated by a congruence of $X^*$ of finite index.
\item[\emph{(3)}] The syntactic congruence $\theta_L$ is of finite index.  \epr
\end{itemize}
\end{thm}

Let $\theta$ be a congruence of $X^*$ saturating an $X$-language $L$. Then $L = (L \theta^\natural){\theta^\natural}^{-1}$, where
\begin{equation*}
\theta^\natural \colon X^* \rightarrow X^* / \theta
\end{equation*}
is the canonical homomorphism, and $X^*/\theta$ is finite iff $\theta$ is of finite index. This applies, in particular, to the syntactic congruence $\theta_L$. The monoid $X^*/\theta_L$ is called the \emph{syntactic monoid}\index{monoid!syntactic} of $L$. On the other hand, if we have a finite monoid $M$, a homomorphism
\begin{equation*}
\varphi \colon X^* \rightarrow M
\end{equation*}
and a subset $H \subseteq M$ for which $L= H \varphi^{-1}$, then $\varphi\varphi^{-1}$ is a congruence of $X^*$ of finite index saturating $L$. It is now clear that Myhill's theorem can be reformulated as follows.

\begin{thm}
\label{Theorem.1.5.8}
For any $X$-language $L$ the following three conditions are equivalent:
\begin{itemize}
\item[\emph{(1)}] $L \in \Rec~X$.
\item[\emph{(2)}] There exist a finite monoid $M$, a homomorphism $\varphi \colon X^* \rightarrow M$ and a subset $H \subseteq M$ such that $L = H \varphi^{-1}$.
\item[\emph{(3)}] The syntactic monoid of $L$ is finite. \epr
\end{itemize}
\end{thm}

An $X$-language $L$ is called \emph{local}\index{language!local}, if there exist sets $H, K \subseteq X$ and $I \subseteq X^2$ such that
\begin{equation*}
L-\{e\} = (HX^* \cap X^*K) - X^*IX^*.
\end{equation*}
The membership of a nonempty word $w$ in such an $L$ can be tested by checking that the first letter of $w$ is in $H$, the last letter of $w$ is in $K$, and that no two consecutive letters of $w$ form a pair belonging to $I$. Note that a local language may, according to our definition, contain the empty word.

A homomorphism $\varphi \colon X^* \rightarrow Y^*$ is called \emph{length-preserving}\index{homomorphism!length-preserving} if $|w\varphi| = |w|$ for all $w \in X^*$. Obviously $\varphi$ is length-preserving iff $X\varphi \subseteq Y$.

In terms of these concepts one more characterization of $\Rec$ can be given.

\begin{thm}
\label{Theorem.1.5.9}
An $X$-language $L$ is recognizable iff $L=U\varphi$ for some alphabet $Y$, local $Y$-language $U$ and length-preserving morphism $\varphi \colon Y^* \rightarrow X^*$. \epr
\end{thm}

An $X$-recognizer $\bA$ is said to be \emph{minimal}\index{recognizer!minimal}, if no $X$-recognizer with fewer states recognizes $L(\bA)$. It is obvious that every regular language has a minimal recognizer. To say more than that, we need a few concepts.

Let $\bA = (A, X, \delta, a_0, A')$ be an $X$-recognizer. It is said to be \emph{connected}\index{recognizer!connected}, if there exists for every $a \in A$ a word $w \in X^*$ such that $a=\delta(a_0, w)$. Two states $a$ and $b$ of $\bA$ are said to be \emph{equivalent}\index{equivalence of states in!recognizer}, and we write $a \sim b$, if
\begin{equation*}
(\forall w \in X^*) \, (\delta(a, w) \in A' \Longleftrightarrow \delta(b, w) \in A').
\end{equation*}
The recognizer $\bA$ is \emph{reduced}\index{recognizer!reduced}, if $a \sim b$ implies $a = b$.

A relation $\theta \in E(A)$ is a \emph{congruence}\index{congruence!of recognizer} of $\bA$, if
\begin{enumerate}[(1)]
\item $a \theta b$ implies $\delta(a,x) \theta \delta(b,x)$ for all $a,b \in A$ and $x \in X$, and
\item $\theta$ saturates $A'$.
\end{enumerate}
Let $C(\bA)$ be the set of all congruences of $\bA$. It is not hard to prove that $\sim$ is a congruence of $\bA$. In fact, it is the greatest congruence of $\bA$.

If $\theta \in C(\bA)$, then one can define a \emph{quotient recognizer}\index{recognizer!quotient}
\begin{equation*}
\bA/\theta = (A/\theta, X, \delta', a_0 \theta, A'/\theta)
\end{equation*}
by putting
\begin{equation*}
\delta'(a \theta, x) = \delta(a,x)\theta \quad \textrm{for all } a \in A \textrm{ and } x \in X.
\end{equation*}
The congruence property (1) guarantees that $\delta'$ is well-defined. An easy induction on $|w|$ shows that
\begin{equation*}
\delta'(a \theta,  w) = \delta(a, w) \theta \quad \textrm{ for all } a \in A  \textrm{ and } w \in  X^*.
\end{equation*}
This implies $L(\bA/\theta)= L(\bA)$. In particular, $L(\bA/\hspace{-.6ex}\sim) = L(\bA)$. It is now obvious that a minimal recognizer should be reduced and, of course, connected.

Let $\bA = (A, X, \delta, a_0, A')$ and $\bB = (B, X, \eta, b_0, B')$ be two $X$-recognizers. A \emph{homomorphism}\index{homomorphism!of recognizer} $\varphi \colon \bA \rightarrow \bB$ is a mapping $\varphi \colon A \rightarrow B$ such that
\begin{enumerate}[(1)]
\item $\delta(a, x)\varphi = \eta(a\varphi, x)$ for all $a \in A$ and $x \in X$,
\item $a_0\varphi = b_0$, and
\item $B' \varphi^{-1} = A'$.
\end{enumerate}
\emph{Epimorphisms}\index{epimorphism!of recognizer} and \emph{isomorphisms}\index{isomorphism of!recognizers} of $X$-recognizers are, respectively, surjective and bijective homomorphisms.

Homomorphisms, congruences and quotients of $X$-recognizers are related to each other the same way as the corresponding concepts in algebra. Hence, for any $\theta \in C(\bA)$, the natural mapping $\theta^\natural$ is an epimorphism $\bA \rightarrow \bA/\theta$. If $\varphi \colon \bA \rightarrow \bB$ is an epimorphism, then $\varphi\varphi^{-1}$ is a congruence of $\bA$ and $\bA/\varphi\varphi^{-1}$ is isomorphic to $\bB$. Moreover,
\begin{equation*}
\delta(a, w)\varphi = \eta(a\varphi, w) \quad \textrm{for all } a \in A, w \in X^*.
\end{equation*}
This implies $L(\bA) = L(\bB)$.

The $X$-recognizer $\bB$ is a \emph{subrecognizer}\index{subrecognizer} of $\bA$ if $B \subseteq A$, $b_0 = a_0$, $B' = A' \cap B$ and $\eta = \delta | B \times X$. The subset $B$ determines such a subrecognizer completely. The \emph{connected part}\index{connected part of recognizer}
\begin{equation*}
A_c = \{\delta(a_0,w) \mid w \in X^*\}
\end{equation*}
of an $X$-recognizer is the state set of a subrecognizer
\begin{equation*}
\bA_c = (A_c, X, \delta_c, a_0, A' \cap A_c)
\end{equation*}
where $\delta_c = \delta | A_c \times X$.

The following theorem summarizes the main facts concerning minimal and reduced recognizers.

\begin{thm}
\label{Theorem.1.5.10}
\begin{itemize}
\item[\emph{(a)}] The minimal recognizer of a regular language is unique up to isomorphism, i.e., if two recognizers are minimal and equivalent to each other, then they are isomorphic.
\item[\emph{(b)}] A recognizer is minimal iff it is connected and reduced.
\item[\emph{(c)}] For any recognizer $\bA$, the quotient $\bA/\hspace{-.6ex}\sim$ is reduced and its connected part $(\bA/\hspace{-.6ex}\sim)_c$ is minimal. The recognizer $\bA_c/\hspace{-.6ex}\sim$ is isomorphic to $(\bA/\hspace{-.6ex}\sim)_c$.
\item[\emph{(d)}] If $\bA$ is minimal, $\bB$ is connected and $L(\bA)= L(\bB)$, then there exists a unique epimorphism $\varphi \colon \bB \rightarrow \bA$. \epr
\end{itemize}
\end{thm}

Theorem~\ref{Theorem.1.5.10} implies that one can find a minimal recognizer for a regular language $L$ by starting with any recognizer $\bA$ of $L$; first one finds the connected part $A_c$ and then one has to determine the equivalent pairs of states in $\bA_c$. For both tasks there are simple algorithms. The order may also be reversed; first form $\bA/\hspace{-.6ex}\sim$ and then find the connected part of this reduced recognizer.

The decidability of the emptiness, finiteness and equality questions for regular languages follows from the following simple observation.

\begin{lm}
\label{Lemma.1.5.11}
Let $\bA$ be an $X$-recognizer with $n$ states.
\begin{itemize}
\item[\emph{(a)}] If $L(\bA)$ contains a word $w$ of length $\geq n$, then one may write $w=uvz$ so that $0<|v|\leq n$ and $uv^kz \in L(\bA)$ for all $k \geq 0$.
\item[\emph{(b)}] $L(\bA)$ is nonempty iff it contains a word of length $<n$.
\item[\emph{(c)}] $L(\bA)$ is infinite iff it contains a word $w$ such that $n \leq |w| < 2n$. \epr
\end{itemize}
\end{lm}
Statement (a) is often referred to as the ``pumping lemma'' for finite recognizers.

To test whether $L(\bA)$ is nonempty it suffices to try all input words of length $<|A|$. Similarly, the finiteness of $L(\bA)$ can be checked by applying all input words $w$ such that $|A|\leq |w| < 2|A|$. From any two $X$-recognizers $\bA$ and $\bB$ one can construct a recognizer for $(L(\bA) - L(\bB))\cup(L(\bB) - L(\bA))$. But this language is empty exactly in case $L(\bA) = L(\bB)$. Hence, the equivalence of $\bA$ and $\bB$ can also be decided.

%% file: Figure.1.1.tex
\centering
\begin{tikzpicture}[node/.style={inner sep=0pt, draw, circle, minimum size=0.65cm, text width=0.2cm, align=center},
every path/.style={->,>=latex},node distance = 1.5cm,
every label/.style={scale=1}]
	\coordinate (initial) {};
	\node[node,right of = initial,node distance=0.6cm,label={[shift={(0.05cm,-0.58cm)}]\small $a_0$}] (q0) {};
	\node[node,below right of = q0] (q2) {\small $c$};
	\node[node,above right of = q0,label={above left:$x$}] (q1) {\small $a$};
	\node[node,right of = q1,label={above left:$y$}] (q3) {\small $b$};
	\node[node,right of = q2] (q4) {\small $d$};
	\node[node,above right of = q4,label={above right:$xy$}] (q5) {\small $q$};
	\coordinate[below of = q1,node distance=0.6cm] (f1) {};
	\coordinate[below of = q2,node distance=0.6cm] (f2) {};
	\coordinate[below of = q3,node distance=0.6cm] (f3) {};
	\path (initial) edge (q0);
	\path (q0) edge node[above,pos=0.3] {$x$} (q1);
	\path (q0) edge node[above,pos=0.6] {$y$} (q2);
	\path (q1) edge node[above=-0.1cm,pos=0.3] {$y$} (q3);
	\path (q2) edge node[above=-0.1cm,pos=0.4] {$x$} (q4);
	\path (q3) edge node[above,pos=0.6] {$x$} (q5);
	\path (q4) edge node[below,pos=0.6] {$y$} (q5);
	
	\path (q1) edge (f1);
	\path (q2) edge (f2);
	\path (q3) edge (f3);
	
	
	\path (q4) edge [bend left] node[below=-0.08cm,pos=0.25] {$x$} (q2);
	\path (q2) edge [bend left=20] node[above,pos=0.25] {$y$} (q5);
	
	\draw (q1) ++ (-0.25cm,0.3cm) arc(185:-5:0.25cm)-- ++(-85:1pt); 
	\draw (q3) ++ (-0.25cm,0.3cm) arc(185:-5:0.25cm)-- ++(-85:1pt);
	\draw (q5) ++ (0.25cm,0.3cm) arc(105:-85:0.25cm)-- ++(190:1pt);
\end{tikzpicture}

%% file: Section.1.6.tex
We shall now consider the most important tools of formal language theory, Chomsky's grammars. A grammar is a device to define a language by showing how to generate the strings of the language. The concept is very flexible, and by imposing various restrictions on grammars several interesting families of languages can be obtained. A good example is provided by the celebrated \emph{Chomsky hierarchy}\index{Chomsky hierarchy} consisting of four families of languages. At the bottom of the hierarchy we find, once more, the recognizable languages. However, most of this section will be devoted to context-free languages. These form the second step in the hierarchy.

\begin{df}
\label{Definition.1.6.1}
\rm
A \emph{grammar}\index{grammar} is a 4-tuple $(N, X, P, a_0)$, where
\begin{enumerate}[(1)]
\item $N$ is a finite nonempty set of \emph{nonterminal symbols}\index{nonterminal symbol of!grammar},
\item $X$ is the \emph{terminal alphabet}\index{alphabet!terminal},
\item $P$ is the finite set of \emph{productions}\index{production of!grammar}, and
\item $a_0 \in N$ is the \emph{initial symbol}\index{initial symbol of!grammar}.
\end{enumerate}
\end{df}
It is required that $N \cap X = \emptyset$. Every production is of the form $\beta \rightarrow \gamma$, where $\beta, \gamma \in (N \cup X)^*$ and $\beta$ contains at least one nonterminal symbol.

Let $G=(N, X, P, a_0)$ be a grammar. For $u,v \in (N \cup X)^*$ we write $u \Rightarrow_G v$ (or just $u \Rightarrow v$, when $G$ is understood) if there exist $u', u'' \in (N \cup X)^*$ and a production $\beta \rightarrow \gamma \in P$ so that $u= u' \beta u''$ and $v= u' \gamma u''$. If $u \Rightarrow_G v$, then $u$ is said to \emph{generate $v$ directly}\index{direct generation in grammar} in $G$. If there exists a \emph{derivation}\index{derivation in!grammar}
\begin{equation*}
u_0 \Rightarrow_G u_1 \Rightarrow_G u_2 \Rightarrow_G \ldots \Rightarrow_G u_n \quad (n \geq 0)
\end{equation*}
such that $u_0 = u$ and $u_n = v$, then we write $u \Rightarrow^*_G v$ (or just $u \Rightarrow^* v$). The \emph{language generated}\index{language!generated by grammar} by $G$ is the $X$-language
\begin{equation*}
L(G) = \{w \in X^* \mid a_0 \Rightarrow^*_G w\}.
\end{equation*}
Two grammars are \emph{equivalent}\index{equivalence of!grammars}, if they generate the same language.

The grammars of Definition~\ref{Definition.1.6.1} are very general and every recursively enumerable language can be generated by such a grammar.

\begin{df}
\label{Definition.1.6.2}
\rm
A grammar $(N, X, P, a_0)$ is called \emph{right linear}\index{grammar!right linear}, if each production is of the form
\begin{equation*}
a \rightarrow xb, \quad a \rightarrow x \quad \textrm{or} \quad  a \rightarrow e,
\end{equation*}
where $a, b \in N$ and $x \in X$. A language is \emph{right linear}\index{language!right linear}, or of \emph{type 3}\index{language!of type} (in the Chomsky hierarchy), if it can be generated by a right linear grammar.
\end{df}

A right linear grammar $G=(N, X, P, a_0)$ can be converted into a nondeterministic $X$-recognizer
\begin{equation*}
\bA = (N \cup \{c\}, X, \delta, \{a_0\}, A') \quad (c \not \in N)
\end{equation*}
which recognizes $L(G)$ as follows. For any $a, b \in N$ and $x \in X$, put
\begin{enumerate}[(i)]
\item $b \in \delta(a, x)$ iff $a \rightarrow xb \in P$,
\item $c \in \delta(a, x)$ iff $a \rightarrow x \in P$, and
\item $\delta(c, x) = \emptyset$.
\end{enumerate}
Finally, let $A' = \{c\} \cup \{a \in N \mid a \rightarrow e \in P\}$. Conversely, every $X$-recognizer $\bA = (A, X, \delta, a_0, A')$ can be replaced by the right linear grammar $G=(A, X, P, a_0)$, where
\begin{equation*}
P = \{a \rightarrow xb \mid \delta(a, x) = b\} \cup \{a \rightarrow e \mid a \in A'\}.
\end{equation*}
These observations lead to one more characterization of $\Rec$:

\begin{thm}
\label{Theorem.1.6.3}
The type 3 languages are exactly the regular languages.  \epr
\end{thm}

Now we proceed to the main topic of this section.

\begin{df}
\label{Definition.1.6.4}
\rm
A grammar $(N, X, P, a_0)$ is \emph{context-free}\index{grammar!context-free} (CF\index{grammar!CF}, for short) if each production is of the form
\begin{equation*}
a \rightarrow \gamma
\end{equation*}
where $a \in N$ and $\gamma \in (N \cup X)^*$. A language is \emph{context-free}\index{language!context-free} (CF)\index{language!CF} if it is generated by a CF grammar. The family of all CF languages is denoted by $\CF$ and the set of CF $X$-languages by $\CF(X)$.
\end{df}

The CF languages are the type 2 languages in Chomsky's hierarchy. Every right linear grammar is CF. Hence $\Rec \subseteq \CF$. If $|X|=1$, then $\Rec\,X = \CF(X)$, but in all other cases the inclusion is proper.

\begin{ex}
\label{Example.1.6.5}
\rm
Suppose $X$ contains two distinct letters $x$ and $y$. Every derivation in the CF grammar
\begin{equation*}
G = (\{a\}, X, \{a \rightarrow xay, a \rightarrow xy\}, a)
\end{equation*}
is of the form
\begin{equation*}
a \Rightarrow xay \Rightarrow xxayy \Rightarrow \ldots \Rightarrow x^{n-1}ay^{n-1} \Rightarrow x^n y^n \quad (n \geq 1).
\end{equation*}
Hence, $L(G)$ is the nonregular language $\{x^n y^n \mid n \geq 1\}$. \epr
\end{ex}

The main fact to connect CF languages with tree automata is that context-free derivations can be represented by \emph{derivation trees}\index{tree!derivation}. A derivation tree is a description of the syntax of a word of the CF language. (Here it would be more natural
to speak about ``sentences'' of a language.) Derivation trees have proved very useful tools in the theory of CF languages. Later we shall define ``trees'' in a way suitable for our purposes, but here there is no need to define the concept too formally.

Let $G=(N, X, P, a_0)$ be a CF grammar. The derivation tree representing a derivation of a word $u \in (X \cup N)^*$ from a symbol $a \in (X \cup N)$ in $G$ is defined by induction on the number $k$ of steps in the derivation:

\begin{itemize}
\item[$1^\circ\,$] If $k=0$, then $u=a$ and the derivation tree consists of a single node labelled by $a$.
\item[$2^\circ\,$] Consider a derivation
\begin{equation*}
\tag{*}
a \Rightarrow u_1 \Rightarrow u_2 \Rightarrow \ldots \Rightarrow u_{k-1} \Rightarrow u
\end{equation*}
where $k \geq 1$. Suppose $u_1 = d_1 \ldots d_m$, where $m \geq 0$ and $d_1, \ldots, d_m \in N \cup X$. At this point the context-freeness of $G$ becomes essential. Every application of a production in (*) rewrites exactly one $d_i$ or a nonterminal derived from exactly one $d_i$. This means that (*) may be decomposed into a number of ``subderivations''
\begin{equation*}
d_i \Rightarrow \ldots \Rightarrow v_i \quad (i = 1,\ldots,m)
\end{equation*}
each of which yields a segment $v_i$ of $u$ and $u= v_1 v_2 \ldots v_m$. If the derivation trees of the subderivations are $t_1, \ldots, t_m$, respectively, then the derivation tree of (*) is that shown in Fig.~\ref{Figure.1.2}.

The possibility $m=0$ was not excluded. Then $k=1$, $u=e$ and the derivation tree reduces to a single node labelled by $a$.
\end{itemize}

The word $xxxyyy$ has the derivation
\begin{equation*}
a \Rightarrow xay \Rightarrow xxayy \Rightarrow xxxyyy
\end{equation*}
in the grammar of Example~\ref{Example.1.6.5}. The corresponding derivation tree is shown in Fig.~\ref{Figure.1.3}.

Consider any derivation
\begin{equation*}
a_0 \Rightarrow \ldots \Rightarrow w
\end{equation*}
\begin{figure}[h]
\input{Figure.1.2}
\caption{\label{Figure.1.2}}
\end{figure}
\begin{figure}[h]
\input{Figure.1.3}
\caption{\label{Figure.1.3}}
\end{figure}
of a terminal word $w \in L(G)$ from the initial symbol. The corresponding derivation tree is also called a derivation tree of $w$, and $w$ can be read from the ``leaves'' of the tree.

The grammar $G$ of Example~\ref{Example.1.6.5} has the rather special property that every word in $L(G)$ has just one derivation in $G$.

\begin{ex}
\label{Example.1.6.6}
\rm
Consider the CF grammar
\begin{equation*}
G = (\{a_0, a, b\}, \{x,y\}, P, a_0)
\end{equation*}
where $P$ consists of the productions
\begin{equation*}
a_0 \rightarrow ab,\;\; a \rightarrow xay,\;\; a \rightarrow xy,\;\;  b \rightarrow ybx \;\textrm{ and }\; b \rightarrow yx.
\end{equation*}
Obviously, $L(G)= \{x^m y^{m+n} x^n \mid m, n \geq 1\}$. The word $xyyx \in L(G)$ has the two derivations
\begin{equation*}
a_0 \Rightarrow ab \Rightarrow xyb \Rightarrow xyyx
\end{equation*}
and
\begin{equation*}
a_0 \Rightarrow ab \Rightarrow ayx \Rightarrow xyyx
\end{equation*}
both of which are represented by the derivation tree shown in Fig.~\ref{Figure.1.4}. In general, the word $x^m y^{m+n} x^n$ has $\binom{m+n}{n}$ different derivations all of which are represented by the same derivation tree. \epr
\end{ex}

\begin{figure}[h]
\input{Figure.1.4}
\caption{\label{Figure.1.4}}
\end{figure}

In Example~\ref{Example.1.6.6} the different derivations of the same word do not represent different syntactic descriptions of the word. In fact, they can all be obtained from each other by changing the order in which the individual steps are carried out. If we agree on some fixed order in which the subderivations are to be carried out, then there would be just one derivation for each derivation tree of a word in the language.

\begin{df}
\label{Definition.1.6.7}
\rm
 A derivation
\begin{equation*}
u_0 \Rightarrow u_1 \Rightarrow u_2 \Rightarrow \ldots \Rightarrow u_k
\end{equation*}
in a CF grammar $G = (N, X, P, a_0)$ is called a \emph{leftmost derivation}\index{leftmost derivation}, if we can write, for every $i= 0,\ldots, k-1$,
\begin{equation*}
u_i = w_iau'_i \quad \textrm{and} \quad u_{i+1} = w_i \gamma u'_i
\end{equation*}
 so that $w_i \in X^*$, $a \in N$ and $a \rightarrow \gamma \in P$. The grammar $G$ is \emph{ambiguous}\index{grammar!ambiguous} if some word $w$ in $L(G)$ has two different leftmost derivations from $a_0$. Otherwise $G$ is \emph{unambiguous}\index{grammar!unambiguous}. A CF language generated by at least one unambiguous CF grammar is said to be \emph{unambiguous}\index{language!unambiguous CF}. If all CF grammars generating a given CF language are ambiguous, then the language is said to be \emph{inherently ambiguous}\index{language!inherently ambiguous CF}.
\end{df}

A CF grammar $G$ is unambiguous if every word $w \in L(G)$ has exactly one derivation tree. It is ambiguous, if at least one word $w \in L(G)$ has more than one derivation tree. The grammars of Examples~\ref{Example.1.6.5} and~\ref{Example.1.6.6} are unambiguous. Every regular language is unambiguous. Of course, a language generated by an ambiguous CF grammar may be unambiguous. The language
\begin{equation*}
\{x^iy^jz^k \mid i = j \textrm{ or } j = k \quad (i, j, k \geq 1)\}
\end{equation*}
is a well-known example of an inherently ambiguous language.

There are many simplifying additional conditions that a CF grammar may always be assumed to satisfy. Some of these are listed below.

\begin{df}
\label{Definition.1.6.8}
\rm
Let $G=(N, X, P, a_0)$ be a CF grammar.
\begin{enumerate}[(a)]
\item $G$ is \emph{reduced}\index{grammar!reduced CF} if either $P=\emptyset$ and $N=\{a_0\}$, or then for every $a \in N$,
\begin{equation*}
a_0 \Rightarrow^* uav \Rightarrow^* w
\end{equation*}
for some $u,v \in (N \cup X)^*$ and $w \in X^*$.
\item $G$ is in \emph{Chomsky normal form}\index{normal form of CF grammar!Chomsky} if each production is of the form
\begin{enumerate}[(i)]
 \item $a \rightarrow bc \quad (a \in N,\, b,c \in  N-a_0)$,
 \item $a \rightarrow x \quad (a \in N,\, x \in X)$, or
 \item $a_0 \rightarrow e$.
\end{enumerate}
\item $G$ is in \emph{Greibach normal form}\index{normal form of CF grammar!Greibach} if each production is of the form
\begin{enumerate}[(i)]
 \item $a \rightarrow xa_1 \ldots a_m \quad (m \geq 0,\, a \in N,\, a_1, \ldots, a_m \in N-a_0,\, x \in X)$, or
 \item $a_0 \rightarrow e$.
\end{enumerate}
\end{enumerate}
If $m \leq k$ for all productions of type (i), then $G$ is said to be in \emph{Greibach $k$-form}\index{Greibach $k$-form} $(k \geq 0)$.
\end{df}

Proofs for the following facts can be found in the references given at the end of the section.
\begin{thm}
\label{Theorem.1.6.9}
\begin{enumerate}
\item[{\rm (a)}] Every CF grammar $(N, X, P, a_0)$ can be converted into an equivalent reduced CF grammar $(N', X, P', a_0)$, where $N' \subseteq N$ and $P' \subseteq P$.
\item[{\rm (b)}] Every CF grammar can be converted into an equivalent CF grammar in any one of the following normal forms: Chomsky normal form, Greibach normal form, and Greibach 2-form. In all cases the grammar can be made reduced. \epr
\end{enumerate}
\end{thm}

We recall now some of the closure properties of the family $\CF$.

\begin{thm}
\label{Theorem.1.6.10} If the languages $U$ and $V$ are CF, then so are $U \cup V$, $UV$ and $U^*$. \epr
\end{thm}

The languages $U = \{x^m y^n z^n \mid m, n \geq 1\}$ and $V=\{x^ny^nz^m \mid m, n \geq 1\}$ are CF, but $U \cap V = \{x^ny^nz^n \mid n \geq 1\}$ is not. This observation implies also that the difference $U-V$ of two CF languages $U$ and $V$ may be noncontext-free. However, the following theorem holds.

\begin{thm}
\label{Theorem.1.6.11} If $U$ is a CF language and $V$ is a regular language, then $U \cap V$ and $U - V$ are CF languages.  \epr
\end{thm}

The following theorem implies, as a special case, that CF is closed under morphisms.

\begin{lm}
\label{Lemma.1.6.12} Let $\varphi \colon \gp X^* \rightarrow \gp Y^*$  be a substitution mapping such that $x\varphi \in \CF (Y)$ for all $x \in X$. If $U \in \CF(X)$, then $U\varphi \in \CF(Y)$. \epr
\end{lm}

The following useful lemma is obtained most naturally by considering derivation trees.

\begin{lm}
\label{Lemma.1.6.13} (Bar-Hillel's pumping lemma). For each CF grammar $G$ one can find two natural numbers $p$ and $q$ such that the following holds for every word $w \in L(G)$: if $|w|>p$, then we may write $w=u_1 v_1 w' v_2 u_2$ so that
\begin{enumerate}
\item[{\rm (i)}] $|v_1w'v_2| \leq q$,
\item[{\rm (ii)}] $v_1 v_2 \not = e$, and
\item[{\rm (iii)}] $u_1 v^i_1 w' v^i_2 u_2 \in L(G)$ for every $i \geq 0$. \epr
\end{enumerate}
\end{lm}

Next we recall some decidability properties of CF languages. A CF language is always assumed to be given by a CF grammar generating it.

\begin{thm}
\label{Theorem.1.6.14} There are algorithms for deciding the following questions:
\begin{enumerate}
\item[{\rm (1)}] Is a given word in a given $\CF$ language?
\item[{\rm (2)}] Is a given $\CF$ language empty?
\item[{\rm (3)}] Is a given $\CF$ language finite? \epr
\end{enumerate}
\end{thm}

The decidability of the finiteness problem follows from Bar-Hillel's lemma. The other two statements can be justified quite directly.

\begin{thm}
\label{Theorem.1.6.15} The following questions are undecidable:
\begin{enumerate}
\item[{\rm (a)}] Are two given $\CF$ languages equal?
\item[{\rm (b)}] Is the intersection of two given $\CF$ languages empty? $|$ finite? $|$ regular? $|$ context-free?
\item[{\rm (c)}] Is the complement $X^* - U$ of a $\CF$ $X$-language $U$ empty? $|$ finite? $|$ regular? $|$ context-free?
\item[{\rm (d)}] Is a given $\CF$ grammar ambiguous?
\item[{\rm (e)}] Is a given $\CF$ language inherently ambiguous? \epr
\end{enumerate}
\end{thm}

In the previous section we noted that every regular language has a minimal recognizer. One might want to find a CF grammar equivalent to a given one with the smallest possible number of nonterminals (\emph{nonterminal minimization problem}\index{problem!nonterminal minimization}) or with a minimum number of productions (\emph{production minimization problem}\index{problem!production minimization}). However, the following theorem holds.

\begin{thm}
\label{Theorem.1.6.16} Both the nonterminal minimization problem and the production minimization problem are unsolvable. \epr
\end{thm}

Let $n$ be a fixed natural number. The sum of two $n$-tuples of nonnegative integers
\begin{equation*}
\ba = (a_1, \ldots, a_n) \textrm{ and } \bb = (b_1, \ldots, b_n)
\end{equation*}
is formed componentwise:
\begin{equation*}
\ba + \bb = (a_1 + b_1, \ldots, a_n + b_n).
\end{equation*}
Similarly, we put
\begin{equation*}
k\ba = (ka_1, \ldots, ka_n)
\end{equation*}
for all $k \in \bN_{\mathbf 0}$ and $\ba \in \bN^n_{\mathbf 0}$.

A subset $K$ of $\bN^n_{\mathbf 0}$ is called \emph{linear}\index{subset!linear}, if there exist an $m \geq 0$ and $n$-tuples $\ba_1, \ldots, \ba_m$, $\bb \in \bN^n_{\mathbf 0}$ such that
\begin{equation*}
K = \{k_1 \ba_1 + \ldots +  k_m\ba_m + \bb \mid k_1, \ldots, k_m \in \bN_{\mathbf 0}\}.
\end{equation*}
A subset of $\bN^n_{\mathbf 0}$ is \emph{semilinear}\index{subset!semilinear} if it is the union of finitely many linear sets.

Let $X$ be an alphabet with $n$ letters $(n \geq 1)$. It is convenient to think that the letters of $X$ are listed in some fixed order, $x_1, \ldots, x_n$.  The \emph{Parikh vector}\index{Parikh vector} of a word $w \in X^*$ is the $n$-tuple
\begin{equation*}
\Par(w) = (a_1, \ldots, a_n)
\end{equation*}
where $a_i$ is the number of occurrences of $x_i$ in $w$ $(i= 1, \ldots, n)$. The resulting \emph{Parikh mapping}\index{mapping!Parikh}
\begin{equation*}
\Par \colon X^* \rightarrow \bN^n_{\mathbf 0}
\end{equation*}
satisfies the conditions
\begin{equation*}
\textrm{(i)~}\Par(e) = (0, \ldots, 0)\phantom{aaaaaaaaaaaaaaaaaa}
\end{equation*}
and
\begin{equation*}
\textrm{(ii)~} \Par (uv) = \Par(u) + \Par(v) \quad (u,v \in X^*).
\end{equation*}
The mapping $\Par$ is extended to $X$-languages in the natural way:
\begin{equation*}
\Par(L) = \{\Par(w) \mid  w \in L\}
\end{equation*}
for all $L \subseteq X^*$.

\begin{thm}
\label{Theorem.1.6.17} For every $\CF$ language $L$, the Parikh set\index{set!Parikh} $\Par(L)$ is semilinear. \epr
\end{thm}

%% file: Figure.1.2.tex
\centering
\begin{tikzpicture}[sibling distance=1.1cm,level distance=0.8cm,
leaf/.style={regular polygon, regular polygon sides=3,draw, shape border rotate=180, minimum height=2cm, scale=0.8}]
\tikzstyle{level 2}=[level distance=0.81cm] 
\begin{scope}[execute at begin node=$, execute at end node=$]

\node[node,white,label={below:a}] {}
child[white]{ node[node] {}
	child[black]{ node[leaf] {t_m} }}
child[white]{}
child[white]{ node[node] {} 
	child[black]{ node[leaf] {t_1} }};
\node[node] {}
child{ node[node,fill=white,label={[left=2pt]left:d_m}] {} }
child[white]{ node[black,scale=0.8] {\dots} }
child{ node[node,fill=white,label={left:d_1}] {} };
\end{scope}
\end{tikzpicture}

%% file: Figure.1.3.tex
\centering
\begin{tikzpicture}[sibling distance=0.7cm,level distance=0.8cm,leaf/.style={level distance=1.1cm}]
\begin{scope}[execute at begin node=$, execute at end node=$]

\node[node,label={left:a}] {}
child[leaf]{ node[node,label={right:y}] {} }
child{ node[node,label={left:a}] {}
	child[leaf]{ node[node,label={right:y}] {} }
	child { node[node,label={left:a}] {} 
		child[leaf]{ node[node,label={right:y}] {} }
		child[draw opacity=0]{}
		child[leaf]{ node[node,label={left:x}] {} }}
	child[leaf]{ node[node,label={left:x}] {} }}
child[leaf]{ node[node,label={left:x}] {}};
\end{scope}
\end{tikzpicture}

%% file: Figure.1.4.tex
\centering
\begin{tikzpicture}
\tikzstyle{level 1}=[sibling distance=2.6cm,level distance=0.8cm]
\tikzstyle{level 2}=[sibling distance=1.7cm,level distance=1.5cm] 
\begin{scope}[execute at begin node=$, execute at end node=$]

\node[node,label={below:a_0}] {}
child{ node[node,label={left:b}] {} 
	child { node[node,label={left:x}] {} }
	child { node[node,label={left:y}] {} }}
child{ node[node,label={left:a}] {}
	child { node[node,label={left:y}] {} }
	child { node[node,label={left:x}] {} }};
\end{scope}
\end{tikzpicture}

%% file: Section.1.7.tex

Automata that produce outputs in response to inputs are generally called sequential machines. The basic example of these is provided by the Mealy-machine which arose as an abstract model of digital circuits with memory. A \emph{Mealy-machine}\index{machine!Mealy} is a system $\bA = (X, A, Y, a_0, \delta, \lambda)$, where 
\begin{enumerate}[(1)]
\item $X$ is the \emph{input alphabet}\index{input alphabet of!Mealy machine}, 
\item $A$ is a finite, nonempty set of \emph{states}\index{state!of Mealy machine},
\item $Y$ is the \emph{output alphabet}\index{output alphabet of!Mealy machine}, 
\item $a_0 \in A$ is the \emph{initial state}\index{initial state of!Mealy machine}, 
\item $\delta \colon A \times X \rightarrow A$ is the \emph{next-state function}\index{next-state function of!Mealy machine}, and 
\item $\lambda \colon A \times X \rightarrow Y$ is the \emph{output function}\index{function!output}. 
\end{enumerate}
In many applications there is no fixed initial state, and $a_0$ is then omitted from the definition. The operation of $\bA$ can be described as follows. If $\bA$ is in state $a~(\in\hspace{-.5ex}A)$ and receives an input $x~(\in\hspace{-.5ex}X)$, then it enters state $\delta (a, x)$ and emits the letter $\lambda(a, x)$. In order to describe the behaviour of $\bA$ under an arbitrary input word $w \in X^*$ we extend $\delta$ and $\lambda$ to mappings 
\begin{equation*}
\hat{\delta} \colon A \times X^* \rightarrow A, \quad \hat{\lambda} \colon A \times X^* \rightarrow Y^*
\end{equation*}
as follows: 
\begin{itemize}
\item[$1^\circ\,$] $\hat{\delta}(a,e) = a$ and $\hat{\lambda}(a,e) = e$ for every $a \in A$.
\item[$2^\circ\,$] $\hat{\delta}(a,wx) = \delta(\hat{\delta}(a,w),x)$ and $\hat{\lambda}(a,wx) = \hat{\lambda}(a,w)\lambda(\hat{\delta}(a,w),x)$ for all $a \in A$, $w \in X^*$, $x \in X$.
\end{itemize}

If $\bA$ receives in state $a$ the input word $w$, it emits the word $\hat{\lambda} (a, w)~(\in\hspace{-.5ex}Y^*)$ and ends up in state $\hat{\delta}(a,w)$. The \emph{translation}\index{translation}\index{translation!induced by Mealy machine} induced by $\bA$ is defined as the relation 
\begin{equation*}
\tau_{\bA} = \{(w, \hat{\lambda}(a_0,w)) \mid w \in X^*\} \quad (\subseteq X^* \times Y^*). 
\end{equation*}
Two Mealy-machines are said to be \emph{equivalent}\index{equivalence of!Mealy machines} if they define the same translation. 

In the case of a Mealy-machine $\bA$ every input word $w$ has exactly one translation $\hat{\lambda}(a_0, w)$ and this has the same length as $w$. Mealy-machines enjoy a number of desirable properties and they have a well-developed theory. For example, the following facts are known: 

\begin{enumerate}[(a)]
\item The translations induced by Mealy-machines have a very simple characterization. 
\item The equivalence problem of Mealy-machines is decidable. 
\item For any Mealy-machine one can find an equivalent minimal Mealy-machine and this is unique up to isomorphism. 
\item Let $\bA$ be the Mealy-machine defined above. If $L \in \Rec~X$, then $L\tau_{\bA} \in \Rec~Y$. If $L \in \Rec~Y$, then $L\tau_{\bA}^{-1} \in \Rec~X$. 
\end{enumerate}

There are several ways to generalize Mealy-machines. First of all, both the next-state and the output behaviour may be nondeterministic. Another generalization allows the sequential machine to emit a word in response to each input letter. Moreover, one may add a set of final states. Then a translation of a word is accepted just in case it leaves the machine in a final state. We shall now define a generalized sequential machine which includes all these features. It is now convenient to use a set of productions which will account both for the next-state behaviour and for the outputs. We arrive at the following concept. 

\begin{df}
\label{Definition.1.7.1}
\rm
A \emph{(nondeterministic) generalized sequential machine\index{machine!generalized sequential}} ($\gsm$)\index{gsm} is a system $\bA =(X, A, Y, a_0, P, A')$ where 
\begin{enumerate}[(1)]
\item $X$ is the \emph{input alphabet}\index{input alphabet of!gsm}, 
\item $A$ is a finite, nonempty set of \emph{states}\index{state!of gsm}, 
\item $Y$ is the \emph{output alphabet}\index{output alphabet of!gsm@$\gsm$}, 
\item $a_0~(\in\hspace{-.5ex}A)$ is the \emph{initial state}\index{initial state of!gsm}, 
\item $P$ is a set of \emph{productions}\index{production of!gsm} of the form $ax \rightarrow wb$ with $a, b \in A$, $x \in X$ and $w \in Y^*$, and 
\item $A' \subseteq A$ is the set of \emph{final states}\index{final state of!gsm}. 
\end{enumerate}
It is assumed that $A \cap (X \cup Y) = \emptyset$. The $\gsm$ $\bA$ is said to be \emph{deterministic}\index{gsm!deterministic} if there exists for each pair $(a,x) \in A \times X$ exactly one production of the form $ax \rightarrow wb$. 
\end{df}

Let $\bA$ be the above $\gsm$. A production $ax \rightarrow wb$ is interpreted as follows. If $\bA$ is in state $a$ and receives the input $x$, $\bA$ may enter state $b$ and simultaneously emit the word $w$. We shall now define the translation performed by $\bA$. For any two words $p,q \in (A \cup X \cup Y)^*$, we write $p \Rightarrow_{\bA} q$ if there exist a production $ax \rightarrow wb$  in $P$ and words $p'$ and $p''$ such that $p=p' axp''$ and $q=p' wbp''$. The reflexive, transitive closure of $\Rightarrow_{\bA}$ is denoted by $\Rightarrow_{\bA}^*$. Thus $p \Rightarrow_{\bA}^* q \; (p,q \in (A \cup  X \cup Y)^*)$ 
holds iff there exists a \emph{derivation}\index{derivation in!gsm} of the form 
\begin{equation*}
p = p_0 \Rightarrow_{\bA} p_1 \Rightarrow_{\bA} \ldots \Rightarrow_{\bA} p_k = q \quad (k \geq 0). 
\end{equation*}
Now, the \emph{translation}\index{translation!induced by gsm} induced by $\bA$ is defined as the relation 
\begin{equation*}
\tau_A = \{(u,v) \mid u \in X^*, v \in Y^*, a_0u \Rightarrow_{\bA}^* vb \textrm{ for some } b \in A' \}. 
\end{equation*}
If $(u,v) \in \tau_{\bA}$, then $v$ is a \emph{translation} of $u$. If $\bA$ is deterministic, then each $X$-word $w$ has at most one translation. Two $\gsm$'s are \emph{equivalent}\index{equivalence of!gsm's} if they induce the same translation. 

The tree transducers, which form the subject matter of Chapter~\ref{Chapter.4}, may be viewed as further generalizations of $\gsm$'s in which trees replace words as inputs and as outputs. The following two theorems may be compared with some of the results to be presented in Chapter~\ref{Chapter.4}. 

\begin{thm}
\label{Theorem.1.7.2} Let $\bA = (X, A, Y, a_0, P, A')$ be a $\gsm$. If $L \in \Rec~X$, then $L\tau_{\bA} \in \Rec~Y$. If $L \in \Rec~Y$, then $L\tau^{-1}_{\bA} \in \Rec~X$. \epr
\end{thm}

\begin{thm}
\label{Theorem.1.7.3} The equivalence problem of deterministic $\gsm$'s is decidable, but the equivalence problem of nondeterministic $\gsm$'s is undecidable.  \epr
\end{thm}

The next-state behaviour of a $\gsm$ is identical to that of a nondeterministic Rabin-Scott recognizer. Thus the following fact, which will be needed in Chapter~\ref{Chapter.4}, is obvious. 

\begin{lm}
\label{Lemma.1.7.4} Let $\bA$ be a $\gsm$ as defined above. For any two states $a,b \in A$, the language 
\begin{equation*}
L(a,b) = \{u \in X^* \mid au \Rightarrow_{\bA}^* bv \textrm{ for some } v \in Y^* \} 
\end{equation*}
is regular.\epr
\end{lm}

%% file: Section.1.8.tex

Extensive treatments of universal algebra can be found in the following two standard references: 

\begin{compactitem}
\item \textsc{P. M. Cohn}, Universal algebra, D. Reidel, Dordrecht (2. ed. 1981). 
\item \textsc{G. Gr\"atzer}, Universal algebra, Springer-Verlag, New York (2. ed. 1979). 
\end{compactitem}

\

The following more concise texts may also be recommended: 
\begin{compactitem}
\item \textsc{H. Lugowski}, Grundz\"uge der universellen Algebra, Teubner, Leipzig (1976). 
\item \textsc{H. Werner}, Einf\"uhrung in die allgemeine Algebra, Bibliographisches Institut, Mannheim (1978). 
\end{compactitem}

\

A good introduction to lattice theory (available in German and in French, too): 
\begin{compactitem}
\item \textsc{G. Sz\'asz}, Introduction to lattice theory, Academic Press, New York (1963). 
\end{compactitem}

\

Two general texts on finite automata and regular expressions: 
\begin{compactitem}
\item \textsc{F. G\'ecseg} and \textsc{I. Pe\'ak}, Algebraic theory of automata, Akad\'emiai Kiad\'o, Budapest (1972). 
\item \textsc{A. Salomaa}, Theory of automata, Pergamon Press, Oxford (1969). 
\end{compactitem}

\

An extensive algebraic treatment of the theory of finite automata can be found in the following two volumes: 
\begin{compactitem}
\item \textsc{S. Eilenberg}, Automata, languages, and machines, Academic Press, New York (Vol. A 1974, Vol. B 1976). 
\end{compactitem}

\

The general area of formal language theory is covered, for example, by the following books: 
\begin{compactitem}
\item \textsc{A. V. Aho} and \textsc{J. D. Ullman}, The theory of parsing, translation, and compiling, Prentice-Hall, Englewood Cliffs, N. J. (1972). 
\item \textsc{M. A. Harrison}, Introduction to formal language theory, Addison-Wesley, Reading, Mass. (1978). 
\item \textsc{J. E. Hopcroft} and \textsc{J. D. Ullmann}, Formal languages and their relation to automata, Addison-Wesley, Reading. Mass. (1969). 
\item \textsc{A. Salomaa}, Formal languages, Academic Press, New York (1973). 
\end{compactitem}

\

A highly recommendable classic on context-free languages is: 
\begin{compactitem}
\item \textsc{S. Ginsburg}, The mathematical theory of context-free languages, McGraw-Hill, New York (1966). 
\end{compactitem}

%% file: Prologue.2.tex
This chapter is devoted to finite-state tree recognizers and the family of forests recognizable by them. Here trees are defined as terms over a finite operator domain, and a forest (or tree language) is just a set of trees. As in the case of formal languages, there are two particularly natural ways to effectively define a forest; a forest can be recognized by an automaton, or it can be generated by a grammar. In Section \ref{Section.2.2} we introduce the tree recognizers which correspond to Rabin--Scott recognizers. It does not make any difference whether Rabin--Scott recognizers are defined to read words from left to right or from right to left, but here we should consider both recognizers that read trees from the leaves down towards the root (frontier-to-root recognizers) and recognizers which work in the opposite direction (root-to-frontier tree recognizers). In both cases the recognizer may be either deterministic or nondeterministic. This gives us four types of finite-state tree recognizers. Three of these define the same family of forests, the family $\Rec$ of recognizable forests. Deterministic root-to-frontier recognizers are essentially weaker and they define a proper subfamily of $\Rec$. In Section \ref{Section.2.3} we define regular tree grammars. After having shown that these can be reduced to a very simple normal form, we prove that regular tree grammars generate exactly the recognizable forests. Often it will be convenient to use regular tree grammars in the study of recognizable forests. In Section \ref{Section.2.4} several operations on forests are considered. Many of these arise as a generalization of some basic language operation. Usually $\Rec$ can be shown to be closed under such operations. However, one should note that there are often many ways to generalize from languages to forests, and a right choice among the alternatives is essential if one wants to generalize the corresponding results, too. For example, there is a natural generalization of the product of languages with respect to which $\Rec$ is not even closed. A related point is demonstrated by the case of tree homomorphisms. Here the greater generality of trees compared with words admits of some entirely new phenomena, such as the copying of subtrees.

In Section \ref{Section.2.5} regular expressions to denote forests are defined, and the appropriate generalized Kleene theorem can then be proved. Section \ref{Section.2.6} contains the minimization theory of deterministic frontier-to-root tree recognizers. In Sections \ref{Section.2.7} to \ref{Section.2.9} the family $\Rec$ is characterized in some further ways. Recognizable forests are described by means of congruences of the term algebra, as solutions of fixed-point equations, and in terms of local forests. Moreover, a Medvedev-type characterization in terms of certain elementary forests and elementary operations is given. In Section \ref{Section.2.10} we show that the emptiness, the finiteness, and the equivalence problems of recognizable forests are decidable. Section \ref{Section.2.11} is devoted to deterministic root-to-frontier recognizers. The forests recognizable by them are characterized by means of a certain closure property. Furthermore, we show that these recognizers have canonical minimal forms. 

In this chapter we try to cover the central parts of what could be called ``the generalized theory of finite automata'', but many topics had to be excluded. Some of these are mentioned in the Notes and references. There we shall also indicate a few other developments not directly related to this chapter as well as some applications of the theory of tree automata.  

%% file: Section.2.1.tex
The ``trees'' which appear in tree automata theory may be visualized as tree-like directed labelled graphs. Such a tree\index{tree} has exactly one node, the root, to which no edge enters. From the root there is exactly one path to every node. Moreover, it is essential that the edges leaving a given node have a specified left-to-right order. This concept has been formalized in several ways, but the variations in the definition are of little or no consequence. We shall choose a definition that suits well an algebraic treatment of the theory.

For the labelling of the nodes of a tree we need two alphabets of different kind, a ranked alphabet and a frontier alphabet. As a rule, these two are assumed to be disjoint. A \emph{ranked alphabet}\index{alphabet!ranked} is a finite nonempty operator domain (cf. Sect. \ref{Section.1.2}). From now on $\Sigma$ always represents a ranked alphabet. Other symbols to be used for ranked alphabets include $\Omega$ and $\Gamma$. The inclusion $\Sigma \subseteq \Omega$ means that $\Sigma_m \subseteq \Omega_m$ for all $m \geq 0$. If $\Sigma_m \cap \Omega_n = \emptyset$ whenever $m \neq n$, then $\Sigma \cup \Omega$ may be defined:
$$(\Sigma \cup \Omega)_m = \Sigma_m \cup \Omega_m \text{  for all  } m \geq 0.$$
A \emph{frontier alphabet}\index{alphabet!frontier} is simply an alphabet in the usual sense, but sometimes we should let it be empty. In fact, in most cases there is no need to exclude this possibility. Our usual symbols for frontier alphabets are $X$, $Y$ and $Z$.

For any $\Sigma$ and $X$, a $\Sigma X$-\emph{tree}\index{SigmaX-tree@$\Sigma X$-tree} is simply a $\Sigma X$-term. Thus the set of $\Sigma X$-trees is $F_{\Sigma}(X)$. In many cases $\Sigma$ or $X$, or both, are either understood or unspecified. In such cases we often speak about $\Sigma$-\emph{trees}\index{Sigma-tree@$\Sigma$-tree}, $X$-\emph{trees}\index{Xtree@$X$-tree} or just \emph{trees}. A similar situation will arise whenever a concept involves a ranked alphabet and a frontier alphabet. We shall not lengthen such definitions by listing the modified names, but they will be used without explanation whenever convenient.

The letters $p,q,r,s$ and $t$ are reserved for trees.

Although trees are defined as strings, they can be visualized as, and are in fact intended as representations of, such tree structures as described above.

\begin{ex}\label{Example.2.1.1}\rm
Let $\Sigma = \Sigma_0 \cup \Sigma_1 \cup \Sigma_2$ be a ranked alphabet, where $\Sigma_0 = \{\gamma \}$, $\Sigma_1 = \{\omega \}$ and $\Sigma_2 = \{\sigma \}$. As the frontier alphabet we take $X= \{x,y \}$. Then $t= \omega \left (\sigma (y, \sigma ( \gamma, x ) ) \right )$ is the $\Sigma X$-tree shown in Fig. \ref{Figure.2.1}. \epr
\end{ex}

\begin{figure}[h]
\input{Figure.2.1}
\caption{{}\label{Figure.2.1}}
\end{figure}

Any other way of writing $\Sigma X$-terms would suit our purpose equally well. For example, in Polish notation the tree $t$ of Example \ref{Example.2.1.1} would be written as $\omega \sigma y \sigma \gamma x$, but it would still be treated in tree automaton theory as the ``tree'' shown in Fig. \ref{Figure.2.1}.

Term induction will now be called \emph{tree induction}.\index{induction!tree} Below some important concepts are defined by tree induction.

\begin{df}\label{Definition.2.1.2}\rm
The \emph{height} $\hg (t)$,\index{height of!tree} the \emph{root} $\rroot (t)$\index{root of tree} and the set of \emph{subtrees}\index{subtree} $\sub (t)$ of a $\Sigma X$-tree $t$ are defined as follows:
\begin{itemize}
\item[$1^\circ$] If $t \in X \cup \Sigma_0$, then $\hg (t) = 0$, $\rroot (t) = t$ and $\sub (t) = \{t\}$.
\item[$2^\circ$]  If $t= \sigma (t_1, \ldots, t_m) \: (m>0)$, then \\
\setlength{\tabcolsep}{1.5pt}
\begin{tabular}{r l l}
  \qquad $\hg (t)$  & = & $\max (\hg (t_i) \mid i=1, \ldots ,m)+1 $, \\
\qquad	$\rroot (t)$ & = & $\sigma$, and  \\
 \qquad  $\sub (t)$ & = & $\bigcup \left ( \sub (t_i) \mid 1 \leq i \leq m \right) \cup t.$
\end{tabular}
\end{itemize}
\end{df}

\noindent For the tree of Example \ref{Example.2.1.1} we get $\hg (t) =3$, $\rroot (t) = \omega$ and $\sub (t)= \{t, \sigma \left ( y, \sigma (\gamma, x)  \right ), \\y, \sigma (\gamma , x), \gamma, x  \}$.

Subtrees of height $0$ are referred to as the \emph{leaves} of the tree. A leaf \index{leaf of tree}is labelled by a letter from the frontier alphabet or by a nullary operator. The \emph{length} $|t|$\index{length of!tree} of a tree $t$ is simply its length as a word. The leaves of tree $t$ of our example are $y, \gamma$ and $x$. Its length is 15 (when parentheses and commas are counted, too). Of course, one can define and prove things about trees by induction on the length; but in practice this mostly reduces to tree induction. Induction on the height $\hg (t)$ is equivalent to tree induction.

We shall use the term \emph{frontier}\index{frontier of tree} in a rather informal way to designate the part of a tree consisting of the leaves. The frontier of the tree of Example \ref{Example.2.1.1} consists of the nodes labelled by $y, \gamma$ and $x$. The same letter or nullary operator could appear several times as a leaf in the frontier. The visual picture of a tree also suggests the notions of a \emph{branch}\index{branch of tree} and that of a \emph{path}\index{path in tree}. In our $t$ there are two main branches leaving the lower $\sigma$. They correspond to the subtrees $y$ and $\sigma( \gamma, x)$. There are three paths from the root to the frontier. They spell out the words $\omega \sigma y$, $\omega \sigma \sigma \gamma$ and $\omega \sigma \sigma x$, respectively. These terms are used in a descriptive manner to aid the intuition and no precise definitions are needed.

\

\textbf{Note.} In the literature the root is often called the ``top'' of the tree, while its frontier is referred to as the ``bottom''. Then ``top-down'' indicates the direction from the root towards the frontier, and ``bottom-up'' means the opposite direction. This terminology is connected with the common practice of drawing trees upside-down.

\

The same tree may occur several times as a subtree of a given tree and one should distinguish between a subtree and an \textit{occurrence of a subtree}\index{occurrence!of subtree}. It is possible to assign coordinates to the nodes of a tree and then indicate a certain occurrence of a subtree by the coordinates of its root. However, the following simple device to specify an occurrence of a subtree will suffice. For any occurrence of a subtree $s$ of a tree $t$, there is a unique way to write $t=usv$. Here $u$ and $v$ are just words and the occurrence of $s$ is uniquely determined by $u$.

We shall now consider some ways to construct new trees from given ones. The very definition of $F_{\Sigma}(X)$ suggests such a construction. If $m \geq 0$, $\sigma \in \Sigma_m$ and $t_1, \ldots, t_m \in F_{\Sigma}(X)$, then $\sigma (t_1, \ldots, t_m)$ is a new $\Sigma X$-tree which could be called the $\sigma$-\emph{catenation}\index{sigma-catenation@$\sigma$-catenation} of $t_1, \ldots, t_m$. It is obtained by connecting the roots of the trees $t_1, \ldots, t_m$ to a new root labelled by $\sigma$. The construction is illustrated by Fig. \ref{Figure.2.2}.

	\begin{figure}[h]
		\centering
		\begin{minipage}{0.49\textwidth}
			\input{Figure.2.2}
			\caption{{}\label{Figure.2.2}}
		\end{minipage}
		\begin{minipage}{0.5\textwidth}
			\input{Figure.2.3}
			\caption{{}\label{Figure.2.3}}
		\end{minipage}
	\end{figure}

Note that the $\sigma$-catenation is the $\sigma$-operation of the $\Sigma X$-term algebra $\cF_{\Sigma}(X)$:
$$\sigma(t_1, \dots, t_m)= \sigma^{\cF_{\Sigma}(X)}(t_1, \dots, t_m).$$

Let $t$ be a $\Sigma X$-tree and suppose we are given a tree $s_x$ for every $x \in X$. The tree denoted by
$$t(x \leftarrow s_x \mid x \in X), \quad \text{or just} \quad t(x \leftarrow s_x), $$
is obtained by substituting in $t$, simultaneously for every $x \in X$, $s_x$ for each occurrence of $x$. The formal definition by tree induction reads as follows:
\begin{itemize}
\item[$1^\circ$] If $t=z \in X$, \hspace{5pt} then   \hspace{5pt} $t(x \leftarrow s_x)=s_z$.
\item [$2^\circ$]  If $t=\sigma \in \Sigma_0$, \hspace{5pt} then   \hspace{5pt} $t(x \leftarrow s_x)= \sigma$.
\item[$3^\circ$] If $t= \sigma(t_1, \ldots, t_m)$, \hspace{5pt} then
\item[] $t(x \leftarrow s_x) = \sigma \left ( t_1(x \leftarrow s_x), \ldots, t_m(x \leftarrow s_x)   \right)$.
\end{itemize}

If the trees $s_x$ are $\Sigma X$-trees, then $t(x \leftarrow s_x)$ is also a $\Sigma X$-tree. However, the construction works also in the more general case where the trees $s_x$ are $\Omega Y$-trees for some $\Omega$ and $Y$ such that $\Sigma_m \cap \Omega_n =\emptyset$ whenever $m\neq n$. Then $t(x \leftarrow s_x) \in F_{\Sigma \cup \Omega}(Y)$.

Suppose $X=\{x_1, \ldots, x_n \}$. One may then write $t(x \leftarrow s_x)$ in the more explicit form
$$t \left (x_1 \leftarrow s_{x_1}, \ldots, x_n \leftarrow s_{x_n}     \right ).$$
If the order $x_1, \ldots, x_n$ is understood, we may write simply $t \left (s_{x_1}, \ldots,  s_{x_n}     \right )$.

A letter $x$ may be left unrewritten by choosing $s_x=x$. The notation $t  (x_1 \leftarrow s_1, \ldots, x_n \leftarrow s_n     )$ is used more generally to indicate a substitution where the letters $x_i$ are rewritten as the corresponding $s_i$ ($i=1, \ldots, n$), but all the other letters of $X$ are left unchanged in the tree $t$.

\begin{ex}\label{Example.2.1.3}\rm
Suppose $\gamma \in \Sigma_0$, $\sigma \in \Sigma_3$ and $x,y,z \in X$. If $t=\sigma \left ( y,\sigma(\gamma, x, y),z \right)$, then
$$t \left ( y \leftarrow x, z \leftarrow \sigma(x,x,z) \right ) = \sigma \left (x,\sigma(\gamma, x,x), \sigma(x,x,z) \right ).$$
The tree is shown in Fig. \ref{Figure.2.3}. \epr
\end{ex}

Often a certain occurrence of a subtree $s$ of a tree $t$ should be replaced by a tree $r$. If the presentation $t=usv$ indicates the particular occurrence of $s$, then the result is $urv$. It is easy to show that $urv$ is also a $\Sigma X$-tree whenever $t,r \in F_{\Sigma}(X)$. The operation may also be described as follows. Let $\xi$ be a new letter. There is a unique tree $t' \in F_{\Sigma}(X \cup \xi)$ with exactly one occurrence of $\xi$ such that $t= t'(\xi \leftarrow s)$. Then $urv = t'(\xi \leftarrow r)$. Other ways to operate on trees will be encountered later on.

Trees define polynomial functions in algebras. These will be very important, and we shall now see how the basic tree operations are reflected in them. Let $\cA =(A, \Sigma)$ be a $\Sigma$-algebra. If $t \in F_{\Sigma}(X)$ is obtained by $\sigma$-catenation from the trees $t_1, \ldots, t_m$ ($m \geq 0, \sigma \in \Sigma_m$), then
$$t^{\cA} = \sigma^{\cA}(t^{\cA}_1, \ldots, t^{\cA}_m)$$
is simply the composition of $t^{\cA}_1, \ldots, t^{\cA}_m$ with $\sigma^{\cA}$. Now we consider the substitution operation. Let $X=\{x_1,\ldots, x_n \}$ and $t,s_1, \ldots, s_n \in F_{\Sigma}(X)$. The polynomial function
$$t(s_1, \ldots, s_n)^{\cA} : A^X \rightarrow A$$
is computed as follows. For any $\alpha :X \rightarrow A$,
$$t(s_1, \ldots, s_n)^{\cA}(\alpha) = t^{\cA}(\beta),$$
where $\beta :X \rightarrow A$ is defined so that $x_i \beta = s^{\cA}_i(\alpha)$ for all $i=1, \ldots, n$.

Finally, consider the replacing of an occurrence of a subtree $s$ of a $\Sigma X$-tree $t$ by a $\Sigma X$-tree $r$. Write $t=t'(\xi \leftarrow s)$ as explained above. For any $\alpha : X \rightarrow A$, we get then
$$t'(\xi \leftarrow r)^{\cA}(\alpha) = t^{\prime \cA}(\alpha ')$$
where $\alpha ' :X \cup \xi \rightarrow A$ is defined so that $\alpha ' |X=\alpha$ and $\xi \alpha = r^{\cA}(\alpha)$.

A $\Sigma X$-forest \index{SigmaX-forest@$\Sigma X$-forest, see also forest} is simply a subset of $F_{\Sigma}(X)$. Many authors call forests \emph{tree languages}.\index{language!tree} In general, we use the letters $R,S$ and $T$ for forests.

If $\Sigma \subseteq \Omega$ and $X \subseteq Y$, then all $\Sigma X$-trees are $\Omega Y$-trees, too. Thus every $\Sigma X$-forest may be viewed as an $\Omega Y$-forest. In most cases this can safely be done. For example, a $\Sigma X$-forest is recognizable (in the sense defined in the next section) as a $\Sigma X$-forest iff it is recognizable as an $\Omega Y$-forest.

Of course, those forests only are of interest that can be defined in some natural way. This chapter is devoted to a family of such forests, the forests recognizable by finite tree automata. In the theory of these forests many concepts and results familiar from the theory of recognizable languages can be perceived. The generalization from words and languages to trees and forests will be considered in the next section.

%% file: Figure.2.1.tex
\centering
\begin{tikzpicture}[sibling distance=1.6cm,level distance=1cm]
\begin{scope}[execute at begin node=$, execute at end node=$]

\node[node,label={left:\omega}] {}
child{ node[node,label={left:\sigma}] {} 
	child { node[node,label={left:\sigma}] {} 
		child { node[node,label={left:x}] {} }
		child { node[node,label={left:\gamma}] {} }}
	child { node[node,label={left:y}] {} }};
\end{scope}
\end{tikzpicture}

%% file: Figure.2.2.tex
\centering
\begin{tikzpicture}[sibling distance=1.1cm,level distance=0.8cm,
leaf/.style={regular polygon, regular polygon sides=3,draw, shape border rotate=180, minimum height=2cm, scale=0.8}]
\tikzstyle{level 2}=[level distance=0.81cm] 
\begin{scope}[execute at begin node=$, execute at end node=$]

\node[node,white,label={below:\sigma}] {}
child[white]{ node[node] {}
	child[black]{ node[leaf] {t_m} }}
child[white]{}
child[white]{ node[node] {} 
	child[black]{ node[leaf] {t_1} }};
\node[node] {}
child{ node[node,fill=white,label={right:d_m}] {} } 
child[white]{ node[black,scale=0.8] {\dots} }
child{ node[node,fill=white,label={left:d_1}] {} };
\end{scope}
\end{tikzpicture}

%% file: Figure.2.3.tex
\centering
\begin{tikzpicture}
\tikzstyle{level 1}=[sibling distance=3cm,level distance=1.1cm]
\tikzstyle{level 2}=[sibling distance=0.7cm,level distance=1.1cm] 
\begin{scope}[execute at begin node=$, execute at end node=$]

\node[node,label={below left:\sigma}] {}
child{ node[node,label={left:\sigma}] {} 
	child { node[node,label={left:z}] {} }
	child { node[node,label={left:x}] {} }
	child { node[node,label={left:x}] {} }}
child{ node[node,label={left:\sigma}] {}
	child { node[node,label={left:x}] {} }
	child { node[node,label={left:x}] {} }
	child { node[node,label={left:\gamma}] {} }}
child[sibling distance=1.8cm]{ node[node,label={left:x}] {}};
\end{scope}
\end{tikzpicture}

%% file: Section.2.2.tex
In this section we introduce tree recognizers, that is, tree automata which define forests. There are four basic types of these recognizers. A tree recognizer may be defined in such a way that it reads its input trees from the frontier towards the root. Then it is called a \emph{frontier-to-root recognizer}, or an $F$-\emph{recognizer} for short. A tree recognizer which reads the trees starting at the root proceeding then towards the frontier is called a \emph{root-to-frontier recognizer}, or simply an $R$-\emph{recognizer}. In both cases the recognizer may be either \emph{deterministic} or \emph{nondeterministic}. As a rule, all tree recognizers considered here are \emph{finite}, i.e., they have a finite number of states.

Our first task will be to compare the families of forests recognizable by these four types of tree recognizers. It turns out that we get just two families. Deterministic $F$-recognizers, nondeterministic $F$-recognizers and nondeterministic $R$-recognizers all have the same recognition power. The forests recognized by them are termed \emph{recognizable}\index{forest!recognizable}.  Deterministic $R$-recognizers are considerably weaker and they yield a rather special subfamily of the recognizable forests.

As stated in the previous section, $\Sigma$ is always a ranked alphabet and $X$ is a frontier alphabet.
\begin{df}\label{Definition.2.2.1}\rm
A \emph{frontier-to-root} $\Sigma X$-\emph{recognizer} \index{SigmaX-recognizer@$\Sigma X$-recognizer!frontier-to-root} or an $(F)\Sigma X$-\emph{recognizer}, for short, $\bA$ consists of
\begin{itemize}
\item[(1)] a finite $\Sigma$-algebra $\cA = (A,\Sigma )$,
\item[(2)] an \emph{initial assignment} $\alpha : X \rightarrow A$ and \index{initial assignment of!Sigma X-recognizer@$\Sigma X$-recognizer}
\item[(3)] a set $A' \subseteq A$ of \emph{final states}. \index{final state of!Sigma X-recognizer@$\Sigma X$-recognizer}
\end{itemize}
\end{df}
We write $\bA =(\cA,\alpha,A')$ or $\bA=(A,\Sigma,X,\alpha,A')$. The \emph{forest recognized} by $\bA$ is the $\Sigma X$-forest
$$T(\bA)=\{t \in F_{\Sigma}(X) \mid t^{\cA}(\alpha) \in A'  \}.$$
A $\Sigma X$-forest $T$ is said to be \emph{recognizable},\index{forest!recognized by Sigma X-recognizer@recognized by $\Sigma X$-recognizer} if there exists a $\Sigma X$-recognizer $\bA$ such that $T=T(\bA)$. The family of recognizable forests is denoted by $\Rec$, and $\Rec(\Sigma ,X)$ denotes the set of all recognizable $\Sigma X$-forests.

The recognizers defined above are finite and deterministic although this has not been emphasized in the name. They are our ``basic'' type of tree recognizer and we shall usually omit the label ``$F$'' which distinguishes them from root-to-frontier tree recognizers. The elements of the underlying algebra $\cA$ are called the \emph{states} of $\bA$\index{state!of Sigma X-recognizer@of $\Sigma X$-recognizer} and $A$ is its \emph{state set}.

If not otherwise specified, $\bA$ will be the $\Sigma X$-recognizer $(\cA, \alpha, A')$. Also $\bB$ and $\bC$ will usually be the $\Sigma X$-recognizers $(\cB, \beta, B')$ and $(\cC, \gamma, C')$, respectively. Here $\cB=(B, \Sigma)$ and $\cC=(C, \Sigma)$ are $\Sigma$-algebras, $\beta:X \rightarrow B$ and $\gamma :X \rightarrow C$ are the initial assignments, and $B'\subseteq B$ and $C' \subseteq C$.

In algebraic terms the operation of the $\Sigma X$-recognizer $\bA$ can be explained as follows. Given an input tree $t \in F_{\Sigma}(X)$ the polynomial function $t^{\cA}$ is evaluated on the initial assignment $\alpha$. The tree is accepted exactly in case the result $t^{\cA}(\alpha)$ is a final state. If
$$\hat{\alpha}:\cF_{\Sigma}(X) \rightarrow \cA$$
is the extension of $\alpha$ to a homomorphism, then
$$t^{\cA}(\alpha)=t \hat{\alpha}  \text{   \ \   for every   \ \  } t \in F_{\Sigma}(X),$$
and we may write
$$T(\bA)=\{t \in F_{\Sigma}(X) \mid t \hat{\alpha} \in A'  \} = A' \hat{\alpha}^{-1}.$$
A more pictorial description of the operation of $\bA$ in automata theoretic terms is also possible. Given an input tree $t$, $\bA$ starts reading it from the leaves in states that depend on the labels of the leaves. If a certain leaf is labelled by a frontier letter $x$, then $\bA$ is in state $x \alpha$ at that leaf. If the label is a nullary operator $\sigma$, then $\bA$ starts from that leaf in state $\sigma^{\cA}$. Now $\bA$ moves down all the branches towards the root step by step as follows. If a given node $v$ is labelled by the $m$-ary operator $\sigma$ ($m>0$), then $\bA$ enters $v$ in state $\sigma^{\cA}(a_1, \ldots, a_m)$, where $a_1, \ldots, a_m$ are the states of $\bA$ at the nodes immediately above $v$, listed in order from left to right. The tree is accepted if $\bA$ enters the root in a final state.

\begin{ex}\label{Example.2.2.2}\rm
Let $\Sigma=\Sigma_1 \cup \Sigma_2$, $\Sigma_1 =\{ \sim \} $, $\Sigma_2 =\{ \wedge, \vee \}$ and $X=\{x,y \}$. Define the operations of the $\Sigma$-algebra $\cA=(\{0,1 \}, \Sigma )$ by the tables below:

\[
\begin{array}{ccccccccccc}
a & \vline & \sim^{\cA}(a) &  & & a & b & \vline & \wedge^{\cA}(a,b) & \vline & \vee^{\cA}(a,b) \\
\cline{1-3}  \cline{6-11}
0 & \vline & 1  & & & 0 & 0 & \vline & 0 & \vline & 0 \\
1 & \vline & 0  & & & 0 & 1 & \vline & 0 & \vline & 1 \\
  &        &    & & & 1 & 0 & \vline & 0 & \vline & 1  \\
  &        &    & &  & 1 & 1 & \vline & 1 & \vline & 1   \\
\end{array}
\]
Define an initial assignment so that $x \alpha =1$ and $y \alpha =0$. To complete the definition of our $\Sigma X$-recognizer $\bA$ we choose $\{1\}$ as the set of final states. The computation of $\bA$ on the tree
$$t= \wedge \left (\sim (\wedge(y,x) ), \vee (\sim (y),x) \right ) $$
is shown in Fig. \ref{Figure.2.4}. The states of $\bA$ at the nodes are shown in parentheses. The tree is accepted since the state at the root is 1. Let $\sim, \wedge$ and $\vee$ have their usual
\begin{figure}[h]
\input{Figure.2.4}
\caption{{}\label{Figure.2.4}}
\end{figure}
meanings as symbols for the logical connectives ``not'', ``and'' and ``or''. Then $\Sigma X$-trees are expressions of propositional logic in the two propositional variables $x$ and $y$. If 0 and 1 are interpreted as the truth values ``false'' and ``true'', respectively, then $\bA$ computes the truth values of propositions, when the truth values of the variables are given. The forest recognized by $\bA$ consists of the propositions (in variables $x$ and $y$) that are true when $x$ is true and $y$ is false.  \epr
\end{ex}

\begin{ex}\label{Example.2.2.3}\rm
Let $\Sigma = \Sigma_2 = \{+, \cdot \}$ and $X=\{x_1, \ldots, x_n\}$ for some $n \geq 1$. The $\Sigma X$-trees may now be interpreted as arithmetic expressions in variables $x_1, \ldots, x_n$. Using the customary infix notation one could write, for example $x_1+x_1 \cdot x_2$ rather than $+(x_1, \cdot (x_1,x_2))$. Let $m>0$ and define the $\Sigma$-algebra $\cA=(\{0,1,\ldots, m-1 \}, \Sigma)$ so that
$$a+^{\cA}b = a+b  \ \ ( {\rm mod} \ m)$$
and
$$a \cdot^{\cA}b = a \cdot b  \ \ ( {\rm mod} \ m)$$
for all $a,b =0,1, \ldots, m-1$. If $t$ is a $\Sigma X$-tree and $\alpha : X \rightarrow A$ is any mapping, then $t^{\cA}(\alpha)$ is the value of the expression $t \ \ ( {\rm mod} \ m)$ when the variables are assigned values according to $\alpha$. Thus any $\Sigma X$-recognizer $\bA =(\cA, \alpha, A')$ based on the algebra $\cA$ recognizes a set of arithmetic expressions which get a value ($ {\rm mod} \ m $) in $A'$ when each variable $x_i$ is given a certain value $x_i \alpha$ ($i=1, \ldots, n$).  \epr
\end{ex}

The examples suggest some useful general observations on tree recognizers. A tree recognizer is a device that evaluates an expression (a tree) for given values of the variables (given by the initial assignment) and decides then on the basis of this value whether the expression belongs to given set or not. Since the state set is finite such an evaluation is always ``modulo something''. For example, we could not construct a tree recognizer which would find out whether the value of an arithmetic expression is a prime or not. Similarly, there is no tree recognizer that recognizes the set of all trees in which two given operators appear the same number of times. The following example discusses another manifestation of the same phenomenon.

\begin{ex}\label{Example.2.2.4}\rm
Let $\Sigma=\Sigma_2=\{\sigma\}$ and let $X$ be an arbitrary nonempty frontier alphabet. Then the forest
$$T= \{\sigma(t,t) \mid t \in F_{\Sigma}(X)  \}$$
is not recognizable. For suppose $T=T(\bA)$ for some $\Sigma X$-recognizer $\bA$. Since $A$ is finite, there must exist two different $\Sigma X$-trees $s$ and $t$ such that $s \hat{\alpha} = t \hat{\alpha}$. But then we would have that
$$\sigma(s,t) \hat{\alpha} =\sigma^{\cA}(s \hat{\alpha}, t \hat{\alpha}) = \sigma^{\cA}(s \hat{\alpha}, s \hat{\alpha} ) = \sigma(s,s) \hat{\alpha} \in A',$$
which implies the contradiction $\sigma(s,t) \in T$.     \epr
\end{ex}

Let us now look how tree recognizers arise as generalizations of the Rabin--Scott recognizers through a universal algebraic interpretation. First, let $\bA =(A, I, \delta, a_0, A')$ be an $I$-recognizer as defined in Sect. \ref{Section.1.5} (to avoid confusion we use $I$ as the input alphabet). Define a ranked alphabet $\Sigma$ such that $\Sigma_1 = I$ and $\Sigma_m= \emptyset$ for all $ m \neq 1$. The next-state mapping of $\bA$ is completely determined by the $\Sigma$-algebra $\cA = (A,\Sigma)$ which is defined so that
$$\sigma^{\cA}(a) = \delta(a,\sigma) \text{ \ \ for all \ \ } a\in A  \text{ \ \ and \ \ } \sigma \in I.$$
If we put $X=\{x\}$, then $I$-words and $\Sigma X$-trees can be identified as follows. The empty word $e$ corresponds to the tree $x$, and a nonempty word $\sigma_1 \ldots \sigma_k$ ($k \geq 1, \sigma_i \in I$) may be interpreted as the tree $\sigma_k (\ldots \sigma_1(x) \ldots )$ (the reverse Polish notation for trees would make the identification even more natural). Define $\alpha : X \rightarrow A$ so that $x \alpha = a_0$. Then
$$\delta (a_0, t)= t ^{\cA}( \alpha) \text{  \ \ for all \ \ } t \in I^*(=F_{\Sigma}(X)!).$$
This implies that the forest recognized by the $\Sigma X$-recognizer $(\cA, \alpha, A')$ is, interpreted as an $I$-language, the language recognized by $\bA$. Hence a Rabin--Scott recognizer may be viewed as a tree recognizer over a unary ranked alphabet and a one-element frontier alphabet. The general $\Sigma X$-recognizers result when one does not require $\Sigma$ to be unary and allows also an arbitrary frontier alphabet $X$.

The nondeterministic frontier-to-root tree recognizers that we soon shall define may be viewed as generalized $F$-tree recognizers in which nondeterminism is allowed both in the assignment of states to the leaves and in the next-state behaviour. First we have to introduce nondeterministic operations and nondeterministic algebras.

An $m$-\emph{ary nondeterministic} (ND) \emph{operation}\index{operation!mary@$m$-ary nondeterministic} on a set $A$ is a mapping from $A^m$ to $\gp A$ ($m \geq 0$). Thus an $m$-ary ND operation
$$f : A^m \rightarrow \gp A$$
assigns to every $m$-tuple of elements from $A$ a subset of $A$. A nullary ND operation
$$f: \{\emptyset \} \rightarrow \gp A $$
fixes a subset of $A$, and $f$ may be identified with this subset $f(\emptyset)$. A \emph{nondeterministic} (ND) $\Sigma$-\emph{algebra} \index{Algebra!nondeterministic} \index{Algebra!nd@ND} $\cA =(A,\Sigma)$ consists of a nonempty set $A$ and a family $\{\sigma^{\cA} \mid \sigma \in \Sigma \} $ of ND operations on $A$ such that for each $\sigma \in \Sigma$, $\sigma^{\cA}$ is $m$-ary if $\sigma \in \Sigma_m$. The ND $\Sigma$-algebra is \emph{finite} \index{Algebra!finite ND} if $A$ is finite. A $\Sigma$-algebra may be viewed as an ND $\Sigma$-algebra when elements $a \in A$ are identified with the corresponding singletons $\{a\}$.

On the other hand, we associate with every ND $\Sigma$-algebra $\cA = (A, \Sigma)$ an ordinary $\Sigma$-algebra, namely the subset algebra
$$\gp \cA = (\gp A, \Sigma) $$
where
$$\sigma ^{\gp \cA}(A_1, \ldots, A_m) = \bigcup \left ( \sigma^{\cA}(a_1, \ldots, a_m) \mid a_1 \in A_1, \ldots, a_m \in A_m   \right)$$
for all $m \geq 0$, $\sigma \in \Sigma_m$ and $A_1, \ldots, A_m \subseteq A$. Now any mapping
$$\alpha : X \rightarrow \gp A $$
may be extended to a homomorphism
$$\hat{\alpha} : \cF_{\Sigma}(X) \rightarrow \gp \cA.$$
Consider a $\Sigma X$-tree $t$. The computation of the set $t \hat{\alpha}$ may be described in automata theoretic terms as follows. If a leaf is labelled by a letter $x$, then the ``automaton'' $\cA$ may start at that leaf in any one of the states in $x \alpha$. If a leaf is labelled by a nullary operator, then $\sigma^{\cA}$ is the set of the possible starting states. Let $v$ be any node in the tree labelled by an $m$-ary symbol $\sigma$ ($m>0$). Let $\sigma(t_1, \ldots, t_m)$ be the subtree of $t$ which has $v$ as its root. Then $t_1 \hat{\alpha}, \ldots, t_m \hat{\alpha}$ are the respective sets of possible states of $\cA$ at the nodes immediately above $v$. Now $\cA$ may enter $v$ in any one of the states from $\sigma^{\gp \cA} (t_1 \hat{\alpha}, \ldots, t_m \hat{\alpha})$. Clearly, $t \hat{\alpha}$ is the set of all states in which $\cA$ may be at the root of~$t$.
\begin{df}\label{Definition.2.2.5}\rm
A \emph{nondeterministic frontier-to-root} $\Sigma X$-\emph{recognizer},\index{SigmaX-recognizer@$\Sigma X$-recognizer!nondeterministic frontier-to-root} or an NDF $\Sigma X$-recognizer\index{SigmaX-recognizer@$\Sigma X$-recognizer!NDF} for short, $\bA$ consists of
\begin{itemize}
\item[(1)] a finite ND $\Sigma$-algebra $\cA = (A, \Sigma)$,
\item[(2)] an \emph{initial assignment} $\alpha : X \rightarrow \gp A$ and \index{initial assignment of!NDF $\Sigma X$-recognizer}
\item[(3)] a set $A' \subseteq A$ of \emph{final states}. \index{final state of!NDF $\Sigma X$-recognizer}
\end{itemize}
\end{df}
We write $\bA = (\cA, \alpha, A')$ or $\bA = (A, \Sigma, X, \alpha, A')$. The \emph{forest recognized} by $\bA$ \index{forest!recognized by NDF $\Sigma X$-recognizer} is the $\Sigma X$-forest
$$T(\bA) = \{t \in F_{\Sigma}(X)  \mid t \hat{\alpha} \cap A' \neq \emptyset \}.$$

The definition of $T(\bA)$ means that a tree $t$ is accepted by $\bA$ iff there is a set of choices of initial states for the leaves and next-states for the other nodes such that $\bA$ enters the root of $t$ in a final state. It is rather obvious that the $\Sigma X$-recognizer
$$\gp \bA  = ( \gp A, \alpha, A''),$$
where
$$A'' = \{A_1 \in\gp A \mid A_1 \cap A' \neq \emptyset \},$$
recognizes the same forest as $\bA$. Indeed, for any $t\in F_{\Sigma}(X)$,
$$t\in T(\gp \mathbf{A}) \text{ \ \ \  iff \ \ \  } t^{\gp \cA}(\alpha) \in A'' \text{ \ \ \ iff \ \ \  }t \hat{\alpha} \in A'' $$
$$\text{ \  \ \  \  \ \ \ \ \ \ \ \ \ \ \ \ \ \  \ iff \ \ \ \  }t \hat{\alpha} \cap A' \neq \emptyset \text{  \ \ \ iff \ \ \  } t\in T(\bA). $$
This is the natural generalization of the usual subset construction as applied to ND Rabin--Scott recognizers, and $\gp \bA$ is the ``subset recognizer'' corresponding to $\bA$. Since every $\Sigma X$-recognizer may be viewed as an equivalent NDF $\Sigma X$-recognizer we have verified the following theorem.
\begin{thm}\label{Theorem.2.2.6}
The forests recognized by nondeterministic frontier-to-root recognizers are exactly the recognizable forests. \epr
\end{thm}

We begin the discussion of root-to-frontier tree recognizers with the nondeterministic version. In a \emph{nondeterministic root-to-frontier} $\Sigma$-\emph{algebra} \index{Algebra!nondeterministic root-to-frontier} (NDR $\Sigma$-\emph{algebra},\index{Algebra!NDR} for short) $\cA = (A, \Sigma)$, $A$ is a nonempty set and every $\sigma \in \Sigma_m$ with $m \geq 1$ is realized as a mapping
$$\sigma^{\cA} : A \rightarrow \gp (A^m).$$
For $\sigma \in \Sigma_0$, $\sigma^{\cA}$ is a subset of $A$. We call $\cA$ \emph{finite}, if $A$ is finite.
\begin{df}\label{Definition.2.2.7}\rm
A \emph{nondeterministic root-to-frontier} $\Sigma X$-\emph{recognizer}\index{SigmaX-recognizer@$\Sigma X$-recognizer!nondeterministic root-to-frontier} $\bA$, or an NDR $\Sigma X$-\emph{recognizer},\index{SigmaX-recognizer@$\Sigma X$-recognizer!NDR} consists of
\begin{itemize}
\item[(1)] a finite NDR $\Sigma$-algebra $\cA = (A,\Sigma)$,
\item[(2)] a set $A' \subseteq A$ of \emph{initial states}, and \index{initial state of!NDR $\Sigma X$-recognizer}
\item[(3)] a \emph{final assignment} \index{final assignment of NDR $\Sigma X$-recognizer} $\alpha : X \rightarrow \gp A$.
\end{itemize}
We write $\bA = (\cA, A', \alpha)$ or $\bA = (A, \Sigma, X, A', \alpha)$. The elements of $A$ are called \emph{states}.\index{state!of NDR $\Sigma X$-recognizer}
\end{df}

In order to make the formal definition of the forest recognized by such an $\bA$ easier to understand, we shall first describe its intended operation. At the root of a given $\Sigma X$-tree $t$, $\bA$ may be in any initial state $a \in A'$. Consider now any node $v$ of $t$ labelled by some $\sigma \in \Sigma_m$ with $m \geq 1$. If $a$ is a possible state of $\bA$ at $v$ and $(a_1, \ldots, a_m) \in \sigma^{\cA}(a)$, then $\bA$ may assume state $a_1$ at the leftmost node immediately above $v$, state $a_2$ at the node immediately to the right of this node etc. For every $m$-tuple in $\sigma^{\cA}(a)$, $\bA$ has such a sequence of possible next-states for the nodes directly above $v$. Note that the possible states at these nodes are connected with each other: $(a_1, \ldots, a_m), (a'_1, \ldots, a'_m) \in \sigma^{\cA}(a)$ does not imply, for example, $(a'_1, a_2, \ldots, a_m) \in \sigma^{\cA}(a)$. The tree $t$ is accepted by $\bA$ if it is possible to choose the initial state for the root and then make the consecutive choices of next-state vectors in such a way that $\bA$ arrives at each leaf labelled by a frontier letter $x$ in a state belonging to $x \alpha$, and at each leaf labelled by a 0-ary symbol $\sigma$ in a state belonging to $\sigma^{\cA}$. It is easier to formalize this recognition process by tracing it from the leaves back to the root. The idea is to see which states at each node can lead to acceptance. For the leaves this is clear. If a leaf is labelled by $x \in X$, then the accepting states for that leaf form the set $x \alpha$. If a leaf is labelled by $\sigma \in \Sigma_0$, then the accepting states are those belonging to $\sigma^{\cA}$. Now one can infer the states that are accepting at the nodes immediately below the leaves. When these have been found, we may determine the states in which $\bA$ should be at nodes one level deeper in a tree. Finally one finds out the accepting states for the root. The tree is accepted iff at least one of these is an initial state.
\begin{df}\label{Definition.2.2.8}\rm
Let $\bA = (\cA, A', \alpha)$ be an NDR $\Sigma X$-recognizer. A mapping
$$\tilde{\alpha} : F_{\Sigma}(X) \rightarrow \gp A$$
is defined as follows:

\begin{itemize}
\item[$1^\circ$] If $x \in X$, then $x\tilde{\alpha} = x \alpha$.
\item[$2^\circ$] If $\sigma \in \Sigma_0$, then $\sigma \tilde{\alpha}=\sigma^{\cA}$.
\item[$3^\circ$] If $t=\sigma (t_1, \ldots, t_m)$ $(m \geq 1)$, then \\
$t\tilde{\alpha} = \{a\in A \mid \sigma^{\cA}(a) \cap \left (t_1 \tilde{\alpha} \times \ldots \times t_m \tilde{\alpha}  \right )  \neq \emptyset \}$.
\end{itemize}
The \emph{forest recognized} by $\bA$ \index{forest!recognized by NDR $\Sigma X$-recognizer} is the $\Sigma X$-forest
$$T(\bA) = \{t \in F_{\Sigma}(X) \mid t\tilde{\alpha} \cap A' \neq \emptyset  \}.$$
\end{df}

\begin{ex}\label{Example.2.2.9}\rm
Let us consider again the arithmetic expressions, defined in Example \ref{Example.2.2.3}. We shall construct an NDR $\Sigma \{x_1,x_2 \}$-recognizer which accepts an expression in variables $x_1$ and $x_2$ iff the value of the expression is divisible by 4 when $x_1=0$ or 2 $(\rm{mod}$ $4)$ and $x_2= 3$ $(\rm{mod}$ $4)$. An obvious choice for a state set is $A = \{0,1,2,3 \}$. The set of initial states is $\{0\}$, and the final assignment is defined by $x_1 \alpha = \{0,2\}$ and $x_2 \alpha = \{3\}$. The next-state behaviour is determined by inferring the possible summands or factors from the sum or product, respectively. We get
$$+^{\cA}(0) = \{(0,0), (1,3), (2,2), (3,1)  \}$$
$$+^{\cA}(1) = \{(0,1), (1,0), (2,3), (3,2)  \}$$
 \  \  \   \   \ \  \ \ \ \ \ \ \ \  \ \  \ \ \ \ \ \ \  \ \ \ \ \ \ \  \ \ \ \ \ etc., and
$$ \text{ \ \ \  } \cdot ^{\cA}(0)=\{0\} \times A \cup A \times \{0\} \cup \{(2,2) \}$$
$$\cdot ^{\cA}(1)=\{(1,1),(3,3)\} \text{ \ \ \ \ \ \ \ \ \ \ \ \ }$$
 \  \  \   \   \ \ \   \ \ \ \ \ \ \ \  \ \  \ \ \ \ \ \ \  \ \ \ \ \ \ \  \ \ \ \ \ etc..

\noindent Note that we would get an equivalent NDF-recognizer by ``inverting'' these operations ($0+^{\cA}0=0$ etc.), and making $\{0\}$ the set of final states and $\alpha$ the initial assignment. \epr
\end{ex}

The concluding observation of Example \ref{Example.2.2.9} can be generalized as follows. We say that the NDF $\Sigma X$-recognizer $\bA = (A,\Sigma, X, \alpha, A')$ and the NDR $\Sigma X$-recognizer $\bB = (B, \Sigma, X, B', \beta)$ are \emph{associated} \index{associated $\Sigma X$-recognizers} if
\begin{itemize}
\item[(1)] $A=B, A'=B'$ and $\alpha = \beta$,
\item[(2)] $(a_1, \ldots, a_m) \in \sigma^{\cB}(a)$ \quad iff \quad $a \in \sigma^{\cA}(a_1, \ldots, a_m)$, \   for all $m \geq 1$, $\sigma \in \Sigma_m$ and $a_1, \ldots, a_m, a \in A$, and
\item[(3)] $\sigma^{\cA}=\sigma^{\cB}$ for every $\sigma \in \Sigma_0$.
\end{itemize}
It easy to see that $\hat{\alpha}=\tilde{\beta}$ if $\bA$ and $\bB$ are associated. Since every NDF tree recognizer has an associated NDR tree recognizer, and conversely, we get
\begin{thm}\label{Theorem.2.2.10}
The forests recognizable by NDR tree recognizers are exactly the recognizable forests. \epr
\end{thm}

A \emph{deterministic root-to-frontier} $\Sigma X$-\emph{recognizer},\index{SigmaX-recognizer@$\Sigma X$-recognizer!deterministic root-to-frontier} or a DR $\Sigma X$-\emph{recognizer}\index{SigmaX-recognizer@$\Sigma X$-recognizer!DR}, is a NDR $\Sigma X$-recognizer $\bA = (\cA, A', \alpha)$ such that $A'$ and all of the sets $\sigma^{\cA}(a)$ ($\sigma \in \Sigma_m, m\geq 1, a \in A$) and $\sigma^{\cA}$ with $\sigma \in \Sigma_0$ contain exactly one element. Thus a DR $\Sigma X$-recognizer $\bA$ has exactly one initial state and in every situation there is exactly one choice of next-state vector. Moreover, there is exactly one final state for each leaf labelled by a nullary symbol. The forest recognized by $\bA$ is defined the same way as in the general case.

That determinism is a real limitation in the case of root-to-frontier recognizers is shown by the following example.

\begin{ex}\label{Example.2.2.11}\rm
Suppose $\sigma \in \Sigma_2$ and $x,y \in X$. If a DR $\Sigma X$-recognizer accepts the trees $\sigma(x,y)$ and $\sigma(y,x)$, then it must accept $\sigma(x,x)$, too. Hence, the forest $T=\{\sigma(x,y), \sigma(y,x)  \}$ cannot be recognized by any DR $\Sigma X$-recognizer. On the other hand, it is obvious that $T \in \Rec (\Sigma, X)$. \epr
\end{ex}

The inability of these recognizers to cope with situations such as that in Example \ref{Example.2.2.11} is due to the fact that they have to read disjoint subtrees separately without any possibility to combine the information gathered from the individual subtrees. In an NDR tree recognizer this handicap is compensated for by their ability to make several guesses about the subtrees jointly before reading them separately.

%% file: Figure.2.4.tex
\centering
\begin{tikzpicture}
\tikzstyle{level 1}=[sibling distance=2cm,level distance=1cm]
\tikzstyle{level 2}=[sibling distance=1.2cm,level distance=1.1cm] 
\begin{scope}[execute at begin node=$, execute at end node=$]

\node[node,label={left:\wedge},label={right:(1)}] {}
child{ node[node,label={left:\vee},label={right:(1)}] {} 
	child { node[node,label={left:x},label={right:(1)}] {} }
	child { node[node,label={left:\sim},label={right:(1)}] {} 
		child { node[node,label={left:y},label={right:(0)}] {} }}
	child[draw opacity=0]{}}
child{ node[node,label={left:\sim},label={right:(1)}] {}
	child { node[node,label={left:\wedge},label={right:(0)}] {} 
		child { node[node,label={left:x},label={right:(1)}] {} }
		child { node[node,label={left:y},label={right:(0)}] {} }}};
\end{scope}
\end{tikzpicture}

%% file: Section.2.3.tex
So far, the recognizable forests have been characterized by means of three types of tree recognizers. Now we shall introduce a class of tree grammars that also defines the family of recognizable forests. These grammars are the natural counterparts to type 3 grammars.
\begin{df}\label{Definition.2.3.1}\rm
A \emph{regular} $\Sigma X$-\emph{grammar}\index{grammar}\index{regular sigmaxgrammar@regular $\Sigma X$-grammar} $G$ consists of
\begin{itemize}
\item[(1)] a finite nonempty set $N$ of \emph{nonterminal symbols},\index{nonterminal symbol of!regular $\Sigma X$-grammar}
\item[(2)] a finite set $P$ of \emph{productions} of the form $a \rightarrow r$, where $a\in N$ and $r \in F_{\Sigma}(N \cup X)$, and\index{production of!regular $\Sigma X$-grammar}
\item[(3)] an \emph{initial symbol} $a_0 \in N$.\index{initial symbol of!regular $\Sigma X$-grammar}
\end{itemize}
It is assumed that $N \cap (\Sigma \cup X) = \emptyset$. We write $G=(N,\Sigma,X,P,a_0)$.
\end{df}

When $\Sigma$ and $X$ are not specified, we speak about \emph{regular tree grammars}\index{regular tree grammar} or just \emph{grammars}, if there is no danger of confusion.

Let $G$ be a regular tree grammar as in the definition above. The right-hand side of a production is a tree in which nonterminal symbols may appear at the leaves only. For $p,q \in F_{\Sigma}(X \cup N)$, we write
$$p \Rightarrow_G q \text{ \ \ \ (or just } p \Rightarrow q)$$
if there exist $a\in N, r \in F_{\Sigma}(X \cup N)$ and words $u,v$ such that $p=uav, q=urv$ and $a\rightarrow r \in P $, i.e., $p \Rightarrow_G q $ means that $q$ is obtained by replacing an occurrence of a nonterminal symbol $a$ by a tree $r$, where $a \rightarrow r$ is a production of the grammar. More generally, we write
$$p \Rightarrow^*_G q \text{ \ \ \ (or just } p \Rightarrow^* q)$$
if $p=q$ or there exists a (nontrivial) derivation
$$ p \Rightarrow_G p_1 \Rightarrow_G \ldots \Rightarrow_G p_{n-1} \Rightarrow_Gq \text{ \ \ }(n \geq 1)$$
of $q$ from $p$. Hence, $\Rightarrow^*$ is the reflexive, transitive closure of $\Rightarrow$, when we view it as a relation in $F_{\Sigma}(X \cup N)$.
\begin{df}\label{Definition.2.3.3}\rm
The \emph{forest generated} by a regular $\Sigma X$-grammar\index{forest!generated by regular $\Sigma X$-grammar} $G=(N,\Sigma,X,P,a_0)$ is the $\Sigma X$-forest
$$T(G)=\{t \in F_{\Sigma}(X) \mid a_0 \Rightarrow^*_G t \}.$$
Two regular $\Sigma X$-grammars $G_1$ and $G_2$ are said to be \emph{equivalent}\index{equivalence of!regular $\Sigma X$-grammars} if $T(G_1)=T(G_2)$.
\end{df}

\begin{ex}\label{Example.2.3.3}\rm
Let $\Sigma = \Sigma_0 \cup \Sigma_2$, $\Sigma_0=\{\omega \}$, $\Sigma_2=\{\sigma\}$ and $X=\{x\}$. Define the regular $\Sigma X$-grammar
$$G=(\{a,b\}, \Sigma,X,P,a),$$
where
$$P=\{a \rightarrow \sigma(x,\sigma(x,b)), a\rightarrow \sigma(\omega,a), b \rightarrow \sigma(x,x)  \}.$$
The tree
$$t= \sigma \left (\omega, \sigma(x,\sigma(x, \sigma(x,x))) \right )$$
is in $T(G)$ and it has the derivation
$$a \Rightarrow \sigma(\omega, a) \Rightarrow \sigma(\omega, \sigma(x,\sigma(x,b))) \Rightarrow t.$$
If the graphical representation of trees is used, this derivation can be written as in Fig.~\ref{Figure.p.69}.
\begin{figure}[h]
\input{Figure.p.69}
\caption{{}\label{Figure.p.69}}
\end{figure}
\epr
\end{ex}

A regular $\Sigma X$-grammar may be viewed as a context-free grammar with a terminal alphabet consisting of $\Sigma,X,$ the parentheses and the comma. Thus, if we treat trees as words, then the forests generated by regular tree grammars are special CF languages. However, we are mainly interested in them as forests, and we shall prove that exactly the recognizable forests can be generated by these grammars. To facilitate the proof first we show that the form of the productions may be restricted considerably without limiting the generative power of regular tree grammars.

To begin with, we note that productions of the form
$$a \rightarrow b \text{ \ \ \ }(a,b \in N)$$
are not needed. All such productions can be deleted if we add to $P$ all productions $a \rightarrow r$ ($a \in N, r \in F_{\Sigma}(X \cup N)-N$) such that $ a \Rightarrow^*b$ and $b\rightarrow r \in P$ for some $b\in N$. (It is easy to see that $a \Rightarrow^*b$ is decidable for $a,b \in N$.)

Call $\hg (r)$ the \emph{height} of the production\index{height of!production} $ a\rightarrow r$. If the height of a production $ a \rightarrow r$ is $>1$, then $r$ is of the form $\sigma(r_1, \ldots, r_m)$, where $m \geq 1, \sigma \in \Sigma_m$ and $\hg(r_i)<\hg(r)$ for each $i=1, \ldots, m$. If we introduce new nonterminal symbols $a_1, \dots, a_m$ and the productions
\begin{equation*}\tag{*}
a \rightarrow \sigma(a_1, \ldots, a_m)
\end{equation*}
and
\begin{equation*} \tag{**}
a_i \rightarrow r_i \quad (i=1,\ldots, m),
\end{equation*}
then the production $a \rightarrow r$ may be deleted without changing the forest generated. Indeed, any application of $a \rightarrow r$ can be replaced by an application of (*) followed by applications of the productions (**). On the other hand, none of the productions (**) can be used unless (*) has first been used, and when (*) has been applied it must be followed by applications of all productions (**) as there is no other way to rewrite the new nonterminals $a_i$. The total effect of these steps is the same as that of a single application of $a \rightarrow r$. Thus every production of height $>1$ can be replaced by productions of lesser height. The process can be repeated until there are no productions of height $>1 $. In (**) there may be productions of the type $a \rightarrow b$, but they can be eliminated. Hence each production of height 0 may be assumed to be of the type
\begin{equation*} \tag{i}
a \rightarrow x \quad (a\in N, x \in X)
\end{equation*}
or of the form
\begin{equation*} \tag{ii}
a \rightarrow \sigma \quad (a \in N, \sigma \in \Sigma_0).
\end{equation*}

A production of height 1 is of the form
$$a \rightarrow \sigma(r_1, \ldots, r_m) \quad (m \geq 1, \sigma \in \Sigma_m, a\in N),$$
where each $r_i$ is a frontier letter, a 0-ary operator or a nonterminal symbol. If $r_i$ is a letter from $X$ or a 0-ary operator, then we may substitute a new nonterminal symbol $d$ for it and introduce the production $d \rightarrow r_i$ of height 0 without changing the forest generated. Thus we may assume that all productions of height 1 are of the form
\begin{equation*} \tag{iii}
a \rightarrow \sigma(a_1, \ldots, a_m) \quad (m \geq 1, \sigma \in \Sigma_m, a, a_1, \ldots, a_m \in N).
\end{equation*}
We say that a regular tree grammar is in \emph{normal form}\index{normal form of regular tree grammar} if each of its productions is of type (i), (ii) or (iii). The previous discussion amounts to the following lemma.

\begin{lm}\label{Lemma.2.3.4}
Every regular tree grammar can be transformed into an equivalent regular tree grammar in normal form. \epr
\end{lm}

\begin{ex}\label{Example.2.3.5}\rm
None of the productions of the grammar considered in Example \ref{Example.2.3.3} is in normal form. The production $a \rightarrow \sigma(x,\sigma(x,b))$ can be replaced by the following set:
$$a \rightarrow \sigma(a_1,a_2), \quad a_1 \rightarrow x, \quad a_2 \rightarrow \sigma(a_1,b).$$
Notice that we could use the new nonterminal symbol $a_1$ twice since in both functions it should be rewritten as $x$. Similarly, the production $a \rightarrow \sigma(\omega, a)$ is replaced by the two productions
$$a \rightarrow \sigma(a_3,a) \quad \text{and} \quad a_3 \rightarrow \omega,$$
and the production $b \rightarrow \sigma(x,x)$ is replaced by $b \rightarrow \sigma(a_1, a_1)$ (we already have $a_1 \rightarrow x$). We have got a grammar in normal form with five nonterminal symbols $a,b,a_1,a_2$ and $a_3$, and the productions
\begin{eqnarray*}
a \rightarrow \sigma(a_1,a_2), \quad &a \rightarrow \sigma(a_3, a), \quad b \rightarrow \sigma(a_1,a_1),\\
 a_1 \rightarrow x, \quad &a_2 \rightarrow \sigma(a_1,b) \quad \text{and} \quad a_3 \rightarrow \omega.
\end{eqnarray*}
\epr
\end{ex}

The following minor generalization of regular tree grammars is introduced as a technical aid. An \emph{extended regular} $\Sigma X$-\emph{grammar}\index{regular sigmaxgrammar@regular $\Sigma X$-grammar!extended}
$$G=(N, \Sigma, X, P, A')$$
is defined otherwise exactly as a regular $\Sigma X$-grammar, but it has a set $A' \subseteq N$ of initial symbols. Also $\Rightarrow^*_G$ is defined the same way as for regular tree grammars. The forest generated by such a $G$ is
$$T(G)=\{t\in F_{\Sigma}(X) \mid a_0 \Rightarrow_G^* t \text{ for some }a_0 \in A'  \}.$$
It is immediately clear that every language generated by an extended regular tree grammar can be generated by an ordinary regular tree grammar, too.

\begin{thm}\label{Theorem.2.3.6}
The forests generated by regular tree grammars are exactly the recognizable forests.
\end{thm}
\pr
We associate with every NDF $\Sigma X$-recognizer $\bA =(A, \Sigma, X, \alpha, A')$ an extended regular $\Sigma X$-grammar
$$G=(A, \Sigma, X, P, A'),$$
where
\begin{gather*}
P=\{a\rightarrow x \mid x\in X, a \in x\alpha  \} \cup \{a \rightarrow \sigma \mid \sigma \in \Sigma_0, a \in \sigma^{\cA} \} \cup \\
 \{  a \rightarrow \sigma(a_1, \ldots, a_m) \mid m \geq 1, \sigma \in \Sigma_m, a,a_1, \ldots, a_m \in A, a\in \sigma^{\cA}(a_1, \ldots, a_m)   \}.
\end{gather*}
The grammar $G$ is in normal form (i.e., the productions are of type (i)--(iii)). It is clear that every extended regular $\Sigma X$-grammar in normal form arises this way from a NDF $\Sigma X$-recognizer. To prove the theorem it suffices now to show that $T(\bA)=T(G)$ for such an associated pair $\bA$ and $G$. To do this we show by tree induction that
\begin{equation*}\tag{*}
a \in t \hat{\alpha} \quad \text{iff} \quad a \Rightarrow^*_G t
\end{equation*}
holds for all $a \in A$ and $t \in F_{\Sigma}(X)$.

$1^\circ$ For $t=x \in X$, $a \in x \hat{a}$ iff $a \rightarrow x \in P$ iff $a \Rightarrow^* x$ (here we needed the fact that $G$ has no productions of the form $ a\rightarrow b$).

$2^\circ$ The case $t=\sigma \in \Sigma_0$ is similar: $a \in \sigma \hat{\alpha}$ iff $a\in \sigma^{\cA}$ iff $a \rightarrow \sigma \in P$ iff $a \Rightarrow^* \sigma$.

$3^\circ$ Let $t=\sigma(t_1, \ldots, t_m) \quad (m \geq 1)$ and suppose that (*) holds for $t_1, \ldots, t_m$ and all states. If $a \Rightarrow^* t$, then there is a derivation of the form
$$a \Rightarrow \sigma(a_1, \ldots, a_m) \Rightarrow^* \sigma(t_1, \ldots, t_m),$$
where $a_1, \ldots, a_m \in N$ and
$$a_i \Rightarrow^* t_i \quad \text{for} \quad i=1, \ldots, m.$$
Then $a \in \sigma^{\cA}(a_1, \ldots, a_m)$ by the definition of $P$, and (*) implies that $a_1 \in t_1 \hat{\alpha}, \ldots, a_m \in t_m \hat{\alpha}$. Hence,
$$a \in \sigma^{\gp \cA}\left (t_1 \hat{\alpha}, \ldots, t_m \hat{\alpha} \right ) = t \hat{\alpha}.$$
Conversely, $a\in t\hat{\alpha}$ means that
$$a \in \sigma^{\cA}(a_1, \ldots, a_m)$$
for some $a_1 \in t_1 \hat{\alpha}, \ldots, a_m \in t_m \hat{\alpha}$. But then (*) implies $a_1 \Rightarrow^* t_1, \ldots, a_m \Rightarrow^* t_m $. Also, $P$ contains the production $a \rightarrow \sigma(a_1, \ldots, a_m)$ and we get the required derivation
$$a \Rightarrow \sigma(a_1, \ldots, a_m) \Rightarrow^*\sigma(t_1, \ldots, t_m)=t.$$
This completes the proof of (*), and we have for every $\Sigma X$-tree $t$,

\begin{center}
\begin{tabular}{ c c l}
$t\in T(\bA)$ & iff & $t\hat{\alpha} \cap A' \neq \emptyset$ \\
& iff & $a\in t\hat{\alpha} $ \ for some \ $a\in A'$ \\
& iff & $a \Rightarrow^*_G t$ \ for some \ $a \in A'$ \\
& iff & $t \in T(G)$.
\end{tabular}
\end{center}
Hence $T(\bA)=T(G)$ as required.
\epr

%% file: Figure.p.69.tex
\centering
\raisebox{-0.34cm}{
\begin{tikzpicture}
	\node [label={$a$}] {};
\end{tikzpicture}} 
$\Rightarrow$
\raisebox{-0.4cm}{
\begin{tikzpicture}[level distance=0.8cm]
	\begin{scope}[execute at begin node=$, execute at end node=$]
		\node[node,label={below:\sigma}] {} 
		child{ node[node,label={left:a}] {}}
		child{ node[node,label={left:\omega}] {}};
	\end{scope}
\end{tikzpicture}}~
$\Rightarrow $
\raisebox{-0.25cm}{
\begin{tikzpicture}[level distance=0.8cm]
	\begin{scope}[execute at begin node=$, execute at end node=$]
		\node[node,label={right:\sigma}] {} 
		child[sibling distance=0.85cm]{ node[node,label={right:\sigma}] {} 
			child{ node[node,label={right:\sigma}] {}
				child{ node[node,label={left:b}] {}}
				child[sibling distance=1.5cm]{ node[node,label={left:x}] {}}}
			child[sibling distance=1.5cm]{ node[node,label={left:x}] {}}}
		child[sibling distance=1.5cm]{ node[node,label={left:\omega}] {}};
	\end{scope}
\end{tikzpicture}}
$\Rightarrow $
\raisebox{-0.25cm}{
\begin{tikzpicture}[level distance=0.8cm]
	\begin{scope}[execute at begin node=$, execute at end node=$]
		\node[node,label={right:\sigma}] {} 
		child[sibling distance=0.85cm]{ node[node,label={right:\sigma}] {} 
			child{ node[node,label={right:\sigma}] {}
				child{ node[node,label={right:\sigma}] {}
					child{ node[node,label={left:x}] {}}
					child[sibling distance=1.5cm]{ node[node,label={left:x}] {}}}
				child[sibling distance=1.5cm]{ node[node,label={left:x}] {}}}
			child[sibling distance=1.5cm]{ node[node,label={left:x}] {}}}
		child[sibling distance=1.5cm]{ node[node,label={left:\omega}] {}};
	\end{scope}
\end{tikzpicture}}

%% file: Section.2.4.tex
In this section some more insight into the family of recognizable forests is 
gained by studying its closure properties with respect to various forest 
operations. In the following definitions and theorems all forests usually have 
the same ranked alphabet and the same frontier alphabet. To show that this is 
no serious limitation, we note the following simple fact.
\begin{lm}\label{Lemma.2.4.1} Let $\Sigma$ and $\Omega$ be ranked alphabets 
such that $\Sigma \subseteq \Omega$, and let $X$ and $Y$ be frontier alphabets 
such that $X \subseteq Y.$ Then 
\[\Rec(\Sigma, X)= \Rec(\Omega, Y) \cap \gp F_\Sigma(X).
\tag*{\epr}\]
\end{lm}

Of course, the lemma presupposes the point of view that every $\Sigma X$-forest
is also an $\Omega Y$-forest. Now let $\Sigma$ and $\Omega$ be any ranked 
alphabets such that $\Sigma_m \cap \Omega_n=\emptyset$ whenever $m\neq n$. 
Also, let $X$ and $Y$ be arbitrary frontier alphabets. The lemma implies that 
if $S \in \Rec(\Sigma, X)$ and $T \in \Rec(\Omega, Y)$, then $S$ and $T$ can 
be regarded as recognizable forests over a common ranked alphabet 
$\Sigma \cup \Omega$ and a common frontier alphabet $X \cup Y$.  

\begin{thm}\label{Theorem.2.4.2}
If $S, T \in \Rec(\Sigma, X)$, then $S \cap T$, $S \cup T$ and $S-T$ are also 
recognizable $\Sigma X$-forests.
\end{thm}
\pr Suppose $S$ and $T$ are recognized by the $\Sigma X$-recognizers $\bA$ and
 $\bB$, respectively. Let ${\cal C}=\cA \times \cB$ and define 
$$\gamma: X \rightarrow C \hbox{\hspace{3mm}}\mbox{ by }\hbox{\hspace{3mm}}  x \mapsto (x\alpha, x\beta).$$
Then 
$$t\hat{\gamma}=(t\hat{\alpha}, t\hat{\beta}) \hbox{\hspace{3mm}}
 \mbox{ for all } t\in 
F_\Sigma(X).$$
This implies that we get from ${\cal C}$ and $\gamma$ $\Sigma X$-recognizers 
for  $S \cap T$, $S \cup T$ and $S-T$ by choosing, respectively, as the set of final states $A'\times B'$,  $A'\times B \cup A \times B'$, and 
 $A'\times (B-B')$. For example, let
$$\bC=({\cal C}, \gamma, A' \times B').$$
For any $t\in F_\Sigma(X)$, 
\begin{eqnarray*}
t\in T(\bC) & \mbox{ iff }  & t\hat{\gamma}=(t\hat{\alpha}, t\hat{\beta})\in 
A'\times B'\\
            & \mbox{ iff }   & t \in T(\bA)\cap T(\bB).
\end{eqnarray*}
That is, $T(\bC)=S\cap T$.
\epr

\

Note that the complement $F_\Sigma(X)-T$ of a recognizable $\Sigma X$-forest
$T$ is recognizable. If $T$ is recognized by a  $\Sigma X$-recognizer $\bA$, 
then the complement is recognized by $(\cA, \alpha, A-A')$. 
\begin{df}\label{Definition.2.4.3}\rm
Let $(T_x\mid x\in X)$ be an $X$-indexed family of $\Sigma X$-forests. For each 
$\Sigma X$-tree $t$ we define a forest $t(x\leftarrow T_x\mid x\in X)$, mostly 
written simply $t(x\leftarrow T_x)$, as follows:
\begin{enumerate}
\item[$1^\circ$] If $t=z\in X$, then $t(x\leftarrow T_x)=T_z$.
\item[$2^\circ$] If $t=\sigma\in \Sigma_0$, then $t(x\leftarrow T_x)=\sigma$.
\item[$3^\circ$] If $t=\sigma(t_1, \ldots, t_m) (m\geq 1)$, then 
 $$t(x\leftarrow T_x)=\{\sigma(s_1, \ldots, s_m)\mid s_i 
\in t_i(x\leftarrow T_x)  \mbox{ for } i=1, \ldots, m\}.$$
\end{enumerate}
The {\it forest product}
\index{product!forest}
 of the family  $(T_x \mid x\in X)$ with the  
$\Sigma X$-forest $T$ is defined as the $\Sigma X$-forest
$$T(x\leftarrow T_x\mid x\in X)= \bigcup ( t(x\leftarrow T_x\mid x\in X)
 \mid t \in T).$$
\end{df}

We shall usually write just $T(x\leftarrow T_x)$. If $T$ consists of a single 
$\Sigma X$ tree $t$, then 
$$T(x\leftarrow T_x)=t(x\leftarrow T_x).$$
The trees $t(x\leftarrow T_x)$ are obtained from $t$ by replacing every occurrence of each letter $x$ by a tree from the corresponding forest $T_x$. Different
occurrences of the same letter $x$ may be rewritten as different trees from $T_x$. 

If $x_1, \ldots, x_n\in X$, then we use the notation
$$T(x_1 \leftarrow T_1, \ldots, x_n \leftarrow T_n)$$
for the forest product $T(x\leftarrow T_x)$, 
where
$$T_x=
\left
\{\begin{array}{lll}
T_i & \mbox{for} & x=x_i \hbox{\hspace{3mm}} (i=1, \ldots, n),\\
x   & \mbox{for} & x\not \in \{ x_1, \ldots, x_n\}. 
\end{array}
\right.
$$
If the letters $x_1, \ldots, x_n$ and their order are understood, then this 
notation may be further simplified to $T(T_1, \ldots, T_n)$.

The comments presented at the beginning of the section show that the 
definition of forest products also includes the cases, where 
$T \subseteq F_\Sigma(X)$ and $T_x \subseteq F_\Omega(Y)$ $(x\in X)$ for any 
such alphabets that $\Sigma_m \cap \Omega_n=\emptyset$ whenever $m\neq n$. 
If $T$ is a $\Sigma X$-forest and the forests $T_x$ are $\Omega Y$-forests,
then $T(x\leftarrow T_x)$ is a $(\Sigma \cup \Omega) Y$-forest.
\begin{ex}\label{Example.2.4.4}\rm
Let $\Sigma=\Sigma_0\cup \Sigma_2$, $\Sigma_0=\{\omega\}$, 
$\Sigma_2=\{\sigma \}$, $X=\{x, y\}$ and $Y=\{y, z\}$. If 
$t=\sigma(x, \sigma(y, x))$, $T_x=\{\sigma(y, z), z\}$ and 
$T_y=\{\sigma(\omega, y), \sigma(z, z)\}$, then $t(x\leftarrow T_x, y\leftarrow T_y)$ contains eight trees, among them the tree $\sigma(\sigma(y, z), \sigma(\sigma(\omega, y), z))$.
\epr
\end{ex}

The following special type of forest products is important.

\begin{df}\label{Definition.2.4.5}\rm
Let $S$ and $T$ be $\Sigma X$-forests and $z \in X$. The $z$-{\it product}
\index{z-product@$z$-product}
of $S$ and $T$ is the forest product 
$$S \cdot_z T=T(x \leftarrow T_x\mid x\in X)$$
where $T_z=S$ and $T_x=x$ for all $x\in X$, $x\neq z$.
\end{df}

The trees in $S \cdot_zT$ are obtained by taking a tree $t$ from $T$ and 
substituting a tree from $S$ for every occurrence of $z$ in $t$. Different 
occurrences of $z$ may be replaced by different trees from $S$. 

\begin{thm}\label{Theorem.2.4.6}
If $T \in \Rec(\Sigma, X)$ and $T_x \in \Rec(\Sigma, X)$ for all $x\in X$, then 
$T(x \leftarrow T_x)\in \Rec(\Sigma, X)$. In particular, $\Rec(\Sigma, X)$ is
closed under all $x$-products $(x \in X)$.
\end{thm}
\pr Here it is convenient to use regular tree grammars. Suppose $T$ and the 
forests $T_x$  $(x \in X)$ are generated by the regular $\Sigma X$-grammars 
$G=(N, \Sigma, X, P, a_0)$ and $G_x=(N_x, \Sigma, X, P_x, a_x)$
$(x \in X)$, respectively. We may assume that the grammars are in normal form 
and that their sets of nonterminal symbols are pairwise disjoint. Construct a 
regular $\Sigma X$-grammar 
$$G'=(N', \Sigma, X, P', a_0)$$
with $N'=N \cup \bigcup (N_x\mid x \in X)$ and 
$$P'=P''\cup \{ a \rightarrow a_x \mid x\in X, \hbox{\hspace{3mm}} a\rightarrow x\in P\} \cup 
\bigcup (P_x\mid x\in X), $$
where $P''$ is $P$ with all productions of the form 
$a\rightarrow x$ $(a\in N, x\in X)$ deleted.

We claim that $T(G')=T(x \leftarrow T_x)$. The idea is that every derivation 
$a_0 \Rightarrow_G \ldots \Rightarrow_G t$ of a tree $t \in T$ can be imitated 
by the productions in $P''$ up to the point where frontier letters $x\in X$ 
are to be generated. Instead of generating a leaf $x$ one transfers then by a 
production $a\rightarrow a_x$ to the beginning of a derivation which generates 
any tree $t_x \in T_x$ in place of the leaf. This means that $G'$ can generate 
all of $T(x\leftarrow T_x)$. On the other hand, every derivation in $G'$ can be
brought into this form by rearranging the applications of the productions 
suitably. Hence, $T(G')\subseteq T(x\leftarrow T_x)$. For a formal proof it 
suffices to show that 
\begin{equation*}\tag{*}
 a \Rightarrow ^*_{G'} p \hbox{\hspace{3mm}} \mbox{ iff } \hbox{\hspace{3mm}}
(\exists q\in F_\Sigma (X)) \hbox{\hspace{1mm}}
a \Rightarrow ^*_G q, \hbox{\hspace{3mm}} p \in q(x \leftarrow T_x) 
 \end{equation*}
holds for all $a\in N$ and $p\in F_\Sigma(X)$. We proceed by tree induction on 
$p$. The fact that the grammars $G$ and $G_x$ are in normal form is used 
without comment.

\begin{itemize}
\item[$1^\circ$] Let $p=y\in X$. Suppose there is a $q\in F_\Sigma(X)$ such that 
$a\Rightarrow ^*_G q$ and $y\in q(x \leftarrow T_x)$. This is possible only in 
case $q=z$ and $y\in T_z$ for some $z\in X$. Then $a\rightarrow z\in P$ and 
hence $a\rightarrow a_z$, $a_z \rightarrow y\in P'$. We get the derivation 
$$a \Rightarrow _{G'} a_z \Rightarrow _{G'} y.$$
On the other hand, all derivations of $y$ from $a$ in $G'$ are of this form.
Hence, if $a\Rightarrow ^*_{G'} y$, then $a\rightarrow a_z$, 
$a_z \rightarrow y\in P'$ for some $z \in X$. This means that 
$a\rightarrow z\in P$ and $a_z\rightarrow y\in P_z$, and thus $z$ is the 
required tree $q$.

\item[$2^\circ$] Let $p=\sigma \in \Sigma_0$. 
\begin{itemize}
\item[(2a)] If there is a $q$ such that $a\Rightarrow ^*_G q$ and 
$\sigma \in q(x\leftarrow T_x)$, then there are two possibilities. The first
one is that $q=\sigma$. Then $P$ and $P'$ both contain $a\rightarrow \sigma$ 
and we get the required derivation $a\Rightarrow^*_ {G'} \sigma$ in one step.
The other possibility is that $q=x\in X$ and $P_x$ contains 
$a_x \rightarrow \sigma$. Then $a \rightarrow a_x$  and $a_x\rightarrow \sigma$ 
are in $P'$ and we get the derivation 
$$a\Rightarrow_{G'} a_x \Rightarrow_{G'} \sigma .$$ 

\item[(2b)] Suppose $a\Rightarrow ^*_{G'} \sigma$. One possibility is that 
$a\rightarrow \sigma \in P'$. Then 
$a\rightarrow \sigma$ is in $P$, too, and we may choose $q=\sigma$. The only 
alternative is that the derivation is of the form 
$a\Rightarrow _{G'} a_x\Rightarrow _{G'} \sigma$ for some $x\in X$. Then 
$a \rightarrow x\in P$ and $\sigma \in T_x$, and we may put $q=x$. 
\end{itemize}

\item[$3^\circ$] Let $p=\sigma(p_1, \ldots, p_m)$ $(m> 0)$.
\begin{itemize}
\item[(3a)] Suppose we have a tree $q$ such that $a\Rightarrow^*_G q$ and 
$p\in q(x \leftarrow T_x)$. Again there are two cases to consider. 
If $q=z \in X$, then $p\in T_z$, $a\rightarrow z\in P$ and 
$a_z \Rightarrow^*_{G_z} p$. Now $a\rightarrow a_z\in P'$ and, since 
$P_z \subseteq P'$, we get 
$$a \Rightarrow _{G'} a_z\Rightarrow^* _{G'} p.$$
The other possibility is that 
$$q=\sigma(q_1, \ldots, q_m)$$
for some $q_1, \ldots, q_m\in F_\Sigma(X)$. Then 
$$p_i \in q_i(x\leftarrow T_x) \hbox{\hspace{3mm}} (i=1, \ldots, m)$$
and the derivation $a \Rightarrow^*_G q$ must begin with 
a step  
$$a \Rightarrow_G \sigma(a_1, \ldots, a_m)$$
such that 
$$a_i \Rightarrow^*_G q_i \hbox{\hspace{3mm}}
\mbox{ for } \hbox{\hspace{3mm}} i=1, \ldots, m.$$
Our silent inductive assumption yields 
$$a_i \Rightarrow^*_{G'} p_i \hbox{\hspace{3mm}}
\mbox{ for } \hbox{\hspace{3mm}} i=1, \ldots, m.$$
Combining these derivations with 
$a\rightarrow \sigma(a_1, \ldots, a_m)\in P'$ we get 
$a \Rightarrow^*_{G'} p$.

\item[(3b)] Suppose $a \Rightarrow^*_{G'} p$. This could mean that 
$a\rightarrow z\in P$ and $a_z \Rightarrow^*_{G_z} p$
for some $z \in X$. Then we may choose $q=z$. The other
possibility is that the derivation takes the form 
$$a  \Rightarrow_{G'}\sigma(a_1, \ldots, a_m) \Rightarrow^*_{G'}
\sigma(p_1, \ldots, p_m).$$
Then there exist $\Sigma X$-trees $q_i$ such that
$$a_i  \Rightarrow^*_G q_i, \hbox{\hspace{3mm}} p_i \in q_i(x \leftarrow T_x) \hbox{\hspace{3mm}} 
(i=1, \ldots, m).$$
Now we may put $q=\sigma(q_1, \ldots, q_m)$. \epr
\end{itemize}
\end{itemize}

\

Next we generalize the iteration operation taking the $x$-products as the 
starting point.
\begin{df}\label{Definition.2.4.7}\rm
Let $T$ be any $\Sigma X$-forest and let $x\in X$. Put 
$T^{0, x}=\{x\}$ and 
$$T^{j+1, x} = T^{j, x} \cdot_x T \cup T^{j, x} $$
for all $j \geq 0$. Then the $x$-{\it iteration} of  $T$ is the
\index{x-iteration@$x$-iteration}
$\Sigma X$-forest
$$T^{* x}=\bigcup ( T^{j, x}\mid j \geq 0).$$
\end{df}

The forest $T^{* x}$ is obtained as follows. First include $x$. New 
members of $T^{* x}$ are obtained by substituting in some $t \in T$ for 
every occurrence of $x$  some tree already known to be in $T^{* x}$. 
Note that 
$T^{1, x}=T \cup x$ and $ T^{j, x} \subseteq T^{j+1, x}$ for every $j \geq 0$.

\begin{thm}\label{Theorem.2.4.8}
If $T \in \Rec(\Sigma, X)$, then $T^{* x} \in \Rec(\Sigma, X)$ for 
each $x\in X$. 
\end{thm}
\pr
Let $G=(N, \Sigma, X, P, a_0)$ be a regular tree grammar in normal form generating the forest 
$T$. Construct an extended regular $\Sigma X$-grammar  
$G'=(N', \Sigma, X, P', A')$, where 
\begin{enumerate}
\item[(1)] $N'= N\cup \{d \}$ $(d \not\in N)$,

\item[(2)] $P'=P \cup \{d \rightarrow x\}\cup 
\{a \rightarrow r \mid a\rightarrow x \in P, \hbox{\hspace{1mm}} a_0\rightarrow r\in P\}$, and

\item[(3)] $A'=\{a_0, d \}$. 
\end{enumerate} 
It is not hard to see that $T(G')=T^{* x}$.
\epr

\

The following operation may be seen as a converse to the $x$-product.
\begin{df}\label{Definition.2.4.9}\rm
Let $S$ and $T$ be $\Sigma X$-forests and let $x\in X$. The $x$-{\it quotient}
\index{x-quotient@$x$-quotient}
of $T$ by $S$ is the forest 
$$S^{-x}T=\{p\in F_\Sigma(X)\mid S\cdot_x \{p\} \cap T\neq \emptyset\}.$$
If $S=\{s \}$ is a singleton, then we write  $S^{-x}T=s^{-x}T$.
\end{df} 

A tree $p$ is in $S^{-x}T$ iff one can convert it into a tree in $T$ by
substituting for every occurrence of $x$ a tree from $S$. If $\Sigma$ is unary and $X=\{x\}$, and if we identify the tree $\sigma_k( \ldots \sigma_1(x)\ldots )$ with the word $\sigma_1\ldots \sigma_k$, then 
$$S^{-x}T=S^{-1}T=\{u \in \Sigma^*\mid Su\cap T\neq \emptyset\}$$
is the usual (left) quotient language.

\begin{thm}\label{Theorem.2.4.10}
If $T \in \Rec(\Sigma, X)$ and $S$ is any $\Sigma X$-forest, then  $S^{-x}T$
is recognizable for every $x\in X$. Moreover, the number of different 
$x$-quotients $S^{-x}T$ for any fixed $T \in \Rec(\Sigma, X)$ is finite.
\end{thm}
\pr Let $\bA$ be a $\Sigma X$-recognizer for $T$. We define an 
NDF $\Sigma X$-recognizer 
$$\bB=(\cA, \beta, A')$$
which is identical to $\bA$ (when states $a\in A$ and singleton sets $\{a\}$
are identified) except for the initial assignment which is defined so that 
$$x\beta=S\hat{\alpha}$$
and
$$z\beta=\{z\alpha\}  \hbox{\hspace{3mm}}\mbox{ for all }  \hbox{\hspace{3mm}}
z\in X, z\neq x.$$
Here $S\hat{\alpha}$ is the set of all states $s\hat{\alpha}$ in which $\bA$ may be after reading a tree $s$ from $S$. By tree induction one verifies that
$$t\hat{\beta}=(S\cdot_xt)\hat{\alpha}$$
for all $t\in F_\Sigma(X)$. Hence 
\begin{eqnarray*}
t \in T(\bB)& \mbox{ iff }  & t\hat{\beta} \cap A'\neq \emptyset \\
            & \mbox{ iff }  & (S\cdot_xt) \hat{\alpha} \cap A'\neq \emptyset\\
            & \mbox{ iff } & S\cdot_xt \cap T\neq \emptyset \\
            & \mbox{ iff } & t \in S^{-x} T
\end{eqnarray*}
for all $t \in F_\Sigma(X)$. This implies $S^{-x} T=T(\bB)$. The second 
statement follows from this construction as the number of possible $\beta$'s is finite. 
\epr

\

Next we introduce the forest operation corresponding to the $\sigma$-catenation of trees which was defined in Section \ref{Section.2.1}.
\begin{df}\label{Definition.2.4.11}
\rm Let $\sigma\in \Sigma$ be an $m$-ary operator and let 
$T_1, \ldots, T_m$ be $m$ $\Sigma X$-forests for some $m\geq 0$. The 
$\sigma$-{\it product} of the forests $T_1, \ldots, T_m$ is the forest 
$$\sigma(T_1, \ldots, T_m)=
\{\sigma(t_1, \ldots, t_m)\mid t_1\in T_1, \ldots, t_m \in T_m\}.$$
\end{df}

If $m=0$, then the $\sigma$-product is always $\{\sigma\}$.
\index{sigma-product@$\sigma$-product}
 In general, 
$$\sigma(T_1, \ldots, T_m)= 
\{\sigma(x_1, \ldots, x_m) \}(x_1\leftarrow T_1, \ldots, x_m \leftarrow T_m).$$

From Theorem \ref{Theorem.2.4.6} we get the following result which could 
easily be proved directly, too.
\begin{cor}\label{Corollary.2.4.12}
If $\sigma\in \Sigma_m$ and $T_1, \ldots, T_m\in \Rec(\Sigma, X)$ 
$(m\geq 0)$, then $\sigma(T_1, \ldots, T_m)\in \Rec(\Sigma, X)$.
\epr
\end{cor} 

We shall now consider some operations in which forests are generally 
transformed into forests over another ranked alphabet. The ranked alphabets 
will be $\Sigma$ and $\Omega$. Moreover, we introduce for every $m\geq 0$, a new alphabet 
$$\Xi_m=\{\xi_1, \ldots, \xi_m\}$$
which is assumed to be disjoint from all other alphabets. 
\begin{df}\label{Definition.2.4.13}\rm
Suppose we are given a mapping
$$h_X: X \rightarrow F_\Omega(Y)$$
and for each $m\geq 0$ a mapping 
$$h_m: \Sigma_m \rightarrow F_\Omega(Y\cup \Xi_m).$$
The {\it tree homomorphism} 
\index{homomorphism!tree}
determined by these mappings is the mapping
$$h: F_\Sigma(X) \rightarrow F_\Omega(Y)$$
defined as follows:
\begin{enumerate}
\item[$1^\circ$] $h(x)=h_X(x)$ for each $x\in X$.
\item[$2^\circ$] $h(\sigma(t_1, \ldots, t_m))=h_m(\sigma)
(\xi_1 \leftarrow h(t_1), \ldots, \xi_m \leftarrow h(t_m))$
for all $m\geq 0$, $\sigma\in \Sigma_m$ and $t_1, \ldots, t_m \in F_\Sigma(X)$. 
\end{enumerate}
The tree homomorphism $h$ is said to  be {\it linear} 
\index{homomorphism!linear tree}
if no letter $\xi_i$ 
appears  more than once in $h_m(\sigma)$ for any $m\geq 0$ and
$\sigma\in \Sigma_m$.
\end{df}

To define such an $h$ it obviously suffices to give $h_X$ and the mappings
$h_m$ for which $\Sigma_m \neq \emptyset$.
\begin{ex}\label{Example.2.4.14}\rm 
Let $\Sigma=\Sigma_2=\{\,|\,\}$, $\Omega= \Omega_1\cup \Omega_2$, 
$\Omega_1=\{'\}$, $\Omega_2=\{\vee\}$ and $X=Y=\{x, y\}$. Define $h_X$ and 
$h_2$ by the conditions 
$$h_X(x)=x, \hbox{\hspace{3mm}}
h_X(y)=y \mbox{ and } h_2(|)=\vee ('(\xi_1), '(\xi_2)).$$
If we interpret $|$ as the Sheffer stroke (i.e., the 2-place NAND), 
$\vee$ as the symbol of disjunction and $'$ as the symbol of negation,
then the tree homomorphism $h$ defined by $h_X$ and $h_2$
transforms
$|$-expressions in variables $x$ and $y$ into equivalent expressions which 
use $\vee$ and $'$ only. If the more customary way to write Boolean 
expressions is used, we get, for example, 
\begin{eqnarray*}
h((x|y)|(x|x)) & = & h(x|y)'\vee h(x|x)'\\
               & = & (x'\vee y')' \vee (x'\vee x')'.
\end{eqnarray*}
This tree homomorphism is linear.
\epr
\end{ex}

Tree homomorphisms are not really  homomorphisms in the sense of algebra. The 
concept is the result of the dual nature of words. When one generalizes from 
languages to forests, words are usually treated as unary terms. On the other 
hand, many concepts in language theory arise from the interpretation of words 
as elements of a free monoid. Here the initial concept was that of a  
homomorphism from the free monoid generated by an alphabet $\Sigma$ to the 
free monoid generated by another alphabet $\Omega$. Such a homomorphism 
rewrites every letter in a word over $\Sigma$ as a word over $\Omega$. When 
$\Sigma$ and $\Omega$ are now viewed as unary ranked alphabets, this means that 
every operator from $\Sigma$ is rewritten as a piece of $\Omega$-tree to
be combined with other such pieces to form the image of a given $\Sigma$-word. The generalization of such mappings to the case of arbitrary ranked alphabets gives tree homomorphisms.

The following example shows that tree  homomorphisms do not always preserve recognizability. 
\begin{ex}\label{Example.2.4.15}\rm Put $\Sigma=\Sigma_1=\{\sigma\}$,
$X=Y=\{x\}$ and $\Omega=\Omega_2=\{\omega\}$. Define $h_X$ and $h_1$ so that 
$$h_X(x)=x \mbox{ and } h_1(\sigma)=\omega(\xi_1, \xi_1).$$ 
All $\Sigma X$-trees are of the type
$$t_k=\sigma(\sigma(\ldots \sigma(x)\ldots))=\sigma^k(x) \hbox{\hspace{3mm}} 
(k\geq 0).$$
Obviously, $h(t_0)=h_X(x)=x$ and, for all $k\geq 0$, 
$$h(t_{k+1})= \omega(h(t_k), h(t_k)).$$
Thus $h(F_\Sigma(X))$ consists of the trees 
$$s_0=x, \hbox{\hspace{3mm}} s_1=\omega(x, x), \ldots, s_{k+1}=\omega(s_k, s_k),
 \ldots\hbox{\hspace{1mm}}.$$
Suppose $\bA=(A, \Omega, Y, \alpha, A')$ is an $\Omega Y$-recognizer such that 
$T(\bA)=h(F_\Sigma (X))$. There must exist two integers $i, j \geq 0$, 
$i\neq j$, such that $s_i \hat{\alpha}=s_j \hat{\alpha}$. But then 
$$\omega(s_i, s_j)\hat{\alpha}= \omega^{\cA}(s_i\hat{\alpha}, s_j\hat{\alpha})=
 \omega^{\cA}(s_i\hat{\alpha}, s_i\hat{\alpha})=s_{i+1}\hat{\alpha}\in A'$$
would imply $\omega(s_i, s_j) \in h(F_\Sigma(X))$. Thus $h(F_\Sigma(X))$ 
cannot be recognizable.
\epr
\end{ex}

The nonpreservation of recognizability in Example \ref{Example.2.4.15} is due 
to the ability of the tree homomorphism to create arbitrarily large identical 
subtrees by copying. No tree recognizer can check whether trees of unbounded 
height are identical or not. Such copying is precluded by linearity, and the 
following closure theorem holds.
\begin{thm}\label{Theorem.2.4.16}
If $h: F_\Sigma(X) \rightarrow F_\Omega(Y)$ is a linear tree homomorphism and 
$T\in \Rec (\Sigma, X)$, then 
 $h(T)\in \Rec (\Omega, Y)$.
\end{thm}
\pr Let $G=(N, \Sigma, X, P, a_0)$ be a regular tree grammar in normal 
form generating $T$. We may assume that $G$ has no superfluous nonterminal 
symbols from which no $\Sigma X$-tree can be generated. Let $\Sigma'$ and 
$\Omega'$ be the ranked alphabets which are obtained by adding all nonterminal 
symbols $a\in N$ to $\Sigma$ and $\Omega$, respectively, as nullary operators.
We extend $h$ to a tree homomorphism
$$h': F_{\Sigma'}(X) \rightarrow F_{\Omega'} (Y)$$ 
by continuing $h_0$ to a mapping
$$h'_0: \Sigma_0\cup N \rightarrow F_{\Omega'} (Y)$$ 
so that $h'_0(a)=a$ for all $a\in N$. Now let  
$$G'=(N, \Omega, Y, P', a_0)$$ 
be the regular $\Omega Y$-grammar, where 
$$P'=\{a \rightarrow h'(p) \mid a \rightarrow p \in P\},$$
i.e., $G'$ is obtained simply by replacing in every production 
$ a \rightarrow p\in P$ the right-hand side by the tree $h'(p)$. The theorem
follows when we show that $T(G')=h(T)$. This again is obvious once we have 
shown that  
\begin{equation*}\tag{*}
a \Rightarrow ^*_{G'} t 
 \hbox{\hspace{3mm}}
\mbox { iff }  \hbox{\hspace{3mm}}(\exists s\in F_\Sigma (X)) 
 \hbox{\hspace{1mm}}
h(s)=t, \hbox{\hspace{3mm}}  a \Rightarrow ^*_G s
\end{equation*}
holds for all $a\in N$ and $t \in F_\Omega(Y)$. 
We prove the two directions of (*) separately.

Suppose  $a \Rightarrow ^*_{G'} t $ for some $a\in N$ and $\Omega Y$-tree $t$. 
We prove the existence of the required $s$ by induction on the length of the 
shortest derivation of $t$ from $a$.

\begin{itemize}
\item[$1^\circ$]
 If $t$ is obtained by a one-step derivation, then $P'$ contains the 
production $a\rightarrow t$. Then $P$ contains a production $a\rightarrow r$ 
such that $h'(r)=t$. If $r$ does not contain any nonterminal symbols, we may 
put $s=r$. Otherwise we choose for every $b\in N$ appearing in $r$ a tree 
$r_b \in F_\Sigma(X)$ such that $b \Rightarrow^*_Gr_b$. Let $s$ be the tree
obtained by substituting in $r$ these trees for the corresponding nonterminal 
symbols. Then $h(s)=h'(r)=t$ since $h'$ deletes all nonterminal symbols from 
$r$.
Moreover,
$$a \Rightarrow_G r \Rightarrow^*_Gs,$$
and $s$ is the required tree.

\item[$2^\circ$] Suppose now that the derivation consists of $k$ steps $(k>1)$
and that (*) holds whenever a shorter derivation exists. The first step 
must be the application of a production $a\rightarrow h'(p)$, where 
$a\rightarrow p \in P$. Since $G$ is in normal form, 
$$p=\sigma(a_1, \ldots, a_m)$$
for some $m> 0$, $\sigma \in  \Sigma_m$ and $a_1, \ldots, a_m \in N$. 
The derivation of $t$ can now be written in the form 
$$a \Rightarrow _{G'} h_m(\sigma) (\xi_1\leftarrow a_1, \ldots, 
\xi_m \leftarrow a_m) \Rightarrow _{G'} \ldots \Rightarrow _{G'} t.$$
For each $\xi_i$ $(i=1, \ldots, m)$ which is present in $h_m(\sigma)$ we have a 
subderivation 
$$a_i \Rightarrow _{G'} \ldots \Rightarrow _{G'} t_i (\in F_\Omega(Y))$$
of length less than $k$. The linearity of $h$ implies that such a $\xi_i$ 
appears in $h_m(\sigma)$ exactly once, and hence $t_i$ is unique. 
For every $t_i$ there is an $s_i\in F_\Sigma(X)$ such that 
$h(s_i)=t_i$ and $a_i \Rightarrow ^*_G s_i$. If a certain $\xi_i$ does not 
appear in $h_m(\sigma)$, then we choose any $s_i \in F_\Sigma(X)$ such that 
$a_i \Rightarrow ^*_G s_i$ and put $t_i=h(s_i)$. 
With these choices we get a tree 
$$s=\sigma(s_1, \ldots, s_m)\in F_\Sigma(X)$$
such that 
$$a  \Rightarrow _G \sigma(a_1, \ldots, a_m)  \Rightarrow ^*_G
\sigma(s_1, \ldots, s_m)=s$$
and 
$$h(s)=h_m(\sigma)(\xi_1\leftarrow h(s_1), \ldots, \xi_m \leftarrow
h(s_m))=t.$$
\end{itemize}

Now we shall prove the converse part of (*). Suppose 
$a\Rightarrow ^*_G s$ and $h(s)=t$ for some 
$a\in N$, $s\in F_\Sigma(X)$ and $t\in F_\Omega(Y)$. To show that this implies
$a\Rightarrow ^*_{G'} t$ we proceed by induction on the length of the shortest 
derivation $a \Rightarrow _G \ldots  \Rightarrow _G s$.

\begin{itemize}
\item[$1^\circ$] If there is a derivation of length one, then it consists simply of 
the application of the production $a\rightarrow s$. But then 
 $a\rightarrow t$ is a production of $G'$ and
$a\Rightarrow _{G'} t$ is the required derivation.

\item[$2^\circ$] Suppose now that the derivation is of the form 
$$a  \Rightarrow _G \sigma(a_1, \ldots, a_m)  \Rightarrow _G \ldots 
\Rightarrow _G \sigma(s_1, \ldots, s_m)=s,$$
where $m>0$, $\sigma\in \Sigma_m$ and $a_1, \ldots, a_m \in N$. For every 
$i=1, \ldots, m$ there is a shorter derivation 
$$a_i \Rightarrow_ G \ldots \Rightarrow_ G s_i.$$
Hence, $a_i \Rightarrow ^*_{G'} h(s_i)$ for each $i=1, \ldots, m$. Moreover, 
$P'$ contains the production 
$$a\rightarrow h_m(\sigma)(\xi_1 \leftarrow a_1, \ldots, \xi_m \leftarrow a_m)$$
corresponding to the production 
$a\rightarrow \sigma(a_1, \ldots, a_m)$ of $G$. Now the required 
derivation is
$$a \Rightarrow_{G'}  
h_m(\sigma)(\xi_1 \leftarrow a_1, \ldots, \xi_m \leftarrow a_m)\Rightarrow_{G'}  
\ldots $$
$$ \Rightarrow_{G'} 
h_m(\sigma)(\xi_1 \leftarrow h(s_1), \ldots, \xi_m \leftarrow h(s_m))$$
$$=h(s)=t.$$
This concludes the proof.
\epr
\end{itemize}

Next we show that arbitrary inverse tree homomorphisms preserve 
recognizability. We need the following technical lemma. Its proof is left as
an exercise.
\begin{lm}\label{Lemma.2.4.17} Consider a $\Sigma$-algebra $\cal A$ and a 
mapping $\alpha: X \rightarrow A$, where $X \cap A=\emptyset$. Let 
$$\overline{\alpha}: \cF_\Sigma (X\cup A) \rightarrow \cA$$
be the unique homomorphism such that $\overline{\alpha}|X=\alpha$ and 
$\overline{\alpha}|A=1_A$. Then $\overline{\alpha}| \cF_\Sigma (X)=
\hat{\alpha}$ and 
$$p(\xi_1\leftarrow p_1, \ldots, \xi_k \leftarrow p_k)\hat{\alpha}=
p(\xi_1\leftarrow p_1\hat{\alpha}, \ldots, 
\xi_k \leftarrow p_k\hat{\alpha})\overline{\alpha}$$
for all $k\geq 0$, $p\in F_\Sigma (X \cup \Xi_k)$ and 
$p_1, \ldots, p_k \in F_\Sigma(X)$. 
\epr
\end{lm}
\begin{thm}\label{Theorem.2.4.18}
Let $h: F_\Sigma(X) \rightarrow F_\Omega(Y)$ be a tree homomorphism. If 
$T \in \Rec(\Omega, Y)$, then $h^{-1}(T)\in \Rec(\Sigma, X)$.
\end{thm}
\pr
Let  $\bA=(A, \Omega, Y, \alpha, A')$ be  an $\Omega Y$-recognizer for $T$.
We construct a $\Sigma X$-recognizer $\bB=(A, \Sigma, X, \beta, A')$
as follows. For any $m\geq 0$, $\sigma \in \Sigma_m$ and
$a_1, \ldots, a_m \in A$, we put
$$\sigma^{\cB}(a_1, \ldots, a_m)=h_m(\sigma)(\xi_1\leftarrow a_1, \ldots, 
\xi_m \leftarrow a_m)\overline{\alpha},$$
where $\overline{\alpha}: F_\Omega(Y \cup A)\rightarrow A$ is the homomorphism 
for which $\overline{\alpha}|X= \alpha$ and  $\overline{\alpha}|A= 1_A$.
In the special case 
$m=0$, we get 
$\sigma^{\cB}=h_0(\sigma)\overline{\alpha}=h_0(\sigma)\hat{\alpha}$.
The initial assignment is defined by putting 
$$x\beta=h(x)\hat{\alpha} \hbox{\hspace{3mm}} \mbox{ for all } 
\hbox{\hspace{3mm}} x\in X.$$
Now a proof by tree induction shows that 
$$s\hat{\beta}=h(s)\hat{\alpha}$$
for all $s\in F_\Sigma(X)$. Hence, $s\in T(\bB)$ iff $h(s)\in 
T(\bA)$. This means that $h^{-1}(T)=T(\bB)$ is recognizable.
\epr

\

As a conclusion we consider a simple, but very important special type
of tree homomorphisms.

\begin{df}\label{Definition.2.4.19} \rm A 
 tree homomorphism $h: F_\Sigma(X) \rightarrow F_\Omega(Y)$ is called 
{\it alphabetic} 
\index{homomorphism!alphabetic tree}
if the defining mappings $h_X$ and $h_m$ $(m\geq 0)$ satisfy 
the following conditions:
\begin{enumerate}
\item[(1)] $h_X(x) \in Y$ for all $x\in X$.
\item[(2)] $h_m(\sigma)=\omega(\xi_1, \ldots, \xi_m)$, where 
$\omega \in \Omega_m$, for all $m\geq 0$, $\sigma \in \Sigma_m$.
\end{enumerate}
\end{df}

An alphabetic tree homomorphism $F_\Sigma(X) \rightarrow F_\Omega(Y)$ can be 
defined only in case $\Omega_m\neq \emptyset$ for all such $m\geq 0$ that 
$\Sigma_m\neq \emptyset$. Alphabetic tree homomorphisms are often  called 
{\it projections}.

Consider the general alphabetic tree homomorphism of the definition. For any
$t\in F_\Sigma(X)$, the image $h(t)$ is obtained simply by rewriting 
every $x$ in $t$ as the letter $h_X(x)$ and every $\sigma \in \Sigma_m$ as the 
operator $\omega$, where $h_m(\sigma)=\omega(\xi_1, \ldots, \xi_m)$.
Hence $h$ preserves completely the ``shape'' of the tree $t$. Obviously, 
$h$ is linear. From Theorems 
\ref{Theorem.2.4.16} and \ref{Theorem.2.4.18} we get 

\begin{cor}\label{Corollary.2.4.20}
Let $h: F_\Sigma(X) \rightarrow F_\Omega(Y)$ be an alphabetic tree
homomorphism. 
\begin{enumerate}
\item[(i)] If $T \in \Rec(\Sigma, X)$, then  $h(T) \in \Rec(\Omega, Y)$.

\item[(ii)] If  $T \in \Rec(\Omega, Y)$,  then  $h^{-1}(T) 
\in \Rec(\Sigma, X)$. \epr
\end{enumerate} 
\end{cor}

%% file: Section.2.5.tex
Kleene's theorem is of central importance in the theory of finite automata
and it is quite natural that it was among the first results to be generalized
to the theory of tree automata. Although the greater generality adds some
technical complications, the standard development of the theory can be
followed quite close here, too, once the right generalizations of the basic
concepts have been found.

We fix again an arbitrary ranked alphabet $\Sigma$ and an arbitrary frontier
alphabet $X$. It turns out that some additional frontier symbols are needed
in the construction of regular forests. Therefore we will operate with an
extended alphabet $Z$ which contains $X$ as a subset.
\begin{df}\label{Definition.2.5.1} \rm
The set of {\it regular} $\Sigma Z$-{\it expressions}
$\RE(\Sigma, Z)$
is defined as the smallest set RE such that the following conditions are
satisfied:
\begin{enumerate}
\item[$1^\circ$] $\emptyset \in \RE$.

\item[$2^\circ$] $\Sigma_0\cup Z\subseteq \RE$.

\item[$3^\circ$] If $\zeta, \eta \in \RE$, then
$(\zeta + \eta)\in \RE$.

\item[$4^\circ$] If $\zeta, \eta \in \RE$ and $z \in Z$, then
$(\zeta \cdot_z \eta)\in \RE$.

\item[$5^\circ$] If $\zeta \in \RE$ and $z \in Z$, then
$(\zeta^{*z})\in \RE$.

\item[$6^\circ$] If $m>0$, $\sigma \in \Sigma_m$,  $\eta_1, \ldots, \eta_m \in
\RE$, then $\sigma(\eta_1, \ldots, \eta_m) \in \RE$.
\end{enumerate}
\end{df}

Thus regular $\Sigma Z$-expressions are strings of symbols from
$\Sigma \cup Z$, of commas etc. Parts $2^\circ$ and $6^\circ$ of the definition
imply that every $\Sigma Z$-tree is a regular  $\Sigma Z$-expression.
Regular expressions are intended as representations of forests.
\begin{df}\label{Definition.2.5.2}\rm The forest $|\eta|$ {\it represented}
by a regular expression
\index{forest!represented by regular expression}
$\eta\in \RE(\Sigma, Z)$ is defined following
 the inductive form of Definition \ref{Definition.2.5.1}:

\begin{enumerate}
\item[$1^\circ$] $|\emptyset |= \emptyset$ (the empty forest).

\item[$2^\circ$] If $\eta \in \Sigma_0\cup Z$, then $|\eta|=\{\eta\}$.

\item[$3^\circ$]
$|(\zeta + \eta)|= |\zeta| \cup |\eta|$.

\item[$4^\circ$] $|(\zeta \cdot_z \eta)|=|\zeta | \cdot_z |\eta|$.

\item[$5^\circ$] $|(\zeta^{*z})|=|\zeta|^{*z}$.

\item[$6^\circ$]  $|\sigma(\eta_1, \ldots, \eta_m)|=
\sigma(|\eta_1|, \ldots, |\eta_m|)$.
\end{enumerate}
\end{df}

Note that the operations in the right-hand sides of $3^\circ-6^\circ$
are forest operations which have been defined in Section \ref{Section.2.4}.
It is easy to see that every tree $t \in F_\Sigma(Z)$ represents, as a regular expression, the one-element forest $\{t\}$.

With this interpretation in mind we may simplify regular expressions by
omitting parentheses that are not needed in order to specify the intended
order of the operations. First of all, the outermost parentheses in
$(\zeta + \eta)$, $(\zeta \cdot_z \eta)$ and $(\zeta^{*z})$ are obviously
superfluous if the expressions do not appear as parts of other expressions.
We may also agree that iterations precede products and that products precede
unions. Then the parentheses around $\zeta^{*z}$ can always be omitted and, for
example,
$$\zeta + \eta \cdot_x \theta^{* y}$$
is interpreted as a short form for
$$(\zeta + (\eta \cdot_x (\theta^{* y}))).$$
\begin{ex}\label{Example.2.5.3}\rm  Let $\Sigma=\Sigma_0\cup \Sigma_2$,
$\Sigma_0=\{\omega\}$ and $\Sigma_2=\{ \sigma\}$ and
$Z=\{ x, y\}$. The forest represented by
$$\eta=\omega\cdot_y \sigma(x, y)^{* y}$$
contains the trees $\omega$, $\sigma(x, \omega)$,
$\sigma( x,  \sigma(x, \omega))$ etc. Note that $y$ has a purely auxiliary
function; it does not appear in any tree of the forest $|\eta|$.
\epr
\end{ex}

In the following definition we make the formal distinction between letters that may appear in trees of the forest represented by a regular expression and
those letters that are used just to mark leaves to be rewritten when
products of forests are formed.

\begin{df}\label{Definition.2.5.4} \rm
Suppose a regular $\Sigma Z$-expression $\zeta$ can be written in the form
$$\zeta=u(\eta \cdot_z \theta)v$$
where $\eta, \theta \in \RE (\Sigma, Z)$ and $z \in Z$. Then every occurrence
of $z$ within the string $\cdot_z \theta$ is said to be {\it bound}.
\index{occurrence!bound}
An occurrence  of a letter $z \in Z$ which is not bound is {\it free}.
\index{occurrence!free}
A letter $z \in Z$ is  {\it bound} in $\zeta\in \RE(\Sigma, Z)$,
if all occurrences of $z$ in $\zeta$ are bound, and it is {\it free}
in $\zeta$ if it has at least one free occurrence in $\zeta$. We denote by
$Z_\zeta$ the set of letters $z \in Z$ free in $\zeta$.
\end{df}

In Example \ref{Example.2.5.3} $Z_\eta=\{x\}$ and $y$ is  bound by the
$y$-product.
\begin{lm}\label{Lemma.2.5.5}
For any $\eta\in \RE(\Sigma, Z)$, $|\eta|\in \Rec(\Sigma, Z_\eta)$.
\end{lm}
\pr
We proceed by induction following the six parts in Definitions
\ref{Definition.2.5.1} and \ref{Definition.2.5.2}.
\begin{itemize}
\item[$1^\circ$] $Z_\emptyset=\emptyset$  and
$|\emptyset|=\emptyset\in \Rec(\Sigma, \emptyset)$.

\item[$2^\circ$] For each $z\in Z$, $Z_z=\{z\}$
and $|z|=\{z\} \in \Rec (\Sigma, \{z\})$.
For $\sigma \in \Sigma_0$, $Z_\sigma=\emptyset$, but still $|\sigma|=
\{\sigma\}\in \Rec(\Sigma, \emptyset)$.

\item[$3^\circ$] If $\eta=\zeta + \theta$, then $Z_\eta=Z_\zeta \cup Z_\theta$ and
$|\eta|=|\zeta|\cup |\theta|\in  \Rec (\Sigma, Z_\eta)$
by Lemma \ref{Lemma.2.4.1} and Theorem \ref{Theorem.2.4.2}.

\item[$4^\circ$] If $\eta=\zeta\cdot_z\theta$, then
(if we omit the trivial case $z\not \in Z_\theta$, $|\eta|=|\theta|)$
$Z_\eta=Z_\zeta \cup (Z_\theta -z)$. There are two cases to consider.
If $z \not \in Z_\zeta$, then $Z_\eta=(Z_\zeta \cup Z_\theta)-z$. From
Theorem \ref{Theorem.2.4.6} we know that
$|\eta|\in \Rec(\Sigma, Z_\zeta \cup Z_\theta)$. Thus it suffices to show that
no tree $t \in |\eta|$ contains any occurrence of $z$. But this is obvious
since every such $t$ is obtained from some $s\in |\theta|$ by replacing every
occurrence of $z$ by a tree from $|\zeta|$, and no tree in $|\zeta|$ contains
$z$. If $z \in Z_\zeta$, then $Z_\eta=Z_\zeta\cup Z_\theta$ and
$|\eta|\in \Rec(\Sigma, Z_\eta)$ follows directly from Theorem
\ref{Theorem.2.4.6}.

\item[$5^\circ$] If $\eta=\zeta^{* z}$ $(z \in Z)$, then $Z_\eta= Z_\zeta \cup z$.
Thus
$|\zeta|\in  Rec(\Sigma, Z_\eta)$ by Lemma \ref{Lemma.2.4.1}. This implies
$|\zeta^{*z}|\in Rec(\Sigma, Z_\eta)$ by Theorem \ref{Theorem.2.4.8}.

\item[$6^\circ$] If $\eta= \sigma(\eta_1, \ldots, \eta_m)$, where $m>0$,
$\sigma \in \Sigma_m$ and $|\eta_i|\in Rec(\Sigma, Z_{\eta_i})$
$(i=1, \ldots, m)$, then $Z_\eta=Z_{\eta_1}\cup \ldots \cup Z_{\eta_m}$
and every $|\eta_i|$ is also a recognizable $\Sigma Z_\eta$-forest.
Corollary \ref{Corollary.2.4.12} yields now $|\eta|\in \Rec(\Sigma, Z_\eta)$.
\epr
\end{itemize}

\

The operations (finite) union, $z$-product and $z$-iteration are called
the {\it regular operations}.
\index{operation!regular}
\index{regular operations}
A forest is {\it regular} if it can be
constructed from finite forests by applying a finite number of regular
operations. In view of the preceding discussion regularity can also be defined
as follows:

\begin{df}\label{Definition.2.5.6}\rm
A $\Sigma X$-forest $T$ is {\it regular}
\index{forest!regular}
if there exist an alphabet
$Z$ $(X \subseteq Z)$ and a regular  $\Sigma Z$-expression $\eta$ such that
$|\eta|=T$.
\end{df}

Note that an unlimited number of auxiliary letters $(z\in Z-X)$
is allowed in a regular expression representing a regular forest,
but that in any particular case just a finite number of them are needed.
Lemma \ref{Lemma.2.5.5} implies now that all regular forests are
recognizable. The next lemma contains the converse statement.
\begin{lm}\label{Lemma.2.5.7}
 For any $\Sigma X$-recognizer $\bA$ one can construct
a regular expression
$\eta\in \RE(\Sigma, X\cup A)$ (we assume $X \cap A=\emptyset$) such that
$|\eta|=T(\bA)$.
\end{lm}
\pr The proof is modelled after the almost standard proof for the
corresponding fact in the language case (due to R. McNaughton
and H. Yamada (1960)).
The notation can be simplified by assuming that
$$A=\{1, 2, \ldots, k\} \hbox{\hspace{3mm}}
\mbox{ for some } \hbox{\hspace{3mm}} k\geq 1.$$
As in Lemma \ref{Lemma.2.4.17} let
$$\overline{\alpha}: \cF_\Sigma(X \cup A) \rightarrow \cA$$
be the homomorphism such that $\overline{\alpha}| X=\alpha$ and
$\overline{\alpha}| A=1_A$. For any $i\in A$, $K \subseteq A$ and $h$,
$0\leq h \leq k$, we denote by $T(K, h, i)$ the set of all
$t\in F_\Sigma(X \cup K)$ such that
\begin{enumerate}
\item[(1)] $t\overline{\alpha}=i$ and

\item[(2)] $s\overline{\alpha}\in \{1, \ldots, h\}$\hbox{\hspace{2mm}}
for all\hbox{\hspace{2mm}}
$s\in \mbox{sub}(t)- (X \cup \Sigma_0\cup t)$.
\end{enumerate}

Thus $t \in T(K, h, i)$ means that the leaves of $t$ may be labelled,
besides frontier letters and nullary symbols, by states from $K$. Moreover,
the computation of $\bA$ on $t$ results in state $i$ and the state
of $\bA$ at any node between the frontier and the root is in the set $\{1, \ldots, h\}$. Obviously,
$$T(\bA)=\bigcup (T(\emptyset, k, i) \mid i\in A').$$
It suffices therefore to show that all sets $T(K, h, i)$ are regular.
To do this we proceed by induction on the number $h$.

\begin{itemize}
\item[$1^\circ$] When $h=0$, no intermediate states between the frontier and
the root are allowed. Every tree $t$ in $T(K, 0, i)$  must hence be of
one of the
following types:
\begin{enumerate}
\item[(i)] $t=x\in X$ and $x\alpha=i$.

\item[(ii)] $t=i\in K$.

\item[(iii)] $t=\sigma \in \Sigma_0$ with $\sigma^{\cA}=i$.

\item[(iv)] $t= \sigma(d_1, \ldots, d_m)$ with
$m> 0$, $d_j \in X \cup \Sigma_0\cup K$
$(j=1, \ldots, m)$ and $t\overline{\alpha}=i$.
\end{enumerate}

In each case a regular expression for $\{t \}$ can be written. The number
of such trees $t$ is finite  and we get a regular expression for
 $T(K, 0, i)$.

\item[$2^\circ$]  Suppose we already have a regular expression for each
$T(K, j, i)$ such that $j \leq h$ for some $h< k$. We show that
\begin{equation*}\tag{*}
T(K, h+1, i)=
\end{equation*}
$$
T(K, h, i)\cup T(K, h, h+1)\cdot _{h+1} T(K\cup h+1, h, h+1)^{* h+1}
\cdot _{h+1} T(K\cup h+1, h, i)
$$
holds for all $K \subseteq A$ and $i \in A$. This will complete the induction
because the right-hand side of (*) is obtained by regular operations from
forests for which we already have regular expressions.

Let $T$ be the right-hand side of  (*). From the construction of $T$ it is
obvious that $T \subseteq T(K, h+1, i)$. If $t \in  T(K, h+1, i)$, then either
$t \in T(K, h, i)$ or $t$ has a proper subtree $s \not \in X \cup \Sigma_0$
such that $s\overline{\alpha}=h+1$. In the former case we get
$t \in T$ directly. In the second case we have
$$t\in \{p_1, \ldots, p_d\}
\cdot_{h+1} \{q_{11}, \ldots, q_{1 e_1}\} \cdot_{h+1} \ldots \cdot_{h+1}
\{q_{j1}, \ldots, q_{j e_j}\} \cdot_{h+1} \{r\},
$$
for some $$p_1, \ldots, p_d\in T(K, h, h+1),
q_{11}, \ldots, q_{1 e_1}, \ldots, q_{j1}, \ldots, q_{j e_j}\in
 T(K\cup h+1,  h, h+1)$$
and $r \in   T(K\cup h+1,  h, i)$. This means that $t$ belongs to the
second part of $T$. \epr
\end{itemize}

\

Combining Lemma \ref{Lemma.2.5.5} and Lemma \ref{Lemma.2.5.7}
we get the following generalized form of Kleene's theorem.

\begin{thm}\label{Theorem.2.5.8}
A forest is recognizable iff it is regular.
\epr
\end{thm}

%% file: Section.2.6.tex
The number of states is a simple and natural measure of the complexity of a
finite automaton. In this section we consider minimal recognizers of forests.
In the case
of a recognizable forest minimality means simply a minimal number of
states, and there is always a minimal recognizer which is unique up
to isomorphism.
All tree recognizers recognizing a nonregular forest must be infinite and
counting the number of states does not make any sense.
Nevertheless, the general definition of minimality is such that the minimal
recognizer of a forest remains unique even in such a case.
The minimal recognizer of a forest can be derived from any recognizer of
this forest. If the forest is recognizable, then the minimalization
procedure is effective. Otherwise, the finiteness of the recognizers is not
needed in this section. Also, some of the concepts and results presented here
will be applied to infinite tree recognizers in the next section.
Thus we will temporarily drop our general assumption that all tree recognizers
dealt with are finite. In all other respects the earlier definitions and
conventions remain valid.

We shall now define homomorphisms, congruences and quotients of
tree recognizers. The reader may find it helpful to review the corresponding material from Section \ref{Section.1.2} before going on.
Algebraic functions and elementary translations (cf. Sect. \ref{Section.1.3})
will also be needed.

\begin{df}\label{Definition.2.6.1} \rm A {\it homomorphism} from a
$\Sigma X$-recognizer $\bA$ to a $\Sigma X$-recognizer $\bB$
\index{homomorphism!of sigmaxrecognizer@of $\Sigma X$-recognizer}
is a mapping $\varphi: A \rightarrow B$ such that

\begin{enumerate}
\item[(1)] $\varphi$ is a homomorphism from the $\Sigma$-algebra $\cA$ to the
$\Sigma$-algebra  $\cB$,

\item[(2)] $\alpha\varphi= \beta$, and

\item[(3)] $B'\varphi^{-1}=A'$.

\end{enumerate}
If $\varphi$ is a homomorphism from $\bA$ to $\bB$, we write
$\varphi: \bA \rightarrow \bB$. A  homomorphism of tree recognizers is an
{\it epimorphism}
\index{monomorphism of!sigmaxrecognizer@$\Sigma X$-recognizer}
if it is surjective, a {\it monomorphism} if it is
injective, and it is called an {\it isomorphism} if it is bijective.
\index{isomorphism of!sigmaxrecognizer@$\Sigma X$-recognizers}
\index{epimorphism!of sigmaxrecognizer@of $\Sigma X$-recognizer}
If there exists an
isomorphism $\varphi: \bA \rightarrow \bB$, then we write $\bA \cong \bB$
and say that $\bA$ and $\bB$ are {\it isomorphic}. If there exists an
epimorphism $\varphi: \bA \rightarrow \bB$, then $\bB$ is said to be an
{\it epimorphic image } of $\bA$.
\index{image!epimorphic}
A  monomorphism is also called an
{\it embedding}.
\end{df}
\index{embedding of!SigmaX-recognizer@$\Sigma X$-recognizer}

Part (3) of Definition \ref{Definition.2.6.1} means that the final states,
and these only, map to final states in a homomorphism. If $\varphi$ is an
epimorphism, then (3) implies $A'\varphi=B'$.

\begin{lm}\label{Lemma.2.6.2} Let $\bA$ and $\bB$ be two $\Sigma X$-recognizers. 
If there exists a homomorphism $\varphi: A \rightarrow B$, then $T(\bA)=T(\bB)$.
\end{lm}
\pr The clauses (1) and (2) of Definition \ref{Definition.2.6.1} imply together
with Lemma \ref{Lemma.1.3.6} that
$$t^\cA(\alpha)\varphi=t^\cB(\alpha\varphi)= t^\cB(\beta)  $$
for every $t \in F_\Sigma(X)$. Now clause (3) shows that
\begin{eqnarray*}
t \in T(\bB) & \mbox{ iff }  & t^\cB(\beta)=t^\cA(\alpha)\varphi\in B' \\
             & \mbox{ iff }  & t^\cA(\alpha)\in A' \\
             & \mbox{ iff }  & t \in T(\bA)
\end{eqnarray*}
for every $t\in F_\Sigma(X)$, and the lemma follows.
\epr
\begin{df}\label{Definition.2.6.3}\rm
A {\it congruence}
\index{congruence!ofSigmaXrecognizer@of $\Sigma X$-recognizer}
of a $\Sigma X$-recognizer $\bA$ is a congruence $\varrho$ of
the algebra $\cA$ saturating $A'$, that is, such that $A'\varrho=A'$. The set of all congruence relations of $\bA$ is denoted by $C(\bA)$.
\end{df}

\begin{lm}\label{Lemma.2.6.4}
 $C(\bA)$ is a principal ideal of the complete lattice $C(\cA)$, and thus
  $(C(\bA),$\\ $\subseteq)$ is a complete lattice itself, too.
\end{lm}
\pr
It suffices to verify the following simple facts:
\begin{enumerate}
\item[(i)] $\delta_\bA \in C(\bA)$ (which implies $ C(\bA)\neq \emptyset$).

\item[(ii)] $\theta \subseteq \varrho \in C(\bA)$ and
 $\theta  \in C(\cA)$ imply $\theta  \in C(\bA)$.

\item[(iii)] $ \vee (\varrho \mid \varrho \in C(\bA))\in C(\bA)$.

\end{enumerate}
In (iii) the supremum is to be formed in $ C(\cA)$.
It is the generating element of the principal ideal.
\epr

\

In Theorem \ref{Theorem.2.6.10} we shall get a more useful description of
the greatest element of
$ C(\bA)$.
\begin{df}\label{Definition.2.6.5}\rm
The {\it quotient $\Sigma X$-recognizer} of a  $\Sigma X$-recognizer $\bA$
\index{SigmaX-recognizer@$\Sigma X$-recognizer!quotient}
with respect to a congruence $\varrho$ is the $\Sigma X$-recognizer
$$\bA/\varrho=(\cA/\varrho, \alpha_\varrho, A'/\varrho),  $$
where $\alpha_\varrho$ is defined so that $x\alpha_\varrho= (x \alpha)\varrho$ for each
$x \in X$.
\end{df}

The usual relations between homomorphisms, congruences and quotients hold for
tree recognizers, too. Some of them are listed in the following theorem.
We omit the proofs since they can be constructed exactly as the corresponding
proofs in algebra.

\begin{thm}\label{Theorem.2.6.6}
\begin{itemize}
\item[{\em (a)}] If $\varrho \in C(\bA)$, the natural mapping
$$ \varrho^\natural : A \rightarrow A/\varrho, \hbox{\hspace{1mm}}  a \mapsto a\varrho
 \hbox{\hspace{3mm}} (a \in A) ,$$
is an epimorphism $ \bA \rightarrow \bA/\varrho $ (called the
{\rm natural epimorphism}).
\index{epimorphism!natural}

\item[{\em (b)}] If $\varphi:  \bA \rightarrow \bB$ is a homomorphism, then the kernel
$\varphi \varphi^{-1}$ is a congruence of $\bA$ and the image
$$\bA \varphi=(\cA\varphi, \beta, A'\varphi)$$
of $\bA$ is isomorphic to $\bA \varphi \varphi^{-1}$. (In
$\bA\varphi$ $\cA\varphi$ is the $\Sigma$-algebra
$(A\varphi, \Sigma)$ such that
$\sigma^{\cA \varphi}=\sigma^{\cB}| A\varphi$ and $\beta$ is to be interpreted as a mapping from $X$ to $A\varphi$.)

\item[{\em (c)}] If $\pi \subseteq \varrho$ for some $\pi, \varrho \in C(\bA)$, then $\bA/\varrho$
is an epimorphic image of $\bA/\pi$.
\epr
\end{itemize}
\end{thm}

From Theorem \ref{Theorem.2.6.6} and Lemma  \ref{Lemma.2.6.2} we get
\begin{cor}\label{Corollary.2.6.7}
If $\varrho \in C(\bA)$, then $T(\bA/\varrho)= T(\bA)$.
\epr
\end{cor}

Thus any congruence of a tree recognizer yields an equivalent recognizer
which is an epimorphic image of the original one. If the recognizer is finite
and the congruence is nontrivial, then a real reduction
in the number   of states is achieved.
Obviously, the greatest congruence gives the smallest quotient recognizer.
The construction of the quotient recognizer involves a merging of states which
are equivalent in the sense that one can be substituted for another in any
computation without affecting the end result. We shall now give a precise
 meaning  to this equivalence of states and show then that the greatest
congruence consists exactly of the pairs of equivalent states.

\begin{df}\label{Definition.2.6.8} \rm Two states $a$ and $b$ of a
$\Sigma X$-recognizer $\bA$ are said to be
{\it equivalent}
\index{equivalence of states in!sigmaxrecognizer@$\Sigma  X$-recognizer}
and we write $a\sim_\bA b$ or just $a\sim b$, iff
$$(\forall f \in \Alg_1(\cA)) \hbox{\hspace{3mm}} f(a) \in A'
\Longleftrightarrow f(b) \in A'.$$
\end{df}

To get a better intuitive grasp of this definition we recall the fact that
for each algebraic function $f \in  \Alg_1(\cA)$ there exists a tree
$t \in F_\Sigma(A \cup \xi )$ such that for all $a\in A$,
$$f(a)=t\hat{\alpha}_a,$$
where $\alpha_a: A \cup \xi\rightarrow A$ is defined by
$\alpha_a|A=1_A$ and $\xi\alpha_a=a$ (Lemma \ref{Lemma.1.3.14}).
This means that $\cA$ computes $f(a)$ from the tree $t$
when one assigns state $a$ to all leaves labelled by $\xi$. On the other
hand, every tree $t \in  F_\Sigma(A \cup \xi )$ defines this way a unary
algebraic function. Such a tree may be thought of as the unprocessed part
of a $\Sigma X$-tree where a leaf labelled by a state $c \in A$ corresponds
to a subtree $s$ such that $s\hat{\alpha}=c$. Once a value $a\in A$ has been
assigned to the leaves labelled by $\xi$ the computation may be completed.
The equivalence of two states $a$ and $b$ means that the assignments
$\xi=a$ and $\xi=b$ give  always the same result $(\mbox{mod }A')$ when such a computation is completed.

\begin{df}\label{Definition.2.6.9}\rm
The $\Sigma X$-recognizer $\bA$ is
\begin{itemize}
\item[(a)] {\it reduced}  if $\sim_\bA=\delta_\bA$,
\index{SigmaX-recognizer@$\Sigma X$-recognizer!reduced}

\item[(b)] {\it connected} if every state of $\bA$ is {\it reachable},
\index{SigmaX-recognizer@$\Sigma X$-recognizer!connected}
\index{reachability of state in!sigmaxrecognizer@$\Sigma X$-recognizer}
i.e., there
exists for every $a\in A$ a tree $t\in F_\Sigma(X)$ such that $t\hat{\alpha}=a$,
and $\bA$ is

\item[(c)] {\it minimal} if it is connected and reduced.
\index{SigmaX-recognizer@$\Sigma X$-recognizer!minimal}
\end{itemize}
\end{df}

That a recognizer is reduced means that no two distinct states  are equivalent.
To be connected means that every state is possible in some computation
performed by the recognizer on some tree. By Lemma \ref{Lemma.1.3.8}, a tree
recognizer $\bA$ is connected iff $X\alpha$ generates $\cA$. In the case of a
finite recognizer minimality really means a minimal number of states among
equivalent recognizers. If a recognizer is not connected, then the
nonreachable states can be discarded without changing the forest recognized.
If $\bA$ is finite and
$\sim_\bA> \delta_\bA$, then $\bA/\hbox{\hspace{-1.5mm}}\sim_\bA$ is a properly smaller recognizer
equivalent to $\bA$. Hence, a finite tree recognizer can be minimal with
respect to the number of states only if it is minimal in the
sense of Definition \ref{Definition.2.6.9}. The converse will be established
later.

\begin{thm}\label{Theorem.2.6.10} For any $\Sigma X$-recognizer $\bA$,
$\sim$ is the greatest congruence of $\bA$ and $\bA/\hbox{\hspace{-1.5mm}}\sim$ is a reduced
$\Sigma X$-recognizer equivalent to $\bA$.
\end{thm}
\pr
It is obvious that $\sim$ is an equivalence relation on $A$. Let
$a\sim b$ $(a, b \in A)$. For any two unary algebraic functions
$f, g \in  \Alg_1(\cA)$,  the composition
$$ f \circ g: \hbox{\hspace{1mm}}  \xi \mapsto
g(f(\xi)) \hbox{\hspace{3mm}} (\xi \in A)$$
is a unary algebraic function. Hence
$$g(f(a))\in A' \hbox{\hspace{3mm}} \mbox{ iff }\hbox{\hspace{3mm}} g(f(b))\in A',$$
and this implies $f(a)\sim f(b)$. By Lemma \ref{Lemma.1.3.16},
$\sim$ is a congruence of $\cA$. If $a\sim b$ and $a\in A'$, then
$b=1_\bA(b)\in A'$. Thus $A'\hbox{\hspace{-1mm}}\sim\,
=A'$
 and $\sim$ is a congruence of $\bA$.
Let $\varrho$ be any congruence of $\bA$. If $a\varrho b$ and $f \in  \Alg_1(\cA)$,
then $\varrho \in C(\cA)$ implies $f(a) \varrho f(b)$. Now $A'\varrho=A'$ implies
$$f(a)\in A' \hbox{\hspace{3mm}} \mbox{ iff } \hbox{\hspace{3mm}} f(b)\in A' .$$
Hence $a\sim b$ and we have shown that $\sim$ is the greatest among the
congruences of $\bA$. Corollary \ref{Corollary.2.6.7} tells us that
$T(\bA)=T(\bA/\hbox{\hspace{-1.5mm}}\sim)$. That $\bA/\hbox{\hspace{-1.5mm}}\sim$ is reduced follows directly from the
fact, well-known in universal algebra, that the lattice
$C(\bA/\hbox{\hspace{-1.5mm}}\sim)$ is
isomorphic to the principal dual ideal $[\sim)$ generated by $\sim$ in
$C(\bA)$. Since $\sim$ is the greatest element of $C(\bA)$,
$[\sim)$ is trivial and thus $\sim _{\bA/\hbox{\hspace{-0.4mm}}\sim}$ must be the diagonal relation of
$A/\hbox{\hspace{-1.5mm}}\sim$. A more direct proof is possible, too. It is not hard to show
that $(a\sim)\sim_{\bA/\hbox{\hspace{-0.4mm}}\sim}(b\sim)$ implies $a\sim b$, and hence
$a\sim=b\sim$.
\epr

\

The quotient recognizer $\bA/\hbox{\hspace{-2mm}}\sim_\bA$ is often called the {\it reduced form}
of $\bA$.
\index{reduced form of $\Sigma X$-recognizer}
 It is clear from Theorem \ref{Theorem.2.6.10} that two tree
recognizers having isomorphic reduced forms are equivalent. We show that the
converse holds for connected recognizers. In other words, equivalent minimal recognizers are shown to be isomorphic.
\begin{thm}\label{Theorem.2.6.11} Let
$\bA$ and $\bB$ be two minimal tree recognizers. If $\bA$ and $\bB$ are equivalent, then they are also isomorphic.
\end{thm}
\pr Define $\varphi: A \rightarrow B$ so that
$$(t\hat{\alpha})\varphi=t\hat{\beta} \hbox{\hspace{3mm}} \mbox{ for all } \hbox{\hspace{3mm}} t\in F_\Sigma(X).$$
We show that $\varphi$ gives the required isomorphism from $\bA$ to
$\bB$. This involves the following seven points:

\begin{itemize}
\item[(i)] $\varphi$ associates with every $a\in A$ a state of $\bB$ since
$\bA$ is connected.

\item[(ii)] To show that $\varphi$ is well-defined we consider the possibility that
$s\hat{\alpha}=t\hat{\alpha}$ for two $\Sigma X$-trees $s$ and $t$.
If $s\hat{\beta}\neq t\hat{\beta}$, then $s\hat{\beta}$ and
$t\hat{\beta}$ are nonequivalent and there exists an algebraic function
$f \in  \Alg_1(\cB)$ such that $f(s\hat{\beta}) \in B'$ and
$f(t\hat{\beta}) \not \in B'$ (or conversely).
By Lemma \ref{Lemma.1.3.14} there exists a tree
 $p\in F_\Sigma(B\cup \xi)$ $(\xi \not \in B\cup X)$
such that for all $b \in B$,
$$f(b)=p^{\cal B}(\beta_b),$$
where $\beta_b: B \cup \xi\rightarrow B$ is defined so that
$\beta_b| B=1_B$ and $\xi\beta_b=b$. Since $\bB$ is connected there exists for
each $b\in B$ a $\Sigma X$-tree $p_b$ such that $p_b \hat{\beta}=b$. Let
$$q=p(b \leftarrow p_b\mid b \in B)(\in F_\Sigma(X \cup \xi)).$$
Consider the $\Sigma X$-trees $q_s=q(\xi \leftarrow s)$ and
$q_t=q(\xi \leftarrow t)$.  Now
$$q_s\hat{\beta}=p^{\cB}(\beta_{s\hat{\beta}})=f(s\hat{\beta} )\in B'$$
and
$$q_t\hat{\beta}=p^{\cB}(\beta_{t\hat{\beta}})=f(t\hat{\beta} )\not\in B'.$$
If we assign in $q$ to every letter $x\in X$ the value $x\alpha$,
we get a function $g \in  \Alg_1(\cA)$ such that for each $a\in A$,
$$g(a) = q^\cA(\alpha_a)$$
where $\alpha_a: X \cup \xi\rightarrow A$ is defined so that
$\alpha_a|X=\alpha$ and $\xi \alpha_a=a$.
Applying Lemma \ref{Lemma.1.3.6} we get now
$$g(s\hat{\alpha})\varphi=q^\cA(\alpha_{s\hat{\alpha}})\varphi=
q_s\hat{\alpha}\varphi=q_s\hat{\beta}\in B'
$$
and
$$g(t\hat{\alpha})\varphi=q^\cA(\alpha_{t\hat{\alpha}})\varphi=
q_t\hat{\alpha}\varphi=q_t\hat{\beta}\not \in B'.
$$
This is in contradiction with our original assumption
that $s\hat{\alpha}=t\hat{\alpha}$.
Hence $q_s\in T(\bB)$, but $q_t \not \in T(\bB)$.
On the other hand, $s\hat{\alpha}=t \hat{\alpha}$ implies
 $q_s\hat{\alpha}=q_t \hat{\alpha}$, and a contradiction with our assumption
that $T(\bA)=T(\bB)$ results.

\item[(iii)] Reversing the roles of $\bA$ and $\bB$ in Part (ii) one sees that
$s\hat{\beta}= t\hat{\beta}$ implies $s\hat{\alpha}=t\hat{\alpha}$ for all
$\Sigma X$-trees $s$ and $t$. This means that $\varphi$ is injective.

\item[(iv)] $\varphi$ is surjective since $\bB$ is connected.

\item[(v)] Let $m\geq 0$, $\sigma \in \Sigma_m$ and $a_1, \ldots, a_m \in A$.
There are trees $t_1, \ldots, t_m \in F_\Sigma(X)$ such that
$a_1=t_1\hat{\alpha},\ldots,  a_m=t_m\hat{\alpha}$. Then

\begin{eqnarray*}
\sigma^\cA(a_1, \ldots, a_m) \varphi  & =   &
\sigma^\cA(t_1\hat{\alpha}, \ldots, t_m\hat{\alpha})\varphi\\
 & =   &
\sigma(t_1, \ldots, t_m)\hat{\alpha}\varphi\\
 & =   &
\sigma(t_1, \ldots, t_m)\hat{\beta}\\
& =   &
\sigma^\cB(t_1\hat{\beta}, \ldots, t_m\hat{\beta})\\
& =   &
\sigma^\cB(a_1\varphi, \ldots, a_m\varphi).
\end{eqnarray*}
Hence $\varphi$ is a homomorphism from $\cA$ to $\cB$.

\item[(vi)] For each $x\in X$, $x\alpha\varphi= x\hat{\alpha}\varphi= x\hat{\beta}=x\beta$.
Thus $\alpha\varphi=\beta$.

\item[(vii)] If $t \hat{\alpha}\in A'$ $(t \in F_\Sigma(X))$, then
$t \hat{\alpha}\varphi=t \hat{\beta}\in B'$ since $t \in T(\bA)=T(\bB)$.
Similarly, $t \hat{\alpha}\varphi\in B'$ implies $t\hat{\alpha}\in A'$. Hence,
$B'\varphi^{-1}=A'$.
\epr
\end{itemize}
\begin{cor}\label{Corollary.2.6.12}
If $\bA$ and $\bB$ are connected $\Sigma X$-recognizers such that
$ T(\bA)=T(\bB)$, then $\bA/\hbox{\hspace{-1.5mm}}\sim_\bA\cong \bB/\hbox{\hspace{-1.5mm}}\sim_\bB$.
\epr
\end{cor}

For every $\Sigma X$-forest $T$ there is at least the infinite
$\Sigma X$-recognizer
$$\bF_T=(\cF_\Sigma(X), 1_X, T)$$
where $\cF_\Sigma(X)=(F_\Sigma(X), \Sigma)$ is the $\Sigma X$-term algebra.
Indeed, for each $t\in F_\Sigma(X)$ we have
$$ t^{\cF_\Sigma(X)}(1_X)=t\in T(\bF_T) \hbox{\hspace{3mm}} \mbox{ iff }
\hbox{\hspace{3mm}} t\in T.   $$
Obviously $\bF_T$ is connected. Hence, $\bF_T/\hbox{\hspace{-1.5mm}}\sim$ is a minimal recognizer
for $T$ (the relation $\sim$ will be examined more closely in the next
section). To show it we shall verify that every quotient recognizer of
a connected tree recognizer is connected.

Let $ \varphi: \bA \rightarrow \bB$ be an epimorphism of $\Sigma X$-recognizers.
If $\bA$ is connected, then so is $\bB$. Indeed, let $b$ be any state of
$\bB$. There exists an $a\in A$ such that $a\varphi=b$. Since $\bA$ is connected
there is a tree $t\in F_\Sigma (X)$ so that $a=t^\cA(\alpha)$. Using Lemma
\ref{Lemma.1.3.6} we get
$$t^\cB(\beta)=t^\cB(\alpha\varphi)=t^\cA(\alpha) \varphi=a\varphi=b.$$
In particular, $\bA/\hbox{\hspace{-1.5mm}}\sim_\bA$ is connected for every tree recognizer $\bA$.

We now have everything needed for the main theorem of the section.
\begin{thm}\label{Theorem.2.6.13}  For every forest $T$ there exists a minimal
tree recognizer, and it is unique up to isomorphism. If $\bA$ is any connected
recognizer of $T$, then the minimal recognizer is an epimorphic image of $\bA$.
In fact, $\bA/\hbox{\hspace{-1.5mm}}\sim_\bA$ is minimal.
\epr
\end{thm}

The theorem is valid for every forest.
It suggests the following two-step procedure for finding the minimal recognizer
for $T$ once any recognizer $\bA$ of $T$ is given:

\begin{itemize}
\item[$1^ \circ$] Discard all nonreachable states from $\bA$. We get a connected
recognizer $\bB$ such that $T(\bB)=T$.

\item[$2^ \circ$] Reduce $\bB$ by finding $\sim_\bB$ and then constructing
$\bB/\hbox{\hspace{-1.5mm}}\sim_\bB$ which is the required minimal recognizer.
\end{itemize}

Both of these steps become effective when $T$ is a recognizable forest and
the given recognizer $\bA$ is finite.

The reachable states of $\bA$ form the subalgebra of $\cA$ generated by
the subset $X \alpha$. This can be found as follows. Let
$H_0=X\alpha \cup \{\sigma ^\cA\mid \sigma \in \Sigma_0\}$ and put
$$H_{i+1}=H_i \cup \{ \sigma^\cA (a_1, \ldots, a_m)\mid m>0, \sigma \in \Sigma_m,
a_1, \ldots, a_m \in H_i\}.$$
Then
$$H_0 \subseteq H_1 \subseteq \ldots \subseteq A$$
and $H_i=[X\alpha]$ $(i\geq 0)$ if $H_{i+1}=H_i$. Such an $i$ must exist
since $A$ is finite.

Suppose now that we have a finite connected $\Sigma X$-recognizer $\bB$
and consider step $2^\circ$. First one should find
$  \Alg_1(\cB)$. It is finite and can be formed repeating the inductive step
of Definition \ref{Definition.1.3.13} a finite number of times. Then
$\sim _\bB$ can be determined directly, using the definition. Although
the minimal recognizer $\bB/\hbox{\hspace{-1.5mm}}\sim_\bB$ certainly can be found this way,
the procedure would be quite tedious in most cases. A computationally
simpler method can be derived from the following lemma. The proof is
left as an exercise.
The crucial aid is Lemma \ref{Lemma.1.3.16}:
an equivalence is a congruence iff it is invariant with respect to all
elementary translations.
\begin{lm}\label{Lemma.2.6.14}
Define a descending sequence $\sim_0\supseteq \sim_1 \supseteq \ldots$
of equivalences on $\bB$ as follows: (i) $B/\hbox{\hspace{-1.5mm}}
\sim_0=\{B', B-B'\}$
and (ii) for all $i\geq 0$ and $a,b \in B$, $a\sim_{i+1} b$
iff $a\sim_i b$ and $f(a) \sim _i f(b)$ for all $f\in \ET(\cB)$. Then
$\sim_i=\sim_\bB$ if
$\sim_{i+1}=\sim_i$, and this holds for some $i< |B|$.
\epr
\end{lm}

%% file: Section.2.7.tex
In this section two strictly algebraic characterizations of the
recognizable forests are presented.  First some ideas from the
previous section are applied to derive a generalization of Nerode's
theorem on regular languages and right congruences of the free monoid
(cf.~Theorem~\ref{Theorem.1.5.6}).  Then we show that the recognizable
forests can be obtained by solving fixed-point equations of a certain
kind.  Again, there is a well-known precursor in the theory of finite
automata.  In fact, in the unary case the equations considered here
reduce to Arden's equations which give the regular languages as their
solutions.

Let $\Sigma$~and~$X$ be fixed and denote the $\Sigma X$-term
algebra~$\cF_\Sigma(X)$ by~$\cF$, for short.  In the previous section
we noted that each $\Sigma X$-forest~$T$ has the (infinite) $\Sigma
X$-recognizer $\bF_T = (\cF, 1_X, T)$.  Consider any $\Sigma
X$-recognizer~$\bA$ such that~$T(\bA) = T$.  It is easy to verify that
the extension of the initial assignment $\alpha \colon X \to A$ to a
homomorphism
\[ \hat \alpha \colon \cF \to \cA \]
is also a homomorphism of $\Sigma X$-recognizers from~$\bF_T$
to~$\bA$.  Indeed, $1_X \hat \alpha = \hat \alpha$ and $A' \hat
\alpha^{-1} = T(\bA) = T$.  The kernel $\hat \alpha \hat\alpha^{-1}$
is a congruence of~$\bF_T$ with a congruence class for each reachable
state of~$\bA$.  If $T$~is recognizable, $\bA$~may be chosen as
finite, and then $\hat \alpha \hat\alpha^{-1}$~is of finite index.
Now, suppose $\bF_T$~has a congruence~$\varrho$ of finite index.  Then
$\bF_T/\varrho$~is a finite $\Sigma X$-recognizer such that
$T(\bF_T/\varrho) = T(\bF_T) = T$ (by
Corollary~\ref{Corollary.2.6.7}).  Hence $T$~is recognizable.  The
congruences of~$\bF_T$ are simply the congruences of~$\cF$ which
saturate~$T$.  Among these there is one of finite index iff the
greatest congruence~$\sim_{\bF_T}$ of~$\bF_T$ is of finite index.  The
congruence~$\sim_{\bF_T}$ ($\sim_T$~for short) is the \emph{Nerode
  congruence of~$T$.}  These observations may be summed up as
\index{Nerode congruence of!forest}

\begin{thm}
  \label{Theorem.2.7.1}
  For every $\Sigma X$-forest~$T$ the following three conditions are
  equivalent:
  \begin{enumerate}[(i)]
  \item $T \in \Rec(\Sigma, X)$.
  \item The term algebra~$\cF_\Sigma(X)$ has a congruence of finite
    index which saturates~$T$.
  \item The index of the Nerode congruence~$\sim_T$ is finite. \epr
  \end{enumerate}
\end{thm}

The recognizer~$\bF_T$ is connected and Theorem~\ref{Theorem.2.6.10}
implies therefore that $\bF_T/\sim_T$ is the minimal
recognizer of the forest~$T$.  To find~$\sim_T$ for a given $\Sigma X$-forest~$T$ 
one could try to apply Definition~\ref{Definition.2.6.8} to~$\bF_T$: for any 
$s, t \in F_\Sigma(X)$,
\[ s \sim_T t \quad \text{iff} \quad \big(\forall p \in F_{\Sigma}(X \cup
\xi)\big) \big(p(\xi \gets s) \in T \Longleftrightarrow p(\xi \gets t) \in
T\big)  \enspace. \]
A part of Theorem~\ref{Theorem.2.7.1} can be restated as follows.

\begin{cor}
  \label{Corollary.2.7.2}
  A $\Sigma X$-forest~$T$ is recognizable iff there exist a finite
  $\Sigma$-algebra~$\cA$, a homomorphism~$\varphi \colon \cF_\Sigma(X)
  \to \cA$ and a subset~$A' \subseteq A$ such that $T =
  A'\varphi^{-1}$. \epr
\end{cor}

The corollary gives, in fact, just an obvious reformulation of the
definition of recognizability.  Without going into the subject any
further here, we note that in this form recognizability may be defined
for subsets of arbitrary algebras (and not just term algebras): a
subset~$T$ of a $\Sigma$-algebra~$\cA$ is said to be recognizable, if
there exist a finite $\Sigma$-algebra~$\cB$, a homomorphism~$\varphi
\colon \cA \to \cB$ and a subset~$H \subseteq B$ such that
$H\varphi^{-1} = T$.  If here $\cA = \cF_\Sigma(X)$, then we get the
recognizable $\Sigma X$-forests, and if $\cA$~is the free
monoid~$X^*$, then we get the recognizable $X$-languages.

As an introduction to the theory of fixed-point equations we first
look at an example of Arden equations.

\begin{ex} \rm
  \label{Example.2.7.3}
  Consider the two-state Rabin-Scott recognizer~$\bA$ defined by the
  state graph shown in Fig.~\ref{Figure.2.5}.  The input alphabet is
  $\Sigma = \{\sigma, \tau\}$.

  \begin{figure}[h]
    \centering
    \input{Figure.2.5}
    \caption{{}\label{Figure.2.5}}
  \end{figure}

  \noindent Let $L_1$~and~$L_2$ be the languages of all words
  taking~$\bA$ from the initial state~$1$ to state $1$~and~$2$,
  respectively.  Then the following equations hold:
  \begin{align}
    \label{Equation.2.7.1}
    \begin{split}
      L_1 &= L_1\sigma \cup L_2\sigma \cup e \\
      L_2 &= L_1\tau \cup L_2\tau \enspace.
    \end{split}
  \end{align}
  If we define a mapping
  \[ \hat\Pi \colon (\gp\Sigma^*)^2 \to (\gp\Sigma^*)^2 \]
  so that for all $U, V \subseteq \Sigma^*$,
  \[ \hat\Pi(U, V) = (U\sigma \cup V\sigma \cup e,\, U\tau \cup V\tau)
  \enspace, \]
  then~\eqref{Equation.2.7.1} means that $(L_1, L_2)$~is a solution of
  the fixed-point equation
  \begin{equation}
    \label{Equation.2.7.2}
    (v_1, v_2) = \hat\Pi(v_1, v_2) \enspace.
  \end{equation}
  Moreover, $(L_1, L_2)$~is the least solution of~\eqref{Equation.2.7.2}
  when $(\gp\Sigma^*)^2$~is partially ordered in the natural way:
  \[ (U_1, V_1) \leq (U_2, V_2) \quad \text{iff} \quad U_1 \subseteq
  U_2 \quad \text{and} \quad V_1 \subseteq V_2 \enspace. \]
  If we view~$\Sigma$ as a unary ranked alphabet and identify
  $\Sigma\{x\}$-trees and $\Sigma$-words as shown in
  Section~\ref{Section.2.2} ($x = e$, $\sigma_k(\dotsm \sigma_1(x)
  \dotsm) = \sigma_1 \dotsm \sigma_k$), then the term
  algebra~$\cF_\Sigma(\{x\})$ may be taken to be
  \[ \cF = (\Sigma^*, \Sigma) \enspace, \]
  where
  \[ \sigma^{\cF}(u) = u\sigma \quad (\sigma \in \Sigma,\, u \in
  \Sigma^*) \enspace. \]
  In the corresponding subset algebra
  \[ \gp\cF = (\gp\Sigma^*, \Sigma) \]
  we have the operations
  \[ \sigma^{\gp\cF}(L) = L\sigma \quad (\sigma \in \Sigma,\, L
  \subseteq \Sigma^*) \enspace. \]
  The mapping~$\hat\Pi$ can be defined in terms of these operations,
  the empty word and unions:
  \[ \hat\Pi(U, V) = \bigl(\sigma^{\gp\cF}(U) \cup \sigma^{\gp\cF}(V) \cup
  x,\, \tau^{\gp\cF}(U) \cup \tau^{\gp\cF}(V)\bigr) \enspace. \]
  Using forest products we may write this as follows:
  \begin{align}
    \label{Equation.2.7.3}
    \begin{split}
      \hat \Pi(U, V) = \bigl(\{\sigma(v_1),\, \sigma(v_2),\, x\} & (v_1
      \gets U,\, v_2 \gets V) \enspace, \\
      \{\tau(v_1),\, \tau(v_2)\} & (v_1 \gets U,\, v_2 \gets V)\bigr)
      \enspace.
    \end{split}
  \end{align}
  Finally, we write~\eqref{Equation.2.7.2} in the more readable form
  \begin{align}
    \label{Equation.2.7.4}
    \begin{split}
      v_1 &= \sigma(v_1) + \sigma(v_2) + x \\
      v_2 &= \tau(v_1) + \tau(v_2)
    \end{split}
  \end{align}
  as a system of equations to be solved in the forest algebra~$\gp\cF$
  which is augmented by union as an operation.  Union is denoted here
  by~$+$.  \epr
\end{ex}

It is obvious that Example~\ref{Example.2.7.3} could be repeated for
any regular language and that the language itself is always the union
of those components of the minimal fixed-point which correspond to
final states.  The interpretation of the equations in terms of forest
operations serves as the starting point for a generalization to
equations for regular forests.

Fix again a ranked alphabet~$\Sigma$ and a frontier alphabet~$X$.  For
any $k \geq 1$, let
\[ F_k = (\gp F_\Sigma(X))^k \]
be the set of $k$-tuples of $\Sigma X$-forests.  We order~$F_k$
partially by componentwise inclusion:
\[ (S_1, \dotsc, S_k) \leq (T_1, \dotsc, T_k) \quad \text{iff} \quad
S_1 \subseteq T_1, \dotsc, S_k \subseteq T_k \enspace. \]
Then~$F_k$ becomes a complete lattice in which least upper bounds and
greatest lower bounds are obtained, respectively, by forming
componentwise unions and intersections, thus
\begin{align*}
  \bigvee \bigl( (S_{i1}, \dotsc, S_{ik}) \mid i \in I \bigr) &= \Bigl(
  \bigcup (S_{i1} \mid i \in I), \dotsc, \bigcup (S_{ik} \mid i \in I)
  \Bigr) \\
  \intertext{and}
  \bigwedge \bigl( (S_{i1}, \dotsc, S_{ik}) \mid i \in I \bigr) &= \Bigl(
  \bigcap (S_{i1} \mid i \in I), \dotsc, \bigcap (S_{ik} \mid i \in I)
  \Bigr)  \enspace.
\end{align*}
The least element is $\pmb{0} = (\emptyset, \dotsc, \emptyset)$. (We
refer the reader to Section~\ref{Section.1.4} for the lattice theory
needed here.)

Let $V_k = \{v_1, \dotsc, v_k\}$ be a set of variables disjoint from
$\Sigma$~and~$X$.  With every $\Sigma(X \cup V_k)$-forest~$P$ we
associate the mapping
\[ \hat P \colon F_k \to \gp F_\Sigma(X) \]
defined so that
\[ \hat P(T_1, \dotsc, T_k) = P(v_1 \gets T_1, \dotsc, v_k \gets
T_k) \]
\index{Sigmaxkpolynomial@$(\Sigma, X, k)$-polynomial}
for all $(T_1, \dotsc, T_k) \in F_k$.  A $k$-tuple $\Pi = (P_1,
\dotsc, P_k)$ of finite $\Sigma(X \cup V_k)$-forests is called a
\emph{$(\Sigma, X, k)$-polynomial} and we associate with it the
mapping
\[ \hat \Pi \colon F_k \to F_k \]
defined so that
\[ \hat \Pi(\bT) = \bigl( \hat P_1(\bT), \dotsc, \hat P_k(\bT) \bigr)
\quad (\bT \in F_k) \enspace. \]

\begin{lm}
  \label{Lemma.2.7.4}
  For any $(\Sigma, X, k)$-polynomial~$\Pi$, the mapping~$\hat\Pi
  \colon F_k \to F_k$ is $\omega$-continuous.
\end{lm}

\pr Let $\Pi = (P_1, \dotsc, P_k)$.  The mapping~$\hat \Pi$ is isotone
as
\[ P(v_1 \gets S_1, \dotsc, v_k \gets S_k) \subseteq P(v_1 \gets T_1,
\dotsc, v_k \gets T_k) \]
obviously holds for all $P \subseteq F_\Sigma(X \cup V_k)$ and $\Sigma
X$-forests $S_1, \dotsc, S_k, T_1, \dotsc, T_k$ such that $S_1
\subseteq T_1, \dotsc, S_k \subseteq T_k$.  Let
\[ \bT_0 \subseteq \bT_1 \subseteq \bT_2 \subseteq \ldots \]
be any ascending $\omega$-sequence of vectors
\[ \bT_i = (T_{i1}, \dotsc, T_{ik}) \in F_k \quad (i \geq 0) \]
of $\Sigma X$-forests.  Now write
\[ \bT = \Bigl( \bigcup (T_{i1} \mid i \geq 0), \dotsc, \bigcup
(T_{ik} \mid i \geq 0) \Bigr) \enspace. \]
In order to prove $\omega$-continuity we should show that
\[ \hat \Pi(\bT) = \Bigl(\bigcup \bigl(\hat P_1(\bT_i) \mid i \geq 0
\bigr), \dotsc, \bigcup \bigl(\hat P_k(\bT_i) \mid i \geq 0 \bigr)
\Bigr) \enspace, \] or equivalently, that
\begin{align}
  \label{Equation.2.7.5}
  \hat P_j(\bT) = \bigcup \bigl(\hat P_j(\bT_i) \mid i \geq 0 \bigr)
  \quad (j = 1, \dotsc, k) \enspace.
\end{align}
Every tree~$t \in \hat P_j(\bT)$ is obtained from some $p \in P_j$ by
substituting a tree from~$\bigcup (T_{im} \mid i \geq 0)$ for every
occurrence of each variable~$v_m$ and each $m = 1, \dotsc, k$.  The
number of occurrences of variables in~$p$ is finite.  Hence there
exists an $i \geq 0$ such that all trees used in this substitution
appear in a component of~$\bT_i$.  Then $t \in \hat P_j(\bT_i)$.  This
shows that the left side of~\eqref{Equation.2.7.5} is included in the
right side of~\eqref{Equation.2.7.5} for each $j = 1, \dotsc, k$.  The
converse inclusions are obvious since $\hat\Pi$ is isotone and $\bT_i
\leq \bT$ for all $i \geq 0$. \epr

\

Now, using Theorem~\ref{Theorem.1.4.8} we get

\begin{cor}
  \label{Corollary.2.7.5}
  For any $(\Sigma, X, k)$-polynomial~$\Pi$, the mapping~$\hat\Pi
  \colon F_k \to F_k$ has the least fixed-point
  \begin{align*}
    [\hat\Pi] = \bigvee (\pmb 0\hat\Pi^i \mid i \geq 0)
    \enspace. \tag*{\epr}
  \end{align*}
\end{cor}

The corollary means that $[\hat\Pi]$~is the least solution of the
fixed-point equation
\begin{align}
  \label{Equation.2.7.6}
  (v_1, \dotsc, v_k) = \hat\Pi(v_1, \dotsc, v_k) \enspace,
\end{align}
where the $v_i$'s are ``unknowns'' that assume $\Sigma X$-forests as
their values.  The equation~\eqref{Equation.2.7.6} can also be written
as a system of equations
\begin{align}
  \label{Equation.2.7.7}
  \begin{cases}
    v_1 = P_1 \\
    \phantom{v_1} \mathrel{\,\vdots} \\
    v_k = P_k \enspace,
  \end{cases}
\end{align}
where the $P$'s are usually expressed as formal sums of their elements
(as we did in Example~\ref{Example.2.7.3}).

The finiteness of the components~$P_i$ was not used in the proof of
Lemma~\ref{Lemma.2.7.4}.  However, it will be essential for obtaining
the main result of this section.  In fact, it will be convenient,
although not necessary, to work with an even more restricted class of
fixed-point equations, which we shall soon introduce.
Example~\ref{Example.2.7.3} provides us with a guideline here, too.

\index{height of!tree}
Let us extend the height function of~$F_\Sigma(X)$ to~$F_\Sigma(X \cup
V_k)$ so that
\[ \hg(v_i) = -1 \quad (i = 1, \dotsc, k) \enspace. \]
Then the $\Sigma(X \cup V_k)$-trees of height~$0$ are
\begin{enumerate}[(i)]
\item the frontier letters $x \in X$,
\item the $0$-ary operators $\sigma \in \Sigma_0$, and
\item the trees of the form $\sigma(v_{i_1}, \dotsc, v_{i_m})$, where
  $m > 0$, $\sigma \in \Sigma_m$ and $v_{i_1}, \dotsc, v_{i_m} \in
  V_k$.
\end{enumerate}

\begin{df} \rm
  \label{Definition.2.7.6}
  \index{Sigmaxkpolynomial@$(\Sigma, X, k)$-polynomial!regular}
  A $(\Sigma, X, k)$-polynomial $\Pi = (P_1, \dotsc, P_k)$ is
  \emph{regular}, if every $\Sigma(X \cup V_k)$-tree of height~$0$
  belongs to exactly one~$P_j$, and the $P_j$'s do not contain any
  other trees.  If $\Pi$~is regular, then $\hat\Pi$~and the
  corresponding fixed-point equation~\eqref{Equation.2.7.6} are also
  said to be \emph{regular}.
  \index{regular sigmaxexpression@regular $\Sigma X$-expression}
  \index{regular fixed-point equation}
  \index{forest!equational}
  A $\Sigma X$-forest~$T$ is called
  \emph{equational} if it can be expressed as the union of some
  components of the least solution of a regular fixed-point equation.
\end{df}

The fixed-point equation in Example~\ref{Example.2.7.3} is regular.
It is easy to see that the same procedure applied to any Rabin-Scott
recognizer will yield a regular fixed-point equation.  Hence, every
regular language is equational when viewed as a unary forest.  It is
also well-known, and easy to prove, that the components of the least
solution of a system of Arden equations are regular.

\begin{ex} \rm
  \label{Example.2.7.7}
  Let $\Sigma = \Sigma_0 \cup \Sigma_2$, $\Sigma_0 = \{\gamma\}$,
  $\Sigma_2 = \{\sigma\}$ and $X = \{x, y\}$.  Then
  \[ \Pi = \bigl(\{x,\, \gamma,\, \sigma(v_1, v_2),\, \sigma(v_2, v_1)\},\;
  \{y,\, \sigma(v_1, v_1),\, \sigma(v_2, v_2)\} \bigr) \]
  is a regular $(\Sigma, X, 2)$-polynomial.  The corresponding regular
  fixed-point equation can be written as the system
  \[
  \begin{cases}
    v_1 = x + \gamma + \sigma(v_1, v_2) + \sigma(v_2, v_1) \\
    v_2 = y + \sigma(v_1, v_1) + \sigma(v_2, v_2) \enspace.
  \end{cases} \]
  The least solution is the pair~$(T_1, T_2)$, where
  \begin{align*}
    T_1 &= \bigl\{x,\, \gamma,\, \sigma(x,y),\, \sigma(\gamma, y),\, \sigma(y,x),\, \sigma(y,\gamma),\,
    \sigma \bigl(x, \sigma(x, x) \bigr), \dotsc \bigr\} \\
    \intertext{and}
    T_2 &= \bigl\{y,\, \sigma(x, x),\, \sigma(\gamma, \gamma),\,
    \sigma(y, y), \dotsc \bigr\} \enspace. \tag*{\epr}
  \end{align*}
\end{ex}

Let $[\hat\Pi] = (T_1, \dotsc, T_k)$ be the least fixed-point for a
given~$(\Sigma, X, k)$-polynomial~$\Pi$.  We define a binary
relation~$\varrho(\Pi)$ in~$F_\Sigma(X)$:
\[ \varrho(\Pi) = \{ (s, t) \mid s, t \in T_i \text{ for some } i = 1,
\dotsc, k\} \enspace. \]

\begin{lm}
  \label{Lemma.2.7.8}
  If $\Pi$~is a regular $(\Sigma, X, k)$-polynomial, then
  $\varrho(\Pi)$ is a congruence of~$\cF_\Sigma(X)$ with at most
  $k$~equivalence classes.  For each congruence~$\varrho$
  of~$\cF_\Sigma(X)$ of index~$k$ ($k \geq 1$) there exists a regular
  $(\Sigma, X, k)$-polynomial~$\Pi$ such that~$\varrho(\Pi) =
  \varrho$.
\end{lm}

\pr Let $\Pi = (P_1, \dotsc, P_k)$ be a regular $(\Sigma, X,
k)$-polynomial and $[\hat\Pi] = (T_1, \dotsc, T_k)$ the corresponding
least fixed-point.  From the definition of~$\varrho(\Pi)$ it is clear
that the relation is symmetric.  To prove that it is reflexive and
transitive, too, we show that every $\Sigma X$-tree~$t$ belongs to
exactly one~$T_i$.  First we note that
\begin{align}
  \label{Equation.2.7.8}
  T_i = P_i(v_1 \gets T_1, \dotsc, v_k \gets T_k) \quad (i = 1,
  \dotsc, k)
\end{align}
as~$[\hat\Pi]$ is a fixed-point of~$\hat\Pi$.  We proceed now by
induction on~$\hg(t)$.
\begin{enumerate}
\item[$1^{\text o}$] If $\hg(t) = 0$, then $t$~is in exactly one of
  the sets~$P_i$ ($i = 1, \dotsc, k$) because $\Pi$~is regular.
  From~\eqref{Equation.2.7.8} we see that $t$~is in the
  corresponding~$T_i$ and that it could belong to some other~$T_j$ ($j
  \neq i$) only in case $v_i \in P_j$.  But $\hg(v_i) = -1$ and
  $v_i$~does not appear in~$\Pi$.
\item[$2^{\text o}$] Consider a tree $t = \sigma(t_1, \dotsc, t_m)$
  ($m > 0$) and assume that all trees of lesser height belong to
  exactly one~$T_i$.  Then there exists for each~$j = 1, \dotsc, m$
  exactly one~$i_j$ ($1 \leq i_j \leq k$) such that $t_j \in
  T_{i_j}$.  Also, there is exactly one~$i$ ($1 \leq i \leq k$) such
  that $p = \sigma(v_{i_1}, \dotsc, v_{i_m}) \in P_i$.  Clearly,
  \[ t \in p(v_1 \gets T_1, \dotsc, v_k \gets T_k) \subseteq T_i
  \enspace. \]
  The uniqueness of the indices~$i_j$ implies that $p$~is the only
  tree of height~$0$ in~$F_\Sigma(X \cup V_k)$ from which $t$~can be
  obtained by the substitutions $v_1 \gets T_1, \dotsc, v_k \gets
  T_k$.  Hence $t$~belongs to~$T_i$ only.
\end{enumerate}

Now we know that $\varrho(\Pi) \in E \bigl(F_\Sigma(X) \bigr)$.  It is
obvious that it has at most $k$~equivalence classes.  (There may be
less than $k$~classes as some $T$'s could be empty.)  To prove that it
is a congruence relation we consider any $m \geq 1$, $\sigma \in
\Sigma_m$ and $s_1, \dotsc, s_m, t_1, \dotsc, t_m \in F_\Sigma(X)$
such that
\[ s_1 \equiv t_1, \dotsc, s_m \equiv t_m \; \bigl(\varrho(\Pi)\bigr)
\enspace. \]
There are indices $i_1, \dotsc, i_m$ such that
\[ s_j, t_j \in T_{i_j} \text{ for } j = 1, \dotsc, m \enspace. \]
Let $\sigma(v_{i_1}, \dotsc, v_{i_m})$ be in~$P_i$.  Then
\[ \sigma(s_1, \dotsc, s_m),\, \sigma(t_1, \dotsc, t_m) \in T_i \]
by~\eqref{Equation.2.7.8}.  Hence
\[ \sigma^{\cF_\Sigma(X)}(s_1, \dotsc, s_m) \equiv
\sigma^{\cF_\Sigma(X)}(t_1, \dotsc, t_m) \; \bigl(\varrho(\Pi) \bigr) \]
as required.

Now, suppose $\varrho \in C \bigl(\cF_\Sigma(X)\bigr)$ and let $S_1,
\dotsc, S_k$ be the equivalence classes of~$\varrho$.  We define a
$(\Sigma, X, k)$-polynomial $\Pi = (P_1, \dotsc, P_k)$ so that
\[ P_i = \{ p \in F_\Sigma(X \cup V_k) \mid \hg(p) = 0,\, p(v_1
\gets S_1, \dotsc, v_k \gets S_k) \subseteq S_i\} \]
for all $i = 1, \dotsc, k$.  The fact that $\varrho$~is a congruence
means that for each~$p$ of height~$0$ there is exactly one~$i$ ($1
\leq i \leq k$) such that
\[ p(v_1 \gets S_1, \dotsc, v_k \gets S_k) \subseteq S_i
\enspace. \]
Hence $\Pi$~is regular.  We claim that $\varrho(\Pi) = \varrho$.
Let $[\hat\Pi] = (T_1, \dotsc, T_k)$.  In order to prove the second
statement of the lemma we show by induction on~$\hg(t)$ that for all
$i = 1, \dotsc, k$,
\[ \big( \forall t \in F_\Sigma(X) \big) \big( t \in S_i
\Longleftrightarrow t \in T_i \big) \enspace. \]
\begin{enumerate}
\item[$1^{\text o}$] If $\hg(t) = 0$, then there is exactly one~$i$
  such that $t \in P_i$.  This means $t \in S_i$.
  From~\eqref{Equation.2.7.8} it follows that $t \in T_i$ for the same~$i$.
\item[$2^{\text o}$] Let $t = \sigma(t_1, \dotsc, t_m)$ ($m > 0$) and
  suppose the claim holds for all trees of height~$< \hg(t)$.  Then
  there are unique indices $i_1, \dotsc, i_m$ such that
  \[ t_j \in S_{i_j} \cap T_{i_j} \quad (j = 1, \dotsc, m)
  \enspace. \]
  Also, there is a unique~$i$ such that
  \[ p = \sigma(v_{i_1}, \dotsc, v_{i_m}) \in P_i \enspace. \]
  Then
  \[ t \in \sigma(S_{i_1},\ldots ,S_{i_m}) =  p(v_1 \gets S_1, \dotsc, v_k \gets S_k) \subseteq S_i \]
  by the definition of~$P_i$.  On the other hand, \eqref{Equation.2.7.8}
  implies $t \in T_i$. \epr
\end{enumerate}

\

If we combine Lemma~\ref{Lemma.2.7.8} and Theorem~\ref{Theorem.2.7.1},
we get

\begin{thm}
  \label{Theorem.2.7.9}
  A forest is equational iff it is recognizable. \epr
\end{thm}

From the first part of this section it is clear that a $\Sigma
X$-forest~$T$ can be recognized by a $k$-state tree recognizer iff
$T$~is saturated by a congruence of~$\cF_\Sigma(X)$ of index~$\leq
k$.  From Lemma~\ref{Lemma.2.7.8} we get a similar connection between
the number of states and the number of variables in a regular
fixed-point equation which defines the forest.

There is also a very close connection between regular tree grammars
and the fixed-point equations considered here.  For example, the
equations of Example~\ref{Example.2.7.7} can be converted into the
following set of productions in which $v_1$~and~$v_2$ are nonterminal
symbols:
\begin{align*}
  v_1 &\to x, & v_1 &\to \gamma, & v_1 &\to \sigma(v_1, v_2), & v_1
  &\to \sigma(v_2, v_1), \\
  v_2 &\to y,  & v_2 &\to \sigma(v_1, v_1), & v_2 &\to \sigma(v_2,
  v_2) \enspace.
\end{align*}
The resulting regular tree grammar generates~$T_1$ if $v_1$~is the
initial symbol, and it generates~$T_2$ if $v_2$~is the initial
symbol.

On the other hand, every regular $\Sigma X$-grammar with
$k$~nonterminal symbols can be converted into a fixed-point system
with $k$~equations.  This system is not necessarily regular, but the
components of the least solution are nevertheless the regular forests
generated by the grammar from the different nonterminal symbols.  For
example, if $\Sigma$~and~$X$ are as in Example~\ref{Example.2.7.7} and
the productions are
\[
  a \to x, \quad a \to \gamma, \quad a \to \sigma(a, b), \quad b \to \sigma(b, b), \quad b \to y ,
\]
then the corresponding equations would be
\begin{align*}
  a &= x + \gamma + \sigma(a, b) \quad \text{and} \\
  b &= y + \sigma(b, b) \enspace,
\end{align*}
where $a$~and~$b$ now are the unknowns.  The least solution
is~$\bigl(T(G_a), T(G_b) \bigr)$, where $G_a$~and~$G_b$ are the
grammars which we obtain by choosing $a$~and~$b$, respectively, as the
initial symbol.

%% file: Figure.2.5.tex
\centering
\begin{tikzpicture}[>=latex,node/.style={draw,circle},node distance = 3cm]
	\node (initial) {};
	\node[node,right of = initial,node distance=0.8cm] (q1) {\tiny $1$};
	\node[node,right of =q1] (q2) {\tiny $2$};
	\node[right of = q2,node distance=0.8cm] (final) {};
	\path[->] (initial) edge (q1);
	\path[->] (q2) edge (final);
	\path[->] (q1) edge node[below,pos=0.1] {\footnotesize $\tau$} (q2);
	\path[->] (q2) edge [bend right] node[above] {\footnotesize $\sigma$} (q1);
	
	\draw [->] (q1) ++ (0.25cm,0.3cm) arc[radius=0.25cm,start angle=-20,end angle =180] --++ (0cm,-1mm);
	\node at (1.2cm,0.65cm) {\footnotesize $\sigma$};
	
	\draw [->] (q2) ++ (0.25cm,0.3cm) arc[radius=0.25cm,start angle=-20,end angle =180] --++ (0cm,-1mm);
 	\node at (4.2cm,0.65cm) {\footnotesize $\tau$};

\end{tikzpicture}

%% file: Section.2.8.tex
Our next description of the recognizable forests is a streamlined
generalization of a well-known characterization of the regular
languages given by J.\@ Medvedev in~1956.  First we define the family
of representable forests.  The theorem states then that the
representable forests are exactly the recognizable forests.  The
representable forests are defined collectively for all ranked
alphabets as the definition involves tree homomorphisms and these may
take us from one alphabet to another.  Recall that $r(\Sigma)$~is the
finite set of nonnegative integers~$m$ for which~$\Sigma_m \neq
\emptyset$.

\begin{df} \rm
  \label{Definition.2.8.1}
  For every pair~$(\Sigma, X)$ we define the ``next-to-root function''
  \[ \mathord{\nroot} \colon F_\Sigma(X) - (\Sigma_0 \cup X) \to
  \bigcup \bigl( (\Sigma \cup X)^m \mid m \in r(\Sigma) \bigr) \]
  so that
  \[ \nroot \bigl(\sigma(t_1, \dotsc, t_m) \bigr) = \bigl( \rroot(t_1),
  \dotsc, \rroot(t_m) \bigr) \]
  for all $m > 0$, $\sigma \in \Sigma_m$ and $t_1, \dotsc, t_m \in
  F_\Sigma(X)$.
\end{df}

\begin{df} \rm
  \label{Definition.2.8.2}
  \index{forest!elementary}
  The \emph{elementary $\Sigma X$-forests} are the forests
  \begin{align*}
    U(d) &= \rroot^{-1}(d) \quad (d \in \Sigma \cup X) \enspace, \text{
      and} \tag{i} \\
    V(d_1, \dotsc, d_m) &= \nroot^{-1}(d_1, \dotsc, d_m) \enspace,
    \tag{ii}
  \end{align*}
  where $m > 0$, $m \in r(\Sigma)$, and $d_1, \dotsc, d_m \in \Sigma
  \cup X$.
\end{df}

Note that the definitions of the $U(d)$-~and~$V(d_1, \dotsc,
d_m)$-forests presume a~$\Sigma$ and an~$X$ although the notations do
not show this.  Clearly, $U(d)$~is the set of all $\Sigma X$-trees
with the root labelled by~$d$, and $V(d_1, \dotsc, d_m)$ consists of
all $\Sigma X$-trees of height $\geq 1$ in which the nodes immediately
above the root are labelled, from left to right, by~$d_1, \dotsc,
d_m$, respectively.  Note also that $U(d) = \{d\}$ when $d \in
\Sigma_0 \cup X$.  We need three more definitions.

\begin{df} \rm
  \label{Definition.2.8.3}
  \index{restriction of!forest}
  The \emph{restriction} of a forest~$T$ is the forest
  \[ \rest(T) = \{t \in T \mid \sub(t) \subseteq T\} \enspace. \]
\end{df}

\begin{df} \rm
  \label{Definition.2.8.4}
  \index{operation!elementary}
  The \emph{elementary operations} on forests are the formation of
  \begin{enumerate}[(i)]
  \item the union of two forests,
  \item the intersection of two forests,
  \item an alphabetic tree homomorphic image of a forest, and
  \item the restriction of a forest.
  \end{enumerate}
\end{df}

\begin{df} \rm
  \label{Definition.2.8.5}
  \index{forest!representable}
  A forest is \emph{representable} if it can be constructed from
  elementary forests by a finite number of applications of elementary
  operations.
\end{df}

Now the theorem can be stated.

\begin{thm}
  \label{Theorem.2.8.6}
  A forest is representable iff it is recognizable.
\end{thm}

\pr To prove that the representable forests are recognizable it
suffices to note that the elementary forests are recognizable and that
the elementary operations preserve recognizability.  Consider any
$\Sigma$~and~$X$.  If $d \in \Sigma_0 \cup X$, then $U(d) = \{d\} \in
\Rec(\Sigma, X)$.  If $d  \in \Sigma_m$ ($m > 0$), then
\[ U(d) = d(y_1, \dotsc, y_m) \bigl(y_1 \gets F_\Sigma(X), \dotsc, y_m
\gets F_\Sigma(X) \bigr) \]
is again recognizable.  Similarly,
\[ V(d_1, \dotsc, d_m) = \bigcup \bigl( \sigma(y_1, \dotsc, y_m)
\bigl(y_1 \gets U(d_1), \dotsc, y_m \gets U(d_m) \bigr) \mid \sigma
\in \Sigma_m \bigr) \]
is recognizable for all $m \in r(\Sigma)$ and $d_1, \dotsc, d_m \in
\Sigma \cup X$.  We have already seen in Section~\ref{Section.2.4}
that unions, intersections and alphabetic tree homomorphisms preserve
recognizability.  Let $T$~be the forest recognized by a $\Sigma
X$-recognizer~$\bA$.  We construct a recognizer for~$\rest(T)$.  First
define a $\Sigma$-algebra $\cB = (A \cup b, \Sigma)$ ($b \notin A$) so
that
\[ \sigma^{\cB}(b_1, \dotsc, b_m) =
\begin{cases}
  \sigma^{\cA}(b_1, \dotsc, b_m) & \text{if } b_1, \dotsc, b_m \in A
  \text{ and } \sigma^{\cA}(b_1, \dotsc, b_m) \in A' \enspace, \\
  b & \text{in all other cases,}
\end{cases} \]
for all $m \geq 0$, $\sigma \in \Sigma_m$ and $b_1, \dotsc, b_m \in A
\cup b$.  The initial assignment $\beta \colon X \to A \cup b$ is
defined so that for each $x \in X$,
\[ x\beta =
\begin{cases}
  x\alpha & \text{ if } x\alpha \in A' \enspace, \\
  b & \text{ if } x\alpha \notin A' \enspace.
\end{cases} \]
Consider any $\Sigma X$-tree~$t$.  It is easy to show that
\[ t\hat\beta =
\begin{cases}
  t\hat\alpha & \text{ if } \sub(t) \subseteq T \enspace, \\
  b & \text{ otherwise.}
\end{cases} \]
Hence, $\bB = (\cB, \beta, A')$ recognizes~$\rest(T)$.

We shall now show that every recognizable forest is representable.
Let $T = T(\bA)$ for some $\Sigma X$-recognizer~$\bA$.  First define a
new ranked alphabet~$\Omega$ such that
\[ \Omega_m = \Sigma_m \times (A \cup X)^m \quad \text{for all} \quad
m \geq 0 \enspace. \]
We construct two representable $\Omega X$-forests $R$~and~$S$ as
follows.  For $c \in A \cup X$ we introduce the notation
\[ \overline c =
\begin{cases}
  c & \text{ if } c \in A \enspace, \\
  c\alpha & \text{ if } c \in X \enspace.
\end{cases} \]
Then
\begin{align*}
  R &= \{x \in X \mid x\alpha \in A'\} \cup {} \\*
  &\phantom{{}={}} \bigcup \bigl( U((\sigma, c_1,
  \dotsc, c_m)) \mid (\sigma, c_1, \dotsc, c_m) \in \Omega,\,
  \sigma^{\cA}(\overline c_1, \dotsc, \overline c_m) \in A' \bigr)
  \enspace.
\end{align*}
The forest~$S$ is the union of all intersections
\[ V(u_1, \dotsc, u_m) \cap U \bigl((\sigma, b_1, \dotsc, b_m) \bigr)
\enspace, \]
where for each $i = 1, \dotsc, m$, either
\begin{enumerate}[(i)]
\item $u_i \in X$ and $\overline b_i = u_i\alpha$, or
\item $u_i = (\tau, c_1, \dotsc, c_k) \in \Omega_k$ ($k \geq 0$) and
  $b_i = \tau^{\cA}(\overline c_1, \dotsc, \overline c_k)$.
\end{enumerate}

Note that the possibility $m = 0$ is included at appropriate places in
the definitions of $R$~and~$S$.

Define the tree homomorphism
\[ h \colon F_\Omega(X) \to F_\Sigma(X) \]
so that
\[ h_m \bigl( (\sigma, b_1, \dotsc, b_m) \bigr) = \sigma \enspace,
\quad \bigl(m \geq 0,\, (\sigma, b_1, \dotsc, b_m) \in \Omega_m
\bigr) \]
and $h_X = 1_X$.  Clearly, $h$~is alphabetic.  We claim that
\[ T = h(P) \]
for the representable forest
\[ P = R \cap \rest(S \cup \Omega_0 \cup X) \enspace. \]
Let $p \in P$.  If $p = (\sigma, e) \in \Omega_0$, then $p \in R$
implies $\sigma^{\cA} \in A'$.  Hence $h(p) = \sigma \in T$.  If $p =
x \in X$, then $p \in R$ implies $h(x)\hat\alpha = x\alpha \in A'$.
Again $h(p) = x \in T$.  Next we show that for every $p \in \rest(S
\cup \Omega_0 \cup X)$ of height~$\geq 1$
\begin{equation}
  \label{Equation.2.8.1}
  h(p)\hat\alpha = \sigma^{\cA}(\overline b_1, \dotsc, \overline b_m)
  \enspace, \quad \text{ where } \quad (\sigma, b_1, \dotsc, b_m) =
  \rroot(p) \enspace.
\end{equation}
We proceed by induction on~$\hg(p)$.
\begin{enumerate}
\item[$1^{\text{o}}$] If $\hg(p) = 1$, then $m \geq 1$ and
  \[ p = (\sigma, b_1, \dotsc, b_m)(u_1, \dotsc, u_m) \]
  for some $u_1, \dotsc, u_m \in \Omega_0 \cup X$.  Since $p \in S$ we
  have $h(u_i)\hat\alpha = \overline b_i$ for all $i = 1, \dotsc, m$.
  But this implies that~\eqref{Equation.2.8.1} holds for~$p$.
\item[$2^{\text{o}}$] Now let
  \[ p = (\sigma, b_1, \dotsc, b_m)(p_1, \dotsc, p_m) \]
  and assume that~\eqref{Equation.2.8.1} holds for the trees~$p_1,
  \dotsc, p_m$.  As~$p$ is in~$S$ and
  \[ h(p)\hat\alpha = \sigma^{\cA} \bigl(h(p_1)\hat\alpha, \dotsc,
  h(p_m)\hat\alpha \bigr) \enspace, \]
  it suffices to show that $h(p_i)\hat\alpha = \overline b_i$ for
  every $i = 1, \dotsc, m$.  We should consider three cases.
  \begin{enumerate}[(a)]
  \item If $p_i$~is of the form $(\tau, c_1, \dotsc, c_k)(r_1, \dotsc,
    r_k)$ ($k > 0$), then the induction hypothesis yields
    \[ h(p_i)\hat\alpha = \tau^{\cA}(\overline c_1, \dotsc, \overline
    c_k) \enspace. \]
    Moreover, $\tau^{\cA}(\overline c_1, \dotsc, \overline c_k) = b_i
    = \overline b_i$ since $p \in S$.
  \item If $p_i = (\sigma, e) \in \Omega_0$, then $h(p_i)\hat\alpha =
    \sigma^{\cA} = b_i = \overline b_i$.
  \item If $p_i = x \in X$, then $h(p_i)\hat\alpha = x\alpha =
    \overline b_i$.
  \end{enumerate}
\end{enumerate}
Now we have completed the proof of~\eqref{Equation.2.8.1}.  Consider
any tree
\[ p = (\sigma, b_1, \dotsc, b_m)(p_1, \dotsc, p_m) \in P \enspace. \]
By using~\eqref{Equation.2.8.1} and the fact that $p \in R$ we get
\[ h(p)\hat\alpha = \sigma^{\cA}(\overline b_1, \dotsc, \overline b_m)
\in A' \enspace. \]
This implies $h(p) \in T$ and we have shown that $h(P) \subseteq T$.

In order to prove the converse inclusion we show first by tree
induction how to construct for each $t \in F_\Sigma(X)$ a tree~$p \in
\rest(S \cup \Omega_0 \cup X)$ such that $h(p) = t$:
\begin{enumerate}
\item[$1^{\text{o}}$] If $t = x \in X$, then we may choose $p = x$.
\item[$2^{\text{o}}$] If $t = \sigma \in \Sigma_0$, put $p = (\sigma,
  e)$.
\item[$3^{\text{o}}$] Let $t = \sigma(t_1, \dotsc, t_m)$ ($m > 0$) and
  suppose we have trees $p_1, \dotsc, p_m \in \rest(S \cup \Omega_0
  \cup X)$ such that $h(p_i) = t_i$ ($i = 1, \dotsc, m$).  If we put
  \[ p = (\sigma, b_1, \dotsc, b_m)(p_1, \dotsc, p_m) \enspace, \]
  where $b_i = t_i\hat\alpha$ for $i = 1, \dotsc, m$, then $h(p) = t$
  and $p \in \rest(S \cup \Omega_0 \cup X)$ as required.
\end{enumerate}

Let $t \in T$ and construct a~$p$ for~$t$ as above.  To prove~$t \in
h(P)$ it suffices to show that~$p \in R$.  This can again be done by
tree induction:
\begin{enumerate}
\item[$1^{\text{o}}$] If $t = x \in X$, then $x\alpha \in A'$ and
  hence $p = x \in R$.
\item[$2^{\text{o}}$] If $t = \sigma \in \Sigma_0$, then $\sigma^{\cA}
  \in A'$ and $p = (\sigma, e) \in U(\sigma, e) \subseteq R$.
\item[$3^{\text{o}}$] Let $t = \sigma(t_1, \dotsc, t_m)$ ($m > 0$).
  If we use~\eqref{Equation.2.8.1} and its notation, we get
  \[ \sigma^{\cA}(\overline b_1, \dotsc, \overline b_m) =
  h(p)\hat\alpha = t\hat\alpha \in A' \enspace. \]
\end{enumerate}
This shows that $p \in R$. \epr 

%% file: Section.2.9.tex
In this section a proper subfamily of the recognizable forests is
introduced.  We will then also get one more characterization of the
recognizable forests, not quite unrelated to that given in the
preceding section.

We need the following auxiliary concept

\begin{df} \rm
  \label{Definition.2.9.1}
  \index{fork of $\Sigma X$-tree}
  The set of \emph{forks} $\fork(t)$ of a $\Sigma X$-tree~$t$ is
  defined as follows:
  \begin{enumerate}
  \item[$1^{\text{o}}$] If $t \in \Sigma_0 \cup X$, then $\fork(t) =
    \emptyset$.
  \item[$2^{\text{o}}$] If $t = \sigma(t_1, \dotsc, t_m)$ ($m > 0$),
    then
    \[ \fork(t) = \fork(t_1) \cup \dotsm \cup \fork(t_m) \cup \bigl\{
    \sigma\bigl( \rroot(t_1), \dotsc, \rroot(t_m) \bigr) \}
    \enspace. \]
  \end{enumerate}
  The set of all forks of $\Sigma X$-trees $\bigcup \bigl( \fork(t)
  \mid t \in F_\Sigma(X) \bigr)$ will be denoted by~$\fork(\Sigma,
  X)$.
\end{df}

\begin{ex} \rm
  \label{Example.2.9.2}
  Let $\Sigma = \Sigma_0 \cup \Sigma_1 \cup \Sigma_2$, $\Sigma_0 =
  \{\gamma\}$, $\Sigma_1 = \{\tau\}$, $\Sigma_2 = \{\sigma\}$ and $X =
  \{x, y\}$.  For the $\Sigma X$-tree
  \[ t = \sigma \bigl(\tau(\gamma),\, \sigma(x,\, \tau(y)) \bigr)
  \enspace, \]
  we have
  \[ \fork(t) = \bigl\{ \sigma(\tau, \sigma),\, \tau(\gamma),\,
  \sigma(x, \tau),\, \tau(y) \bigr\} \enspace. \]
  Graphically these forks are represented as in Fig.~\ref{Figure.p.106} respectively.  Obviously, $\fork(\Sigma, X)$ is always
  finite and here it consists of 30~forks. \epr
  \begin{figure}[h]
    \centering
    \input{Figure.p.106}
    \caption{{}\label{Figure.p.106}}
  \end{figure}
\end{ex}

Local forests may now be defined.

\begin{df} \rm
  \label{Definition.2.9.3}
  \index{forest!local}
  A $\Sigma X$-forest~$T$ is \emph{local} if there are sets $R
  (\subseteq \Sigma \cup X)$ and $F \bigl(\subseteq \fork(\Sigma, X)
  \bigr)$ such that, for each $t \in F_\Sigma(X)$,
  \[ t \in T \quad \text{ iff } \quad \rroot(t) \in R \quad \text{ and
  } \quad \fork(t) \subseteq F \enspace. \]
  Then we write $T = \Loc(R, F)$.
\end{df}

Hence the membership of a $\Sigma X$-tree~$t$ in the local
forest~$\Loc(R, F)$ can be decided by testing for the local properties
$\rroot(t) \in R$ and $\fork(t) \subseteq F$.

A $\Sigma X$-recognizer for~$\Loc(R, F)$ can be constructed as
follows.  First we define a $\Sigma$-algebra $\cA = (A, \Sigma)$.  Let
$A = \Sigma \cup X \cup 0$ ($0 \notin \Sigma \cup X$).  For every
$\sigma \in \Sigma_0$, put $\sigma^{\cA} = \sigma$.  For $m > 0$,
$\sigma \in \Sigma_m$ and $a_1, \dotsc, a_m \in A$ let
\[ \sigma^{\cA}(a_1, \dotsc, a_m) =
\begin{cases}
  \sigma & \text{if } \sigma(a_1, \dotsc, a_m) \in F \enspace, \\
  0 & \text{otherwise.}
\end{cases} \]
Let $\alpha \colon X \to A$ be the embedding $x \mapsto x$ ($x \in
X$). It is easy to show that for all $t \in F_\Sigma(X)$,
\[ t\hat\alpha =
\begin{cases}
  \rroot(t) & \text{if } \fork(t)  \subseteq F \enspace, \\
  0 & \text{otherwise.}
\end{cases} \]
This readily implies $T(\bA) = \Loc(R, F)$ for $\bA = (\cA, \alpha,
R)$.  Hence we have

\begin{thm}
  \label{Theorem.2.9.4}
  Every local forest is recognizable. \epr
\end{thm}

The converse of Theorem~\ref{Theorem.2.9.4} does not hold.  For
example, the forest consisting of the single tree of
Example~\ref{Example.2.9.2} is not local as there are many other trees
with the same root and the same forks.  However, the following fact
can be proved.

\begin{thm}
  \label{Theorem.2.9.5}
  For every recognizable $\Sigma X$-forest~$T$ there exist a ranked
  alphabet~$\Omega$, a frontier alphabet~$Y$, a local $\Omega
  Y$-forest~$S$ and an alphabetic tree homomorphism
  \[ h \colon F_\Omega(Y) \to F_\Sigma(X) \]
  such that $T = h(S)$.
\end{thm}

\pr Let $G = (N, \Sigma, X, P, a_0)$ be a regular $\Sigma X$-grammar
generating~$T$.  We assume that $G$~is in normal form.  A new ranked
alphabet~$\Omega$ is defined so that
\[ \Omega_m = \{ [a \to \sigma(a_1, \dotsc, a_m)] \mid a \to
\sigma(a_1, \dotsc, a_m) \in P,\, \sigma \in \Sigma_m\} \]
for all $m \geq 0$.  Also, let
\[ Y = \{[a \to x] \mid a \to x \in P,\, x \in X\} \enspace. \]
The local $\Omega Y$-forest $S = \Loc(R, F)$ is defined by the sets
\[ R = \{[a_0 \to p] \mid a_0 \to p \in P\} \]
and
\begin{align*}
  F &= \bigl\{ [a \to \sigma(a_1, \dotsc, a_m)]([a_1 \to p_1], \dotsc,
  [a_m \to p_m]) \mid m > 0,\, \\
  &\phantom{{}= \bigl\{ \quad} a \to \sigma(a_1, \dotsc, a_m),\, a_1
  \to p_1, \dotsc, a_m \to p_m \in P \bigr\} \enspace.
\end{align*}
Finally, define an alphabetic tree homomorphism
\[ h \colon F_\Omega(Y) \to F_\Sigma(X) \]
by the mappings
$$  h_Y \colon Y \to F_\Sigma(X), \quad [a \to x] \mapsto x ,$$
and
$$h_m \colon \Omega_m \to F_\Sigma(X \cup \Xi_m), \quad [a
  \to \sigma(a_1, \dotsc, a_m)] \mapsto \sigma(\xi_1, \dotsc, \xi_m)
  \enspace.$$
Now $h(S) = T$, and thereby the theorem, follows from
\eqref{Item.2.9.5.1}~and~\eqref{Item.2.9.5.2}:
\begin{enumerate}[(1)]
\item \label{Item.2.9.5.1} If $a \Rightarrow_G^* t$, for some $a \in
  N$ and $t \in F_\Sigma(X)$, then there is a tree~$s \in F_\Omega(Y)$
  such that $h(s) = t$, $\fork(s) \subseteq F$ and $\rroot(s)$ is of
  the form~$[a \to p]$.
\item \label{Item.2.9.5.2} If $s \in F_\Omega(Y)$ is such that
  $\fork(s) \subseteq F$ and $\rroot(s) = [a \to p]$ for some $p \in
  F_\Sigma(N \cup X)$, then $a \Rightarrow_G^* h(s)$.
\end{enumerate}

Part~\eqref{Item.2.9.5.1} can be proved by induction on the length of
the derivation of~$t$ and \eqref{Item.2.9.5.2}~by tree induction
on~$s$. \epr

\

Note that $h(S)$~is always recognizable when $S$~is a local forest and
$h$~an alphabetic tree homomorphism (Theorem~\ref{Theorem.2.9.4} and
Corollary~\ref{Corollary.2.4.20}). 

%% file: Figure.p.106.tex
\centering
\raisebox{-0.5cm}{
\begin{tikzpicture}[sibling distance=1.6cm,level distance=1.5cm]
	\begin{scope}[execute at begin node=$, execute at end node=$]
		\node[node,label={below:\sigma}] {} 
		child{ node[node,label={left:\sigma}] {}}
		child{ node[node,label={left:\tau}] {}};
	\end{scope}
\end{tikzpicture}}
\text{,} 
\raisebox{-0.5cm}{
\begin{tikzpicture}[sibling distance=1.4cm,level distance=1.5cm]
	\begin{scope}[execute at begin node=$, execute at end node=$]
		\node[node,label={below:\tau}] {} 
		child{ node[node,label={left:\gamma}] {}};
	\end{scope}
\end{tikzpicture}}
\text{,} 
\raisebox{-0.5cm}{
\begin{tikzpicture}[sibling distance=1.6cm,level distance=1.5cm]
	\begin{scope}[execute at begin node=$, execute at end node=$]
		\node[node,label={below:\sigma}] {} 
		child{ node[node,label={left:\tau}] {}}
		child{ node[node,label={left:x}] {}};
	\end{scope}
\end{tikzpicture}}
\text{and} 
\raisebox{-0.5cm}{
\begin{tikzpicture}[sibling distance=1.4cm,level distance=1.5cm]
	\begin{scope}[execute at begin node=$, execute at end node=$]
		\node[node,label={below:\tau}] {} 
		child{ node[node,label={left:y}] {}};
	\end{scope}
\end{tikzpicture}}

%% file: Section.2.10.tex
In this section we shall show that some of the first questions one
might ask about given tree recognizers are algorithmically decidable.
To begin with, we have the \emph{emptiness problem:}
\index{problem!emptiness} Is the forest recognized by a given tree
recognizer empty? Or one may ask whether this forest is finite or
infinite.  This is the \emph{finiteness
  problem}. \index{problem!finiteness} Finally, we have the important
\emph{equivalence problem:} \index{problem!equivalence} Do two given
tree recognizers recognize the same forest? \index{equivalence of!tree
  recognizers} In fact, the more general \emph{inclusion problem:}
\index{problem!inclusion} ``$T(\bA) \subseteq T(\bB)$?'' is shown to
be decidable.  The problems are quite easy and the proofs follow the
strategy familiar from finite automata theory with a ``pumping lemma''
as the key result.  We have seen in Section~\ref{Section.2.2} that any
nondeterministic frontier-to-root, or root-to-frontier, tree
recognizer can be converted into an equivalent deterministic
F-recognizer.  Hence we may again restrict ourselves to our basic type
of tree recognizers.

We need the following special notation.  Let $\Sigma$~and~$X$ be
given.  Introduce a new letter~$\xi$ and let $T_\xi$~be the set of all
$\Sigma(X \cup \xi)$-trees in which $\xi$~appears exactly once.  For
any $q \in T_\xi$ and $p \in F_\Sigma(X) \cup T_\xi$ we denote~$q(\xi
\gets p)$ by~$p \cdot q$.  Also, we define the powers~$q^k$ as
follows:
\begin{enumerate}
\item[$1^{\text{o}}$] $q^0 = \xi$,
\item[$2^{\text{o}}$] $q^{n+1} = q \cdot q^n$ \quad ($n \geq 0$).
\end{enumerate}

Using these notations we may formulate the pumping lemma of tree
recognizers as follows.

\begin{lm}
  \label{Lemma.2.10.1}
  Let $\bA$~be a $k$-state $\Sigma X$-recognizer.  If $t \in T(\bA)$
  and $\hg(t) \geq k$, then there are trees $p \in F_\Sigma(X)$ and
  $q, r \in T_\xi$ such that
  \begin{enumerate}[(a)]
  \item $t = p \cdot q \cdot r$,
  \item $\hg(q) \geq 1$ and
  \item $p \cdot q^i \cdot r \in T(\bA)$ for all $i = 0, 1, 2,
    \dotsc$.
  \end{enumerate}
\end{lm}

\pr Suppose $t \in T(\bA)$ and $\hg(t) \geq k(=\lvert A \rvert)$.
Then we can write $t = \sigma(t_1, \dotsc, t_m)$ ($m > 0$, $\sigma \in
\Sigma_m$).  Choose some $j$ ($1 \leq j \leq m$) such that $\hg(t_j) =
\hg(t) - 1$.  Then
\[ t = t_j \cdot s_1 \enspace, \]
where
\[ s_1 = \sigma(t_1, \dotsc, t_{j-1}, \xi, t_{j+1}, \dotsc, t_m) \in
T_\xi \enspace. \]
If $\hg(t_j) > 0$, we may decompose~$t_j$ the same way.  Since $\hg(t)
\geq k$ the process can be repeated $k$~times and finally we obtain a
representation
\[ t = t' \cdot s_k \cdot \ldots \cdot s_2 \cdot s_1 \enspace, \]
where $t' \in F_\Sigma(X)$ and $s_1, \dotsc, s_k \in T_\xi$.
Moreover, $\hg(s_i) \geq 1$ for every $i = 1, \dotsc, k$.  Let
\[ u_{k+1} = t' \enspace, \quad u_k = t' \cdot s_k \enspace, \dotsc,
\quad u_1 = t' \cdot s_k \cdot \ldots \cdot s_1 = t \enspace. \]
There must be indices $h$~and~$j$, $k + 1 \geq h > j \geq 1$, such
that 
\[ u_h\hat\alpha = u_j\hat\alpha \enspace. \]
Now let $p = u_h$, $q = s_{h-1} \cdot \ldots \cdot s_j$ and $r =
s_{j-1} \cdot \ldots \cdot s_1$ (if~$j = 1$, then~$r = \xi$).  Then $t
= p \cdot q \cdot r$ and $\hg(q) \geq 1$.  Also, our choice of
$p$~and~$q$ implies
\begin{equation}
  \label{Equation.2.10.1}
  p\hat\alpha = (p \cdot q)\hat\alpha \enspace.
\end{equation}
We assume that $A \cap X = \emptyset$, and extend~$\hat\alpha$ to a
homomorphism
\[ \overline \alpha \colon \cF_\Sigma(X \cup A) \to \cA \]
so that $\hat\alpha|A = 1_A$.  By Lemma~\ref{Lemma.2.4.17} $s\overline
\alpha = s\hat\alpha$ whenever $s \in F_\Sigma(X)$.  We verify now by
induction on~$i$ that
\begin{equation}
  \label{Equation.2.10.2}
  (p \cdot q^i)\hat\alpha = (p \cdot q)\hat\alpha
\end{equation}
for every $i \geq 0$.  From~\eqref{Equation.2.10.1} we know that this
is true for $i = 0$.  Suppose~\eqref{Equation.2.10.2} holds for a
given~$i$.  This assumption and~\eqref{Equation.2.10.1} imply
\begin{align*}
  (p \cdot q^{i+1})\hat \alpha &= q \bigl(\xi \gets (p \cdot
  q^i)\hat\alpha \bigr)\overline \alpha = q\bigl(\xi \gets (p \cdot
  q)\hat\alpha \bigr)\overline \alpha \\
  &= q \bigl(\xi \gets p\hat\alpha \bigr)\overline \alpha = (p\cdot
  q)\hat\alpha \enspace.
\end{align*}
Using~\eqref{Equation.2.10.2} we get for each~$i \geq 0$,
\begin{align*}
  (p \cdot q^i \cdot r)\hat\alpha &= r \bigl(\xi \gets (p \cdot
  q^i)\hat\alpha \bigr) \overline \alpha \\
  &= r \bigl(\xi \gets (p \cdot q)\hat \alpha \bigr) \overline \alpha
  \\
  &= (p \cdot q \cdot r)\hat\alpha \enspace.
\end{align*}
Hence, $p \cdot q^i \cdot r \in T(\bA)$ for all $i \geq 0$. \epr

\begin{thm}
  \label{Theorem.2.10.2}
  Let $\bA$~be a $k$-state $\Sigma X$-recognizer.  Then $T(\bA)$~is
  nonempty iff it contains a tree of height less than~$k$.  Hence the
  emptiness problem of recognizable forests is decidable.
\end{thm}

\pr Suppose $T(\bA)$~is nonempty.  Let $t$~be a tree in~$T(\bA)$ of
minimal length.  If $\hg(t) \geq k$, we apply the pumping lemma and
write $t = p \cdot q \cdot r$.  But then $T(\bA)$~would contain the
tree $p \cdot r$ which is properly shorter than~$t$ as $\hg(q) \geq
1$.  Hence $\hg(t) < k$ must hold.  The converse part is trivial.  The
emptiness of~$T(\bA)$ can always be decided by going through the
finite set of trees of height~$< k$. \epr

\

Suppose two $\Sigma X$-recognizers $\bA$~and~$\bB$ are given.
Clearly, $T(\bA) \subseteq T(\bB)$ iff $T(\bA) - T(\bB) = \emptyset$.
But $T(\bA) - T(\bB)$ is recognized by
\[ \bC = \bigl(\cA \times \cB, \gamma, A' \times (B - B') \bigr)
\enspace, \]
where $x\gamma = (x\alpha,x\beta)$ for $x \in X$.  Thus the question
``$T(\bA) \subseteq T(\bB)$?'' can be answered by deciding whether
$T(\bC)$~is empty or not.  The equivalence problem can similarly be
reduced to the emptiness problem.  Of course, its decidability follows
also from the decidability of the inclusion problem.  We have
justified

\begin{thm}
  \label{Theorem.2.10.3}
  The inclusion problem and the equivalence problem of tree
  recognizers are decidable. \epr
\end{thm}

Finally we consider the finiteness problem.

\begin{thm}
  \label{Theorem.2.10.4}
  It is decidable whether the forest recognized by a given tree
  recognizer is finite or infinite.
\end{thm}

\pr Let $\bA$~be a $k$-state $\Sigma X$-recognizer and write
\[ T = T(\bA) - \{t \in F_\Sigma(X) \mid \hg(t) < k\} \enspace. \]
We claim that $T(\bA)$ is finite iff $T = \emptyset$.  Obviously the
condition is sufficient since the set of $\Sigma X$-trees of
height~$< k$ is finite.  If $T \neq \emptyset$ and $t \in T$, then
$\hg(t) \geq k$ and we may apply the pumping lemma and write $t = p
\cdot q \cdot r$ so that
\[ p \cdot q^i \cdot r \in T(\bA) \quad \text{for all } i \geq 0
\enspace. \]
These trees are pairwise distinct since $\hg(q) \geq 1$.  Hence
$T(\bA)$~is infinite.  The forest~$T$ is recognizable and one can
easily construct a recognizer for it.  This means that the
condition~$T = \emptyset$ is effectively testable. \epr

\

The decidability of the finiteness problem may also be deduced from
the following corollary of the pumping lemma.  The proof is an
exercise.

\begin{lm}
  \label{Lemma.2.10.5}
  Let $\bA$~be a $k$-state tree recognizer.  Then $T(\bA)$~is infinite
  iff it contains a tree~$t$ such that
  \[ k \leq \hg(t) < 2k \enspace. \tag*{\epr} \]
\end{lm}

%% file: Section.2.11.tex
In Section~\ref{Section.2.2} it was shown that NDR-recognizers recognize exactly the family $\Rec$, but that there are recognizable forests that cannot be recognized by any deterministic R-recognizer. The limited recognition power of DR-recognizers is due to the fact that they have no way of combining the information gathered from disjoint subtrees. This implies that a DR-recognizer will accept any tree in which every path from the root to the frontier appears in some tree accepted by the recognizer. It will turn out that this closure property characterizes the forests recognizable by DR-recognizers. Here a ``path'' contains, not only a list of the labels of the nodes traversed, but also the information about the directions taken at the nodes. In the later part of this section we shall consider the minimization of DR-recognizers. It will be shown that every DR-recognizer can be reduced to a canonical minimal form which is unique up to isomorphism.

Let $\Sigma$ be a fixed ranked alphabet. In order to avoid some troublesome technicalities, we shall assume that $\Sigma_0=\emptyset$. We associate with $\Sigma$ a unary ranked alphabet
\[
\Gamma=\Gamma_1=\bigcup(\Gamma(\sigma)\mid \sigma\in\Sigma),
\]
where for all $\sigma,\tau\in\Sigma$,
\begin{enumerate}
\item[(i)] $\Gamma(\sigma)=\{\sigma_1,\dots,\sigma_m\}$ if $\sigma\in\Sigma_m$ ($m\geq 1$), and
\item[(ii)] $\Gamma(\sigma)\cap\Gamma(\tau)=\emptyset$ if $\sigma\neq\tau$.
\end{enumerate}
The paths in $\Sigma$-trees can now be defined as $\Gamma$-trees.

\begin{df}\label{Definition.2.11.1}\normalfont
Let $X$ be any frontier alphabet. For each $x\in X$ the set $g_x(t)$ of \emph{$x$-paths}\index{x-path@$x$-path of $\Sigma X$-tree} of a $\Sigma X$-tree $t$ is defined as follows:
\begin{enumerate}
\item[1$^{\circ}$] $g_x(x)=\{x\}$, and $g_x(y)=\emptyset$ for all $y\neq x$, $y\in X$.
\item[2$^{\circ}$] If $t=\sigma(t_1,\dots,t_m)$ ($\sigma\in\Sigma_m$, $m>0$), then $g_x(t)=\sigma_1(g_x(t_1))\cup\dots\cup\sigma_m(g_x(t_m))$.
\end{enumerate}
We extend $g_x$ to a mapping from $\gp F_\Sigma(X)$ to $\gp F_\Gamma(X)$ in the natural way. Moreover, we put
\[
g(T)=\bigcup(g_x(T)\mid x\in X)
\]
for each $T\subseteq F_\Sigma(X)$.
\end{df}

Label the edges of the graph representing a tree $t\in F_\Sigma(X)$ so that the $i^{\textrm{th}}$ edge (counted from the left) leaving a node labelled by a symbol $\sigma$ always gets the label $\sigma_i$. Then the elements of $g_x(t)$ ($x\in X$) are spelled out by the paths leading from the root to a leaf labelled by $x$ when we interpret a word $\sigma_{1i_1}\dots\sigma_{ki_k}x$ ($k\geq 0,\sigma_{1i_1},\dots,\sigma_{ki_k}\in\Gamma$) as the $\Gamma X$-tree $\sigma_{1i_1}(\dots\sigma_{ki_k}(x)\dots)$. Moreover, every such path gives an element of $g_x(t)$.

\begin{lm}\label{Lemma.2.11.2}
If $T\in\Rec(\Sigma,X)$, then $g(T)\in\Rec(\Gamma,X)$.
\end{lm}

\pr
Let $G=(N,\Sigma,X,P,a_0)$ be a regular $\Sigma X$-grammar in normal form generating $T$. The case $T=\emptyset$ being trivial, we may assume that every $G_a=(N,\Sigma,X,P,a)$ ($a\in N$) generates a nonempty forest. Let $G'=(N,\Gamma,X,P',a_0)$ be the regular $\Gamma X$-grammar, where
\[\begin{array}{c}
P'=\{a\rightarrow\sigma_i(a_i)\mid a\rightarrow\sigma(a_1,\dots,a_m)\in P,\ m>0,\ 1\leq i\leq m\}\cup{}\\\{a\rightarrow x\mid a\rightarrow x\in P,x\in X\}.
\end{array}\]
We claim that $T(G')=g(T)$. This follows when we show that, for every tree
\[
p=\sigma_{1i_1}(\dots\sigma_{ki_k}(x)\dots)\in F_\Gamma(X)
\]
and every $a\in N$,
\begin{equation}\tag{*}
p\in T(G_a')\text{\quad iff\quad} p\in g(T(G_a)),
\end{equation}
where $G_a'=(N,\Gamma,X,P',a)$.

We proceed by induction on $\hg(p)$.
\begin{enumerate}
\item[1$^{\circ}$] If $\hg(p)=0$, then $p=x$. In this case (*) obviously holds as $a\rightarrow x$ is in $P'$ iff it is in $P$.
\item[2$^{\circ}$] Suppose $\hg(p)>0$ and that (*) holds for all trees of lesser height.
\end{enumerate}

If $p\in T(G_a')$, then $a\Rightarrow_{G'}^*\sigma_{1i_1}(a_{i_1})$ and $a_{i_1}\Rightarrow_{G'}^*\sigma_{2i_2}(\dots\sigma_{ki_k}(x)\dots)$ for some $a_{i_1}\in N$, and $P$ contains a production $a\to\sigma_1(a_1,\dots,a_m)$ such that $i_1\leq m$. By the inductive assumption there exists a tree $t_{i_1}\in T(G_{a_{i_1}})$ such that $\sigma_{2i_2}(\dots\sigma_{ki_k}(x)\dots)\in g_x(t_{i_1})$. Moreover, we may choose for every $i\neq i_1$, $1\leq i\leq m$, a tree $t_i\in T(G_{a_i})$. Then $t=\sigma_1(t_1,\dots,t_m)\in T(G_a)$ and $p\in g_x(t)\subseteq g(T(G_a))$.

Conversely, let $p\in g(T(G_a))$. Then $p\in g_x(t)$ for some $t\in T(G_a)$. Obviously, $t$ is of the form $\sigma_1§(t_1,\dots,t_m)$, where $i_1\leq m$, and it has a derivation
\[
a\Rightarrow_G\sigma_1(a_1,\dots,a_m)\Rightarrow_G^* t.
\]
This means that $P'$ contains the production $a\to\sigma_{1i_1}(a_{i_1})$. Moreover, $t_{i_1}\in T(G_{a_{i_1}})$ and $\sigma_{2i_2}(\dots\sigma_{ki_k}(x)\dots)\in g_x(t_{i_1})$. Hence, we get a derivation
\[
a\Rightarrow_G^*\sigma_{1i_1}(a_{i_1})\Rightarrow^*_{G'}p,
\]
which shows that $p\in T(G'_a)$.
\epr

\

Let $g$ be the mapping of Definition~\ref{Definition.2.11.1} associated with a given frontier alphabet $X$. Then we write $\tau_X=gg^{-1}$. It is clear that $\tau_X$ is a closure operation in $F_\Sigma(X)$, i.e., for all $S,T\subseteq F_\Sigma(X)$,
\begin{enumerate}
\item[(i)] $S\subseteq S\tau_X$,
\item[(ii)] $S\subseteq T$ implies $S\tau_X\subseteq T\tau_X$, and
\item[(iii)] $S\tau_X\tau_X=S\tau_X$.
\end{enumerate}
For any $T\subseteq F_\Sigma(X)$, $T\tau_X$ is the \emph{closure}\index{closure!of forest} of $T$, and $T$ is said to be \emph{closed}\index{forest!closed} if $T\tau_X=T$.

Now, consider an arbitrary NDR $\Sigma X$-recognizer $\bA=(\cA,A',\alpha)$. For each $a\in A$, let
\[
T(\bA,a)=\{t\in F_\Sigma(X)\mid a\in t\tilde\alpha\}.
\]
A state $a\in A$ is a \emph{$0$-state},\index{zero-state@0-state} if $T(\bA,a)=\emptyset$. We say that $\bA$ is \emph{normalized}\index{normalized NDR $\Sigma X$-recognizer} if for all $m>0$, $\sigma\in\Sigma_m$ and $a\in A$ one of the following two alternatives holds:
\begin{enumerate}
\item[(1)] Each component of every vector in $\sigma^{\cA}(a)$ is a 0-state.
\item[(2)] No component of any vector of $\sigma^{\cA}(a)$ is a 0-state.
\end{enumerate}

A normalized NDR $\Sigma X$-recognizer $\bA$ has the following important property. Let $p\in g_x(s)$ ($x\in X$) for some $\Sigma X$-tree $s$ such that $\bA$ has a computation on $s$ which begins at the root in an initial state and ends at the leaf corresponding to $p$ in a state which belongs to $x\alpha$. Then there exists a tree $t$ in $T(\bA)$ such that $p\in g_x(t)$. Such a $t$ can be built around the $x$-path $p$ by completing it with trees from appropriate $T(\bA,a)$-forests.

An NDR $\Sigma X$-recognizer $\bA$ becomes normalized if we omit from each set $\sigma^{\cA}(a)$ every vector which contains a 0-state. This does not change $T(\bA)$ because the use of a vector containing a 0-state cannot lead to an accepting computation. Hence, we have

\begin{lm}\label{Lemma.2.11.3}
For every NDR-recognizer there is an equivalent normalized NDR-recog\-nizer.
\epr
\end{lm}

We associate with each NDR $\Sigma X$-recognizer $\bA$ a DR $\Sigma X$-recognizer $\gp\bA=(\gp\cA,A',\beta)$ defined as follows:
\begin{enumerate}
\item[(i)] $\gp\cA=(\gp A,\Sigma)$ is the deterministic root-to-frontier algebra such that
\[
\sigma^{\gp\cA}(H)=\left(\bigcup(\pi_1(\sigma^\cA(a))\mid a\in H),\dots,\bigcup(\pi_m(\sigma^\cA(a))\mid a\in H)\right)
\]
for all $H\in\gp A$, $m>0$ and $\sigma\in\Sigma_m$. Here $\pi_i$ ($1\leq i\leq m$) is the $i^{\textrm{th}}$ projection.
\item[(ii)] For each $x\in X$, $x\beta=\{H\in\gp A\mid H\cap x\alpha\neq\emptyset\}$.
\end{enumerate}

\begin{lm}\label{Lemma.2.11.4}
For every normalized NDR $\Sigma X$-recognizer $\bA$, $T(\gp\bA)=T(\bA)\tau_X$.
\end{lm}

\pr
In order to prove the inclusion $T(\gp\bA)\subseteq T(\bA)\tau_X$, we consider an arbitrary tree $s\in T(\gp\bA)$ and an $x$-path $p\in g_x(s)$ ($x\in X$). We should show that  $p\in g(T(\bA))$. Let $p=\sigma_{1i_1}(\dots(\sigma_{ki_k}(x))\dots)$. By the definition of $\gp\bA$ there are states $a_0,a_1,\dots,a_k\in A$ such that
\begin{enumerate}
\item[(i)] $a_0\in A'$ and $a_k\in x\alpha$, and
\item[(ii)] $a_j\in\pi_{i_j}(\sigma_j^{\cA}(a_{j-1}))$ for $j=1,\dots,k$.
\end{enumerate}
Since $\bA$ is normalized, this implies that there is a tree $t\in T(\bA)$ such that $p\in g_x(t)$. Hence $p\in g(T(\bA))$. Now, let $s\in T(\bA)\tau_X$ and consider any $x$-path
\[
p=\sigma_{1i_1}(\dots\sigma_{ki_k}(x)\dots)\in g_x(s)\quad(x\in X).
\]
Then $p\in g_x(t)$ for some $t\in T(\bA)$ and there are states $a_0,a_1,\dots,a_k\in A$ such that the above conditions (i) and (ii) hold. But the definition of $\gp\bA$ implies that the state of $\gp\bA$ at the leaf corresponding to $p$ includes $a_k$ for any tree in which $p$ is an $x$-path. Hence $\gp\bA$ arrives at the leaf of $s$ corresponding to $p$ in a state belonging to $x\alpha$. This holds for every leaf of $s$ and therefore $s\in T(\gp\bA)$.
\epr

\begin{cor}\label{Corollary.2.11.5}
If $T\in\Rec(\Sigma,X)$, then $T\tau_X\in\Rec(\Sigma,X)$.
\epr
\end{cor}

Lemmas~\ref{Lemma.2.11.3} and~\ref{Lemma.2.11.4} also imply that every closed recognizable forest is recognized by a DR recognizer. But it is easy to see that $T(\gp\bA)=T(\bA)$ if $\bA$ is deterministic. Hence we may state the following result.

\begin{thm}\label{Theorem.2.11.6}
A recognizable forest can be recognized by a DR recognizer iff it is closed.
\epr
\end{thm}

The rest of this section deals with the minimization of DR-recognizers. First two general remarks. When $\bA=(\cA,a_0,\alpha)$ is a DR $\Sigma X$-recognizer, then the NDR algebra $\cA=(A,\Sigma)$ is deterministic and we may view each $\sigma^{\cA}$ ($\sigma\in\Sigma_m$, $m>0$) as a mapping
\[
\sigma^{\cA}\colon A\to A^m.
\]
Hence we write $\sigma^{\cA}(a)=(a_1,\dots,a_m)$ rather than $\sigma^{\cA}(a)=\{(a_1,\dots,a_m)\}$. The second remark concerns normalized DR recognizers. If the DR $\Sigma X$-recognizer $\bA$ is normalized, one of the following conditions holds for each pair $(a,\sigma)\in A\times\Sigma$:
\begin{enumerate}
\item[(1)] Every component of $\sigma^{\cA}(a)$ is a 0-state.
\item[(2)] No component of $\sigma^{\cA}(a)$ is a 0-state.
\end{enumerate}
Of course, Lemma~\ref{Lemma.2.11.3} and the construction which led to it are valid here, too, but we define a ``standard'' normalized form $\bA^*=(\cA^*,a_0,\alpha)$ of $\bA$ as follows:

\begin{enumerate}
\item[(i)] If $\bA$ has no 0-state, then put $\bA^*=\bA$.
\item[(ii)] If $\bA$ has a 0-state, choose one of them, say $d$, and define then for all $m>0$, $\sigma\in\Sigma_m$, and $a\in A$,
\[
\sigma^{\cA^*}(a)=\left\{\begin{array}{ll}
(d,\dots,d)\ (\in A^m)&\text{if $\sigma^{\cA}(a)$ contains a 0-state},\\
\sigma^{\cA}(a)&\text{otherwise.}
\end{array}\right.
\]
\end{enumerate}
It is easy to prove that $\bA^*$ is normalized and deterministic, and that $T(\bA^*)=T(\bA)$.

Normalized DR recognizers have also the following useful property.

\begin{lm}\label{Lemma.2.11.7}
Let $\bA$ and $\bB$ be normalized  DR $\Sigma X$-recognizers, and let $a\in A$, $b\in B$, $m>0$, $\sigma\in\Sigma_m$, $\sigma^\cA(a)=(a_1,\dots,a_m)$ and $\sigma^\cB(b)=(b_1,\dots,b_m)$. If $T(\bA,a)=T(\bB,b)$, then $T(\bA,a_i)=T(\bB,b_i)$ for all $i=1,\dots,m$.
\end{lm}

\pr
If one of the states $a_i$ ($1\leq i\leq m$) is a 0-state, then all of them are. Moreover, $T(\bA,a)=T(\bB,b)$ does not contain any tree of the form $\sigma(t_1,\dots,t_m)$. Hence, one of the forests $T(\bB,b_i)$ ($1\leq i\leq m£$), and therefore every one of them, is empty. Thus $T(\bA,a_i)=T(\bB,b_i)=\emptyset$ for all $i=1,\dots,m$.

Suppose now that $T(\bA,a_i)\neq\emptyset$ and $T(\bB,b_i)\neq\emptyset$ for all $i=1,\dots,m$. Consider any $i$ ($1\leq i\leq m$) and $t_i\in T(\bA,a_i)$. Choose any $t_1\in T(\bA,a_1),\dots,t_{i-1}\in T(\bA,a_{i-1}),t_{i+1}\in T(\bA,a_{i+1}),\dots,t_m\in T(\bA,a_m)$. Then $\sigma(t_1,\dots,t_m)\in T(\bA,a)=T(\bB,b)$ implies $t_i\in T(\bB,b_i)$. By a symmetrical argument, $T(\bB,b_i)\subseteq T(\bA,a_i)$ holds for every $i=1,\dots,m$. Hence, $T(\bA,a_i)=T(\bB,b_i)$ for every $i=1,\dots,m$, as required.
\epr

\

We shall now define a few algebraic concepts for DR recognizers. Let $\bA=(\cA,a_0,\alpha)$ and $\bB=(\cB,b_0,\beta)$ be DR $\Sigma X$-recognizers.

A \emph{homomorphism}\index{homomorphism!of DR $\Sigma X$-recognizer} from $\bA$ to $\bB$ is a mapping $\varphi\colon A\to B$ such that
\begin{enumerate}
\item[(i)] for all $m>0$, $\sigma\in\Sigma_m$ and $a\in A$, $\sigma^\cB(a\varphi)=(a_1\varphi,\dots,a_m\varphi)$, where $(a_1,\dots,a_m)=\sigma^\cA(a)$,
\item[(ii)] $a_0\varphi=b_0$, and
\item[(iii)] for every $x\in X$, $x\beta\varphi^{-1}=x\alpha$.
\end{enumerate}
If $\varphi$ is a homomorphism from $\bA$ to $\bB$, we write $\varphi\colon\bA\to \bB$. If such a $\varphi$ is surjective, it is called an \emph{epimorphism}.\index{epimorphism!of DR $\Sigma X$-recognizer} For an epimorphism condition (iii) implies $x\alpha\varphi=x\beta$, too. If there exists an epimorphism $\varphi$ from $\bA$ onto $\bB$, then $\bB$ is an \emph{epimorphic} image\index{image!epimorphic} of $\bA$. If $\varphi\colon \bA\to \bB$ is bijective, then $\bA$ and $\bB$ are \emph{isomorphic},\index{isomorphism of!DR $\Sigma X$-recognizers} and we write $\bA\cong\bB$.

A \emph{congruence}\index{congruence!of DR $\Sigma X$-recognizer} on $\bA$ is an equivalence relation $\varrho$ on $A$ such that
\begin{enumerate}
\item[(i)] for all $m>0$, $\sigma\in\Sigma_m$ and $a,a'\in A$, $a\varrho=a'\varrho$ implies $\sigma^{\cA}(a)/\varrho=\sigma^{\cA}(a')/\varrho$ (recall the notation from Section~\ref{Section.1.1}), and
\item[(ii)] $\varrho$ saturates every set $x\alpha$ ($x\in X$).
\end{enumerate}

If $\varrho$ is a congruence on $\bA$, then the \emph{quotient recognizer}\index{SigmaX-recognizer@$\Sigma X$-recognizer!quotient DR} determined by $\varrho$ is the DR $\Sigma X$-recognizer
\[
\bA/\varrho=(\cA/\varrho,a_0\varrho,\alpha_\varrho),
\]
where $\cA/\varrho=(A/\varrho,\Sigma)$ is defined by
\[
\sigma^{\cA/\varrho}(a\varrho)=\sigma^{\cA}(a)/\varrho\quad (\sigma\in\Sigma_m,m>0,a\in A),
\]
and $\alpha_\varrho\colon X\to A/\varrho$ is defined by $x\alpha_\varrho=x\alpha/\varrho$ ($x\in X$). It is easy to see that $\bA/\varrho$ is well-defined.

The following theorem is easily obtained by modifying the proofs of the corresponding facts from algebra.

\begin{thm}\label{Theorem.2.11.8}
Let $\bA$ and $\bB$ be DR $\Sigma X$-recognizers.
\begin{enumerate}
\item[(a)] If $\varrho$ is a congruence of $\bA$, then the natural mapping $\varrho^\sharp\colon A\to A/\varrho$ defines an epimorphism of $\bA$ onto $\bA/\varrho$.
\item[(b)] If $\varphi\colon\bA\to\bB$ is an epimorphism, then $\varrho=\varphi\varphi^{-1}$ is a congruence on $\bA$, and $\bA/\varrho\cong\bB$.
\epr
\end{enumerate}
\end{thm}

The following fact will be needed later.

\begin{thm}\label{Theorem.2.11.9}
If $\bB$ is an epimorphic image of $\bA$, then $T(\bA)=T(\bB)$.
\end{thm}

\pr
Let $\varphi\colon\bA\to\bB$ be an epimorphism. We verify by tree induction that
\begin{equation}\tag{*}
t\tilde\alpha=t\tilde\beta\varphi^{-1}\text{ and }t\tilde\alpha\varphi=t\tilde\beta,
\end{equation}
for every $t\in F_\Sigma(X)$.

\begin{enumerate}
\item[1$^{\circ}$] For $t=x\in X$, (*) follows directly from the fact that $\varphi$ is an epimorphism.
\item[2$^{\circ}$] Let $t=\sigma(t_1,\dots,t_m)$ and assume that (*) holds for $t_1,\dots,t_m$. Suppose $a\in t\tilde\alpha$. If $\sigma^{\cA}(a)=(a_1,\dots,a_m)$, this means that $a_1\in t_1\tilde\alpha,\dots,a_m\in t_m\tilde\alpha$. Hence, $a_1\varphi\in t_1\tilde\beta,\dots,a_m\varphi\in t_m\tilde\beta$. This implies
\[
\sigma^{\cB}(a\varphi)=(a_1\varphi,\dots,a_m\varphi)\in t_1\tilde\beta\times\dots\times t_m\tilde\beta.
\]
Hence, $a\varphi\in t\tilde\beta$. Suppose now that $a\varphi\in t\tilde\beta$, and let $\sigma^{\cA}(a)=(a_1,\dots,a_m)$. Then $a_1\varphi\in t_1\tilde\beta,\dots,a_m\varphi\in t_m\tilde\beta$, which implies $a_1\in t_1\tilde\alpha,\dots,a_m\in t_m\tilde\alpha$. Hence, $a\in t\tilde\alpha$. The equality $t\tilde\alpha=t\tilde\beta\varphi^{-1}$ implies $t\tilde\alpha\varphi=t\tilde\beta$ as $\varphi$ is surjective.

Now, (*) implies that for every $t\in F_\Sigma(X)$,
\[
\parbox{\linewidth}{\hspace*{\fill}%
$\begin{array}[b]{lll}
t\in T(\bA)&\text{iff}& a_0\in t\tilde\alpha\\
&\text{iff}& a_0\varphi\text{($=b_0$)}\in t\tilde\alpha\varphi\text{($=t\tilde\beta$)}\\
&\text{iff}& t\in T(\bB).\\
\end{array}$%
\hspace*{\fill}\llap{$\Box$}}
\]
\end{enumerate}

We call two states $a$ and $a'$ of a DR $\Sigma X$-recognizer $\bA$ \emph{equivalent},\index{equivalence of states in!DR recognizer} and we write $a\sim_{\bA}a'$ (or just $a\sim a'$), if $T(\bA,a)=T(\bA,a')$. Obviously, $\sim_{\bA}$ is an equivalence relation on $A$. We say that $\bA$ is \emph{reduced},\index{SigmaX-recognizer@$\Sigma X$-recognizer!reduced DR} if $\sim_{\bA}=\delta_A$.

\begin{lm}\label{Lemma.2.11.10}
If $\bA$ is a normalized DR $\Sigma X$-recognizer, then $\sim$ is a congruence on $\bA$ and $\bA/{\sim}$ is reduced.
\end{lm}

\pr
First we show that $\sim$ is a congruence relation.
\begin{enumerate}
\item[(i)] Consider any $m>0$, $\sigma\in\Sigma_m$ and $a,a'\in A$ such that $a\sim a'$. Let
\[
\sigma^{\cA}(a)=(a_1,\dots,a_m) \text{\quad and\quad} \sigma^{\cA}(a')=(a_1',\dots,a_m').
\]
But $a\sim a'$ means that $T(\bA,a)=T(\bA,a')$, and Lemma~\ref{Lemma.2.11.7} implies that
\[
T(\bA,a_i)=T(\bA,a_i') \text{\quad for all\quad} i=1,\dots,m.
\]
Hence, $a_i\sim a_i'$ for all $i=1,\dots,m$.

\item[(ii)] If $a\in x\alpha$ and $a\sim a'$, for some $x\in X$ and $a,a'\in A$, then $x\in T(\bA,a)=T(\bA,a')$ implies $a'\in x\alpha$. Hence, $\sim$ saturates $x\alpha$.

Now we know that the quotient recognizer $\bA/{\sim}$ can be defined. It is reduced as $(a{\sim})\sim_{\bA/{\sim}}(a'{\sim})$ implies $a{\sim}=a'{\sim}$ ($a,a'\in A$) because, by Theorem~\ref{Theorem.2.11.9},
\[
\parbox{\linewidth}{\hspace*{\fill}%
$T(\bA,a)=T(\bA/{\sim},a{\sim})=T(\bA/{\sim},a'{\sim})=T(\bA,a').$%
\hspace*{\fill}\llap{$\Box$}}
\]
\end{enumerate}

\

Let $a,a'\in A$. We write $a\vdash a'$ if there exist an $m>0$ and a $\sigma\in\Sigma_m$ such that $a'$ appears in $\sigma^{\cA}(a)$. The reflexive, transitive closure of $\vdash$ is denoted by $\vdash^*$. If $a\vdash^* a'$, we say that $a'$ is \emph{reachable}\index{reachability of state in!drsimgaxrecognizer@DR $\Sigma X$-recognizer} from $a$. The DR recognizer $\bA$ is said to be \emph{connected}\index{SigmaX-recognizer@$\Sigma X$-recognizer!connected DR} if every state is reachable from the initial state.

The \emph{connected component}\index{connected component of DR $\Sigma X$-recognizer}
\[
\bA^c=(\cA^c,a_0,\alpha^c)
\]
of a DR $\Sigma X$-recognizer $\bA$ is defined as follows:
\begin{enumerate}
\item[(i)] $\cA^c=(A^c,\Sigma)$, where $A^c=\{a\in A\mid a_0\vdash^* a\}$ and $\sigma^{\cA^c}(a)=\sigma^{\cA}(a)$ for all $\sigma\in\Sigma$ and $a\in A^c$.
\item[(ii)] $x\alpha^c=x\alpha\cap A^c$ for each $x\in X$.
\end{enumerate}
Clearly, the operations $\sigma^{\cA^c}\colon A^c\to(A^c)^m$ are completely defined ($m>0, \sigma\in\Sigma_m$).

The proof of Lemma~\ref{Lemma.2.11.11} is quite straightforward and we shall omit it.

\begin{lm}\label{Lemma.2.11.11}
Let $\bA$ be any DR $\Sigma X$-recognizer. Then
\begin{enumerate}
\item[(a)] $\bA^c$ is connected and deterministic,
\item[(b)] $\bA^c=\bA$ iff $\bA$ is connected,
\item[(c)] $T(\bA^c)=T(\bA)$, and
\item[(d)] if $\bA$ is normalized, then so is $\bA^c$.
\epr
\end{enumerate}
\end{lm}

We are now ready to present the main theorem of the minimization theory of DR recognizers.

\begin{thm}\label{Theorem.2.11.12}
Let $\bA$ and $\bB$ be connected, normalized DR $\Sigma X$-recognizers. Then $T(\bA)=T(\bB)$ iff $\bA/{\sim_{\bA}}\cong\bB/{\sim_{\bB}}$.
\end{thm}

\pr
If $\bA/{\sim_{\bA}}$ and $\bB/{\sim_{\bB}}$ are isomorphic, then
\[
T(\bA)=T(\bA/{\sim_{\bA}})=T(\bB/{\sim_{\bB}})=T(\bB)
\]
by Theorems~\ref{Theorem.2.11.8} and~\ref{Theorem.2.11.9}.

Assume now that $T(\bA)=T(\bB)$. We define a mapping
\[
\varphi\colon A/{\sim_{\bA}}\to B/{\sim_{\bB}}
\]
by the condition that
\[
(a{\sim_{\bA}})\varphi=b{\sim_{\bB}}\text{\quad if\quad} T(\bA,a)=T(\bB,b)\text{\quad($a\in A, b\in B$).}
\]
The following steps (i)--(v) show that $\varphi$ is the required isomorphism.
\begin{enumerate}
\item[(i)] $(a{\sim_{\bA}})\varphi$ is defined for all $a{\sim_{\bA}}\in A/{\sim_{\bA}}$. Since $\bA$ is connected, there exist for every $a\in A$ a $k\geq 0$ and states $a_1,\dots,a_k\in A$ such that
\[
a_0\vdash a_1\vdash a_2\vdash\dots\vdash a_k=a.
\]
Using Lemma~\ref{Lemma.2.11.7} one shows by induction on the smallest $k$ (corresponding to the given $a$) that there is a $b$ such that $T(\bA,a)=T(\bB,b)$.
\item[(ii)] $\varphi$ is well-defined. If $a \sim_{\bA} a'$, $T(\bA,a)=T(\bB,b)$ and $T(\bA,a')=T(\bB,b')$ for some $a,a' \in A$
and $b,b'\in B$, then $b{\sim_{\bB}}=b'{\sim_{\bB}}$.
\item[(iii)] $\varphi$ is injective. Similarly as (ii).
\item[(iv)] $\varphi$ is surjective. If we exchange the roles of $\bA$ and $\bB$ in (i), we see that there exists for every $b\in B$ an $a\in A$ such that $T(\bA,a)=T(\bB,b)$.
\item[(v)] $\varphi$ is a homomorphism. That $\varphi$ preserves the operations follows from Lemma~\ref{Lemma.2.11.7}. If $a{\sim_{\bA}}\in x\alpha/{\sim_{\bA}}$ ($x\in X$) and $(a{\sim_{\bA}})\varphi=b{\sim_{\bB}}$, then $x\in T(\bA,a)=T(\bB, b)$ implies $b{\sim_{\bB}}\in x\beta/{\sim_{\bB}}$. Likewise, $(a{\sim_{\bA}})\varphi=b{\sim_{\bB}}\in x\beta/{\sim_{\bB}}$ implies $a{\sim_{\bA}}\in x\alpha/{\sim_{\bA}}$. Thus $x\beta{\sim_{\bB}}\varphi^{-1}=x\alpha{\sim_{\bA}}$ for every $x\in X$.
\epr
\end{enumerate}

\

A DR recognizer $\bA$ is said to be \emph{minimal}\index{SigmaX-recognizer@$\Sigma X$-recognizer!minimal DR} if no DR recognizer with fewer states recognizes $T(\bA)$. If $\bA$ is minimal, then it is connected by Lemma~\ref{Lemma.2.11.11}. As $T(\bA^*)=T(\bA)$ we may also assume that $\bA$ is normalized. Then $T(\bA)=T(\bA/{\sim_{\bA}})$ implies that $\bA$ should be reduced, too. Conversely, if $\bA$ is connected, normalized and reduced, then it is minimal and every normalized minimal DR recognizer is isomorphic to it (Theorem~\ref{Theorem.2.11.12}). These facts imply that the following three steps yield for any DR recognizer $\bA$ an equivalent minimal DR recognizer $\bB$. Moreover, this $\bB$ is normalized.

\begin{enumerate}
\item[]Step 1. Form $\bA^*$.

\item[]Step 2. Form ${\bA^*}^c$.

\item[]Step 3. Form $\sim$ for ${\bA^*}^c$, and put $\bB={\bA^*}^c/{\sim}$.
\end{enumerate}

It is not hard to see that these steps are effectively realizable.

%% file: Section.2.12.tex
\begin{enumerate}
\item Let $\mathrm{leaf}(t)$ denote the set of symbols which label the leaves of a given $\Sigma X$-tree $t$. Define $\mathrm{leaf}(t)$ by tree induction.
\item
\begin{enumerate}
\item[(a)] Define the length $|t|$ of a $\Sigma X$-tree $t$ (as a word) by tree induction.
\item[(b)] For the sake of simplicity, let $\Sigma=\Sigma_2$. Derive an upper bound for $|t|$ in terms of $\hg(t)$. Give also a lower bound for $|t|$ in terms of $\hg(t)$.
\end{enumerate}
\item Let $\Sigma=\Sigma_0\cup\Sigma_2$, $\Sigma_0=\{\omega\}$, $\Sigma_2=\{\sigma\}$, and let $X=\{x,y\}$. Construct a CF grammar which generates the set $F_\Sigma(X)$ of all $\Sigma X$-trees (when these are viewed as words). Is the set of all $\Sigma X$-trees still a CF language if we use the Polish notation for $\Sigma X$-terms?
\item Let $\Sigma$ and $X$  be as in the previous exercise. Decide which ones of the $\Sigma X$-forests, $R$, $S$, and $T$ are recognizable, when these are defined as follows:
\begin{enumerate}
\item[(i)] $t\in R$ iff the number of $\sigma$'s in $t$ is odd.
\item[(ii)] $t\in S$ iff all paths from the root to a leaf are of the same length.
\item[(iii)] $t\in T$ iff no leaf labelled by $y$ appears to the left of a leaf labelled by $x$.
\end{enumerate}
\item Let $\bA$ be an NDF $\Sigma X£$-recognizer and $\bB$ an NDR $\Sigma X$-recognizer which are associated in the sense of Section~\ref{Section.2.2}. Prove the equality $\hat\alpha=\tilde\beta$ by tree induction.
\item Use regular tree grammars to prove directly that $\Rec(\Sigma, X)$ is closed under $\sigma$-products (Corollary~\ref{Corollary.2.4.12}).
\item Let us change the definition of the forest product $T(x\leftarrow T_x)$ (cf.~Definition~\ref{Definition.2.4.3}) in such a way that every occurrence of each letter $x\in X$ should be rewritten as the same tree $t_x\in T_x$. Then we get the new product
\[
T[x\leftarrow T_x\mid x\in X]=\{t(x\leftarrow t_x\mid x\in X)\mid t\in T, t_x\in T_x\ (x\in X)\}.
\]
Is $\Rec(\Sigma, X)$ closed under this product?
\item Let $T$ be a $\Sigma X$-forest and let $x\in X$. Describe the forests $T\cdot_x\emptyset$ and $\emptyset\cdot_x T$.
\item Do the following laws hold for $x$-products?
\begin{enumerate}
\item[(a)] $R\cdot_x(S\cup T)=(R\cdot_xS)\cup(R\cdot_xT)$.
\item[(b)] $(R\cup S)\cdot_xT=(R\cdot_xT)\cup(S\cdot_xT)$.
\item[(c)] $R\cdot_x(S\cdot_yT)=(R\cdot_xS)\cdot_yT$.
\end{enumerate}
\item Let us change Definition~\ref{Definition.2.4.7} so that $T^{j+1,x}=T\cdot_xT^{j,x}\cup T^{j,x}$ for all $j\geq 0$. Does the new $x$-iteration coincide with the original one? If not, does it preserve recognizability?
\item Let $x\neq y$ ($x,y\in X$). Is it possible that $(T^{*x})^{*y}\neq(T^{*y})^{*x}$ for some $\Sigma X$-forest $T$?
\item Show that the construction of the tree recognizer for the forest $S^{-x}T$ given in the proof of Theorem~\ref{Theorem.2.4.10} is effective when $S$ is recognizable (and given by a tree recognizer).
\item Prove Lemma~\ref{Lemma.2.4.17}.
\item Prove Corollary~\ref{Corollary.2.4.20} directly without using Theorems~\ref{Theorem.2.4.16} and~\ref{Theorem.2.4.18}.
\item Let $\varphi\colon\cF_\Sigma(X)\to\cF_\Sigma(X)$ be a homomorphism of $\Sigma$-algebras. Prove that if $T\in\Rec(\Sigma,X)$, then (a)~$T\varphi\in\Rec(\Sigma,X)$ and (b)~$T\varphi^{-1}\in\Rec(\Sigma,X)$.
\item The set of \emph{atomic}\index{SigmaX-tree@$\Sigma X$-tree!atomic} $\Sigma X$-trees is defined as
\[
A(\Sigma,X)=\{\sigma(x_{i_1},\dots, x_{i_m})\mid m\geq 0,\ \sigma\in\Sigma_m,\ x_{i_1},\dots,x_{i_m}\in X\}.
\]
For the sake of definiteness, let $X=\{x_1,\dots,x_n\}$ ($n\geq 1$). Prove that
\[
(\dots(A(\Sigma,X)^{*x_1})^{*x_2}\dots)^{*x_n}=F_\Sigma(X)
\]
(cf.~\textsc{Thatcher} and \textsc{Wright}~\cite{thawri68}).
\item Let $\Sigma=\Sigma_2=\{\sigma\}$ and $X=\{x\}$. Write a regular expression for the forest of all $\Sigma X$-trees which contain an even number of $\sigma$'s.
\item Let $\Sigma$ and $X$ be as in Exercise~3. Construct a $\Sigma X$-recognizer for the forest represented by the regular expression $\sigma(x,y)\cdot_z\sigma(\omega,\sigma(\omega,z))^{*z}$.
\item Prove Theorem~\ref{Theorem.2.6.6}.
\item If $\bA$ is a $\Sigma X$-recognizer and $T(\bA)=T$, then $\hat\alpha$ is a homomorphism from $\bF_T$ to $\bA$. Prove Lemma~\ref{Lemma.2.6.2} using this observation.
\item Prove Lemma~\ref{Lemma.2.6.14}.
\item In Section~\ref{Section.2.7} we noted that one may define recognizability for subsets of algebras. We call $T$ ($\subseteq A$) a \emph{recognizable subset}\index{subset!recognizable} of the $\Sigma$-algebra $\cA=(A,\Sigma)$, if there exists a congruence $\theta$ of finite index which saturates $T$. Denote by $\Rec\,\cA$ the set of all recognizable subsets of $\cA$. Prove the following facts:
\begin{enumerate}
\item[(a)] If $S,T\in\Rec\,\cA$, then $S\cup T,S\cap T,S-T\in\Rec\,\cA$.
\item[(b)] If $\varphi\colon\cA\to\cB$ is a homomorphism and $T\in\Rec\,\cB$, then $T\varphi^{-1}\in\Rec\,\cA$.
\end{enumerate}
 (Note. $T\in\Rec\,\cA$ does not imply $T\varphi\in\Rec\,\cB$. A counterexample where $\cA$ and $\cB$ are monoids can be found in Eilenberg's book (Vol.~A) mentioned among the references of Chapter~\ref{Chapter.1}.)
 \item Let $\Sigma=\Sigma_2=\{\sigma\}$ and $X=\{x,y\}$, and let $(U,V)$ be the least fixed-point of the system
\begin{align*}
u&=x+\sigma(\sigma(u,v),y)\\
v&=\sigma(y,u).
\end{align*}
Find a regular $(\Sigma,X,k)$-polynomial $\Pi$ ($k\geq 2$) such that $U$ and $V$ can be represented as unions of some components of $[\hat\Pi]$. (For a general treatment of such questions see~\textsc{Mezei} and \textsc{Wright}~\cite{mezwri67}.)
\item Show that every local $\Sigma X$-forest $\Loc(R,F)$ can be represented in terms of the elementary forests and the elementary operations intersection, union, and restriction. Note the resulting connection between the Theorems~\ref{Theorem.2.8.6} and~\ref{Theorem.2.9.5}.
\item Show that the decidability of the equivalence problem of tree recognizers follows from the results of Section~\ref{Section.2.6}.
\item Prove Lemma~\ref{Lemma.2.10.5}.
\item Prove that it is decidable whether a recognizable forest can be recognized by a DR-recognizer.
\item Are all local forests recognizable by DR-recognizers?
\item Present algorithms for carrying out Steps~2 and~3 of the minimization algorithm for DR-recognizers which was outlined in Section~\ref{Section.2.11}.
\end{enumerate}

%% file: Section.2.13.tex
The observation (made about 1960) that finite automata may be defined as unary algebras is attributed to J.~R. B\"uchi and J.~B. Wright (see \textsc{Mezei} and \textsc{Wright}~\cite{mezwri67}, \textsc{Thatcher}~\cite{tha73}). The generalization to tree automata was suggested independently by \textsc{Doner}~\cite{don65,don70} and by \textsc{Thatcher} and \textsc{Wright}~\cite{thawri65,thawri68}. Many of the basic results presented in this chapter were obtained in various forms by several authors, and often it would be hard to establish any priorities. Most of the important early contributions can be found in \textsc{Mezei} and \textsc{Wright}~\cite{mezwri67}, \textsc{Eilenberg} and \textsc{Wright}~\cite{eilwri67}, \textsc{Thatcher} and \textsc{Wright}~\cite{thawri68}, \textsc{Doner}~\cite{don70}, \textsc{Thatcher}~\cite{tha70}, \textsc{Pair} and \textsc{Quere}~\cite{paique68}, \textsc{Brainerd}~\cite{bra68,bra69a}, \textsc{Arbib} and \textsc{Give'on}~\cite{arbgive68}, and \textsc{Magidor} and \textsc{Moran}~\cite{magmor69}.

Already in many of these papers trees were defined as terms, and this formalism is now very common. However, most authors use no separate frontier alphabet. Also, often operators may have more than one rank. The original reason for our use of frontier alphabets was to keep the character of the algebras independent of the number of frontier symbols. Another popular formalism defines a tree as a pair $(D,\lambda)$ consisting of a \emph{``tree domain''}\index{domain!tree} $D$ and a labelling mapping $\lambda$. Each element $d$ of $D$ specifies a node of the tree and $\lambda(d)$ is the label of this node. This definition is quite convenient for discussing concepts and operations which involve specific occurrences of subtrees. Tree domains were introduced by S.~Gorn in 1965 (for a reference, see \textsc{Brainerd}~\cite{bra69a}).

Deterministic and nondeterministic frontier-to-root tree recognizers were defined, and their equivalence was established, by \textsc{Thatcher} and \textsc{Wright}~\cite{thawri68}, \textsc{Doner}~\cite{don70}, and \textsc{Magidor} and \textsc{Moran}~\cite{magmor69}. Root-to-frontier tree recognizers were introduced by \textsc{Rabin}~\cite{rab69}, and \textsc{Magidor} and \textsc{Moran}~\cite{magmor69}. Magidor and Moran showed the equivalence of NDF and NDR recognizers, and they also studied DR recognizers.

Regular tree grammars and the results of Section~\ref{Section.2.3} are due to \textsc{Brainerd}~\cite{bra69a}. In Brainerd's grammars the form of the productions is quite general, but he shows that they can be reduced to, what we call, regular tree grammars.

The Boolean closure properties of $\Rec(\Sigma,X)$ were noted in many of the early papers mentioned above. The Kleene theorem (Theorem~\ref{Theorem.2.5.8}) was proved by \textsc{Thatcher} and \textsc{Wright}~\cite{thawri68} and by \textsc{Magidor} and \textsc{Moran}~\cite{magmor69}. A simplified proof was given by \textsc{Arbib} and \textsc{Give'on}~\cite{arbgive68}. Alphabetic tree homomorphisms (called there projections\index{projection}) and Corollary~\ref{Corollary.2.4.20} appear in \textsc{Thatcher} and \textsc{Wright}~\cite{thawri68}. General tree homomorphisms arose as special cases of finite-state tree transductions (see \textsc{Thatcher}~\cite{tha70,tha73} and \textsc{Engelfriet}~\cite{eng75b}). Tree transductions and tree homomorphisms will be considered in Chapter~\ref{Chapter.4}. Forest products (or \emph{``substitutions''}\index{substitution}) were also introduced in this context. \textsc{Ito} and \textsc{Ando}~\cite{itoand74} present a complete axiom system for the equality of regular expressions (cf.~also \textsc{\'Esik}~\cite{esi81}).

Minimal tree recognizers and Nerode congruences are discussed in \textsc{Brainerd}~\cite{bra68}, \textsc{Arbib} and \textsc{Give'on}~\cite{arbgive68}, and  \textsc{Magidor} and \textsc{Moran}~\cite{magmor69}.

The theory of equational forests is from \textsc{Mezei} and \textsc{Wright}~\cite{mezwri67}. We have simplified the exposition by considering only regular fixed-point equations. Mezei and Wright considered also equational and recognizable subsets of general algebras (cf.~Exercise~22). They proved that the equational subsets of an algebra (of finite type) are the homomorphic images of the recognizable subsets of term algebras. Applied to term algebras this result gives our Theorem~\ref{Theorem.2.7.9}. \textsc{Eilenberg} and \textsc{Wright}~\cite{eilwri67} present these results in a category theoretic form. For various classes of subsets in general algebras we refer also to \textsc{Wagner}~\cite{wag71}, \textsc{Lescanne}~\cite{les76}, \textsc{Marchand}~\cite{mar81}, \textsc{Shepard}~\cite{she69}, and \textsc{Steinby}~\cite{ste81}. \textsc{Dubinsky}~\cite{dub75} discusses equational and recognizable subsets of nondeterministic algebras. \textsc{Maibaum}~\cite{mai74}, and \textsc{Engelfriet} and \textsc{Schmidt}~\cite{engscc77-80} extend the subject into another direction by considering many-sorted algebras.

The material of Section~\ref{Section.2.8} is from \textsc{Costich}~\cite{cos72}. Local forests, or similar concepts, and results related to Theorems~\ref{Theorem.2.9.4} and~\ref{Theorem.2.9.5} can be found in \textsc{Doner}~\cite{don70}, \textsc{Thatcher}~\cite{tha67,tha70}, and \textsc{Takahashi}~\cite{tak75a}.

The characterization of the forests recognizable by DR recognizers is from \textsc{Vir\'agh}~\cite{vir80}, although the basic idea is discernible already in \textsc{Magidor} and \textsc{Moran}~\cite{magmor69} (cf.~also \textsc{Thatcher}~\cite{tha73}). The minimization theory of DR recognizers appears in \textsc{G\'ecseg} and \textsc{Steinby}~\cite{gesste78a}.

We should also mention an alternative approach, originating with \textsc{Pair} and \textsc{Quere}~\cite{paique68} and popular among French writers, in which the basic objects are tuples of trees rather than trees. The usual tree operations are then augmented by operations which catenate tuples of trees or form a tree from an $m$-tuple by creating a new root labelled by an $m$-ary operator. As an abstract framework for their study Pair and Quere introduced \emph{``binoids''},\index{binoid} the tuples of trees form such a binoid. Their results include the basic closure properties and a Kleene Theorem. This formalism has been developed further by \textsc{Arnold} and \textsc{Dauchet}~\cite{arndau78d-79} to a theory of \emph{``magmoids''}\index{magmoid} which also embodies many of the ideas of \textsc{Eilenberg} and \textsc{Wright}~\cite{eilwri67}. \textsc{Arnold}~\cite{arn77a,arn77b} discusses many topics relevant to this chapter within the framework of magmoids.

We shall now discuss briefly some topics and applications of the theory not covered by this book. The survey is by no means complete, and in many cases the choices were dictated by personal preference. Some more remarks will be made at the end of Chapters~\ref{Chapter.3} and~\ref{Chapter.4}.

The category theoretic treatment of recognizable and equational subsets by \textsc{Eilenberg} and \textsc{Wright}~\cite{eilwri67} was already mentioned. It is based on Lawvere's \emph{``theories''}.\index{theories} This approach was developed further by \textsc{Give'on} and \textsc{Arbib}~\cite{givarb68}, and others. The theory of magmoids has also evolved from the same ideas. We have avoided the use of category theory altogether, but the bibliography contains a sample from the extensive and highly diversified literature on the subject. The items of interest include \textsc{Alagi\'c}~\cite{ala75a,ala75b}, \textsc{Arbib} and \textsc{Manes}~\cite{arbman74}, \textsc{Bobrow} and \textsc{Arbib}~\cite{bobarb}, \textsc{Goguen}~\cite{gog75}, \textsc{Goguen} et al~\cite{gogtha74,gogthawagwri77}, \textsc{Horv\'ath}~\cite{hor79,hor81}, and \textsc{Trnkov\'a} and \textsc{Ad\'amek}~\cite{trnada79}.

The \emph{structure theory} of tree automata has received little attention although some initial steps were taken already by \textsc{Magidor} and \textsc{Moran}~\cite{magmor69}. \textsc{Ricci}~\cite{ric73} considered cascade products of tree automata. Iterative realizations\index{realization of!tree automaton} and general products\index{product!of tree automata} of tree automata are studied in \textsc{Steinby}~\cite{ste77b}. Two sections of \textsc{G\'ecseg} and \textsc{Steinby}~\cite{gecste78b-79} are devoted to the subject. It is evident that generalizations from the unary case will usually not be easy in this area.

Transition monoids have proved very useful in finite automaton theory and some equivalents of them for tree automata have been suggested. The \emph{``$m$-ary monoids''}\index{monoid!mary@$m$-ary} of \textsc{Give'on}~\cite{giv71} and the \emph{``substitution algebras''} of \textsc{Yeh}~\cite{yeh71} \index{Algebra!substitution}are in fact special Menger algebras. The same idea reappears in the \emph{``clone\index{Algebra!clone} algebras''} of \textsc{Turner}~\cite{tur75}. \textsc{Sommerhalder}~\cite{som74} develops the concept further and associates with an algebra a sequence $M_1,M_2,\dots$ of monoids. Here $M_n$ consists of all $n$-tuples of $n$-ary polynomial functions of the algebra. It would be easy to define syntactic monoids of forests along these lines, but no such theory seems to have evolved yet. Another variant of the transition semigroup concept has been studied by \textsc{Helton}~\cite{hel76}.

We shall mention some other algebraic topics of potential interest. A $\Sigma X$-forest $T$ is said to be recognizable by a $\Sigma$-algebra $\cA=(A,\Sigma)$ if one may choose $\alpha\colon X\to A$ and $A'(\subseteq A)$ in such a way that $(\cA,\Sigma,A')$ recognizes $T$. Families of forests recognizable by algebras belonging to a given variety (equational class) were considered by \textsc{Steinby}~\cite{ste77a} and by \textsc{G\'ecseg} and \textsc{Horv\'ath}~\cite{gechor76}. For a further study in this direction it would  probably be advantageous to follow the example of Eilenberg's theory of $M$-varieties and varieties of recognizable languages and consider \emph{``$\omega$-varieties''}\index{omega-variety@$\omega$-variety} (usually called \emph{pseudovarieties}\index{pseudovariety}) of algebras and the families of forests corresponding to them; an $\omega$-variety is a class of finite algebras closed under the construction of subalgebras, homomorphic images and finite direct products. In \textsc{Steinby}~\cite{ste79} it was shown that Eilenberg's basic variety theorem can be extended to $\omega$-varieties and varieties of recognizable subsets of free algebras (suitably defined). A specialization of this result to term algebras gives a correspondence between $\omega$-varieties and varieties of recognizable forests. A $\Sigma X$-forest $T$ is said to be \emph{rationally represented} by an $\Omega X$-recognizer $\bA$ if there exists an embedding $\varphi\colon F_\Sigma(X)\to F_\Omega(X)$ of a certain kind such that $T\varphi=T(\bA)$. A variety $\cK$ of algebras is said to be \emph{rationally complete} if every recognizable forest can be rationally represented by a recognizer based on a finite algebra belonging to $\cK$. \textsc{G\'ecseg}~\cite{gec77} studies the rational completeness\index{rational completeness} of varieties and the equivalence of tree recognizers with respect to rational representation.\index{rational representation} Further results can be found in \textsc{Mar\'oti}~\cite{maro77}, and \textsc{Marchand}~\cite{mar79} also contains some related ideas.

We shall now list a few references to some more topics. \emph{Probabilistic tree automata}\index{probabilistic tree automaton} and related topics have been discussed by \textsc{Magidor} and \textsc{Moran}~\cite{magmor69,magmor70}, \textsc{Ellis}~\cite{ell71} and \textsc{Karpi\'nski}~\cite{kar74b,kar75}. Forests of \emph{infinite trees}\index{tree!infinite} appear in \textsc{Rabin}~\cite{rab69}, \textsc{Engelfriet}~\cite{eng72}, \textsc{Casteran}~\cite{cas78} and \textsc{Courcelle}~\cite{cou78}. An alternative way to generate forests is provided by the \emph{tree adjunct grammars}\index{grammar!tree adjunct} studied by \textsc{Joshi,} \textsc{Levy} and \textsc{Takahashi}~\cite{joslevtak73,joslevtak75}, \textsc{Levy}~\cite{levy73}, and \textsc{Levy} and \textsc{Joshi}~\cite{levjos73}. Also \emph{Lindenmayer systems}\index{Lindenmayer system} (L-systems) for trees have been considered; see \textsc{\v{C}ulik}~\cite{cul74}, \textsc{\v{C}ulik} and \textsc{Maibaum}~\cite{culmai74}, \textsc{Engelfriet}~\cite{eng76a,eng77}, \textsc{Karpi\'nski}~\cite{kar77}, \textsc{Steyart}~\cite{stey78}, and \textsc{Szilard}~\cite{szi74}.

Although we present our subject as a part of pure automata and formal language theory, it should be clear that it has many connections to the more applied aspects of language specification, translation and semantics. As a conclusion we would like to point out some less obvious areas of application.

When \textsc{Doner}~\cite{don65,don70} and \textsc{Thatcher} and \textsc{Wright}~\cite{thawri65,thawri68} introduced tree automata their goal was to prove the decidability of the weak second order theory of multiple successors. Further applications to logic can be found in \textsc{Rabin}~\cite{rab69,rab70}.

In \emph{syntactic pattern recognition}\index{syntactic pattern recognition} patterns are decomposed into simple basic elements which are represented by letters of an alphabet. A pattern is then represented, for example, as a word. However, essential information about the relations between the basic elements may be lost if the corresponding letters are simply concatenated to form a word. It is possible that these can be described adequately by representing the pattern as a tree, and then tree automata theory may be used. For example, the considered class of patterns may be generated by a tree grammar or recognized by a tree recognizer. One specific problem prompted by syntactic pattern recognition is the \emph{inference of forests}\index{inference of forests} from samples. The interested reader may consult the books by \textsc{Fu}~\cite{fu82} and \textsc{Gonzalez} and \textsc{Thomason}~\cite{gontho78}. Some papers from this area are \textsc{Berger} and \textsc{Pair}~\cite{berpai78}, \textsc{Brayer} and \textsc{Fu}~\cite{brafu77}, \textsc{Fu} and \textsc{Bhargava}~\cite{fubha73}, \textsc{Gonzalez,} \textsc{Edwards} and \textsc{Thomason}~\cite{gonedwtho76}, \textsc{Lu} and \textsc{Fu}~\cite{lufu78}, \textsc{Pair}~\cite{pai76a}, \textsc{Tai}~\cite{tai79}, and \textsc{Williams}~\cite{wil75}.

%% file: Prologue.3.tex
The words generated by a context-free grammar can be read from derivation trees. The connection between forests and languages implied by this fact is the subject matter of this chapter. In the first section we define the yield-function by means of which a word is extracted from a tree. In Section~\ref{Section.3.2} the basic relations between recognizable forests and context-free grammars are established. The usual definition of derivation trees must be modified slightly as to make them ``trees'' in our sense of the term, but the difference is inessential. The forest of derivation trees of any CF grammar is shown to be recognizable. On the other hand, we shall see that the yield of any recognizable forest is a CF language. Hence tree recognizers may also be viewed as recognizers of CF languages. The section is concluded by showing that every CF language is the yield of a local forest recognizable by a deterministic R-recognizer.

The inverse image of a CF language under the yield-function is not always a recognizable forest, but we show in the beginning of Section~\ref{Section.3.3} that the inverse image of a regular language is a recognizable forest. Also, a slightly restricted converse of this fact is presented. Then we show that every CF language can be obtained from a recognizable forest with a fixed and very simple ranked alphabet. Section~\ref{Section.3.3} is concluded by some examples which show how facts about context-free languages can be proved using the theory of recognizable forests. 

In Section~\ref{Section.3.4} another, less well-known, way to obtain the context-free languages from recognizable forests is presented.

%% file: Section.3.1.tex
We shall now formally define the function that extracts a word from the frontier of a tree. This will also give a function that associates a language with every forest.

\begin{df}\label{Definition.3.1.1}\normalfont
The \emph{yield}\index{yield of!tree} $\yd(t)$ of a $\Sigma X$-tree $t$ is defined inductively as follows:
\begin{enumerate}
\item[1$^{\circ}$] $\yd(x)=x$ for all $x\in X$.
\item[2$^{\circ}$] If $t=\sigma(t_1,\dots,t_m)$ ($m\geq 0$, $\sigma\in\Sigma_m$), then $\yd(t)=\yd(t_1)\dots\yd(t_m)$.
\end{enumerate}
The \emph{yield}\index{yield of!forest} of a $\Sigma X$-forest $T$ is the $X$-language $\yd(T)=\{\yd(t)\mid t\in T\}$.
\end{df}

To obtain the yield of a tree $\sigma(t_1,\dots,t_m)$ one concatenates the yields of the subtrees $t_1,\dots,t_m$. In particular, $\yd(\sigma)=e$ for all $\sigma\in\Sigma_0$. More generally, $\yd(t)=e$ iff $t\in F_\Sigma(\emptyset)$. The mapping
\[
\yd\colon F_\Sigma(X)\to X^*
\]
is not injective; in general, a word is the yield of several trees.

We use the same symbol $\yd$ for its extension to forests. Of course, $\yd$ presupposes a $\Sigma$ and an $X$ although our notation does not show this.

\begin{ex}\label{Example.3.1.2}\normalfont
Let $\omega\in\Sigma_0$, $\sigma\in\Sigma_3$, and $x,y\in X$. For $s=\sigma(x,\sigma(y,\omega,y),\omega)$ and $t=\sigma(\omega,x,\sigma(y,y,\omega))$ we have $\yd(s)=\yd(t)=xyy$.
\epr
\end{ex}

Whether or not a given word $w\in X^*$ is the yield of some $\Sigma X$-tree depends on the length of $w$ and the arities of the operators in $\Sigma$.

\begin{lm}\label{Lemma.3.1.3}
Let $r(\Sigma)=\{m_1,\dots,m_k\}$. For a word $w\in X^*$ there exists a tree $t\in ´F_\Sigma(X)$ such that $\yd(t)=w$ iff the length of $w$ can be expressed in the form
\[
|w|=h_1(m_1-1)+\dots+h_k(m_k-1)+1
\]
for some (integers) $h_1,\dots,h_k\geq 0$.
\epr
\end{lm}

The proof of the lemma is an exercise. It is easy to see that $\yd(F_\Sigma(X))=X^*$ iff $\Sigma_0\neq\emptyset$ and $\Sigma-(\Sigma_1\cup\Sigma_0)\neq\emptyset$. When this is the case, there exists for every $X$-language $L$ a $\Sigma X$-forest $T$ such that $\yd(T)=L$. The greatest among these is the forest
\[
\yd^{-1}(L)=\{t\in F_\Sigma(X)\mid\yd(t)\in L\}.
\]
In general, we know just that $\yd(\yd^{-1}(L))\subseteq L$. From Lemma~\ref{Lemma.3.1.3} one easily gets

\begin{cor}\label{Corollary.3.1.4}
For a given $L\subseteq X^*$, there exists a forest $T\subseteq F_\Sigma(X)$ such that $\yd(T)=L$ iff
\[
\{|w|\mid w\in L\}\subseteq\{h_1(m_1-1)+\dots+h_k(m_k-1)+1\mid h_1,\dots,h_k\geq 0\},
\]
where $\{m_1,\dots,m_k\}=r(\Sigma)$.
\epr
\end{cor}

In the following lemma we list some obvious properties of $\yd$ and $\yd^{-1}$.

\begin{lm}\label{Lemma.3.1.5}
Let $S$ and $T$ be $\Sigma X$-forests, and $K$ and $L$ $X$-languages. Then
\begin{enumerate}
\item[(a)] $\yd(S\cup T)=\yd(S)\cup\yd(T)$,
\item[(b)] $\yd(S\cap T)\subseteq\yd(S)\cap\yd(T)$,
\item[(c)] $\yd^{-1}(K\cup L)=\yd^{-1}(K)\cup\yd^{-1}(L)$,
\item[(d)] $\yd^{-1}(K\cap L)=\yd^{-1}(K)\cap\yd^{-1}(L)$, and
\item[(e)]$\yd^{-1}(K-L)=\yd^{-1}(K)-\yd^{-1}(L)$.
\epr
\end{enumerate}
\end{lm}

%% file: Section.3.2.tex
In the customary definition of derivation trees the inner nodes are labelled by nonterminal symbols and a nonterminal may appear at nodes with different numbers of outgoing edges. Since we allowed a symbol of a ranked alphabet to have just one rank, the definition of derivation trees should be modified accordingly.

Let $G=(N,X,P,a_0)$ be a CF grammar as defined in Section~\ref{Section.1.6}. We associate with $G$ a ranked alphabet $\Sigma^G$ thus: for each $m\geq 0$,
\[
\Sigma_m^G=\{(a,m)\mid(\exists a\to \eta\in P)|\eta|=m\}.
\]

\begin{df}\label{Definition.3.2.1}\normalfont
Let $G$ and $\Sigma^G$ be as above. For every $d\in N\cup X$ the set $D(G,d)$ of \emph{derivation trees}\index{tree!derivation} with $d$ as the root is defined by the following conditions:
\begin{enumerate}
\item[1$^\circ$] $D(G,x)=\{x\}$ for each $x\in X$.
\item[2$^\circ$] For $a\in N$, $(a,0)\in D(G,a)$ iff $a\to e\in P$.
\item[3$^\circ$] Suppose $a\to d_1\dots d_m\in P$, with $m>0$, $a\in N$ and $d_1,\dots,d_m\in N\cup X$. If $t_1\in D(G,d_1),\dots,t_m\in D(G,d_m)$, then $(a,m)(t_1,\dots,t_m)\in D(G,a)$.
\item[4$^\circ$] Nothing is in any $D(G,d)$ unless this follows from a finite number of applications of the rules 1$^\circ$, 2$^\circ$ and 3$^\circ$.
\end{enumerate}
The \emph{derivation forest}\index{forest!derivation} of $G$ is the $\Sigma^G X$-forest $D(G)=D(G,a_0)$.
\end{df}

Exactly as in the case of conventional derivation trees, every $t$ in $D(G,d)$ ($d\in N\cup X$) corresponds to a unique leftmost derivation in $G$ of the word $\yd(t)$ from $d$. Also, every derivation
\[
d\Rightarrow_G u_1\Rightarrow_G\dots\Rightarrow_Gu_{k-1}\Rightarrow_G w,
\]
with $d\in N\cup X$ and $w\in X^*$, can be described by a tree $t\in D(G,d)$ such that $\yd(t)=w$. This is easily shown by induction on the length of the derivation. Hence, $L(G)=\yd(D(G))$.

\begin{thm}\label{Theorem.3.2.2}
The derivation forests of CF grammars are local and, therefore, recognizable.
\end{thm}

\pr
Let $G=(N,X,P,a_0)$ be a CF grammar. It is obvious that $D(G)$ is the local $\Sigma^G X$-forest $L(R,F)$ (in the notation of Section~\ref{Section.2.9}), where
\[
R=\{(a_0,m)\mid m\geq0, (a_0,m)\in\Sigma^G_m\}
\]
and the set $F$ of the allowed forks is defined as follows. If $m>0$ and $a\to d_1\dots d_m\in P$, then we include in $F$ every fork $(a,m)(c_1,\dots,c_m)$ such that for all $i=1,\dots,m$,
\[
c_i=\left\{\begin{array}{ll}
d_i&\text{if $d_i\in X$},\\
(d_i,k)&\text{with $k\geq 0$ and $(d_i,k)\in\Sigma^G_k$, if $d_i\in N$.}
\end{array}\right.
\]
Nothing is in $F$ unless this follows from the construction described above.
\epr

\

It is also easy to see that $D(G)$ is generated by the regular $\Sigma^G X$-grammar $G_D=(N,\Sigma^G,X,P_D,a_0)$, where
\[
P_D=\{a\to(a,m)(d_1,\dots,d_m)\mid m\geq0,\ a\to d_1\dots d_m\in P,\ d_1,\dots,d_m\in N\cup X\}.
\]

\begin{ex}\label{Example.3.2.3}\normalfont
Consider the CF grammar
\[
G=(\{a_0,b\},\{x,y\},\{a_0\to xa_0b,a_0\to e,b\to xyb,b\to y\},a_0).
\]
In this case $\Sigma^G=\Sigma^G_0\cup\Sigma^G_1\cup\Sigma^G_3$, where $\Sigma^G_0=\{(a_0,0)\}$, $\Sigma^G_1=\{(b,1)\}$ and $\Sigma^G_3=\{(a_0,3),(b,3)\}$. The productions of the grammar $G_D=(N,\Sigma^G,X,P_D,a_0)$ generating $D(G)$ are $a_0\to(a_0,3)(x,a_0,b)$, $a_0\to(a_0,0)$, $b\to(b,3)(x,y,b)$ and $b\to (b,1)(y)$. The allowed roots of the local forest $D(G)$ are $(a_0,0)$ and $(a_0,3)$, and the possible forks are $(a_0,3)(x,(a_0,0),(b,1))$, $(a_0,3)(x,(a_0,0),(b,3))$, $(a_0,3)(x,(a_0,3),(b,1))$, $(a_0,3)(x,(a_0,3),\allowbreak(b,3))$, $(b,3)(x,y,(b,1))$, $(b,3)(x,y,(b,3))$ and $(b,1)(y)$.
\epr
\end{ex}

Theorem~\ref{Theorem.3.2.2} yields immediately

\begin{cor}\label{Corollary.3.2.4}
Every CF language is the yield of a recognizable forest.
\epr
\end{cor}

The converse is also true:

\begin{thm}\label{Theorem.3.2.5}
The yield of any recognizable forest is a context-free language.
\end{thm}

\pr
Let $G=(N,\Sigma,X,P,a_0)$ be a regular $\Sigma X$-grammar generating the given recognizable $\Sigma X$-forest $T$. To simplify matters we assume that $G$ is in normal form. Now we construct the CF grammar $G_1=(N,X,P_1,a_0)$ with
\[
P_1=\{a\to\yd'(p)\mid a\to p\in P\}.
\]
Here $\yd'$ is the yield-function corresponding to the extended frontier alphabet $X\cup N$. Inductions on the lengths of the derivations show that
\begin{enumerate}
\item[(1)] $a\Rightarrow^*_Gt$ implies $a\Rightarrow^*_{G_1}\yd(t)$, for all $a\in N$, $t\in F_\Sigma(X)$, and that
\item[(2)] for all $w\in X^*$ and $a\in N$, $a\Rightarrow^*_{G_1} w$ only in case there exists a tree $t\in F_\Sigma(X)$ such that $a\Rightarrow^*_Gt$ and $\yd(t)=w$.
\end{enumerate}
These two facts imply that $\yd(T)=L(G_1)$ is CF.
\epr

\

In view of Theorem~\ref{Theorem.3.2.5} any tree recognizer may be seen as a device which recognizes a CF language by checking the possible syntaxes of given words; a word is accepted iff it is the yield of at least one tree accepted by the tree recognizer.

\begin{df}\label{Definition.3.2.6}\normalfont
The \emph{language recognized}\index{language!recognized by $\Sigma X$-recognizer} by a $\Sigma X$-recognizer $\bA$ is the $X$-language $L(\bA)=\yd(T(\bA))$.
\end{df}

The previous results can now be expressed as follows.

\begin{thm}\label{Theorem.3.2.7}
A language is recognized by a tree recognizer iff it is context-free.
\epr
\end{thm}

The equivalence expressed in Theorem~\ref{Theorem.3.2.7} is effective both ways; for any CF language given by a CF grammar we can construct a tree recognizer, and for any tree recognizer $\bA$ we can construct a CF grammar generating $L(\bA)$.

By Theorem~\ref{Theorem.3.2.2} every CF language is the yield of a local forest. We shall now show that even a smaller class of forests will suffice. To this end we replace derivation trees by trees in which the inner nodes are labelled by complete productions.

With every CF grammar $G=(N,X,P,a_0)$ we associate another ranked alphabet $\Sigma^P$ defined as follows. For each $m\geq0$, let
\[
\Sigma^P_m=\{(a\to\eta)\mid\text{$a\to\eta$ is in $P$ and $|\eta|=m$}\},
\]
i.e., the $m$-ary symbols correspond to the productions with right-hand sides of length $m$.

\begin{df}\label{Definition.3.2.8}\normalfont
Let $G$ and $\Sigma^P$ be as above. For every $d\in N\cup X$ the set $P(G,d)$ of \emph{production trees}\index{tree!production} with $d$ at the root is defined by the following conditions:
\begin{enumerate}
\item[1$^\circ$] $P(G,x)=\{x\}$ for each $x\in X$.
\item[2$^\circ$] For $a\in N$, $(a\to e)\in P(G,a)$ iff $a\to e\in P$.
\item[3$^\circ$] Suppose $a\to d_1\dots d_m\in P$ ($m>0$, $a\in N$ and $d_1,\dots,d_m\in N\cup X)$. If $p_1\in P(G,d_1),\dots,p_m\in P(G,d_m)$, then $(a\to d_1\dots d_m)(p_1,\dots,p_m)\in P(G,a)$.
\item[4$^\circ$] Nothing is in any $P(G,d)$ unless this follows from a finite number of applications of 1$^\circ$, 2$^\circ$ and 3$^\circ$.
\end{enumerate}
The \emph{production forest}\index{forest!production} of $G$ is the $\Sigma^P X$-forest $P(G)=P(G,a_0)$.
\end{df}

In our previous discussion of DR-recognizers we excluded nullary symbols, but since the ranked alphabets $\Sigma^P$ may contain such symbols, we now extend the definition of a DR $\Sigma X$-recognizer $\bA = (\cA,a_0,A')$ by setting $\sigma^\cA \in A$ and $\sigma \tilde{\alpha} = \{\sigma^\cA\}$ for any $\sigma \in \Sigma_0$.

\begin{thm}\label{Theorem.3.2.9}
The production forest $P(G)$ of any CF grammar $G$ is local and it is also recognizable by a deterministic R-recognizer.
\end{thm}

\pr
Let $G=(N,X,P,a_0)$ be a CF grammar. The presentation of $P(G)$ as a local forest is similar to that of $D(G)$. We construct a DR $\Sigma^P X$-recognizer $\bA=(A,\Sigma^P,X,A',\alpha)$ as follows. Put $A=N\cup X\cup\{d\}$ ($d\notin N\cup X$), $A'=\{a_0\}$, and for each $x\in X$, $x\alpha=\{x\}$. Next, the underlying root-to-frontier algebra $\cA=(A,\Sigma^P)$ is defined. If $\sigma=(a\to e)\in\Sigma^P_0$, then $\sigma^\cA=a$. Let $\sigma=(a\to c_1\dots c_m)\in\Sigma^P_m$ with $m>0$. Then we put $\sigma^\cA(a)=(c_1,\dots,c_m)$, and $\sigma^\cA(b)=(d,\dots,d)$ for all $b\neq a$. It is easy to show by tree induction that for all $t\in F_{\Sigma^P}(X)$ and $a\in N\cup X$,
\[
a\in t\tilde\alpha\text{\quad iff\quad} t\in P(G,a).
\]
This implies that $\bA$ recognizes $P(G)$.
\epr

\

The language recognized by an R-recognizer is defined in the natural way. As it is obvious that $\yd(P(G))=L(G)$ for every CF grammar $G$, we may state

\begin{cor}\label{Corollary.3.2.10}
Every CF language is recognized by a deterministic R-recognizer.
\epr
\end{cor}

%% file: Section.3.3.tex

Every $\CF$ language $L$ is the yield of many different forests.
Such a forest is not necessarily recognizable. In particular,
the greatest of them (for a given $\Sigma$)
$\yd^{-1}(L)$ may be nonrecognizable.

\begin{ex}
\label{Example.3.3.1}
{\rm
Let $\Sigma = \Sigma_2 = \{ \sigma \}$ and $X = \{ x, y \}$. Consider the
(minimal linear) $\CF$ language $L = \{ x^n y^n \mid n \geq 1 \}$. If
$\yd^{-1}(L)$ were recognized by a $\Sigma X$-recognizer
$\bA$, then $\bA$ would accept all trees $\sigma(s_i, t_i)$ $(i \geq 1)$, where
(i) $s_1 = x$, $t_1 = y$ and (ii) $s_{i+1} = \sigma(s_k, x)$ and
$t_{k+1} = \sigma(y, t_k)$ for all $k \geq 1$. As $\bA$ is finite, it would
then also accept some tree $\sigma(s_i, t_j)$ with $i \neq j$. But this is a
contradiction, because $\yd(\sigma(s_i, t_j)) = x^i y^j \not\in L$. \epr
} 
\end{ex}

In contrast to Example \ref{Example.3.3.1} we have

\begin{thm}
\label{Theorem.3.3.2}
If $L$ is a regular $X$-language, then $\yd^{-1}(L) \in \Rec(\Sigma, X)$
for any ranked alphabet $\Sigma$.
\end{thm}

\pr
Let $M$ be a finite monoid, $\varphi : X^* \rightarrow M$ a
 homomorphism and $H$ a subset of $M$ such that $L = H\varphi^{-1}$.
Let $\cA = (M, \Sigma)$ be the $\Sigma$-algebra defined so that
$$
\sigma^{\cA}(a_1, \ldots, a_m) = a_1 \cdot a_2 \cdot \ldots
\cdot a_m \;\; (\mbox{product in } M)
$$
for all $m \geq 0$, $\sigma \in \Sigma_m$ and $a_1, \ldots, a_m \in M$.
In particular, $\sigma^{\cA} = 1$ when $\sigma \in \Sigma_0$.
If we put
$$
\alpha = \varphi | X \colon X \rightarrow M,
$$
then
$$
t\hat{\alpha} = \yd(t) \varphi \; \mbox{ for all } t \in F_{\Sigma}(X).
$$
This implies that $\yd^{-1}(L) = T(\bA)$ for the
$\Sigma X$-recognizer $\bA = (\cA, \alpha, H)$. Indeed, for all
$t \in F_{\Sigma}(X)$,
\begin{eqnarray*}
t \in T(\bA) & \mbox{ iff } & t \hat{\alpha} = \yd(t) \varphi \in H \\
 & \mbox{ iff } & \yd(t) \in L \\
 & \mbox{ iff } & t \in \yd^{-1}(L).
\end{eqnarray*}

\epr

\

The full converse of Theorem \ref{Theorem.3.3.2} is not valid, but the
following result will be proven in Exercises~6 and~7.

\begin{thm}
\label{Theorem.3.3.3}
\sloppy Let $L \; (\subseteq X^*)$ be a language and $\Sigma$ a ranked alphabet
such that $\yd(\yd^{-1}(L)) = L$. Then $\yd^{-1}(L) \in \Rec(\Sigma, X)$
implies $L \in \Rec X$. \epr
\end{thm}

The ranked alphabets $\Sigma^G$ and $\Sigma^P$ depend on the given
$\CF$ grammar. We shall now show that every $\CF$ language is the yield of
a recognizable forest over a fixed ranked alphabet. In fact, a very
simple alphabet will suffice.

\begin{thm}
\label{Theorem.3.3.4}
Let $\Sigma$ be a ranked alphabet which contains a binary operator and
a nullary operator. Then every $\CF$ language is recognized by a
$\Sigma$-recognizer. For e-free $\CF$ languages the binary symbol alone
is sufficient.
\end{thm}

\pr
Let us consider the e-free case first. Every $\CF$ language
$L \subseteq X^+$ is generated by a $\CF$ grammar
$G = (N, X, P, a_0)$ in Chomsky normal form, where each production
is of the form $a \rightarrow bc$ or $a \rightarrow x$ ($a, b, c \in N$,
$x \in X$). By Lemma~\ref{Lemma.2.4.1} we may assume that
$\Sigma = \Sigma_2 = \{ \sigma \}$. Let $G_1 = (N, \Sigma, X, P_1, a_0)$ be
the regular $\Sigma X$-grammar, where
$$
P_1 = \{ a \rightarrow \sigma(b, c) \mid a \rightarrow bc \in P \}
\cup \{ a \rightarrow x \mid a \rightarrow x \in P \}.
$$
Adjoin $N$ to the frontier alphabet and let
$$
\yd' \colon F_\Sigma(X \cup N) \rightarrow (X \cup N)^*
$$
be the corresponding yield-function. By induction on the length of
the derivation one can verify that for every derivation
$$
a \Rightarrow_G u_1 \Rightarrow_G \ldots \Rightarrow_G u_k \;\;
(a \in N, k \geq 1)
$$
there is a derivation
\begin{equation*}
\tag{*}
a \Rightarrow_{G_1} p_1 \Rightarrow_{G_1} \ldots \Rightarrow_{G_1} p_k
\;\; (p_1, \ldots, p_k \in F_\Sigma(X \cup N))
\end{equation*}
such that $\yd'(p_i) = u_i$ for $i = 1, \ldots, k$. This implies
$L(G) \subseteq \yd(T(G_1))$ as $\yd'|F_\Sigma(X) = \yd$.
The converse inclusion follows from the fact that for every
derivation (*) we have a derivation
$$
a \Rightarrow_G \yd'(p_1) \Rightarrow_G \ldots \Rightarrow_G \yd'(p_k).
$$

If $L \subseteq X^*$ and $e \in L$, then we find, as above, a recognizable
$\Sigma X$-forest $T$ such that $\yd(T) = L - \{ e \}$. Now add a nullary
operator $\omega$ to $\Sigma$ and let $T' = T \cup \omega$.
Then $T'$ is recognizable and $\yd(T') = L$.
\epr

\

The connections established above suggest the possibility of developing,
or just interpreting, the theory of context-free languages in terms
of tree automata and recognizable forests. We shall illustrate this
by a few examples. The results themselves are well known.

\begin{thm}
\label{Theorem.3.3.5}
The intersection of a context-free language with a regular language
is context-free.
\end{thm}

\pr
Consider a $\CF$ language $L \subseteq X^*$ and a regular language $U$
over the same alphabet. Choose any ranked alphabet $\Sigma$ and a
recognizable $\Sigma X$-forest $R$ such that $\yd(R) = L$. Then
$$
L \cap U = \yd(R \cap \yd^{-1}(U)).
$$
Since $R \cap \yd^{-1}(U) \in \Rec(\Sigma, X)$ by
Theorem~\ref{Theorem.3.3.2} and Theorem~\ref{Theorem.2.4.2},
this means that $L \cap U$ is context-free.
\epr

\

The next example shows how the regular forest operations relate to
language operations.

\begin{df}
\label{Definition.3.3.6}
{\rm
Let $U$ and $V$ be $X$-languages and $x \in X$. The
{\em $x$-substitution\/}\index{x-substitution@$x$-substitution}
of $U$ into $V$ is the language $U \cdot_x V$ of all words
$$
w_0 u_1 w_1 u_2 \ldots w_{k-1} u_k w_k,
$$
where $k \geq 0$, $u_1, \ldots, u_k \in U$,
$w_0 x w_1 x \ldots x w_{k-1} x w_k \in V$ and $x$ does not appear
in the word $w_0 w_1 \ldots w_k$.

The {\em $x$-substitution closure\/} of $U$ is the language \index{closure!x@$x$-substitution}
$$
U^{*x} = \bigcup (U^{i,x} \mid i \geq 0 ),
$$
where $U^{0,x} = \{ x \}$ and $U^{i,x} = U^{i-1,x} \cdot_x U
\cup U^{i-1,x}$ for $i > 0 $.
} 
\end{df}

Consider two $\Sigma X$-forests $S$ and $T$ and a symbol $x \in X$.
Every tree $p \in S \cdot_x T$ is obtained from some tree $t \in T$
by replacing each occurrence of $x$ by some tree from $S$. Suppose
$x$ appears $k$ times ($k \geq 0$) in $t$ and that we get $p$ by replacing
these occurrences, from left to right, by the trees
$s_1, \ldots, s_k \in S$. If
$$
\yd(t) = w_0 x w_1 x \ldots x w_k,
$$
then
$$
\yd(p) = w_0 \yd(s_1) w_1 \yd(s_2) \ldots \yd(s_k) w_k \in \yd(S) \cdot_x
\yd(T).
$$
Conversely, if $w \in  \yd(S) \cdot_x
\yd(T)$, then we may write $w$ in the form
$$
w = w_0 u_1 w_1 u_2 \ldots w_{k-1} u_k w_k
$$
so that $k \geq 0$, $w_0 x w_1 x \ldots x w_k \in \yd(T)$ and
$u_1, \ldots, u_k \in \yd(S)$. Then there are trees $t \in T$ and
$s_1, \ldots, s_k \in S$ such that $\yd(t) = w_0 x w_1 x \ldots x w_k$
and $\yd(s_1) = u_1$, \ldots, $\yd(s_k) = u_k$. If we replace the occurrences
of $x$ in $t$ by the trees $s_1, \ldots, s_k$, then we get a tree
$p \in S \cdot_x T$ such that $\yd(p) = w$. An easy induction on $i$
shows now that
$$
\yd(T^{i, x}) = \yd(T)^{i, x} \;\; \mbox{ for all } i \geq 0.
$$
Using these observations we get

\begin{lm}
\label{Lemma.3.3.7}
For any two $\Sigma X$-forests $S$ and $T$, and any letter
$x \in X$,
\begin{enumerate}
\item[{\rm (a)}] $\yd(S \cdot_x T) = \yd(S) \cdot_x \yd(T)$
\end{enumerate}
and
\begin{enumerate}
\item[{\rm (b)}] $\yd(T^{*x}) = \yd(T)^{*x}$. \epr
\end{enumerate}
\end{lm}

Now we can derive the following well-known description of the family
of context-free languages.

\begin{thm}
\label{Theorem.3.3.8}
The context-free languages form the smallest family of languages
which contains the finite languages and is closed under (finite)
union, $x$-substitutions and $x$-substitution closures.
\end{thm}

\pr
Clearly, all finite languages are context-free. Let $U, V \subseteq X^*$
be $\CF$ and $x \in X$. There exist recognizable forests $S, T \subseteq
F_\Sigma(X)$ such that $\yd(S) = U$, $\yd(T) = V$.
Now $U \cup V = \yd(S \cup T)$, $U \cdot_x V = \yd(S \cdot_x T)$
and $V^{*x} = \yd(T^{*x})$ are all seen to be context-free.
On the other hand, the Kleene theorem
(Theorem~\ref{Theorem.2.5.8}) together with
Corollary~\ref{Corollary.3.2.4} and Lemma~\ref{Lemma.3.3.7} shows
that every $\CF$ language can be obtained from finite languages
by forming unions, $x$-substitutions and
$x$-substitution closures.
\epr

Note that when a $\CF$ $X$-language is expressed in terms of
finite languages, unions, substitutions and substitution closures,
symbols not in $X$ may be used as auxiliary symbols in substitutions.

As an example we consider the language $L = \{ x^n y^n \mid n \geq 0 \}$.
Let $ \omega \in \Sigma_0$ and $\sigma \in \Sigma_3$. Then $L$ is the
yield of, for example, the recognizable $\Sigma X$-forest
$$
T = \{ \omega, \sigma(x, \omega, y), \sigma(x, \sigma(x, \omega, y), y),
\ldots \}
$$
which has the regular expression $\omega \cdot_z \sigma(x, z, y)^{*z}$.
From this we get for $L$ the representation
$$
L = \{ e \} \cdot_z \{ xzy \}^{*z}.
$$
Here $z$ is an auxiliary letter which does not appear in the language
represented.


%% file: Section.3.4.tex

If an ordinary finite automaton is viewed as a unary algebra,
then its input symbols form a ranked alphabet. There is a way
to interpret $\Sigma X$-trees as words over $\Sigma$ in the general
case, too. When this is done, recognizable forests become
$\CF$ languages. Moreover, every $\CF$ language can be obtained
this way as a recognizable forest once its alphabet is
suitably ranked.

We consider the unary case as an introduction. The word
$$
t \eta = \sigma_1 \ldots \sigma_k \in \Sigma^*
$$
can be obtained from the corresponding
$\Sigma \{ x \}$-tree
$$
t = \sigma_k( \ldots \sigma_1(x) \ldots )
$$
recursively as follows:
\begin{enumerate}
\item[{\rm 1$^\circ$}] $x \eta = e$ for all $x \in X$.
\end{enumerate}
\begin{enumerate}
\item[{\rm 2$^\circ$}] $t \eta = s \eta \sigma$ if
$t = \sigma(s)$ ($\sigma \in \Sigma$).
\end{enumerate}
Another way to get $t \eta$ would be to erase the parentheses and
$x$ and then reverse the resulting word. Both of these constructions
can serve as a basis for the generalization to the case of an arbitrary
ranked alphabet. The reversing of the order of the word
is an inessential step due to our way of writing trees,
and it will be omitted in the generalization.

Let $\Sigma$ be an arbitrary ranked alphabet and $X$ any frontier
alphabet. We shall treat $\Sigma$ as an ordinary alphabet, too.
We assume that $\Sigma$ and $X$ are disjoint and that they do not
contain (, ) or the comma. Let
$$
Y = \Sigma \cup X \cup \{ (, ), , \}
$$
and define
$$
\eta : Y^* \rightarrow \Sigma^*
$$
as the monoid homomorphism such that
$$
y \eta =
\begin{cases}
y \mbox{ for } y \in \Sigma,\\
e \mbox{ for } y \in Y - \Sigma.
\end{cases}
$$

Applied to a $\Sigma X$-tree $t$ $\eta$ erases all frontier letters
$x \in X$, the parentheses and the commas leaving the symbols
$\sigma \in \Sigma$ intact.
It is easy to see that this can be carried out as follows, too.

\begin{lm}
\label{Lemma.3.4.1}
The words $t \eta$ ($t \in F_\Sigma(X)$) can be found recursively
as follows:
\begin{enumerate}
\item[{\rm 1$^\circ$}] $x \eta = e$ for all $x \in X$.
\end{enumerate}
\begin{enumerate}
\item[{\rm 2$^\circ$}]
If $t = \sigma(t_1, \ldots, t_m)$
($m \geq 0$, $\sigma \in \Sigma_m$), then
$t \eta = \sigma t_1 \eta \ldots t_m \eta$. \epr
\end{enumerate}
\end{lm}

We have already noted that every regular $\Sigma X$-grammar may
also be viewed as a $\CF$ grammar generating a $Y$-language. Moreover,
it is well-known that the family of context-free languages is closed
under homomorphisms. Hence we have

\begin{lm}
\label{Lemma.3.4.2}
If $T \in \Rec(\Sigma, X)$, then $T \eta \in \CF(\Sigma)$. \epr
\end{lm}

Next we prove the following converse of Lemma~\ref{Lemma.3.4.2}.

\begin{lm}
\label{Lemma.3.4.3}
Let $\Sigma$ and $X$ be alphabets. If $\Sigma$ is ranked so that
$\Sigma_2 = \Sigma$, then there exists for each $\CF$ language
$L \subseteq \Sigma^*$ a recognizable $\Sigma X$-forest $T$ such that
$T \eta = L$.
\end{lm}

\pr
First, let $L$ be $e$-free. Then $L$ is generated by a $\CF$ grammar
$G = (N, \Sigma, P, a_0)$ in Greibach 2-form, where each production
is of the form (i) $a \rightarrow \sigma bc$,
(ii) $a \rightarrow \sigma b$ or (iii) $a \rightarrow \sigma$
($a, b, c \in N$, $\sigma \in \Sigma$). We convert $G$ into a
regular $\Sigma X$-grammar $G_1 = (N, \Sigma, X, P_1, a_0)$, where
the set $P_1$ of productions is defined as follows. Fix any
$x \in X$ and put then
\begin{eqnarray*}
P_1 = &\{ a \rightarrow \sigma(b, c)  \mid   a \rightarrow \sigma bc \in P \} \cup 
\{ a \rightarrow \sigma(b, x) \mid a \rightarrow \sigma b \in P \} \cup\\
 &\{ a \rightarrow \sigma(x, x) \mid a \rightarrow \sigma \in P \}.
\end{eqnarray*}
In order to show that $T(G_1)$ is the required recognizable forest we
extend $\eta$ to a homomorphism
$$
\eta_1 | (Y \cup N)^* \rightarrow (\Sigma \cup N)^*
$$
so that $\eta_1 | Y = \eta$ and $\eta_1 | N = 1_N$. It is easy
to see that to every derivation
$$
a \Rightarrow_G u_1 \Rightarrow_G \ldots \Rightarrow_G u_k \;\;
(a \in N, k \geq 1)
$$
there corresponds a derivation
\begin{equation*}
\tag{*}
a \Rightarrow_{G_1} v_1 \Rightarrow_{G_1} \ldots \Rightarrow_{G_1}
v_k
\end{equation*}
such that $v_i \eta_1 = u_i$ ($i = 1, \ldots, k$). Conversely, every
derivation (*) is matched by the derivation
$$
a \Rightarrow_G v_1 \eta_1 \Rightarrow_G \ldots \Rightarrow_G v_k \eta_1.
$$
Since $\eta_1 | Y^* = \eta$, this implies $T(G_1) \eta = L(G) = L$.
If $e \in L$, we apply this construction to $L - e$ and add then
the tree $x$ to $T(G_1)$.
\epr

In the representation of Lemma~\ref{Lemma.3.4.3} the frontier alphabet
$X$ can be fixed in advance independently of $\Sigma$ and the
language $L$. A one-element alphabet $X = \{ x \}$ always suffices.

We say that a $\Sigma X$-recognizer $\bA$ {\em $\eta$-accepts\/}
a word $w \in \Sigma^*$, \index{word!eta@$\eta$-accepted} if it accepts at least one
$\Sigma X$-tree $t$ such that $t \eta = w$. The $\Sigma$-language
$\eta(\bA)$
{\em $\eta$-recognized\/}\index{language!etarecognized@$\eta$-recognized}
 by $\bA$ is the
set of all words $\eta$-accepted by $\bA$. In this terminology
the previous results may be summed up as follows.

\begin{thm}
\label{Theorem.3.4.4}
A language is $\eta$-recognized by some tree recognizer iff it is
a context-free language. \epr
\end{thm}


%% file: Section.3.5.tex

\begin{enumerate}
\item Is is possible that $\yd^{-1}(w)$ is infinite
for some word $w$?
\item Prove Lemma~\ref{Lemma.3.1.3}.
\item Find an example of a nonrecognizable forest $T$ such that
$\yd(T)$ is a recognizable language.
\item Show that for every $\CF$ grammar $G$, $D(G)$ is the image
of $P(G)$ under an alphabetic tree homomorphism.
\item Recall that a {\em groupoid\/}\index{groupoid}
 is an algebra with one binary
operation (and no other operations). For
$\Sigma = \Sigma_2 = \{ \sigma \}$, $F_\Sigma(X)$ is the free groupoid
generated by $X$. Verify that $\yd \colon F_\Sigma(X) \rightarrow X^+$
is a groupoid epimorphism. Then prove that a language $L \subseteq
X^+$ is context-free iff it is the homomorphic image of a recognizable
subset of the free groupoid generated by $X$
(cf. Exercise 2.22, and \textsc{Mezei} and \textsc{Wright} \cite{mezwri67}).
\item The set $\Comb(\Sigma, X)$ of ``comb-like'' $\Sigma X$-trees is defined
as the smallest set $S$ satisfying the conditions $1^\circ$ and $2^\circ$:
\begin{description}
\item[{\rm $1^\circ$}] $X \cup \Sigma_0 \subseteq S$.
\item[{\rm $2^\circ$}] If $m > 0$, $\sigma \in \Sigma_m$, $x_1, \ldots, x_{m-1}
\in X$ and $t \in S$, then $\sigma(x_1, \ldots, x_{m-1}, t) \in S$.
\begin{description}
\item[{\rm (a)}] Prove that  $\Comb(\Sigma, X) \in \Rec(\Sigma, X)$.
\item[{\rm (b)}] Let $T$ be a recognizable forest such that $T \subseteq
\Comb(\Sigma, X)$.\\
Show that $T$ is generated by a regular $\Sigma X$-grammar
$(N, \Sigma, X, P, a_0)$ in which each production has the form
$a \rightarrow \sigma(x_1, \ldots, x_{m-1}, b)$, $a \rightarrow \omega$
or $a \rightarrow x$ ($a, b \in N$, $m > 0$, $\sigma \in \Sigma_m$,
$x_1, \ldots, x_{m-1} \in X$, $\omega \in \Sigma_0$, $x \in X$).
\item[{\rm (c)}] Infer from (b) that $\yd(T) \in \Rec X$ for every
recognizable $T \subseteq \Comb(\Sigma, X)$.
\item[{\rm (d)}] Prove that for every $\Sigma X$-tree $t$ there exists
a comb-like $\Sigma X$-tree $s$ such that $\yd(s) = \yd(t)$. Deduce from
this fact that if $\yd(\yd^{-1}(L)) = L$ for some $L \subseteq X^*$, then
$$
\yd(\yd^{-1}(L) \cap \Comb(\Sigma, X)) = L.
$$
\end{description}
\end{description}
\item Prove Theorem~\ref{Theorem.3.3.3} using the results of the
previous exercise.
\item Give another proof for Theorem~\ref{Theorem.3.3.4} using the fact
that every $\CF$ language can be generated by an invertible $\CF$ grammar
in Chomsky normal form.
\end{enumerate}
In Exercises 9--12 the theory of recognizable forests should be
applied.
\begin{enumerate}
\setcounter{enumi}{8}
\item Prove that the languge $U - V$ is $\CF$ if $U$ is $\CF$ and $V$ is
a regular language.
\item Let $\varphi \colon X^* \rightarrow Y^*$ be a homomorphism of monoids.
Prove that $L \varphi^{-1} \in \CF(X)$ for every $L \in \CF(Y)$.
\item Let $h(t)$ denote the tree which is obtained from a given
tree by rewriting every operator $\sigma$ as its rank $r(\sigma)$.
Obviously $\yd(h(t)) = \yd(t)$. Show that $h$ can be defined,
for any given $\Sigma$ and $X$, as an alphabetic tree homomorphism.
Two $\CF$ grammars $G_1$ and $G_2$ are said to be
{\em structurally 
equivalent\/}\index{structural equivalence of CF grammars} 
if $h(D(G_1)) = h(D(G_2))$. Prove that
there is an algorithm to determine whether or not two
$\CF$ grammars are structurally equivalent.
\item Prove Bar-Hillel's pumping lemma (Lemma~\ref{Lemma.1.6.13}).
\item Let $G$ be a regular $\Sigma X$-grammar. Construct a
$\CF$ grammar $G'$ such that $L(G') = T(G) \eta$. Note that
Lemma~\ref{Lemma.3.4.2} follows as a result.
\end{enumerate}


%% file: Section.3.6.tex

The basic connection between recognizable forests and
context-free languages has been established in various ways.
\textsc{Mezei} and \textsc{Wright} \cite{mezwri67} proved that the
equational subsets of an algebra of finite type
(in the monoid $X^*$ these are the $\CF$ languages) are the
homomorphic images of the recognizable subsets of term algebras,
i.e., recognizable forests. Applied to groupoids this theorem
gives the result of Exercise~5 (credited to D. Muller). It also
implies Theorem~\ref{Theorem.3.3.4} which was explicitly formulated by
\textsc{Magidor} and \textsc{Moran} \cite{magmor69}.
The proof using derivation forests goes back to
\textsc{Thatcher} \cite{tha67,tha70} and \textsc{Doner} \cite{don70}.
Various forms of production trees have been used in this context
by \textsc{Engelfriet} \cite{eng75a}, and
\textsc{Steinby} \cite{ste77a}. Theorem~\ref{Theorem.3.3.2} appears,
for example, in \textsc{Rounds} \cite{rou70b}. It is a special instance
of the fact that the inverse homomorphic images of recognizable subsets
of algebras are recognizable
(cf. Exercise 2.22). Theorem~\ref{Theorem.3.3.3} appears to be well-known.
The proof outlined in Exercises~6 and~7 is from
\textsc{Steyart} \cite{stey77b}. The idea to use tree automata
in the theory of $\CF$ languages was proposed by \textsc{Rounds} \cite{rou70a}.
More examples of such applications can be found in \textsc{Thatcher}
\cite{tha73} and \textsc{Engelfriet} \cite{eng75a}. The results
of Section~\ref{Section.3.4} are due to \textsc{Ferenci} \cite{fer76}.
The interested reader may consult \textsc{Ferenci} \cite{fer80} for
further work in this direction.

As a conclusion we mention a few other topics. Using a ranked
nonterminal alphabet it is possible to define {\em context-free tree
grammars.}\index{grammar!context-free tree}
 \textsc{Rounds} \cite{rou69,rou70a,rou70b} shows that
the yield-languages of $\CF$ forests are exactly the indexed languages.
\textsc{Arnold} and \textsc{Dauchet}
\cite{arndau76d,arndau77,arndau78a}, and \textsc{Engelfriet} and
\textsc{Schmidt} \cite{engscc77-80} are some further references.

Possibilities to extend some of the results of this chapter to type 0 or
context-sensitive languages by generalizing the tree-concept have been
investigated by \textsc{Benson} \cite{ben75}, \textsc{Buttelman}
\cite{but75a,but75b}, \textsc{Hart} \cite{har74,har76},
and \textsc{R\'{e}v\'{e}sz} \cite{rev77}. Hierarchies of term languages
obtained by iteration of the yield-forming process have been studied
by \textsc{Maibaum} \cite{mai74}, \textsc{Engelfriet} and
\textsc{Schmidt} \cite{engscc77-80}, and \textsc{Turner}
\cite{tur73,tur75}. Families of languages defined by tree recognizers
based on algebras belonging to a given variety of algebras were
considered by \textsc{Steinby} \cite{ste77a}. \textsc{G\'{e}cseg} and
\textsc{Horv\'{a}th} \cite{gechor76} showed that a proper variety
may be {\em complete\/}\index{complete variety}
 in the sense that every $\CF$ language is
recognizable by a finite algebra of the variety
(cf.~the Notes and references section of Chapter~\ref{Chapter.2}).


%% file: Prologue.4.tex

In this chapter we shall deal with systems transforming trees into
trees similarly as generalized sequential machines transform strings
into strings. There are two main categories of such systems:
frontier-to-root tree transducers which process a tree from the leaves
down towards the root, and root-to-frontier tree transducers which work
in the opposite direction. Special classes of tree transducers will play
a basic part in decomposing tree 
transformations\index{composition of!tree transformations} into simpler ones.


%% file: Section.4.1.tex

Throughout this chapter, $\Sigma$, $\Omega$ and $\Delta$ will stand
for ranked alphabets. It will be assumed that whenever an operator
belongs to more than one ranked alphabet, then it has the same rank
in all of them. Moreover, $X$, $Y$ and $Z$ will always stand for (finite,
nonvoid) frontier alphabets.

Let us recall that $F_\Sigma(S)$ as defined in Section~\ref{Section.2.1}
denotes the set of $\Sigma$-trees over the frontier alphabet $S$.
Here we shall allow $S$ to be a possibly infinite set of trees and
then use the notation $F_\Sigma[S]$ for $F_\Sigma(S)$. One can easily
see that in such a case there always exist a ranked alphabet $\Omega$
and a frontier alphabet $Y$ such that $F_{\Sigma}[S] \subseteq
F_\Sigma(Y)$.

Binary relations $\tau \subseteq F_\Sigma(X) \times F_\Omega(Y)$
will be called {\em tree
transformations.}\index{tree transformation} An inclusion
$(p, q) \in \tau$ is interpreted to mean that $\tau$ may transform
$p$ into $q$. Because tree transformations are binary relations,
we can speak about {\em compositions,}
{\em inverses,}\index{inverse of tree transformation}
{\em domains\/}\index{domain!of tree transformation}
 and {\em ranges\/}\index{range of!tree transformation}
 of tree transformations
as defined in Section~\ref{Section.1.1}.

With each tree transformation
 $\tau \subseteq F_\Sigma(X) \times F_\Omega(Y)$
we associate the
{\em translation\/}\index{translation!induced by tree transformation}
$\{ (\yd(p), \yd(q)) \mid (p, q) \in \tau \}$ from
$X^*$ into $Y^*$.

The important tree transformations are those which can be given
in an effective way. Next we define two general systems (tree transducers)
inducing such transformations. We shall need a countably infinite set
$$
\Xi = \{ \xi_1, \xi_2, \ldots \}
$$
of auxiliary variables. The subset of $\Xi$ consisting of its first
$n \geq 0$ elements will be denoted by $\Xi_n$, i.e.,
$\Xi_n = \{ \xi_1, \ldots, \xi_n \}$. The role of an auxiliary variable
is to indicate an occurrence of a subtree in a tree.

If all variables occurring in a tree $q$ are among
$\xi_1$, \ldots, $\xi_n$, then the notation $q(\xi_1, \ldots, \xi_n)$
may be also  used for $q$. Moreover, if $q_1$, \ldots, $q_n$ are
arbitrary trees, then we generally write $q(q_1, \ldots, q_n)$ for
$q(\xi_1 \leftarrow q_1, \ldots, \xi_n \leftarrow q_n)$.

\begin{df}
\label{Definition.4.1.1}
{\rm
A {\em frontier-to-root tree
transducer\/}\index{tree transducer!frontier-to-root}
(F-{\em transducer\/})\index{ftransducer@F-transducer}
is a system $\gA = (\Sigma, X, A, \Omega, Y, P, A')$ where
\begin{enumerate}
\item[{\rm (1)}] $\Sigma$ and $\Omega$ are ranked alphabets,
\item[{\rm (2)}] $X$ and $Y$ are frontier alphabets,
\item[{\rm (3)}] $A$ is a ranked alphabet consisting of unary operators,
the {\em state set\/}\index{state!of F-transducer} of $\gA$,

(It will be assumed that $A$ is disjoint with all other sets in the
definition of $\gA$, except $A'$.)
\item[{\rm (4)}] $A' \subseteq A$ is the set of {\em final
states,}\index{final state of!F-transducer} and
\item[{\rm (5)}] $P$ is a finite set of
{\em productions\/}\index{production of!F-transducer}
(or {\em rewriting rules\/})\index{rewriting rule of!F-transducer}
 of the following two types:
\begin{description}
\item[{\rm (i)}] $x \rightarrow a(q)$ ($x \in X$, $a \in A$, $q \in
F_\Omega(Y)$),
\item[{\rm (ii)}] $\sigma(a_1(\xi_1), \ldots, a_m(\xi_m)) \rightarrow
a(q(\xi_1, \ldots, \xi_m))$ ($\sigma \in \Sigma_m$, $m \geq 0$,
$a_1, \ldots, a_m, a \in A$, $q(\xi_1, \ldots, \xi_m) \in
F_\Omega(Y \cup \Xi_m)$).
\end{description}
(In the sequel we shall write simply $\sigma(a_1, \ldots, a_m)$
for $\sigma(a_1(\xi_1), \ldots, a_m(\xi_m))$.)
\end{enumerate}
} 
\end{df}

We shall use also the notation $(p, q)$ for a production
$p \rightarrow q$. Moreover, if $a \in A$
is a state and $t$ is a tree, then we generally write $at$ for
$a(t)$. Similarly, if $T$ is a forest, then
$AT$ will denote the forest $\{ at \mid a \in A, t \in T \}$.
Furthermore, for any
$a \in A$, we put $\gA(a) = (\Sigma, X, A, \Omega, Y, P, a)$.

Let us note that in the above definition it would be more
exact to speak about production schemes instead of productions.
Indeed, soon we shall see that they define patterns for rewriting
trees.

Next we define the
transformations induced by F-transducers.
Consider the F-transducer $\gA$ of Definition~\ref{Definition.4.1.1}
and, for every $p \in F_\Sigma[X \cup A \Xi]$, let
$p \tau_{\gA}^*$ be the subset of
$A F_\Omega(Y \cup \Xi)$ given as follows:
\begin{enumerate}
\item[{\rm (1)}] if $p = a \xi$ ($a \in A$, $\xi \in \Xi$), then
$a \xi \in p \tau_{\gA}^*$,
\item[{\rm (2)}] if $p \in X \cup \Sigma_0$, then
$aq \in p \tau_{\gA}^*$ for all $(p, aq) \in P$,
\item[{\rm (3)}] \sloppy if $p = \sigma(p_1, \ldots, p_m)$ ($\sigma \in \Sigma_m$,
$m > 0$), then $aq(q_1, \ldots, q_m) \in p \tau_{\gA}^*$ for all
$(\sigma(a_1, \ldots, a_m), aq) \in P$ and
$a_i q_i \in p_i \tau_{\gA}^*$ ($a, a_i \in A$, $i = 1, \ldots, m$), and
\item[{\rm (4)}] nothing is in any $p \tau_{\gA}^*$ unless this follows
from (1)--(3).
\end{enumerate}

\begin{df}
\label{Definition.4.1.2}
{\rm
Take an F-transducer $\gA = (\Sigma, X, A, \Omega, Y, P, A')$.
Then the relation
$$
\tau_{\gA} = \{ (p, q) \mid p \in F_\Sigma(X), q \in F_{\Omega}(Y),
a q \in p  \tau_{\gA}^* \mbox{ for some } a \in A' \}
$$
is called the
{\em transformation\/}\index{transformation induced by!F-transducer}
 induced by $\gA$.
} 
\end{df}

For Definition~\ref{Definition.4.1.2} it would be enough to apply
$ \tau_{\gA}^*$ to trees from $F_\Sigma(X)$. The above more general
case will be needed later.

\label{page.140}
Sometimes in our proofs we should know how an input tree is transformed
step by step into an output tree. Again, let $\gA$ be the
F-transducer of Definition~\ref{Definition.4.1.1}, and consider
two trees $p, q \in F_\Sigma[X \cup A F_\Omega(Y \cup \Xi)]$. It is
said that $p$ {\em directly derives\/} $q$ in $\gA$ if $q$ can
be obtained from $p$ by
\begin{enumerate}
\item[{\rm (i)}] replacing an occurrence of an $x \in X$ in $p$ by
the right side $a \overline{q}$ of a production
$x \rightarrow a \overline{q}$ from $P$, or by
\item[{\rm (ii)}] replacing an occurrence of a subtree
$\sigma(a_1 q_1, \ldots, a_m q_m)$ ($\sigma \in \Sigma_m$,
$a_1, \ldots, a_m \in A$, $q_1, \ldots, q_m \in F_\Omega(Y \cup \Xi)$
in $p$ by $a \overline{q}(q_1, \ldots, q_m)$, where
$\sigma(a_1, \ldots, a_m) \rightarrow a \overline{q}$ is a production from $P$.
\end{enumerate}

Each application of rule (i) or rule (ii) is called a
{\em direct derivation\/}\index{direct derivation in!F-transducer}
 in $\gA$. If $q$ is obtained
from $p$ by a direct derivation in $\gA$ (i.e., $p$ directly derives
$q$ in $\gA$), then we write $p \Rightarrow_{\gA} q$. Therefore,
$\Rightarrow_{\gA}$ is a binary relation in
$ F_\Sigma[X \cup A F_\Omega(Y \cup \Xi)]$. If there is no danger of
confusion, we generally omit $\gA$ in $\Rightarrow_{\gA}$.

By finitely many consecutive applications of direct derivations we
get derivations. Accordingly, for any two trees
$p, q \in  F_\Sigma[X \cup A F_\Omega(Y \cup \Xi)]$ we say that
\begin{equation*}
\tag{1}
p = p_0 \Rightarrow p_1 \Rightarrow \ldots \Rightarrow p_i \Rightarrow
\ldots \Rightarrow p_j \Rightarrow \ldots \Rightarrow p_k = q
\end{equation*}
$$
(k \geq 0, p_\ell \in  F_\Sigma[X \cup A F_\Omega(Y \cup \Xi)], \;
\ell = 1, \ldots, k, \; 0 \leq i < j \leq k)
$$
is a {\em derivation\/} of $q$ from $p$ in $\gA$, $k$ is the
{\em length\/}\index{length of!derivationinftransducer@derivation in F-transducer}
 of this derivation and
$p_i \Rightarrow \ldots \Rightarrow p_j$ is a
{\em subderivation\/}\index{subderivation in!F-transducer}
of (1). In this case we write $p \Rightarrow_{\gA}^* q$, or
$p \Rightarrow^* q$ if $\gA$ is understood, and say that $p$ {\em derives\/}
$q$ in $\gA$. Therefore, $\Rightarrow^*$ is the reflexive-transitive
closure of $\Rightarrow$. Obviously, when $p \Rightarrow^* q$, there
could be several (but finitely many) derivations of $q$ from $p$.
However, when we write $p \Rightarrow^* q$, we usually have in mind,
at least implicitly, a certain well-defined derivation of $q$ from $p$.
Consequently, we may say that $p \Rightarrow^* q$ is a derivation.

\sloppy Using the notation $\Rightarrow^*$ the transformation $\tau_{\gA}$
induced by an F-transducer $\gA = (\Sigma, X, A, \Omega, Y, P, A')$
can also be given thus:
$$
\tau_{\gA} = \{ (p, q) \mid p \in F_\Sigma(X), q \in F_\Omega(Y),
\; p \Rightarrow^* aq \; \mbox{ for some } a \in A' \}.
$$

As $\gA$ may have different productions with the same left side,
there could be more than one $q \in F_\Omega(Y)$ such that
$(p, q) \in \tau_{\gA}$ for a given $p \in F_\Sigma(X)$, i.e.,
$\gA$ is in general nondeterministic. However, at each step
of a transformation we have only finitely many choices.
Therefore, $p \tau_{\gA}$ is finite for every $p \in F_\Sigma(X)$.

A tree transformation is an
F-{\em transformation\/}\index{ftransformation@F-transformation}
if it can be induced by an F-transducer. The class of all
F-transformations will be denoted by $\cF$.

Take an arbitrary set $A$. The $i$th component of a vector
$\ba \in A^n$ will be denote by $a_i$; i.e.,
$\ba = (a_1, \ldots, a_n)$. If $a_1 = \ldots = a_n = a$ then
for $\ba$ we write $a^n$. If $\ba \in A^n$ and
$\bb \in B^m$ are arbitrary two vectors, then
${\bf (a, b)}$ will stand for
$(a_1, \ldots, a_n, b_1, \ldots, b_m)$. Assume that $k = \min (m, n)$.
Then $\ba \bb$ stands for $(a_1 b_1, \ldots, a_k b_k)$ or
$((a_1, b_1), \ldots, (a_k, b_k))$, depending on the context.

Consider a $p \in F_\Sigma(X \cup \Xi_n)$, and let
$\bp = (p_1, \ldots, p_n)$ be a vector of trees. Then we shall
write $p(\bp)$ for $p(p_1, \ldots, p_n)$. Moreover,
if $\bp \in F_\Sigma(X \cup \Xi_n)^m$ and
$\bq = (q_1, \ldots, q_n)$ is a vector of trees, then
$\bp (\bq)$ will stand for $(p_1(\bq), \ldots, p_m(\bq))$.

Consider the homomorphism $\varphi \colon (X \cup \Xi)^* \rightarrow \Xi^*$
given by $x \varphi = e$ ($x \in X$) and $\xi \varphi = \xi$
($\xi \in \Xi$)\label{varphiDefinition}. Set
$$
\hat{F}_\Sigma(X \cup \Xi_n) = \{ p \in F_\Sigma(X \cup \Xi_n) \mid
\yd(p) \varphi \mbox{ is a permutation of } \xi_1, \ldots, \xi_n \}
$$
and
$$
\hat{\hat{F}}_\Sigma(X \cup \Xi_n) = \{ p \in F_\Sigma(X \cup \Xi_n)
\mid \yd(p) \varphi = \xi_1 \ldots \xi_n \}.
$$
Moreover, if $m > 0$ then let
\begin{eqnarray*}
\hat{F}_\Sigma^m(X \cup \Xi_n) & = & \{ p \in F_\Sigma(X \cup \Xi_n)^m
\mid \yd(p_1) \varphi \ldots \yd(p_m) \varphi \mbox{ is a } \\
& & \mbox{ permutation of } \; \xi_1, \ldots, \xi_n \}.
\end{eqnarray*}

Now let $\gA = (\Sigma, X, A, \Omega, Y, P, A')$ be an F-transducer,
and consider a derivation
$$
\alpha \colon p \Rightarrow^* q \;\; (p, q \in F_\Sigma[X \cup A F_\Omega(Y)]).
$$
Let
\begin{equation*}
\tag{2}
r(p_1, p_2) \Rightarrow r(p_{1_1}, p_2) \Rightarrow \ldots \Rightarrow
r(p_{1_k}, p_2) \Rightarrow r(p_{1_k}, p_2')
\end{equation*}
$$
(r \in \hat{F}_\Sigma[X \cup A F_\Omega(Y) \cup \Xi_2])
$$
be a subderivation of $\alpha$, where the first $k$ direct derivation
steps apply to the subtree $p_1$, and then the $(k+1)$th step concerns
the subtree $p_2$. Replacing the subderivation (2) in $\alpha$ by
\begin{equation*}
\tag{3}
r(p_1, p_2) \Rightarrow r(p_1, p_2') \Rightarrow r(p_{1_1}, p_2')
\Rightarrow \ldots \Rightarrow
r(p_{1_k}, p_2')
\end{equation*}
we obviously get a new derivation
$$
\beta \colon p \Rightarrow^* q.
$$

The replacement of (2) in $\alpha$ by (3) is called an
{\em inversion\/}\index{inversion of direct derivations in!F-transducer}
of direct derivations. Finitely many inversions of direct derivations
is a
{\em reordering\/}\index{reordering of direct derivations
in!F-transducer} of direct derivations.

In the sequel we do not distinguish between derivations obtained from
each other by reorderings of direct derivations.

Again, consider the above F-transducer $\gA$ and a tree
$p \in F_\Sigma(X)$. Then by
$$
p = \overline{p}(p_1, \ldots, p_m) \Rightarrow^*
\overline{p}(a_1 q_1, \ldots, a_m q_m) \Rightarrow^* aq (q_1, \ldots, q_m)
$$
$$
(\overline{p} \in \hat{F}_\Sigma(X \cup \Xi_m), \;
p_i \Rightarrow^* a_i q_i, \; i = 1, \ldots, m, \;
\overline{p}(a_1 \xi_1, \ldots, a_m \xi_m) \Rightarrow^* aq)
$$
we mean the derivation\index{derivation in!F-transducer}
$$
\overline{p}(p_1, \ldots, p_m) \Rightarrow
\overline{p}(p_{1_1}, \ldots, p_m) \Rightarrow \ldots \Rightarrow
\overline{p}(p_{1_{k_1}}, \ldots, p_m) \Rightarrow \ldots
$$
$$
\overline{p}(p_{1_{k_1}}, \ldots, p_{m_1}) \Rightarrow \ldots \Rightarrow
\overline{p}(p_{1_{k_1}}, \ldots, p_{m_{k_m}}) =
$$
$$
\overline{p}(a_1 q_1, \ldots, a_m q_m) \Rightarrow^* aq (q_1, \ldots, q_m)
$$
if $p_i \Rightarrow^* a_i q_i$ is the derivation
$ p_i \Rightarrow p_{i_1} \Rightarrow \ldots \Rightarrow p_{i_{k_i}} = a_i q_i$
($a_i \in A$, $q_i \in F_\Omega(Y)$, $i = 1, \ldots, m$), and
$\overline{p}(a_1 q_1, \ldots, a_m q_m) \Rightarrow^* aq(q_1, \ldots, q_m)$
is obtained by replacing $\xi_i$ in
$\overline{p}(a_1 \xi_1, \ldots, a_m \xi_m) \Rightarrow^* aq$
by $q_i$ ($i = 1, \ldots, m$).

If we say that we write the derivation
$$
\alpha \colon p \Rightarrow^* aq \;\; (a \in A, p \in F_\Sigma(X),
q \in F_\Omega(Y))
$$
in the (more detailed) form
$$
\beta \colon p = \overline{p}(p_1, \ldots, p_m) \Rightarrow^*
\overline{p}(a_1 q_1, \ldots, a_m q_m) \Rightarrow^* a \overline{q}
(q_1, \ldots, q_m)
$$
$$
 (\overline{p} \in \hat{F}_\Sigma(X \cup \Xi_m), \;
p_i \Rightarrow^* a_i q_i, \; i = 1, \ldots, m, \;
\overline{p}(a_1 \xi_1, \ldots, a_m \xi_m) \Rightarrow^* a\overline{q})
$$
this also generally means that $\beta$ is a reordering of $\alpha$.
Of course, such a reordering always exists.

In the special case $\overline{p} = \sigma(\xi_1, \ldots, \xi_m)$
($\sigma \in \Sigma_m$) we write $\beta$ in the form
$$
\beta \colon \sigma(p_1, \ldots, p_m) \Rightarrow^* \sigma(a_1 q_1, \ldots,
a_m q_m) \Rightarrow a \overline{q}(q_1, \ldots, q_m)
$$
$$
(p_i \Rightarrow^* a_i q_i, \; i = 1, \ldots, m, \;
(\sigma(a_1, \ldots, a_m), a\overline{q}) \in P).
$$

We illustrate the concepts of F-transducers and F-transformations
by

\begin{ex}
\label{Example.4.1.3}
{\rm
Let $\gA = (\Sigma, \{ x \}, \{ a_0, a_1 \}, \Omega, \{ y \}, P,
\{ a_0 \})$, where $\Sigma = \Sigma_2 = \{ \sigma \}$,
$\Omega = \Omega_1 = \{ \omega \}$ and $P$ consists of the productions
$x \rightarrow a_1 y$ and $\sigma(a_1, a_1) \rightarrow a_0 \omega(\xi_1)$.

Consider the tree $\sigma(x, x)$. One of the possible derivations
$$
\sigma(x, x) \Rightarrow \sigma(a_1 y, x) \Rightarrow
\sigma(a_1 y, a_1 y) \Rightarrow a_0 \omega(y)
$$
is illustrated by Fig.~\ref{Figure.4.1}.
\begin{figure}[h]
\input{Figure.4.1}
\caption{\label{Figure.4.1}}
\end{figure}

Thus $(\sigma(x, x), \omega(y))$ is in $\tau_{\gA}$. In fact,
$\tau_{\gA}$ consists of this single pair
 $(\sigma(x, x), \omega(y))$. Indeed, the only $\Sigma X$-tree of
height 0 is $x$, which obviously is not in $\dom(\tau_{\gA})$. If
$p \in F_\Sigma(X)$ is a tree with height greater than 1, then it should
contain at least one of the following trees as a subtree:
$$
\sigma(\sigma(x, x), \sigma(x, x)), \;\;\;
\sigma(\sigma(x, x), x) \;\; \mbox{ and } \;\;
\sigma(x, \sigma(x, x)).
$$
One can easily see that none of these subtrees can be transformed
by $\gA$. \epr
} 
\end{ex}

F-transducers transform a tree from the leaves of the tree towards the
root of the tree. Now we define a system which works in the
opposite direction.

\begin{df}
\label{Definition.4.1.4}
{\rm
A {\em root-to-frontier tree
transducer\/}\index{root-to-frontier tree transducer}
\index{tree transducer!root-to-frontier}
 (R-{\em transducer\/})\index{rtransducer@R-transducer}
is a system
$\gA = (\Sigma, X, A, \Omega, Y, P, A')$, where
\begin{enumerate}
\item[{\rm (1)}] $\Sigma$, $X$, $A$, $\Omega$, $Y$ and $A'$ are
specified the same way as in Definition~\ref{Definition.4.1.1}, but
here $A'$ is called the set of
{\em initial states,}\index{initial state of!R-transducer}
\item[{\rm (2)}] $P$ is a finite set of
{\em productions\/}\index{production of!Rt-ransducer@R-transducer}
(or {\em rewriting rules\/})\index{rewriting rule of!Rtransducer@R-transducer}
 of the following two types:
\begin{description}
\item[{\rm (i)}] $ax \rightarrow q$ ($a \in  A$, $x \in X$,
$q \in F_\Omega(Y)$),
\item[{\rm (ii)}] $a \sigma (\xi_1, \ldots, \xi_m) \rightarrow q$
($a \in A$, $\sigma \in \Sigma_m$, $m \geq 0$,
$q \in F_\Omega[Y \cup A \Xi_m]$).
\end{description}
\end{enumerate}
} 
\end{df}

In the sequel we shall write simply
$a \sigma$ for $a \sigma(\xi_1, \ldots, \xi_m)$.
Moreover, for a production $p \rightarrow q$ we shall use the
notation $(p, q)$, too.

Obviously, a production of type (ii) in Definition~\ref{Definition.4.1.4}
can be written in the form
$$
a \sigma \rightarrow q(\ba_1 \xi_1^{n_1}, \ldots, \ba_m \xi_m^{n_m})
$$
where $\ba_i \in A^{n_i}$, $n_i \geq 0$, $i = 1, \ldots, m$,
$n_1 + \ldots + n_m = n$, and
$q \in \hat{F}_\Omega(X \cup \Xi_n)$. In the sequel we shall assume that
whenever $1 \leq i \leq m$ and
$n_1 + \ldots + n_{i-1} + 1 \leq i_i < i_2 \leq n_1 + \ldots + n_i$,
$\xi_{i_1}$ precedes $\xi_{i_2}$ in $\yd(q) \varphi$.
Here $\varphi$ is the homorphism defined on p.~\pageref{varphiDefinition}.

Next we define the transformations induced by R-transducers.
Let $\gA$ be the R-transducer of Definition~\ref{Definition.4.1.4}.
For any $a \in A$ and $p \in F_\Sigma(X)$ we define the subsets
$p \tau_{\gA, a}$ as follows:
\begin{enumerate}
\item[{\rm (i)}] if $p \in \Sigma_0 \cup X$ and
$(ap, q) \in P$ then $q \in p \tau_{\gA, a}$,
\item[{\rm (ii)}] if $p = \sigma(p_1, \ldots, p_m)$
($\sigma \in \Sigma_m$, $m > 0$), then for any
$(a \sigma, q(\ba_1 \xi_1^{n_1}, \ldots, \ba_m \xi_m^{n_m})) \in P$
and $q_{i_j} \in p_i \tau_{\gA, a_{i_j}}$ ($1 \leq i \leq m$,
$1 \leq j \leq n_i$),
$q (\bq_1, \ldots, \bq_m) \in p \tau_{\gA, a}$ where
$\bq_i = (q_{i_1}, \ldots, q_{i_{n_i}})$ ($i = 1, \ldots, m$),
\item[{\rm (iii)}] nothing is in any $p \tau_{\gA, a}$ unless this
follows from (i) and (ii).
\end{enumerate}

\begin{df}
\label{Definition.4.1.5}
{\rm
Let $\gA = (\Sigma, X, A, \Omega, Y, P, A')$ be an R-transducer.
Then the
{\em transformation\index{transformation induced by!rtransducer@R-transducer}
induced
by\/}
$\gA$ is the relation
$$
\tau_{\gA} = \{ (p, q) \mid p \in F_\Sigma(X), q \in F_\Omega(Y),
\; q \in p \tau_{\gA, a} \mbox{ for some } a \in A' \}.
$$

A tree transformation is an
R-{\em transformation\/}\index{rtransformtion@R-transformation} if it can be
induced by an R-transducer. The class of all R-transformations
will be denoted by $\cR$.
} 
\end{df}

For R-transformations we also give another definition which shows
how a transformation is carried out step by step.

Let $p, q \in F_\Omega[Y \cup AF_\Sigma(X \cup \Xi)]$ be trees,
and consider the R-transducer of Definition~\ref{Definition.4.1.4}.
It is said that {\em $p$ directly derives
$q$\/}\index{direct derivation in!Rtransducer@R-transducer} in $\gA$ if $q$
can be obtained from $p$ by
\begin{enumerate}
\item[{\rm (i)}] replacing an occurrence of a subtree $a x$
($a \in A$, $x \in X$) in $p$ by the right side
$\overline{q}$ of a production $a x \rightarrow \overline{q}$
in $P$, or by
\item[{\rm (ii)}] replacing an occurrence of a subtree
$a \sigma(p_1, \ldots, p_m)$ ($a \in A$, $\sigma \in \Sigma_m$,
$m \geq 0$, $p_1, \ldots, p_m \in F_\Sigma(X \cup \Xi)$) in $p$
by $\overline{q}(p_1, \ldots, p_m)$ where $a \sigma \rightarrow
\overline{q}$ is in $P$.
\end{enumerate}

Each application of steps (i) and (ii) is called a {\em direct
derivation\/} in $\gA$. The relation expressing the direct derivation
will be denoted by $\Rightarrow_{\gA}$, i.e., we write
$p \Rightarrow_{\gA} q$ if $q$ is obtained from $p$ by a direct
derivation in $\gA$. Frequently, $\gA$ will be omitted in
$\Rightarrow_{\gA}$. Any finite sequence of consecutive direct
derivations defines a derivation\index{derivation in!Rtransducer@R-transducer}.
More precisely,
\begin{equation*}
\tag{4}
p = p_0 \Rightarrow p_1 \Rightarrow \ldots \Rightarrow p_i \Rightarrow
\ldots \Rightarrow p_j \Rightarrow \ldots \Rightarrow p_k = q
\end{equation*}
$$
(k \geq 0, \; p_\ell \in F_\Omega[Y \cup A F_\Sigma(X \cup \Xi)], \;
\ell = 0, \ldots, k, \; 0 \leq i < j \leq k)
$$
is a {\em derivation\/} of $q$ from $p$ in $\gA$, $k$ is the
{\em length\/}\index{length of!derivationinrtransducer@derivation in R-transducer}
 of this derivation and
$p_i \Rightarrow \ldots \Rightarrow p_j$ is a
{\em subderivation\/}\index{subderivation in!R-transducer}
of (4). If $q$ can be obtained from $p$ by a derivation, then we
write $p \Rightarrow_{\gA}^* q$, or simply $p \Rightarrow^* q$ if
$\gA$ is understood from the context. Thus, $\Rightarrow^*$
is the reflexive-transitive closure of $\Rightarrow$. Similarly
as in the case of an F-transducer, we suppose that the notation
$p \Rightarrow^* q$ implies a certain derivation of $q$ from $p$
in $\gA$.

Using the notation $\Rightarrow^*$, the transformation $\tau_{\gA}$ induced
by an R-transducer
$\gA = (\Sigma, X, A, \Omega, Y, P, A')$ can equivalently
be defined thus:
$$
\tau_{\gA} = \{ (p, q) \mid p \in F_\Sigma(X), q \in F_\Omega(Y), \;
ap \Rightarrow^* q \; \mbox{ for some } a \in A' \}.
$$

Let us note that although an R-transducer $\gA$ is generally
a nondeterministic system, $p \tau_{\gA}$ is finite for every input
tree $p$ of $\gA$.

Let $\gA = (\Sigma, X, A, \Omega, Y, P, A')$ be an R-transducer.
Consider some $n > 0$,
$\ba \in A^n$, $\bp \in F_\Sigma(X)^n$, $\bq \in F_\Omega(Y)^n$
and derivations $a_i p_i \Rightarrow^* q_i$ ($i = 1, \ldots, n$). Then
$\ba \bp \Rightarrow^* \bq$ will denote the vector of these
derivations. Moreover, we assume that
$\ba \bp \Rightarrow^* \bq$ implicitly expresses the $n$ derivations
$a_i p_i \Rightarrow^* q_i$ ($i = 1, \ldots, n$).

Take the above R-transducer $\gA$ and a derivation
$$
\alpha \colon p \Rightarrow^* q \;\; (p, q \in F_\Omega[Y \cup A F_\Sigma(X)]).
$$
Let
\begin{equation*}
\tag{5}
r(p_1, p_2) \Rightarrow r(p_{1_1}, p_2) \Rightarrow \ldots \Rightarrow
r(p_{1_k}, p_2) \Rightarrow r(p_{1_k}, p_2')
\end{equation*}
$$
(r \in \hat{F}_\Omega[Y \cup A F_\Sigma(X \cup \Xi_2)])
$$
be a subderivation of $\alpha$, where the first $k$ direct
derivation steps are carried out in the subtree $p_1$, and then in
the $(k+1)$th step we apply a production in the subtree $p_2$.
Replacing the subderivation (5) in $\alpha$ by
\begin{equation*}
\tag{6}
r(p_1, p_2) \Rightarrow r(p_1, p_2') \Rightarrow r(p_{1_1}, p_2')
\Rightarrow \ldots \Rightarrow
r(p_{1_k}, p_2')
\end{equation*}
we get a derivation
$$
\beta \colon p \Rightarrow^* q.
$$

The replacement  of (5) in $\alpha$ by (6) is called an
{\em inversion\/}\index{inversion of direct derivations in!R-transducer}
of direct derivations. By finitely many applications of inversions
we get a
{\em reordering\/}\index{reordering of direct derivations
in!R-transducer} of direct derivations. We shall not
distinguish between derivations in an R-transducer if they are
reorderings of each other.

Again, take the above R-transducer $\gA$, a state $a \in A$ and
a tree $p \in F_\Sigma(X)$. Then by
$$
a p = a \overline{p} (p_1, \ldots, p_m) \Rightarrow^*
q (\ba_1 p_1^{n_1}, \ldots, \ba_m p_m^{n_m}) \Rightarrow^*
q ( \bq_1, \ldots, \bq_m)
$$
$$
(\overline{p} \in \hat{F}_\Sigma(X \cup \Xi_m), \;
a \overline{p} \Rightarrow^* q(\ba_1 \xi_1^{n_1}, \ldots, \ba_m \xi_m^{n_m}),
\; \ba_i \in A^{n_i},
$$
$$
n_i \geq 0, \; i = 1, \ldots, m, \; n_1 + \ldots + n_m = n,
\; q \in \hat{F}_\Omega(Y \cup \Xi_n),
$$
$$
\ba_j p_j^{n_j} \Rightarrow \bq_j, \; j = 1, \ldots, m)
$$
we mean the derivation
$$
a \overline{p}(p_1, \ldots, p_m) \Rightarrow^*
q(\ba_1 p_1^{n_1}, \ldots, \ba_m p_m^{n_m}) \Rightarrow
$$
$$
q(p_{1_1(1)}, a_{1_2} p_1, \ldots, a_{1_{n_1}} p_1, \ldots,
a_{m_1} p_m, \ldots, a_{m_{n_m}} p_m) \Rightarrow \ldots
$$
$$
q(p_{1_1(k_1)}, a_{1_2} p_1, \ldots, a_{1_{n_1}} p_1, \ldots,
a_{m_1} p_m, \ldots, a_{m_{n_m}} p_m) \Rightarrow \ldots
$$
$$
q(p_{1_1(k_1)},  \ldots,  p_{1_{n_1}(k_{1_{n_1}})}, \ldots,
a_{m_1} p_m, \ldots, a_{m_{n_m}} p_m) \Rightarrow \ldots
$$
$$
q(p_{1_1(k_1)},  \ldots,  p_{{1_{n_1}(k_{1_{n_1}})}}, \ldots,
p_{{m_1}(k_{m_1})},  \ldots,  a_{m_{n_m}} p_m) \Rightarrow \ldots
$$
$$
q(p_{1_1(k_1)},  \ldots,  p_{{1_{n_1}(k_{1_{n_1}})}}, \ldots,
p_{{m_1}(k_{m_1})},  \ldots,  p_{m_{n_m}(k_{m_{n_m}})} =
$$
$$
q (q_{1_1}, \ldots, q_{1_{n_1}}, \ldots,
q_{m_1}, \ldots, q_{m_{n_m}}), \;\; \mbox{ assuming }
$$
that $\ba_i p_i^{n_i} \Rightarrow^* \bq_i$ ($1 \leq i \leq m$) has
its component derivations
$$
a_{i_j} p_i \Rightarrow p_{{i_j}(1)} \Rightarrow \ldots \Rightarrow
 p_{{i_j}(k_{i_j})} = q_{i_j} \;\;
(q_{i_j} \in F_\Omega(Y), \; j = 1, \ldots, n_i),
$$
and $a \overline{p}(p_1, \ldots, p_m) \Rightarrow^*
q(\ba_1 p_1^{n_1}, \ldots, \ba_m p_m^{n_m})$ is obtained
by replacing $\xi_i$ ($i = 1, \ldots, m$)
in $a \overline{p} \Rightarrow^*
q(\ba_1 \xi_1^{n_1}, \ldots, \ba_m \xi_m^{n_m})$ by $p_i$.

When we say that we write the derivation
$$
\alpha \colon ap \Rightarrow^* q \;\;\; (a \in A,
p \in F_\Sigma(X), q \in F_\Omega(Y))
$$
in the more detailed form
$$
\beta \colon  ap = a \overline{p}(p_1, \ldots, p_m) \Rightarrow^*
\overline{q}( \ba_1 p_1^{n_1}, \ldots, \ba_m p_m^{n_m}) \Rightarrow^*
\overline{q} (\bq_1, \ldots, \bq_m)
$$
$$
(\overline{p} \in \hat{F}_\Sigma(X \cup \Xi_m), \;
a \overline{p} \Rightarrow^* \overline{q} (\ba_1 \xi_1^{n_1}, \ldots,
\ba_m \xi_m^{n_m}), \; \ba_i \in A^{n_i}, \; n_i \geq 0,
$$
$$
i = 1, \ldots, m, \; n_1 + \ldots + n_m = n, \;
\overline{q} \in \hat{F}_\Omega(Y \cup \Xi_n), \;
\ba_j p_j^{n_j} \Rightarrow \bq_j, \; j = 1, \ldots, m),
$$
it generally also means that $\beta$ is a reordering of $\alpha$.
Obviously, such a reordering always exists.

In case $\overline{p} = \sigma(\xi_1, \ldots, \xi_m)$
($\sigma \in \Sigma_m$), we write $\beta$ in the form
$$
\beta \colon a \sigma(p_1, \ldots, p_m) \Rightarrow^*
\overline{q}( \ba_1 p_1^{n_1}, \ldots, \ba_m p_m^{n_m})
\Rightarrow^* \overline{q}(\bq_1, \ldots, \bq_m)
$$
$$
(( a \sigma, \overline{q}(\ba_1 \xi_1^{n_1}, \ldots, \ba_m \xi_m^{n_m}))
\in P, \; \ba_i \in A^{n_i}, \; n_i \geq 0, \; i = 1, \ldots, m, \;
n_1 + \ldots + n_m = n,
$$
$$
\overline{q} \in \hat{F}_\Omega(Y \cup \Xi_n), \;
\ba_j p_j^{n_j} \Rightarrow^* \bq_j, \; j = 1, \ldots, m).
$$

\begin{ex}
\label{Example.4.1.6}
{\rm
Let $\gA = (\Sigma, \{ x \}, \{ a_0, a_1, a_2 \}, \Omega, \{ y_1, y_2 \},
P, a_0)$ be the R-transducer, where $\Sigma = \Sigma_1 = \{ \sigma \}$,
$\Omega = \Omega_1 \cup \Omega_2$, $\Omega_1 = \{ \omega_1 \}$ and
$\Omega_2 = \{ \omega_2 \}$ and $P$ consists of the productions
$$
a_0 \sigma \rightarrow \omega_2( a_1 \xi_1, a_2 \xi_1),
$$
$$
a_1 \sigma \rightarrow \omega_1 (a_1 \xi_1), \;
a_2 \sigma \rightarrow \omega_1 (a_2 \xi_1),
$$
$$
a_1 x \rightarrow y_1, \; a_2 x \rightarrow y_2 .
$$

Consider the trees $p = \sigma(\sigma(\sigma(x)))$ and
$q = \omega_2(\omega_1(\omega_1(y_1)), \omega_1(\omega_1(y_2)))$.
Then a derivation of $q$ from $a_0 p$ is illustrated
in Fig.~\ref{Figure.4.2}.
\begin{figure}[h]
\input{Figure.4.2}
\caption{\label{Figure.4.2}}
\end{figure}

By induction on the heights of input trees one can easily
prove that
$$
\tau_{\gA} = \{ (\sigma^n(x), \omega_2(\omega_1^{n-1}(y_1),
\omega_1^{n-1}(y_2))) \mid n = 1, 2, \ldots \},
$$
where $\sigma^0(\xi) = \xi$ and $\sigma^n(\xi) =
\sigma(\sigma^{n-1}(\xi))$ if $n > 0$. \epr
} 
\end{ex}

Both F-transducers and R-transducers generalize generalized sequential
machines from strings to trees (or from unary polynomial symbols
to polynomial symbols of arbitrary finite type if strings are
interpreted as unary polynomial symbols, as we did in
Section~\ref{Section.2.2}). At the same time there are the
following main differences between F-transducers and R-transducers:
\begin{enumerate}
\item[{\rm (1)}] An F-transducer first processes an input subtree
nondeterministically and then makes copies of the resulting
output subtree.
\item[{\rm (2)}] An R-transducer can first make copies of an input subtree
and then process each copy independently in a nondeterministic fashion.
\item[{\rm (3)}] F-transducers should process even those subtrees
which are deleted afterwards.
\end{enumerate}

Before ending this section we state and prove some simple
general results.

The concept of tree homomorphism was introduced in
Section~\ref{Section.2.4}. It is easy to see that the tree
homomorphism $h \colon F_\Sigma(X) \rightarrow F_\Omega(Y)$,
given by the mappings
$$
h_m \colon \Sigma_m \rightarrow F_\Omega(Y \cup \Xi_m) \;\; (m \geq 0)
$$
and
$$
h_X \colon X \rightarrow F_\Omega(Y),
$$
can be induced by the one-state F-transducer
$\gA = (\Sigma, X, \{ a \}, \Omega, Y, P, a)$ where
$$
P = \{ x \rightarrow a h_X (x) \mid x \in X \} \cup
\{ \sigma(a, \ldots, a) \rightarrow
a h_m(\sigma) \mid \sigma \in \Sigma_m, m \geq 0 \}.
$$

\begin{df}
\label{Definition.4.1.7}
{\rm
A one-state F-transducer $\gA = (\Sigma, X, \{ a \}, \Omega, Y, P, a)$
is an HF-{\em transducer\/}\index{HF-transducer} if for every $x \in X$, resp.
$\sigma \in \Sigma$, in $P$ there is exactly one production
with left side $x$, resp. $\sigma(a, \ldots, a)$.
} 
\end{df}

We have seen that every tree homomorphism can be induced by an
HF-transducer. The converse is also true: transformations induced
by HF-transducers are tree homomorphisms.

We now introduce the R-transducer counterpart of
HF-transducers.

\begin{df}
\label{Definition.4.1.8}
{\rm
A one-state R-transducer
 $\gA = (\Sigma, X, \{ a \}, \Omega, Y, P, a)$ is an
HR-{\em transducer\/}\index{HR-transducer}
if for each $d \in X \cup \Sigma$ in $P$ there is exactly one
production with the left side $ad$.
} 
\end{df}

Next we prove that the class of all tree homomorphisms coincides with
the class of all transformations induced by HR-transducers.

\begin{thm}
\label{Theorem.4.1.9}
The class of transformations induced by
{\rm HF}-transducers coincides with the class of all transformations
induced by {\rm HR}-transducers.
\end{thm}

\pr
Let $\gA = (\Sigma, X, \{ a \}, \Omega, Y, P, a)$
be an HF-transducer. Consider the R-transducer
$\gB = (\Sigma, X, \{ a \}, \Omega, Y, P', a)$, where
$P'$ is given in the following way:
$$
(ax, q) \in P' \;\; \Longleftrightarrow \;\; (x, aq) \in P \;\; (x \in X)
$$
and
$$
(a \sigma, q(a \xi_1, \ldots, a \xi_m))
 \in P'
\;\; \Longleftrightarrow \;\;  (\sigma(a, \ldots, a), a q) \in P \;
(\sigma \in \Sigma_m, m \geq 0,
q \in F_\Omega(Y \cup \Xi_m)).
$$
It is obvious that $\gB$ is an HR-transducer.

By induction on $\hg(p)$, we show that for an arbitrary
$p \in F_\Sigma(X)$ and $q \in F_\Omega(Y)$ the equivalence
\begin{equation*}
\tag{7}
ap \Rightarrow^*_{\gB} q \;\; \Longleftrightarrow \;\; p \Rightarrow^*_{\gA} aq
\end{equation*}
holds. This obviously implies $\tau_{\gA} = \tau_{\gB}$.

If $\hg(p) = 0$, then (7) holds by the definition of $P'$.

Let $p = \sigma(p_1, \ldots, p_m)$ ($\sigma \in \Sigma_m$, $m > 0 $),
and assume that (7) has been proved for all trees in
$F_\Sigma(X)$ with heights less than $\hg(p)$.

\sloppy Suppose that the left side of (7) holds, i.e., we have
$ap = a \sigma(p_1, \ldots, p_m)$ $\Rightarrow_{\gB}$
$\overline{q}(a p_1, \ldots, a p_m)$ $\Rightarrow_{\gB}^*$
$\overline{q}(q_1, \ldots, q_m) = q$, where
$(a \sigma, \overline{q}( a \xi_1, \ldots, a \xi_m)) \in P'$ and
$a p_i \Rightarrow_{\gB}^* q_i$ ($i = 1, \ldots, m$). Then,
by the definition of $P'$, the production
$\sigma(a, \ldots, a) \rightarrow a \overline{q}( \xi_1, \ldots, \xi_m)$
is in $P$. Moreover, by the induction hypothesis,
$p_i \Rightarrow_{\gA}^* a q_i$ is valid for each $i$
($1 \leq i \leq m$). Therefore, we have a desired derivation
$$
p = \sigma(p_1, \ldots, p_m) \Rightarrow_{\gA}^*
\sigma(a q_1, \ldots, a q_m) \Rightarrow_{\gA}
a \overline{q}(q_1, \ldots, q_m) = a q.
$$

The fact that $p \Rightarrow_{\gA}^* aq$ implies
$ap \Rightarrow_{\gB}^* q$ can be shown by reversing the above
argument.

To see that every HR-transformation is induced by an HF-transducer,
it suffices to observe that every HR-transducer $\gB$ arises from
an HF-transducer $\gA$ by the above construction. Hence HR- and
HF-transducers appear in equivalent ``associated'' pairs.
\mbox{\ \ }\epr

\

We prove two more results.

\begin{thm}
\label{Theorem.4.1.10}
The following statements hold.
\begin{enumerate}
\item[{\rm (i)}] For every {\rm F-}transformation
$\tau \subseteq F_\Sigma(X) \times F_\Omega(Y)$,
$\dom(\tau) \in \Rec(\Sigma, X)$.
\item[{\rm (ii)}] There exists a tree homomorphism
$h \colon F_\Sigma(X) \rightarrow F_\Omega(Y)$ such that
${\rm range}(h) \not\in \Rec(\Omega, Y)$.
\end{enumerate}
\end{thm}

\pr
In order to show (i) consider an F-transducer
$\gA = ( \Sigma, X, A, \Omega, Y, P, A')$.
Construct an NDF $\Sigma X$-recognizer $\bB =
(\cB, \beta, B')$, where $\cB = (A, \Sigma)$,
$B' = A'$, and, for all $m \geq 0$, $\sigma \in \Sigma_m$ and
$a_1 \ldots, a_m \in A$,
$$
\sigma^{\cB}(a_1, \ldots, a_m) =
\{ a \mid (\exists q \in F_\Omega(Y \cup \Xi_m))
((\sigma(a_1, \ldots, a_m), aq) \in P) \}.
$$
Finally let
$$
x \beta = \{ a \in A \mid (\exists q \in F_\Omega(Y))
((x, aq) \in P) \} \;\; (x \in X).
$$
We end the proof of (i) by the observation that for all $a \in A$ and
$p \in F_\Sigma(X)$ the equivalence
$$
a \in p \hat{\beta} \;\; \Longleftrightarrow \;\;
(\exists q \in F_\Omega(Y)) (p \Rightarrow^* aq)
$$
holds. This can be shown by induction on $\hg(p)$.

For a proof of (ii), see Example~\ref{Example.2.4.15}.
\epr

\

 Example~\ref{Example.2.4.15} shows also that the translation of a
context-free language by a tree transducer is not always context-free.
In fact, in this example the finite language $\{ x \}$ is translated
into the non-$\CF$ language
$\{ x^{2^n} \mid n \geq 0 \}$.

\begin{lm}
\label{Lemma.4.1.11}
For each $T \in \Rec(\Sigma, X)$ there exists an {\rm F}-transducer
$\gA$ such that $\dom(\tau_{\gA}) = {\rm range}(\tau_{\gA}) = T$
and $\tau_{\gA}$ is the identity mapping of $T$.
\end{lm}

\pr
Let $\bB = (\cB, \beta, B')$ be an F $\Sigma X$-recognizer
with $\cB = (B, \Sigma)$ and $T(\bB) = T$. Take the F-transducer
$\gA = (\Sigma, X, B, \Sigma, X, P, B')$ where
\begin{eqnarray*}
P & = & \{  x \rightarrow \beta(x) x \mid x \in X \} \cup
\{ \sigma(b_1, \ldots, b_m) \rightarrow b \sigma(\xi_1, \ldots, \xi_m) \mid \\
& & m \geq 0, \; \sigma \in \Sigma_m, \;\; b, b_1, \ldots, b_m \in B, \;
\sigma^{\cB}(b_1, \ldots, b_m) = b \}.
\end{eqnarray*}
Obviously, $\gA$ has the desired properties.
\epr

\

We end off this Section with

\begin{df}
\label{Definition.4.1.12}
{\rm
Two R- or F-transducers $\gA$ and $\gB$ are
{\em equivalent\/}\index{equivalence of!R- and F-transducers}
if $\tau_{\gA} = \tau_{\gB}$ holds.
}
\end{df}


%% file: Figure.4.1.tex
\begin{align*} 
	\raisebox{-0.5\height}{
		\begin{tikzpicture}[level distance=1cm]
		\begin{scope}[execute at begin node=$, execute at end node=$]
		\node[node,label={left:\sigma}] {} 
		child{ node[node,label={left:x}] {}}
		child{ node[node,label={left:x}] {}};
		\end{scope}
		\end{tikzpicture}}&&
	\Rightarrow &&
	\raisebox{-0.5\height}{
		\begin{tikzpicture}[level distance=1cm]
		\begin{scope}[execute at begin node=$, execute at end node=$]
		\node[node,label={left:\sigma}] {} 
		child{ node[node,label={left:x}] {}}
		child{ node[node,label={left:a_1}] {}
			child{ node[node,label={left:y}] {}}};
		\end{scope}
		\end{tikzpicture}}&&
	\Rightarrow &&
	\raisebox{-0.5\height}{
		\begin{tikzpicture}[level distance=1cm]
		\begin{scope}[execute at begin node=$, execute at end node=$]
		\node[node,label={left:\sigma}] {} 
		child{ node[node,label={left:a_1}] {}
			child{ node[node,label={left:y}] {}}}
		child{ node[node,label={left:a_1}] {}
			child{ node[node,label={left:y}] {}}};
		\end{scope}
		\end{tikzpicture}}&&
	\Rightarrow &&
	\raisebox{-0.5\height}{
		\begin{tikzpicture}[level distance=1cm]
		\begin{scope}[execute at begin node=$, execute at end node=$]
		\node[node,label={left:a_0}] {}
		child{ node[node,label={left:\omega}] {}
			child{ node[node,label={left:y}] {}}};
		\end{scope}
		\end{tikzpicture}}
\end{align*}

%% file: Figure.4.2.tex
\tikzset{level distance=0.8cm,sibling distance=1.25cm}
\begin{align*} 
\raisebox{-0.5\height}{
	\begin{tikzpicture}
	\begin{scope}[execute at begin node=$, execute at end node=$]
	\node[node,label={left:a_0}] {} 
	child{ node[node,label={left:\sigma}] {}
		child{ node[node,label={left:\sigma}] {}
			child{ node[node,label={left:\sigma}] {}
				child{ node[node,label={left:x}] {}}}}};
	\end{scope}
	\end{tikzpicture}}&&
\Rightarrow &&
\raisebox{-0.5\height}{
	\begin{tikzpicture}
	\begin{scope}[execute at begin node=$, execute at end node=$]
	\node[node,label={left:\omega_2}] {} 
	child{ node[node,label={left:a_2}] {}
		child{ node[node,label={left:\sigma}] {}
			child{ node[node,label={left:\sigma}] {}
				child{ node[node,label={left:x}] {}}}}}
	child{ node[node,label={left:a_1}] {}
		child{ node[node,label={left:\sigma}] {}
			child{ node[node,label={left:\sigma}] {}
				child{ node[node,label={left:x}] {}}}}};
	\end{scope}
	\end{tikzpicture}}&&
\Rightarrow &&
\raisebox{-0.5\height}{
	\begin{tikzpicture}
	\begin{scope}[execute at begin node=$, execute at end node=$]
	\node[node,label={left:\omega_2}] {} 
	child{ node[node,label={left:a_2}] {}
		child{ node[node,label={left:\sigma}] {}
			child{ node[node,label={left:\sigma}] {}
				child{ node[node,label={left:x}] {}}}}}
	child{ node[node,label={left:\omega_1}] {}
		child{ node[node,label={left:a_1}] {}
			child{ node[node,label={left:\sigma}] {}
				child{ node[node,label={left:x}] {}}}}};
	\end{scope}
	\end{tikzpicture}}&&
\Rightarrow &&
\raisebox{-0.5\height}{
	\begin{tikzpicture}
	\begin{scope}[execute at begin node=$, execute at end node=$]
	\node[node,label={left:\omega_2}] {} 
	child{ node[node,label={left:a_2}] {}
		child{ node[node,label={left:\sigma}] {}
			child{ node[node,label={left:\sigma}] {}
				child{ node[node,label={left:x}] {}}}}}
	child{ node[node,label={left:\omega_1}] {}
		child{ node[node,label={left:\omega_1}] {}
			child{ node[node,label={left:a_1}] {}
				child{ node[node,label={left:x}] {}}}}};
	\end{scope}
	\end{tikzpicture}} && 
\Rightarrow &
\end{align*}
\begin{align*} 
\raisebox{-0.5\height}{
	\begin{tikzpicture}
	\begin{scope}[execute at begin node=$, execute at end node=$]
	\node[node,label={left:\omega_2}] {} 
	child{ node[node,label={left:a_2}] {}
		child{ node[node,label={left:\sigma}] {}
			child{ node[node,label={left:\sigma}] {}
				child{ node[node,label={left:x}] {}}}}}
	child{ node[node,label={left:\omega_1}] {}
		child{ node[node,label={left:\omega_1}] {}
			child{ node[node,label={left:y_1}] {}}}};
	\end{scope}
	\end{tikzpicture}}&&
\Rightarrow &&
\raisebox{-0.5\height}{
	\begin{tikzpicture}
	\begin{scope}[execute at begin node=$, execute at end node=$]
	\node[node,label={left:\omega_2}] {} 
	child{ node[node,label={left:\omega_1}] {}
		child{ node[node,label={left:a_2}] {}
			child{ node[node,label={left:\sigma}] {}
				child{ node[node,label={left:x}] {}}}}}
	child{ node[node,label={left:\omega_1}] {}
		child{ node[node,label={left:\omega_1}] {}
			child{ node[node,label={left:y_1}] {}}}};
	\end{scope}
	\end{tikzpicture}}&&
\Rightarrow &&
\raisebox{-0.5\height}{
	\begin{tikzpicture}
	\begin{scope}[execute at begin node=$, execute at end node=$]
	\node[node,label={left:\omega_2}] {} 
	child{ node[node,label={left:\omega_1}] {}
		child{ node[node,label={left:\omega_1}] {}
			child{ node[node,label={left:a_2}] {}
				child{ node[node,label={left:x}] {}}}}}
	child{ node[node,label={left:\omega_1}] {}
		child{ node[node,label={left:\omega_1}] {}
			child{ node[node,label={left:y_1}] {}}}};
	\end{scope}
	\end{tikzpicture}}&&
\Rightarrow &&
\raisebox{-0.5\height}{
	\begin{tikzpicture}
	\begin{scope}[execute at begin node=$, execute at end node=$]
	\node[node,label={left:\omega_2}] {} 
	child{ node[node,label={left:\omega_1}] {}
		child{ node[node,label={left:\omega_1}] {}
			child{ node[node,label={left:y_2}] {}
				child[opacity=0]{ }}}}
	child{ node[node,label={left:\omega_1}] {}
		child{ node[node,label={left:\omega_1}] {}
			child{ node[node,label={left:y_1}] {}}}};
	\end{scope}
	\end{tikzpicture}} &
\end{align*}

%% file: Section.4.2.tex
In this section we shall define several classes of F- and R-transformations and then compare them with each other with respect to set theoretic inclusion. It will turn out that in most cases the classes to be investigated are incomparable.

\begin{df}\label{Definition.4.2.1}\rm
Let $\gA=(\Sigma, X, A, \Omega, Y, P,  A')$ be an F-transducer. Then:
\begin{itemize}
\item[(1)] A production of $\gA$ is {\it linear}\index{linear production of!F-transducer} if each auxiliary variable occurs at most once in its right-hand side. Moreover, $\gA$ is a {\it linear}\index{ftransducer@F-transducer!linear} F-{\it transducer} (LF-{\it transducer}\index{LF-transducer}) if all of its productions are linear.
\item[(2)] $\gA$ is a {\it totally defined}\index{ftransducer@F-transducer!totally defined} F-{\it transducer} (TF-{\it transducer})\index{TF-transducer} if
\item[(i)] for each $x\in X$ there is a production in $P$ with left-hand side $x$ and
\item[(ii)] for all $m\geq 0, \sigma\in\Sigma_{m}$ and $a_{1}, \ldots, a_{m}\in A$ there is a production in $P$ with left-hand side $\sigma(a_{1},\ldots,a_{m})$.
\item[(3)] $\gA$ is a {\it nondeleting}\index{ftransducer@F-transducer!nondeleting} F-{\it transducer} (NF-{\it transducer}\index{NF-transducer}) if for every production\linebreak	 $\sigma(a_{1},\ldots, a_{m})\rightarrow aq$ $(\sigma\in\Sigma_{m}, m\geq 0)$ from $P$ each $\xi_{i}\in\Xi_{m}$ occurs at least once in $q$.
\item[(4)] $\gA$ is a {\it deterministic}\index{ftransducer@F-transducer!deterministic} F-{\it transducer} (DF-{\it transducer})\index{DF-transducer} if there are no two distinct productions in $P$ with the same left-hand side.
\item[(5)] $\gA$ is an F-{\it relabeling}\index{frelabeling@F-relabeling} if each of its productions is of the form
\item[(i)] $x\rightarrow ay$ $(x\in X, a\in A, y\in Y)$ or
\item[(ii)] $\sigma(a_{1},\ldots,a_{m})\rightarrow a\omega(\xi_{1},\ldots,\xi_{m})$, where $\sigma\in\Sigma_{m}, a_{1}, \ldots, a_{m}, a\in A, \omega\in\Omega_{m}$.
\end{itemize}
Transformations induced by $\mathrm{F}$-relabelings are also called $\mathrm{F}$-{\it relabelings}.
\end{df}

To illustrate the above concepts, let us take the following example.

\begin{ex}\label{Example.4.2.2}\rm
Let $\gA=(\Sigma, \{x\}, \{a_{0}, a_{1}\}, \Omega, \{y\}, P, \{a_{1}\})$ be the F-transducer with $\Sigma=\Sigma_{2}=\{\sigma\}$ and $\Omega=\Omega_{2}=\{\omega\}$, where $P$ consists of the productions
\begin{gather*}
x\rightarrow a_{0}y,\\
\sigma(a_{0}, a_{0})\rightarrow a_{1}\omega(\xi_{1}, \xi_{2}), \sigma(a_{0}, a_{1})\rightarrow a_{0}\omega(\xi_{1}, \xi_{2}), \sigma(a_{1}, a_{0})\rightarrow a_{1}\omega(\xi_{1}, \xi_{2}),\\
\sigma(a_{1}, a_{1})\rightarrow a_{1}\omega(\xi_{1}, \xi_{2}).
\end{gather*}

Then $\gA$ is a linear, totally defined, nondeleting and deterministic F-transducer. Moreover, $\gA$ is an F-relabeling.\epr
\end{ex}

Example \ref{Example.4.1.3} gives an F-transducer which is linear and deterministic, but it is neither totally defined nor nondeleting.

Let us note that F-relabelings are always linear and nondeleting F-transducers.

We now define the R-transducer counterparts of the above classes of F-transducers.

\begin{df}\label{Definition.4.2.3}\rm
Let $\gA=(\Sigma, X, A, \Omega, Y, P, A')$ be an R-transducer. Then:

\begin{enumerate}
\item[(1)] A production of $\gA$ is {\it linear}\index{linear production of!R-transducer} if each auxiliary variable occurs at most once in its right-hand side. Moreover, $\gA$ is a {\it linear} R-{\it transducer}\index{rtransducer@R-transducer!linear} (LR-{\it transducer}\index{LR-transducer}) if all of its productions are linear.
\item[(2)] $\gA$ is {\it a totally defined} $R$-{\it transducer}\index{rtransducer@R-transducer!totally defined} (TR-{\it transducer})\index{TR-transducer} if
\item[(i)] for all $a\in A$ and $x\in X$ there is a production in $P$ with left-hand side $ax$, and
\item[(ii)] for all $a\in A$ and $\sigma\in\Sigma_{m}$ $(m\geq 0)$ there is a production in $P$ with left-hand side $a\sigma$.
\item[(3)] $\gA$ is a {\it nondeleting} R-{\it transducer}\index{rtransducer@R-transducer!nondeleting} (NR-{\it transducer}\index{NR-transducer}) if for every production $a\sigma\rightarrow q$ $(\sigma\in\Sigma_{m}, m>0)$ from $P$ each $\xi_{i}\in\Xi_m$ occurs at least once in $q$.
\item[(4)] $\gA$ is a {\it deterministic} R-{\it transducer}\index{rtransducer@R-transducer!deterministic} (DR-{\it transducer})\index{drtransducer@DR-transducer} if $A'$ is a singleton and there are no distinct productions in $P$ with the same left-hand side.
\item[(5)] $\gA$ is an R-{\it relabeling}\index{Rre-labeling@R-relabeling} if each of the productions of $\gA$ has the form
\item[(i)] $ax\rightarrow y$ $(a\in A, x\in X, y\in Y)$ or
\item[(ii)] $a\sigma\rightarrow\omega(a_{1}\xi_{1},\ldots,a_{m}\xi_{m})$, where $a, a_{1}, \ldots, a_{m}\in A, \sigma\in\Sigma_{m}, \omega\in\Omega_{m}$. Transformations induced by R-relabelings will also be called R-{\it relabelings}.
\end{enumerate}
\end{df}

\begin{ex}\label{Example.4.2.4}\rm
Let $\gA=(\Sigma, \{x\}, \{a_{0}, a_{1}\}, \Omega, \{y_{1}, y_{2}\}, P, \{a_{0}\})$ be an R-transducer with $\Sigma=\Sigma_{2}=\{\sigma\}$ and $\Omega=\Omega_{2}=\{\omega\}$. Moreover, $P$ consists of the productions
\begin{align*}
a_{0}x\rightarrow y_{1},&\ \ a_{1}x\rightarrow y_{2},\\
a_{0}\sigma\rightarrow\omega(a_{1}\xi_{1}, a_{1}\xi_{2}),&\ \ a_{1}\sigma\rightarrow\omega(a_{0}\xi_{1}, a_{0}\xi_{2}).
\end{align*}

Then $\gA$ is a linear, totally defined, nondeleting and deterministic R-transducer. Moreover, $\gA$ is an R-relabeling. \epr
\end{ex}

The R-transducer of \ref{Example.4.1.6} is deterministic and nondeleting, but it is neither linear nor totally defined.

Let us note that R-relabelings are linear and nondeleting R-transducers.

The abbreviations introduced above for classes of tree transducers can be combined to indicate further subclasses. For instance, an LNF-transducer is a linear nondeleting F-transducer. Moreover, a transformation is a K-{\it transformation}\index{ktransformation@K-transformation} if it can be induced by a K-transducer. The class of all K-transformations will be denoted by $\cK$. Thus, for example, $\cLNF$ is the class of all LNF-transformations, i.e., the class of all transformations induced by linear nondeleting F-transducers. By Theorem \ref{Theorem.4.1.9}, we shall write simply $\cH$ instead of $\cH\cF$ and $\cH\cR$.
Moreover, $\cF$rel, resp. $\cR$rel, will denote the class of F-relabelings, resp. R-relabelings.

We now prove

\begin{thm}\label{Theorem.4.2.5}
$\cF$ and $\cR$ are incomparable.
\end{thm}

\pr In order to prove Theorem \ref{Theorem.4.2.5}, we give (i) an F-transformation which is not in $\cR$ and (ii) an R-transformation which cannot be induced by any F-transducer.

(i) Consider the LDF-transducer $\gA$ of Example \ref{Example.4.1.3}. If for an R-transducer $\gB=(\Sigma, \{x\}, B, \Omega, \{y\}, P', B')$ we have $(\sigma(x, x), \omega(y))\in\tau_{\gB}$, then at the first step of a derivation $b\sigma(x, x)\Rightarrow_{\gB}^{*}\omega(y)$ $(b\in B')$ we should apply a production of the form $b\sigma\rightarrow b'\xi_{1}, b\sigma\rightarrow b'\xi_{2}, b\sigma\rightarrow\omega(b'\xi_{1}), b\sigma\rightarrow\omega(b'\xi_{2})$ or $b\sigma\rightarrow\omega(y)$, where $b'\in B$. In each of the above cases one of the auxiliary variables $\xi_{1}$ and $\xi_{2}$ is deleted. Therefore, $\dom(\tau_{\gB})$ is infinite.

(ii) Take the DR-transducer $\gA$ of Example \ref{Example.4.1.6}. Assume that an F-transducer $\gB=(\Sigma,\{x\},B,\Omega, \{y_{1}, y_{2}\}, P', B')$ induces $\tau_{\gA}$. Obviously, $P'$ should then contain a production of the form
$$
\sigma(b)\rightarrow b_{1}\omega_{2}(q_{1}, q_{2})\ (b, b_{1}\in B).
$$
We may confine ourselves to the following cases:
\begin{itemize}
\item[(I)] $q_{1}=\sigma^{k}(y_{1})$ and $q_{2}=\sigma^{k}(y_{2})$,
\item[(II)] $q_{1}=\sigma^{l}(\xi_{1})$ and $q_{2}=\sigma^{k}(y_{2})$,
\item[(III)] $q_{1}=\sigma^{k}(y_{1})$ and $q_{2}=\sigma^{l}(\xi_{1})$,
\item[(IV)] $q_{1}=\sigma^{m}(\xi_{1})$ and $q_{2}=\sigma^{n}(\xi_{1})$.
\end{itemize}
Obviously, in a derivation $\sigma^{r}(x)\Rightarrow_{\gB}^{*}b'\omega_{2}(\omega_{1}^{r-1}(y_{1}), \omega_{1}^{r-1}(y_{2}))$ $(r>1, b'\in B')$ the last application of the above productions can be followed by applications of productions of the form $\sigma(\bar b)\rightarrow \bar b_{1}\xi_{1}$ $(\bar b,\bar b_{1}\in B)$ only. Let $t$ be the maximum of exponents in (I)--(IV). If $r>t+1$ and $\tau_{\gB}(\sigma^{r}(x))=\omega_{2}(\omega_{1}^{r-1}(y_{i}), \omega_{1}^{r-1}(y_{j}))\ (1\leq i,j\leq 2)$ then $i=j$. \epr

\

From the proof of Theorem \ref{Theorem.4.2.5} we directly get

\begin{cor}\label{Corollary.4.2.6}
$\cDF$ and $\cDR$ are incomparable and so are $\cDF$ and $\cR$, and $\cF$ and $\cDR$. \mbox{\ \ \ \ }\epr
\end{cor}

As we have mentioned one of the main differences between F- and R-transducers is that while F-transducers first process an input subtree and then copy the resulting output subtree, R-transducers first copy an input subtree and then treat these copies independently. In the case of an LR-transducer none of the input subtrees of a tree is copied during the translation of the tree. This property leads to

\begin{thm}\label{Theorem.4.2.7}
$\cLR$ is a proper subclass of $\cLF$.
\end{thm}

\pr By (i) in the proof of Theorem \ref{Theorem.4.2.5}, $\cLF$ is not a subclass of $\cLR$. Thus, it is enough to show the validity of $\cLR\subseteq \cLF$.

Let $\gA=(\Sigma, X, A, \Omega, Y, P,  A')$ be an LR-transducer. Then the productions from $P$ can be written in the form
\begin{itemize}
\item[(i)] $ax\rightarrow q$ $(a\in A, x\in X, q\in F_{\Omega}(Y))$, or
\item[(ii)] $a\sigma(\xi_{1}, \ldots, \xi_{m})\rightarrow q(a_{1}\xi_{1}, \ldots, a_{m}\xi_{m})$ $(a, a_{1}, \ldots, a_{m}\in A, m\geq 0, \sigma\in\Sigma_{m},  q\in F_{\Omega}[Y\cup A\Xi_m])$.
\end{itemize}

Now take the following R-transducer $\overline \gA$. If $\gA$ is nondeleting, then $\overline\gA=\gA$.
In the opposite case $\overline\gA=(\Sigma, X,\overline{A},\Omega, Y,\overline{P}, A')$ is given as follows. Let $\overline{A}=A\cup \{*\}$ $(*\not\in A)$. Fix any $\overline{y}\in Y$ and enlarge $P$ by all productions $*x\rightarrow\overline{y}$ $(x\in X)$ and $*\sigma\rightarrow \overline{y}$ $(m\geq 0, \sigma\in\Sigma_{m})$. Denote by $\overline{P}$ the resulting set of productions. Obviously, $\overline\gA$ is linear and equivalent to $\gA$. The only difference between $\overline{\gA}$ and $\gA$ is that $\overline{\gA}$ transforms (in  state $*$) even those subtrees of a tree $p\in F_{\Sigma}(X)$ which are deleted during the corresponding derivation of $p$ in $\gA$.

Next, construct the F-transducer $\gB=(\Sigma, X, B, \Omega, Y, P', B')$, where $ B=\overline{A}$ and $B'=A'$. Moreover, given any $x\in X, b\in B$ and $q\in F_{\Omega}(Y), x\rightarrow bq$ is in $P'$ iff $bx\rightarrow q$ is in $\overline{P}$. Furthermore, the production
$$
\sigma(b_{1}, \ldots, b_{m})\rightarrow bq(\xi_{1}, \ldots, \xi_{m})\ (\sigma\in\Sigma_{m}, m\geq 0, b_{1}, \ldots, b_{m}, b\in B, q\in F_{\Omega}(Y\cup \Xi_m))$$
is in $P'$ iff $\overline{P}$ contains a production
$$
b\sigma\rightarrow q(c_{1}\xi_{1}, \ldots, c_{m}\xi_{m}),
$$
such that for each $i=1, \ldots, m$,
$$
b_{i}=\left\{\begin{array}{ll}
c_{i}&\text{if } \xi_{i} \text{ occurs in } q,\\
*&\text{otherwise.}
\end{array}\right.
$$
Obviously $\gB$ is linear.

In order to complete the proof of Theorem \ref{Theorem.4.2.7}, it is enough to show that the equivalence
\begin{equation*}\label{Equation.4.2.1}\tag{1}p\Rightarrow_{\gB}^{*}bq \;\; \Longleftrightarrow \;\; bp\Rightarrow_{\gA}^{*}q
\end{equation*}
holds for all $b\in B, p\in F_{\Sigma}(X)$ and $ q\in F_{\Omega}(Y)$. We shall proceed by induction on $\hg(p)$.

If $\hg(p)=0$, then (\ref{Equation.4.2.1}) obviously holds by the definition of $P'$.

Now let $p=\sigma(p_{1}, \ldots, p_{m})$ $(\sigma\in\Sigma_{m}, m>0)$, and assume that (\ref{Equation.4.2.1}) has been proved for all trees in $F_{\Sigma}(X)$ of lesser height.

(I) Let $p\Rightarrow_{\gB}^{*}bq$ hold. More in detail, let
$$
p=\sigma(p_{1}, \ldots, p_{m})\Rightarrow_{\gB}^{*}\sigma(b_{1}q_{1}, \ldots, b_{m}q_{m})\Rightarrow_{\gB}b\overline{q}(q_{1}, \ldots, q_{m})=bq
$$
where $p_{i}\Rightarrow_{\gB}^{*}b_{i}q_{i}$ $(i=1, \ldots,  m)$. Then by the induction hypothesis, we have $b_{i}p_{i}\Rightarrow_{\overline\gA}^*q_{i}\ (i=1,\ldots,m)$. Moreover, by the definition of $P'$, $b\sigma\rightarrow\overline{q}(b_{1}\xi_{1},\ldots,b_{m}\xi_{m})$ is in $\overline{P}$. Therefore,
$$
bp=b\sigma(p_{1},\ldots,p_{m})\Rightarrow\overline{q}(b_{1}p_{1},\ldots,b_{m}p_{m})\Rightarrow^{*}\overline{q}(q_{1},\ldots,q_{m})=q
$$
also exists in $\overline\gA$.

(II) Assume that in $\overline\gA$ we have a derivation
$$
bp=b\sigma(p_{1},\ldots,p_{m})\Rightarrow\overline{q}(b_{1}p_{1},\ldots,b_{m}p_{m})\Rightarrow^{*}\overline{q}(q_{1},\ldots,q_{m})=q
$$
where each $q_{i}$ $(i=1,\ldots,m)$ is obtained by a derivation $b_{i}p_{i}\Rightarrow^{*}q_{i}$ in $\overline\gA$. Moreover, let $b_{i}=*$ and $q_{i}=\overline{y}$ if $\xi_{i}$ does not occur in $\overline{q}$. Then $\sigma(b_{1},\ldots,b_{m})\rightarrow b\overline{q}$ is in $P'$.
Furthermore, by the induction hypothesis, there are derivations $p_{i}\Rightarrow_{\gB}^{*}b_{i}q_{i}$ $(i=1,\ldots,m)$. Therefore, the derivation
$$
p=\sigma(p_{1},\ldots,p_{m})\Rightarrow_{\gB}^{*}\sigma(b_{1}q_{1},\ldots,b_{m}q_{m})\Rightarrow_{\gB}b\overline{q}(q_{1},\ldots,q_{m})=bq
$$
is also valid.\epr

\

For linear nondeleting tree transformations we have the following stronger result.

\begin{thm}\label{Theorem.4.2.8}
$\cLNR=\cLNF$.
\end{thm}

\pr The LF-transducer $\gB$ constructed to the LNR-transducer $\gA$ in the proof of the previous Theorem is obviously nondeleting.

Conversely, let $\gC=(\Sigma,X,C,\Omega,Y,P'',C')$ be an arbitrary LNF-transducer.
Construct the R-transducer $\gA=(\Sigma,X,C,\Omega,Y,P,C')$, where $P$ is defined as follows:
$$
(ax,q)\in P \;\; \Longleftrightarrow \;\; (x,aq)\in P''
$$
and
\begin{gather*}
(a\sigma,\ q(a_{1}\xi_{1}, \ldots, a_{m}\xi_{m}))\in P \;\; \Longleftrightarrow \;\;
(\sigma(a_{1}, \ldots, a_{m}), aq(\xi_{1}, \ldots, \xi_{m}))\in P'',
\end{gather*}
where $x\in X, a, a_{1}, \ldots, a_{m}\in A, \sigma\in\Sigma_{m}$ $(m\geq 0)$ and $q\in F_{\Omega}(Y\cup\Xi_{m})$. Obviously,
$\gA$ is an LNR-transducer.
Now to $\gA$ construct the F-transducer $\gB$ as in the proof of Theorem \ref{Theorem.4.2.7}. Then
$\gB=\gC$. \epr

\

The LF-transducer $\gB$ constructed to an R-relabeling in the proof of Theorem \ref{Theorem.4.2.7} is obviously an F-relabeling. Moreover, the R-transducer $\gA$ given to an F-relabeling $\gC$ in the proof of Theorem \ref{Theorem.4.2.8} is an R-relabeling. Thus, we have

\begin{cor}\label{Corollary.4.2.9}
$\cF\mathrm{rel}=\cR\mathrm{rel}.$\epr
\end{cor}

According to Corollary \ref{Corollary.4.2.9}, we may speak simply about relabelings.

One can easily show the existence of an LNF-transformation which is not a relabeling.

Our comparison results can be summarized by the diagram in Fig.~\ref{Figure.p.155}.
\begin{figure}[h]
\input{Figure.p.155}
\caption{{}\label{Figure.p.155}}
\end{figure}

%% file: Figure.p.155.tex
\centering
\begin{tikzpicture}[circ/.style={draw, circle,minimum size=0.7cm},
rect/.style={rounded corners=7pt, draw, text width=2.5cm, align=center,text height=0.22cm},
level distance=1.1cm]
\begin{scope}[execute at begin node=\tiny$, execute at end node=$]

\node[rect] {\cF_\rel=\cR_\rel}
child{ node[rect] {\cLNF=\cLNR} 
	child{ node[circ] {\cLR}
		child[sibling distance=3.2cm,level distance=2.2cm]{ node[circ] {\cR}
			child[sibling distance=2.2cm,level distance=-1.5cm]{ node[circ] {\cDR}}
			child[draw opacity=0]{}}
		child[sibling distance=2.6cm,level distance=0.8cm]{ node[circ] {\cLF}
			child[draw opacity=0]{}
			child[sibling distance=2.2cm,level distance=1.5cm]{ node[circ] {\cF}
				child[draw opacity=0]{}
				child[sibling distance=2.2cm,level distance=-1.5cm]{ node[circ] {\cDF}}
			}
		}
	}
};
\end{scope}
\end{tikzpicture}

%% file: Section.4.3.tex
Let $\cK$ be a class of tree transformations. We say that $\cK$ is {\it closed under composition}\index{class!of tree transformations closed under composition} if $\tau_{1}\circ\tau_{2}\in \cK$ whenever $\tau_{1}, \tau_{2}\in \cK$. As we shall see, some of our classes of tree transformations are closed under composition while others are not. On the other hand, in many cases it is possible to decompose a tree transformation into
a composition of simpler ones.

For any two classes $\cK_{1}$ and $\cK_{2}$ of tree transformations, we introduce the notation $\cK_1\circ \cK_{2}=\{\tau_{1}\circ\tau_{2}\mid\tau_{1}\in \cK_{1},\tau_{2}\in \cK_{2}\}$. Using this notation, the closure of a class $\cK$ of tree transformations under composition can be expressed by the
inclusion $\cK\circ \cK\subseteq \cK$. Similarly, the fact that all transformations in $\cK$ can be given as compositions of a transformation in $\cK_{1}$ by a transformation from $\cK_2$
can be expressed by $\cK\subseteq \cK_{1}\circ \cK_{2}$. Finally, if $\cK$ is a class of tree transformations, then let $\cK^{1}=\cK$ and $\cK^{n}=\cK\circ \cK^{n-1}$ $(n>1)$. All of the classes defined in the previous section $(\cR, \cF, \cLF,\cH\ \mathrm{etc}.)$ include all identity transformations
$\{(t,\ t)\mid t\in F_{\Sigma}(X)\}$. Hence, if $\cK$ is any one of these classes, then we know that
$$
\cK\subseteq \cK^{2}\subseteq \cK^{n}\subseteq\ldots.
$$

First we prove a decomposition theorem concerning F-transformations.

\begin{lm}\label{Lemma.4.3.1}
$\cF\subseteq \cLF\circ \cH$  and $\cF\subseteq \cLR\circ\cH$.
\end{lm}

\pr Let $\gA=(\Sigma, X, A, \Delta, Z, P, A')$ be an arbitrary F-transducer. Arrange the productions from $P$ in a fixed order and number them from 1 to $|P|$. For all $i(=1,\ldots,|P|)$, if the left side of the $i$th production is $x\in X$, then let $x^{(i)}$ be a new letter. Denote by $Y$ the set of all such $x^{(i)}$. Moreover, for all $i(=1, \ldots, |P|)$, if the symbol $\sigma\in\Sigma_{m}$ $(m\geq 0)$ occurs in the left-hand side of the $i$th production, then $\sigma^{(i)}$ will be a new $m$-ary operator. The set of all such operators will be denoted by $\Omega.$

Now we introduce the F-transducer $\gB=(\Sigma,X,A,\Omega,Y,P',A')$, where $P'$ is defined as follows:
\begin{itemize}
\item[(i)] $x\rightarrow ax^{(i)}$ $(x\in X, a\in A)$ is in $P'$ iff the $i$th production in $P$ is $x\rightarrow ar$ for some $r$,

\item[(ii)] $\sigma(a_{1},\ldots,a_{m})\rightarrow a\sigma^{(i)}(\xi_{1},\ldots,\xi_{m})$ $(\sigma\in\Sigma_{m}, m\geq 0,\ a_{1}, \ldots, a_{m}\in A)$ is in $P'$ iff the $i$th production in $P$ is $\sigma(a_{1}, \ldots, a_{m})\rightarrow ar$ for some $r$.
\end{itemize}

\noindent Obviously, $\gB$ is linear and nondeleting. Thus, by Theorem \ref{Theorem.4.2.8}, $\tau_{\gB}$ is a linear nondeleting R-transformation, as well.

Next define the F-transducer $\gC=(\Omega,Y, \{c_{0}\}, \Delta, Z, P'', c_{0})$ in the following way:
\begin{itemize}
\item[(i)] $x^{(i)}\rightarrow c_{0}r$ is in $P''$ iff the $i$th production in $P$ is $x\rightarrow ar$,

\item[(ii)] $\sigma^{(i)}(c_{0}, \ldots, c_{0})\rightarrow c_{0}r$ is in $P''$ iff the $i$th production in $P$ is $\sigma(a_{1}, \ldots, a_{m})\rightarrow ar$.
\end{itemize}
Then $\gC$ is an HF-transducer.

We prove that $\tau_{\gA}=\tau_{\gB}\circ\tau_{\gC}$. For this it is enough to show that, for all $p\in F_{\Sigma}(X)$, $r\in F_{\Delta}(Z)$ and $a\in A$, the equivalence
\begin{equation*}\label{Equation.4.3.1}\tag{1}
p\Rightarrow_{\gA}^{*}ar \;\; \Longleftrightarrow \;\; (\exists q\in F_{\Omega}(Y))(p\Rightarrow_{\gB}^{*}aq\wedge q\Rightarrow_{\gC}^{*}c_{0}r)
\end{equation*}
holds. We proceed by induction on $\hg(p)$.

If $\hg(p)=0$, then (\ref{Equation.4.3.1}) obviously holds.

Assume that $p=\sigma(p_{1},\ldots,p_{m})$ $(\sigma\in\Sigma_{m}, m>0)$ and that (\ref{Equation.4.3.1}) has been proved for all trees from $F_{\Sigma}(X)$ of lesser height.

(I) Let
\begin{equation*}\label{Equation.4.3.2}\tag{2}
p\Rightarrow_{\gA}^{*}\sigma(a_{1}r_{1}, \ldots, a_{m}r_{m})\Rightarrow_{\gA}a\overline{r}(r_{1}, \ldots, r_{m})=ar,
\end{equation*}
where $p_{i}\Rightarrow_{\gA}^{*}a_{i}r_{i}$ $(r_{i}\in F_{\Delta}(Z))$ holds for each $i(=1, \ldots, m)$. Then, by the induction hypothesis, there are trees $q_{i}\in F_{\Omega}(Y)$ $(i=1, \ldots, m)$ such that $p_{i}\Rightarrow_{\gB}^{*}a_{i}q_{i}$ and $q_{i}\Rightarrow_{\gC}^{*}c_{0}r_{i}$ hold. Assume that the production $\sigma(a_{1}, \ldots, a_{m})\rightarrow a\overline{r}$ last applied in (\ref{Equation.4.3.2}) is the $i$th one in $P$. Then
$$(\sigma(a_{1}, \ldots, a_{m}), a\sigma^{(i)}(\xi_{1}, \ldots, \xi_{m}))\in P' \text{ and } (\sigma^{(i)}(c_{0}, \ldots, c_{0}), c_{0}\overline{r})\in P''.$$

Therefore, taking $q=\sigma^{(i)}(q_{1},\ldots,q_{m})$, we have the desired derivations
$$
p\Rightarrow_{\gB}^{*}\sigma(a_{1}q_{1}, \ldots, a_{m}q_{m})\Rightarrow_{\gB}a\sigma^{(i)}(q_{1}, \ldots, q_{m})=aq
$$
and
$$
q\Rightarrow_{\gC}^{*}\sigma^{(i)}(c_{0}r_{1}, \ldots, c_{0}r_{m})\Rightarrow_{\gC}c_{0}\overline{r}(r_{1}, \ldots, r_{m})=c_{0}r.
$$

(II) The fact that the right side of (\ref{Equation.4.3.1}) implies its left side can be proved by inverting the above computation.\epr

\begin{lm}\label{Lemma.4.3.2} $\cF\circ \cH\subseteq \cF$.
\end{lm}

\pr Let $\gA=(\Sigma, X, A, \Omega, Y, P, A')$ be an F-transducer and $\gB=(\Omega, Y, \{b_{0}\}, \Delta, Z, P', b_{0})$ an HF-transducer. We shall construct an F-transducer $\gC$ whose productions will be composed of productions of $\gA$ and derivations in $\gB$. For this, using the fact that derivations in $\gB$ can be started from trees in $F_{\Omega}[Y\cup b_{0}\Xi]$ (see p. \pageref{page.140}), we define derivations in $\gB$ for trees in $F_{\Omega}(Y\cup\Xi)$. Take two trees $q\in F_{\Omega}(Y\cup\Xi_m)$ and $r\in F_{\Delta}(Z\cup\Xi_m)$. We write $q\Rightarrow_{\gB}^{*}b_{0}r$ if
$$
q(b_{0}\xi_{1}, \ldots, b_{0}\xi_{m})\Rightarrow_{\gB}^{*}b_{0}r
$$
holds. Now define an F-transducer $\gC=(\Sigma, X, A,\Delta,Z,P'',A')$, where $P''$ is given as follows:
\begin{itemize}
\item[(i)] $x\rightarrow ar$ $(x\in X, a\in A, r\in F_{\Delta}(Z))$ is in $P''$ iff there is a production $x\rightarrow aq$ in $P$ such that $q\Rightarrow_{\gB}^{*}b_{0}r$ holds,

\item[(ii)] $\sigma(a_{1}, \ldots, a_{m})\rightarrow ar$ $(\sigma\in\Sigma_{m}, m\geq 0, a_{1}, \ldots, a_{m}, a\in A, r\in F_{\Delta}(Z\cup\Xi_{m}))$ is in $P''$ iff there is a production $\sigma(a_{1},\ldots, a_{m})\rightarrow aq$ in $P$ such that $q\Rightarrow_{\gB}^{*}b_{0}r$ holds. Since at each step of the transformation of a tree the number of applications is finite, $P''$ is finite.
\end{itemize}

We prove that for all $a\in A, p\in F_{\Sigma}(X)$ and $r\in F_{\Delta}(Z)$ the equivalence

\begin{equation*}\label{Equation.4.3.3}\tag{3}
p\Rightarrow_{\gC}^{*}ar \;\; \Longleftrightarrow \;\; (\exists q\in F_{\Omega}(Y))(p\Rightarrow_{\gA}^{*}aq\wedge q\Rightarrow_{\gB}^{*}b_{0}r)
\end{equation*}
holds. We proceed by induction on $\hg(p)$.

If hg $(p)=0$ then (\ref{Equation.4.3.3}) obviously holds.

Assume that $p=\sigma(p_{1}, \ldots, p_{m})$ $(\sigma\in\Sigma_{m}, m>0)$ and that (\ref{Equation.4.3.3}) has been proved for all trees from $ F_{\Sigma}(X)$ of lesser height.

(I) First we show that the right side of (\ref{Equation.4.3.3}) implies its left side. For this assume that the derivations
$$
p\Rightarrow_{\gA}^{*}\sigma(a_{1}q_{1}, \ldots, a_{m}q_{m})\Rightarrow_{\gA}a\overline{q}(q_{1},\ldots,q_{m})=aq
$$
$$
(p_{i}\Rightarrow_{\gA}^{*}a_{i}q_{i},\ i=1, \ldots, m)
$$
and
$$
q\Rightarrow_{\gB}^{*}\overline{q}(b_{0}r_{1}, \ldots, b_{0}r_{m})\Rightarrow_{\gB}^{*}b_{0}\overline{r}(r_{1}, \ldots, r_{m})=b_{0}r
$$
$$
(q_{i}\Rightarrow_{\gB}^{*}b_{0}r_i, i=1, \ldots,\ m)
$$
are given. Then, by the induction hypothesis, the relations $p_{i}\Rightarrow_{\gC}^{*}a_{i}r_{i}$ $(i=1, \ldots, m)$ also hold. Moreover, by the definition of $P''$, $\sigma(a_{1}, \ldots, a_{m})\to a\overline{r}$ is in $P''$. Thus, we have the derivation
\begin{equation*}\label{Equation.4.3.4}\tag{4}
p\Rightarrow_{\gC}^{*}\sigma(a_{1}r_{1}, \ldots, a_{m}r_{m})\Rightarrow_{\gC}a\overline{r}(r_{1},\ldots,r_{m})=ar.
\end{equation*}

(II) Suppose that (\ref{Equation.4.3.4}) and the derivations $p_{i}\Rightarrow_{\gC}^{*}a_{i}r_{i}$ $(i=1, \ldots, m)$ are valid. Then, by the induction hypothesis, there are trees $q_{i}\in F_{\Omega}(Y)$ $(i=1, \ldots, m)$ such that $p_{i}\Rightarrow_{\gA}^{*}a_{i}q_{i}$ and $q_{i}\Rightarrow_{\gB}^{*}b_{0}r_{i}$ hold. Moreover, by the definition of $P''$, there exists a $\overline{q}\in F_{\Omega}(Y\cup\Xi_m)$ with $(\sigma(a_{1}, \ldots, a_{m}), a\overline{q})\in P$ and $\overline{q}\Rightarrow_{\gB}^{*}b_{0}\overline{r}$. Therefore, for $q=\overline{q}(q_{1}, \ldots, q_{m})$
$$
p\Rightarrow_{\gA}^{*}\sigma(a_{1}q_{1}, \ldots, a_{m}q_{m})\Rightarrow_{\gA}a\overline{q}(q_{1}, \ldots, q_{m})=aq
$$
and
$$
q\Rightarrow_{\gB}^{*}\overline{q}(b_{0}r_{1}, \ldots, b_{0}r_{m})\Rightarrow_{\gB}^{*}b_{0}\overline{r}(r_{1}, \ldots, r_{m})=b_{0}r
$$
hold. \epr

\

From Theorem \ref{Theorem.4.2.7} and the Lemmas \ref{Lemma.4.3.1} and \ref{Lemma.4.3.2} we directly obtain

\begin{thm}\label{Theorem.4.3.3}
$\cF=\cLF\circ\cH=\cLR\circ\cH$.\epr
\end{thm}

The constructions in the proofs of Lemma \ref{Lemma.4.3.1} and \ref{Lemma.4.3.2} preserve determinism. Thus, we have

\begin{cor}\label{Corollary.4.3.4}
$\cDF=\cLDF\circ \cH$.\epr
\end{cor}

Now we investigate some special classes of F-transformations for closure under composition.

\begin{lm}\label{Lemma.4.3.5}
Let $\gA=(\Sigma, X, A, \Omega, Y, P, A')$ be an \emph{F}-transducer. Then there exists a totally defined \emph{F}-transducer $\gB=(\Sigma, X, B, \Omega, Y, P', B')$ such that $\tau_{\gA}=\tau_{\gB}.$ Moreover, if $\gA$ is linear, then $\gB$ can be chosen linear, too.
\end{lm}

\pr Let $B=A\cup \{*\}$ and $B'=A'$. The required $\gB$ results if we put
$$
P'=P\cup\{x\rightarrow*y\mid x\in X, y\in Y\}\cup\{\sigma(b_{1}, \ldots, b_{m})\rightarrow*y\mid\sigma\in\Sigma_{m},
$$
$$
m\geq 0, b_{1}, \ldots, b_{m}\in B, y\in Y\}.
$$

If $\gA$ is linear, then so is $\gB$. \epr

\begin{thm}\label{Theorem.4.3.6}
The following equalities hold:
\begin{enumerate}\rm
\item[(i)] $\cLF\circ \cLF=\cLF$,
\item[(ii)] $\cLR\circ \cLR=\cLF$.
\end{enumerate}
\end{thm}

\pr In order to show (i), take two LF-transducers $\gA=(\Sigma,X, A, \Omega, Y, P, A')$ and $\gB=(\Omega, Y, B, \Delta, Z, P', B')$. In view of Lemma \ref{Lemma.4.3.5}, we may assume that
$\gB$ is totally defined. Construct an F-transducer $\gC=(\Sigma, X, C, \Delta, Z, P'',  C')$ with $C=A\times B$ and $C'=A'\times B'$. Furthermore, $P''$ is defined as follows:
\begin{itemize}
\item[(I)] $x\rightarrow(a, b)r$ $(x\in X, (a, b)\in C, r\in F_{\Delta}(Z))$ is in $P''$ iff there is a production $x\rightarrow aq$ in $P$ such that $q\Rightarrow_{\gB}^{*}br$ holds,
\item[(II)] $\sigma((a_{1}, b_{1}), \ldots, (a_{m}, b_{m}))\rightarrow(a, b) r$
$$
(\sigma\in\Sigma_{m}, m\geq 0, (a_{1}, b_{1}), \ldots, (a_{m}, b_{m}), (a, b)\in C, r\in F_{\Delta}(Z\cup\Xi_m))
$$
\end{itemize}
is in $P''$ iff there is a production $\sigma(a_{1}, \ldots, a_{m})\rightarrow aq$ in $P$ such that $q(b_{1}\xi_{1}, \ldots, b_{m}\xi_{m})\Rightarrow_{\gB}^{*}br$ holds.

We shall prove that for arbitrary $p\in F_{\Sigma}(X), r\in F_{\Delta}(Z)$ and $(a, b)\in C$ the equivalence
\begin{equation*}\label{Equation.4.3.5}\tag{5}
p\Rightarrow_{\gC}^{*}(a, b)r \;\; \Longleftrightarrow \;\; (\exists q\in F_{\Omega}(Y))(p\Rightarrow_{\gA}^{*}aq\wedge q\Rightarrow_{\gB}^{*}br)
\end{equation*}
holds. We proceed by induction on $\hg(p)$.

If $\hg(p)=0$, then (\ref{Equation.4.3.5}) obviously holds.

Now let $p=\sigma(p_{1}, \ldots,p_{m})$ $(\sigma\in\Sigma_{m}, m>0)$, and assume that (\ref{Equation.4.3.5}) has been proved for all trees of lesser height.

First we show that the right side of (\ref{Equation.4.3.5}) implies the left side. Suppose we are given derivations
$$
p\Rightarrow_{\gA}^{*}\sigma(a_{1}q_{1}, \ldots, a_{m}q_{m})\Rightarrow_{\gA}a\overline{q}(q_{1}, \ldots, q_{m})=aq
$$
and
$$
q\Rightarrow_{\gB}^{*}\overline{q}(b_{1}r_{1}, \ldots, b_{m}r_{m})\Rightarrow_{\gB}^{*}b\overline{r}(r_{1}, \ldots, r_{m})=br
$$
where $p_{i}\Rightarrow_{\gA}^{*}a_{i}q_{i}$ and $q_{i}\Rightarrow_{\gB}^{*}b_{i}r_{i}$ $(i=1, \ldots, m)$. (Observe that for each $i$ $(1\leq i\leq m)$ there exists an $r_{i}$ such that $q_{i}\Rightarrow_{\gB}^{*}b_{i}r_{i}$ holds since $\gB$ is totally defined.) Then, by the induction hypothesis, the derivations $p_{i}\Rightarrow_{\gC}^{*}(a_{i}, b_{i})r_{i}$ $(i=1, \ldots, m)$ are also valid. Furthermore, by the definition of $P''$, the production
$$
\sigma((a_{1}, b_{1}), \ldots, (a_{m}, b_{m}))\rightarrow(a, b)\overline{r}
$$
is in $P''$. Therefore, we get the derivation
$$
p\Rightarrow_{\gC}^{*}\sigma((a_{1}, b_{1})r_{1}, \ldots, (a_{m}, b_{m}) r_{m})\Rightarrow_\gC(a, b)\overline{r}(r_{1}, \ldots, r_{m})=(a, b)r.
$$

The fact that the left side of (\ref{Equation.4.3.5}) implies its right side can be shown by reversing the above argument.

In order to prove (ii) it is enough to note that the HF-transducer $\gC$ constructed to the LF-transducer $\gA$ in the proof of Lemma \ref{Lemma.4.3.1} is also linear. Moreover, by Theorem~\ref{Theorem.4.2.7}, the inclusion $\cLR\subseteq \cLF$ holds. \epr

\

Using an argument similar to that used in the proof of Theorem \ref{Theorem.4.3.6} (i), one can prove

\begin{thm}\label{Theorem.4.3.7} The classes $\cDF$ and $\cH$ are closed under composition.\epr
\end{thm}

From Theorem \ref{Theorem.4.3.7}, by Theorem \ref{Theorem.4.3.6} (i), we get

\begin{cor}\label{Corollary.4.3.8}
The class $\cLDF$ is closed under composition.\epr
\end{cor}

Using our decomposition results, one can prove

\begin{thm}\label{Theorem.4.3.9}
$\cF\circ \cDF=\cF$. \epr
\end{thm}

Now we turn to decomposition of R-transducers.

\begin{lm}\label{Lemma.4.3.10}
$\cR\subseteq \cH\circ \cLR.$
\end{lm}

\pr Let $\gA=(\Sigma, X, A, \Delta, Z, P,  A')$ be an arbitrary R-transducer. Let $n$ be the greatest integer with $\Sigma_{n}\neq\emptyset$. For any production $d\in P$ and natural number $i$ $(1\leq i\leq n)$, denote by $k(d,i)$ the number of occurrences of $\xi_{i}$ in the right-hand side of $d$. Set $k= \max\{k(d, i)\mid d\in P, i=1, \ldots, n\}$. Furthermore, take the ranked alphabet $\Omega$ given by $\Omega=\bigcup(\Omega_{m\cdot k}\mid m\geq 0)$ and $\Omega_{m\cdot k}=\{\sigma'\mid\sigma\in\Sigma_{m}\}$ $(m\geq 0)$.

Let $\gB=(\Sigma, X, \{b_{0}\}, \Omega, X, P', b_{0})$ be the HR-transducer where $P'$ consists of all productions
$$
b_{0}x\rightarrow x\ (x\in X)
$$
and
$$
b_{0}\sigma\rightarrow\sigma'(b_{0}^{k}\xi_{1}^{k}, \ldots, b_{0}^{k}\xi_{m}^{k})\ (\sigma\in\Sigma_{m}, m\geq 0).
$$

Next define an LR-transducer $\gC=(\Omega,X,A,\Delta,Z,P'',A')$, where $P''$ is given as follows:
\begin{itemize}
\item[(i)] $ax\rightarrow r$ $(x\in X)$ is in $P''$ iff it is in $P$.
\item[(ii)] Let $\sigma\in\Sigma_{m}$ $(m\geq 0)$ and $\boldsymbol\xi_{i}\in\Xi^k$ with $\xi_{i_{j}}=\xi_{(i-1)k+j}$ $(i=1, \ldots, m, j=1, \ldots, k)$. Then $a\sigma'\rightarrow r(\ba_{1}\boldsymbol\xi_{1}, \ldots,\ba_{m}\boldsymbol\xi_{m})$ is in $P''$ iff $a\sigma\rightarrow r(\ba_{1}\xi_{1}^{n_{1}}, \ldots, \ba_{m}\xi_{m}^{n_m})$ is in $P$ (for some $n_{1}, \ldots, n_{m})$.
\end{itemize}

For each $p\in F_{\Sigma}(X)$ let us denote by $p'\in F_{\Omega}(X)$ the tree given as follows:
\begin{enumerate}
\item[(I)] if $p=x\in X$, then $p'=x$,
\item[(II)] if $p=\sigma(p_{1}, \ldots,p_{m})$ $(\sigma\in\Sigma_{m}, m\geq 0)$, then $p'=\sigma'(p_{1}^{\prime {k}}, \ldots,p_{m}^{\prime k})$.
\end{enumerate}

It is easy to show that the transformation $\tau_{\gB}$ is exactly the mapping $p\rightarrow p'$ $(p\in F_{\Sigma}(X))$.

In order to prove $\tau_{\gA}=\tau_{\gB}\circ\tau_{\gC}$ it is enough to show that for all $a\in A, p\in F_{\Sigma}(X)$ and $r\in F_{\Delta}(Z)$ the equivalence
\begin{equation*}\label{Equation.4.3.6}\tag{6}
ap\Rightarrow_{\gA}^{*}r \;\; \Longleftrightarrow \;\;  ap'\Rightarrow_{\gC}^{*}r
\end{equation*}
holds. We proceed by induction on $\hg(p)$.

If $\hg(p)=0$ then, by the choice of $P''$, (\ref{Equation.4.3.6}) is obviously valid.

Now let $p=\sigma(p_{1},\ \ldots,p_{m})$ $(\sigma\in\Sigma_{m}, m>0)$, and assume that (\ref{Equation.4.3.6}) has been proved for all trees of lesser height.

First we prove that the left side of (\ref{Equation.4.3.6}) implies its right side. Assume that
$$
ap\Rightarrow_{\gA}\overline{r}(\ba_{1}p_{1}^{n_1}, \ldots, \ba_{m}p_{m}^{n_m})\Rightarrow_{\gA}^{*}\overline{r}(\br_{1},\ldots,\br_{m})=r
$$
where $\ba_{i}p_{i}^{n_i}\Rightarrow_{\gA}^{*}\br_{i}$ $(i=1,\ldots,m)$. Then, by the definition of $P''$, the production $a\sigma'\rightarrow\overline{r}(\ba_{1}\boldsymbol{\xi}_{1},\ldots,\ba_{m}\boldsymbol{\xi}_{m})$ is in $P''$. Moreover, by the induction hypothesis, there are derivations $\ba_{i}p_{i}^{\prime n_{i}}\Rightarrow_{\gC}^* \br_{i}$ for all $i(=1, \ldots, m)$. Therefore, we have the desired derivation
$$
ap'\Rightarrow_{\gC}\overline{r}(\ba_{1}p_{1}^{\prime n_{1}},\ldots,\ba_{m}p_{m}^{\prime n_{m}})\Rightarrow_{\gC}^{*}\overline{r}(\br_{1},\ldots,\br_{n})=r.
$$

The fact that the right side of (\ref{Equation.4.3.6}) implies its left side can be proved by the converse of the computation above.\epr

\begin{lm}\label{Lemma.4.3.11}
$\cH\circ\cR\subseteq \cR.$
\end{lm}

\pr Let $\gA=(\Sigma, X, \{a_{0}\}, \Omega, Y, P,  a_{0})$ be an HR-transducer and $\gB=(\Omega, Y, B, \Delta, Z, P',\linebreak B')$ an arbitrary R-transducer. Take the R-transducer $\gC=(\Sigma,X,B,\Delta, Z, P'', B')$, where $P''$ is given in the following way:
\begin{itemize}
\item[(i)] $bx\to r$ $(b\in B, x\in X, r\in F_{\Delta}(Z))$ is in $P''$ iff there is a production $a_{0}x\rightarrow q$ in $P$ such that $bq\Rightarrow_{\gB}^{*}r$ holds;
\item[(ii)] $b\sigma\rightarrow r$ $(b\in B, \sigma\in\Sigma_{m}, m\geq 0,  r\in F_{\Delta}[Z\cup B\Xi_m])$ is in $P''$ iff there is a production $a_{0}\sigma\rightarrow q(a_{0}\xi_{1},\ldots,a_{0}\xi_{m})$ $(q\in F_{\Omega}(Y\cup \Xi_m))$ in $P$ such that $bq\Rightarrow_{\gB}^{*}r$ holds.
\end{itemize}

To show $\tau_{\gA}\circ\tau_{\gB}=\tau_{\gC}$ it is enough to prove that for arbitrary $b\in B, p\in F_{\Sigma}(X)$ and $r\in F_{\Delta}(Z)$ the equivalence
$$
bp\Rightarrow_{\gC}^{*}r \;\; \Longleftrightarrow \;\; (\exists q\in F_{\Omega}(Y))(a_{0}p\Rightarrow_{\gA}^{*}q\wedge bq\Rightarrow_{\gB}^{*}r)
$$
holds. This can be carried out by induction on $\hg(p)$.
\epr

\

From Lemmas \ref{Lemma.4.3.10} and \ref{Lemma.4.3.11} we directly get

\begin{thm}\label{Theorem.4.3.12} $\cR=\cH\circ \cLR.$ \epr
\end{thm}

Using Theorems \ref{Theorem.4.3.3} and \ref{Theorem.4.3.12} we obtain

\begin{thm}\label{Theorem.4.3.13} For each $n\geq 1$ the inclusions $\cF^{n}\subseteq \cR^{n+1}$ and $\cR^{n}\subseteq \cF^{n+1}$ hold. \epr
\end{thm}

Taking $n=1$ in Theorem \ref{Theorem.4.3.13}, we see that every F-transformation can be given as the composition of two R-transformations, and each R-transformation can be obtained as the composition of two F-transformations. Thus, taking Theorem \ref{Theorem.4.2.5} into account, we get

\begin{cor}\label{Corollary.4.3.14} Neither $\cF$ nor $\cR$ is closed under composition. \epr
\end{cor}

One can show that $\cF$ is not closed under composition by LNF-transformations either. For $\cR$, we have

\begin{thm}\label{Theorem.4.3.15} $\cR\circ \cLNR=\cR.$
\end{thm}

\pr By Theorem \ref{Theorem.4.3.12}, it suffices to show that $\cLR$ is closed under compositions by LNR-transformations.

Let $\gA=(\Sigma,X,A,\Omega,Y,P,A')$ be an LR-transducer and $\gB=(\Omega, Y, B, \Delta, Z, P', B')$ an LNR-transducer. Take the R-transducer $\gC=(\Sigma, X, C, \Delta, Y, P'', C')$ with $C=A\times B$ and $C'=A'\times B'$. Moreover, $P''$ is given as follows:
\begin{itemize}
\item[(i)] $(a, b)x\rightarrow r$ $((a, b)\in C, x\in X, r\in F_{\Delta}(Z))$ is in $P''$ iff there is a production $ax\rightarrow q$ in $P$ such that $bq\Rightarrow_{\gB}^{*}r$ holds.
\item[(ii)]$(a, b)\sigma\to r((a_{1}, b_{1})\xi_{1}, \ldots, (a_{m}, b_{m})\xi_{m})$
$$
((a, b), (a_{1}, b_{1}), \ldots, (a_{m}, b_{m})\in C, \sigma\in\Sigma_{m}, m\geq 0, r\in F_{\Delta}[Z\cup C\Xi_m])
$$
is in $P''$ iff there is a production $a\sigma\rightarrow q(a_{1}\xi_{1}, \ldots, a_{m}\xi_{m})$ $(q\in F_{\Omega}(Y\cup \Xi_m))$ in $P$ such that $bq\Rightarrow_{\gB}^{*}r(b_{1}\xi_{1}, \ldots,b_{m}\xi_{m})$ holds.
\end{itemize}

In order to show $\tau_{\gC}=\tau_{\gA}\circ\tau_{\gB}$ it is enough to prove that for arbitrary $(a, b)\in C, p\in F_{\Sigma}(X)$ and $q\in F_{\Delta}(Z)$ the equivalence
$$
(a,b)p\Rightarrow_{\gC}^{*}r \;\; \Longleftrightarrow \;\; (\exists q\in F_{\Omega}(Y))(ap\Rightarrow_{\gA}^{*}q\wedge bq\Rightarrow_{\gB}^{*}r)
$$
holds. This can be done by induction on $\hg(p)$.
\epr

\

Later on we need the following results.

\begin{lm}\label{Lemma.4.3.16} Let $\tau\subseteq F_{\Sigma}(X)\times F_{\Omega}(Y)$ be an arbitrary F-transformation and $T\in \Rec(\Omega,Y)$. Then $T\tau^{-1}\in \Rec(\Sigma,X)$.
\end{lm}

\pr By Lemma \ref{Lemma.4.1.11}, there exists an F-transducer $\gA$ with $\dom(\tau_{\gA})=\range(\tau_{\gA})=T$ and $\tau_{\gA}$ is the identity mapping on $T$. Moreover, by the proof of Lemma \ref{Lemma.4.1.11}, we may suppose that $\gA$ is deterministic. Furthermore, by Theorem \ref{Theorem.4.3.9}, $\cF\circ \cDF=\cF$. Thus, since $T\tau^{-1}=\dom(\tau \circ\tau_{\gA})$, in order to prove Lemma \ref{Lemma.4.3.16}, it is enough to show that the domain of an F-transformation is recognizable. But this is true by (i) of Theorem \ref{Theorem.4.1.10}.\epr

\

From Theorem \ref{Theorem.4.1.10} and Lemma \ref{Lemma.4.3.16}, using the inclusion $\cR\subseteq \cF^{2}$ (see Theorem \ref{Theorem.4.3.13}), we get

\begin{cor}\label{Corollary.4.3.17} Let $\tau\subseteq F_{\Sigma}(X)\times F_{\Omega}(Y)$ be an arbitrary \emph{R}-transformation. If $T\in\Rec(\Omega,Y)$, then $T\tau^{-1}\in\Rec(\Sigma,X)$. In particular, $\dom(\tau)\in\Rec(\Sigma,X)$. \epr
\end{cor}

%% file: Section.4.4.tex
Consider an F-transducer $\gA=(\Sigma, X, A, \Omega, Y, P, A')$. Take a tree $p=\sigma(p_{1},\ldots,p_{m})\in F_{\Sigma}(X)$ $(\sigma\in\Sigma_{m}, m>0)$ and a derivation $\sigma(p_{1}, \ldots, p_{m})\Rightarrow^{*} \sigma(a_{1}q_{1}, \ldots,a_{m}q_{m})$ $(a_{i}\in A,q_{i}\in F_{\Omega}(Y), p_{i}\Rightarrow^{*}a_{i}q_{i}, i=1, \ldots, m)$. Then, knowing the states $a_{1}, \ldots, a_{m}$, our transducer can decide which production $\sigma(a_{1}, \ldots, a_{m})\rightarrow q$ to apply next. In other words, after inspecting the properties of the subtrees $p_{1}, \ldots, p_{m}$, the F-transducer $\gA$ can select the production to be applied in the next step of the translation of $p$. Moreover, these properties of subtrees are regular in the sense that $\dom(\tau_{\gA(a_i)})$ is a regular forest for each $i(=1, \ldots, m)$. Obviously, R-transducers lack this possibility. This observation leads to the idea to provide R-transducers with regular look-ahead as follows.

\begin{df}\label{Definition.4.4.1}\rm A {\it root-to-frontier tree transducer with regular look-ahead}\index{root-to-frontier tree transducer!with regular look-ahead} $(\mathrm{R}_{\mathrm{R}}$-{\it trans\-ducer}\index{RR-transducer@$\mathrm{R}_{\mathrm{R}}$-transducer}) is a system $\gA=(\Sigma, X, A, \Omega, Y, P, A')$, where
\begin{itemize}
\item[(1)] $\Sigma, X, A, \Omega, Y$ and $A'$ have the same meanings as in Definition \ref{Definition.4.1.4},
\item[(2)] $P$ is a finite set of {\it productions}\index{production of!RR-transducer@$\mathrm{R}_{\mathrm{R}}$-transducer} (or {\it rewriting rules}\index{rewriting rule of!RRtransducer@$\mathrm{R}_{\mathrm{R}}$-transducer}) of the form $(p\rightarrow q,D)$, where $p\rightarrow q$ is an R-transducer production and $D$ is a mapping of the set of all auxiliary variables occurring in $p$ into $\Rec(\Sigma, X)$.
\end{itemize}

If $p$ is of the form $ax$ $(x\in X)$ or $a\sigma$ with $\sigma\in\Sigma_{0}$, then the domain of $D$ is empty. We write such rules generally as $ax\rightarrow q$ and $a\sigma\rightarrow q$, respectively. Moreover, for any $a\in A$, we put $\gA(a)=(\Sigma, X, A, \Omega, Y, P, a)$.
\end{df}

\begin{df}\label{Definition.4.4.2}\rm Let $\gA$ be the $\mathrm{R}_{\mathrm{R}}$-transducer of Definition \ref{Definition.4.4.1}. $\gA$ is called {\it deterministic}\index{RR-transducer@$\mathrm{R}_{\mathrm{R}}$-transducer!deterministic} if the following conditions are satisfied:
\begin{itemize}
\item[(i)] $A'$ is a singleton.
\item[(ii)] If $(p_{1}\rightarrow q_{1}, D_{1})$ and $(p_{2}\rightarrow q_{2}, D_{2})$ are two productions in $P$ with $p_{1}=p_{2},$ and $q_{1}\neq q_{2}$, then there exists an $i$ $(1\leq i\leq m)$ such that $D_{1}(\xi_{i})\cap D_{2}(\xi_{i})=\emptyset,$ where $m$ is the number of auxiliary variables in $p_{1}(=p_{2})$.
\end{itemize}

{\it Linear}\index{RR-transducer@$\mathrm{R}_{\mathrm{R}}$-transducer!linear} and {\it nondeleting}\index{RR-transducer@$\mathrm{R}_{\mathrm{R}}$-transducer!nondeleting} $\mathrm{R}_{\mathrm{R}}$-transducers are defined in the same way as their $\mathrm{R}$-transducer counterparts.
\end{df}

\begin{df}\label{Definition.4.4.3}\rm Take an $\mathrm{R}_{\mathrm{R}}$-transducer $\gA=(\Sigma, X, A, \Omega, Y, P, A')$, and let $p, q\in F_{\Omega}[Y\cup AF_{\Sigma}(X)]$ be two trees. It is said that $p$ {\it directly derives} $q$ in $\gA$\index{direct derivation in!RRtransducer@$\mathrm{R}_{\mathrm{R}}$-transducer} (in notation, $p\Rightarrow_{\gA}q$) if $q$ can be obtained from $p$
\begin{itemize}
\item[(i)] by replacing an occurrence of an $ax$ $(a\in A, x\in X)$ in $p$ by the right side $\overline{q}$ of a production $ax\rightarrow\overline{q}$ in $P$, or
\item[(ii)] by replacing an occurrence of a subtree $a\sigma(p_{1}, \ldots, p_{m})$ $(a\in A, \sigma\in\Sigma_{m},  m\geq 0, p_{1}, \ldots,p_{m}\in F_{\Sigma}(X))$ in $p$ by $\overline{q}(p_{1}, \ldots, p_{m})$, where $(a\sigma\rightarrow\overline{q}, D)$ is in $P$ and
$p_{i}\in D(\xi_{i})$ for each $i(=1, \ldots,\ m)$.
\end{itemize}

A sequence
$$
p=p_{0}\Rightarrow_{\gA}p_{1}\Rightarrow_{\gA}\ldots\Rightarrow_{\gA}p_{k}=q\ (k\geq 0)
$$
obtained by consecutive applications of direct derivations is a {\it derivation}\index{derivation in!RRtransducer@$\mathrm{R}_{\mathrm{R}}$-transducer} of $q$ from $p$ in $\gA$. When such a derivation exists, we write $p\Rightarrow_{\gA}^{*}q$. Again, this notation will also be used to indicate a certain derivation.

If there is no danger of confusion, then we generally omit $\gA$ in $\Rightarrow_{\gA}$ and $\Rightarrow_{\gA}^{*}.$
\end{df}

According to Definition \ref{Definition.4.4.3}, the difference between derivations in $\mathrm{R}$-transducers and $\mathrm{R}_{\mathrm{R}}$-transducers is that in case of an $\mathrm{R}_{\mathrm{R}}$-transducer $\gA$ a production $a\sigma\rightarrow q$ can be applied to a tree $a\sigma(p_{1},\ldots,p_{m})$ if and only if there is a production $(a\sigma\rightarrow q, D)$ of $\gA$ such that each subtree $p_{i}$ $(1\leq i\leq m)$ is in the recognizable forest $D(\xi_{i})$.

\begin{df}\label{Definition.4.4.4}\rm Let $\gA=(\Sigma, X, A, \Omega, Y, P, A')$ be an $\mathrm{R}_{\mathrm{R}}$-transducer. Then the relation
$$\tau_{\gA}=\{(p, q)\mid p\in F_{\Sigma}(X), q\in F_{\Omega}(Y),\ ap\Rightarrow^{*}q \text{ for some } a\in A'\}$$
is called the {\it transformation induced} by $\gA.$\index{transformation induced by!rrtransducer@$\mathrm{R}_{\mathrm{R}}$-transducer}

A relation $\tau$ is an $\mathrm{R}_{\mathrm{R}}$-{\it transformation}\index{RR-transformation@$\mathrm{R}_{\mathrm{R}}$-transformation} if there exists an $\mathrm{R}_{\mathrm{R}}$-transducer $\gA$ such that $\tau=\tau_{\gA}.$

{\it Linear\index{RR-transformation@$\mathrm{R}_{\mathrm{R}}$-transformation!linear}, nondeleting\index{RR-transformation@$\mathrm{R}_{\mathrm{R}}$-transformation!nondeleting}} and {\it deterministic\index{RR-transformation@$\mathrm{R}_{\mathrm{R}}$-transformation!deterministic}} $\mathrm{R}_{\mathrm{R}}$-{\it transformations} are defined in an obvious way.

The class of all $\mathrm{R}_{\mathrm{R}}$-transformations will be denoted by $\cR_{R}.$
\end{df}

Let us note that there exists a recursive definition of transformations induced by $\mathrm{R}_{\mathrm{R}}$-transducers. This can be obtained by an obvious modification of the corresponding definition of transformations induced by $\mathrm{R}$-transducers.

Moreover, for $\mathrm{R}_{\mathrm{R}}$-transducers the notion of a reordering of direct derivations can be defined in the same way as in the case of $\mathrm{R}$-transducers. Furthermore, the remarks concerning different forms of derivations in $\mathrm{R}$-transducers are valid for $\mathrm{R}_{\mathrm{R}}$-transducers, too.

To illustrate the concepts of $\mathrm{R}_{\mathrm{R}}$-transducers and $\mathrm{R}_{\mathrm{R}}$-transformations, consider

\begin{ex}\label{Example.4.4.5}\rm Let $X=\{x\}$ and $\Sigma=\Sigma_{1}\cup\Sigma_{2}$, where $\Sigma_{i}=\{\sigma_{i}\}$ $(i=1,2)$. Take the forests $T_{1}=\{\sigma_{1}(x)\}^{*x}$ and $T_{2}=\{\sigma_{1}(x)\}$. Let $\gA=(\Sigma, X, \{a_{0}, a_{1}\}, \Omega, Y, P,  a_{0})$ be the $\mathrm{R}_{\mathrm{R}}$-transducer where $\Omega=\Omega_{1}=\{\omega\}, Y=\{y\}$ and $P$ consists of the productions
$$
(a_{0}\sigma_{2}\rightarrow\omega(a_{1}\xi_{1}), D_{1})\ \ (D_{1}(\xi_{1})=T_{1}, D_{1}(\xi_{2})=T_{2}),
$$
$$
(a_{1}\sigma_{1}\rightarrow\omega(a_{1}\xi_{1}), D_{2})\ \ (D_{2}(\xi_{1})=T_{1}),
$$
$$
a_{1}x\rightarrow y.
$$
Then $\tau_{\gA}=\{(\sigma_{2}(\sigma_{1}^{n}(x), \sigma_{1}(x)), \omega^{n+1}(y))\mid n=0,1, \ldots\}$. Observe that (without regular look-ahead) the corresponding $\mathrm{R}$-transducer would induce the transformation \linebreak$\{(\sigma_{2}(\sigma_{1}^{n}(x), p), \omega^{n+1}(y))\mid p\in F_{\Sigma}(X), n=0,1, \ldots\}.$ \epr
\end{ex}

Obviously $\mathrm{R}$-transducers are special cases of $\mathrm{R}_{\mathrm{R}}$-transducers. On the other hand, $\mathrm{R}_{\mathrm{R}}$-transducers can restrict the domain of possible subtrees of input trees even if these are deleted. In fact, no $\mathrm{R}$-transducer could induce the $\tau_{\gA}$ considered in the above example. Assume that such an $\mathrm{R}$-transducer
$$
\gB=(\Sigma, X, B, \Omega, Y, P', B')
$$
exists. Then for every $n(\geq 0)$, the production applied first in a derivation $b_{0}\sigma_{2}(\sigma_{1}^{n}(x),$ $\sigma_{1}(x))\Rightarrow_{\gB}^{*}\omega^{n+1}(y)$ $(b_{0}\in B')$ should be of the form
\begin{itemize}
\item[(i)] $b_{0}\sigma_{2}\rightarrow q(b\xi_{1})$ or
\item[(ii)] $b_{0}\sigma_{2}\rightarrow q(b\xi_{2})$ $(b\in B, q=\omega^{m}(\xi_{1}), m\geq 0)$.
\end{itemize}
Let $k$ be the maximum of the heights of right sides of productions from $P'$ and $n\geq 3k$. Then the considered production should be of the form (i). But in this case all pairs $(\sigma_{2}(\sigma_{1}^{n}(x),p), \omega^{n+1}(y))$ $(p\in F_{\Sigma}(X))$ are in $\tau_{\gB}$, which is a contradiction.\epr

\begin{thm}\label{Theorem.4.4.6} The following inclusions hold:
\begin{itemize}
\item[\emph{(i)}]$\cR_{R}\subseteq \cDF\rel\circ \cR,$
\item[\emph{(ii)}] $\cLR_{R}\subseteq \cDF\rel\circ \cLR,$
\item[\emph{(iii)}] $\cDR_{R}\subseteq \cDF\rel\circ \cDR,$
\item[\emph{(iv)}]$\cLDR_{R}\subseteq \cDF\rel\circ \cLDR.$
\end{itemize}
\end{thm}

\pr Let $\gA=(\Sigma, X, A, \Delta, Y, P, A')$ be an arbitrary $\mathrm{R}_{\mathrm{R}}$-transducer. Let $T_{1}, \ldots, T_{k}\ (\subseteq F_{\Sigma}(X))$ be all regular forests which appear as images in the $D$-mappings of the productions in $P$. Denote by $V$ the set of all $k$-dimensional vectors with components $0$ or $1$. Now take a ranked alphabet $\Omega$, where $\Omega_{0}=\Sigma_{0}$, and for each $m>0, \Omega_{m}=\Sigma_{m}\times V^{m}$. Thus, the elements from $\Omega_{m}$ $(m>0)$ can be given in the form $(\sigma, (\mathbf{v}_{1}, \ldots, \mathbf{v}_{m}))$, where $\sigma\in\Sigma_{m}$ and $\mathbf{v}_{1}, \ldots, \mathbf{v}_{m}\in V.$

Let $\bA_{i}=(\cA_{i}, \alpha_{i}, A_{i}')$ be $\Sigma X$-recognizers with $\cA_{i}=(A_{i}, \Sigma)$ and $T(\bA_{i})=T_{i}$ $(i=1, \ldots, k)$. We introduce the $\mathrm{F}$-transducer $\gB=(\Sigma, X, B, \Omega, X, P', B')$ where $B=B'=A_{1}\times\ldots\times A_{k}$ and $P'$ consists of the following productions:
\begin{itemize}
\item[(I)] $x\rightarrow(x\alpha_{1}, \ldots, x\alpha_{k})x$ $(x\in X)$,
\item[(II)] $\sigma\rightarrow(\sigma^{\cA_{1}}, \ldots, \sigma^{\cA_{k}})\sigma$ $(\sigma\in\Sigma_{0})$,
\item[(III)] $\sigma(\ba_{1}, \ldots, \ba_{m})\rightarrow \ba(\sigma, (\mathbf{v}_{1}, \ldots, \mathbf{v}_{m}))(\xi_{1}, \ldots, \xi_{m})$
$$
(\sigma\in\Sigma_{m}, m>0; \ba, \ba_{i}\in B, \mathbf{v}_{i}\in V, i=1, \ldots, m),
$$
\end{itemize}

where
$$
\ba=(\sigma^{\cA_{1}}(a_{1_{1}}, \ldots, a_{m_{1}}), \ldots, \sigma^{\cA_{k}}(a_{1_{k}}, \ldots, a_{m_{k}}))
$$
and $v_{i_{j}}=1$ iff $a_{i_{j}}\in A_{j}'$. Obviously, $\gB$ is a deterministic $\mathrm{F}$-relabeling.

One can easily show that $\gB$ relabels every $\Sigma X$-tree $p$ in the following way:
\begin{itemize}
\item[$(\alpha)$] if $p\in X\cup\Sigma_{0}$, then $\tau_{\gB}(p)=p$,
\item[$(\beta)$] if $p=\sigma(p_{1}, \ldots,p_{m})$ $(\sigma\in\Sigma_{m}, m>0)$ then $\tau_{\gB}(p)=(\sigma, (\mathbf{v}_{1}, \ldots, \mathbf{v}_{m}))(\tau_{\gB}(p_{1}), \ldots, \linebreak\tau_{\gB}(p_{m}))$, where $v_{i_{j}}=1$ iff $p_{i}\in T_{j}$ $(1\leq i\leq m, 1\leq j\leq k)$.
\end{itemize}

Next construct the $\mathrm{R}$-transducer $\gC=(\Omega, X, A, \Delta, Y, P'', A')$ where $P''$ consists of the productions below:
\begin{enumerate}
\item[$(\alpha')$] $ap\rightarrow r$ $(a\in A, p\in X\cup\Omega_{0}, r\in F_{\Delta}(Y))$ is in $P''$ iff it is in $P,$
\item[$(\beta')$] $a(\sigma, (\mathbf{v}_{1}, \ldots, \mathbf{v}_{m}))\rightarrow r$ $(a\in A;\sigma\in\Sigma_{m}, m>0;\mathbf{v}_{i}\in V, i=1, \ldots, m;r\in F_{\Delta}[Y\cup A\Xi_m])$ is in $P''$ iff $(\sigma, (\mathbf{v}_{1}, \ldots , \mathbf{v}_{m}))$ occurs in a tree $\tau_{\gB}(p)$ $(p\in F_{\Sigma}(X))$ and $P$ contains a production $(a\sigma\rightarrow r, D)$ such that $v_{i_{j}}=1$ whenever $D(\xi_{i})=T_{j}$ $(1\leq i\leq m, 1\leq j\leq k)$.
\end{enumerate}

In order to prove $\tau_{\gA}=\tau_{\gB}\circ\tau_{\gC}$ it is enough to show that for arbitrary $a\in A, p\in F_{\Sigma}(X)$ and $r\in F_{\Delta}(Y)$ the equivalence
$$
ap\Rightarrow_{\gA}^{*}r \;\; \Longleftrightarrow \;\; a\tau_{\gB}(p)\Rightarrow_{\gC}^{*}r
$$
holds. This can be carried out by induction on $\hg(p)$.

It is also easy to show that $\gC$ is deterministic (linear) if $\gA$ is deterministic (linear). \epr

\

Theorem \ref{Theorem.4.4.6} (iii) shows that $\mathrm{DR}_{\mathrm{R}}$-transducers induce (partial) mappings.

Next we show that $\cR_{R}$ is closed under certain special $\mathrm{F}$-transformations.

\begin{thm}\label{Theorem.4.4.7} The following inclusions hold:
\begin{itemize}
\item[\emph{(i)}] $\cR_{R}\circ \cLF\subseteq \cR_{R},$
\item[\emph{(ii)}] $\cDR_{R}\circ \mathcal{DLF}\subseteq \cDR_{R},$
\item[\emph{(iii)}] $\cDR_{R}\circ\mathcal{DLR}\subseteq \cDR_{R},$
\item[\emph{(iv)}] $\cDR_{R}\circ\cH\subseteq \cDR_{R}.$
\end{itemize}
\end{thm}

\pr Let $\gA=(\Sigma,X,A,\Omega,Y,P,A')$ be an $\mathrm{R}_{\mathrm{R}}$-transducer, and take an $\mathrm{LF}$- transducer $\gB=(\Omega, Y, B, \Delta, Z, P', B')$.

We want to treat cases (i) and (ii) together. Since the set of initial states of a $\mathrm{DR}_{\mathrm{R}}$-transducer should be a singleton we shall use the $\mathrm{LF}$-transducer $\overline{\gB}=(\Omega,Y,\overline{B},\Delta,Z,\overline{P}',b_{0})$ instead of $\gB$, where $\overline{B}=B\cup b_{0}$ $(b_{0}\not\in B)$ and $\overline{P}'$ is obtained by enlarging $P'$ by the following productions: if $y\rightarrow bq$ $(y\in Y)$, is in $P'$ and $b\in B'$, then $y\rightarrow b_{0}q$ is in $\overline{P}'$. Similarly, if $\sigma(b_{1}, \ldots, b_{m})\rightarrow bq$ $(\sigma\in\Sigma_{m}, m\geq 0)$ is in $P'$ and $b\in B'$ then the production $\sigma(b_{1}, \ldots, b_{m})\rightarrow b_{0}q$ is in $\overline{P}'$. It is obvious that $\tau_{\overline{\gB}}=\tau_{\gB}.$

Construct the $\mathrm{R}_{\mathrm{R}}$-transducer $\gC=(\Sigma,X,A\times\overline{B},\Delta,Z,P'',A'\times\{b_{0}\})$, where $P''$ is given as follows:
\begin{itemize}
\item[(I)] $(a, b)p\rightarrow r$ $(a\in A, b\in\overline{B}, p\in X\cup\Sigma_{0}, r\in F_{\Delta}(Z))$ is in $P''$ iff there exists a production $ap\rightarrow q$ in $P$ such that $q \Rightarrow^{*}_{\overline\gB}br$ holds.
\item[(II)] Assume that the production $(a\sigma\rightarrow q(\ba_{1}\xi_{1}^{n_{1}}, \ldots, \ba_{m}\xi_{m}^{n_m}), D)$ $(a\in A$; $\sigma\in\Sigma_{m}, m>0;\ba_{i}\in A^{n_{i}}, i=1, \ldots, m; n_{1}+\ldots+n_{m}=n, q\in\hat{F}_{\Omega}(Y\cup\Xi_n))$ is in $P$ and that there is a derivation $q(\bb_{1}\boldsymbol\xi_{1}, \ldots, \bb_{m}\boldsymbol\xi_{m})\Rightarrow^{*}_{\overline\gB}br(\boldsymbol\xi_{1}, \ldots, \boldsymbol\xi_{m})$ with $b\in B;\bb_{i}\in B^{n_{i}}, \boldsymbol\xi_{i}\in\Xi^{n_{i}}, \xi_{i_{j}}=\xi_{n_{1}+\ldots+n_{i-1}+j}, 1\leq j\leq n_{i}, i=1, \ldots, m$ and $r \in F_{\Delta}(Z\cup\Xi_{n})$. Then $P''$ contains the production $((a, b)\sigma\to r(\ba_{1}\bb_{1}\xi_{1}^{n_{1}}, \ldots, \ba_{m}\bb_{m}\xi_{m}^{n_m}), D')$, where $D'(\xi_{i})=\bigcap(\tau_{\gA(a_{i_{j}})}^{-1}(\dom(\tau_{\gB(b_{i_{j}})}))\mid j=1, \ldots, n_{i})\cap D(\xi_{i})$ $(i=1, \ldots,  m)$. If $b\in B'$, then $((a, b_{0})\sigma\rightarrow r(\ba_{1}\bb_{1}\xi_{1}^{n_{1}}, \ldots, \ba_{m}\bb_{m}\xi_m^{n_m}), D')$ is also in $P''.$
\end{itemize}

By Corollary \ref{Corollary.4.3.17}, the domain of an $\mathrm{R}$-transformation is regular. Moreover, also by Corollary \ref{Corollary.4.3.17}, the inverse of an $\mathrm{R}$-transformation preserves regularity. Thus, by Corollary \ref{Corollary.4.2.9} and Theorems \ref{Theorem.4.4.6} and \ref{Theorem.2.4.2}, $D'(\xi_{i})$ $(1\leq i\leq m)$ is regular.

In order to show $\tau_{\gA}\circ\tau_{\overline{\gB}}=\tau_{\gC}$ it is enough to prove that for all $(a, b)\in A\times\overline{B}, p\in F_{\Sigma}(X)$ and $r\in F_{\Delta}(Z)$ the equivalence
$$
(a, b)p\Rightarrow_{\gC}^{*}r \;\; \Longleftrightarrow \;\; (\exists q\in F_{\Omega}(Y))(ap\Rightarrow^{*}_{\overline\gA}q\wedge q\Rightarrow_{\gB}^{*}br)
$$
holds. This can be done by induction on $\hg(p).$

One can easily check that if $\gA$ and $\gB$ are deterministic, then so is $\gC$. Thus, (i) and (ii) are valid.

For (iii), take a $\mathrm{DR}_{\mathrm{R}}$-transducer $\gA=(\Sigma, X, A, \Omega, Y, P,  a_{0})$ and a DLR-transducer $\gB=(\Omega, Y, B, \Delta, Z, P', b_{0})$.

Consider the $\mathrm{R}_{\mathrm{R}}$-transducer $\gC=(\Sigma, X, A\times B, \Omega, Y, P'', (a_{0}, b_{0}))$, where $P''$ is given in the following way:
\begin{itemize}
\item[(I)] If $ap\rightarrow q$ $(a\in A, p\in X\cup\Sigma_{0}, q\in F_{\Omega}(Y))$ is in $P$ and $bq\Rightarrow_{\gB}^{*}r$ $(b\in B, r\in F_{\Delta}(Z))$ holds, then $(a, b)p\rightarrow r$ is in $P''.$
\item[(II)] Suppose that $(a\sigma\rightarrow q(\ba_{1}\xi_{1}^{n_{1}}, \ldots, \ba_{m}\xi_{m}^{n_m}), D)$ $(a\in A, \sigma\in\Sigma_{m}, m>0, \ba_{i}\in A^{n_{i}}, i=1, \ldots, m, n_{1}+\ldots+n_{m}=n, q \in\hat{F}_{\Omega}(Y\cup\Xi_n))$ is in $P$ and there is a derivation $bq\Rightarrow_{\gB}^{*}r(\bb_{1}\boldsymbol\xi_{1}, \ldots, \bb_{m}\boldsymbol\xi_{m})$ with $b\in B, \bb_{i}\in B^{n_{i}}, \boldsymbol\xi_{i}\in\Xi^{n_i}, \xi_{i_{j}}=\xi_{n_{1}+\ldots+n_{i-1}+j}, 1\leq j\leq n_{i}, i=1, \ldots, m$ and $r\in F_{\Delta}(Z\cup\Xi_n)$. Then the production
$$
((a, b)\rightarrow r(\ba_{1}\bb_{1}\xi_{1}^{n_{1}}, \ldots, \ba_{m}\bb_{m}\xi_{m}^{n_m}), D')
$$
is in $P''$, where for every $i(=1,\ \ldots,\ m)$,
$$ D'(\xi_{i})=\bigcap(\dom(\tau_{\gA(a_{i_{j}})})\mid\xi_{i_{j}}\ (1\leq j\leq n_{i}) \text{ does not occur in }r)\cap D(\xi_i).$$
\end{itemize}

Obviously, $\gC$ is a $\mathrm{DR}_{\mathrm{R}}$-transducer. Moreover, for all $a\in A, b\in B, p\in F_{\Sigma}(X)$ and $r\in F_{\Delta}(Z)$ the equivalence
$$
(a, b)p\Rightarrow_{\gC}^{*}r\Longleftrightarrow(\exists q\in F_{\Omega}(Y))(ap\Rightarrow_{\gA}^{*}q\wedge bq\Rightarrow_{\gB}^{*}r)
$$
holds. This can be proved by induction on $\hg(p)$. Therefore, $\tau_{\gC}=\tau_{\gA}\circ\tau_{\gB}$. Thus we have shown that $\cDR_{R}\circ \mathcal{DLR}\subseteq \cDR_{R}.$

To show (iv), let $\gA=(\Sigma, X,A,\Omega,Y,P,a_{0})$ be a $\mathrm{DR}_{\mathrm{R}}$-transducer and $\gB=(\Omega, Y, \{b_{0}\}, \Delta, Z, P',  b_{0})$ an $\mathrm{HF}$-transducer.

Construct an $\mathrm{R}_{\mathrm{R}}$-transducer $\gC=(\Sigma, X, A, \Delta, Z, P'', a_{0})$,  where $P''$ is given as follows:
\begin{itemize}
\item[(I)] $ap\rightarrow r$ $(a\in A, p\in\Sigma_{0}\cup X, r\in F_{\Delta}(Z))$ is in $P''$ iff there is a production $ap\rightarrow q$ in $P$ such that $q\Rightarrow_{\gB}^{*}b_{0}r$ holds.
\item[(II)] Suppose that the production $(a\sigma\rightarrow q(\ba_{1}\xi_{1}^{n_1}, \ldots, \ba_{m}\xi_{m}^{n_m}), D)$ $(a\in A, \sigma\in\Sigma_{m}, m>0, \ba_{i}\in A^{n_{i}}, i=1, \ldots, m, n_{1}+\ldots+n_{m}=n, q \in\hat{F}_{\Omega}(Y\cup\Xi_n))$ is in $P$ and there is a derivation $q(b_{0}^{n_{1}}\boldsymbol\xi_{1}, \ldots,\ b_{0}^{n_{m}}\boldsymbol\xi_{m})\Rightarrow_{\gB}^{*}b_{0}r(\xi_{1_{1}}^{k_{11}}, \ldots, \xi_{1_{n_{1}}}^{k_{1n_{1}}}, \ldots, \xi_{m_{1}}^{k_{m1}}, \ldots, \xi_{m_{n_{m}}}^{k_{mn_{m}}})$ where $\boldsymbol\xi_{i}\in\Xi^{n_i}, \xi_{i_{j}}=\xi_{n_{1}+\ldots+n_{i-1}+j}, 1\leq j\leq n_{i}, i=1, \ldots, m, k_{11}+\ldots+k_{1n_{1}}+\ldots+k_{m1}+\ldots+k_{mn_{m}}=k, r \in\hat{F}_{\Delta}(Z\cup\Xi_k)$. Then the production
$$
(a\sigma\rightarrow r(a_{1_{1}}^{k_{11}}\xi_{1}^{k_{11}}, \ldots, a_{1n_{1}}^{k_{1n_{1}}}\xi_{1}^{k_{1n_{1}}}, \ldots, a_{m_{1}}^{k_{m1}}\xi_{m}^{k_{m1}}, \ldots, a_{m_{n_{m}}}^{k_{mn_{m}}}\xi_{m}^{k_{mn_{m}}}), D')
$$
is in $P''$, where for every $i(=1, \ldots, m), D'(\xi_i)=\bigcap(\dom(\tau_{\gA(a_{i_{j}})})\mid\xi_{i_{j}}$ occurs in $q$ but it does not occur in $r)\cap D(\xi_{i})$.
\end{itemize}

Using a similar argument as in the proof of (ii), we get that $D'(\xi_{i})$ is a regular forest. It is obvious that $\gC$ is deterministic.

Finally, to show $\tau_{\gA}\circ\tau_{\gB}=\tau_{\gC}$ it is enough to prove that for all $a\in A, p\in F_{\Sigma}(X)$ and $r\in F_{\Delta}(Z)$ the equivalence
$$
ap\Rightarrow_{\gC}^{*}r \;\; \Longleftrightarrow \;\;  p\tau_{\gA(a)}\Rightarrow_{\gB}^{*}b_{0}r
$$
holds. This can be done by induction on $\hg(p)$. \epr

\

From Theorem \ref{Theorem.4.4.7} we get

\begin{cor}\label{Corollary.4.4.8} The inclusions
\begin{itemize}
\item[\emph{(i)}] $\cR_{R}\circ\cF\rel\subseteq \cR_{R},$
\item[\emph{(ii)}] $\cDR_{R}\circ \cDF\rel\subseteq \cDR_{R}$, and
\item[\emph{(iii)}] $\cDR_{R}\circ \cDR\rel\subseteq \cDR_{R}$
\end{itemize}
hold.\epr
\end{cor}

Next we show that the classes of $\mathrm{LF}$-transformations and $\mathrm{LR}_{\mathrm{R}}$-transformations coincide.

\begin{thm}\label{Theorem.4.4.9} $\cLR_{R}=\cLF$.
\end{thm}

\pr Since $\cDF\rel \subseteq \cLNF$, the inclusion $\cLR_{R}\subseteq \cLF$ is implied by Theorems \ref{Theorem.4.4.6} (ii), \ref{Theorem.4.2.8} and \ref{Theorem.4.3.6} (ii).

In order to prove $\cLF\subseteq \cLR_{R}$, take an $\mathrm{LF}$-transducer $\gA=(\Sigma, X, A, \Omega, Y, P, A')$. Consider the $\mathrm{R}_{\mathrm{R}}$-transducer $\gB=(\Sigma, X, A, \Omega,Y, P', A')$, where $P'$ is given as follows:
\begin{itemize}
\item[(i)] If $x\rightarrow aq$ $(x\in X, a\in A, q\in F_{\Omega}(Y))$ is in $P$, then $ax\rightarrow q$ is in $P'.$
\item[(ii)] If $\sigma(a_{1}, \ldots, a_{m})\rightarrow aq$ $(\sigma\in\Sigma_{m}, m\geq 0, a_{1}, \ldots, a_{m}, a\in A, q\in F_{\Omega}(Y\cup\Xi_m))$ is in $P$, then $(a\sigma\rightarrow q(a_{1}\xi_{1}, \ldots, a_{m}\xi_{m}), D)$ is in $P'$, where for every $i(=1, \ldots, m)$,
$$D(\xi_{i})=\left\{\begin{array}{l}
\dom(\tau_{\gA(a_{i})}) \text{ if $\xi_i$ does not occur in } q,\\
F_{\Sigma}(X)\text{ otherwise}.
\end{array}\right.$$
\end{itemize}

Obviously, $\gB$ is an $\mathrm{LR}_{\mathrm{R}}$-transducer. To prove $\tau_{\gA}=\tau_{\gB}$ it is enough to show that for each $a\in A,  p\in F_{\Sigma}(X)$ and $q\in F_{\Omega}(Y)$ the equivalence
$$
p\Rightarrow_{\gA}^{*}aq \;\; \Longleftrightarrow \;\; ap\Rightarrow_{\gB}^{*}q
$$
holds. Again, we omit the straightforward inductive proof. \epr

\

In the proof of the above theorem we used look-ahead to ensure that the $\mathrm{LR}_{\mathrm{R}}$-transducer will not transform any tree which contains a subtree for which the $\mathrm{LF}$-transducer has no transform but which it would later delete.

From Theorem \ref{Theorem.4.4.9}, by Theorem \ref{Theorem.4.3.6} (i), we get

\begin{cor}\label{Corollary.4.4.10} $\cLR_{R}$ is closed under composition. \epr
\end{cor}

Next we show that $\cR_{R}$ is closed under $\mathrm{LR}_{\mathrm{R}}$-transformations and $\cDR_{R}$ is closed under composition.

\begin{thm}\label{Theorem.4.4.11} The following equations hold:
\begin{itemize}
\item[\emph{(i)}] $\cR_{R}\circ\cLR_{R}=\cR_{R},$
\item[\emph{(ii)}] $\cDR_{R}\circ \cDR_{R}=\cDR_{R}.$
\end{itemize}
\end{thm}

\pr $\cR_{R}\circ\cLR_{R}=\cR_{R}$ follows from Theorem \ref{Theorem.4.4.7} by Theorem \ref{Theorem.4.4.9}.

Since, for each $\Sigma$ and $X$, the identity mapping on $F_{\Sigma}(X)$ is in $\cDR_{R}$, in order to prove (ii) it is enough to show the validity of the inclusion $\cDR_{R}\circ \cDR_{R}\subseteq \cDR_{R}.$

By Theorem \ref{Theorem.4.4.6} (iii), the inclusion $\cDR_{R}\circ\cDR_{R}\subseteq \cDR_{R}\circ \cDF\rel\circ \cDR$ holds from which, using Corollary \ref{Corollary.4.4.8} (ii), we get $\cDR_{R}\circ\cDR_{R}\subseteq \cDR_{R}\circ \cDR$. This latter inclusion, by the proof of Lemma \ref{Lemma.4.3.10}, implies $\cDR_{R}\circ \cDR_{R}\subseteq\cDR_{R}\circ\cH\circ\mathcal{DLR}$. Now, using Theorem \ref{Theorem.4.4.7} (iv), we get $\cDR_{R}\circ \cDR_{R}\subseteq \cDR_{R}\circ \mathcal{DLR}$, from which by Theorem \ref{Theorem.4.4.7} (iii), we arrive at the desired inclusion $\cDR_{R}\circ\cDR_{R}\subseteq \cDR_{R}$. \epr

\

To end this section we prove the analogue of Theorem \ref{Theorem.4.3.12}.

\begin{thm}\label{Theorem.4.4.12} $\cR_{R}=\cH\circ \cLR_{R}.$
\end{thm}

\pr The inclusion $\cH\circ \cLR_{R}\subseteq \cR_{R}$ directly follows from Theorem \ref{Theorem.4.4.11} (i). To show $\cR_{R}\subseteq \cH\circ\cLR_{R}$, consider an $\mathrm{R}_{\mathrm{R}}$-transducer $\gA=(\Sigma,X,A,\Delta,Z,P,A')$. Omit regular look-ahead in $\gA$ and for the resulting $\mathrm{R}$-transducer consider the $\mathrm{H}$-transducer $\gB$ and $\mathrm{LR}$-transducer $\gC$ given in the proof of Lemma \ref{Lemma.4.3.10}. Now it is impossible to provide $\gC$ with a suitable regular look-ahead in an obvious way since $\mathrm{H}$-transducers do not preserve regularity. We shall solve this problem in the following way.

Take the tree homomorphism $h:F_{\Omega}(X)\rightarrow F_{\Sigma}(X)$ given as follows:
\begin{itemize}
\item[(i)] $h_{X}(x)=x$ $(x\in X)$,
\item[(ii)] $h_{mk}(\sigma')=\sigma(\xi_{1}, \xi_{k+1}, \ldots, \xi_{(m-1)k+1})$ $(\sigma\in\Sigma_{m}, m\geq 0)$.
\end{itemize}

One can easily verify that for every $p\in F_{\Sigma}(X)$ the equality $h(\tau_{\gB}(p))=p$ holds, i.e., $h(p')=p$ (for $p'$, see the proof of Lemma \ref{Lemma.4.3.10}).

Now replacing each production $a\sigma'\rightarrow r(\ba_{1}\boldsymbol\xi_{1}, \ldots, \ba_{m}\boldsymbol\xi_{m})$ $(\sigma\in\Sigma_{m}, m>0, (a\sigma\rightarrow r(\ba_{1}\xi_{1}^{n_{1}}, \ldots, \ba_{m}\xi_{m}^{n_m}), D)\in P)$ in $P''$ by $(a\sigma'\rightarrow r(\ba_{1}\boldsymbol\xi_{1}, \ldots, \ba_{m}\boldsymbol\xi_{m}), D')$, where $D'(\xi_{i_j})=h^{-1}(D(\xi_{i}))$ $(i=1, \ldots, m, j=1, \ldots, k)$, from $\gC$ we get an $\mathrm{LR}_{\mathrm{R}}$-transducer since, by Theorem \ref{Theorem.2.4.18}, $h^{-1}$ preserves recognizability. Let us denote the resulting $\mathrm{LR}_{\mathrm{R}}$-transducer also by $\gC.$

Using tree induction, it is easy to prove that $\tau_{\gA}=\tau_{\gB}\circ\tau_{\gC}$. \epr

%% file: Section.4.5.tex
In studying certain properties of tree transformations it is technically useful to consider systems that translate trees into strings. Such systems are also of interest as mathematical models of syntax directed translations of context-free languages.

\begin{df}\label{Definition.4.5.1}\rm A {\it generalized syntax directed translator}\index{generalized syntax directed translator} (GSDT\index{GSDT}) is a system $\gA=(\Sigma,X,A,Y,P,A')$, where
\begin{itemize}
\item[(1)] $\Sigma$ is a ranked alphabet,
\item[(2)] $A$ is a unary ranked alphabet (the {\it state set})\index{state!of GSDT},
\item[(3)] $X$ and $Y$ are alphabets,
\item[(4)] $A'\subseteq A$ is the set of {\it initial states}\index{initial state of!GSDT}, and
\item[(5)] $P$ is a finite set of {\it productions}\index{production of!GSDT} (or {\it rewriting rules}\index{rewriting rule of!GSDT}) of the following two types:
\begin{itemize}
\item[(i)] $ax\rightarrow w$ $(a\in A, x\in X, w\in Y^{*}),$
\item[(ii)] $a\sigma\rightarrow w$ $(a\in A, \sigma\in\Sigma_{m}, m\geq 0, w\in(Y\cup A\Xi_{m})^{*})$. (Here $A\Xi_{m}$ is treated as an alphabet; the elements of it are the trees of the form $a\xi_{i}$ with $a\in A$ and $\xi_{i}\in\Xi_{m}.$)
\end{itemize}
\end{itemize}
\end{df}

For $ap\rightarrow w$ we shall use the notation $(ap, w)$, too. Moreover, for any $a\in A,$ we put $\gA(a)=(\Sigma, X, A, Y, P, a)$.

Next we define translations induced by a GSDT $\gA$. To this end, we associate with each $a\in A$ and $p\in F_{\Sigma}(X)$ a subset $p\tau_{\gA, a}$ as follows:
\begin{itemize}
\item[(i)] if $p\in(X\cup\Sigma_{0})$, then $p\tau_{\gA,a}=\{w\mid (ap, w)\in P\}$;
\item[(ii)] if $p=\sigma(p_{1}, \ldots,p_{m})$ $(\sigma\in\Sigma_{m}, m>0)$, then for all
$$
(a\sigma, w_{1}a_{i(1)}\xi_{i_{1}}w_{2}\ldots w_{k}a_{i(k)}\xi_{i_{k}}w_{k+1})\in P
$$
$(1\leq i_{j}\leq m, j=1, \ldots, k, w_{1}, \ldots, w_{k+1}\in Y^{*})$ and $v_{i_{j}}\in p_{i_{j}}\tau_{\gA,a_{i(j)}}$ $(j=1, \ldots, k)$ the word $w_{1}v_{i_{1}}w_{2}\ldots w_{k}v_{i_{k}}w_{k+1}$ is in $p\tau_{\gA,a}$, and
\item[(iii)] nothing is in any $p\tau_{\gA,a}$ unless this follows from (i) and (ii).
\end{itemize}

\begin{df}\label{Definition.4.5.2}\rm Let $\gA=(\Sigma, X, A, Y, P,  A')$ be a GSDT. Then the {\it translation induced} by $\gA$\index{translation!induced by GSDT} is the relation $\tau_{\gA}=\{(p, w)\mid p\in F_{\Sigma}(X), w\in Y^{*}, w\in p\tau_{\gA,a}\text{ for some }a\in A'\}.$

The class of all translations induced by GSDTs will be denoted by $\cG.$
\end{df}

For translations induced by GSDTs we give another definition showing how a translation is carried out step by step.

Let $\gA$ be the GSDT of Definition \ref{Definition.4.5.1}. Take two words $v, w \in(Y\cup AF_{\Sigma}(X\cup\Xi))^{*}.$ (Here again each element of $AF_{\Sigma}(X\cup\Xi)$ is considered a symbol, i.e., we ignore the fact that these elements are composed of simpler objects.) We say that $v$ {\it directly derives} $w$ in $\gA$, and write $v\Rightarrow_{\gA}w$, if $w$ can be obtained from $v$ by
\begin{itemize}
\item[(i)] replacing an occurrence of $ax$ $(a\in A, x\in X)$ in $v$ by the right side $\overline{w}$ of a production $ax\rightarrow\overline{w}$ from $P$, or

\item[(ii)] replacing an occurrence of an $a\sigma(p_{1}, \ldots,p_{m})$ $(a\in A, \sigma\in\Sigma_{m}, m\geq 0, p_{1}, \ldots,p_{m}\in F_{\Sigma}(X\cup\Xi))$ in $v$ by $w_{1}a_{i(1)}p_{i_{1}}w_{2}\ldots w_{k}a_{i(k)}p_{i_{k}}w_{k+1}$ where
$$a\sigma\rightarrow w_{1}a_{i(1)}\xi_{i_{1}}w_{2}\ldots w_{k}a_{i(k)}\xi_{i_{k}}w_{k+1}\ (1\leq i_{j}\leq m, j=1, \ldots, k, w_{1}, \ldots, w_{k+1}\in Y^{*})$$ is a production in $P.$
\end{itemize}

Each application of a step (i) or (ii) is called a {\it direct derivation} in $\gA$\index{direct derivation in!GSDT}. A sequence $$v=v_{0} \Rightarrow_{\gA}v_{1}\Rightarrow_{\gA}\ldots\Rightarrow_{\gA}v_{k}=w\ \ (k\geq 0, v_{i}\in(Y\cup AF_{\Sigma}(X\cup\Xi))^{*}, i=0, \ldots,  k)$$
of consecutive direct derivations is a {\it derivation} of $w$ from $v$ in $\gA$\index{derivation in!GSDT}, and $k$ is the {\it length}\index{length of!derivation in GSDT} of this derivation. If $w$ can be obtained from $v$ by a derivation in $\gA$, then we write $v\Rightarrow_{\gA}^{*}w$. Thus $\Rightarrow_{\gA}^{*}$ is the reflexive-transitive closure of $\Rightarrow_{\gA}$. Again, we suppose that the notation $v\Rightarrow_{\gA}^{*}w$ implicitly includes a given derivation of $w$ from $v.$

Using the notation $\Rightarrow_{\gA}^{*}$, the translation $\tau_{\gA}$ induced by a GSDT $\gA=(\Sigma, X,A, Y, P, A')$ can be given by
$$
\tau_{\gA}=\{(p, w)\mid p\in F_{\Sigma}(X), w\in Y^{*}, ap\Rightarrow_{\gA}^{*}w\text{ for some }a\in A'\}.$$


The concept of a reordering of direct derivations in GSDTs can be defined in a similar way as in the case of an $\mathrm{R}$-transducer. Moreover, different forms of derivations can be introduced in an obvious manner.

{\it Deterministic\index{GSDT!deterministic}, linear\index{GSDT!linear}, totally defined\index{GSDT!totally defined}} and {\it nondeleting\index{GSDT!nondeleting}} GSDTs are defined in a natural way. Moreover, a one-state totally defined deterministic GSDT is a GSDH-{\it translator}\index{GSDH-translator}. The translation induced by a GSDH-translator is called a {\it generalized syntax directed homomorphism}\index{generalized syntax directed homomorphism} (GSD {\it homomorphism}\index{GSD homomorphism}). The class of all GSD homomorphisms will be denoted by $\cG_{\hom}.$

\begin{ex}\label{Example.4.5.3}\rm
Let $\gB=(\Sigma, \{x\}, \{b_{0},b_{1},b_{2}\}, \{y_{1},y_{2}\}, P', b_{0})$ be a GSDT, where $\Sigma=\Sigma_{1}=\{\sigma\}$ and $P'$ consists of the productions
\begin{gather*}
b_{0}\sigma\rightarrow b_{1}\xi_{1}b_{2}\xi_{1},\\
b_{1}\sigma\rightarrow b_{1}\xi_{1},\ \ b_{2}\sigma\rightarrow b_{2}\xi_{1},\\
b_{1}x\rightarrow y_{1},\ \ b_{2}x\rightarrow y_{2}.
\end{gather*}
Then $\gB$ is deterministic, totally defined and nondeleting, but it is not linear.

Take the tree $p=\sigma(\sigma(\sigma(x)))$ and the word $w=y_{1}y_{2}$. Moreover, consider the derivation
\begin{gather*}
p\Rightarrow_{\gB}b_{1}\sigma(\sigma(x))b_{2}\sigma(\sigma(x))\Rightarrow_{\gB}b_{1}\sigma(x) b_{2}\sigma(\sigma(x))\Rightarrow_{\gB}b_{1}xb_{2}\sigma(\sigma(x))\Rightarrow_\gB\\
y_{1}b_{2}\sigma(\sigma(x))\Rightarrow_{\gB}y_{1}b_{2}\sigma(x)\Rightarrow_{\gB}y_{1}b_{2}x\Rightarrow_{\gB}y_{1}y_{2}=w,
\end{gather*}
i.e., $\tau_{\gB}(p)=\yd(\tau_{\gA}(p))$, where $\gA$ is the $\mathrm{R}$-transducer of Example \ref{Example.4.1.6}. One can easily show that the previous equality holds for every $p\in F_{\Sigma}(\{x\})$. \epr
\end{ex}

The above relation generally holds between GSDTs and $\mathrm{R}$-transducers as it is shown by

\begin{thm}\label{Theorem.4.5.4} For each \emph{GSDT} $\gA=(\Sigma,X,A,Y,P,A')$ there exist a ranked alphabet $\Omega$ and an $\mathrm{R}$-transducer $\gB=(\Sigma, X, A, \Omega,Y, P', A')$ such that $\tau_{\gA}=\{(p, \yd(q))\mid (p,q)\in\tau_{\gB}\}$. Moreover, if $\gA$ is linear, deterministic, nondeleting or a \emph{GSDH}-transducer, then $\gB$ can also be chosen, correspondingly, as a linear, deterministic, nondeleting or an $\mathrm{RH}$-transducer.

Conversely, for every $\mathrm{R}$-transducer $\gB$ there exists a \emph{GSDT} $\gA$ such that $\{(p, \yd(q))\mid (p, q)\in\tau_{\gB}\}=\tau_{\gA}$. If $\gB$ is, respectively linear, deterministic, nondeleting or an $\mathrm{RH}$- transducer, then $\gA$ is linear, deterministic, nondeleting or a \emph{GSDH}-translator.
\end{thm}

\pr Let $\gA=(\Sigma,X,A,Y,P,A')$ be a GSDT. To define $\gB$, for each production $ap\rightarrow w$ $(a\in A, p\in X\cup\Sigma, w\in(Y\cup A\Xi)^{*})$ in $P$, let $\omega_{(ap,w)}$ be an operator with rank $|w|$. Let $\Omega$ be the resulting ranked alphabet. Moreover, $P'$ is defined as follows:
\begin{itemize}
\item[(i)] If $ap\rightarrow w$ $(a\in A, p\in X\cup\Sigma_{0}, w\in Y^{*})$ is in $P$ and $|w|=k$, then the production $ap\rightarrow \omega_{(ap,w)}(q_{1},\ldots,q_{k})$ $(q_{i}\in Y, i=1, \ldots, k)$ with $\yd(\omega_{(ap,w)}(q_{1}, \ldots, q_{k}))=w$ is in $P'.$

\item[(ii)] If $a\sigma\rightarrow w$ $(a\in A, \sigma\in\Sigma_{m}, m>0, w\in(Y\cup A\Xi_{m})^{*})$ is in $P$ with $|w|=k$, then the production $a\sigma\rightarrow\omega_{(a\sigma,w)}(q_{1}, \ldots, q_{k})$ $(q_{i}\in Y\cup A\Xi_{m}, i=1, \ldots, k)$ satisfying $\yd(\omega_{(a\sigma,w)}(q_{1}, \ldots, q_{k}))=w$ is in $P'$, where $\yd$ is taken over the frontier alphabet $Y\cup A\Xi_{m}.$
\end{itemize}

In order to prove $\tau_{\gA}=\{(p,\yd(q))\mid (p, q)\in\tau_{\gB}\}$ it is enough to show that, for all $a\in A, p\in F_{\Sigma}(X)$ and $w\in Y^{*}$, the equivalence
$$
ap\Rightarrow_{\gA}^{*}w \;\; \Longleftrightarrow \;\; (\exists q\in F_{\Omega}(Y))(ap\Rightarrow_{\gB}^{*}q\wedge \yd(q)=w)
$$
holds. This can be done in an obvious way by induction on $\hg(p)$.

It is also obvious from the construction of $\gB$ that the remaining conclusions of the first part of Theorem \ref{Theorem.4.5.4} hold, too.

Conversely, consider an $\mathrm{R}$-transducer $\gB=(\Sigma, X, B, \Omega, Y, P', B')$. The productions of the desired GSDT $\gA=(\Sigma, X, B, Y, P, B')$ are given as follows:
\begin{itemize}
\item[(I)] For all $b\in B, p\in X\cup\Sigma_{0}$ and $q\in F_{\Omega}(Y)$, if $bp\rightarrow q$ is in $P'$, then $ bp\rightarrow \yd(q)$ is in $P.$

\item[(II)] For all $b\in B, \sigma\in\Sigma_{m}$ $(m>0)$ and $q\in F_{\Omega}(Y\cup B\Xi_{m})$, if $b\sigma\rightarrow q$ is in $P',$ then $ b\sigma\rightarrow\yd(q)$ is in $P$, where $\yd$ is again taken over the alphabet $Y\cup B\Xi_{m}.$
\end{itemize}

To prove $\tau_{\gA}=\{(p,\yd(q))\mid (p, q)\in\tau_{\gB}\}$ it is enough to show that the equivalence
$$
bp\Rightarrow_{\gA}^{*}w \;\; \Longleftrightarrow \;\; (\exists q\in F_{\Omega}(Y))(bp\Rightarrow_{\gB}^{*}q\wedge \yd(q)=w)
$$
holds for arbitrary $b\in B, p\in F_{\Sigma}(X)$ and $w\in Y^{*}$. This can be carried out by induction on $\hg(p)$. Moreover, the remaining conclusions of the second part of Theorem \ref{Theorem.4.5.4} are obviously valid.\epr

%% file: Section.4.6.tex
The images of regular forests under tree transformations are called surface forests. In this section we compare classes of surface forests belonging to different classes of tree transformations.

\begin{df}\rm\label{Definition.4.6.1}
Let $\cK$ be a class of tree transformations. A forest $S\subseteq F_\Omega(Y)$ is called a {\em $\cK$-surface forest} \index{forest!ksurface@$\cK$-surface} if there exist a ranked alphabet $\Sigma$, a frontier alphabet $X$, a forest $R\in \Rec(\Sigma,X)$, and a $\cK$-transformation $\tau\subseteq F_\Sigma(X)\times F_\Omega(Y)$ such that $S=R\tau$. The class of all $\cK$-surface forests is denoted by $\Surf(\cK)$.
\end{df}

The following lemma is obvious.

\begin{lm}\label{Lemma.4.6.2} If $\cK$ is a class of tree transformations which contains all identity transformations, then $\Rec$ is included as a subclass in $\Surf(\cK)$. \epr
\end{lm}

Of course, this lemma applies to all of the classes of tree transformations which we have considered ($\cF, \cR, \cLF, \cH$ etc.).

Next we characterize F-transformations  preserving regularity. For this we should introduce some more terminology.

\begin{df}\rm\label{Definition.4.6.3} A tree transformation $\tau\subseteq F_\Sigma(X)\times F_\Omega(Y)$ is said to {\em preserve regularity} \index{tree transformation!preserving regularity}
if $R\tau \in \Rec(\Omega,Y)$ whenever $R \in \Rec(\Sigma,X)$. Moreover, a class $\cK$ of tree transformations {\em preserves regularity} if every
$\tau$ in $\cK$ preserves regularity. \index{class!of tree transformations preserving regularity}
\end{df}

We say that an F-transducer $\gA=(\Sigma,X,A,\Omega,Y,P,A')$ is {\em connected} if for each $a\in A$ there are $p\in F_\Sigma(X)$ and
$q\in F_\Omega(Y)$ such that $p\Rightarrow^*aq$ holds. \index{ftransducer@F-transducer!connected}

\begin{df}\rm\label{Definition.4.6.4} For each $p\in F_\Sigma(X\cup \Xi_n)$, $\ppath_i(p)$ $(1\leq i\leq n)$ \index{path in tree} is given in the following way:
\begin{enumerate}
\item[(i)] if $p\in \Sigma_0\cup X$, then $\ppath_i(p)=\emptyset$,
\item[(ii)] if $p=\xi_i$, then $\ppath_i(p)=\{e\}$,
\item[(iii)] if $p=\xi_j$ $(j\neq i)$, then $\ppath_i(p)=\emptyset$,
\item[(iv)] if $p=\sigma(p_1,\ldots,p_m)$ $(\sigma\in \Sigma_m, m>0)$, then
$$\ppath_i(p)=\{jw_j\mid w_j \in  \ppath_i(p_j), j=1,\ldots,m  \}.$$
\end{enumerate}
\end{df}

Thus, $\ppath_i(p)$ is a language over the alphabet $\{1,\ldots,m \}$, where $m$ is the maximal integer with $\Sigma_m\neq \emptyset$.

Obviously, the elements of $\ppath_i(p)$ describe paths leading from the root of $p$ to a leaf labelled by $\xi_i$.

If $\ppath_i(p)$ consists of a single word, then $l\big(\ppath_i(p) \big)$ denotes the length of this word.

\begin{lm}\label{Lemma.4.6.5} $\cLF$ preserves regularity.
\end{lm}
\pr Since the F-transducer  given in the proof of Lemma \ref{Lemma.4.1.11} is linear, by Theorem \ref{Theorem.4.3.6} (i), it is enough to show that for each LF-transducer $\gA=(\Sigma,X,A,\Omega,Y,P,A')$, $\range(\tau_{\gA})$ is regular. Without loss of generality, we may assume that $\gA$ is connected.

Consider the regular $\Omega Y$-grammar $G=(A,\Omega,Y,P',A')$, where $P'$ is given as follows:
\begin{itemize}
\item[(i)] if $x\rightarrow aq$ $\big(x\in X, a\in A, q\in F_\Omega(Y)\big)$ is in $P$, then $a\rightarrow q$ is in $P'$,
\item[(ii)] if $\sigma(a_1,\ldots,a_m)\rightarrow aq$ $\big(\sigma\in \Sigma_m, m\geq 0, a_1,\ldots,a_m,a\in A, q\in F_\Omega(Y\cup \Xi_m)\big)$ is in $P$, then $a\rightarrow q(a_1,\ldots,a_m)$ is in $P'$.
\end{itemize}

In order to prove the lemma it is enough to show that the equivalence
\begin{equation*}
\tag{1}\label{Equation.4.6.5.1}a\Rightarrow^*_G q \;\; \Longleftrightarrow \;\;  \big(\exists p\in F_\Sigma(X)\big)(p\Rightarrow^*_{\gA} aq)
\end{equation*}
holds for all $a\in A$ and $q\in F_\Omega(Y)$.
\begin{itemize}
\item[(I)] First we prove that the left side of (\ref{Equation.4.6.5.1}) implies its right side. For this, assume that $a\Rightarrow^*_G q$ is valid. We shall proceed by
induction on the length $l$ of $a\Rightarrow^*_G q$.

Let $l=1$. Then $a\rightarrow q$ is in $P'$, and the following two cases are possible:
\begin{itemize}
\item[(Ia)] There is a production $x\rightarrow aq$ $\big(x\in X, a\in A, q\in F_\Omega(Y)\big)$.
\item[(Ib)] There is a production $\sigma(a_1,\ldots,a_m) \rightarrow aq$ $(\sigma\in \Sigma_m, m\geq 0, a_1,\ldots,a_m,a\in A)$ such that in $q$ no auxiliary variables occur, i.e.,
$q\in F_\Omega(Y)$.
\end{itemize}

In case (Ia) take $p=x$.

In case (Ib), since $\gA$ is connected, there are $p_i\in F_\Sigma(X)$ and $q_i\in F_\Omega(Y)$ $(i=1,\ldots,m)$ such that
$p_i\Rightarrow^*_{\gA} a_iq_i$ hold. Now taking $p=\sigma(p_1,\ldots,p_m)$ we have $p=\sigma(p_1,\ldots,p_m) \Rightarrow^*_{\gA}\sigma(a_1q_1,\ldots,a_mq_m) \Rightarrow_{\gA} aq(q_1,\ldots,q_m)=aq$.

Next, assume that $l>1$ and that our statement has been proved for derivations of length less than $l$. Then $a\Rightarrow^*_G q$ can be written in the form $a \Rightarrow_G \overline{q}(a_1,\ldots,a_m) \Rightarrow^*_G \overline{q}(q_1,\ldots,q_m)=q$, where $\sigma(a_1,\ldots,a_m) \rightarrow a\overline{q}$ is in $P$ for some $\sigma \in \Sigma_m$ $(m>0)$ and $a_i\Rightarrow^*_G q_i$ $(1\leq i\leq m)$ if $\xi_i$ occurs in $\overline{q}$. By the induction hypothesis, for all such $i$ there exists a $p_i\in F_\Sigma(X)$ with $p_i\Rightarrow^*_{\gA} a_iq_i$. In the remaining cases, i.e., if $\xi_i$ does not occur in  $\overline{q}$, let $p_i\in F_\Sigma(X)$ and $q_i\in F_\Omega(Y)$ $(1\leq i\leq m)$ be arbitrary such that $p_i\Rightarrow^*_{\gA} a_iq_i$. Then $p=\sigma(p_1,\ldots,p_m)$ satisfies $p\Rightarrow^*_{\gA} aq$.
\item[(II)] Assume that $p\Rightarrow^*_{\gA} aq$ holds. We shall show by induction on $\hg(p)$ that the left side of (\ref{Equation.4.6.5.1}) is also valid.
If $\hg(p)=0$, then, by the choice of  $P'$, the right side of (\ref{Equation.4.6.5.1}) obviously implies its left side.

Now let $p=\sigma(p_1,\ldots,p_m)$ $(\sigma \in \Sigma_m, m>0)$, and assume that our statement has been proved for all trees from $F_\Sigma(X)$ with height less than $\hg(p)$. Moreover, let us write $p\Rightarrow^*_{\gA} aq$ in the form
$p \Rightarrow^*_{\gA}\sigma(a_1q_1,\ldots,a_mq_m) \Rightarrow_{\gA} a\overline{q}(q_1,\ldots,q_m)$, where $\sigma(a_1,\ldots,a_m) \rightarrow a\overline{q}$ is in $P$ and $p_i\Rightarrow^*_{\gA} a_iq_i$ $(i=1,\ldots, m)$. Then, by the definition of $P'$ and the induction hypothesis, we have $a \Rightarrow_G \overline{q}(a_1,\ldots,a_m) \Rightarrow^*_G \overline{q}(q_1,\ldots,q_m)=q$.
\epr
\end{itemize}

\

From Lemma \ref{Lemma.4.6.5}, using Theorems \ref{Theorem.4.2.7} and \ref{Theorem.4.4.9}, respectively, we get the following results.

\begin{cor}\label{Corollary.4.6.6} $\cLR$ preserves regularity.
\epr
\end{cor}

\begin{cor}\label{Corollary.4.6.7} $\cLR_R$ preserves regularity.
\epr
\end{cor}

A state $a\in A$ of an F-transducer $\gA=(\Sigma,X,A,\Omega,Y,P,A')$ is {\em nondeleting} \index{state!nondeleting} if there exist two trees $p\in \hat F_\Sigma(X\cup \Xi_1)$ and $q\in F_\Omega(Y\cup \Xi_1)$ such that $p(a\xi_1) \Rightarrow^* a'q$ for some $a'\in A'$ and $\xi_1$ occurs in $q$. Otherwise $a$ is {\em deleting}. \index{state!deleting} The state $a$ is {\em copying} \index{state!copying} if there are two trees $p\in \hat F_\Sigma(X\cup \Xi_1)$ and $q\in F_\Omega(Y\cup \Xi_1)$ such that $p(a\xi_1) \Rightarrow^* a'q$ for some $a'\in A'$ and $\xi_1$ occurs at least twice in $q$.

\begin{lm}\label{Lemma.4.6.8} Let $\gA=(\Sigma,X,A,\Omega,Y,P,A')$ be a connected F-transducer. If $\tau_{\gA}$ preserves regularity and $a\in A$ is copying, then $\range(\tau_{\gA(a)})$ is finite.
\end{lm}
\pr Assume that $\tau_{\gA}$ preserves regularity. Let $a\in A$ be a copying state, and take two trees $p\in \hat F_\Sigma(X\cup \Xi_1)$ and $q\in \hat F_\Omega(Y\cup \Xi_n)$ such that $p(a\xi_1) \Rightarrow^* a'q(\xi_1^n)$ where $a'\in A'$  and $n>1$. Suppose that $\range(\tau_{\gA(a)})$ is infinite. Then there is an $s\in \range(\tau_{\gA(a)})$ with $\hg(s)>k\cdot|A|$, where $k$ is the maximum of the heights of the right-hand sides of the productions in $P$. Let $r\in F_\Sigma(X)$ be a tree such that $r\Rightarrow^* as$. Since $\hg(s)>k\cdot|A|$, there are trees $r_1,r_2\in
\hat F_\Sigma(X\cup \Xi_1)$ and $r_3\in F_\Sigma(X)$ such that the following conditions are satisfied:
\begin{itemize}
\item[(i)] $r_1\big(r_2(r_3)\big)=r$,
\item[(ii)] $r_3\Rightarrow^* bs_3$, $r_2(b\xi_1)\Rightarrow^* bs_2$ and $r_1(b\xi_1)\Rightarrow^* as_1$ for some $b\in A$, $s_1,s_2\in
F_\Omega(Y\cup \Xi_1)$ and $s_3\in F_\Omega(Y)$,
\item[(iii)] $\hg(s_2)> 0$, and $\xi_1$ occurs in $s_1$ and $s_2$,
\item[(iv)] $s_1(s_2(s_3))=s$.
\end{itemize}
Therefore, for each $i(=0,1,\ldots)$, there is a derivation $p_i=p\big(r_1(r_2^i(r_3))\big)\Rightarrow^* a'q(t_i^n)=q_i$ where $t_i=s_1\big(s_2^i(s_3)\big)$ (the powers $t^i$ of any tree $t\in  F_\Sigma(X\cup \Xi_1)$ are defined thus: $t^0=\xi_1$, and $t^{i+1}=t(t^i)$ for each $i\geq 0$). Obviously,
$\hg(q_i)$ increases with $i$ when $i$ is large enough.

Now consider the forest $T=\{p_i\mid i=0,1,\ldots\}$. Obviously, $T$ is regular. Since $\tau_{\gA}$ preserves regularity, this implies that $T'=T\tau_{\gA}$ is also regular. Take an $\Omega Y$-recognizer $\bB=(B,\Omega,Y,\beta,B')$ with $T'=T(\bB)$. Choose an
$$i\geq \big(2k(\hg(p(r))+1)+2|B|\big)k\big(\hg(p(r))+1\big).$$
Then there exists a tree $t\in F_\Omega(Y)$ with $k\big(\hg(p(r))+1\big)+|B| \le \hg(t) < k\big(\hg(p(r))+1\big)+2|B|$ such that
\begin{equation*}
\tag{2}\label{Equation.4.6.8.2} \overline{q}=q(t,t_i^{n-1})
\end{equation*}
is also in $T'$. To prove the lemma it is enough to show that there exist no $j$ and $a''\in A'$ such that $p_j\Rightarrow^* a''\overline{q}$.
Suppose
$$p_j\Rightarrow^* a''q'({t'}^m)=a''\overline{q}$$
holds, where $a''\in A'$, $q'\in \hat F_\Omega(Y\cup \Xi_m)$, $r_3\Rightarrow^*b_1s_1'$, $r_2(b_l\xi_1)\Rightarrow^*b_{l+1}s_{l+1}'$ $\big(b_1,b_{l+1}\in A$,  $s_{l+1}\in  F_\Omega(Y\cup \Xi_1)$, $l=1,\ldots,j$, $s_1'\in F_\Omega(Y)\big)$, $r_1(b_{j+1}\xi_1)\Rightarrow^*b_{j+2}s_{j+2}'$
$\big(b_{j+2}\in A$,  $s'_{j+2}\in  F_\Omega(Y\cup \Xi_1)\big)$, $p(b_{j+2}\xi_1)\Rightarrow^*a''s_{j+3}'=a''q'$ and $t'=s'_{j+2}\big(s'_{j+1}(\ldots(s_1')\ldots)\big)$.

By the choice of $i$, there exists a $u$ $(2\leq u \leq j+3)$ such that $\xi_1$ occurs in $s'_u,s'_{u+1},\ldots,s'_{j+3}$ but $\xi_1$ does not occur in $s'_{u-1}$. Moreover, let $u-1\leq u_1<\ldots<u_v\leq j+3$ be a maximal sequence with $1\leq \hg(\overline{s}_{u_1})< \ldots < \hg(\overline{s}_{u_v})$, where $\overline{s}_l=s'_l\big(s'_{l-1}(\ldots(s'_1)\ldots)\big)$ $(l=1,\ldots,j+3)$.
Then $v\geq 2k\big(\hg(p(r))+1\big)+2|B|$. Taking into consideration that $\hg(t)\geq k(\hg(p(r))+1)+|B|$ (and $|B|\geq 1$), for an $l$
($2\leq l\leq v$), the word $w$ forming $\ppath_1(q)$ is a subword of a word in $\ppath_1\big(s'_{j+3}(s'_{j+2}(\ldots(s'_{u_l})\ldots))\big)$. (Informally speaking, this means that there is a word in  $\ppath_1\big(s'_{j+3}(s'_{j+2}(\ldots(s'_{u_l})\ldots))\big)$  going through the root of $t$.) Therefore, we have $l\big(\ppath_1(q)\big)+\hg(t) \geq 2k\big(\hg(p(r))+1\big)+2|B|$. But, by (\ref{Equation.4.6.8.2}) and the choice of $t$, $l\big(\ppath_1(q)\big)+\hg(t) < 2k\big(\hg(p(r))+1\big)+2|B|$, which is a contradiction. \epr

\begin{lm}\label{Lemma.4.6.9} Let $\gA=(\Sigma,X,A,\Omega,Y,P,A')$ be a connected F-transducer such that for every copying state $a\in A$, $\range(\tau_{\gA(a)})$ is finite. Then $\gA$ is equivalent to a linear F-transducer.
\end{lm}
\pr Suppose that $a_1,\ldots,a_k$ are all the copying states of $\gA$. Let $T_i=\range(\tau_{\gA(a_i)})$ $(i=1,\ldots,k)$.  Moreover, set $T=\bigcup(T_i\mid i=1,\ldots,k)$. By our assumptions, $T$ is finite.

Define an F-transducer $\gB=(\Sigma,X,B,\Omega,Y,P',B')$, where
$$B=(A-\{a_i\mid i=1,\ldots,k\})\cup \bigcup(\{a_i\}\times T_i\mid i=1,\ldots,k)$$
and
$$B'=(A'\cup A'\times T)\cap B.$$
Moreover, $P'$ is given as follows:
\begin{itemize}
\item[(i)] If $p\rightarrow aq$ ($p\in \Sigma_0\cup X$) is in $P$ and $a=a_i$ for some $i$ $(1\le i\le k)$, then $p\rightarrow (a,q)q$ is in $P'$. If $a\not\in \{ a_1,\ldots,a_k\}$, then $p\rightarrow aq$ itself is in $P'$.
\item[(ii)] Let
$$\sigma(b_1,\ldots,b_m) \rightarrow aq(\xi_1,\ldots,\xi_m)$$
$(\sigma \in \Sigma_m, m>0, b_1,\ldots,b_m,a\in A, q\in F_\Omega(Y\cup \Xi_m))$ be in $P$. We distinguish the following cases:
\begin{itemize}
\item[(iia)] The state $a$ is deleting. Fix any $\overline{q}\in F_\Omega(Y\cup \Xi_m)$ such that every $\xi_i$ occurs at most once in $\overline{q}$. Then $P$ contains every linear production $\sigma(c_1,\ldots,c_m) \rightarrow a\overline{q}(\xi_1,\ldots,\xi_m)$
such that
$$c_j
=\left\{
\begin{array}{ll}
 (b_j,q_j) (q_j\in T_l) & \text{ if $b_j$ is copying and $b_j=a_l$},\\
b_j & \text{  otherwise.}
\end{array}
\right.
$$
\item[(iib)] The state $a$ is nondeleting but not copying. Then all productions
$$\sigma(c_1,\ldots,c_m) \rightarrow aq(\eta_1,\ldots,\eta_m)$$
are in $P'$ where for each $j(=1,\ldots,m),$
$$c_j
=\left\{
\begin{array}{ll}
 (b_j,q_j)\mbox{\hspace{2mm}} (q_j\in T_l) & \text{ if $b_j$ is copying and $b_j=a_l$},\\
b_j & \text{  otherwise}
\end{array}
\right.
$$
and
$$\eta_j
=\left\{
\begin{array}{ll}
 \xi_j & \text{ if $\xi_j$ occurs at most once in  $q$},\\
q_j (=\pi_2(c_j))& \text{  otherwise.}
\end{array}
\right.
$$
(Observe that if $\xi_j$ occurs at least twice in $q$, then $b_j$ is copying.)
\item[(iic)] The state $a$ is copying. Then $P'$ contains all productions
$$\sigma(c_1,\ldots,c_m) \rightarrow (a,\overline{q})\overline{q}$$
where $\overline{q}=q(\eta_1,\ldots,\eta_m)$ and for each $j(=1,\ldots,m),$
$$c_j
=\left\{
\begin{array}{ll}
 (b_j,q_j)\mbox{\hspace{2mm}} (q_j\in T_l) & \text{ if $b_j$ is copying and $b_j=a_l$},\\
b_j & \text{  otherwise}
\end{array}
\right.
$$
and
$$\eta_j
=\left\{
\begin{array}{ll}
 q_j & \text{ if $\xi_j$ occurs in  $q$},\\
\text{any fixed tree from $F_\Omega(Y)$} & \text{  otherwise.}
\end{array}
\right.
$$
\end{itemize}
(Note that $b_j$ is copying if $\xi_j$ occurs in $q$.)
\end{itemize}
This ends the construction of $P'$. Obviously, $\gB$ is an LF-transducer.

We show that $\gA$ is equivalent to $\gB$.
\begin{itemize}
\item[(I)] Assume that $p\Rightarrow^*_{\gA} aq$ $(p\in F_\Sigma(X), q\in F_\Omega(Y), a\in A)$ holds. We prove that
\begin{itemize}
\item[(Ia)] $p\Rightarrow^*_{\gB} aq$ if $a$ is nondeleting but not copying,
\item[(Ib)] $p\Rightarrow^*_{\gB} (a,q)q$ if $a$ is  copying,
\item[(Ic)] $p\Rightarrow^*_{\gB} a\overline{q}$ for some $\overline{q}\in F_\Omega(Y)$ if $a$ is  deleting.
\end{itemize}

We shall proceed by induction on $\hg(p)$. If $\hg(p)=0$ then, by (i), (Ia), (Ib) and (Ic) obviously hold.

Next let $p=\sigma(p_1,\ldots,p_m)$ $(\sigma\in\Sigma_m, m>0)$, and write $p\Rightarrow^*_{\gA} aq$ in the more detailed form
$$\sigma(p_1,\ldots,p_m)\Rightarrow^*_{\gA}\sigma(b_1q'_1,\ldots,b_mq'_m) \Rightarrow_{\gA} aq'(q'_1,\ldots,q'_m)=aq$$
where $\sigma(b_1,\ldots,b_m)\rightarrow aq'$ is in $P$ and for each $j$ $(1\le j\le m)$, $p_j\Rightarrow^*_{\gA} b_jq'_j$. Then, by the induction hypothesis, for all $j(=1,\ldots,m)$, we have $p_j\Rightarrow^*_{\gB} c_jq_j$, where
\begin{itemize}
\item[(Ia$'$)] $c_j=b_j$ and $q_j=q'_j$ if $b_j$ is nondeleting and not copying,
\item[(Ib$'$)] $c_j=(b_j,q_j)$ and $q_j=q'_j$ if $b_j$ is copying,
\item[(Ic$'$)] $c_j=b_j$ and $q_j=\overline{q}_j$ for some $\overline{q}_j\in F_\Omega(Y)$ if $b_j$ is  deleting.
\end{itemize}

Therefore:
\begin{itemize}
\item[(Ia$''$)] If $a$ is nondeleting but not copying, then the production
$$\sigma(c_1,\ldots,c_m) \rightarrow aq'(\eta_1,\ldots,\eta_m)$$
is in $P'$, were $\eta_j$ $(j=1,\ldots,m)$ is given by (iib).
\item[(Ib$''$)] If $a$ is  copying then the production
$$\sigma(c_1,\ldots,c_m) \rightarrow (a,\overline{q}')\overline{q}'$$
with $\overline{q}'=q'(\eta_1,\ldots,\eta_m)$
is in $P'$, were $\eta_j$ $(j=1,\ldots,m)$ is given by (iic).
\item[(Ic$''$)] If $a$ is  deleting then the production
$$\sigma(c_1,\ldots,c_m) \rightarrow a\overline{q}'$$
given by (iia) is in $P'$.
\end{itemize}

Thus, in all three cases the required derivations in $\gB$ exist.\\[2mm]
\item[(II)] Assume that one of the following relations hold:
\begin{itemize}
\item[(IIa)] $p\Rightarrow^*_{\gB} aq$ or
\item[(IIb)] $p\Rightarrow^*_{\gB} (a,q)q$
\end{itemize}
where $p\in F_\Sigma(X)$, $q\in F_\Omega(Y)$ and $a\in A$.

Then, by reversing the above computation, one can show that the desired derivations
\begin{itemize}
\item[(IIc)]  $p\Rightarrow^*_{\gA} aq$ if $a$ is nondeleting,
\item[(IId)] $p\Rightarrow^*_{\gA} a\overline{q}$ for some $\overline{q}\in F_\Omega(Y)$ if $a$ is  deleting
\end{itemize}
exist. Since the final states are nondeleting, this ends the proof of the lemma.
\epr
\end{itemize}

\

We can now state and prove

\begin{thm}\label{Theorem.4.6.10} Let $\gA=(\Sigma,X,A,\Omega,Y,P,A')$ be an arbitrary F-transducer. Then $\tau_{\gA}$ preserves regularity iff $\gA$ is equivalent to an LF-transducer.
\end{thm}
\pr If $\gA$ is equivalent to an LF-transducer then, by Lemma \ref{Lemma.4.6.5}, $\tau_{\gA}$ preserves regularity.

Conversely, let $\tau_{\gA}$ preserve regularity. We may assume that $\gA$ is connected. Then by Lemmas \ref{Lemma.4.6.8} and \ref{Lemma.4.6.9},
$\gA$ is equivalent to an LF-transducer.\epr

\

From Example \ref{Example.2.4.15}, we directly obtain

\begin{thm}\label{Theorem.4.6.11} Neither $\cF$ nor $\cR$ preserves regularity. \epr
\end{thm}

The following result shows that $\Surf(\cF)\subset \Surf(\cR)$. More precisely, we have

\begin{thm}\label{Theorem.4.6.12} $\Surf(\cF)=\Surf(\cH)$ and $\Surf(\cH)$ is a proper subclass of $\Surf(\cR)$.
\end{thm}
\pr The first statement of Theorem \ref{Theorem.4.6.12} follows from Theorem \ref{Theorem.4.3.3} and Lemma \ref{Lemma.4.6.5}.

It is obvious that $\Surf(\cH) \subseteq \Surf(\cR)$. We show that the inclusion is proper. For this, consider Example \ref{Example.4.1.6}.
Moreover, let $S= \{\omega_2(\omega_1^n(y_1),\omega_1^n(y_2))\mid n=0,1,\ldots \}$. If $R$ denotes the regular forest
$\{\sigma(x)\}\cdot_x\{\sigma(x)\}^{*x}$, then $R\tau_{\gA}=S$. Therefore, $S\in \Surf(\cR)$.

Assume that for an HR-transducer $\gB=(\Delta,Z,\{b_0\},\Omega,Y,P',b_0)$ and regular forest $T\subseteq F_\Delta(Z)$, we have $S=T\tau_\gB$. Then $\gB$ can be chosen linear since in the opposite case in $T\tau_\gB$ there is a tree with at least two occurrences of a subtree. Therefore, by
Theorem \ref{Theorem.2.4.16}, $S$ is regular. But one can show similarly as in Example \ref{Example.2.4.15} that $S$ is not regular. \epr

\

Next we show some closure properties of surface forests which will be needed also in Section \ref{Section.4.7}.

\begin{thm}\label{Theorem 4.6.13} Let $S\in\Surf(\cF)$ and let $T$ be a recognizable forest. Then $S\cap T\in \Surf(\cF)$.
\end{thm}
\pr Let $\tau_1\subseteq F_\Sigma(X)\times F_\Omega(Y)$ be an F-transformation and $S=R\tau_1$ where $R\in\Rec(\Sigma,X)$. Take an arbitrary regular forest $T\subseteq F_\Omega(Y)$. Denote by $\tau_2\subseteq F_\Omega(Y)\times F_\Omega(Y)$ the DF-transformation given in the proof of Lemma \ref{Lemma.4.1.11} which corresponds to $T$. Then $R\tau_1\circ\tau_2=S\cap T$. But, by Theorem \ref{Theorem.4.3.9}, $\tau_1\circ\tau_2$ is an F-transformation.\epr

\

For R-surface forests we have a similar result.

\begin{thm}\label{Theorem.4.6.14} The intersection of an {\rm R}-surface forest with a regular forest is again an {\rm R}-surface forest.
\end{thm}
\pr The proof is similar to that of the previous theorem, but now we shall use the fact that the transformation given in the proof of Lemma \ref{Lemma.4.1.11} is an LNR-transformation. Moreover, by Theorem \ref{Theorem.4.3.15}, the composition of an R-transformation by an LNR-transformation is again an R-transformation. \epr

\

By Theorem \ref{Theorem.4.3.7}, $\cDF$ is closed under composition. Therefore, $\Surf(\cDF)$ is closed under DF-transformations. Although $\cDR$ is not closed under composition, we shall show that $\Surf(\cDR)$ is closed under DR-transformations. For this, we need

\begin{thm}\label{Theorem.4.6.15} Let $\gA=(\Sigma,X,A,\Omega,Y,P,a_0)$ and $\gB=(\Omega,Y,B,\Delta,Z,P',b_0)$ be any {\rm DR}-transducers. Then there exists a {\rm DR}-transducer $\gC=(\Sigma,X,C,\Delta,Z,P'',c_0)$ such that for every $R\subseteq F_\Sigma(X)$, $S\tau_{\gC}=R\tau_{\gA}\circ \tau_{\gB}$, where
$S=R\cap \dom(\tau_{\gA}\circ \tau_{\gB})$.
\end{thm}
\pr Let $C=A\times B$ and $c_0=(a_0,b_0)$. We want to define  $P''$ in such a way that whenever $ap\Rightarrow^*_{\gA} q$ $\big(a\in A, p\in F_\Sigma(X), q\in F_\Omega(Y)\big)$ and $bq\Rightarrow^*_{\gB} r$ $\big(b\in B, r\in F_\Delta(Z)\big)$ hold, then $(a,b)p\Rightarrow^*_{\gC} r$. If $p\in \Sigma_0\cup X$, then $(ap,q)\in P$. If we put the production $(a,b)p\rightarrow r$ in $P''$, $\gC$ will have the desired property for these $a,b,p,q$ and $r$.

Now let $p=\sigma(p_1,\ldots,p_m)$ $(\sigma\in\Sigma_m, m>0)$ and suppose
$$ap=a\sigma(p_1,\ldots,p_m)\Rightarrow_\gA \overline{q}(\ldots, a_{ij}p_i,\ldots)\Rightarrow^*_\gA \overline{q}(\ldots, q_{ij},\ldots)=q,$$
where $(a\sigma,\overline{q}(\ldots, a_{ij}\xi_i,\ldots))\in P$ ($\overline{q}\in \hat F_\Omega(Y\cup \Xi_n)$ for some $n$) and $a_{ij}p_i\Rightarrow^*_{\gA} q_{ij}$, i.e., the considered copy of $p_i$ is translated by $\gA$ starting in state $a_{ij}$ into $q_{ij}$. Furthermore, suppose that applying to $q$ the transducer $\gB$ starting in $b$, we get
$$bq =b\overline{q}(\ldots, q_{ij},\ldots)\Rightarrow^*_\gB\overline{r}(\ldots,b_{ij1}q_{ij},\ldots,b_{ijk}q_{ij},\ldots)\Rightarrow^*_\gB\overline{r}(\ldots,r_{ij1},\ldots,r_{ijk},\ldots)=r$$
$$(b\overline{q}\Rightarrow^*_\gB\overline{r}, \overline{r}\in F_\Delta(Z\cup \Xi_n), b_{ijl}q_{ij}\Rightarrow^*_\gB r_{ijl}, l=1,\ldots,k)$$ (meaning that the given occurrence of $q_{ij}$ in $\overline{q}$
has $k$ translations by $\gB$ starting the translations in states $b_{ij1},\ldots, b_{ijk}$). Thus, if we have the production
$$(a,b)\sigma \rightarrow \overline{r}\big(\ldots,(a_{ij},b_{ij1})\xi_i,\ldots,(a_{ij},b_{ijk})\xi_i,\ldots\big)$$
in $P''$ and suppose that $\gC$ has the required property for trees with height less than $\hg(p)$, then
$(a,b)p\Rightarrow^*_{\gC} r$ also holds. Accordingly, the formal definition of $P''$ reads as follows:
\begin{itemize}
\item[(i)] The production $(a,b)x\rightarrow r$ $\big((a,b)\in C, x\in X, r\in F_\Delta(Z)\big)$ is in $P''$ if there is an $ax\rightarrow q$ in $P$ such that $bq \Rightarrow^*_\gB r$.
\item[(ii)] If the production $a\sigma \rightarrow q(\ba_1\xi_1^{n_1},\ldots,\ba_m\xi_m^{n_m})$ $\big(a\in A, \sigma\in\Sigma_m, m\geq 0, \ba_i\in A^{n_i}, i=1,\ldots, m, \, n_1+\ldots +n_m=n, q\in \hat F_\Omega(Y\cup\Xi_n)\big)$ is in $P$ and
$$bq\Rightarrow^*_{\gB} r\big(\bb_{11}\xi_{11}^{n'_{11}},\ldots,\bb_{1n_1}\xi_{1n_1}^{n'_{1n_1}},\ldots,
\bb_{m1}\xi_{m1}^{n'_{m1}},\ldots,\bb_{mn_m}\xi_{mn_m}^{n'_{mn_m}}\big)$$
$\big(\bb_{ij}\in B^{n'_{ij}},\, \xi_{ij}=\xi_{n_1+\ldots +n_{i-1}+j},\, i=1,\ldots,m, \, j=1,\ldots,n_i, \, n'_{11}+\ldots + n'_{mn_m}=n', r\in \hat F_\Delta(Z\cup \Xi_{n'})\big)$ holds, then the production
$$(a,b)\sigma\rightarrow r\big((a_{1_1}^{n'_{11}}\bb_{11},\ldots,a_{1_{n_1}}^{n'_{1n_1}}\bb_{1n_1})\xi_1^{k_1},\ldots,(a_{m_1}^{n'_{m1}}\bb_{m1},\ldots,a_{m_{n_m}}^{n'_{mn_m}}\bb_{mn_m})\xi_m^{k_m}\big)$$
in in $P''$, where $k_i=n'_{i1}+\ldots +n'_{in_i}$ $(i=1,\ldots,m)$.
\end{itemize}

Obviously, $\gC$ is a DR-transducer. Moreover, to prove the theorem it is enough to show that for arbitrary $(a,b)\in C, p\in F_\Sigma(X), q\in F_\Omega(Y)$ and $r\in F_\Delta(Z)$, $ap\Rightarrow^*_\gA q$
and $bq\Rightarrow^*_\gB r$ jointly imply $(a,b)p\Rightarrow^*_\gC r$. This can be proved by induction on $\hg(p)$. \epr

\

Let us note that the $\gC$ constructed above may delete certain subtrees of input trees so that $\dom(\tau_\gC)$ becomes larger than $\dom (\tau_\gA\circ \tau_\gB)$.

If $R$ in Theorem \ref{Theorem.4.6.15} is regular then, by Corollary \ref{Corollary.4.3.17} and Theorem \ref{Theorem.2.4.2}, $S$ is also regular. Thus we have

\begin{cor}\label{Corollary.4.6.16} $\Surf(\cDR)$ is closed under {\rm DR}-transformations.
\epr
\end{cor}

\

%% file: Section.4.7.tex
In Section \ref{Section.4.3} it has been shown that neither $\cF$ nor $\cR$ is closed under composition.
In the next section we shall prove that compositions of $n$ F-transformations or $n$ R-transformations
lead to proper hierarchies when $n$ assumes the values $0,1,2,\ldots$.

The purpose of this section is to introduce concepts and present results needed in Section \ref{Section.4.8}.

Let $K$ be a class of forests and $\cS$ a class of tree transformations. Then $\cS(K)$
denotes the class $\{T\tau \mid T\in K, \tau \in \cS\}$. Moreover, $\yd\,\cS(K)$ will stand for
$\{\yd(T) \mid T \in \cS(K)\}$.

\begin{df}\rm\label{Definition.4.7.1} Let $\Sigma$ be a ranked alphabet and $X$ an alphabet. Let $f$ be a mapping which associates with each $d\in\Sigma\cup X$ a nonvoid recognizable forest $T_d\subseteq F_{\Omega(d)}(\Xi_1)$ where $\Omega(d)$ is a ranked alphabet consisting of unary operational symbols only. It is also supposed that $\Omega(d)$ is disjoint with $\Sigma$.

Now define the mapping $\overline{f}$ from the set of all $\Sigma X$-forests into the set of subsets of $F_{\Sigma\cup\Omega}(X)$ $\big(\Omega=\bigcup(\Omega(d)\mid d\in \Sigma\cup X)\big)$ in the following way:
\begin{enumerate}
\item[(i)] if $p\in \Sigma_0\cup X$, then $\overline{f}(p)=\{q(p)\mid q\in T_p\}$,
\item[(ii)] if $p=\sigma(p_1,\ldots,p_m)$ $(\sigma\in \Sigma_m, m>0, p_1,\ldots,p_m\in F_\Sigma(X))$, then
$$\overline{f}(p)=\{q(\sigma(q'_1,\ldots,q'_m))\mid q\in T_\sigma, q'_i\in \overline{f}(p_i), i=1,\ldots,m\},\mbox{\hspace{5mm}}\text{and}$$
\item[(iii)] if $T\subseteq F_\Sigma(X)$, then $\overline{f}(T)=\bigcup\big(\overline{f}(p)\mid p\in T\big).$
\end{enumerate}
\end{df}

The mapping $\overline{f}$ is called a {\em regular insertion}.\index{regular insertion}

In the sequel we shall write simply $f$ for $\overline{f}$.

The above regular insertion can be interpreted as follows: $f$ inserts directly below each node of a tree $p\in F_\Sigma(X)$ a unary tree from the regular forest $T_d$ if the label of the node in question is $d$. The insertion of $\xi_1$ means that the given node is unchanged. The name ``regular insertion'' is more expressive if trees are given in Polish prefix form. In this case $f$ inserts a word from $T_d$ directly before an occurrence of $d$ in the word $p$.

\begin{lm}\label{Lemma.4.7.2} $\Rec$ is closed under regular insertion.
\end{lm}
\pr Let $T\subseteq F_\Sigma(X)$ be a regular forest and $f$ a regular insertion given by $f(d)=T_d$ $(d\in \Sigma\cup X, T_d\subseteq F_\Omega(\Xi_1))$. Consider a regular tree grammar $G=(N,\Sigma,X,P,a_0)$ given in normal form such that $T(G)=T$. Moreover, for every $T_d$ $(d\in \Sigma\cup X)$ let $G^d=(N^d,\Omega,\Xi_1,P^d,a^d_0)$ be a regular tree grammar in normal form generating $T_d$. For each $d\in \Sigma\cup X$ and $a\in N$ consider the tree grammar $G^d_a=(N^d_a,\Omega,\Xi_1,P^d_a,(a^d_0,a))$, where $N^d_a=N^d\times \{a\}$ and
$$P^d_a=\{(a^d,a)\rightarrow \omega((b^d,a))\mid a^d\rightarrow \omega(b^d)\in P^d \}\cup \{ (a^d,a)\rightarrow\xi_1 \mid a^d\rightarrow \xi_1\in P^d\}.$$
Obviously, $T(G^d_a)=T_d$ holds for each $d\in \Sigma\cup X$ and $a\in N$.

Assume that the sets of nonterminal symbols of the grammar $G^d$ $(d\in \Sigma\cup X)$ are pairwise disjoint and also disjoint with $N$ and $N\times (\Sigma\cup X)$. Construct the tree grammar $G'=(N',\Sigma\cup \Omega, X, P',a_0)$, where $N'=\bigcup(N^d_a \mid d\in \Sigma\cup X, a\in N)\cup N\cup N\times (\Sigma\cup X)$ and $P'$ is given as follows:
\begin{align*}
P' & = \{a\rightarrow (a^d_0,a)\mid a\in N, d\in \Sigma\cup X\}\\
& \cup \bigcup\big(P^d_a-\{(a^d,a)\rightarrow\xi_1\mid a^d\in N^d\}\mid a\in N, d\in \Sigma\cup X\big)\\
& \cup \{ (a^d,a)\rightarrow (a,d) \mid a^d\rightarrow \xi_1 \in P^d_a, a^d\in N^d, a\in N, d\in \Sigma\cup X\}\\
& \cup \{(a,\sigma)\rightarrow\sigma(a_1,\ldots,a_m)\mid a\rightarrow \sigma(a_1,\ldots,a_m)\in P, \sigma\in\Sigma_m, m>0,a,a_1,\ldots,a_m\in N\}\\
& \cup \{(a,d)\rightarrow d\mid a\rightarrow d\in P, a\in N, d\in \Sigma_0\cup X\}.
\end{align*}

From the construction of $G'$ it is obvious that the following statements are valid:
\begin{itemize}
\item[(ia)] For any production $a\rightarrow \sigma(a_1,\ldots,a_m)\in P$ $(\sigma \in \Sigma_m, m>0)$ and tree $q\in T_\sigma$ there exists a derivation in $G'$
$$a\Rightarrow(a_0^\sigma,a)\Rightarrow ^*q((a^\sigma,a))\Rightarrow q((a,\sigma))\Rightarrow q(\sigma(a_1,\ldots,a_m))\mbox{\hspace*{10mm}}(a^\sigma\in N^\sigma).$$
\item[(ib)] For any production $a\rightarrow d\in P$ $(d \in \Sigma_0\cup X)$ and tree $q\in T_d$ there exists a derivation in $G'$, $a\Rightarrow(a_0^d,a)\Rightarrow ^*q((a^d,a))\Rightarrow q((a,d))\Rightarrow d \quad (a^d\in N^d)$.
\end{itemize}
Conversely,
\begin{itemize}
\item[(ii)] for any $a\in N$ and $p\in F_{\Sigma\cup \Omega}(X)$ each derivation $a\Rightarrow^*_{G'}p$ should have the form
\begin{itemize}
\item[(iia)] $a\Rightarrow  (a_0^\sigma,a) \Rightarrow q_1((a_1^\sigma,a))\Rightarrow \ldots\Rightarrow q_n((a_n^\sigma,a))
\Rightarrow q_n((a,\sigma))\Rightarrow$\\[2mm]
\mbox{\hspace{35mm}}$q_n(\sigma(a_1,\ldots,a_m))\Rightarrow^* p$\\[2mm]
for some $a\rightarrow \sigma(a_1,\ldots,a_m)\in P, q_n\in T_\sigma, \sigma\in \Sigma_m, m>0$ and $a_0^\sigma,\ldots,a_n^\sigma\in N^\sigma$,
or the form
\item[(iib)] $a\Rightarrow  (a_0^d,a) \Rightarrow q_1((a_1^d,a))\Rightarrow \ldots\Rightarrow q_n((a_n^d,a))
\Rightarrow q_n((a,d))\Rightarrow q_n(d)$ for some $a\rightarrow d\in P, q_n\in T, d\in\Sigma_0\cup X$ and $a_0^d,\ldots,a_n^d\in N^d$.
\end{itemize}
\end{itemize}

Properties (ia), (ib), and (ii) obviously imply that $T(G')=f(T)$. \epr

\begin{lm}\label{Lemma.4.7.3} Let $K$ be a class of forests closed under regular insertion. Then $\cR(K)$ is also closed under regular insertion.
\end{lm}
\pr Let $R\in K$ be an arbitrary $\Sigma X$-forest and take an R-transducer $\gA= (\Sigma, X, A, \Omega, Y, P, A')$. Set $S=R\tau_{\gA}$. Moreover, for every $d\in\Sigma\cup X$ take a unary operator $\#_{d}$, and let $f$ be the regular insertion given by
$f(d)=\{\#_{d}(\xi_{1})\}^{*\xi_{1}}.$

First we shall show that if $g$ is a regular insertion for which $g(d)=\{\#(\xi_{1})\}^{*\xi_{1}}$ $(d\in\Omega \cup Y)$, then
$g(S) \in \cR(K)$.

Construct the R-transducer $\gB=(\Sigma \cup \{\#_{d}\mid d\in\Sigma \cup X\}, X, B, \Omega \cup\{\#\},Y,P',A')$ with $B=A\cup C$, where $C=\{\overline{p}\mid p\in \big(\bigcup (\sub(q)\mid q \text{ is the right-hand side of a rule in } P)-\Xi\big)\}$. Moreover, $P'$ is the union of the following ten sets of productions:

\begin{align*}
 P_{1}= & \{a\#_{d}\rightarrow\#(a\xi_{1}), a\#_d\rightarrow a\xi_{1}\mid a\in A, d\in\Sigma \cup X\}, \\[2mm]
 P_{2}= & \{a\#_{d}\rightarrow\omega(\overline{q}_{1}\xi_{1}, \ldots,\overline{q}_{m}\xi_{1})\mid ad\rightarrow q  \text{ is in $P$ for some } \\
&  d\in\Sigma\cup X, a\in A, q=\omega(q_{1}, \ldots, q_m), \omega\in\Omega_{m},\ m>0\}, \\[2mm]
P_3=& \{a\#_d\rightarrow \overline{q}\xi_{1}\mid ad\rightarrow q \text{ is in $P$ for some } d\in\Sigma,
 q=a'\xi_{i}, a,a'\in A\}, \\[2mm]
 P_{4}=& \{a\#_{d}\rightarrow\overline{q}\xi_{1}\mid ad\rightarrow q \text{ is in $P$ for some } d\in\Sigma \cup X, a\in A,  q=\omega\in \Omega_{0}\}, \\[2mm]
 P_{5}=& \{a\#_{d}\rightarrow\overline{q}\xi_{1}\mid ad\rightarrow q \text{ is in $P$ for some } d\in\Sigma \cup X, a\in A, q=y\in Y\},\\[2mm]
 P_{6}=& \{\overline{q}\#_{d}\rightarrow \#(\overline{q}\xi_{1}),\, \overline{q}\#_d \rightarrow\omega(\overline{q}_{1}\xi_{1},
 \ldots,\overline{q}_{m}\xi_{1}),\, \overline{q}\#_{d}\rightarrow\overline{q}\xi_{1}\mid \\
 & q=\omega(q_{1}, \ldots,q_{m}), \omega\in\Omega_{m},\ m>0\}, \\[2mm]
 P_7=& \{\overline{a\xi_{i}}\#_{d}\rightarrow\#(\overline{a\xi_{i}}\xi_{1}),\, \overline{a\xi_i }\#_{d}\rightarrow\overline{a\xi_{i}}\xi_{1},\,
\overline{\omega}\#_{d}\rightarrow \#(\overline{\omega}\xi_{1}),\, \overline{\omega}\#_{d}\rightarrow\overline{\omega}\xi_{1}, \\
 & \overline{y}\#_{d}\rightarrow\#(\overline{y}\xi_{1}),\,\overline{y}\#_{d}\rightarrow\overline{y}\xi_{1}\mid 1\leq i\leq r(P),
r(P) \text{ is the maximum of ranks} \\
 & \text{of the operators appearing in the left-hand sides of productions from } P, \\
 & a\in A, \omega\in\Omega_{0}, y\in Y\},\\[2mm]
 P_{8}=& \{\overline{a\xi_{i}}\sigma\rightarrow a\xi_{i}\mid a\in A, \sigma\in\Sigma_{m}, m>0, 1\leq i\leq m\}, \\[2mm]
 P_9 =& \{\overline{\omega}d\rightarrow \omega \mid \omega\in\Omega_{0},  d\in\Sigma \cup X\} \text{ and} \\[2mm]
P_{10}=& \{\overline{y}d\rightarrow y\mid y\in Y,\ d\in\Sigma \cup X\}.
\end{align*}

One can easily see that $\gB$ works as follows: assume that for some $ a\in A,p\in F_\Sigma(X)$ and $q\in F_{\Omega}(Y)$ a derivation $ap\Rightarrow_{\gA}^{*}q$ exists. Let $q'$ be a tree obtained by inserting in $q$ arbitrary trees from $\{\#(\xi_1)\}^{*\xi_{1}}$ below symbols from $\Omega \cup Y$. Then for a $p'\in f(p)$, $ap'\Rightarrow^*_\gB q'$ holds. Conversely, if for some $a\in A, p\in F_\Sigma(X), p'\in f(p)$ and
$q'\in F_{\Omega\cup\{\#\}}(Y)$ a derivation $ap'\Rightarrow^*_\gB q'$ holds then there is a $q\in F_\Omega(Y)$ such that $q'\in g(q)$ and $ap\Rightarrow_{\gA}^{*}q$.

Now, consider an arbitrary regular insertion $h$ (into $\Omega Y$-trees). For each $d\in \Omega\cup Y$, there is a regular tree grammar $G_d=(N_d,\Omega(d),\Xi_1,P_d,\{a_{d_0}\})$ such that $h(d)=T(G_d)$. We may assume
that every $G_d$ is in normal form. Since $\Omega(d)$ is unary, this means that the productions of $G_d$ are of the form $a_d\rightarrow \omega_d(a'_d)$ or $a_d\rightarrow \xi_1$ $(a_d,a'_d\in N_d,\omega_d\in \Omega(d))$. Furthermore we may assume that the sets $N_d$ are pairwise disjoint. Now construct the R-transducer
\[\gC=(\Omega\cup\{\#\},Y,C,\Delta,Y,P'',C')\]
with
\[C=\bigcup(N_d\mid d\in \Omega\cup Y),\mbox{\hspace{10mm}}C'=\{ a_{d_0}\mid d\in \Omega\cup Y\}\]
and\[ \Delta=\bigcup\big(\Omega(d)\mid d \in \Omega\cup Y\big)\cup\Omega\mbox{\hspace{5mm}}
\big(\Delta_1=\bigcup\big(\Omega(d)\mid d \in \Omega\cup Y\big)\cup\Omega_1,\Delta_m=\Omega_m\,\, (m\neq 1)\big).\]
Furthermore, $P''$ is given as follows:
\begin{itemize}
\item[(I)] $a_d\#\rightarrow \omega_d(a'_d\xi_1)$ $(a_d,a'_d\in N_d, \omega_d\in \Omega(d), d\in \Omega\cup Y)$ 
is in $P''$ if $a_d\rightarrow \omega_d(a'_d)$ is in $P_d$.
\item[(II)] $a_\omega\omega\rightarrow \omega(a_{d_{1_0}}\xi_1,\ldots,a_{d_{m_0}}\xi_m)$ is in $P''$ for $\omega\in \Omega_m, m\geq 0, d_1,\ldots,d_m\in\Omega\cup Y$ \\[2mm]
and $a_\omega\in N_\omega$ if $a_\omega\rightarrow \zeta_1$ is in $P_\omega$.
\item[(III)] For each $y\in Y$ and $a_y\in N_y, a_yy\rightarrow y$ is in $P''$ if $ßa_y\rightarrow \xi_1$ is in $P_y$.
\end{itemize}
Obviously, $\gC$ is an R-relabeling. Therefore, by Theorem \ref{Theorem.4.3.15}, $\tau_\gB\circ\tau_\gC = \tau$ is an R-transformation. Moreover, by the constructions of $\gB$ and $\gC$, it is clear that the equality $h(S)=f(R)\tau$ holds. \epr

\

In the next section we shall need

\begin{thm}\label{Theorem.4.7.4} Let $\tau:X^* \rightarrow Y^*$ be a mapping induced by a deterministic $\rm gsm$ and $\Sigma$ a ranked alphabet. Then there exist a ranked alphabet $\Omega$ and a $\mathrm{DR}_\mathrm{R}$-transducer $\gB=(\Sigma,X,B,\Omega,Y,P',b_0)$ such that the equality
$\yd(T)\tau=\yd(T\tau_\gB)$ holds for every $T\subseteq F_\Sigma(X)$.
\end{thm}
\pr Consider the deterministic gsm $\bA=(X,A,Y,a_0,P,A')$ inducing $\tau$. We shall show the existence of a ranked alphabet $\Omega$ and that of a $\mathrm{DR}_\mathrm{R}$-transducer $\gB=(\Sigma,X,B,\Omega,Y,P',b_0)$ such that for any $p\in F_\Sigma(X)$,
\begin{enumerate}
\item[(i)] $\yd(p\tau_\gB)=\yd(p)\tau$ if $\yd(p)\in\dom(\tau)$, and
\item[(ii)] $p\in \dom(\tau_\gB)$ implies $\yd(p)\in \dom(\tau)$.
\end{enumerate}
These obviously will imply the validity of Theorem \ref{Theorem.4.7.4}.

For each $a_1,a_2\in A$, let $T(a_1,a_2)$ denote the set of all such trees $p\in F_\Sigma(X)$ that $a_1\yd(p)\Rightarrow^*_\bA wa_2$ holds for some $w\in Y^*$. By Lemma \ref{Lemma.1.7.4} and Theorem \ref{Theorem.3.3.2}, every $T(a_1,a_2)=\yd^{-1}\big(L(a_1,a_2)\big)$ is a regular forest. Now let $B=(A\times A)\cup \{b_0\}$ $(b_0\not\in A)$ and $\Omega=\Sigma\cup\{\omega_{ax}\mid a\in A, x\in X\}$, where $r(\omega_{ax})$ equals the length of the word $w$ obtained from the production $ax\rightarrow wa'\in P$ $(a\in A')$. (The ranks of symbols from $\Sigma$ are unchanged.) Moreover, $P'$ is given as follows:
\begin{itemize}
\item[(I)] For arbitrary $m>0,\sigma\in \Sigma_m$ and $a_1,a_2,\ldots,a_{m+1}\in A$, $P'$ contains the production $\big((a_1,a_{m+1})\sigma\rightarrow \sigma((a_1,a_2)\xi_1,\ldots,(a_m,a_{m+1})\xi_m),D\big)$ where $D(\xi_i)=T(a_i,a_{i+1})$ $(i=1,\ldots,m)$.
\item[(II)] If $\sigma\in\Sigma_0$ and $a\in A$, then the production $(a,a)\sigma\rightarrow \sigma$ is in $P'$.
\item[(III)] For arbitrary $x\in X$ and $(a_1,a_2)\in A\times A$, $P'$ contains the production $(a_1,a_2)x\rightarrow q$, where $a_1x\Rightarrow_\bA wa_2$ $(w\in Y^*)$ and $q\in F_\Omega(Y)$ is a fixed tree with $\yd(q)=w$ (such $q$ exists by the definition of $\omega_{a_1x}$).
\item[(IV)] For arbitrary $m>0, \sigma\in\Sigma_m$ and $a_1,\ldots,a_{m+1}\in A$, if $a_1=a_0$ and $a_{m+1}\in A'$, then the production $\big(b_0\sigma\rightarrow \sigma((a_1,a_2)\xi_1,\ldots,(a_m,a_{m+1})\xi_m),D\big)$ is in $P'$, where
$D(\xi_i)=T(a_i,a_{i+1})$ $(i=1,\ldots,m)$.
\item[(V)] For arbitrary $x\in X$, if $a_0x\Rightarrow_\bA wa_1$ $(w\in Y^*)$ and $a_1\in A'$, then the production $b_0x\rightarrow q$ is in $P'$, where $q\in F_\Omega(Y)$ is a fixed tree with $\yd(q)=w$ (again, by the definition of $\omega_{a_0x}$, such $q$ exists).
\item[(VI)] If $a_0\in A'$ and $\sigma\in\Sigma_0$, then the production $b_0\sigma\rightarrow \sigma$ is in $P'$.
\end{itemize}

In order to prove Theorem \ref{Theorem.4.7.4} it is enough to show that for arbitrary $(a_1,a_2)\in A\times A, p\in F_\Sigma(X)$ and $q\in F_\Omega(Y)$ the implication
\[(a_1,a_2)p \Rightarrow^*_\gB q \, \;\; \Longrightarrow \;\; a_1\yd(p)\Rightarrow^*_\bA\yd(q)a_2\]
holds. This can be carried out by induction on $\hg(p)$. \epr

\

We shall now introduce some more concepts that will be needed in the next section.

Let $\gA=(\Sigma,X,A,\Omega,Y,P,A')$ be an R-transducer. Take a tree $p\in F_\Sigma(X)$ and a node $d$ of $p$. Denote by $s$ the subtree of $p$ at this node $d$. Consider  a state $a$ and a derivation $\alpha:ap\Rightarrow^* q$ $(q\in F_\Omega(Y))$. Suppose exactly $k$ copies of this occurrence of $s$ are created during $\alpha$ and that these are translated into the trees $t_1,\ldots,t_k$ $(\in F_\Omega(Y))$
starting the translations, respectively, in states $a_1,\ldots,a_k$. In the next definition we distinguish a sequence of these states which will be called the state-sequence of $\alpha$ at $d$.

\begin{df}\rm\label{Definition.4.7.5} Let $\gA=(\Sigma,X,A,\Omega,Y,P,A')$ be an R-transducer. Take a derivation
\[\alpha:ap\Rightarrow^* q\mbox{\hspace*{10mm}}(a\in A, p\in F_\Sigma(X),q\in F_\Omega(Y)).\]
Let $d$ be a node of $p$ and $s$ the subtree at this node $d$. Replace the given occurrence of $s$ in $p$ by $\xi_1$ and denote by $r$ the resulting tree. Write $\alpha$ in the form
\[ap=ar(s)\Rightarrow^*\overline{q}(\ba s^n)\Rightarrow^*\overline{q}(\bt),\]
where $\overline{q}\in \hat{\hat F}_\Omega(Y\cup \Xi_n), \ba\in A^n, ar\Rightarrow^*\overline{q}(\ba \xi_1^n),\ba s^n\Rightarrow^* \bt$ and $\bt\in F_\Omega(Y)^n$. Denote by $a_id_i\rightarrow q_i$ $(a_i\in A, d_i\in \Sigma\cup X)$ the production applied first in the derivation $a_is \Rightarrow^* t_i$ $(i=1,\ldots,n)$. Then $\ba=(a_1,\ldots,a_n)$ is the {\em state-sequence} \index{state-sequence of!R-transducer} and
\[(a_1d_1\rightarrow q_1,\ldots,a_nd_n\rightarrow q_n)\]
is the {\em production-sequence} of $\alpha$ at $d$. \index{production-sequence}
\end{df}

Often we shall speak about the state-sequence and production-sequence of $\alpha$ at a subtree $s$. In such cases the node to which the given occurrence of $s$ belongs will be clear from the context.

We now define state-sequences for derivations in GSDTs.

\begin{df}\rm\label{Definition.4.7.6} Let $\gA=(\Sigma,X,A,Y,P,A')$ be a GSDT. Take a derivation
\[\alpha:\,ap\Rightarrow^* w\mbox{\hspace*{10mm}}(a\in A, p\in F_\Sigma(X),w\in Y^*).\]
Let $d$ be a node of $p$ and $s$ the subtree of $p$ at $d$. Replace the given occurrence of $s$ in $p$ by $\xi_1$ and denote by $r$ the resulting tree. Write $\alpha$ in the form
\[ap=ar(s)\Rightarrow^*w_1a_1sw_2\ldots w_na_nsw_{n+1}\Rightarrow^*w_1v_1w_2\ldots w_nv_nw_{n+1},\]
where $ar\Rightarrow^* w_1a_1\xi_1w_2\ldots w_na_n\xi_1w_{n+1}$ $(w_i\in Y^*, i=1,\ldots,n+1, a_1,\ldots,a_n\in A)$ and $a_is\Rightarrow^* v_i$ $(v_i\in Y^*, i=1,\ldots,n)$.
Then $\ba=(a_1,\ldots,a_n)$ is the {\em state-sequence of} $\alpha$ at $d$. \index{state-sequence of!GSDT}
\end{df}

Like in the case of R-transducers, we shall also speak about the state-sequence of $\alpha$ at the subtree $s$.

\begin{df}\rm\label{Definition.4.7.7} Let $\gA$ be an R-transducer $\gA=(\Sigma,X,A,\Omega,Y,P,A')$
[a GSDT $\gA=(\Sigma,X,A,Y,P,A')$]. Then a derivation $\alpha:\,ap\Rightarrow^* q$ $(a\in A, p\in F_\Sigma(X),q\in F_\Omega(Y))$ [$\beta:\, ap\Rightarrow^* w$ $(a\in A, p\in F_\Sigma(X),w\in Y^*)]$ is {\em $k$-copying}
if for every node $d$ of $p$ the length of the state sequence of $\alpha$ $[\beta]$ at $d$ is at most $k$.
Moreover, $\gA$ is {\em $k$-copying} if every derivation
$\alpha:\,ap\Rightarrow^* q$ $(p\in F_\Sigma(X),q\in F_\Omega(Y))$ [$\beta:\,ap\Rightarrow^* w$ $(p\in F_\Sigma(X),w\in Y^*)]$ with $a\in A'$ is $k$-copying. \index{GSDT!kcopying@$k$-copying} \index{rtransducer@R-transducer!kcopying@$k$-copying} \index{kc@$k$-copying derivation in!GSDT}\index{kc@$k$-copying derivation in!R-transducer}
Finally, $\gA$ is {\em finite-copying} if it is $k$-copying for some $k$. \index{GSDT!finite copying}
\index{rtransducer@R-transducer!finite copying}
\end{df}

We shall use the notation $\cR_k$ for the class of all transformations induced by $k$-copying R-transducers. Similarly, $\cG_k$ denotes the class of all transformations induced by $k$-copying GSDT's. Moreover, $\cR_f$
and $\cG_f$ will stand for the classes of transformations induced by finite-copying R-transducers and finite copying GSDT's, respectively. Corresponding notations will be used for the classes $\cDR, \cDG$ etc.

The next result shows that R-transformational languages can be studied through generalized syntax directed translations.

\begin{thm}\label{Theorem.4.7.8} For every $k$-copying {\rm GSDT} $\gA=(\Sigma,X,A,Y,P,A')$ there exist a ranked alphabet $\Omega$ and a $k$-copying {\rm R}-transducer $\gB=(\Sigma,X,A,\Omega,Y,P',A')$ such that
$\tau_\gA=\{(p,\yd(q))\mid (p,q)\in \tau_\gB \}$.

Conversely, for every $k$-copying {\rm R}-transducer $\gB$ there exists a $k$-copying {\rm GSDT} $\gA$ such that
$\tau_\gA=\{(p,\yd(q))\mid (p,q)\in \tau_\gB \}$.
\end{thm}
\pr The R-transducer and GSDT constructed in the proof of Theorem \ref{Theorem.4.5.4} obviously have the required properties. \epr

\

The following theorem gives sufficient conditions under which $\cR_k(K)=\cDR_k(K)$ holds for a given class $K$ of forests.

\begin{thm}\label{Theorem.4.7.9} Let $K$ be a class of forests closed under relabeling and regular insertion. Take an {\rm R}-transducer $\gA=(\Sigma,X,A,\Omega,Y,P,A')$, an $R\in K$ and a positive integer $k$. Then
\[S=\{q\in F_\Omega(Y)\mid \text{there is a $k$-copying derivation $ap\Rightarrow^* q$ for some $a\in A'$ and $p\in R$}\}\]
is in $\cDR_k(K)$.
\end{thm}
\pr Since $K$ is closed under regular insertion, we may assume that $A'$ is a singleton. Indeed, in the opposite case enlarge $A$ by a new state $a_0$, $\Sigma$ by a new unary operational symbol $\sigma$ and $P$ by all productions $a_0\sigma\rightarrow a\xi_1$ $(a\in A')$. Let $\overline{\gA}$ be the resulting R-transducer with initial state $a_0$, and let $\overline{R}=f(R)$ , where $f$ is a regular insertion given by $f(d)=\{\sigma(\xi_1)\}$ $(d\in X\cup \Sigma)$. Then $\overline{R}\in K$ and $\tau_{\overline{\gA}}(\overline{R})=\tau_\gA(R)$. Furthermore, a derivation $ap\Rightarrow^*_\gA q$ $(a\in A', p\in R, q\in F_\Omega(Y))$ is $k$-copying if the corresponding derivation   $a_0\sigma(p)\Rightarrow^*_{\overline{\gA}} q$ is $k$-copying, and conversely. Thus, we shall assume that $A'=\{ a_0\}$. Now we introduce the alphabet
\[\overline{X}=\{((a_1x,q_1),\ldots,(a_tx,q_t))\mid t\leq k, x\in X, a_ix\rightarrow q_i\in P
\mbox{\hspace*{3mm}}(i=1,\ldots,t)\}\]
and the ranked alphabet $\Delta$ with
\[\Delta_m=\{((a_1\sigma,q_1),\ldots,(a_t\sigma,q_t))\mid t\leq k, \sigma\in \Sigma_m, a_i\sigma\rightarrow q_i\in P
\mbox{\hspace*{3mm}}(i=1,\ldots,t)\}\]
$(m=0,1,\ldots)$. Consider the R-transducer $\gB=(\Sigma,X,\{b_0\},\Delta,\overline{X},P',b_0)$ where $P'$ consists of the productions
\[b_0x\rightarrow ((a_1x,q_1),\ldots,(a_tx,q_t))\mbox{\hspace*{5mm}}(x\in X,((a_1x,q_1),\ldots,(a_tx,q_t))\in\overline{X})\]
and
\[b_0\sigma\rightarrow ((a_1\sigma,q_1),\ldots,(a_t\sigma,q_t))(b_0\xi_1,\ldots,b_0\xi_m)\]
\[(\sigma\in\Sigma_m,((a_1\sigma,q_1),\ldots,(a_t\sigma,q_t))\in\Delta_m, m=0,1,\ldots).\]

Obviously, $\gB$ is an R-relabeling which relabels trees in the following way: if $\sigma\in \Sigma$
[resp. $x\in X$] is a label at a node $d$ of a tree $p\in F_\Sigma(X)$, then $\gB$ relabels $d$ by a sequence of productions $((a_1\sigma,q_1),\ldots,(a_t\sigma,q_t))$ [resp. $((a_1x,q_1),\ldots,(a_tx,q_t))$] from $P$ with length at most $k$.

Next define an R-transducer $\gC=(\Delta,\overline{X},C,\Omega,Y,P'',c_0)$ with
\[C=\{(u;a_1,\ldots,a_t)\mid 1 \leq u\leq t\leq k, a_i\in A \,\,(i=1,\ldots,t)\}\]
and $c_0=(1;a_0)$. Moreover, $P''$ is defined as follows:
\begin{itemize}
\item[(i)] For each $(u;a_1,\ldots,a_t)\in C$ and $((a_1p,q_1),\ldots,(a_tp,q_t))\in \Delta_0\cup \overline{X}$, \\
$(u;a_1,\ldots,a_t) ((a_1p,q_1),\ldots,(a_tp,q_t))\rightarrow q_u$ is in $P''$.
\item[(ii)] Let $(u;a_1,\ldots,a_t)\in C$ and $((a_1\sigma,q_1),\ldots,(a_t\sigma,q_t))\in\Delta_m$ $(m>0)$. Write
$(a_i\sigma,q_i)$ in the more detailed form $a_i\sigma\rightarrow q_i(\ba_{i1}\xi_1^{n_{i1}},\ldots,\ba_{im}\xi_m^{n_{im}})$ $(\ba_{ij}\in A^{n_{ij}}, j=1,\ldots, m, n_{i1}+\ldots+n_{im}=n_i, q_i\in \hat F_\Omega(Y\cup \Xi_{n_i}), i=1,\ldots, t$). Then the production
\[(u;a_1,\ldots,a_t) ((a_1\sigma,q_1),\ldots,(a_t\sigma,q_t))\rightarrow \]
\[q_u\big(((u_{11};\bb_1),\ldots,(u_{1n_{u1}};\bb_1))\xi_1^{n_{u1}},\ldots,((u_{m1};\bb_m),\ldots,(u_{mn_{um}};\bb_m))\xi_m^{n_{um}}\big)\]
in is $P'$, provided that $n_{1j}+\ldots+n_{tj}\leq k$ $(j=1,\ldots,m)$, where $u_{jl}=n_{1j}+\ldots+n_{u-1j}+l, \bb_j=(\ba_{1j},\ldots,\ba_{tj})$ and $j=1,\ldots,m$.
\end{itemize}

Obviously, $\gC$ is a deterministic R-transducer. Furthermore, one can easily see the following connection between derivations in $\gA$ and $\gC$:

Let $p\in F_\Sigma(X)$ and $q\in F_\Omega(Y)$ be arbitrary trees, and take a $k$-copying derivation
\[\alpha:\,a_0p\Rightarrow^*_\gA q.\]
Consider the tree $\overline{p}$ with $(p,\overline{p})\in\tau_\gB$ which is the result of relabeling each node $d$ of $p$ by the production-sequence of $\alpha$ at $d$. Then in $\gC$ we have a derivation
\[\beta:(1;a_0)\overline{p}\Rightarrow^*_\gC q\]
such that if $\ba=(a_1,\ldots,a_n)$ $(n\leq k)$ is the state-sequence of $\alpha$ at $d$ then $((1;\ba),\ldots,(n;\ba))$ is the state-sequence of $\beta$ at $d$. Conversely, if for a $\overline{p}'\in F_\Delta(\overline{X})$ and $q'\in F_\Omega(Y)$ there is a derivation
\[\beta':(1;a_0)\overline{p}'\Rightarrow^*_\gC q',\]
then for the (uniquely determined) tree $p'\in F_\Sigma(X)$ with $(p',\overline{p}')\in\tau_\gB$ we have the derivation
\[\alpha':a_0p'\Rightarrow^*_\gA q'.\]
Moreover, the state-sequence of $\beta'$ at a node $d$ of $\overline{p}'$ is of the form
$((1;\ba'),\ldots,(m;\ba'))$  $(\ba'=(a'_1,\ldots,a'_m))$ with $m\leq k$, and $\ba'$ is the state-sequence of $\alpha'$ at $d$. Therefore, $\gC$ is $k$-copying and $S=R\tau_\gB\circ \tau_\gC$ holds. Since $K$ is closed under relabelings, this implies $S\in \cDR_k(K)$. \epr

\

From  Theorem \ref{Theorem.4.7.9}, by Theorem \ref{Theorem.4.7.8}, we get

\begin{cor}\label{Corollary.4.7.10} Let $K$ be a class of forests closed under relabeling and regular insertion. Take a {\rm GSDT} $\gA=(\Sigma,X,A,Y,P,A')$, a $T\in K$ and a positive integer $k$. Then the language
\[L=\{w\in Y^*\mid \text{there is a $k$-copying derivation $ap\Rightarrow^* w$ for some $a\in A'$ and $p\in T$} \}\]
is in $\cDG_k(K)$.\epr
\end{cor}

Three more language operations will be needed.

\begin{df}\rm\label{Definition.4.7.11} Let $X$ be an alphabet and $\#\not\in X$ a symbol. For each $L\subseteq X^*$, $\rres(L,\#)$ ({\em re}gular {\em s}ubstitution) denotes the language defined as follows:
\begin{enumerate}
\item[(i)] if $L=\{e\}$, then $\rres(L,\#)=\#^*$,
\item[(ii)] if $L=\{x\}$ $(x\in X)$, then $\rres(L,\#)=\#^*x\#^*$,
\item[(iii)] if $L=\{ux\}$ $(u\in X^*, x\in X)$, then $\rres(L,\#)=\rres(u,\#)\rres(x,\#)$,
\item[(iv)] if $L$ is arbitrary, then $\rres(L,\#)=\bigcup(\rres(w,\#)\mid w\in L)$.
\end{enumerate}
\end{df}

\begin{thm}\label{Theorem.4.7.12} Let $K$ be a class of forests closed under regular insertion. For each $R\in K$ there exist a linear nondeleting {\rm GSDT} $\gA$ and a forest $S\in K$ such that $\rres(\yd(R),\#)=S\tau_\gA$.
\end{thm}
\pr Let $R\subseteq F_\Sigma(X)$, $R\in K$, and denote $\yd(R)$ by $L$. Let $\Delta=\Delta_1=\{\overline{d}\mid d\in\Sigma\cup X\}$ and let $f$ be the regular insertion defined by $f(d)=\{\overline{d}(\xi_1)\}^{*\xi_1}$ $(d\in \Sigma\cup X)$. Define the GSDT $\gA=(\Omega,X,\{a_0\},X\cup\{\#\},P,a_0)$ with $\Omega=\Sigma\cup\Delta$ $(\Omega_1=\Sigma_1\cup \Delta,\Omega_m=\Sigma_m, m\neq 1)$ so that
\[P=\{a_0\overline{x}\rightarrow \#a_0\xi_1, a_0\overline{x}\rightarrow a_0\xi_1\# \mid x\in X\} \cup
\{a_0\overline{\sigma}\rightarrow \#a_0\xi_1\mid\sigma\in\Sigma_0\} \cup \]
\[\{a_0x\rightarrow x \mid x\in X\}\cup \{a_0\sigma\rightarrow a_0\xi_1\ldots a_0\xi_m\mid \sigma \in\Sigma_m, m\geq 0\}.\]
Obviously, $\gA$ is a linear nondeleting GSDT satisfying $\rres(L,\#)=f(R)\tau_\gA$. Moreover, by our assumptions, $f(R)=S\in K$. \epr

\begin{thm}\label{Theorem.4.7.13} Let $Y$ be an alphabet and $\#\not\in Y$ a symbol. Take a language $L\subseteq Y^*$ and a class $K$ of forests closed under relabeling and regular insertion. If $\rres(L,\#)\in \cDG(K)$, then $L\in \cDG_f(K)$.
\end{thm}
\pr Let $\rres(L,\#)=T\tau_\gA$ where $\gA=(\Sigma,X,A,Y\cup\{\#\},P,a_0)$ is a deterministic GSDT and $T\subseteq F_\Sigma(X)$ is a forest from $K$. Moreover, let $A=\{a_1,\ldots,a_k\}$. A word
$y_{i_1}\#^{n_1}y_{i_2}\#^{n_2}\ldots y_{i_{r-1}}\#^{n_{r-1}}y_{i_r}$ $(\in\rres(L,\#), y_{i_1},\ldots,y_{i_r}\in Y)$ is called {\em proper} if $n_1,n_2,\ldots,n_{r-1}$ are pairwise distinct. \index{word!proper}

Consider a derivation
\[\alpha:\, a_0p\Rightarrow^*w_1b_1p_1w_2b_2p_1w_3\ldots w_sb_sp_1w_{s+1}\Rightarrow^*
w_1v_1w_2v_2w_3\ldots w_sv_sw_{s+1}=w,\]
where $p\in T$, $p_1$ is a subtree of $p$, $(b_1,b_2,\ldots,b_s)$ is the state-sequence of $\alpha$ at $p_1$, $b_ip_1\Rightarrow^* v_i$ $(i=1,\ldots,s)$ and $w_1,\ldots,w_{s+1},v_1,\ldots,v_s \in (Y\cup\{\#\})^*$. If $w$ is proper and $b_i=b_j$ $(i\neq j)$, then in $v_i$ (and thus in $v_j$) at most one symbol from $Y$ may occur.

Now for each $\sigma\in \Sigma_m$ $(m>0)$ take all pairs $(\sigma,M)$, where $M$ is a matrix of type $k\times m$ whose elements are from $Y\cup A\Xi_m\cup\{e\}$. Moreover, let $\Omega$ be a ranked alphabet with $\Omega_0=\Sigma_0$ and $\Omega_m=\{(\sigma,M)\mid\sigma \in \Sigma_m\}$ $(m>0)$.

Let $Y=\{y_1,\ldots,y_l\}$ and denote by $T_{ij}$ $(i=1,\ldots,k, j=1,\ldots,l)$ the set of all trees
$p\in F_\Sigma(X)$ for which $v\in\#^*y_j\#^*$, where $v$ is the word obtained from the derivation $a_ip\Rightarrow^* v$. Moreover, let $T_{il+1}$ $(i=1,\ldots,k)$ be the forest of all trees $p\in F_\Sigma(X)$ satisfying $v\in\#^*$, where $v$ is obtained again by the derivation $a_ip\Rightarrow^* v$.

By Theorems \ref{Theorem.4.5.4} and \ref{Theorem.3.3.2} and Corollary \ref{Corollary.4.3.17}, the
$T_{ij}$ $(i=1,\ldots,k, j=1,\ldots,l+1)$ are recognizable forests. Therefore, there are $\Sigma X$-recognizers $\bA_{ij}=(\cA_{ij},\alpha_{ij},A'_{ij})$ $(i=1,\ldots,k, j=1,\ldots,l+1)$ with  $\cA_{ij}=(A_{ij},\Sigma)$ such that $T(\bA_{ij})=T_{ij}$. Consider the DF-relabeling $\gB=(\Sigma,X,B,\Omega,X,P',B)$ where
\[B=\{(p\hat\alpha_{11},\ldots,p\hat\alpha_{1l+1},\ldots,p\hat\alpha_{k1},\ldots,p\hat\alpha_{kl+1})\mid
p\in F_\Sigma(X)\},\]
and $P'$ is given as follows:
\begin{itemize}
\item[(i)] For each $x\in X$, the production
\[x\rightarrow (x\alpha_{11},\ldots,x\alpha_{1l+1},\ldots,x\alpha_{k1},\ldots,x\alpha_{kl+1})x\]
is in $P'$.
\item[(ii)] For every $\sigma\in\Sigma_0$, the production
\[\sigma\rightarrow (\sigma^{\cA_{11}},\ldots,\sigma^{\cA_{1l+1}},\ldots,\sigma^{\cA_{k1}},\ldots,\sigma^{\cA_{kl+1}})\sigma\]
is in $P'$.
\item[(iii)] For each $\sigma\in\Sigma_m$ $(m>0)$ the productions
\[\sigma(\bb_1,\ldots,\bb_m)\rightarrow \bb(\sigma,M)(\xi_1,\ldots,\xi_m)\]
are in $P'$, where $\bb_t=(b_{11}^{(t)},\ldots,b_{1l+1}^{(t)},\ldots,b_{k1}^{(t)},\ldots,b_{kl+1}^{(t)})$, $\bb=(b_{11},\ldots,b_{1l+1},\ldots,b_{k1},\ldots,b_{kl+1})\in B$ $(t=1,\ldots,m)$,
$b_{ij}=\sigma^{\cA_{ij}}(b_{ij}^{(1)},\ldots,b_{ij}^{(m)})$ $(i=1,\ldots,k, j=1,\ldots, l+1)$ and the element $m_{it}$ $(i=1,\ldots,k, t=1,\ldots,m)$ of matrix $M$ is given by
$$m_{it}
=\left\{
\begin{array}{ll}
 e & \text{ if } b_{il+1}^{(t)}\in A'_{il+1},\\
y_u & \text{ if } b_{iu}^{(t)}\in A'_{iu}\,(1\leq u\leq l),\\
a_i\xi_t & \text{  otherwise.}
\end{array}
\right.
$$
Obviously, $m_{it}$ is well-defined since there are no two components $b_{ij_1}^{(t)}$ and $b_{ij_2}^{(t)}$
 $(1\leq i\leq k, 1\leq j_1,j_2 \leq l+1, j_1\neq j_2)$ such that $b_{ij_1}^{(t)}\in A'_{ij_1}$ and
$b_{ij_2}^{(t)}\in A'_{ij_2}$ both hold.
\end{itemize}

By the definition of $\gB$, it relabels trees in the following way: take a tree $p\in F_\Sigma(X)$, and let
$\sigma(p_1,\ldots,p_m)$ $(m>0)$ be the subtree of $p$ at a node $d$. The $\gB$ provides us with the information about which of the subtrees $p_1,\ldots,p_m$ is translated by $\gA(a_i)$ $(i=1,\ldots,k)$ into a word from $(Y\cup\{\#\})^*$ with
\begin{itemize}
\item[(I)] no occurrence of letters from $Y$,
\item[(II)] exactly one occurrence of letters from $Y$,
\item[(IIIa)] at least two occurrences of letters from $Y$, or
\item[(IIIb)] the given subtree is not in $\dom(\tau_{\gA(a_i)})$.
\end{itemize}
Next take the GSDT $\gC=(\Omega,X,A,Y,P'',a_0)$ where $P''$ is given as follows:
\begin{itemize}
\item[(a)] If $ap\rightarrow w$ $\big(a\in A, p\in X\cup\Sigma_0, w\in (Y\cup\{\#\})^*\big)$ is in $P$, then the production obtained from $ap\rightarrow w$ by replacing all occurrences of $\#$ in $w$ by $e$ will be in $P''$.
\item[(b)] Let $a\sigma\rightarrow w$ $\big(a\in A, \sigma\in\Sigma_m, m>0, w\in (Y\cup\{\#\}\cup A\Xi_m)^*\big)$ be in $P$.
Then all productions $a(\sigma,M)\rightarrow w'$ are in $P''$ where $w'$ is the result of replacing all occurrences of $a_i\xi_j$ in $w$ by $m_{ij}$ $(1\leq i\leq k, 1\leq j\leq m)$ and all occurrences of $\#$ by $e$.
\end{itemize}

It is clear that $\gC$ is deterministic. Moreover, one can show by induction on $\hg(p)$ for arbitrary $a\in A, p\in F_\Sigma(X)$ and $w\in (Y\cup\{\#\})^*$ the implication
\[ap\Rightarrow^*_{\gA}w \;\; \Longrightarrow\;\; a\tau_\gB(p)\Rightarrow^*_{\gC}\varphi(w)\]
holds, where $\varphi: (Y\cup\{\#\})^*\rightarrow Y^*$ is the homomorphism given by $\varphi(y)=y$ $(y\in Y)$ and $\varphi(\#)=e$. Thus
\begin{align*}
\tag{1}\label{Equation.4.7.13.1}L=\{& w'\in Y^*\mid a_0\tau_\gB(p)\Rightarrow^*_{\gC}w', a_0 p\Rightarrow^*_{\gA}w,\\
& p\in T, w\in (Y\cup\{\#\})^* \text{ and $w$ is proper if } |w'|>2\}.
\end{align*}
Furthermore, by our remark concerning state-sequences of derivations yielding proper words and the construction of $\gC$, the elements of a state-sequence of a derivation $a_0\tau_\gB(p)\Rightarrow^*_{\gC}w'$ from (\ref{Equation.4.7.13.1}) are different at any node of $\tau_\gB(p)$. Therefore, since $\gC$ has $ßk$ elements, each element of $L$ can be obtained by a $k$-copying derivation in $\gC$. Finally, since by our assumptions $T\tau_\gB\in K$, using Corollary \ref{Corollary.4.7.10} we get $L\in \cDG_k(K)$. \epr


\begin{df}\label{Definition.4.7.14}\rm Let $X$ be an alphabet and $\# \not\in X$ a symbol. For each language $L \subseteq X^*$, the language $c_\ast(L, \#)$ is defined by
\[
c_\ast(L, \#) = \{(w  \#)^n  \mid w \in L,\ n= 1,2,\ldots \}.
\]
\end{df}

\begin{thm}\label{Theorem.4.7.15} Let $K$ be a class of forests closed under regular insertion. For each $R \in K$ there exist a $\DGSDT$ $\gA$ and a forest $S \in K$ such that $c_\ast\big(\yd(R),\#\big) = S \tau_\gA$.
\end{thm}
\pr Suppose $R \subseteq F_\Sigma(X)$ and let $L = \yd(R)$. We introduce the ranked alphabet \mbox{$\Delta = \Delta_1 = \{\overline{d} \mid d \in \Sigma \cup X\}$} and define a regular insertion $f$ by $f(d) = \{\overline{d}(\xi_1)\}^{\ast \xi_1}$ \mbox{$(d \in \Sigma \cup X)$}. Moreover, let $\Omega$ be the ranked alphabet for which $\Omega_1 = \Sigma_1 \cup \Delta$ and $\Omega_m = \Sigma_m\ (m\ge 0,\ m\not=1)$. Consider the $\GSDT$
\[
\gA = (\Omega,X,\{a_1,a_2\}, X \cup \{\#\},P,a_1)
\]
where
\begin{align*}
P = \; &\{ a_1\overline{d} \rightarrow a_1 \xi_1 a_2 \xi_1 \# \mid d \in \Sigma \cup X\}\\
& \cup \{a_2\overline{d} \rightarrow a_2\xi_1 \mid d \in \Sigma \cup X\}\\
& \cup \{a_1 x \rightarrow e \mid x \in X\} \cup  \{a_1 \sigma \rightarrow e \mid \sigma \in \Sigma_m,\ m \ge 0\}\\
& \cup \{a_2x \rightarrow x \mid x \in X\}  \cup  \{a_2\sigma \rightarrow a_2 \xi_1 \ldots a_2 \xi_m \mid \sigma \in \Sigma_m,\ m\ge 0\}.
\end{align*}
It is obvious that $\gA$ is a deterministic $\GSDT$ satisfying $c_\ast(L,\#) = S\tau_\gA$, where $S = f(R)$. Moreover, by our assumptions $S \in K$.
\epr

\begin{thm}\label{Theorem.4.7.16} Let $U \subseteq c_\ast(L,\#)\ (L \subseteq Z^\ast,\ \# \not\in Z)$ be a language containing infinitely many words $(w\#)^n$ for each $w \in L$. Furthermore, let $K$ be a class of forests closed under relabeling and regular insertion. If $U \in \cDG_f\big(\cR(K)\big)$, then $L \in \cDG(K)$.
\end{thm}
\pr \sloppy Let $\gA = (\Sigma,X,A,\Omega,Y,P,A')$ be an {\rm R}-transducer and $\gB = (\Omega,Y,B,Z\cup\{\#\},P',b_0)$ a $k$-copying deterministic $\GSDT$. Moreover, take a forest $R \subseteq F_\Sigma(X)$ from $K$ satisfying $U = (R\tau_\gA)\tau_\gB$.
 Since $K$ is closed under regular insertion, we may, without any loss of generality, assume that $A'$ is a singleton, say $A' =\{a_0\}$. First we shall construct an {\rm R}-transducer $\overline{\gA} = ( \Sigma,X,\overline{A}, \Omega,Y,\overline{P},\overline{a}_0)$ which translates every $p \in F_\Sigma(X)$ into a tree $q \in F_\Omega(Y)$ in the same way as $\gA$ provided that $q \in \dom(\tau_\gB)$. In addition, if during the translation of $p$ into $q$ by $\gA$, an occurrence of a subtree $p'$ in $p$ is translated starting in a state $a$ into a tree $q'$, then during the corresponding translation of $p$ by $\overline{\gA}$, $p'$ will be translated starting in a state consisting of $a$ and the state-sequence of the derivation of $q$ in $\gB$ at the subtree $q'$.  Thus, $\overline{\gA}$ will have the property that if during the above translation of $p$ by $\overline{\gA}$, two copies of an occurrence of $p'$ are translated starting in states $\overline{a}_1$ and $\overline{a}_2$, respectively, into the trees $q_1$ and $q_2$ such that $\overline{a}_1 = \overline{a}_2$, then the state-sequences of the derivation of $q$ in $\gB$ at $q_1$ and $q_2$ coincide.

Let $\tau_\gB(q) = (w\#)^m\ (w \in Z^\ast)$. If $m$ is large enough, then the properties of $\overline{\gA}$ will make it possible to replace in a derivation $\overline{a}_0 p \Rightarrow^\ast_{\overline{\gA}}q$ different derivations of $p'$ starting from the same state by one of them such that for the resulting output tree $\overline{q}$ we shall have $\tau_{\gB}(\overline{q}) = (w\#)^{m'}$ with $m' \ge m$. By prescribing the applications of productions of $\overline{\gA}$ in this manner we shall arrive at a {\rm DR}-transducer $\gA_1$ such that $(S\tau_{\gA_1})\tau_{\gB}$ contains infinitely many words $(w\#)^m$ for each $w \in L$ and $S$ is obtained from $R$ by a relabeling. Afterwards applying a deterministic $\gsm$ to $(S\tau_{\gA_1})\tau_{\gB}$, we shall get $L$.

Thus construct the {\rm R}-transducer $\overline{\gA} = (\Sigma,X,\overline{A},\Omega,Y,\overline{P},\overline{a}_0)$ where
\[
\overline{A} = \{(a,\bb) \mid a \in A,\ \bb \in B^n,\ n=0,1,\ldots,k\}
\]
and $\overline{a}_0 = \big(a_0,(b_0)\big)$. Moreover, $\overline{P}$ is given in the following way:
\begin{itemize}
\item[(i)] Let $ap \rightarrow q\ \big(a \in A,\ p \in X \cup \Sigma_0,\ q \in F_\Omega(Y)\big)$ be in $P$ and take a vector \mbox{$\bb \in B^n$} $(0 \le n \le k)$. Then the production $(a,\bb)p \rightarrow q$ is in $\overline{P}$.

\item[(ii)] Let $a \sigma \rightarrow q(\ba_1 \xi_1^{n_1}, \ldots, \ba_m \xi_m^{n_m})$ $\big(a \in A,$ $\sigma \in \Sigma_m$, $m > 0$, $\ba_i \in A^{n_i}$, $i=1,\ldots,m$, $n_1 + \ldots + n_m=n$, $q \in \hat{F}_\Omega(Y \cup \Xi_n)\big)$ be in $P$ and $\bb = (b_1,\ldots,b_s) \in B^s$. Moreover, for every $u\ (1 \le u \le s)$, and every $j\ (1 \le j \le n)$ take the derivation
\[
\begin{array}{c}
b_u q \Rightarrow_\gB^\ast w_{uj_1} b_{uj_1} \xi_j w_{uj_2} \ldots w_{uj_{u_j}} b_{uj_{u_j}} \xi_j w_{uj_{u_j+1}}\\
\big(w_{uj_1}, \ldots, w_{uj_{u_j+1}} \in (Z \cup \{\#\} \cup B(\Xi_n - \{\xi_j\}))^\ast, \; b_{uj_1},\ldots, b_{uj_{u_j}} \in B\big).
\end{array}
\]
Set $\bb_j = (b_{1j_1}, \ldots ,b_{1j_{1_j}}, \ldots, b_{sj_1}, \ldots, b_{sj_{s_j}}) \ (j=1,\ldots,n)$. Then the production
\[
\begin{array}{c}
(a,\bb)\sigma \rightarrow q\big(((a_{11},\bb_1),\ldots,(a_{1n_1},\bb_{n_1})) \xi_1^{n_1}, ((a_{21},\bb_{n_1+1}),\ldots\\
\ldots , (a_{2n_2}, \bb_{n_1+n_2})) \xi_2^{n_2},\ldots, ((a_{m_1},\bb_{n_1+\ldots + n_{m-1} +1}),\ldots,(a_{mn_m},\bb_n)) \xi_m^{n_m} \big)
\end{array}
\]
is in $\overline{P}$, provided that for each $j=1,\ldots,n$ the length of the sequence $\bb_j$ is not greater than $k$.
\end{itemize}

From the construction of $\overline{\gA}$, one can easily see the following connection between $\gA$ and $\overline{\gA}$. Take a tree $p \in F_\Sigma(X)$, a node $d$ of $p$ and let $p'$ be the subtree of $p$ at $d$. Moreover, write $p = r(p')\ \big(r \in \hat{F}_\Sigma(X \cup \Xi_1)\big)$, and consider a derivation
\[
\begin{array}{c}
\alpha: \;  a_0 r(p') \Rightarrow_\gA^\ast \overline{q}(\ba p'^n) \Rightarrow _\gA^\ast \overline{q}(\bt) = q\\[1mm]
\big(q \in F_\Omega(Y),\ a_0 r \Rightarrow_\gA^\ast \overline{q}(\ba \xi_1^n),\ \overline{q} \in \hat{\hat{F}}_\Omega(Y \cup \Xi_n), \ \ba p'^n \Rightarrow_\gA^\ast \bt, \ \bt \in F_\Omega(Y)^n\big)
\end{array}
\]
with $q \in \dom(\tau_\gB)$. Then in $\overline{\gA}$ we have a derivation
\[
\begin{array}{c}
\beta: \; \big(a_0,(b_0)\big)r(p') \Rightarrow^\ast \overline{q}\big(((a_1,\bb_1),\ldots,(a_n,\bb_n)) p'^n \big) \Rightarrow^\ast \overline{q}(\bt) = q,
\end{array}
\]
where $\bb_i\ (1 \le i \le n)$ is the state-sequence of the derivation
\[
\gamma: \; b_0q \Rightarrow_\gB^\ast w \ \big(\in (Z \cup \{\#\})^\ast \big)
\]
at the subtree $t_i$. Therefore, if $(a_i,\bb_i) = (a_j,\bb_j)\ (1\le i,j\le n)$, then the state-sequences of $\gamma$ at the subtrees $t_i$ and $t_j$ coincide.  We can assume that $\gA$ itself has this property, because the equality $\tau_\gA \circ \tau_\gB = \tau_{\overline{\gA}} \circ \tau_\gB$ obviously holds.

Consider a word $(w \#)^m \in (R\tau_\gA)\tau_\gB$ with $m > 2k+1$. More exactly, let $p \in R$ be a tree for which under the derivation $a_0p \Rightarrow_\gA^\ast q \; \big(\in F_\Omega(Y)\big)$ the equality $\tau_\gB(q)=(w\#)^m$ holds. Let $r \in \hat{F}_\Sigma(X \cup \Xi_1)$ and $p' \in F_\Sigma(X)$ with $r(p')=p$. Moreover, write the above derivation in the form
\[
\begin{array}{c}
\alpha': \; a_0 r(p') \Rightarrow_\gA^\ast \overline{q}(\ba p'^n) \Rightarrow_\gA^\ast \overline{q}(\bt) = q\\
\big(q \in F_\Omega(Y), \ a_0 r \Rightarrow_\gA^\ast \overline{q}(\ba \xi_1^n), \ \overline{q} \in \hat{\hat{F}}_\Omega(Y \cup \Xi_n), \ \ba p'^n \Rightarrow_\gA^\ast \bt, \ \bt \in F_\Omega(Y)^n\big).
\end{array}
\]
Assume that a state $a \in A$ occurs more than once in $\ba$, and let $a_{i_1},\ldots,a_{i_j} \ (1 \le i_1 < \ldots < i_j \le n)$ be all occurrences of $a$ in $\ba$. Then the state-sequences of
\[
\beta': \; b_0q \Rightarrow_\gA^\ast (w\#)^m \ \big(\in (Z \cup \{\#\})^\ast \big)
\]
at the subtrees $t_{i_1},\ldots,t_{i_j}$ coincide. Let $(b_1,\ldots,b_s)$ be this common state-sequence.

Among $t_{i_1},\ldots,t_{i_j}$ let $t_{i_1}$ be the tree for which $\tau_{\gB(b_1)}(t_{i_1}) \ldots \tau_{\gB(b_s)}(t_{i_1})$ has a maximal number of occurrences of $\#$.  Replace the considered occurrences of $t_{i_1},\ldots, t_{i_j}$ in $q$ by $t_{i_1}$, and denote by $q'$ the resulting tree. We claim that for $q'$ we have $\tau_\gB(q') = (w\#)^{m'}$ with $m' \ge m$. To prove it let us distinguish the following two cases:

\begin{enumerate}
\item[(I)] There exists an $r \ (1 \le r \le s)$ such that $\#$ occurs at least twice in the word $\tau_{\gB(b_r)}(t_{i_1})$. Then our claim obviously holds.
\item[(II)] $\#$ occurs at most once in each word $\tau_{\gB(b_1)}(t_{i_1}),\ldots,\tau_{\gB(b_s)}(t_{i_1})$. Take a fixed $r$ $(1 < r \le j)$, and write $\beta'$ in the form
\[
\begin{array}{c}
b_0q \Rightarrow_\gB^\ast w_1 b_1 t_{i_r} w_2 \ldots w_s b_s t_{i_r} w_{s+1} \Rightarrow_\gB^\ast \\
w_1 v_1 w_2 \ldots w_s v_s w_{s+1} = (w\#)^m.
\end{array}
\]
Since $m>2k+1$ and $s\le k$, there exists a $w_u\ (1 \le u \le s+1)$ such that $\#$ occurs at least twice in $w_u$. This also implies our claim.
\end{enumerate}

Thus we have got the following result. If we replace in $\alpha'$ every subderivation \mbox{$a_rp' \Rightarrow_\gA^\ast t_r$} $(a_r=a, r=i_1,\ldots,i_j)$  by $ap' \Rightarrow_\gA^\ast t_{i_1}$, then $b_0q' \Rightarrow_\gB^\ast (w\#)^{m'}$ with $m' \ge m$ holds for the resulting output tree $q'$. Therefore, prescribing the applications of the productions of $\gA$ in this way, we arrive at  a deterministic {\rm R}-transformation whose composition by $\tau_\gB$, applied to a suitable forest from $K$, for each $w \in L$ yields infinitely many words $(w\#)^m \ (m\ge 1)$, and only such words. Next we show how this can be carried out. First we define a deterministic {\rm R}-transducer $\gA_1$.

Let $A = \{a_1\ldots,a_s\}$, and define a set $\overline{X}$ of variables by
\[
\overline{X} = \{\big(x,(c_1,\ldots,c_s)\big) \mid x\in X,\ c_i=(a_ix,q_i) \in P \text{ or } c_i= \ast,\ i=1,\ldots,s\}
\]
where $\ast$ is a new symbol. Moreover, define the ranked alphabet $\Delta$, where for each $m \ (\ge 0)$,
\[
\Delta_m = \{\big(\sigma,(c_1,\ldots,c_s)\big) \mid \sigma \in \Sigma_m,\ c_i = (a_i\sigma, q_i) \in P \text{ or } c_i = \ast,\ i=1,\ldots,s\}.
\]
Now take the {\rm R}-transducer $\gA_1 = (\Delta,\overline{X},A,\Omega,Y,P_1,a_0)$ for which $P_1$ is given as follows:
\begin{itemize}
\item[$(\alpha)$] For each $a_i \in A$ and $\big(x,(c_1,\ldots,c_s)\big) \in \overline{X}, \text{ if } c_i = (a_i x,q_i)$, then the production
\[
a_i\big(x,(c_1,\ldots,c_s)\big) \rightarrow q_i
\]
is in $P_1$.

\item[$(\beta)$] For each $a_i \in A$ and $\big(\sigma,(c_1,\ldots,c_s)\big) \in \Delta_m, \text{ if } c_i=(a_i\sigma,q_i)$, then the production
\[
a_i\big(\sigma,(c_1,\ldots,c_s)\big) \rightarrow q_i
\]
is in $P_1$.
\end{itemize}

Obviously, $\gA_1$ is a deterministic {\rm R}-transducer.

Next, let $\gD = (\Sigma,X,\{d_0\},\Delta,\overline{X},P'',d_0)$ be the F-relabeling where
\[
\begin{array}{c}
P'' = \{x \rightarrow d_0\big(x,(c_1,\ldots,c_s)\big) \mid x \in X,\ \big(x,(c_1,\ldots,c_s)\big) \in \overline{X}\} \cup \\
\{\sigma(d_0,\ldots,d_0) \rightarrow d_0\big(\sigma,(c_1,\ldots,c_s)\big)(\xi_1,\ldots,\xi_m) \mid \sigma \in \Sigma_m,\\
\big(\sigma,(c_1,\ldots,c_s)\big) \in \Delta_m,\, m \ge 0\}.
\end{array}
\]
Put $S=R\tau_\gD$. Since $K$ is closed under relabeling, $S \in K$. Moreover, taking into consideration the remarks preceding the construction of $\gA_1$, one can easily see that, for each $w \in L$, $(S\tau_{\gA_1})\tau_\gB$ contains infinitely many words of the form $(w \#)^m \ (m\ge 1)$, and only such words.

Finally, take the deterministic $\gsm$ $\bC = (Z \cup \{\#\}, \{c_0,c_1\}, Z,c_0,P_{\bC},\{c_1\})$ where
\[
P_{\bC} = \{c_0z \rightarrow zc_0 \mid z \in Z\} \cup \{c_0\# \rightarrow ec_1\} \cup \{c_1\overline{z} \rightarrow ec_1 \mid \overline{z} \in Z \cup \{\#\}\}.
\]
Obviously, $(w\#)^m\tau_{\bC} = w$ for all $w \in Z^\ast$ and $m\ge 1$.

Denote by $\gB_1$ the deterministic $k$-copying {\rm R}-transducer obtained from $\gB$ by Theorems \ref{Theorem.4.5.4} and \ref{Theorem.4.7.8}. Moreover, let $\gC_1$ be the $\mathrm{DR}_{\mathrm{R}}$-transducer given to $\bC$ by Theorem \ref{Theorem.4.7.4}. Then the equality $L =  \yd(S\tau_{\gA_1} \circ \tau_{\gB_1} \circ \tau_{\gC_1})$ holds. Thus, by a repeated application of Theorem \ref{Theorem.4.4.6} (iii) and Corollary \ref{Corollary.4.4.8} (ii) and using Theorem \ref{Theorem.4.6.15} and Corollary \ref{Corollary.4.3.17}, we get for a suitable deterministic {\rm R}-transformation $\tau$ and a suitable $T \in K$ the equality $T\tau=S\tau_{\gA_1} \circ \tau_{\gB_1} \circ \tau_{\gC_1}$. (Observe that the {\rm F}-transducer $\gA$ given in Lemma \ref{Lemma.4.1.11} is an F-relabeling. Hence, closure under relabeling implies closure under intersection with regular forests.) Finally, again by Theorem \ref{Theorem.4.5.4}, we have $L \in \cDG(T)$.
\epr

\begin{df}\label{Definition.4.7.17}\rm  Let $X$ be an alphabet and $\# \not\in X$ a symbol. Then for $L \subseteq X^\ast$ the language $c_2(L,\#)$ is defined by $c_2(L,\#) = \{w\# w\mid w \in L\}$.
\end{df}

\begin{thm}\label{Theorem.4.7.18} Let $K$ be a class of forests closed under relabeling and regular insertion. If $R \in K$, then there exist a 2-copying $\GSDH$-transducer $\gA$ and a forest $T\in K$ such that $c_2(\yd(R),\#) = T\tau_\gA$.
\end{thm}
\pr Suppose $R \subseteq F_\Sigma(X)$ and let $L = \yd(R)$. Moreover, take the ranked alphabet $\Delta= \Delta_1=\{\overline{d} \mid d \in \Sigma \cup X\}$, and consider the regular insertion defined by $f(d)=\{\overline{d}(\xi_1)\}^{\ast \xi_1}$
$(d \in \Sigma \cup X)$, and set $S=f(R)$. Then $S \in K$. Finally, let $\Omega = \Sigma \cup \Delta$ be the ranked alphabet with $\Omega_1 = \Sigma_1 \cup \Delta$ and $\Omega_m=\Sigma_m \ (m\ge 0,\ m\not=1)$.

Now consider the R-relabeling $\gB = (\Omega,X,\{b_0,b_1\},\Omega,X,P,b_0)$, where
\[
\begin{array}{c}
P = \{b_0 \overline{d} \rightarrow \overline{d}(b_1\xi_1) \mid d \in \Sigma \cup X\} \cup\\
\{b_1\sigma \rightarrow \sigma(b_1\xi_1,\ldots,b_1\xi_m) \mid \sigma \in \Sigma_m,\ m\ge 0\} \cup\\
\{b_1x \rightarrow x \mid x \in X\}.
\end{array}
\]

Obviously, $T=S\tau_\gB$ consists of all trees of the form $\overline{d}(r)$, where $r \in R$ and \mbox{$d = \rroot(r)$}. Since $\gB$ is a relabeling, $T\in K$. Now we construct the required $\GSDT$ $\gA =(\Omega,X,\{a_0\},X \cup \{\#\},P',a_0)$, where
\[
\begin{array}{c}
P' = \{a_0\overline{d} \rightarrow a_0\xi_1\#a_0\xi_1 \mid d \in \Sigma \cup X\} \cup \\
\{a_0\sigma \rightarrow a_0\xi_1 \ldots a_0\xi_m \mid \sigma \in \Sigma_m,\ m\ge 0\} \cup \{a_0x \rightarrow x \mid x \in X\}.
\end{array}
\]

It is clear that $\gA$ is a 2-copying $\GSDH$-transducer and that $c_2(L,\#) = T\tau_\gA$ holds.
\epr

\begin{thm}\label{Theorem.4.7.19} Let $Y$ be an alphabet and $\# \not\in Y$ a symbol. Take a language $L \subseteq Y^\ast$ and a class $K$ of forests closed under relabeling and regular insertion. If $c_2(L,\#) \in \cG(K)$, then $L \in \cDG(K)$.
\end{thm}
\pr The idea behind the proof is similar to that of Theorem \ref{Theorem.4.7.16}, but this is much simpler.

Let $\gA=(\Sigma,X,A,Y\cup\{\#\},P,A')$ be a $\GSDT$ and $R \in K$ a $\Sigma X$-forest such that $R\tau_\gA = c_2(L,\#)$. Since $K$ is closed under regular insertion, we may assume that $A'$ is a singleton, say $A'=\{a_0\}$.

Take a tree $p \in R$, a subtree $p'$ of $p$ and let $p=r(p')\ \big(r \in \hat{F}_\Sigma(X\cup \Xi_1)\big)$. Consider a derivation
\[
\alpha: \; a_0r(p') \Rightarrow^\ast w_1 a_1 p' w_2
\ldots w_k a_k p' w_{k+1} \Rightarrow^\ast w_1 v_1 w_2 \ldots w_k v_k w_{k+1} = w\#w,
\]
where $a_0r(\xi_1) \Rightarrow^\ast w_1 a_1\xi_1 w_2 \ldots w_k a_k \xi_1 w_{k+1}$, $w_1,\ldots,w_{k+1}, v_1,\ldots,v_k \in (Y\cup\{\#\})^\ast$ and $a_i p' \Rightarrow^\ast v_i \ (i=1,\ldots,k)$. Then $(a_1,\ldots,a_k)$ is the state-sequence of $\alpha$ at $p'$. Assume that a state $a \in A$ occurs at least twice in $(a_1,\ldots,a_k)$, and let $a_{i_1}$ and $a_{i_2}\ (1 \le i_1 < i_2 \le k)$ be two such occurrences of $a$. Then, taking the relevant occurrences of $v_{i_1}$ and $v_{i_2}$ in $w\#w$, we have the decomposition $w\# w = u_1 v_{i_1}u_2 v_{i_2} u_3$. On the other hand the words $u_1 v_{i_j}u_2 v_{i_j} u_3 \ (j = 1,2)$ are also in $R\tau_\gA$. Hence, $v_{i_1} = v_{i_2}$ must hold. This implies that if we replace for each $t\ (1 \le t \le k)$ such that $a_t=a$, $a_tp' \Rightarrow^\ast v_t$ by $a_tp' \Rightarrow^\ast v_{i_1}$, we get the same word $w\# w$. Therefore, prescribing accordingly the applications of productions from $P$, we arrive at a deterministic $\GSDT$ yielding $c_2(L,\#)$. This can be carried out in the same way as in the proof of Theorem \ref{Theorem.4.7.16}, but here the resulting $\gA_1$ is a $\DGSDT$. Thus, taking the F-relabeling $\gD$ defined in the proof of Theorem \ref{Theorem.4.7.16}, for $S=R\tau_\gD$, we have $S\in K$ and $S\tau_{\gA_1} = c_2(L,\#)$. Moreover, by Theorem \ref{Theorem.4.5.4}, there exists a {\rm DR}-transducer $\gB_1$ with $c_2(L,\#) = \yd(S\tau_{\gB_1})$. Finally, consider the deterministic $\gsm$ $\bC$ of the proof of Theorem \ref{Theorem.4.7.16} with $Y$ instead of $Z$, and let $\gC_1$ be the corresponding $\mathrm{DR}_{\mathrm{R}}$-transducer. Then the equality $L = \yd(S\tau_{\gB_1} \circ \tau_{\gC_1})$ holds. Thus, by \mbox{Theorem \ref{Theorem.4.4.6} (iii)}, Corollary \ref{Corollary.4.4.8} (ii), Theorem \ref{Theorem.4.6.15} and Corollary \ref{Corollary.4.3.17}, for suitable {\rm DR}-transformation $\tau$ and a $T \in K$, we get $T\tau = S\tau_{\gB_1} \circ \tau_{\gC_1}$. This, by Theorem \ref{Theorem.4.5.4}, implies $L \in \cDG(T)$.
 \epr

%% file: Section.4.8.tex
In this section we prove that the compositions of $n$ {\rm F}-transformations or $n$ {\rm R}-transformations form proper hierarchies when $n=0,1,2,\ldots$. Similar results will be shown for the classes of forests ($n$-surface forests) which can be obtained from regular forests by compositions of $n$ F- or $n$ {\rm R}-transformations. All these results will follow from the fact that the classes of languages ($n$-transformational languages) obtained by taking the yields of $n$-surface forests form a proper hierarchy.

\begin{df}\label{Definition.4.8.1}\rm A forest $T$ is an $(n,\mathrm{R})$-{\em surface forest}
\index{forest!nrsurface@$(n,\mathrm{R})$-surface}
if $T\in \Surf(\cR^n)$.  $(n,\mathrm{F})$- and $(n,\mathrm{R}_{\mathrm{R}})$-{\em surface forests} are defined in a similar way.
\index{forest!nfsurface@$(n,\mathrm{F})$-surface}
\index{forest!nrrsurface@$(n,\mathrm{R}_{\mathrm{R}})$-surface}
\end{df}

\begin{df}\label{Definition.4.8.2}\rm A (string) language $L$ is an $(n,\mathrm{R})$-{\em transformational language}
\index{language!nrtransformational@$(n,\mathrm{R})$-transformational}
if \mbox{$L=\yd(T)$} for some $(n,\mathrm{R})$-surface forest $T$. $(n,\mathrm{F})$- and  $(n,\mathrm{R}_{\mathrm{R}})$-{\em transformational languages}
\index{language!nftransformational@$(n,\mathrm{F})$-transformational}
\index{language!nrrtransformational@$(n,\mathrm{R}_{\mathrm{R}})$-transformational}
are defined similarly.

If $n=1$ then we shall speak about {\rm R}-, {\rm F}- and $\mathrm{R}_{\mathrm{R}}$-{\em transformational languages},
\index{language!Rt-ransformational@{\rm R}-transformational}
\index{language!ftransformational@{\rm F}-transformational}
\index{language!RR-transformational@$\mathrm{R}_{\mathrm{R}}$-transformational}
as well.
\end{df}

The following results show that in studying $(n,\mathrm{R})$-surface forests and {\rm $(n,\mathrm{R})$}-transformational languages we can use $\mathrm{R}_{\mathrm{R}}$-transformations, too.

\begin{thm}\label{Theorem.4.8.3} For each natural number $n$, the equality $\Surf(\cR^n) = \Surf(\cR_{\mathrm{R}}^n)$ holds.
\end{thm}
\pr This follows from Theorems \ref{Theorem.4.4.6} (i) and \ref{Theorem.4.3.15} and Lemma \ref{Lemma.4.6.5}.
\epr

\

From Theorem \ref{Theorem.4.8.3} we directly get

\begin{cor}\label{Corollary.4.8.4} For every natural number $n$, the class of {\rm $(n,\mathrm{R})$}-transformational languages coincides with the class of $(n,\mathrm{R}_\mathrm{R})$-transformational languages. \epr
\end{cor}

Using Theorems \ref{Theorem.4.4.7} (i) and \ref{Theorem.4.2.7}, from Theorem \ref{Theorem.4.8.3} we obtain

\begin{cor}\label{Corollary.4.8.5} For every natural number $n$, $\Surf(\cR^n)$ is closed under {\rm LF}-transformations and {\rm LR}-transformations. \epr
\end{cor}

Now we can state and prove a result giving a recursive procedure by which the hierarchy theorems can be proved easily. The procedure will be based on the ``bridge theorems'' of the previous section which concern the operations $\rres, c_2$ and $c_\ast$. These associate with each language which is not in a given class another language which is not in another, larger class.

\begin{thm}\label{Theorem.4.8.6}Let $K$ be a class of forests closed under relabeling and regular insertion. If $\yd \cDR_f(K) \subset \yd \cR(K)$, then for each integer $n\ge 1$,
\[
\yd \cR^n(K) \subset \yd \cDR_f \big( \cR^n(K) \big) \subset \yd \cDR \big( \cR^n(K) \big) \subset \yd \cR^{n+1}(K).
\]
\end{thm}
\pr By Theorem \ref{Theorem.4.3.15} and Lemma \ref{Lemma.4.7.3}, $\cR^n(K)$ is closed under relabeling and regular insertion, for every $n \ge 1$. In the sequel these facts will be used without further mention.

We shall proceed by induction on $n$. Let $n=1$. Take a forest $R$ such that $R \in \cR(K)$ and $\yd(R) \not\in \yd \cDR_f(K)$. Then by Theorems \ref{Theorem.4.7.12}, \ref{Theorem.4.5.4} and \ref{Theorem.4.2.8} there exist an {\rm LNF}-transformation $\tau$ and a forest $S \in \cR(K)$ such that $\rres\big(\yd(R),\# \big) = \yd(S\tau)$. Moreover, by Theorem \ref{Theorem.4.3.15}, $S\tau \in \cR(K)$. On the other hand, since $\yd(R) \not\in \yd \cDR_f(K)$, by Theorems \ref{Theorem.4.7.13} and \ref{Theorem.4.5.4}, $\rres \big(\yd(R),\# \big) \not\in \yd \cDR(K)$. Thus, the proper inclusion $\yd \cDR(K) \subset \yd \cR(K)$ holds.

Next take an $R \in \cR(K)$ with $\yd(R) \not\in \yd \cDR(K)$. Then, by Theorems \ref{Theorem.4.7.18} and \ref{Theorem.4.7.8}, there exist a 2-copying homomorphism $\tau$ and a forest $S \in \cR(K)$ such that $c_2 \big(\yd(R),\# \big) = \yd(S\tau)$. On the other hand, since $\yd(R) \not\in \yd \cDR(K)$, by Theorems \ref{Theorem.4.5.4} and \ref{Theorem.4.7.19}, $c_2 \big(\yd(R),\# \big) \not\in \yd\cR(K)$. Therefore, the inclusion $\yd \cR(K) \subset \yd \cDR_f(\cR(K))$ is valid.

Again take an $R \in \cR(K)$ with $\yd(R) \not\in \yd \cDR(K)$. By Theorems \ref{Theorem.4.7.15} and \ref{Theorem.4.5.4} there exist a {\rm DR}-transformation $\tau$ and a forest $S \in \cR(K)$ such that, $c_\ast \big(\yd(R),\# \big) = \yd(S\tau)$. Moreover, since $\yd(R) \not\in \yd \cDR(K)$, by Theorems \ref{Theorem.4.7.16} and \ref{Theorem.4.7.8}, $c_\ast \big(\yd(R),\# \big) \not\in \yd \cDR_f \big(\cR(K) \big)$. Thus we have got that
\[
\yd \cDR_f \big(\cR(K) \big) \subset \yd \cDR \big(\cR(K) \big).
\]

Finally, take an $R \in \cR^2(K)$ with $\yd(R) \not\in \yd \cDR_f \big(\cR(K)\big)$. Then again by Theorems \ref{Theorem.4.7.12} and \ref{Theorem.4.5.4}, there exist an {\rm LNF}-transformation $\tau$ and a forest $S \in \cR^2(K)$ such that $\rres\big(\yd(R),\#\big) = \yd(S\tau)$. Moreover, by Theorem \ref{Theorem.4.3.15}, $S\tau \in \cR^2(K)$. On the other hand, since $\yd(R) \not\in \yd\cDR_f\big(\cR(K)\big)$, by Theorems \ref{Theorem.4.7.13} and \ref{Theorem.4.7.8}, $\rres\big(\yd(R),\#\big) \not\in \yd \cDR\big(\cR(K)\big)$. Therefore, $\yd \cDR\big(\cR(K)\big) \subset \yd \cR^2(K)$.

Summarizing our results, we have
\[
\yd \cR(K) \subset \yd \cDR_f\big(\cR(K)\big) \subset \yd \cDR\big(\cR(K)\big) \subset \yd \cR^2(K)
\]
which completes the proof for $n=1$.

The transition from $n$ to $n+1$ is illustrated by Fig. \ref{Figure.4.3}.
\epr

\begin{figure}[h]
\input{Figure.4.3}
\caption{{}\label{Figure.4.3}}
\end{figure}
According to Theorem \ref{Theorem.4.8.6}, to show that the classes of {\rm ($n$,R)}-transformational languages form a proper hierarchy it is enough to prove the properness of the inclusion $\yd \cDR_f(\Rec) \subset \yd \cR(\Rec)$. For this we need

\begin{lm}\label{Lemma.4.8.7} For each $k$-copying $\DGSDT$ $\gA = (\Sigma,X,A,Y,P,a_0)$ there exists a linear $\DGSDT$ $\gB=(\Sigma,X,B,Y,P',b_0)$ such that $\Par(T\tau_\gB) = \Par(T\tau_\gA)$, for every forest \mbox{$T \subseteq F_\Sigma(X)$}.
\end{lm}
\pr For each $w \in (Y \cup A\Xi)^\ast$, let $\overline{w}$ denote the word obtained from $w$ by erasing all $a\xi$'s $(a \in A,\ \xi \in \Xi)$.

Let $B = \{(a_1,\ldots,a_n) \mid n \le k,\ a_i \in A (i=1,\ldots,n)\}$ and $b_0=(a_0)$. Moreover, $P'$ is defined in the following way:
\begin{itemize}
\item[\rm (i)] Let $\ba =(a_1,\ldots,a_n) \in B$ and $x \in X$ be arbitrary. Assume that the productions $a_ix \rightarrow v_i$ $(a_i \in A,\ v_i \in Y^\ast,\ i=1,\ldots,n)$ are in $P$. Then the production \mbox{$\ba x \rightarrow v_1\ldots v_n$} is in $P'$.

\item[\rm (ii)] Take an arbitrary $\ba = (a_1,\ldots,a_n) \in B$ and $\sigma \in \Sigma_m\ (m\ge 0)$. Suppose $P$ contains, for each $i=1,\ldots,n$, a production
\[
\begin{array}{c}
a_i\sigma \rightarrow w_{ij_1} a_{ij_1} \xi_j w_{ij_2} \ldots w_{ij_{i_j}} a_{ij_{i_j}}  \xi_j w_{ij_{i_j+1}} = w_i \\
\big(w_{ij_1},\ldots, w_{ij_{i_j+1}} \in (Y \cup A(\Xi_m - \{\xi_j\}))^\ast,\ a_{ij_1}, \ldots, a_{ij_{i_j}} \in A,\ 1 \le j \le m \big).
\end{array}
\]
Then the production
\[
\begin{array}{c}
\ba \sigma \rightarrow (a_{11_1},\ldots,a_{11_{1_1}},\ldots,a_{n1_1}, \ldots, a_{n1_{n_1}}) \xi_1 \ldots\\[1mm]
\ldots (a_{1m_1},\ldots,a_{1m_{1_m}},\ldots,a_{nm_1},\ldots,a_{nm_{n_m}})\xi_m \overline{w}_1 \ldots \overline{w}_n
\end{array}
\]
is in $P'$, provided that $1_j + \ldots + n_j \le k\ (j=1,\ldots,m$).
\end{itemize}

Obviously, $\gB$ is a linear $\DGSDT$. Moreover, the derivations in $\gA$ and in $\gB$ are related as follows. Take a vector $\ba \in A^n\ (n\le k)$ and a tree $p \in F_\Sigma(X)$. Consider the derivations $\alpha: \; \ba p^n \Longrightarrow^\ast_\gA w$, where $w=w_1\ldots w_n \in Y^\ast$ and $\alpha_i: \; a_i p \Longrightarrow_\gA^\ast w_i\ (i=1,\ldots,n)$. By the state-sequence of $\alpha$ at a node $d$ of $p$ we mean $(\ba_1,\ldots,\ba_n)$, where $\ba_i\ (1 \le i \le n)$ is the state-sequence of $\alpha_i$ at $d$. Furthermore, we say that $\alpha$ is $k$-copying if the length of the state-sequence of $\alpha$ at any node of $p$ is at most $k$. Assume that $\alpha$ is $k$-copying. Then for some $w' \in Y^\ast$, $\beta \colon \ba p^n \Longrightarrow_\gB^\ast w'$ exists. One can easily show by induction on $\hg(p)$ that the state-sequence of $\beta$ at any node $d$ of $p$ is of length one (if it exists) and coincides, as a sequence of states of $\gA$, with the state-sequence of $\alpha$ at $d$. Finally, $w$ is a permutation of $w'$. Therefore, the equality $\Par(T\tau_\gA) = \Par(T\tau_\gB)$ holds.
\epr

\

From Lemma \ref{Lemma.4.8.7}, by Theorems \ref{Theorem.1.6.17} and \ref{Theorem.4.5.4} and Corollary \ref{Corollary.4.6.6}, we get

\begin{cor}\label{Corollary.4.8.8} Let $T \subseteq F_\Sigma(X)$ be a recognizable forest and $\gA = (\Sigma,X,A,Y,P,a_0)$ a finite-copying $\DGSDT$. Then $\Par(T\tau_\gA)$ is semilinear. \epr
\end{cor}

We now can state and prove that the hierarchy of {\rm ($n$,R)}-transformational languages is infinite.

\begin{thm}\label{Theorem.4.8.9} For every natural number $n$, the inclusions
\[
\yd \cR^n(\Rec) \subset \yd \cDR_f\big(\cR^n(\Rec)\big) \subset \yd \cDR\big(\cR^n(\Rec)\big) \subset \yd \cR^{n+1}(\Rec)
\]
hold.
\end{thm}
\pr By Lemma \ref{Lemma.4.7.2} and Corollary \ref{Corollary.4.6.6}, $\Rec$ is closed under regular insertion and relabeling. Thus, by Theorems \ref{Theorem.4.8.6}, \ref{Theorem.4.5.4}, and \ref{Theorem.4.7.8}, and Corollary \ref{Corollary.4.8.8}, it is enough to show that there exist a regular forest $T \subseteq F_\Sigma(X)$ and a $\GSDT$ $\gA=(\Sigma,X,A,Y,P,a_0)$ such that $\Par(T\tau_\gA)$ is not semilinear. For this let $\Sigma=\Sigma_1=\{\sigma\}$, $A=\{a_0\}$, $X=\{x\}$, $Y=\{y\}$ and $P = \{a_0\sigma \rightarrow a_0\xi_1a_0\xi_1, \ a_0x\rightarrow y\}$. Moreover, let $T=\{\sigma(x)\}^{\ast x}$. Then $T\tau_\gA=\{y^{2^n} \mid n=0,1,\ldots\}$. Thus, $\Par(T\tau_\gA)=\{\langle 2^n\rangle \mid n=0,1,\ldots\}$, which is obviously not semilinear.
\epr

\

From Theorem \ref{Theorem.4.8.9} we directly get

\begin{cor}\label{Corollary.4.8.10} For every natural number $n$ the inclusions
\begin{itemize}
\item[\rm(i)] $\yd \cR^n(\Rec) \subset \yd \cR^{n+1}(\Rec)$,
\item[\rm(ii)] $\cR^n(\Rec) \subset \cR^{n+1}(\Rec)$,
\item[\rm(iii)] $\cR^n \subset \cR^{n+1}$
\end{itemize}
hold. \epr
\end{cor}

Finally, we give two more hierarchies of transformational languages, surface forests and tree transformations.

\begin{thm}\label{Theorem.4.8.11} For every natural number $n$ the inclusions
\[
\yd \cR^n(\Rec) \subset \yd \, \cF^{n+1}(\Rec) \subset \yd \cR^{n+1}(\Rec)
\]
are valid.
\end{thm}

\pr By Theorems \ref{Theorem.4.3.3} and \ref{Theorem.4.3.12} and Corollary \ref{Corollary.4.6.6}, the inclusions $\yd \cR^n(\Rec) \subseteq \yd\, \cF^{n+1}(\Rec) \subseteq \yd \cR^{n+1}(\Rec)$ hold. By the proofs of Theorems \ref{Theorem.4.8.6} and \ref{Theorem.4.8.9}, $\yd \cR^n(\Rec)$ is a proper subclass of $\yd \cH\big(\cR^n(\Rec)\big)$. Moreover, by Theorems \ref{Theorem.4.3.3} and \ref{Theorem.4.3.12} and Corollary \ref{Corollary.4.6.6}, the equality $\cH\big(\cR^n(\Rec)\big) = \cF^{n+1}(\Rec)$ holds. Thus, the inclusion $\yd \cR^n(\Rec) \subset \yd \cF^{n+1}(\Rec)$ is valid. Finally, by Theorem \ref{Theorem.4.8.9}, $\yd \cH\big(\cR^n(\Rec)\big) \subseteq \yd \cDR\big(\cR^n(\Rec)\big) \subset \yd \cR^{n+1}(\Rec)$. Therefore, the inclusion $\yd \, \cF^{n+1}(\Rec) \subset \yd \cR^{n+1}(\Rec)$ is also valid.
\epr

\

From Theorem \ref{Theorem.4.8.11}, using Theorems \ref{Theorem.4.3.3} and \ref{Theorem.4.3.12} and Corollary \ref{Corollary.4.6.6}, we get the following results.

\begin{cor}\label{Corollary.4.8.12} For every natural number $n$ the inclusions
\[
\cR^n(\Rec)\subset \cF^{n+1}(\Rec) \subset \cR^{n+1}(\Rec)
\]
hold. \epr
\end{cor}

\begin{cor}\label{Corollary.4.8.13} For every natural number $n$ the inclusions
\begin{itemize}
\item[\rm(i)] $\yd \cF^n(\Rec) \subset \yd \cF^{n+1}(\Rec)$,
\item[\rm(ii)] $\cF^n(\Rec) \subset \cF^{n+1}(\Rec)$,
\item[\rm(iii)] $\cF^n \subset \cF^{n+1}$
\end{itemize}
are valid. \epr
\end{cor}

%% file: Figure.4.3.tex
\begin{tikzpicture}[>=latex]
	\begin{scope}[execute at begin node=$, execute at end node=$]
	
		\node [below,scale=0.9] at (0.1\textwidth,0) {yd\cDR(\cR^n(K))} ;
		\node [below,scale=0.9] at (0.3\textwidth,0) {yd\cR^{n+1}(K)} ;
		\node [below,scale=0.9] at (0.5\textwidth,0) {yd\cDR_f(\cR^{n+1}(K))} ;
		\node [below,scale=0.9] at (0.7\textwidth,0) {yd\cDR(\cR^{n+1}(K))} ;
		\node [below,scale=0.9] at (0.9\textwidth,0) {yd\cR^{n+2}(K)} ;
		
		\draw [-] (0,0) -- (\textwidth,0); 
		
		\draw [-] (0.2\textwidth,0) -- (0.2\textwidth,-0.2); 
		\draw [-] (0.4\textwidth,0) -- (0.4\textwidth,-0.2); 
		\draw [-] (0.6\textwidth,0) -- (0.6\textwidth,-0.2); 
		\draw [-] (0.8\textwidth,0) -- (0.8\textwidth,-0.2); 
		\draw [-] (1.0\textwidth,0) -- (1.0\textwidth,-0.2);
		
		\draw[<-] (9.5,0) arc [start angle=15,end angle=165,radius=3.2cm];
		\node[scale=0.9] at (7.3,2.4){c_\ast};
		
		\draw[<-] (7,0) arc [start angle=15,end angle=165,radius=1.4cm];
		\node[scale=0.9] at (6.4,1.1){c_2};
		
		\draw[<-] (13,0) arc [start angle=15,end angle=165,radius=1.4cm];
		\node[scale=0.9] at (12.8,0.9){\rres};
	\end{scope}
\end{tikzpicture}

%% file: Section.4.9.tex
Since the equivalence problem for (nondeterministic) generalized sequential machines is undecidable, there exists no algorithm to decide for arbitrary two tree transducers whether or not they are equivalent. In this section we show that there is an algorithm for deciding the equivalence of two tree transducers when at least one of them induces a partial mapping. Moreover, we shall prove that it is decidable whether the tree transformation induced by a given tree transducer is a partial mapping when restricted to a given recognizable forest.

We start by introducing a concept.

\begin{df}\rm \label{Definition.4.9.1} Let $p \in F_\Sigma(X)$. A tree $p' \in \hat{\hat{F}}_\Sigma(X \cup \Xi^n)$ is called a {\em supertree}
\index{supertree}
of $p$ if there are trees $p_1,\ldots,p_n \in F_\Sigma(X)$ such that $p=p'(p_1,\ldots,p_n)$. 
\end{df}

To prove the decidability results we shall give five reduction rules formulated in the following five lemmas. In these lemmas $\gA=(\Sigma,X,A,\Omega,Y,P,A')$ will be a fixed {\rm R}-transducer and $\bB=(\cB,\beta,B')$ will be a fixed $\Sigma X$-recognizer with $\cB=(B,\Sigma)$ and $T(\bB) =T$. Furthermore, set $Q=\{p\in T\mid |p\tau_\gA|\ge 2\}$, i.e., $Q$ consists of all trees from $T$ which are translated into at least two different output trees by $\gA$. 

\begin{lm}\label{Lemma.4.9.2} Let $p_1, p_2 \in \hat{F}_\Sigma(X\cup \Xi_1)$, $p_3 \in F_\Sigma(X),\ n_1,n_1',n_2,n_2'  \ge 0$, 
\mbox{$q_1 \in \hat{F}_\Omega(Y \cup \Xi_{n_1})$}, 
$q_1' \in \hat{F}_\Omega(Y \cup \Xi_{n_1'})$, $
\bq_2 \in \hat{F}_\Omega^{n_1}(Y \cup \Xi_{n_2})$, $
\bq_2' \in \hat{F}_\Omega^{n_1'}(Y \cup \Xi_{n_2'})$, $
\bq_3 \in F_\Omega(Y)^{n_2}$, $
\bq_3' \in F_\Omega(Y)^{n_2'}$, $
a_0, a_0' \in A'$ and $\ba_i \in A^{n_i}$, $\ba_i' \in A^{n_i'}\ (i=1,2)$.
Moreover, set $A_i=\{a_{i_j} \mid j=1,\ldots,n_i\}$ and $A_i'=\{a_{i_j}' \mid j=1,\ldots,n_i'\}\ (i=1,2)$. Assume that the following conditions are satisfied:
\begin{enumerate}
\item[\rm (i)] $p_1\big(p_2(p_3)\big) \in T$,
\item[\rm(ii)] $a_0 p_1 \Rightarrow^\ast q_1(\ba_1 \xi_1^{n_1})$, $a_0'p_1 \Rightarrow^\ast q_1'(\ba_1' \xi_1^{n_1'})$,
\item[\rm(iii)] $\ba_1 p_2^{n_1} \Rightarrow^\ast \bq_2(\ba_2 \xi_1^{n_2})$, $\ba_1' p_2^{n_1'} \Rightarrow^\ast \bq_2'(\ba_2' \xi_1^{n_2'})$,
\item[\rm(iv)] $\ba_2 p_3^{n_2} \Rightarrow^\ast \bq_3$, $\ba_2' p_3^{n_2'} \Rightarrow^\ast \bq_3'$,
\item[\rm(v)] $p_3\hat{\beta} = p_2(p_3)\hat{\beta}$, $A_1 \subseteq A_2$, $A_1' \subseteq A_2'$,
\item[\rm(vi)] for all $\br \in F_\Omega(Y)^{n_1}$ and $\br' \in F_\Omega(Y)^{n_1'}$, $q_1(\br) \not= q_1'(\br')$.
\end{enumerate}
Then $p_1(p_3) \in Q$.
\end{lm}

\pr First let us note that the conditions of Lemma \ref{Lemma.4.9.2} imply $p_1\big(p_2(p_3)\big) \in Q$. 

Next take two mappings $f\colon \{1,\ldots,n_1\} \rightarrow \{1,\ldots,n_2\}$ and $g \colon \{1,\ldots,n_1'\} \rightarrow \{1, \ldots, n_2'\}$ such that $a_{1_i} = a_{2_{f(i)}}\ (i=1,\ldots,n_1)$ and $a_{1_i}' = a_{2_{g(i)}}'\ (i=1,\ldots,n_1')$. By (v), there are such mappings $f$ and $g$. Thus, by (iv), we have $\ba_1p_3^{n_1}\Rightarrow^\ast\br$ and $\ba_1'p_3^{n_1'}\Rightarrow^\ast\br'$ with $\br = (q_{3_{f(1)}},\ldots,q_{3_{f(n_1)}})$ and $\br' = (q_{3_{g(1)}}',\ldots,q_{3_{g(n_1')}}')$. This, by (ii) implies $a_0p_1(p_3) \Rightarrow^\ast q_1(\br)$ and  $a_0'p_1(p_3) \Rightarrow^\ast q_1'(\br')$. 
By (vi), $q_1(\br) \not= q_1'(\br')$. Moreover by (v), $p_1(p_3) \in T$. Therefore, $p_1(p_3) \in Q$.
\epr 

\begin{lm}\label{Lemma.4.9.3} Let $p_1 \in \hat{F}_\Sigma(X \cup \Xi_1)$, $p_2 \in F_\Sigma(X)$, $n,n' > 0$, $q_1 \in \hat{F}_\Omega(Y \cup \Xi_n)$, $q_1' \in \hat{F}_\Omega(Y \cup \Xi_{n'})$, $\bq_2 \in F_\Omega(Y)^n$, $\bq_2' \in F_\Omega(Y)^{n'}$, $a_0,a_0' \in A'$, $\ba \in A^n$ and $\ba' \in A^{n'}$. Furthermore, let $K$ be the maximum of the heights of the right-hand sides of the productions from $P$. Assume that the following conditions are satisfied:
\begin{enumerate}
\item[\rm (i)] $p_1(p_2) \in T$,
\item[\rm(ii)] $a_0 p_1 \Rightarrow^\ast q_1(\ba \xi_1^{n})$, $a_0'p_1 \Rightarrow^\ast q_1'(\ba' \xi_1^{n'})$,
\item[\rm(iii)] $\ba p_2^{n} \Rightarrow^\ast \bq_2$, $\ba' p_2^{n'} \Rightarrow^\ast \bq_2'$,
\item[\rm(iv)] $\ppath_1(q_1)$ is an initial segment of $\ppath_1(q_1')$, 
and\\
\mbox{$l\big(\ppath_1(q_1')\big) - l\big(\ppath_1(q_1)\big) > |\gp A|^2 |B| K$}, $\hg(p_2) \ge |\gp A|^2 |B|$. 
\end{enumerate}
Then there exists an $r \in F_\Sigma(X)$ with $|r| < |p_2|$ such that $p_1(r) \in Q$.
\end{lm}
\pr Set $R=\{r \in F_\Sigma(X) \mid p_1(r) \in T$, $|r| \le |p_2|$, $\ba r^n \Rightarrow^\ast \bs$, \mbox{$\ba' r^{n'} \Rightarrow^\ast \bs'$} for some $\bs \in F_\Omega(Y)^n$  and $\bs' \in F_\Omega(Y)^{n'}\}$. Obviously, $R$ is nonvoid. Denote by $r$ an element from $R$ with minimal length. We prove that $p_1(r) \in Q$ and $\hg(r) < |\gp A|^2 |B|$.

First assume that $\hg(r) \ge |\gp A|^2 |B|$. Then there are 
\[
\begin{array}{c}
r_1,r_2 \in \hat{F}_\Sigma(X \cup \Xi_1),\ r_3  \in F_\Sigma(X),\ m_1,m_1',m_2,m_2' \ge 0,\  \bs_1 \in \hat{F}_\Omega^n(Y \cup \Xi_{m_1}),\\[1mm]
\bs_1' \in \hat{F}_\Omega^{n'}(Y\cup \Xi_{m_1'}),\ 
\bs_2 \in \hat{F}_\Omega^{m_1}(Y\cup \Xi_{m_2}),\  
\bs_2' \in \hat{F}_\Omega^{m_1'}(Y\cup \Xi_{m_2'}),\\[1mm]
\bs_3 \in F_\Omega(Y)^{m_2},\ \bs_3' \in F_\Omega(Y)^{m_2'},\ \bb_i \in A^{m_i},\ \bb_i' \in A^{m_i'}\ (i=1,2) \text{ such } 
\end{array}
\]
that
\begin{itemize}
\item[(I)] $r=r_1\big(r_2(r_3)\big)$, $r_2 \not= \xi_1$,
\item[(II)] $\ba r_1^n \Rightarrow^\ast \bs_1(\bb_1 \xi_1^{m_1})$, $\ba' r_1^{n'} \Rightarrow^\ast \bs_1'(\bb_1' \xi^{m_1'})$,
\item[(III)]  $\bb_1 r_2^{m_1} \Rightarrow^\ast \bs_2(\bb_2 \xi_1^{m_2})$, $\bb_1' r_2^{m_1'} \Rightarrow^\ast \bs_2'(\bb_2' \xi_1^{m_2'})$,
\item[(IV)]   $\bb_2 r_3^{m_2} \Rightarrow^\ast \bs_3$, $\bb_2' r_3^{m_2'} \Rightarrow^\ast \bs_3'$,
\item[(V)] $r_3 \hat{\beta} = r_2(r_3)\hat{\beta}$, $B_1 \subseteq B_2$ and $B_1' \subseteq B_2'$, where $B_i=\{b_{i_j} \mid 1 \le j \le m_i\}$, \\
$B_i'=\{b_{i_j}' \mid 1 \le i \le m_i'\}\ (i=1,2)$. 
\end{itemize}

Take two mappings $f\colon \{1,\ldots,m_1\} \rightarrow \{1,\ldots,m_2\}$ and $g\colon \{1,\ldots,m_1'\} \rightarrow \{1,\ldots,m_2'\}$ such that  $b_{1_i} = b_{2_{f(i)}}\ (1 \le i \le m_1)$ and $b_{1_i}'=b_{2_g(i)}'\ (1\le i \le m_1')$. Obviously, $\ba t^n \Rightarrow^\ast \bs_1(s_{3_{f(1)}},\ldots,s_{3_{f(m_1)}})$ and $\ba' t^{n'} \Rightarrow^\ast  \bs_1'(s_{3_{g(1)}}',\ldots,s_{3_{g(m_1')}}')$, where $t=r_1(r_3)$. Moreover, $r_1(r_3) \hat{\beta} = r \hat{\beta}$ also holds. Therefore, $r_1(r_3) \in R$, which is a contradiction since \mbox{$|r_1(r_3)| < |r|$}.

Thus, we got that $\hg(r) < |\gp A|^2 |B|$. Therefore, for arbitrary vectors $\bs \in F_\Omega(Y)^n$ and $\bs' \in F_\Omega(Y)^{n'}$ satisfying $\ba r^n \Rightarrow^\ast \bs$ and $\ba' r^{n'} \Rightarrow^\ast \bs'$, the inequalities $\hg(s_1),\hg(s_1') \le |\gp A|^2 |B|K$ hold. This, by (iv), obviously implies the conclusion of Lemma \ref{Lemma.4.9.3}.  
\epr

\begin{lm}\label{Lemma.4.9.4} Let $p_1,p_2,p_3 \in \hat{F}_\Sigma(X \cup \Xi_1)$, $p_4 \in F_\Sigma(X)$, $n_i, n_i', m_i \ge 0\ (i=1,2,3)$,
\[
\begin{array}{c}
q_1 \in \hat{F}_\Omega(Y \cup \Xi_{n_1+1}),\ q_1' \in \hat{F}_\Omega(Y \cup \Xi_{n_1'+1}),\ r_1 \in \hat{F}_\Omega(Y \cup \Xi_{m_1}),\\[1mm]
\bq_2 \in \hat{F}_\Omega^{n_1}(Y \cup \Xi_{n_2}),\ \bq_2' \in \hat{F}_\Omega^{n_1'}(Y \cup \Xi_{n_2'}),\ \br_2 \in \hat{F}_\Omega^{m_1}(Y \cup \Xi_{m_2}),\\[1mm]
\bq_3 \in \hat{F}_\Omega^{n_2}(Y \cup \Xi_{n_3}),\ \bq_3' \in \hat{F}_\Omega^{n_2'}(Y \cup \Xi_{n_3'}),\ \br_3 \in \hat{F}_\Omega^{m_2}(Y \cup \Xi_{m_3}),\\[1mm]
\bq_4 \in F_\Omega(Y)^{n_3},\ \bq_4' \in F_\Omega(Y)^{n_3'},\ \br_4 \in F_\Omega(Y)^{m_3},\\[1mm]
a_0,a_0' \in A',\ a \in A,\ \ba_i \in A^{n_i},\ \ba_i' \in A^{n_i'},\ \bb_i \in A^{m_i}\ (i=1,2,3).
\end{array}
\]
Moreover, take an $r \in F_\Omega(Y)$, and let $r' = r_1\big(\br_2(\br_3(\br_4))\big)$.
Finally, set $A_i=\{a_{i_j} \mid j=1,\ldots,n_i\}$, $A_i'=\{a_{i_j}' \mid j=1,\ldots, n_i'\}$ and $B_i=\{b_{i_j}' \mid j=1,\ldots,m_i\}\ (i=1,2,3)$. Assume that the following conditions are satisfied:
\begin{enumerate}
\item[\rm (i)] $p_1\big(p_2(p_3(p_4))\big) \in T$,
\item[\rm(ii)] $a_0 p_1 \Rightarrow^\ast q_1(a \xi_1, \ba_1 \xi_1^{n_1})$, $
a_0'p_1 \Rightarrow^\ast q_1'\big(r_1(\bb_1 \xi_1^{m_1}), \ba_1' \xi_1^{n_1'}\big)$,
\item[\rm(iii)] $\ba_1  p_2^{n_1} \Rightarrow^\ast \bq_2(\ba_2 \xi_1^{n_2})$, $
\ba_1' p_2^{n_1'} \Rightarrow^\ast \bq_2'(\ba_2' \xi_1^{n_2'})$,\\[1mm]
$a p_2 \Rightarrow^\ast a \xi_1$, $
\bb_1 p_2^{m_1} \Rightarrow^\ast \br_2(\bb_2 \xi_1^{m_2})$,
\item[\rm(iv)] $\ba_2  p_3^{n_2} \Rightarrow^\ast \bq_3(\ba_3 \xi_1^{n_3})$, $
\ba_2' p_3^{n_2'} \Rightarrow^\ast \bq_3'(\ba_3' \xi_1^{n_3'})$,\\[1mm]
$a p_3 \Rightarrow^\ast a \xi_1$, $
\bb_2 p_3^{m_2} \Rightarrow^\ast \br_3(\bb_3 \xi_1^{m_3})$,
\item[\rm(v)] $\ba_3 p_4^{n_3} \Rightarrow^\ast \bq_4$, $
\ba_3' p_4^{n_3'} \Rightarrow^\ast \bq_4'$, $
a p_4 \Rightarrow^\ast r$, $
\bb_3p_4^{m_3} \Rightarrow^\ast \br_4$,
\item[\rm(vi)] $p_4 \hat{\beta} = p_3(p_4)\hat{\beta} = p_2\big(p_3(p_4)\big)\hat{\beta}$, $
A_1 \subseteq A_2 \subseteq A_3$,\\[1mm]
$A_1' \subseteq A_2' \subseteq A_3'$, $
B_1=B_2 \subseteq B_3$,
\item[\rm(vii)] $r\not=r'$, $\ppath_1(q_1) = \ppath_1(q_1')$.
\end{enumerate}
Then at least one of the trees $p_1\big(p_2(p_4)\big)$, $p_1\big(p_3(p_4)\big)$ and  $p_1(p_4)$ is in $Q$. 
\end{lm}
\pr First note that the conditions of Lemma \ref{Lemma.4.9.4} imply  $p_1\big(p_2(p_3(p_4))\big) \in Q$. Indeed, let $t=q_1\big(\xi_1,\bq_2(\bq_3(\bq_4))\big)$ and  $t'=q_1'\big(\xi_1,\bq_2'(\bq_3'(\bq_4'))\big)$. Then 
\[
a_0 p_1\big(p_2(p_3(p_4))\big) \Rightarrow^\ast t(r), \; \; \;
a_0' p_1\big(p_2(p_3(p_4))\big) \Rightarrow^\ast t'(r')
\]
and $t(r) \not= t(r')$. 

Take six mappings $f_i\colon \{1,\ldots,n_i\} \rightarrow   \{1,\ldots,n_{i+1}\}$, $
g_i\colon \{1,\ldots,n_i'\} \rightarrow   \{1,\ldots,n_{i+1}'\}$ and
\[
h_i\colon \{1,\ldots,m_i\} \rightarrow   \{1,\ldots,m_{i+1}\} \; (i=1,2)
\]
such that 
\[
\begin{array}{c}
a_{i_j} = a_{i+ 1_{f_i(j)}}\ (i=1,2,\ 1 \le j \le n_i)\  a_{i_j}'=a_{i+1_{g_i(j)}}'\ (i=1,2,\ 1\le j\le n_i'),\\
b_{i_j} = b_{i+1_{h_i(j)}} \ (i=1,2, \ 1 \le j \le m_i).
\end{array}
\]
Furthermore, set $f_3= f_1\circ f_2$, $g_3=g_1\circ g_2$ and $h_3=h_1 \circ h_2$. Moreover, introduce the notations 
\begin{align*}
&\bs_1 = (q_{3_{f_1(1)}}, \ldots, q_{3_{f_1(n_1)}})(\bq_4), \ 
\bs_1' = (q_{3_{g_1(1)}}', \ldots, q_{3_{g_1(n_1')}}')(\bq_4'),\\
&\bt_1 = (r_{3_{h_1(1)}}, \ldots, r_{3_{h_1(m_1)}})(\br_4),\\
&\bs_2 = \bq_2(q_{4_{f_2(1)}}, \ldots, q_{4_{f_2(n_2)}}), \ 
\bs_2' = \bq_2'(q_{4_{g_2(1)}}', \ldots, q_{4_{g_2(n_2')}}'),\\
&\bt_2 = \br_2(r_{4_{h_2(1)}}, \ldots, r_{4_{h_2(m_2)}}),\\
&\bs_3 = (q_{4_{f_3(1)}}, \ldots, q_{4_{f_3(n_1)}}), \ 
\bs_3' = (q_{4_{g_3(1)}}', \ldots, q_{4_{g_3(n_1')}}'),\\
&\bt_3 = (r_{4_{h_3(1)}}, \ldots, r_{4_{h_3(m_1)}}).
\end{align*}
Then the following derivations obviously hold:
\[
\begin{array}{c}
a_0p_1\big(p_3(p_4)\big) \Rightarrow^\ast q_1(r,\bs_1),\  
a_0'p_1\big(p_3(p_4)\big) \Rightarrow^\ast q_1'\big(r_1(\bt_1),\bs_1'\big),\\[1mm]
a_0p_1\big(p_2(p_4)\big) \Rightarrow^\ast q_1(r,\bs_2),\  
a_0'p_1\big(p_2(p_4)\big) \Rightarrow^\ast q_1'\big(r_1(\bt_2),\bs_2'\big),\\[1mm]
a_0p_1(p_4) \Rightarrow^\ast q_1(r,\bs_3),\  
a_0'p_1(p_4) \Rightarrow^\ast q_1'\big(r_1(\bt_3),\bs_3'\big).
\end{array}
\]
It is also obvious that $p_1\big(p_3(p_4)\big), p_1\big(p_2(p_4)\big), p_1(p_4) \in T$. 

Now assume that $p_1\big(p_2(p_4)\big) \not\in Q$. Then, by (vi) and (vii), $m_1,m_2,m_3 >0$ and there exists an $i\ (1 \le i \le m_2)$ such that $r_{3_i}(\br_4)\not=r_{4_{h_2(i)}}$. We can choose $h_1$ in such a way that for some $j\ (1 \le j \le m_1)$ $h_1(j)=i$ holds. Now assume that, under the latter choice of $h_1$, none of $p_1\big(p_3(p_4)\big)$ and $p_1(p_4)$ are in $Q$. Then we get $r_1(\bt_1)=r_1(\bt_3)=r$. But this is impossible since $t_{1_j} \not= t_{3_j}$.
\epr

\begin{lm}\label{Lemma.4.9.5} Let $p_1,p_2,p_3 \in \hat{F}_\Sigma(X \cup \Xi_1)$, $p_4 \in F_\Sigma(X)$, $n_i, n_i', m_i \ge 0\ (i=1,2,3)$,
\[
\begin{array}{c}
q_1 \in \hat{F}_\Omega(Y \cup \Xi_{n_1+1}),\ 
q_1' \in \hat{F}_\Omega(Y \cup \Xi_{n_1'+1}),\ 
r_1 \in \hat{F}_\Omega(Y \cup \Xi_{m_1}),\\[1mm]
\bq_2 \in \hat{F}_\Omega^{n_1}(Y \cup \Xi_{n_2}),\ 
\bq_2' \in \hat{F}_\Omega^{n_1'}(Y \cup \Xi_{n_2'}),\ 
\br_2 \in \hat{F}_\Omega^{m_1}(Y \cup \Xi_{m_2}),\\[1mm]
\bq_3 \in \hat{F}_\Omega^{n_2}(Y \cup \Xi_{n_3}),\ 
\bq_3' \in \hat{F}_\Omega^{n_2'}(Y \cup \Xi_{n_3'}),\ 
\br_3 \in \hat{F}_\Omega^{m_2}(Y \cup \Xi_{m_3}),\\[1mm]
\bq_4 \in F_\Omega(Y)^{n_3},\ 
\bq_4' \in F_\Omega(Y)^{n_3'},\ 
\br_4 \in F_\Omega(Y)^{m_3},\\[1mm]
a_0,a_0' \in A',\ 
\ba_i \in A^{n_i},\ 
\ba_i' \in A^{n_i'},\ 
\bb_i \in A^{m_i}\ 
(i=1,2,3).
\end{array}
\]
Moreover, take an $r' \in F_\Omega(Y)$, and let $r = r_1\big(\br_2(\br_3(\br_4))\big)$.
Finally, set $A_i=\{a_{i_j} \mid j=1,\ldots,n_i\}$, $A_i'=\{a_{i_j}' \mid j=1,\ldots, n_i'\}$ and $B=\{b_{i_j} \mid j=1,\ldots,m_i\}\ (i=1,2,3)$. Assume that the following conditions are satisfied:
\begin{enumerate}
\item[\rm (i)] $p_1\big(p_2(p_3(p_4))\big) \in T$,
\item[\rm(ii)] $a_0 p_1 \Rightarrow^\ast q_1\big(r_1(\bb_1\xi_1^{m_1}), \ba_1 \xi_1^{n_1}\big)$, $
a_0'p_1 \Rightarrow^\ast q_1'(r', \ba_1' \xi_1^{n_1'})$,
\item[\rm(iii)] $\ba_1  p_2^{n_1} \Rightarrow^\ast \bq_2(\ba_2 \xi_1^{n_2})$, $
\ba_1' p_2^{n_1'} \Rightarrow^\ast \bq_2'(\ba_2' \xi_1^{n_2'})$, $
\bb_1 p_2^{m_1} \Rightarrow^\ast \br_2(\bb_2 \xi_1^{m_2})$,
\item[\rm(iv)] $\ba_2  p_3^{n_2} \Rightarrow^\ast \bq_3(\ba_3 \xi_1^{n_3})$, $
\ba_2' p_3^{n_2'} \Rightarrow^\ast \bq_3'(\ba_3' \xi_1^{n_3'})$, $
\bb_2 p_3^{m_2} \Rightarrow^\ast \br_3(\bb_3 \xi_1^{m_3})$,
\item[\rm(v)] $\ba_3 p_4^{n_3} \Rightarrow^\ast \bq_4$, $
\ba_3' p_4^{n_3'} \Rightarrow^\ast \bq_4'$, $
\bb_3 p_4^{m_3} \Rightarrow^\ast \br_4$,
\item[\rm(vi)] $p_4 \hat{\beta} = p_3(p_4)\hat{\beta} = p_2\big(p_3(p_4)\big)\hat{\beta}$,\\ 
$A_1 \subseteq A_2 \subseteq A_3$, $
A_1' \subseteq A_2' \subseteq A_3'$, $
B_1=B_2 \subseteq B_3$,
\item[\rm(vii)] $r\not=r'$, $
\ppath_1(q_1) = \ppath_1(q_1')$.
\end{enumerate}
Then at least one of the trees $p_1\big(p_2(p_4)\big)$, $p_1\big(p_3(p_4)\big)$ and  $p_1(p_4)$ is in $Q$. 
\end{lm}
\pr The proof of this lemma is similar to that of Lemma \ref{Lemma.4.9.4}.
\epr

\begin{lm}\label{Lemma.4.9.6}
Let
\[
\begin{array}{c}
p_1,p_2 \in \hat{F}_\Sigma(X \cup \Xi_1),\ 
p_3 \in F_\Sigma(X),\ 
k,l,m,k',l',m' \ge 0,\\[1mm]
q_1 \in \hat{F}_\Omega(Y \cup \Xi_{k+1}),\ 
q_1' \in \hat{F}_\Omega(Y \cup \Xi_{k'+1}),\ 
q_2 \in \hat{F}_\Omega(Y \cup \Xi_{l+1}),\ 
q_2' \in \hat{F}_\Omega(Y \cup \Xi_{l'+1}), \\[1mm]
\br \in \hat{F}_\Omega^k (Y \cup \Xi_m),\ 
\br' \in \hat{F}_\Omega^{k'}(Y \cup \Xi_{m'}),\ 
q_3 \in \hat{F}_\Omega (Y \cup \Xi_1),\ 
q_3',r \in F_\Omega(Y), \\[1mm]
\bs \in F_\Omega(Y)^l,\ 
\bs' \in F_\Omega(Y)^{l'},\ 
\bt \in F_\Omega(Y)^m,\ 
\bt' \in F_\Omega(Y)^{m'},\ 
a_0, a_0' \in A',\ 
a,a' \in A, \\[1mm]
\ba \in A^k,\ 
\ba' \in A^{k'},\ 
\bb \in A^l,\ 
\bb' \in A^{l'},\ 
\bc \in A^m \text{ and } \bc' \in A^{m'}.
\end{array}
\]
Moreover, set $ A_1=\{ a_i \mid i=1, \ldots ,k \}$, $
B_1=\{b_i \mid i=1, \ldots, l\}$, $
C_1=\{c_i \mid i=1, \ldots, m\}$, $
A_1'=\{a_i' \mid i=1, \ldots, k'\}$, $
B_1'=\{b_i' \mid i=1, \ldots, l'\}$ and $C_1'=\{c_i' \mid i=1, \ldots m'\}$. Assume that the following conditions are satisfied:

\begin{enumerate}
\item[\rm (i)] $p_1\big(p_2(p_3)\big) \in T$,
\item[\rm (ii)] $a_0p_1\Rightarrow^\ast q_1(a\xi_1,\ba \xi_1^k)$, $
a_0'p_1\Rightarrow^\ast q_1'(a'\xi_1, \ba' \xi_1^{k'})$,
\item[\rm (iii)] $ap_2 \Rightarrow^\ast q_2(a \xi_1, \bb\xi_1^l)$, $
a'p_2 \Rightarrow^\ast q_2'(a' \xi_1, \bb'\xi_1^{l'})$,\\[1mm]
$\ba p_2^k \Rightarrow^\ast \br(\bc \xi_1^m)$, $
\ba'p_2^{k'} \Rightarrow^\ast \br'(\bc'\xi_1^{m'})$,
\item[\rm (iv)] $ap_3 \Rightarrow^\ast q_3(r)$, $
a'p_3 \Rightarrow^\ast q_3'$, $
\bb p_3^l \Rightarrow^\ast \bs$, $
\bb'p_3^{l'} \Rightarrow^\ast \bs', \\[1mm]
\bc p_3^m \Rightarrow^\ast \bt$, $
\bc' p_3^{m'} \Rightarrow^\ast \bt'$,
\item[\rm (v)] $A_1 \subseteq B_1 \cup C_1$, $
A_1' \subseteq B_1' \cup C_1'$, $
p_3 \hat{\beta} = p_2(p_3) \hat{\beta}$,
\item[\rm (vi)] $\ppath_1(q_1') = \ppath_1(q_1)\ppath_1(q_3) \text{ and } r\not=q_3'$.
\end{enumerate}
Then $p_1(p_3) \in Q$.
\end{lm}

\pr Introduce the notation $\bd=(\bb,\bc)$, $\bd'=(\bb', \bc')$, $\bu=(\bs, \bt)$ and $\bu'=(\bs', \bt')$. Moreover, take two mappings $f \colon \{1, \ldots, k\} \rightarrow \{1, \ldots, l+m\}$ and $g \colon \{1, \ldots, k'\} \rightarrow \{1, \ldots, l'+m'\}$ satisfying the equalities $a_i=d_{f(i)}\ (1 \le i \le k)$ and $a'_i=u_{g(i)}$ \mbox{$(1 \le i \le k')$}. Obviously, there are derivations $a_0p_1(p_3) \Rightarrow^\ast q_1\big(q_3(r),u_{f(1)}, \ldots, u_{f(k)}\big)$ and $a_0'p_1(p_3) \Rightarrow^\ast q_1'\big( q_3', u_{g(1)}', \ldots, u_{g(k')}'\big)$. Moreover, $p_1(p_3) \in T$. Since
\[
\ppath_1 \big( q_1(q_3(\xi_1), u_{f(1)}, \ldots, u_{f(k)}) \big) = \ppath_1 \big( q_1'(\xi_1, u_{g(1)}', \ldots, u_{g(k')}') \big)
\]
and $q_3' \not= r, q_1\big( q_3(r)$, $u_{f(1)}, \ldots, u_{f(k)}) \not= q_1'(q_3', u_{g(1)}', \ldots, u_{g(k')}')$. Hence, $p_1(p_3) \in Q$.
\epr

\

Now we are ready to state a theorem from which the main decidability results of this section easily follow. 

\begin{thm}\label{Theorem.4.9.7}
There exists an algorithm to decide whether $Q$ is empty.
\end{thm}
\pr
Let $K$ denote the maximum of the heights of the right-hand sides of the productions from $P, \| A \| = 2^{|A|}$ and let $L$ be the number of all words over $\{ 1,\ldots,r_\Sigma \}$ with length at most $\|A\|^2|B|K$, where $r_\Sigma$ is the maximal $m$ for which $\Sigma_m \not= \emptyset$. Moreover, let $k=\|A\|^2|A|^2 |B|2L+1$, $l=k+(2\|A\|^3 |A| |B|) ( \|A\|^2 |B| K+1)$ and $m=l+2\|A\|^3 |B|$.

We shall show that $Q$ is nonvoid iff it contains a tree with height less than $m$. The case $K=0$ being obvious, we assume that $K \not= 0$.

Let $p$ be an element of $Q$ with minimal length, and $q, q' \in F_\Omega(Y)$ trees such that $q \not= q'$ and $(p,q),(p,q') \in \tau_\gA$. Assume that $\hg(p) \ge m$. Then there are $a_0, a_0' \in A'$, $p_0, \ldots, p_m \in \hat{F}_\Sigma(X \cup \Xi_1)$, $p_{m+1} \in F_\Sigma(X)$, $n_i, n_i' \ge 0\ (i = 0, \ldots, m)$, $q_0 \in \hat{F}_\Omega(Y \cup \Xi_{n_0})$, $q_0' \in \hat{F}_\Omega (Y \cup \Xi_{n_0'})$, $\bq_i \in \hat{F}_\Omega^{n_{i-1}} (Y \cup \Xi_{n_i})$, $\bq_i' \in \hat{F}_\Omega^{n_{i-1}'}(Y \cup \Xi_{n_i'})\ (i=1, \ldots, m)$, $\bq_{m+1} \in F_\Omega(Y)^{n_m}$, $\bq_{m+1}' \in F_\Omega(Y)^{n_m'}$, $\ba_i \in A^{n_i}$, $\ba_i' \in A^{n_i'}\ (i=0, \ldots, m)$ such that the following conditions are satisfied:

\begin{enumerate}
\item[(1)] $p=p_0 \big( p_1(\ldots (p_{m+1}) \ldots) \big),\ 
p_i \not= \xi_1$ $(i=1, \ldots, m)$,
\item[(2)] $q = q_0\big( \bq_1(\ldots (\bq_{m+1}) \ldots) \big),\ 
q' = q_0'\big( \bq_1'( \ldots (\bq_{m+1}') \ldots) \big)$,
\item[(3)] $a_0p_0 \Rightarrow^\ast q_0(\ba_0 \xi_1^{n_0}),\ 
a_0'p_0 \Rightarrow^\ast q_0'(\ba_0' \xi_1^{n_0'})$,\\[1mm]
$\ba_ip_{i+1}^{n_i} \Rightarrow^\ast \bq_{i+1}(\ba_{i+1} \xi_1^{n_{i+1}}),\ 
\ba_i'p_{i+1}^{n_i'} \Rightarrow^\ast \bq_{i+1}'(\ba_{i+1}' \xi_1^{n_{i+1}'})$\\[1mm]
$(i=0,\ldots, m-1),\ 
\ba_mp_{m+1}^{n_m} \Rightarrow^\ast \bq_{m+1},\ 
\ba_m'p_{m+1}^{n_m'} \Rightarrow^\ast \bq_{m+1}'$.
\end{enumerate}

For $i=0,\ldots,m$, introduce the notations $\check{p}_i = p_0 \big(p_1(\ldots(p_i)\ldots) \big)$, $\check{q}_i = q_0 \big(\bq_1(\ldots(\bq_i)\ldots)\big)$ and $\check{q}_i'=q_0'(\bq_1'(\ldots(\bq_i')\ldots)\big)$. Moreover, let $\hat{p}_i=p_{i+1}\big(\ldots(p_{m+1})\ldots\big)$, $\hat{q}_i=\bq_{i+1} \big(\ldots(\bq_{m+1})\ldots \big)$ and $\hat{q}_i'=\bq_{i+1}'\big( \ldots(\bq_{m+1}')\ldots\big)\ (i=0,\ldots,m)$.
Finally, set $A_i=\{a_{i_j} \mid 1 \le j \le n_i\}$ and $A_i'=\{a_{i_j}' \mid 1 \le j \le n_i'\}\ (i=0, \ldots, m)$.

If $\check{q}_l(\br) \not= \check{q}_l'(\br')$ holds for all $\br \in F_\Omega(Y)^{n_l}$ and $\br' \in F_\Omega(Y)^{n_l'}$, then the fact that $m-l+1 > |\gp A|^2|B|$ makes Lemma \ref{Lemma.4.9.2} applicable and hence there are $i$ and $j$ with $l \le i < j \le m$ such that $\check{p}_i(\hat{p}_j) \in Q$. This is obviously a contradiction since $|\check{p}_i(\hat{p}_j)| < |p|$.

Thus, we may assume that at least one of $n_l$ and $n_l'$, say $n_l$, is greater than $0$. Moreover, it can also be supposed that there are an $i_l\ (1 \le i_l \le n_l)$, an $r' \in \hat{F}_\Omega(Y \cup \Xi_1)$ and an $s' \in F_\Omega(Y)$ such that $q'=r'(s')$, $\ppath_1(r')=\ppath_{i_l}(\check{q}_l)$ and $s' \not= \hat{q}_{l_{i_l}}$. Then for each $j<l,\ n_j>0$. Now let $i_j\ (0 \le j <l,\ 1 \le i_j \le n_j)$ be those uniquely determined integers for which $\ppath_{i_j}(\check{q}_j)$ are initial segments of $\ppath_{i_l}(\check{q}_l)$. Without loss of generality, we may assume that $i_0=\ldots=i_l=1$.

Now suppose that there exists no $w \in \{\ppath_i(\check{q}_l') \mid 1 \le i \le n_l'\}$ such that $\ppath_1(\check{q}_l)$ is an initial segment of $w$ or $w$ is an initial segment of $\ppath_1(\check{q}_l)$. Then for each $i\ (l \le i \le m)$, set
\[
B_i=\{a_{i_j} \mid \ppath_1(\check{q}_l) \text{ is an initial segment of } \ppath_j(\check{q}_i)\}
\]
and
\[
C_i = \{ a_{i_j} \mid \ppath_1(\check{q}_l) \text{ is not an initial segment of } \ppath_j(\check{q}_i)\}.
\]

Since the cardinality of $\{l, \ldots, m\}$ is $2 \|A\|^3|B|+1$, there are $i_1, i_2, i_3\ (l\le i_1<i_2<i_3\le m)$ such that the following conditions are satisfied: $\hat{p}_{i_1}\hat{\beta} =\hat{p}_{i_2}\hat{\beta} = \hat{p}_{i_3}\hat{\beta}$, $B_{i_1}=B_{i_2} \subseteq B_{i_3}$, $C_{i_1} \subseteq C_{i_2} \subseteq C_{i_3}$ and $A_{i_1}' \subseteq A_{i_2}' \subseteq A_{i_3}'$. From this, by Lemma \ref{Lemma.4.9.5} we get that at least one of the trees $\check{p}_{i_2}(\hat{p}_{i_3}), \check{p}_{i_1}(\hat{p}_{i_2})$ and $\check{p}_{i_1}(\hat{p}_{i_3})$ is in $Q$, which is again a contradiction.

Therefore, for an $i_l\ (1\le i_l\le n_l')$, $\ppath_{i_l}(\check{q}_l')$ is an initial segment of $\ppath_1(\check{q}_l)$ or $\ppath_1(\check{q}_l)$ is an initial segment of $\ppath_{i_l}(\check{q}_{l}')$. Let $i_j\ (0\le j <l,\ 1\le i_j \le n_j')$ be those uniquely determined integers for which $\ppath_{i_j}(\check{q}_j')$ are initial segments of $\ppath_{i_j}(\check{q}_l')$. Without loss of generality we may assume that $i_0=\ldots =i_l=1$. We can also assume that $\ppath_1(\check{q}_l)$ is an initial segment of $\ppath_1(\check{q}_l')$.

Now let us distinguish the following two cases:
\begin{enumerate}
\item[a)] $\ppath_1(\check{q}_k')$ is an initial segment of $\ppath_1(\check{q}_l)$. If in addition for some \mbox{$i\ (0\le i \le k)$}, $\abs\big( l(\ppath_1(\check{q}_i))-l(\ppath_1(\check{q}_i'))\big) > \|A\|^2|B|K$ then, by Lemma \ref{Lemma.4.9.3}, there exists an $r \in F_\Omega(Y)$ such that $\check{q}_i(r) \in Q$ and $|r|<|\hat{p}_i|$. (Here $\mathrm{abs}$ stands for absolute value.) This obviously is a contradiction. Therefore, for each $i\ (0\le i \le k)$, $\mathrm{abs}\big(l(\ppath_1(\check{q}_i))-l(\ppath_1(\check{q}_i'))\big) \le ||A||^2|B|K$. Then, since the cardinality of $\{1,\ldots,k\}$ is $\|A\|^2|A|^2|B|2L+1$, for some integers $i$ and $j\ (1\le i < j \le k)$, we have:
\begin{enumerate}
\item[(I)] $\ppath_1(\check{q}_i)$ is an initial segment of $\ppath_1(\check{q}_i')$, $\ppath_1(\check{q}_j)$ is an initial segment of $\ppath_1(\check{q}_j')$, $\ppath_1(\check{q}_i')/ \ppath_1(\check{q}_i)=\ppath_1(\check{q}_j')/ \ppath_1(\check{q}_j)$, or 

\item[(II)] $\ppath_1(\check{q}_i')$ is an initial segment of $\ppath_1(\check{q}_i)$, $\ppath_1(\check{q}_j')$ is an initial segment of $\ppath_1(\check{q}_j)$, $\ppath_1(\check{q}_i)/\ppath_1(\check{q}_i') = \ppath_1(\check{q}_j) / \ppath_1(\check{q}_j')$. (Here $uv/u=v$ for any two words $u$ and $v$.) Moreover, $\hat{p}_j\hat{\beta} = \hat{p}_i\hat{\beta}$, $a_{i_1}=a_{j_1}$, $a_{i_1}' = a_{j_1}'$, $B_i \subseteq B_j$ and $B_i' \subseteq B_j'$, where $B_s=\{a_{s_t} \mid 2 \le t \le n_s\}$ and $B_s'=\{a_{s_t}' \mid 2 \le t \le n_s'\}\ (s=i,j)$. Then, by Lemma \ref{Lemma.4.9.6}, $\check{p}_i(\hat{p}_j) \in Q$, which is a contradiction since $|\check{p}(\hat{p}_{i_j})| < |p|$. 
\end{enumerate}
\item[b)] $\ppath_1(\check{q}_l)$ is an initial segment of $\ppath_1(\check{q}_k')$. We shall show that
\[
l\big(\ppath_1(\check{q}_l)\big)-l\big(\ppath_1(\check{q}_k)\big)>\|A\|^2|B|K.
\] 
Then $l\big(\ppath_1(\check{q}_k')\big)-l\big(\ppath_1(\check{q}_k)\big)>\|A\|^2|B|K$ will also hold, which, by Lemma \ref{Lemma.4.9.3}, will be a contradiction.

\hspace*{3mm}Thus, assume that $l\big(\ppath_1(\check{q}_l)\big)-l\big(\ppath_1(\check{q}_k)\big) \le \|A\|^2|B|K$. Then, since the cardinality of $\{k+1, \ldots, l\}$ is $(2\|A\|^3|A| |B|)(\|A\|^2 |B| K+1)$, there are $i_1$ and $i_2\ (k \le i_1 < i_2 \le l)$ such that $i_2-i_1=2\|A\|^3|A| |B|$ and $\ppath_1(\check{q}_{i_1}) = \ldots = \ppath_1(\check{q}_{i_2})$, i.e., $q_{(i_1+1)_1}=\ldots = q_{i_{2_1}}=\xi_1$. Now for each $j\ (i_1 \le j \le i_2)$ set
\[
B_j=\{ a_{j_t}' \mid 1 \le t \le n_j',\ \ppath_1(\check{q}_{i_1}') \text{ is an initial segment of } \ppath_1(\check{q}_j')\}
\]
and
\[
C_j = \{a_{j_t}' \mid 1 \le t \le n_j',\ \ppath_1(\check{q}_{i_1}') \text{ is not an initial segment of } \ppath_1(\check{q}_j')\}.
\]
Since the cardinality of $\{i_1, \ldots, i_2\}$ is $2\|A\|^3|A||B|+1$, there are integers $j_1, j_2$ and $j_3\ (i_1 \le j_1 < j_2 < j_3 \le i_2)$ such that $\hat{p}_{j_1} \hat{\beta} = \hat{p}_{j_2} \hat{\beta} = \hat{p}_{j_3} \hat{\beta}$, $a_{j_{1_1}} = a_{j_{2_1}} = a_{j_{3_1}}$, $\overline{A}_{j_1} \subseteq \overline{A}_{j_2} \subseteq \overline{A}_{j_3}$, $B_{j_1} = B_{j_2} \subseteq B_{j_3}$ and $C_{j_1} \subseteq C_{j_2} \subseteq C_{j_3}$, where $\overline{A}_{j_t}= \{a_{j_{t_s}} \mid 2 \le s \le n_{j_t} \}\ (t=1,2,3)$. Therefore, by Lemma \ref{Lemma.4.9.4}, at least one of the trees $\check{p}_{j_2}(\hat{p}_{j_3})$, $\check{p}_{j_1}(\hat{p}_{j_2})$ and $\check{p}_{j_1}(\hat{p}_{j_3})$ is in $Q$ which is again a contradiction.\epr
\end{enumerate}

Now we are ready to prove

\begin{thm}\label{Theorem.4.9.8} \sloppy For any two {\rm R}-transducers $\gA = (\Sigma, X, A, \Omega, Y, P, A')$ and $\gB = (\Sigma, X, B, \Omega, Y, P', B')$ and any recognizable $\Sigma X$-forest $T$ it is decidable
\begin{enumerate}
\item[\rm (i)] whether $\tau_\gA | T$ is a (partial) mapping,
\item[\rm (ii)] whether $\tau_\gA | T \subseteq \tau_\gB | T$, provided  that $\tau_\gB | T$ is a (partial) mapping,
\item[\rm (iii)] whether $\gA$ is equivalent to $\gB$, provided that $\tau_\gA$ or $\tau_\gB$ is a (partial) mapping,
\end{enumerate}
and
\begin{enumerate}
\item[\rm (iv)] whether $\gA$ is equivalent to $\gB$, provided that at least one of them is deterministic.
\end{enumerate}
\end{thm}

\pr
By Theorem \ref{Theorem.4.9.7}, (i) is true. Moreover, (iii) and (iv) follow from (ii) since the domain of an {\rm R}-transformation is regular and, by Theorem \ref{Theorem.2.10.3}, it is decidable for two regular forests whether one of them contains the other one. Therefore, it is enough to prove (ii).

\sloppy We may assume that $A \cap B = \emptyset$. Let us construct an {\rm R}-transducer $\gC = (\Sigma, X, C, \Omega, Y, P'', C')$ with $C= A \cup B$, $C'=A' \cup B'$ and $P''= P \cup P'$. Obviously, $\tau_\gC | T = \tau_\gA | T \cup \tau_\gB | T$. Thus $\tau_\gA | T \subseteq \tau_\gB | T$ holds iff $\dom(\tau_\gA) \cap T \subseteq \dom(\tau_\gB) \cap T$ and $\tau_\gC | T$ is a partial mapping.
\epr

\

Before stating the analogous result for {\rm F}-transducers we prove a lemma.

\begin{lm}\label{Lemma.4.9.9}
For any {\rm F}-transducer $\gA = (\Sigma, X, A, \Delta, Y, P, A')$ and $R \in \Rec(\Sigma, X)$ one can effectively give an {\rm R}-transducer $\gB = (\Omega, X, B, \Delta, Y, P', B')$ and a forest \mbox{$S \in \Rec(\Omega, X)$} such that $\tau_\gA | R$ is a partial mapping iff $\tau_\gB | S$ is a partial mapping.
\end{lm}

\pr Construct an $\mathrm{R}_{\mathrm{R}}$-transducer $\overline{\gA} = (\Sigma, X, A, \Delta, Y, \overline{P}, A')$ where $\overline{P}$ is given as follows:
\begin{enumerate}
\item[\rm (i)] If $x \rightarrow ar\ \big(x \in X,\ a \in A,\ r \in F_\Delta(Y)\big)$ is in $P$, then $ax \rightarrow r$ is in $\overline{P}$.
\item[\rm(ii)] If $\sigma(a_1, \ldots, a_m) \rightarrow ar\ \big(\sigma \in \Sigma_m,\ m \ge 0,\ a_1, \ldots, a_m, a \in A,\ r \in F_\Delta(Y \cup \Xi_m) \big)$ is in $P$, then $\big(a\sigma \rightarrow r (a_1\xi_1, \ldots, a_m\xi_m), D \big)$ is in $\overline{P}$, where $D(\xi_i) = \dom(\tau_{\gA(a_i)})\ (i=1, \ldots, m)$. Since, by Theorem \ref{Theorem.4.1.10} (i), $\dom(\tau_{\gA(a)})\ (a \in A)$ is regular, $\overline{\gA}$ is an $\mathrm{R}_{\mathrm{R}}$-transducer. Observe that $\tau_{\gA(a)} \subseteq \tau_{\overline{\gA}(a)}$ holds for every $a \in A$.
\end{enumerate}

We shall show that for all $\{a, a'\} \subseteq A$ and $p \in F_\Sigma(X)$ the equivalence

\begin{equation*}\tag{1}
| \tau_{\gA(a)}(p) \cup \tau_{\gA(a')}(p)| >1 \;\; \Longleftrightarrow \;\;  | \tau_{\overline{\gA}(a)}(p) \cup \tau_{\overline{\gA}(a')}(p)| > 1
\end{equation*}
holds. (Note that $a$ and $a'$ are not necessarily distinct.)

Since $\tau_{\gA(a)} \subseteq \tau_{\overline{\gA}(a)}$, the left side of (1) implies its right side.

The converse will be proved by induction on $\hg(p)$. If $\hg(p)=0$, then our statement obviously holds. Now let $p=\sigma(p_1, \ldots, p_m)\ \big(\sigma \in \Sigma_m, m>0, p \in F_\Sigma(X)\big)$ and $r, r' \in F_\Delta(Y)$ be such that $ap \Rightarrow_{\overline{\gA}}^\ast r$, $a'(p) \Rightarrow_{\overline{\gA}}^\ast r'$ and $ r \not= r'$. Moreover, assume that the right side of (1) implies its left side for every state and every $\Sigma X$-tree of height less than $\hg(p)$.

Let us write the above derivations in the form
\[
a\sigma \Rightarrow_{\overline{\gA}} \overline{r}(a_1^{n_1}\xi_1^{n_1}, \ldots, a_m^{n_m} \xi_m^{n_m}),\ a_i^{n_i}p_i^{n_i} \Rightarrow_{\overline{\gA}} \br_i \ (i=1, \ldots, m)
\]
and
\[
\begin{array}{c}
a'\sigma \Rightarrow_{\overline{\gA}} \overline{r}'(b_1^{n_1'}\xi_1^{n_1'}, \ldots, b_m^{n_m'}\xi_m^{n_m'}),\ b_i^{n_i'}p_i^{n_i'} \Rightarrow_{\overline{\gA}} \br_i'\ (i=1, \ldots, m),\\[1mm]
\text{where } a, a', a_i, b_i \in A,\ i=1, \ldots, m,\ n_1+ \ldots +n_m=n,\ n_1'+\ldots + n_m' = n',\\[1mm]
\overline{r} \in \hat{F}_\Delta(Y \cup \Xi_n),\ \overline{r}' \in \hat{F}_\Delta(Y \cup \Xi_{n'}),\ \overline{r}(\br_1, \ldots, \br_m) = r \text{ and}\\[1mm]
\overline{r}'(\br_1, \ldots, \br_m) = r'. \text{ Moreover, } \big(\sigma(a_1, \ldots, a_m),\ a \overline{r}(\xi_1^{n_1}, \ldots, \xi_m^{n_m})\big),\\[1mm]
\big( \sigma(b_1, \ldots, b_m),\ a'\overline{r}'(\xi_1^{n_1'}, \ldots, \xi_m^{n_m'}) \big) \in P.
\end{array}
\]

Now distinguish the following two cases:
\begin{enumerate}
\item[(I)] There exists an $i\ (1 \le i \le m)$ with $n_i >0$ and $|\tau_{\overline{\gA}(a_i)}(p_i)| > 1$ or there exists a $j\ (1 \le j \le m)$ with $n_j'>0$ and $|\tau_{\overline{\gA}(b_j)}(p_j)|>1$. Then, by the induction hypothesis, $|\tau_{\gA(a_i)}(p_i)|>1$ or $|\tau_{\gA(b_j)}(p_j)|>1$. Therefore, by the definition of $\overline{P}, |\tau_{\gA(a)}(p)| >1$ or $|\tau_{\gA(a')}(p)| > 1$ also holds.

\item[(II)] Assume that there are no $i$ and $j$ satisfying (I). Then, $r_{i_1}=\ldots=r_{i_{n_i}}=r_i$ $(1 \le i \le m)$ if $n_i>0$. For all such $i$, by $\tau_{\gA(a_i)} \subseteq \tau_{\overline{\gA}(a_i)}$ and the choice of $D$, we have $p_i\Rightarrow_\gA^\ast a_ir_i$. Moreover, again by the choice of $D$, if $n_i=0$ then also there exists an $r_i \in F_\Delta(Y)$ such that $p_i \Rightarrow_\gA^\ast a_i r_i$ holds. Thus, we have the derivation $p \Rightarrow_\gA^\ast ar$. Using similar arguments, one can show that $p \Rightarrow_\gA^\ast a'r'$ is also valid. Therefore, $|\tau_{\gA(a)}(p) \cup \tau_{\gA(a')}(p)| > 1$.
\end{enumerate}

Thus, we have proved that $\tau_\gA|R$ is a partial mapping iff $\tau_{\overline{\gA}}|R$ is a partial mapping. By Theorem \ref{Theorem.4.4.6} (i), there exist a deterministic F-relabeling $\tau \colon F_\Sigma(X) \rightarrow F_\Omega(X)$ and an {\rm R}-transducer $\gB=(\Omega, X, B, \Delta, Y, P'', B')$ such that $\tau_{\overline{\gA}} = \tau \circ \tau_\gB$. Moreover, by Lemma~\ref{Lemma.4.6.5}, $R\tau = S$ is in $\Rec(\Omega, X)$ and $S$ can be obtained effectively from $R$. Therefore, $\tau_{\overline{\gA}}|R$ is a partial mapping iff $\tau_\gB|S$ is a partial mapping.
\epr

\

Now we state and prove

\begin{thm} \label{Theorem.4.9.10}
For any two {\rm F}-transducers $\gA=(\Sigma, X, A, \Omega, Y, P, A')$ and $\gB=(\Sigma, X, B, \Omega, Y, P', B')$ and recognizable $\Sigma X$-forest $T$, it is decidable

\begin{enumerate}
\item[\rm (i)] whether $\tau_\gA |T$ is a partial mapping,
\item[\rm (ii)] whether $\tau_\gA |T \subseteq \tau_\gB |T$, provided that $\tau_\gB |T$ is a partial mapping,
\item[\rm (iii)] whether $\gA$ is equivalent to $\gB$, provided that $\tau_\gA$ or $\tau_\gB$ is a partial mapping, and
\item[\rm (iv)] whether $\gA$ is equivalent to $\gB$, provided that at least one of them is deterministic.
\end{enumerate}
\end{thm}

\pr
Obviously, (i) follows from Theorem \ref{Theorem.4.9.8} by Lemma \ref{Lemma.4.9.9}. Moreover, (ii) implies (iii) and (iv) since, by Theorem \ref{Theorem.4.1.10} (i), the domain of an {\rm F}-transformation is recognizable. Thus, it suffices to prove (ii).

Assume that $A \cap B = \emptyset$, and construct the {\rm F}-transducer
\[
\gC = (\Sigma, X, C, \Omega, Y, P'', C')
\]
with $C=A \cup B$, $C' = A' \cup B'$ and $P''=P \cup P'$. Obviously, $\tau_\gC = \tau_\gA \cup \tau_\gB$. Therefore, $\tau_\gA|T \subseteq \tau_\gB|T$ iff $\dom(\tau_\gA) \cap T \subseteq \dom(\tau_\gB) \cap T$ and $\tau_\gC|T$ is a partial mapping.
\epr

%% file: Section.4.10.tex
\begin{enumerate}
\item Define generalized sequential machines as tree transducers when strings are interpreted as unary trees in the usual way.
\item Let $\tau$ be a {\rm DR}-transformation. Then $\dom(\tau)$ can be recognized by a {\rm DR}-recognizer.
\item Show that the classes $\cLDF$ and $\cLDR$, and similarly the classes $\cLNDF$ and $\cLNDR$, are incomparable.
\item Let us call a {\rm DR}-transducer $\gA=(\Sigma, X, A, \Omega, Y, P, A')$ {\em simple}, 
\index{drtransducer@DR-transducer!simple} if for every $a\sigma \rightarrow q \in P$, whenever $a_1\xi_i$ and $a_2\xi_i$ occur in $q$, then $a_1=a_2$. If $\gA$ is a simple {\rm DR}-transducer, then $\tau_\gA$ can be induced by an {\rm F}-transducer.
\item Prove that $\cDR$ is not closed under composition.
\item The composition of a totally defined {\rm DR}-transformation by an {\rm R}-transformation is an {\rm R}-transformation.
\item Is $\cR$ closed under composition with {\rm LR}-transformations from the right?
\item Show that $\cF$ is not closed under composition with {\rm LNF}-transformations from the right.
\item Prove Theorems \ref{Theorem.4.3.7} and \ref{Theorem.4.3.9}.
\item Find two {\rm R}-transformations $\tau_1$ and $\tau_2$ such that $\tau_1 \circ \tau_2$ is the {\rm F}-transformation given in Example \ref{Example.4.1.3}.
\item Give two {\rm F}-transformations whose composition is the {\rm R}-transformation of Example~\ref{Example.4.1.6}.
\item Show that $\cF$ and $\cR_R$ are incomparable.
\item Prove that $\cDR_R$ is closed under {\rm DF}-transformations.
\item An {\rm F}-transformation (or an {\rm R}-transformation) is a partial mapping iff it can be induced by a $\mathrm{DR}_{\mathrm{R}}$-transducer.
\item Find a $\mathrm{DR}_{\mathrm{R}}$-transducer which is not equivalent to any {\rm DR}-transducer.
\item The equivalence problem of two $\mathrm{R}_{\mathrm{R}}$-transducers is decidable, provided that at least one of them induces a partial mapping.
\item Find an algorithm to decide for an {\rm F}-transducer whether it is equivalent to an {\rm LF}-transducer.
\item Let $\gA=(\Sigma, X, A, Y, P, A')$ be a $\GSDT$ and $\Omega$ a ranked alphabet. Let $\{n_1, \ldots, n_r\}$ be the set of lengths of right-hand sides of all rules from $P$ (each element of $A\Xi$ is counted as one symbol). Moreover, let $r(\Omega) = \{m_1, \ldots, m_s\}$. Assume that there exists a mapping $f \colon \{n_1, \ldots, n_r\} \rightarrow r(\Omega)$ such that the equality
\[
n_k=m_{f(k)} + l_1(m_1-1)+ \ldots + l_s(m_s-1)
\]
holds for every $k(=1, \ldots, r)$, where $l_1, \ldots, l_s \ge 0$. Then there is an {\rm R}-transducer $\gB=(\Sigma, X, B, \Omega, P', B')$ with $\tau_\gA=\{\big(p, \yd(q)\big) \mid (p,q) \in \tau_\gB\}$.
\item Find an {\rm R}-transducer $\gA$ such that $\tau_\gA$ preserves recognizability, but $\gA$ is not equivalent to any {\rm LF}-transducer.
\item An {\rm R}-transducer $\gA=(\Sigma, X, A, \Omega, Y, P, a_0)$ is called {\em k-metalinear}
\index{rtransducer@R-transducer!$k$-metalinear}
if the following conditions are satisfied:
\begin{enumerate}
\item[(1)] $a_0$ does not appear in the right-hand sides in rules from $P$,
\item[(2)] for each rule $a_0\sigma \rightarrow q\ (\sigma \in \Sigma_m)$ in $P$ every $\xi_i\ (1 \le i \le m)$ can occur in $q$ at most $k$ times, and
\item[(3)] for each rule $a\sigma \rightarrow q\ (a \not= a_0,\ \sigma \in \Sigma_m)$ in $P$ the number of occurences of each $\xi_i\ (1 \le i \le m)$ in $q$ is $0$ or $1$.
\end{enumerate}
Let $\gA$ be a $k$-metalinear {\rm R}-transducer. Does $\tau_\gA$ preserve recognizability?
\item For a ranked alphabet $\Sigma$ let $\tilde{\Sigma} = \tilde{\Sigma}_0 \cup \tilde{\Sigma}_1$ be the ranked alphabet with $\tilde{\Sigma}_0 = \Sigma_0$ and $\tilde{\Sigma}_1 = \{ \tilde{\sigma} \mid \sigma \in \Sigma_m,\ m>0 \}$. Define the mapping $\ph \colon F_\Sigma(X) \rightarrow \gp F_{\tilde{\Sigma}}(X)$

by $\ph(d) = \{d\}\ (d \in \Sigma_0 \cup X)$ and 
\[
\ph\big(\sigma(p_1, \ldots, p_m)\big) = \{\tilde{\sigma}(t) \mid t \in \ph(p_1) \cup \ldots \cup \ph(p_m) \}\ 
\]

$\big(\sigma \in \Sigma_m,\ m>0, \ p_1, \ldots, p_m \in F_\Sigma(X)\big)$. Show that if $T \in \Surf(\cR)$ then $\ph(T) = \bigcup \big(\ph(t) \mid t \in T \big)$ is recognizable.
\item Is $\Surf(\cR)$ closed under intersection?
\item Give a recursive definition of the concepts of state-sequence and production-sequence.
\item For every {\rm F}-transducer there is an equivalent totally defined {\rm F}-transducer with a single final state.
\item For every {\rm DF}-transducer ({\rm DR}-transducer) one can effectively give an equivalent {\rm DF}-transducer ({\rm DR}-transducer) with a minimal number of states.

\end{enumerate}

%% file: Section.4.11.tex
The concept of the {\rm R}-transducer was introduced by \textsc{Rounds} \cite{rou70b} and \textsc{Thatcher} \cite{tha70} thus extending generalized sequential machines from strings to trees and to give a tree automaton formalism for parts of mathematical linguistics (in particular, for the theory of syntax directed compilation). The {\rm F}-transducer is due to \textsc{Thatcher} \cite{tha73}. As in the case of tree recognizers, many of the authors dealing with tree transducers allow a symbol from a ranked alphabet to have more than one rank, and most of them use no separate frontier alphabets.

The results of Section \ref{Section.4.2} can be found in \textsc{Engelfriet} \cite{eng75b}, and most results of Section \ref{Section.4.3} are also from this work. Theorems \ref{Theorem.4.3.3}, \ref{Theorem.4.3.12}, \ref{Theorem.4.3.13} were obtained by \textsc{Baker} \cite{bak73}.

Tree transducers with regular look-ahead are defined and investigated in \textsc{Engelfriet} \cite{eng76-77}. Generalized syntax directed translations were introduced by \textsc{Aho} and \textsc{Ullman} \cite{ahoull1971} in the special case where the domain of the translation is the forest of all {\em parse trees}
\index{tree!parse}
of a given context-free grammar. (Parse trees are almost the same as our production trees.) Applying a generalized syntax directed translation in the sense of Aho and Ullman is equivalent to applying a $\DGSDT$ of Section \ref{Section.4.5} which, by Theorem \ref{Theorem.4.5.4}, is equivalent to applying a {\rm DR}-transducer and then taking the yield of the resulting tree. The more general concept of a $\GSDT$ was introduced in \textsc{Baker} \cite{bak78b}. In the same work she proved that for each $n, \yd \Surf(\cR^n)$ and $\yd \Surf (\cF^n)$ are properly contained in the family of deterministic context-sensitive languages.

The results of Section \ref{Section.4.6} are from \textsc{Engelfriet} \cite{eng75b}, \textsc{G\'ecseg} \cite{gec81} and \textsc{Rounds} \cite{rou70b}.

The first result about the $\Surf(\cR^n)$-hierarchy can be found in \textsc{Ogden} and \textsc{Rounds} \cite{ogdrou72}, where they proved that $\Surf(\cR)$ is a proper subclass of $\Surf(\cR^2)$ and conjectured the properness of the hierarchy. It was \textsc{Engelfriet} \cite{eng78a, eng82} who succeeded in proving that the $\cR^n$-, $\Surf(\cR^n)$-, and $\yd \Surf(\cR^n)$-hierarchies (and their {\rm F}-transducer counterparts) are proper. Section \ref{Section.4.7} and \ref{Section.4.8} are based on his work.

The decidability results of Section \ref{Section.4.9} are from \textsc{\'Esik} \cite{esi80}. Using a different technique \textsc{Zachar} \cite{zac79} also proved the decidability of the equivalence problem of {\rm DF}-transducers.

As a conclusion we mention some other topics relevant to the subject matter of Chapter~\ref{Chapter.4}.

A {\em sequential program machine}
\index{machine!sequential program} \index{sp-machine}
(sp-{\em machine}) introduced by \textsc{Buda} \cite{bud79} is such a generalization of a $\gsm$ whose inputs are strings and whose outputs are $n$-tuples of $n$-ary trees. Buda showed that the equivalence problem of sp-machines is solvable and that this implies that the equivalence of certain program schemes is also decidable.
 
Engelfriet and Fil\`e introduced a new type of tree transducer called {\em macro tree transducer}
\index{tree transducer!macro}
 which is a combination of the {\rm R}-transducer and the context-free tree grammar (see \textsc{Engelfriet} \cite{eng80}). They propose to use macro tree transducers to model \index{grammar!attribute} {\em attribute grammars} of D. E. Knuth (Math. Systems Theory 2 (1968), 127--145: Correction: ibid 5 (1971), 95--96). For tree transformations in terms of magmoids we refer the reader to \textsc{Arnold} and \textsc{Dauchet} \cite{arndau76b, arndau76e}, \textsc{Dauchet} \cite{dau77a, dau77b}, and \textsc{Lilin} \cite{lil78a, lil78b}.

Finally, we note that much of the category theoretic work mentioned in the Notes and References of Chapter \ref{Chapter.2} deal with tree transductions.

%% file: bib.tex
\renewcommand{\bibpreamble}{
We hope that most of the literature dealing with tree automata, tree grammars, forests, tree transductions, or their applications (published by the end of 1982) is listed in this bibliography. It also includes some more general works which devote at least a part to our subject, as well as a few items on closely related topics. As to the latter category the decision on inclusion or exclusion has sometimes been difficult. Of a paper published more than once in almost identical form, just the more complete, or the more widely available, version is mentioned. Preliminary reports and unpublished theses are not included except for a few cases. Items published by the same author(s) in the same year are distinguished for reference by a letter after the year. For some of the most often recurring journals and proceedings we use the following abbreviations:

\

$n$. Ann. ACM STC = Proceedings of the $n$\textsuperscript{th} Annual ACM Symposium on Theory of Computing

$n$.  Coll. Lille = Les Arbres en Alg\'ebre et en Programmation, $n$\textsuperscript{th} Colloque du Lille, Universit\'e de Lille I

IC = Information and Control

$n$. IEEE Symp. ($n \le 15$) = $n$\textsuperscript{th} Annual Symposium on Switching and Automata Theory

$n$. IEEE Symp ($n> 15$) = $n$\textsuperscript{th} Annual Symposium on Foundations of Computer Science

J. ACM = J. Assoc. Comput. Mach.

J. CSS = J. Comput. System Sci.

LN in CS = Lecture Notes in Computer Science (Springer-Verlag)

MST = Mathematical Systems Theory

S-C-C = Systems-Computers-Controls

}

%% file: Appendix.tex
{\Large\bf SOME FURTHER TOPICS AND REFERENCES}\\[2mm]
by Magnus Steinby

\vskip0.5cm

\noindent The purpose of this Appendix is to supplement the original book with notes on some further topics and a selection of more recent references. The choice of topics and references is partly influenced by personal preferences, but I trust that the areas included deserve to be mentioned, and that the general expositions, surveys and research papers appearing in the bibliography are useful. Hence, I hope that these notes may serve as an initial guide to the subjects discussed, and that they give an idea of the continuing vitality of the theory and of its applications.

Before considering any specific areas, let me note some works of a general nature published after \emph{Tree Automata} was written. J.~R.~B\"uchi's posthumous book \emph{Finite Automata, Their Algebras and Grammars} \cite{buc89} appeared  in 1989 (edited by D.~Siefkes). The main part of it treats unary algebras, finite acceptors, regular languages and production systems, but in a manner that suggests tree automata and tree languages as natural generalizations. The last two chapters deal with terms, trees, algebras as tree automata, tree grammars, and connections between context-free languages and pushdown automata. Especially this latter part of the book appears quite unfinished, but the author's grand design, a theory that would encompass algebras, automata, formal languages and rewriting systems, is clearly discernible. The terminology and notation is often nonstandard, sometimes even confusing, but a patient reader is rewarded by original insights and interesting historical remarks.

The book \emph{Tree Automata and Languages} \cite{nivpod92} edited M.~Nivat and A.~Podelski, which appeared in 1992, is a collection of papers that discuss a  variety of topics involving trees. The survey paper \cite{gecste97} by F.~G\'ecseg and M.~Steinby  may be viewed as a condensed and somewhat  modernized  version of \emph{Tree Automata}, but it also takes up some further topics and its bibliography includes many additional items.

In their book \emph{Syntax-Directed Semantics. Formal Models Based on Tree Transducers} \cite{fulvog98}, Z.~F\"ul\"op and H.~Vogler consider formal models of syntax-directed semantics based on tree transducers. They also develop a fair amount of the general theory of total deterministic top-down, macro, attributed, and macro attributed tree transducers. In particular, they compare with each other the classes of tree transformations defined by the different types of tree transducers, and they present several composition and decomposition results for these  tree transformations.

The internet book \emph{Tree Automata Techniques and Applications} \cite{tata08}, to be referred to as TATA, is a joint enterprise of several authors. First launched in 1997, it has already been revised and extended a few times. The presentation is often rather informal, but the ideas are richly illustrated by examples and many interesting facts are also given as exercises. The first two chapters review some basic material about finite tree recognizers, regular tree languages, and regular tree grammars, but also mention context-free tree languages. Chapter 6 contains a brief account of tree transducers (without proofs). The remaining five chapters deal with topics not covered by our book. The tutorial \cite{lod12} by C.~L\"oding focuses on applications of tree automata and emphasizes algorithmic aspects.

\emph{Automata on infinite trees} and the connections between \emph{tree automata and logic} were the most important topics excluded from \emph{Tree Automata}. The two are strongly linked with each other and have been studied intensively ever since tree automata were introduced, and by now they form an extensive theory with important applications to logic and computer science. Although mainly concerned with the word case,  the survey papers \cite{tho90} and \cite{tho97} by W.~Thomas offer very readable introductions to this area, and they also include extensive bibliographies.  Chapter 3 of TATA \cite{tata08} is a further useful general reference, and some of the papers in \cite{nivpod92} deal with this topic. The book \emph{Automata, Logics, and Infinite Games} \cite{grathowil02} edited by E.~Gr\"adel, Thomas and T.~Wilke contains twenty tutorial papers that form an excellent overview of the study of automata, logics and games.  About half of them concern trees and tree automata. Besides MSO logics, they elucidate the uses and properties of various modal logics, fixed-point logics and guarded logics, and demonstrate the usefulness of alternating tree automata.

The continual development of the theory of \emph{tree transformations} is also largely driven by applications, and tree transducers will be mentioned also in connection with some the other themes to be discussed below. Here I shall note separately a few  important topics. The study of compositions of tree transformation classes initiated by B.~S.~Baker (1973, 1979)\footnote{The references can be found in the original bibliography of \emph{Tree Automata}} and J.~Engelfriet (1975) has been pursued further especially by F\"ul\"op and S.~V\'agv\"olgyi \cite{fulvag87,fulvag90,fulvag91,ful91}. In particular, they have considered semigroups of the compositions of some given tree transformation classes, and presented rewriting systems by which the equality of two composition classes can be decided. They have also considered some variants of Engelfriet's (1977) important idea of regular look-ahead for top-down tree transducers (\cite{fulvag89}, for example). Recently, Engelfriet, S.~Maneth and H.~Seidl \cite{engmansei14} have shown that in certain cases it can be decided whether a deterministic top-down tree transducer with regular look-ahead is equivalent to  a deterministic top-down tree transducer, and that such a transducer without look-ahead can be constructed if the answer is positive. Macro tree transducers were first defined by \mbox{Engelfriet} (1980) but, as noted in \cite{fulvog98} for example, the primitive recursive program schemes independently introduced by B.~Courcelle and P.~Franchi-Zannettacci \cite{coufra82} amount to many-sorted versions of them. Macro and other higher-level tree transducers have been studied in depth by Engelfriet and Vogler \cite{engvog85,engvog86,engvog88,engvog91} (cf. also \cite{engman99, engman03}). For further information about these matters, I recommend the bibliographic notes in \cite{fulvog98}. The work \cite{bozfulrah11} on equational tree transformations by S.~Bozapalidis, F\"ul\"op and G.~Rahonis is a natural extension of a classical theme.

The decidability of the question whether the image of a given regular tree language under a given tree homomorphism is regular, has been a relatively long-standing open problem, but recently an affirmative solution was presented by G.~Godoy and O.~Gim\'enez \cite{godgim13}. Their approach uses tree automata with equality or disequality tests, and their work contains also some results of independent interest concerning such automata. Moreover, it has some applications to term rewriting and XML theory.  F\"ul\"op and P.~Gyenizse \cite{fulgye93} have shown that injectivity is undecidable for tree homomorphisms while it is decidable for linear deterministic top-down tree transformations. Furthermore, in \cite{ful94} F\"ul\"op proves that several questions concerning the ranges of deterministic top-down tree transformations are undecidable. The decidability of the equivalence of deterministic top-down tree transducers  was proved by \'Esik already in 1980. More recently, Engelfriet, Maneth and Seidl \cite{engmansei09} showed that the equivalence of total deterministic top-down tree transducers can be decided in polynomial time by reducing the transducers to a certain canonical form, and their method can be applied also to deterministic top-down tree transducers with regular look-ahead. In \cite{friseiman11}, S.~Friese, Seidl and Maneth present a corresponding equivalence checking algorithm based on normal forms for bottom-up tree transducers.
In \cite{engman06}, Engelfriet and Maneth prove that the equivalence of deterministic MSO tree transducers is decidable. These results, as well as many other decidability questions for tree transducers are discussed in the recent survey paper \cite{man14} by Maneth. Finally, two quite recent contributions should be mentioned. Firstly, Seidl, Maneth and G.~Kemper \cite{seimankem15} prove the decidability of the equivalence of deterministic top-down tree-to-string transducers. In \cite{filmanreytal15}, E.~Filiot, Maneth, P.-A.~Reynier and J.-M.~Talbot introduce tree transducers for which every output tree is augmented with information about the origin of each of its nodes, and prove several decidability results concerning the equivalence or injectivity of such transducers.

Since terms can be seen as syntactic representations of trees over ranked alphabets, it is to be expected that there are some connections between \emph{tree automata and term rewriting systems} (TRSs). Indeed,  various tree automata and tree grammars are often defined as special term rewriting systems. On the other hand,  tree automata can be used for solving problems concerning TRSs and such applications have, in turn, inspired new developments in the theory tree automata. In the mid-1980s it was noted that the set $Red(\mathcal{R})$ of terms reducible by a finite left-linear TRS $\mathcal{R}$, as well as its complement, the set $Irr(\mathcal{R})$ of irreducible terms, are regular tree languages. Since this means that many questions concerning reducibility and normal forms are decidable for such TRSs, the observation was quickly followed by several studies of related matters. Thus it was shown that a finite TRS $\mathcal{R}$ for which $Red(\mathcal{R})$ is regular can be ``linearized'' and that the regularity of $Red(\mathcal{R})$ is decidable, the regular sets $Red(\mathcal{R})$ were characterized in terms of a new class of finite tree automata, and questions of ground reducibility were considered.  So-called monadic and semi-monadic TRSs were studied using tree pushdown automata. For extending such applications to TRSs that are not  left-linear,  new classes of tree automata are needed. The problem here is that automata that are able to recognize also non-regular sets $Red(\mathcal{R})$ or the sets of all ground instances of a given non-linear term, tend to be too powerful to be manageable themselves. An example of increased power combined with good decidability properties is provided by the automata with comparisons between brothers introduced in the 1990s. The ground tree transducer is another important tree automaton sprung from the theory of term rewriting. Much material  concerning these matters can be found in TATA \cite{tata08}, and introductions to this subject and many references are provided also by the surveys \cite{giltis95}, \cite{ott99} and \cite{ste03}. For some recent work on this theme, cf. \cite{vag13}, for example.

\emph{Weighted tree automata, tree series and weighted tree transformations} have been studied quite extensively in recent years. Most aspects of this work (up to around 2009) are reviewed in the handbook chapter \cite{fulvog09} by F\"ul\"op and Vogler, and a broad introduction is provided also by the survey paper \cite{esikui03} by Z.~\'Esik and W.~Kuich. \emph{Weighted logics} for weighted tree automata have been studied by M.~Droste, Vogler and others, cf. \cite{drovog06,fulstuvog12}, for example. Equational weighted tree transformations are considered by Bozapalidis, F\"ul\"op and Rahonis \cite{bozfulrah12}. In \cite{rah07} Rahonis introduces weighted Muller-automata on infinite trees and a corresponding weighted  MSO-logic. The dissertation \cite{mat09} of C.~Mathissen  contains, among other matters, also much interesting material belonging to this area as well as a useful bibliography.

In an \emph{unranked tree} a node labeled with a given symbol may have any number of children. Languages of such trees were considered already in the 1960s in two notable papers. J.~W.~Thatcher (1967) introduced finite unranked tree recognizers and showed that the yields of the recognizable unranked tree languages are precisely the context-free languages. C.~Pair and A.~Quere (1968) created an algebraic framework for the study of regular unranked tree languages that also incorporated \emph{hedges}, i.e. finite sequences of unranked trees, and they proved many of the usual properties of regular sets for recognizable unranked tree languages. Nevertheless, the topic received little attention before it was discovered that it is natural to represent XML documents by unranked trees and that unranked tree automata may be useful for handling questions concerning them. The revival of the theory of unranked tree and hedge languages by M.~Murata et al. \cite{mur95,mur00,brumurwoo01} initiated a lively activity in the area. TATA \cite{tata08} devotes a chapter to unranked tree languages and their applications. As a sample from the extensive literature, let us mention just the papers \cite{crilodtho05,marnevsch08,marnie07,nev02} and the survey \cite{sch07} by T.~Schwentick. Since this work is mostly quite application-oriented, algorithmic and complexity issues are much to the fore. X.~Piao and K.~Salomaa \cite{piasal11,piasal12} have considered state complexity questions connected with conversions between different types of unranked automata as well as lower bounds for the size of unranked tree automata. An overview of logics for unranked trees is given by L.~Libkin \cite{lib06}. Weighted unranked tree automata are studied in \cite{drovog11} and \cite{drohen15} by Droste, Vogler,  and  D.~Heusel.

\emph{Natural language description and processing} has become an important area of application of the theory of tree automata and tree languages. Of course, parse trees of natural languages have always been prime examples of `trees' and some of the early works on tree automata explicitly refer to linguistic motivations, but the current activity took really off much later. In his book \cite{mor03} F.~Morawietz discusses \emph{formalizations of natural language syntax} that are based on monadic second-order (MSO) logic on trees and tree language theory. A key fact here is the effective correspondence between weak MSO logic and finite tree automata established already by Thatcher and J.~B.~Wright (1965, 1968) and J.~Doner (1965, 1970), but actually a whole array of tree language-related notions are utilized or noted as potentially useful. These include tree walking automata \cite{bloeng97,bojcol06,bojcol08} macro tree transducers \cite{engvog85,engman99}, and tree-adjoining grammars (cf. \cite{jossch97}, for a survey). Recently, the theory of tree automata has attracted the attention of linguists especially because of the promise shown by tree-based  approaches to \emph{machine translation}. Besides classical notions and results appearing already in our book, work in this area draws also upon some newer developments. In particular, it has both utilized and inspired work on unranked and weighted tree languages as well as weighted tree transducers. Furthermore, it has revived the interest  in the generalized top-down tree transducers studied much earlier by E.~Lilin (1978). Also compositions and decompositions of various tree transformations are used in machine translation systems. The papers \cite{kni07,knigra05} expose some of the relevant questions from a linguist's point of view, while the papers \cite{englilmal09,graknimay08,mal11a,mal11b} form a sample of theoretical work in the area.

Almost all papers on \emph{varieties of tree languages}, and \emph{classes of special regular tree languages} in general, have appeared after 1984. Most of  the work in this area published before 2005 is at least mentioned in the survey \cite{ste05}, and all the references pointed to (by author and year) below can be found there. Eilenberg-like variety theories for tree languages were presented by Steinby (1979, 1992, 1998) and J.~Almeida (1990, 1995).  \'Esik (1999) has set forth a variety theory in which finitary algebraic theories take the place of finite algebras, and later he together with P.~Weil \cite{esiwei05} formulated a similar theory in terms of preclones. Syntactic monoids of tree languages were introduced by Thomas (1982, 1984) and studied further by Salomaa (1983). A similar notion for binary trees has been used by Nivat and Podelski (1989, 1992). The families of regular tree languages considered in the literature include those of the finite and co-finite tree languages (G\'ecseg and B.~Imreh 1988), definite,   reverse definite and generalized definite tree languages (U.~Heuter 1989, 1992), $k$-testable tree languages (Heuter 1989, T.~Knuutila 1992), and aperiodic tree languages (Thomas 1984).  All of them are varieties of tree languages (cf. Steinby 1992, 1998), and in some cases the corresponding varieties of finite algebras are also known.

Although Thomas (1984) could characterize the aperiodic tree languages by their syntactic monoids, it was obvious that such a characterization is not possible for all varieties of tree languages. This was confirmed when S.~Salehi \cite{sal05} described the (generalized) varieties definable by syntactic monoids or semigroups. His result shows, for example, that the definite tree languages cannot be characterized by syntactic semigroups (as claimed in an earlier paper). However, in \cite{canste11} A.~Cano Gomez and Steinby introduce generalized syntactic semigroups (and monoids) in terms of which the definite tree languages can be characterized. Wilke (1996) gave an effective characterization of the reverse definite binary tree languages in terms of so-called tree algebras. Salehi and Steinby \cite{salste07b} studied the tree algebra formalism in some detail and presented a variety theorem for it.
Noticing that the well-known equivalence of aperiodicity, star-freeness, and first-order definability of string languages fails for trees, Thomas (1984) introduced logics in which set quantifications are limited to chains or to antichains of nodes. He proved then, for example, that a regular tree language is star-free iff it is antichain-definable. This line of research has been pursued further by Heuter (1989, 1991) and A.~Potthoff (1994, 1995), for example.

Some families of tree languages have been introduced by first defining a  class of finite algebras. For example, the monotone tree languages studied by G\'ecseg and Imreh (2002) were defined as the languages recognized by monotone algebras. Similarly, \'Esik and Sz.~Iv\'an \cite{esiiva07} introduce a hierarchy of aperiodicity notions for finite algebras and consider then, besides the properties of the obtained varieties of finite algebras, the corresponding families of tree languages.
There are a few different extensions of the variety theory of tree languages: positive varieties of tree languages  by T.~Petkovi\'c and Salehi \cite{petsal05}, varieties of many sorted sets (with tree languages as a special case) by Salehi and Steinby \cite{salste07a}, and varieties of recognizable tree series by F\"ul\"op and Steinby \cite{fulste11}.

A section of \emph{Tree Automata} is devoted to deterministic root-to-frontier (DR) recognizers and DR tree languages, but the topic has been studied quite extensively also later. In her thesis E.~Jurvanen (1995) considers closure properties and the variety generated by DR tree languages as well as ways of strengthening DR recognizers. The latter include, in particular, the  regular frontier check mechanism introduced by Jur\-va\-nen, Potthoff and Thomas (1994). The thesis is also a good general introduction and a reference for work done before 1995. In the synchronized deterministic top-down automata of Salomaa \cite{sal94,sal96} a limited communication between the computations in different branches is allowed. G\'ecseg and Steinby (2001) introduced syntactic monoids for DR tree languages, and these were used by G\'ecseg and Imreh (2002, 2004) for characterizing monotone, nilpotent and definite DR tree languages. In \cite{marnevsch08} W.~Martens, F.~Neven and Schwentick discuss several aspects of DR-recognition. In particular, motivated by applications to schema languages for XML, they study DR recognizers of unranked tree languages.

The book \emph{Grammatical Picture Generation. A Tree-Based Approach} \cite{dre06} by F.~Drewes is a comprehensive treatment of \emph{tree-based picture generation}. The picture generating systems considered consist, roughly speaking, of a device for producing a tree language and a picture algebra that interprets trees as pictures. The devices used for producing the tree languages include regular tree grammars, ET0L tree grammars, branching tree grammars, and tree transducers. The needed tree language theory is given in several inserts in the main text and in a separate appendix. Thus this fascinating book offers also a  general introduction to tree languages.

A great number of concepts and results from several branches of mathematics are used in the theory of tree automata. However, as a conclusion of this appendix, I shall mention some introductions to just  two subjects most intimately connected with tree automata: universal algebra and term rewriting. Besides the texts listed at the end of Chapter I of \emph{Tree Automata}, there are several other good books on universal algebra. As general introductions, I recommend the classic \cite{bursan81} by S.~Burris and H.~P.~Sankappanavar and the more recent textbook by C.~Bergman \cite{ber12}. The book \cite{wec92} by W.~Wechler, written expressly for computer scientists, is also very useful. The books \cite{ave95} by J.~Avenhaus and \cite{baanip98} by F.~Baader and T.~Nipkow offer two good introductions to term rewriting systems.

\renewcommand{\bibname}{Bibliography of the Appendix}
\renewcommand{\bibpreamble}{}